\documentclass[11pt,oneside]{book}
\usepackage{a4wide}
\usepackage{caption,amsmath}
\usepackage[x11names]{xcolor}
\usepackage{hyperref}
\hypersetup{urlcolor=blue!50!green,linkcolor=DodgerBlue2,citecolor=SpringGreen4,colorlinks, linktocpage}

\usepackage{cite}
\usepackage{makeidx,amssymb,epsfig,euscript,amsfonts,dsfont,stmaryrd,subfigure,slashed,mathrsfs}

\usepackage[nottoc]{tocbibind}

\usepackage{enumitem}
\newlist{nitems}{enumerate}{1}

\setlist[nitems, 1]
{label={\bf\thechapter.\arabic{nitemsi}\;},
ref={\thechapter.\arabic{nitemsi}},
leftmargin=*,
rightmargin=0pt
}

\newenvironment{priklady}[1]
{
\vspace{1mm}\setcounter{secnumdepth}{-1} \section{Problems}
\vspace{1mm}
\begin{nitems}
}
{
\end{nitems}
\setcounter{secnumdepth}{2} 
}

\usepackage[english]{babel}
\usepackage[utf8]{inputenc}
\usepackage[OT2,OT1]{fontenc}

\makeatletter\def\cleardoublepage{\clearpage\if@twoside \ifodd\c@page\else\hbox{} \thispagestyle{empty}\newpage\fi\fi}
\makeatother
\makeatletter\def\jmenokap{\@chapapp}\makeatother
\usepackage[Bjarne]{fncychap}
\ChTitleVar{\ifodd\thepage\raggedleft\else\raggedright\fi\LARGE\rm\bfseries}
\ChNameVar{\ifodd\thepage\raggedleft\else\raggedright\fi\normalsize\rm}
\ChNumVar{\ifodd\thepage\raggedleft\else\raggedright\fi \bfseries\Large}
\usepackage{fancyhdr}

\renewcommand{\chaptermark}[1]{\markboth{\MakeUppercase{\thechapter\quad#1}}{}}
\renewcommand{\sectionmark}[1]{\markright{\MakeUppercase{\thesection\quad#1}}}
\newcommand{\helv}{\fontseries{bx}\fontshape{sc}\fontsize{7}{9}\selectfont}
\newcommand{\m}{\text{-}\!}
\newcommand{\widebar}[1]{\overline{\hspace{-1pt}#1\hspace{-1pt}}\hspace{2pt}}

\newcommand{\ccdot}{\!\cdot\!}

\newcommand{\appA}[1]{Appendix~\ref{appenA}}
\newcommand{\appB}[1]{Appendix~\ref{appenB}}
\newcommand{\appC}[1]{Appendix~\ref{appenC}}
\newcommand{\appD}[1]{Appendix~\ref{appenD}}
\newcommand{\appE}[1]{Appendix~\ref{appenE}}
\newcommand{\qq}[1]{\textquotedblleft#1\/\textquotedblright} % to jsou uvozovky \qq{aaa} = "aaa"
\newcommand{\qs}[1]{\textquoteleft#1\/\textquoteright} % apostrofy \qs{aaa}='aaa'
\newcommand{\J}{\mathds{1}}     % jedn.matice
\newcommand{\ti}[1]{\text{\it #1}}   % index v matematice
\newcommand{\tiz}[1]{{(\!\text{{\it #1}})}} % index v matematice + zavorky
\newcommand{\bm}[1]{\begin{pmatrix}#1\end{pmatrix}} % pro matice, vektory... napr.\bm{a&b\\c&d}
\renewcommand{\U}{\mathcal{U}}
\newcommand{\V}{\mathcal{V}}
\newcommand{\M}{M}
\newcommand{\Mt}{\hspace{.5mm}\widetilde{\hspace{-.5mm}\M}}
\newcommand{\Mp}{\mathfrak{M}}
\newcommand{\Mpt}{\hspace{.2mm}\widetilde{\hspace{-.2mm}\Mp}}
\newcommand{\Ut}{\widetilde{\mathcal{U}}}
\newcommand{\Vt}{\widetilde{\mathcal{V}}}
\newcommand{\hv}{\tilde{h}}
\newcommand{\Phit}{\widetilde{\Phi}}
\newcommand{\LIPS}{\text{LIPS}}
\newcommand{\BR}{\text{BR}}
\newcommand{\lagr}{\mathscr{L}}
\newcommand{\partialv}{\partial\hspace{-4pt}\raisebox{9pt}[0pt]{$\scriptscriptstyle\shortrightarrow$}}
\newcommand{\partialvb}{\partial\hspace{-5pt}\raisebox{9pt}[0pt]{$\scriptscriptstyle\shortleftarrow$}}
\newcommand{\partialvob}{\partial\hspace{-6pt}\raisebox{9pt}[0pt]{$\scriptscriptstyle\leftrightarrow$}}
\newcommand{\dmd}{\partialvob_\mu}

\newcommand{\dmm}{\partialvob^\mu}
\newcommand{\OO}{O}
\newcommand{\s}{\scalebox{.8}}
\newcommand{\eV}{\text{eV}}
\newcommand{\MeV}{\text{MeV}}
\newcommand{\GeV}{\text{GeV}}
\newcommand{\TeV}{\text{TeV}}
\newcommand{\psib}{\bar{\psi}}
\newcommand{\Tr}{\text{Tr}}

\makeindex

\begin{document}

%\input{title}
%%%%%%%%%%%%%%%%%%%%%%%%%%%%%%%%%%%%%%%%%%%%%%%%%%%%%%%%%%%%%%%%%%%%%%%%%%%%%%%%%%%%%%%%%%%%%%%%%%%%%%%%%%%%%%%%%%%%%%%%%%%%%%%%%%%%%%%%
%%%%%%%%%%%%%%%%%%%%%%%%%%%%%%%%%%%%%%%%%%%%%%%%%%%%%%%%%%%%%%%%%%%%%%%%%%%%%%%%%%%%%%%%%%%%%%%%%%%%%%%%%%%%%%%%%%%%%%%%%%%%%%%%%%%%%%%%

%\thispagestyle{empty}\phantom{a}\newpage\thispagestyle{empty}
%\thispagestyle{empty}\vspace*{1cm}
%\begin{center}
%\Large FUNDAMENTALS OF ELECTROWEAK THEORY
%\end{center}
%\newpage \thispagestyle{empty}\vspace*{1cm}\newpage
\thispagestyle{empty}\vspace*{1cm}
\begin{center}
{\Huge FUNDAMENTALS\\OF ELECTROWEAK THEORY\\[0.5cm]}
SECOND EDITION\\[2cm]
{\large Ji\v{r}\'i Ho\v{r}ej\v{s}\'{i}} \\
\href{mailto:jiri.horejsi@mff.cuni.cz}{jiri.horejsi@mff.cuni.cz} \\[1cm]
{\it Institute of Particle and Nuclear Physics\\
Faculty of Mathematics and Physics\\
Charles University}\\[2.4cm]
\includegraphics[width=4cm]{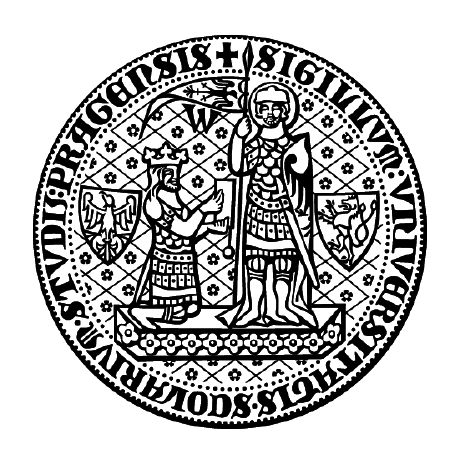}
%\hspace{-0.2cm}
\vfill
%Charles University in Prague\\
Prague\\
2022
\end{center}
\newpage\thispagestyle{empty}
\noindent
\begin{center}
\large \bf Abstract
\end{center}

\large
The present text is an updated version of an earlier author’s book on the electroweak theory (published originally in 2002, ISBN 80-246-0639-9). It reflects the ultimate completion of the standard model by the long-awaited discovery of the Higgs boson (ten years after the first edition) and incorporates also some minor corrections of the previous text, removing typos, etc. As regards an immediate motivation to come up with such an update of the original rather old book, the 10th anniversary of the Higgs boson discovery seems to be an opportune moment for doing it now. The publication of the current upgraded text within the e-print arXiv is aimed at its best possible availability for any interested reader.
\vfill

%%%%%%%%%%%%%%%%%%%%%%%%%%%%%%%%%%%%%%%%%%%%%%%%%%%%%%%%%%%%%%%%%%%
\setcounter{footnote}{0}
\tableofcontents %\mainmatter

%%%%%%%%%%%%%%%%%%%%%%%%%%%%%%%%%%%%%%%%%%%%%%%%%%%%%%%%%%%%%%%%%%%%%%%%%%%%%%%%%%%%%%%%%%%%%%%%%%%%%%%%%%%%%%%%%%%%%%%%%%%%%%%%%%%%%%%%
%%%%%%%%%%%%%%%%%%%%%%%%%%%%%%%%%%%%%%%%%%%%%%%%%%%%%%%%%%%%%%%%%%%%%%%%%%%%%%%%%%%%%%%%%%%%%%%%%%%%%%%%%%%%%%%%%%%%%%%%%%%%%%%%%%%%%%%%
%%%%%%%%%%%%%%%%%%%%%%%%%%%%%%%%%%%%%%%%%%%%%%%%%%%%%%%%%%%%%%%%%%%%%%%%%%%%%%%%%%%%%%%%%%%%%%%%%%%%%%%%%%%%%%%%%%%%%%%%%%%%%%%%%%%%%%%%

%\input{title}
%\thispagestyle{empty}\phantom{a}\newpage\thispagestyle{empty}

%\newcommand{\hhh}{\fontfamily{phv}\fontseries{b}\fontsize{30}{11}\selectfont}
%\newcommand{\hhh}{\fontfamily{uhv}\fontsize{30}{20}\selectfont}

%%%%%%%%%%%%%%%%%%%%%%%%%%%%%%%%%%%%%%%%%%%%%%%%%%%%%%%%%%%%%%%%%%%
%%%%%%%%%%%%%%%%%%%%%%%%%%%%%%%%%%%%%%%%%%%%%%%%%%%%%%%%%%%%%%%%%%%%%%%%%%%%%%%%%%%%%%%%%%%%%%%%%%%%%%%%%%%%%%%%%%%%%%%%%%%%%%%%%%%%%%%%

\setcounter{footnote}{0}
%\setcounter{page}{5}

%%%%%%%%%%%%%%%%%%%%%%%%%%%%%%%%%%%%%%%%%%%%%%%%%%%%%%%%%%%%%%%%%%%
%%%%%%%%%%%%%%%%%%%%%%%%%%%%%%%%%%%%%%%%%%%%%%%%%%%%%%%%%%%%%%%%%%%%%%%%%%%%%%%%%%%%%%%%%%%%%%%%%%%%%%%%%%%%%%%%%%%%%%%%%%%%%%%%%%%%%%%%

%%%%%%%%%%%%%%%%%%%%%%%%%%%%%%%%%%%%%%%%%%%%%%%%%%%%%%%%%%%%%%%%%%%
%%%%%%%%%%%%%%%%%%%%%%%%%%%%%%%%%%%%%%%%%%%%%%%%%%%%%%%%%%%%%%%%%%%%%%%%%%%%%%%%%%%%%%%%%%%%%%%%%%%%%%%%%%%%%%%%%%%%%%%%%%%%%%%%%%%%%%%%
%\newpage\fancyhead[CO]{\helv PREFACE} \fancyhead[CE]{\helv PREFACE}
\newpage\markboth{\MakeUppercase{Preface to second edition}}{\MakeUppercase{Preface to second edition}}

\setcounter{secnumdepth}{-1} 
\chapter{Preface to second edition}%\addcontentsline{toc}{chapter}{Preface to second edition}

In 2012, ten years after the first edition of my book {\it Fundamentals of electroweak theory\/}, the long-awaited Higgs boson was discovered, and the edifice of the standard model (SM) was thereby completed. The discovery came despite the skepticism and doubts of many prominent physicists (in this context, see e.g. the essay \cite{refWel}), and it certainly represented one of the most important milestones in the development of particle physics. On the other hand, during the past two decades, there was no other breakthrough discovery that would reveal clearly a new physics beyond SM (though some well-known open problems persist, which apparently cannot be solved within SM). Taking all this into account, I have found it appropriate to refurbish the relatively old original text of my book, since it is still commonly in use (however, the first edition is currently out of stock). Note also that the present update of the original version has been scheduled to appear, symbolically, just 10 years after the Higgs boson discovery. Instead of striving for a conventional book edition of the upgraded text, I have decided to publish it within the e-print arXiv in order to make it available most easily for any interested reader at any place. Of course, apart from incorporating the Higgs boson as a physical reality, I have also corrected some misprints and other minor mistakes found in the original version. Furthermore, as a bonus for true enthusiasts, I have added some new problems to be solved (or at least contemplated); these extend the corresponding sections of chapters~\ref{chap2}, \ref{chap5} and~\ref{chap7}.
As in the first edition of this work, the main emphasis is put here on fundamentals rather than the current phenomenology of electroweak physics. Anyway, the phenomenological aspects of SM are extremely important, and so it is very helpful that several other new books by other authors appeared during the past twenty years, where one may find, among other things, a detailed treatment of various aspects of electroweak physics that are not covered by the present text. At least some of them are duly included in the updated bibliography; they are marked as \cite{Alt}, \cite{Lan}, \cite{Pal} and \cite{Pas} respectively. Some other new items have also been added explicitly to the list of references. Needless to say, the relevant literature on SM physics (and beyond) is vast; a rather comprehensive list can be found e.g. in the book \cite{Lan}.  
Last but not least, I would like to thank Karol Kampf for his help with the preparation of the \LaTeX{} file of the current text.

\vfil \noindent Prague, October 2022 \hfill Ji\v{r}\'{\i} Ho\v{r}ej\v{s}\'{\i}

%\input{preface}
%%%%%%%%%%%%%%%%%%%%%%%%%%%%%%%%%%%%%%%%%%%%%%%%%%%%%%%%%%%%%%%%%%%
%%%%%%%%%%%%%%%%%%%%%%%%%%%%%%%%%%%%%%%%%%%%%%%%%%%%%%%%%%%%%%%%%%%%%%%%%%%%%%%%%%%%%%%%%%%%%%%%%%%%%%%%%%%%%%%%%%%%%%%%%%%%%%%%%%%%%%%%
%\newpage\fancyhead[CO]{\helv PREFACE} \fancyhead[CE]{\helv PREFACE}
\newpage\markboth{\MakeUppercase{Preface to first edition}}{\MakeUppercase{Preface to first edition}}
\chapter{Preface to first edition}%\addcontentsline{toc}{chapter}{Preface to first edition}

This work is an extended version of a one-semester course of
lectures on the theory of electroweak interactions that I taught
regularly at the Faculty of Mathematics and Physics of the Charles
University in Prague during the 1990s. I have also included here some
selected material from my earlier courses, delivered  at the same
school in the second half of the 1980s. Throughout those years, I
could benefit from the feedback provided by the students who
attended my lectures; thus, I believe that the contents of the
present text is properly tuned to the needs of an uninitiated
reader who wants to understand the basic principles of the
electroweak Standard Model (SM), as well as their origin and
meaning. For pedagogical reasons, I have adopted partly a
historical approach -- starting from a Fermi-type theory of weak
interactions\index{weak!interaction|(}, explaining subsequently
the theoretical motivation for intermediate vector bosons, and
only then proceeding to the basic concepts of the gauge theory of
electroweak interactions.

The main point of the discussion of the old weak interaction
theory (Chapter~\ref{chap1} and~\ref{chap2}) is to demonstrate that an effective
Fermi-type Lagrangian can in fact be \qq{measured} in a series of
appropriate experiments and one is thus led to the universal $V-A$
theory, providing a key input for the construction of SM. In many
other treatments of the electroweak theory it has become quite
common to start directly with an a priori knowledge of the $V-A$
theory; nevertheless, I believe that one should not take lightly
the fact that establishing the $V-A$ structure of charged weak
currents took more than twenty years (between 1934 and 1958), and
this development was rather dramatic interplay between experiment
and theory. Besides that, the first two chapters can help the
students of {\it nuclear\/} physics -- who usually do not exploit
the full SM -- to understand the origins of our present-day
knowledge concerning \qq{ordinary} weak interactions.

In Chapter~\ref{chap3} and also later, in connection with the successive
construction of the electroweak SM, the issue of \qq{good
high-energy behaviour} of scattering amplitudes, or -- in common
technical parlance -- the \qq{tree unitarity}\index{tree
unitarity}, is often emphasized (note that this is a necessary
condition for perturbative renormalizability\index{perturbative
renormalizability}). In my experience, such a kind of
argumentation is very helpful and natural when explaining the
construction of SM and an \qq{inevitability} of its essential
ingredients. In this regard, the reader may find it useful to
consult occasionally a companion work, namely my earlier book {\it
Introduction to electroweak unification: SM from tree unitarity\/}
published some years ago (and quoted here as \cite{Hor}) -- this
supplies a lot of technical details omitted in the present text.
Of course, an impatient reader, who wishes to arrive at a
formulation of the electroweak theory as soon as possible, can
start immediately with Chapter~\ref{chap4} devoted to the basics of
non-Abelian gauge theories. Finally, a most pragmatic student may
jump immediately into the Section~\ref{sec7.10} (that contains a succinct
overview of the SM), skipping the rest of the book contents
altogether.

One more remark concerning the contents is in order here. In
accordance with the book title, the emphasis is put on the
\qq{fundamentals}, which means that a discussion of applications
of the electroweak theory is rather suppressed in favour of a
thorough elucidation of the basic principles and their origin.
This may be disappointing for a more phenomenologically oriented
reader, but I am deeply convinced that for mastering a theory, the
understanding of its genesis is as important as a precise
formulation of the theory itself. Moreover, there are many other
textbooks, specialized monographs and review articles devoted to
the practical applications and phenomenology, where the interested
reader can find the required information; some of these sources
are quoted in our bibliography.

The bulk of the present text is devoted to the electroweak
Standard Model, but after going through its exposition, one should
keep in mind the remark made at the end of the synoptic Section~\ref{sec7.10}: despite the stunning phenomenological success of the SM, the prevailing opinion now is that it cannot be the whole story. The
SM should be viewed as an effective theory valid (with remarkable
accuracy) within a limited energy domain, explored till the end of
20th century. The contours of a deeper electroweak theory are to
be unveiled in the forthcoming decades -- to this end,
experimental input provided by the new accelerator facilities
(such as the Large Hadron Collider at CERN, etc.) will be of
crucial importance.

For understanding of the presented material with all technical
details, a preliminary knowledge of quantum field theory at the
level of Feynman diagrams is necessary. In order to make the
reader's life easier, several appendices have been included, which
contain a lot of important formulae and/or describe some special
techniques employed in the main body of the text. The list of
quoted or recommended literature is divided into two parts:
\qq{References} represent mostly (with several exceptions) the
original articles that are particularly important in the
considered context, while \qq{Bibliography} contains books and
review articles. Needless to say, the list is far from complete
and I apologize in advance to all authors whose important work was
not mentioned here.

Each chapter of the main text is supplemented with a set of
problems, or exercises, to be solved. Some of them are not
entirely trivial and may require long and tedious calculations. In
any case, a diligent reader should not be discouraged by finding
out that an appropriate answer cannot be obtained within less than
half an hour or so. On the other hand, some exercises should
stimulate the student's appetite for further reading; in fact, we
thus also partly make up for a broader discussion of applications
of the electroweak theory -- in particular, this can be said about
the computation of the $Z$ boson production cross section (see the
Problem \ref{pro7Compute} in Chapter~\ref{chap7}).

In the course of writing this book I was helped, in various ways,
by many people. I would like to thank all students who read the
preliminary versions of the manuscript when preparing for their
exams, and pointed out to me errors and numerous misprints; in
this respect, I am particularly grateful to Jarom\'{\i}r
Ka\v{s}par, who came through a substantial portion of the text and
also checked most of the formulae. For technical assistance in the
early stages of the whole process I am indebted to Marie
Navr\'{a}tilov\'{a}. My special thanks are due to Karol Kampf, for
preparing the final \LaTeX version of the manuscript, as well as
for ultimate proofreading of the complete text. Last but not
least, let me add that this work was partially supported by the
Centre for Particle Physics, the Czech Ministry of Education
project No. LN00A006.
\\
\\
\\ Prague, December 2002 \hfill Ji\v{r}\'{\i} Ho\v{r}ej\v{s}\'{\i}

%\input{conventions}
%%%%%%%%%%%%%%%%%%%%%%%%%%%%%%%%%%%%%%%%%%%%%%%%%%%%%%%%%%%%%%%%%%%
%%%%%%%%%%%%%%%%%%%%%%%%%%%%%%%%%%%%%%%%%%%%%%%%%%%%%%%%%%%%%%%%%%%%%%%%%%%%%%%%%%%%%%%%%%%%%%%%%%%%%%%%%%%%%%%%%%%%%%%%%%%%%%%%%%%%%%%%
\newpage\markboth{\MakeUppercase{Conventions and notation}}{\MakeUppercase{Conventions and notation}}
\chapter{Conventions and notation}%\addcontentsline{toc}{chapter}{Conventions and notation}

Some of the conventions employed in this book are given in the
main text and, in particular, in the Appendix~\ref{appenA}. For reader's
convenience, and to avoid any misunderstanding, we summarize the
most important items here.

Unless stated otherwise, we always use the natural system of units
in which $\hbar = c = 1$\index{natural units}. Numerical values of
observable quantities (such as decay rates or scattering cross
sections) are converted into ordinary units by setting
\begin{align*}
&1\,\MeV^{-1} \doteq 6.58 \times 10^{-22}\,\text{s} \intertext{or}
&1\,\MeV^{-1} \doteq 197\,\text{fm}
\end{align*}
where $1\,\text{fm} = 10^{-13}\text{cm}$ (fm stands for \qq{fermi}
or \qq{femtometer}).

Most of the other conventions correspond to the textbook
\cite{BjD}. The indices of any Lorentz four-vector take on values
0, 1, 2, 3. The metric is defined by\index{metric tensor}
$$
g_{\mu\nu}=g^{\mu\nu}=\text{diag}(+1,\,-1,\,-1,\,-1)
$$
so that e.g. the scalar product $k\cdot p$ is
$$
k\cdot p = k_0 p_0 - \vec{k}\cdot\vec{p}
$$
Dirac matrices $\gamma^\mu$, $\mu=0,1,2,3$ are defined by means of
the standard representation \cite{BjD}. We also employ the usual
symbol $\slashed{p} = p_\mu\gamma^\mu$ for an arbitrary
four-vector $p$. We should particularly stress the definition of
the $\gamma_5$ matrix,
$$
\gamma_5 = i \gamma^0 \gamma^1 \gamma^2 \gamma^3
$$
that coincides with \cite{BjD} (see also \cite{ItZ}, \cite{PeS},
and \cite{Ryd}), but differs e.g. from \cite{Wei}. Further, the
fully antisymmetric Levi-Civita tensor is fixed by
$$
\epsilon_{0123} = +1
$$
(let us remark that this convention differs in sign e.g. from that
used in \cite{ItZ} and \cite{PeS}).

Our conventions for Dirac spinors are described in Appendix~\ref{appenA}. Let
us emphasize that the normalization employed here differs from
\cite{BjD} (it coincides e.g. with \cite{LaL}, \cite{PeS}).

Finally, the Lorentz invariant transition (scattering) amplitude,
or simply \qq{matrix element} $\mathcal{M}_{fi}$ has an opposite
sign with respect to \cite{BjD}; the convention adopted here
coincides e.g. with that of \cite{LaL}.

\setcounter{secnumdepth}{2} 

%\input{kniha11}  %kapitola 1  1.1
%%%%%%%%%%%%%%%%%%%%%%%%%%%%%%%%%%%%%%%%%%%%%%%%%%%%%%%%%%%%%%%%%%%
%%%%%%%%%%%%%%%%%%%%%%%%%%%%%%%%%%%%%%%%%%%%%%%%%%%%%%%%%%%%%%%%%%%%%%%%%%%%%%%%%%%%%%%%%%%%%%%%%%%%%%%%%%%%%%%%%%%%%%%%%%%%%%%%%%%%%%%%
\chapter{Beta decay}\label{chap1}
\section{Kinematics}

\fancyhead[CE]{\helv\leftmark} \index{propagator!of the
photon|see{photon propagator}}\index{Kobayashi--Maskawa
matrix|see{CKM matrix}} \index{U(1) group@$U(1)$
group|seealso{Abelian group}} \index{triangle anomaly|see{ABJ
anomaly}}\index{electrodynamics|see{QED}}\index{chromodynamics|see{QCD}}\index{GF@$G_F$|see{Fermi
constant}}\index{currents|seealso{charged, neutral, hadronic or
leptonic current}} \index{beta decay!of neutron|ff}
\index{high-energy behaviour|see{tree
unitarity}}\index{hypercharge|see{weak hypercharge}}\index{phase
space|seealso{LIPS}}\index{pion|seealso{decay of the
pion}}\index{axial anomaly|see{ABJ anomaly}} \index{high-energy
divergences|seealso{tree unitarity}}

The oldest and best-known example of a process caused by weak
interaction is the nuclear beta decay, i.e. the spontaneous
emission of electrons (or positrons) from an atomic nucleus.
Experimentally observed for the first time at the end of the 19th
century, it played subsequently a very important role in
establishing the present-day theory of weak interactions. Thus, we
will start our road towards the famous $V-A$ theory\index{V-A
theory@$V-A$ theory} by discussing some essential features of the
beta-decay processes.

   As we know now, a process of that kind can be described as the
decay of a neutron into proton, electron and an electrically
neutral particle called (anti)neutrino
\begin{equation}
\label{eq1.1} n \rightarrow p+e^-+\bar{\nu}_e
\end{equation}
(throughout this chapter we will write simply $\nu$ instead of
$\nu_e$). Let us remind the reader that a third particle in the
final state of (\ref{eq1.1}) is necessary (though very difficult
to detect) to explain the observed continuous spectrum of the
beta-electron energies; in fact, the existence of such a neutral
elusive particle was postulated  by W.~Pauli in the early days of
the beta-decay theory to save -- in a natural way -- the
fundamental law of energy conservation. It is also well known that
the neutrino must be very light -- the current upper bound for its
mass is a few electronvolts, i.e. five orders of magnitude less
than the electron mass. While the $m_\nu$ can be safely neglected
for most practical purposes, an ultimate resolution of the puzzle
of neutrino mass\index{neutrino!mass} (and of possible related
phenomena) constitutes one of the most challenging experimental
goals of particle physics.\footnote{In fact, until 1998, even the
possibility of a strictly massless neutrino was acceptable. Since
then, the accumulating experimental evidence concerning the
so-called oscillation phenomena\index{oscillation
phenomena}\index{oscillation phenomena|seealso{neutrino
oscillations}} (cf. e.g. \cite{Bil}) made it clear that neutrinos
must have non-vanishing masses.}

At least a rough estimate of the $m_\nu$ value can be obtained
from  simple kinematical characteristics of the decay process
(\ref{eq1.1}). In particular, one can calculate the maximum energy
of the emitted electron (which of course depends on the masses of
the particles involved) and  compare it with a measured value;
this can in principle give a desired bound. The endpoint value for
the electron energy spectrum\index{energy spectrum of the
electron} will occur frequently in our future considerations, so
let us now calculate it explicitly. Denoting the four-momenta of
the neutron, proton, electron and antineutrino in (\ref{eq1.1}) by
$P, p, k$ and $k'$ respectively, the energy-momentum conservation
requires that
\begin{equation}
\label{eq1.2}
P=p+k+k'
\end{equation}
For our purpose it is most helpful to utilize a simple
consequence of (\ref{eq1.2}) for suitable kinematical invariants,
namely
\begin{equation}
\label{eq1.3}
(P-k)^2=(p+k')^2
\end{equation}
One may observe that the quantity on the right-hand side of  eq.
(\ref{eq1.3}) cannot be less than $(m_p+m_\nu)^2$: indeed, taking
into account Lorentz invariance\index{Lorentz!invariance}, it can
be calculated in the proton -- antineutrino c.m. system, where one
obviously gets
\begin{equation}
\label{eq1.4}
(p+k')^2=(E^{c.m.}_p +E^{c.m.}_\nu)^2\geq(m_p+m_\nu)^2
\end{equation}
Of course, because of the Lorentz invariance, the lower bound
(\ref{eq1.4}) is actually independent of the reference frame.
Thus, expressing the invariant $(P-k)^2$ in terms of the variables
corresponding to the {\it laboratory system\/} (the neutron rest
frame), one gets an inequality
\begin{equation}
\label{eq1.5}
m^2_n-2m_n E_e +m^2_e\geq (m_p+m_\nu)^2
\end{equation}
which immediately yields the desired upper bound for the electron
energy;  the endpoint of the electron energy spectrum obviously
corresponds to the value

\begin{equation}
\label{eq1.6} E^\ti{max}_e=\frac{m^2_n-(m_p+m_\nu)^2+m^2_e}{2m_n}
\end{equation}

Analogous bounds for the proton and antineutrino energies can be
obtained by means of  the same method. These are
\begin{align}
\label{eq1.7}
E^\ti{max}_p&=\frac{m^2_n+m^2_p-(m_e+m_\nu)^2}{2m_n}\\
\label{eq1.8}
E^\ti{max}_\nu&=\frac{m^2_n-(m_p+m_e)^2+m^2_\nu}{2m_n}
\end{align}
The formula (\ref{eq1.6}) makes it obvious that from the position
of the endpoint of the electron energy spectrum one may infer information about the value of the neutrino rest mass. Since the
relevant experimental data are still compatible with zero, we will
set $m_\nu=0$ for the time being; we shall comment on the some
effects of $m_\nu\neq 0$ later on. Taking now into account the
known values of the relevant masses
\begin{equation}
\label{eq1.9} m_e=0.51\ \MeV,\qquad m_p=938.27\ \MeV,\qquad
m_n=939.56\ \MeV
\end{equation}
the kinematical bound (\ref{eq1.6}) can be approximately written as
\begin{align}
\label{eq1.10} E^\ti{max}_e&=\frac{m^2_n-m^2_p+m^2_e}{2m_n}\doteq
\frac{m^2_n-m^2_p}{2m_n}\\
&\doteq m_n-m_p=1.29\ \MeV \notag
\end{align}
The value (\ref{eq1.10}) indicates that an electron emitted in the
beta decay can be -- at least near the endpoint of the spectrum
-- highly relativistic. To see this clearly, let us calculate the
maximum electron velocity; using (\ref{eq1.10}) and the familiar
kinematical formulae, one gets
\begin{align}
\label{eq1.11}
\beta^\ti{max}_e&=\sqrt{1-\frac{m^2_e}{(E^\ti{max}_e)^2}}\doteq
\sqrt{1-\frac{m^2_e}{(\Delta m)^2}}\\
&\doteq 1-\frac{1}{2}\frac{m^2_e}{(\Delta m)^2} \doteq0.92 \notag
\end{align}
where we have denoted  $\Delta m=m_n-m_p$. On the other hand, the
maximum proton velocity is of the order $\Delta m/m_p$, as one can
deduce easily from (\ref{eq1.7}): indeed, neglecting $m_e$, one
has
\begin{align}
\label{eq1.12} \beta^\ti{max}_p&\doteq \sqrt
{1-{m^2_p}/\Bigl(\frac{m^2_n+m^2_p}
{2m_n}\Bigr)^2}=\frac{m^2_n-m^2_p}{m^2_n+m^2_p}\\
&= \frac{\Delta m}{m_p}+O\bigl((\frac{\Delta
m}{m_p})^2\bigr)\doteq 1.37\times 10^{-3} \notag
\end{align}
which means that the recoil proton is certainly non-relativistic
over the whole kinematical range.\footnote{However, the
reader should keep in mind that the maximum proton velocity
(\ref{eq1.12}) amounts to about 400 km s$^{-1}$, i.e. such a particle
is pretty fast by everyday standards.}

   The lesson one can learn from these simple considerations is
essentially twofold. First, in the beta-decay process
(\ref{eq1.1}) a particle (neutron) is annihilated and three new
particles are created -- this obviously calls for employing the
framework of quantum field theory, which is able to incorporate
such processes in a very natural way (in other words, the ordinary
quantum mechanics is not quite adequate for such a purpose).
Second, the final-state electron (to say nothing of the
quasi-massless neutrino) is relativistic, at least for a certain
part of its energy spectrum -- it means that one should use a {\it
relativistic quantum field theory model\/} to achieve a
satisfactory treatment of the dynamics of this decay process. At
the same time, one may expect some technical simplifications in
connection with the non-relativistic nature of the recoil proton.

%\input{kniha12}  %            1.2
%%%%%%%%%%%%%%%%%%%%%%%%%%%%%%%%%%%%%%%%%%%%%%%%%%%%%%%%%%%%%%%%%%%
%%%%%%%%%%%%%%%%%%%%%%%%%%%%%%%%%%%%%%%%%%%%%%%%%%%%%%%%%%%%%%%%%%%%%%%%%%%%%%%%%%%%%%%%%%%%%%%%%%%%%%%%%%%%%%%%%%%%%%%%%%%%%%%%%%%%%%%%
\section{Fermi theory}

\index{Fermi!theory|(}The first quantitative theory of beta decay
was formulated in 1934 by E.~Fermi. In his pioneering work
\cite{ref1}, he suggested a direct interaction of four
spin-$\frac{1}{2}$ quantum fields, corresponding to the particles
involved in the process (\ref{eq1.1}). The interaction Hamiltonian
density\index{Hamiltonian density} proposed by Fermi can be
written as
\begin{equation}
\label{eq1.13}
\mathscr{H}^\tiz{Fermi}_\ti{int}=G(\psib_p\gamma^\mu\psi_n)
(\psib_e\gamma_\mu\psi_\nu)+\text{h.c.}
\end{equation}
where the $\psi$'s stand for the relevant four-component spinor
(i.e. fermionic) fields, the $\gamma^\mu$ are standard Dirac
matrices and $G$ is a coupling constant. Alternatively, one may
write Lagrangian density\index{Lagrangian density|ff}, which in
this case corresponds simply to changing the sign in
(\ref{eq1.13}), i.e.\index{coupling constants!Fermi}\index{gamma
matrices}
\begin{equation}
\label{eq1.14}
\lagr^\tiz{Fermi}_\ti{int}=-G(\psib_p\gamma^\mu\psi_n)
(\psib_e\gamma_\mu\psi_\nu)+\text{h.c.}
\end{equation}

It is easy to see that the first term of the interaction
Lagrangian describes, in a straightforward way (in the first order
of perturbation theory), the neutron decay (\ref{eq1.1}) (and
related processes like e.g. $e^+ +n\rightarrow p+\bar\nu$, etc.),
while its Hermitean conjugate incorporates the nuclear transition
of a proton into neutron, positron and neutrino (and related
reactions). Now it should be also clear why the third particle
produced in the neutron beta decay is called {\it anti\/}neutrino.
First of all, to describe the process in question one needs an
annihilation operator for the neutron and creation operators for
proton and electron -- this in turn means that the $\psib_p$ and
$\psi_e$ must occur in the interaction Lagrangian. The Lorentz
invariance\index{Lorentz!invariance} then requires that $\psib_e$
be paired with $\psi_\nu$ into a bilinear covariant form; however,
according to the conventional terminology, the $\psi_\nu$ contains
a creation operator for antiparticle (along with an annihilation
operator for particle).

The form (\ref{eq1.14}) reflects the original Fermi's idea that
the \qq{weak nuclear force}\index{weak!nuclear force|ff}
responsible for beta decay has essentially zero range, i.e. that
-- unlike e.g. the electromagnetism -- there is no relevant
bosonic particle mediating the weak interaction. Although we know
now that an intermediary does exist (this is the famous $W$
boson\index{W boson@$W$ boson}), the original \qq{conservative}
assumption of the contact character of the weak force is, in fact,
a very good approximation to reality {\it at sufficiently low
energies\/} (the reason is, of course, that the $W$ boson is very
heavy). Thus, in this chapter we will stick to the framework of a
direct four-fermion interaction\index{four-fermion interaction},
using the paradigm of (\ref{eq1.14}) and its subsequent
generalizations.

A note of historical character is perhaps in order here. In spite
of some clear differences between the weak and electromagnetic
forces, the original Fermi's form (\ref{eq1.14}) has certainly
been inspired by electrodynamics -- the bilinear combinations
(\qq{currents})\index{currents} of the fermion fields appearing in
(\ref{eq1.14}) are Lorentz {\it four-vectors}, similarly to the
electromagnetic current (coupled to vector four-potential)
familiar from QED\index{quantum!electrodynamics (QED)}. (Needless
to say, in the early 1930s there was no previous empirical
evidence that would support such a theoretical construction for
beta decay.) In this sense, the original Fermi's ideas clearly
constitute the first step towards an electro-weak unification, and
represent thus a rather fortunate conjecture indeed.

Before examining the phenomenological consequences of  the Fermi
theory, let us add one more remark concerning  general structural
aspects of  the Lagrangian (\ref{eq1.14}). A generic feature of
any four-fermion Lagrangian of the type (\ref{eq1.14}) is a
specific dimensionality of the corresponding coupling constant
$G$; in our natural system of units this is
\begin{equation}
\label{eq1.15}
[G]=M^{-2}
\end{equation}
where $M$ is an arbitrary mass scale. As we shall see later,
such a fact plays an important technical role in the development
of weak interaction theory, so let us now show how the result
(\ref{eq1.15}) can be inferred directly from the structure of the
relevant Lagrangian. To this end, one should first realize that
the action integral of any Lagrangian density $\lagr$
\begin{equation}
\label{eq1.16} \mathscr{I}=\int\lagr d^4 x
\end{equation}
is dimensionless in natural units (remember that an action has
dimension of $\hbar$ in general); this in turn implies
\begin{equation}
\label{eq1.17}
[\lagr ]=M^4
\end{equation}
since the four-dimensional volume element in (\ref{eq1.16})  has
dimension of $M^{-4}$ (a length unit is $M^{-1}$). Using the
general result (\ref{eq1.17}) one can determine easily the
dimensionality of a fermion field: the kinetic term of  Dirac
Lagrangian density has the usual form
$i\psib\slashed{\partial}\psi$ and the derivative has dimension of
$M$ (namely that of an inverse length), so one gets immediately
\begin{equation}
\label{eq1.18}
[\psi]=M^{3/2}
\end{equation}
From (\ref{eq1.14}), (\ref{eq1.17}) and (\ref{eq1.18}) the result
(\ref{eq1.15}) is then obvious.\footnote{In a similar way, it
is easy to see that for a bosonic field $B$ (e.g. scalar or vector)
one has $[B]=M$. Proving this is left to the reader as a simple
but instructive exercise. Then it is also obvious that e.g. in QED
the relevant coupling constant is dimensionless.} A note on terminology
is in order here. For an interaction Lagrangian density a corresponding
\qq{scale dimension} is usually introduced, which can be defined as
the dimension of the corresponding field monomial alone (i.e. of
the considered interaction term with the coupling constant removed);
more precisely, it is taken to be the relevant exponent of the
arbitrary mass referred to above. Thus, we will write e.g.
\begin{equation}
\label{eq1.19} \dim\lagr^\tiz{Fermi}_\ti{int}=6\quad(=4\cdot
\frac{3}{2})
\end{equation}
(for the individual fields we write similarly
$\dim\psi=\frac{3}{2}$ and $\dim B=1$ resp.). Taking into account
(\ref{eq1.17}), it is obvious that the scale dimension of an
interaction Lagrangian fixes uniquely the dimensionality of the
corresponding coupling constant and vice versa.

\begin{figure}[h]
\centering \s{\includegraphics{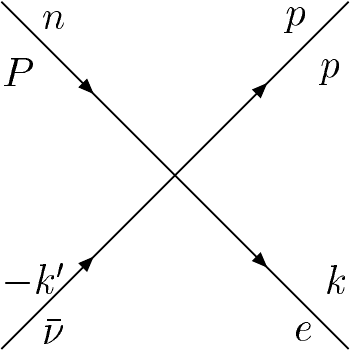}}
\caption{Lowest-order Feynman graph for the neutron beta decay in
a model of direct four-fermion interaction\index{four-fermion
interaction}.} \label{fig1}\index{Feynman diagrams!for neutron
beta decay}
\end{figure}

Let us now discuss some specific physical implications of the
original Fermi Lagrangian (\ref{eq1.14}). In the first order of
perturbation expansion of the $S$-matrix, the neutron beta decay
can be represented by a simple Feynman diagram shown in
Fig.\,\ref{fig1}.
The corresponding matrix element reads
\begin{equation}
\label{eq1.20}
{\cal M}_{fi}=-G[{\bar u}_p(p)\gamma^\mu u_n (P)][{\bar u}_e(k)
\gamma_\mu v_\nu (k')]
\end{equation}
where the $u$ and $v$ are momentum-space wave functions (Dirac
spinors)\index{Dirac!spinors} for the particles involved. From
(\ref{eq1.20}) one can easily calculate the corresponding decay
rate and fit the value of the coupling constant $G$ to the
measured neutron lifetime\index{lifetime!of the neutron}; one thus
gets roughly $G\doteq 10^{-5}\ \GeV^{-2}$. We will discuss the
decay rate calculation and the determination of the Fermi constant
later in this chapter, within the framework of a more general
(improved) four-fermion interaction Lagrangian. Now we are going
to examine more closely the structure of the matrix element
(\ref{eq1.20}) to get information about possible limitations of
the original Fermi model (\ref{eq1.14}).

We consider the neutron decay in its rest system. As we have
seen in the preceding section, the final-state proton is then
safely non-relativistic; in a first approximation, we will
therefore neglect the proton momentum altogether. Using the
well-known general formulae for solutions of Dirac equation in
the standard representation (see Appendix~\ref{appenA}), the nucleon spinors
can be then approximately written as
\begin{equation}
\label{eq1.21} u_n= \bm{U^{(r)}_n \\ 0},\qquad u_p\doteq
\bm{U^{(r)}_p \\ 0}
\end{equation}
where the $U^{(r)}_{p,n},\: r=1,2$ are two-component objects
\begin{equation}
\label{eq1.22} U^{(1)}_{p,n}=\sqrt{2M}\bm{1\\ 0},\qquad
U^{(2)}_{p,n}=\sqrt{2M}\bm{0\\ 1}
\end{equation}
and we have denoted by $M$ the nucleon average mass,
\begin{equation}
\label{eq1.23}
M=\frac{1}{2}(m_p+m_n)
\end{equation}
Note that the neglected proton momentum is actually of the order
of the nucleon mass difference (cf.(\ref{eq1.12})), so writing the
$M$ instead of a nucleon mass (whenever it does not lead to an
inconsistency) fits precisely into our \qq{quasi-static}
approximation scheme. Of course, the lower components of the $u_n$
are exactly zero, since the decaying neutron is at rest by
definition. Taking now into account the explicit form of the
standard Dirac gamma matrices, it is easy to see that in the
non-relativistic approximation (\ref{eq1.21}) the nucleon part of
the matrix element (\ref{eq1.20}) becomes
\begin{align}
\label{eq1.24} {\bar u}_p \gamma^0 u_n &= u^\dagger_p u_n \doteq
U^\dagger_p U_n\nonumber\\
{\bar u}_p\gamma^j u_n &= u^\dagger_p\alpha^j u_n \doteq 0\,,\qquad
\quad \text{for}\; j=1,2,3
\end{align}
Note that the last implication obviously holds as the matrices
$\alpha^j$
\begin{equation}
\label{eq1.25} \alpha^j=\bm{0 & \sigma_j \\ \sigma_j & 0}
\end{equation}
(with $\sigma_j$ being the Pauli matrices\index{Pauli matrices})
only connect \qq{large} and \qq{small} components of Dirac
spinors. The matrix element (\ref{eq1.20}) thus can be written, in
our quasi-static approximation for the recoil proton, as
\begin{equation}
\label{eq1.26} M_{fi}\doteq -G(U^\dagger_p U_n)(\bar u_e\gamma_0
v_\nu)
\end{equation}
The result (\ref{eq1.26}) shows that within the Fermi model, an
\qq{effective transition operator} for nucleons is actually the
{\it unit matrix\/} and thereby a nucleon spin flip is not
possible. At the level of atomic nuclei this means that the
original Fermi Lagrangian can only account for beta decay
processes with no change of the nucleonic spin. However, with the
development of nuclear spectroscopy it has become clear that beta
transitions involving a spin change ($\Delta J=1$ in particular)
do occur, with intensity comparable to the $\Delta J=0$ case.
Examples are, e.g., $\text{He}^6(0^+)\rightarrow
\text{Li}^6(1^+)+e^-+\bar\nu$ or $\text{B}^{12}(1^+)\rightarrow
\text{C}^{12}(0^+)+e^- +\bar\nu$ etc. (For historical reasons, the
spin-changing beta-decay processes are called Gamow--Teller
transitions\index{Gamow--Teller transitions|ff} while those caused
by the effective unit operator as in (\ref{eq1.26}) are Fermi
transitions.)\footnote{It should be emphasized that here and in
what follows we always have in mind only the so-called {\bf
allowed} transitions\index{allowed transitions} -- these occur in
the lowest order even when the quasi-static approximation for
nucleons is adopted. For a more detailed discussion of allowed and
forbidden\index{forbidden transition} beta decay processes see
e.g. \cite{CoB}, Chapter 5.}

One may thus conclude that the original Fermi model, though
conceptually correct (and applicable at least in a limited
sense), is certainly incomplete and must therefore be generalized
if one wants to get a realistic effective theory of weak nuclear
force.

%\input{kniha13}  %            1.3
%%%%%%%%%%%%%%%%%%%%%%%%%%%%%%%%%%%%%%%%%%%%%%%%%%%%%%%%%%%%%%%%%%%
%%%%%%%%%%%%%%%%%%%%%%%%%%%%%%%%%%%%%%%%%%%%%%%%%%%%%%%%%%%%%%%%%%%%%%%%%%%%%%%%%%%%%%%%%%%%%%%%%%%%%%%%%%%%%%%%%%%%%%%%%%%%%%%%%%%%%%%%
\section[Generalization of Fermi theory and parity violation]{Generalization of Fermi theory\\and parity violation}
%\def\rightmark{\thesection}
%\markboth{\thepage\hfil\slshape\leftmark}{{\slshape\rightmark}\hfil\thepage}
\index{parity!violation|ff}\index{couplings $S, V, A, T, P$}

 In fact, there is a straightforward way how to
generalize the simple Fermi model. Along with the vector-like
currents appearing in (\ref{eq1.14}), other possible covariant
bilinear combinations of the relevant spinor fields may be
included as well, i.e. one can construct a four-fermion
interaction\index{four-fermion interaction} Lagrangian using the
whole set of scalar ($S$), vector ($V$), tensor ($T$), axial
vector ($A$)\index{axial vector} and pseudoscalar ($P$) bilinear
Dirac forms (cf. Appendix~\ref{appenA}). Such an extension of the original
Fermi model was suggested first by G.~Gamow and E.~Teller
\cite{ref2}, and it is not difficult to realize that one is thus
indeed capable to describe both the spin-conserving and the
spin-changing nuclear beta transitions (as we shall see later in
this section, interaction terms of the type $A$ and $T$ are those
which can account for the Gamow--Teller $\Delta J=1$ transitions).

While the above-mentioned construction represents a rather
straightforward and natural step in building a realistic
theoretical framework for description of weak nuclear force, a
real breakthrough came in 1956, when T.~D.~Lee and C.~N.~Yang in
their fundamental paper \cite{ref3} suggested that one could
abandon the traditional assumption of parity symmetry (i.e. the
invariance under spatial inversion) of the weak interaction
Lagrangian. They observed that such a \qq{mirror symmetry}, though
naively taken for granted (e.g. in analogy with electrodynamics),
actually had no support in the available data and  proposed
therefore a set of experiments which could truly test this
fundamental issue.\footnote{Note that Lee and Yang came up with
their radical idea in order to solve a conundrum concerning the
strange meson $(K^+)$\index{meson} decays into pions -- a problem
that is usually quoted as the \qq{$\tau - \theta$
puzzle}\index{tau - theta@$\tau-\theta$ puzzle} in the literature.
For a detailed discussion of this important piece of particle
physics history, see e.g. \cite{Adv} or \cite{CaG}.} A series of
subsequent experiments revealed clearly the envisaged
parity-breaking phenomena (some of these effects will be discussed
explicitly later on) and the parity violation in weak interactions
has thus become one of the most dramatic discoveries of the 20th
century physics (Lee and Yang received the Nobel prize in 1957).

Let us now see how these aspects of the weak nuclear force can be
described formally. A general four-fermion
interaction\index{four-fermion interaction} Lagrangian for
beta-decay processes, including all the algebraic structures
mentioned above and taking into account a possible parity
violation, can be written as\index{bilinear covariant forms|(}
\begin{equation}
\label{eq1.27}
\lagr^{(\beta)}_{int}=\sum_{j=S,V,A,T,P}C_j(\bar\psi_p\Gamma_j
\psi_n)[\bar\psi_e(1+\alpha_j\gamma_5)\Gamma^j\psi_\nu]
\end{equation}
(tacitly assuming the presence of the h.c. term), where
\begin{eqnarray}
\label{eq1.28}
\Gamma_j&=&1,\gamma_\mu,\gamma_5\gamma_\mu,
\sigma_{\mu\nu},\gamma_5\nonumber \\
\Gamma^j&=&1,\gamma^\mu,\gamma_5\gamma^\mu,
\sigma^{\mu\nu},\gamma_5
\end{eqnarray}
for $j=S,V,A,T,P$ consecutively; the symbol $\sigma_{\mu\nu}$ means
\begin{equation}
\label{eq1.29}
\sigma_{\mu\nu}=\frac{i}{2}[\gamma_\mu,\gamma_\nu]
\end{equation}
and $1$ denotes the $4\times 4$ unit matrix. The parameters $C_j$
in (\ref{eq1.27}) have dimension of $M^{-2}$ in analogy with the
original Fermi coupling constant $G$. The $\alpha_j$ are
dimensionless and provide a measure of parity violation, as it
should be clear from the familiar transformation properties of the
fermion bilinear forms under space inversion $\cal P$. Indeed, for
each $j=S,V,A,T,P$ there are two terms in the Lagrangian,
descending from the factor of $1+\alpha_j\gamma_5$: the term
corresponding to the unity is $\cal P$-even (i.e. true Lorentz
scalar) while that involving $\alpha_j\gamma_5$ is $\cal P$-odd
(Lorentz pseudoscalar). The point is that the presence of an extra
$\gamma_5$ always changes the parity of a Lorentz-covariant
bilinear form under $\cal P$ (cf. Appendix~\ref{appenA}). Note, however, that
both terms coming from $1+\alpha_j\gamma_5$ must be present if the
Lagrangian is designed to describe parity-violating effects: if
one drops the $\cal P$-even terms in (\ref{eq1.27}) and keeps only
those involving $\alpha_j \gamma_5$, then the remaining Lagrangian
is in fact parity-conserving (though naively $\cal P$-odd) since
one can redefine the neutrino field by means of a unitary
transformation $\psi'_\nu=\gamma_5\psi_\nu$ without changing the
physical contents of the theory.

The generalized four-fermion interaction (\ref{eq1.27}) is
described in terms of ten arbitrary parameters $\alpha_j,C_j$. For
the sake of simplicity, we take all these parameters to be {\it
real}, which in fact means that invariance of (\ref{eq1.27}) under
time reversal\index{time reversal} is tacitly assumed (we will
discuss this issue in more detail within the framework of the
standard model of electroweak interactions). Note also that in
principle one could add to (\ref{eq1.27}) infinitely many other
terms involving derivatives of the fermion fields (i.e.
interaction terms of dimension higher than six). The form
(\ref{eq1.27}) represents, in this sense, a minimal model
involving non-derivative (i.e. lowest-dimensional) four-fermion
couplings only. Even so, introducing as many as ten arbitrary
parameters into our \qq{realistic} beta-decay Lagrangian certainly
makes it much less elegant than the original Fermi model. In
subsequent sections we will see that the number of relevant
parameters can in fact be significantly reduced when the Ansatz
(\ref{eq1.27}) is confronted with experimental data. At the end of
the day, a rather simple and elegant interaction Lagrangian
emerges, which in certain sense is quite similar to the old Fermi
model. In other words, it turns out that Fermi was \qq{almost
right} when writing his provisional theory of weak nuclear force
{\it a priori\/} in terms of vectorial
currents\index{Fermi!theory|)}.

Next, we discuss the properties of the relevant transition
amplitude. Obviously, the lowest-order matrix element for neutron
beta decay  (corresponding to the Feynman graph in
Fig.\,\ref{fig1}) now becomes
\begin{equation}
\label{eq1.30} {\cal M}^{(\beta)}_{fi}=\sum_{j=S,V,A,T,P}C_j(\bar
u_p\Gamma_j u_n)[\bar u_e(1+\alpha_j\gamma_5)\Gamma^j
v_\nu]\phantom{\J}
\end{equation}
Let us examine how the last expression is simplified if one
employs the non-relativistic approximation for nucleons. Using
the standard representation
\begin{equation}\label{eq1.31}
\gamma^0= \bm{\J& 0\\0&-\J},\quad
\gamma^j=\bm{0&\sigma_j\\-\sigma_j&0},\quad \gamma_5=\bm{0&\J\\
\J&0}
\end{equation}
it is not difficult to find that the non-relativistic (static)
reduction of the nucleon part of the matrix element (\ref{eq1.30})
follows the pattern shown in Table \ref{table1} (the result for
the $V$ term has already
been discussed in the preceding section).\\
\begin{table}[htb]
\begin{center}
\begin{tabular}{|c|c|c|}\hline
\bf {Algebraic type} & \multicolumn{2}{|c|}{\bf
Nucleon matrix elements}\\
\cline{2-3}
\bf of coupling &
\bf Covariant form &
\bf Static approximation\\ \hline
$S$ & $\bar u_p u_n$ & $ U^\dagger_p U_n $\\ \hline
$V$ & $\bar u_p\gamma_\mu u_n$ & $U^\dagger_p U_n$\\ \hline
$A$ & $\bar u_p\gamma_5\gamma_\mu u_n$ &
$U^\dagger_p\vec{\sigma}U_n$\\ \cline{1-3}
$T$ & $\bar u_p\sigma_{\mu\nu}u_n $ &
$U^\dagger_p\vec{\sigma}U_n $\\ \cline{1-3}
$P$ & $\bar u_p\gamma_5 u_n $ & 0\\ \hline
\end{tabular}
\end{center}
\caption{Scheme of the reduction of nucleonic part of a beta-decay
matrix element in non-relativistic (static) approximation.}
\label{table1}
\end{table}\index{bilinear covariant forms|)}
\noindent
\\
More precisely, such a scheme means that the matrix element
(\ref{eq1.30}) can be recast as
\begin{equation}
\label{eq1.32}
{\cal M}^{(\beta)}_{fi}={\cal M}_S+{\cal M}_V+{\cal M}_A+{\cal M}
_T+{\cal M}_P
\end{equation}
where
\begin{align}
\label{eq1.33} {\cal M}_S&\doteq C_S(U^\dagger_p U_n)[\bar
u_e(1+\alpha_S\gamma_5) v_\nu]\notag\\
{\cal M}_V&\doteq C_V(U^\dagger_p U_n)[\bar
u_e(1+\alpha_V\gamma_5) \gamma_0 v_\nu]\notag\\
{\cal M}_A&\doteq C_A(U^\dagger_p\sigma_j U_n)[\bar
u_e(1+\alpha_A\gamma_5)\gamma_5\gamma^j v_\nu]\notag\\
{\cal M}_T&\doteq 2C_T(U^\dagger_p\sigma_j U_n)[\bar
u_e(1+\alpha_T\gamma_5)\Sigma^j v_\nu]\notag\\
{\cal M}_P&\doteq 0
\end{align}
with
\begin{equation}
\label{eq1.34} \Sigma^j=\frac{1}{2}\epsilon^{jkl}\sigma^{kl}=
\bm{\sigma_j&0\\0&\sigma_j}
\end{equation}
We thus see that the pseudoscalar ($P$) term does not contribute
at all in the considered approximation and the remaining algebraic
structures come in pairs with similar properties: the $S$ and $V$
couplings are effectively represented by the $2\times2$ unit
matrix (i.e. a spin-zero transition operator), while the $A$ and
$T$ couplings both lead to Pauli matrices and constitute thereby
an effective transition operator carrying spin 1; these can
therefore account for spin-changing Gamow--Teller processes
(remember the good old Wigner--Eckart theorem). Let us remark that
the traditional terminology, mentioned briefly in previous
section, can now be made more precise: processes due to $S$ and/or
$V$ couplings (i.e. effectively mediated by unit matrix) are
called Fermi (F) transitions\index{Fermi!transitions} and those
caused by $A,T$ couplings (i.e. effectively mediated by Pauli
matrices) are Gamow--Teller (GT) transitions. This rather
technical definition can be translated into a more physical
language as follows. With regard to the spin of the initial and
final nucleon system $J_{i,f}$, there are essentially three types
of beta-decay processes. If $J_i=J_f=0$, only the $S$ and/or $V$
couplings can contribute and such a process is therefore {\bf pure
F transition}. If $\Delta J=| J_f-J_i | =1$, then only the $A$
and/or $T$ terms contribute and we have a {\bf pure GT
transition}. For $J_i=J_f\neq0$ one can in principle get a
contribution from both type of couplings and such a process may be
naturally called {\bf mixed transition}\index{mixed transition}.
We have already given examples of pure GT transitions in the
preceding section. A well-known case of a pure F process is the
$0^+\rightarrow 0^+$ transition O$^{14}\rightarrow$
N$^{14^{\ast}}+e^+ +\nu$, while the free neutron decay or the
tritium decay H$^3\rightarrow$ He$^3+e^-+\bar\nu$ can serve as
examples of mixed transitions\index{tritium decay}.

Now we have the necessary technical prerequisites at hand and
we can employ the matrix elements (\ref{eq1.33}) to calculate some
observable dynamical characteristics of beta-decay processes,
that will help us to determine the values of  the free parameters
in the Lagrangian (\ref{eq1.27}). This will be the subject
of subsequent sections.

%\input{kniha14}  %            1.4
%%%%%%%%%%%%%%%%%%%%%%%%%%%%%%%%%%%%%%%%%%%%%%%%%%%%%%%%%%%%%%%%%%%
%%%%%%%%%%%%%%%%%%%%%%%%%%%%%%%%%%%%%%%%%%%%%%%%%%%%%%%%%%%%%%%%%%%%%%%%%%%%%%%%%%%%%%%%%%%%%%%%%%%%%%%%%%%%%%%%%%%%%%%%%%%%%%%%%%%%%%%%
\section{The electron energy spectrum}\label{sec1.4}

\index{energy spectrum of the electron|(}Observable quantities for
the considered processes are expressed in terms of appropriate
decay rates. The starting point of our calculations will be the
differential decay rate for a free neutron in its rest frame,
involving the element of the corresponding three-particle phase
space\index{phase space}
\begin{equation}
\label{eq1.35}
dw=\frac{1}{2m_n}|{\cal M}|^2\frac{d^3 k}
{(2\pi)^3 2E(k)}\frac{d^3k'}{(2\pi)^3 2E(k')}\frac{d^3p}
{(2\pi)^3 2E(p)}(2\pi)^4 \delta^4(P-k-k'-p)
\end{equation}
where we have denoted the relevant momenta in accordance with
(\ref{eq1.2}). Various interesting quantities can be then obtained by
integrating (\ref{eq1.35}) over some kinematical variables (in other
words, over the phase-space volume elements). We will discuss the
phase-space integrations later on and calculate first the matrix
element squared, which is the object of central importance,
reflecting the weak interaction dynamics.

To begin with, let us consider the situation where the particles
are unpolarized, i.e. one does not care about a particular spin
(projection) of a decay product and the decaying neutron is
supposed to have spin \qq{up} or \qq{down} with equal probability. In
such a case, the $\overline{|{\cal M}|^2}$ is to be summed
over the spins of final-state particles and averaged over the
initial neutron spin, i.e. the relevant  quantity  is
\begin{equation}
\label{eq1.36} \overline{|{\cal M}|^2}=\frac{1}{2} \sum_{\text{\it
spins} \atop n,p,e,\bar\nu} |{\cal M}|^2
\end{equation}
This brings about some simplifications, especially within our
quasi-static approximation for nucleons. In particular, as a
result of the summation over nucleon spins, there is no
interference between the Fermi ($S, V$) and Gamow--Teller ($A, T$)
parts of the matrix element (\ref{eq1.32}), (\ref{eq1.33}). Let us
prove this simple statement for the reader's convenience. Obviously,
such an interference term in $|{\cal M}|^2=\cal{M~M}^\ast$ would
certainly contain a nucleonic factor
\begin{equation}
\label{eq1.37}
U^\dagger_pU_n\;U^\dagger_n\sigma_jU_p
\end{equation}
that can be identically recast as the trace
\begin{equation}
\label{eq1.38}
\Tr(U^{(r)}_pU^{(r)\dagger}_pU^{(r')}_nU^{(r')\dagger}_n
\sigma_j)
\end{equation}
Spin sums for the two-component Pauli spinors (\ref{eq1.22}) are
proportional to the unit matrix:
\begin{align}
\label{eq1.39} \sum^2_{r=1}U^{(r)}_{p,n}U^{(r)\dagger}_{p,n}&=
(\sqrt{2M})^2\left[\bm{1\\0}\bm{1 & 0}+\bm{0\\
1}\bm{0&1}\right]
\notag\\
&= 2M\left[\bm{1&0\\ 0& 0}+\bm{0& 0\\ 0& 1}\right]=
2M\bm{1&0\\ 0& 1}
\end{align}
The expression (\ref{eq1.38}) summed over $r,r'$ thus becomes
proportional to $\Tr\ \sigma_j$, which of course vanishes for any
$j=1,2,3$.

The quantity (\ref{eq1.36}) can thus be written (within the
non-relativistic approximation (\ref{eq1.33})) as
\begin{equation}
\label{eq1.40}
\overline{|{\cal M}|^2}=\overline{|{\cal M}_S+{\cal
M}_V|^2}+\overline{|{\cal M}_A+{\cal M}_T|^2}=
\overline{|{\cal M}_F|^2} +\overline{|{\cal
M}_{GT}|^2}
\end{equation}
Let us now work out the Fermi part of the last expression. Using
the explicit form of ${\cal M}_S$ and ${\cal M}_V$ as given in
(\ref{eq1.33}), some familiar properties of gamma matrices
and the usual trick of introducing traces of matrix products,
one gets first
\begin{eqnarray}
\label{eq1.41}
\lefteqn {{|{\cal M}_F|^2}={|{\cal M}_S|^2}+
{|{\cal M}_V|^2}+{\cal M}_S{\cal M}^\ast_V+
{\cal M}^\ast_S{\cal M}_V}\\
&=&C^2_S \Tr(U_pU^\dagger_pU_nU^\dagger_n)\Tr[u_e
\bar u_e(1+\alpha_S\gamma_5)v_\nu\bar
v_\nu(1-\alpha_S\gamma_5)]\nonumber\\
&+&C^2_V \Tr(U_pU^\dagger_pU_nU^\dagger_n)\Tr[u_e
\bar u_e(1+\alpha_V\gamma_5)\gamma_0 v_\nu\bar
v_\nu(1+\alpha_V\gamma_5)\gamma_0]\nonumber\\
&+&C_SC_V \Tr(U_pU^\dagger_pU_nU^\dagger_n)\Tr[u_e \bar
u_e(1+\alpha_S\gamma_5)v_\nu\bar
v_\nu(1+\alpha_V\gamma_5)\gamma_0]+{\text{c.c.}}\nonumber
\end{eqnarray}
(of course, the complex conjugation refers only to the last line
in (\ref{eq1.41})). After the summation over spins (cf.
(\ref{eq1.39}) and (\ref{eqA.66})) this becomes
\begin{eqnarray}
\label{eq1.42}
\lefteqn{\overline{|{\cal M}_F|^2}=4C^2_S M^2 \Tr[(\slashed{k} +m_e)
(1+\alpha_S\gamma_5)\slashed{k}'(1-\alpha_S\gamma_5)]}\nonumber\\
&+&4C^2_V M^2 \Tr[(\slashed{k} +m_e)(1+\alpha_V\gamma_5)\gamma_0\slashed{k}'
(1+\alpha_V\gamma_5)\gamma_0]\nonumber\\
&+&4C_SC_VM^2\Tr[(\slashed{k} +m_e)(1+\alpha_S\gamma_5)\slashed{k}'
(1+\alpha_V\gamma_5)\gamma_0]+ \text{c.c.}
\end{eqnarray}
The leptonic traces can be simplified to
\begin{eqnarray}
\overline{|{\cal M}_F|^2}&=&4C^2_S(1+\alpha^2_S)
M^2 \Tr{(\slashed{k}}\slashed{k}')\nonumber\\
&+&4C^2_V(1+\alpha^2_V)
M^2 \Tr{(\slashed{k}}\gamma_0\slashed{k}'\gamma_0)\nonumber\\
&+&4C_S C_V(1-\alpha_S\alpha_V)
m_e M^2 \Tr
(\slashed{k}'\gamma_0)+{\text{c.c.}}\nonumber
\end{eqnarray}
and a straightforward calculation then yields
\begin{eqnarray}
\label{eq1.43}
\overline{|{\cal M}_F|^2}&=&16C^2_S(1+\alpha^2_S)
M^2 (k\cdot k')\nonumber\\
&+&16C^2_V(1+\alpha^2_V)M^2 (2E_e E_{\bar\nu}-k\cdot
k')\nonumber\\
&+&32C_S C_V(1-\alpha_S\alpha_V)m_e M^2 E_{\bar\nu}
\end{eqnarray}
The Lorentz scalar product $k\cdot k'$ can be
expressed as
\begin{equation}
\label{eq1.44}
k\cdot k'=E_e E_{\bar\nu}-|\vec k|
|\vec k'| \cos{\vartheta}=E_e E_{\bar\nu}
(1-\beta_e \cos \vartheta)
\end{equation}
and (\ref{eq1.43}) thus finally becomes
\begin{eqnarray}
\label{eq1.45}
\overline{|{\cal M}_F|^2}&=&16M^2 E_e E_{\bar\nu}
\Bigl[C^2_S(1+\alpha^2_S)(1-\beta_e \cos \vartheta)\nonumber\\
&+&C^2_V(1+\alpha^2_V)(1+\beta_e \cos \vartheta)+2C_S C_V
(1-\alpha_S\alpha_V)\frac{m_e}{E_e}\Bigr]
\end{eqnarray}
The Gamow--Teller part of (\ref{eq1.40}) can be evaluated in a
similar way, using the standard trace techniques. We defer the
calculation to the next section, in order not to clutter the
present section with too many technicalities. The result has a
form analogous to (\ref{eq1.45}); it reads
\begin{eqnarray}
\label{eq1.46}
\lefteqn{\overline{|{\cal M}_{GT}|^2}=\overline{|{\cal M}_A+
{\cal M}_T|^2}=}\nonumber\\
&=&16M^2 E_e E_{\bar\nu}\Bigl[3C^2_A(1+\alpha^2_A)(1-\frac{1}{3}
\beta_e \cos \vartheta)+12C^2_T(1+\alpha^2_T)(1+{\frac{1}{3}
\beta_e \cos \vartheta)}\nonumber\\
&-&12C_A C_T(1-\alpha_A\alpha_T)\frac{m_e}{E_e}\Bigr]
\end{eqnarray}
Notice that while there is no interference between ${\cal M}_F$
and ${\cal M}_{GT}$, the $S-V$ or $A-T$ interference terms can in
principle occur, depending on the values of the relevant
parameters $C_j,\alpha_j$.

Let us now turn to the calculation of some interesting
differential decay rates defined with respect to the leptonic
kinematical variables. Our ultimate goal will be the electron
energy spectrum (we still have in mind the case of unpolarized
particles). To this end, we start from the basic formula
(\ref{eq1.35}) and integrate first over the proton momentum. Such
an integration is essentially trivial -- we simply use up the
three-dimensional delta function corresponding to the momentum
conservation, i.e. replace the $\vec{p}$ by $-(\vec{k}+\vec{k}')$.
This immediately yields
\begin{eqnarray}
\label{eq1.47}
dw&\doteq&\frac{1}{2m_n}\overline{|{\cal M}|^2}\frac
{d^3k}{(2\pi)^3 2E_e}\frac{d^3 k'}{(2\pi)^3 2E_{\bar\nu}}
\frac{1}{2m_p}\nonumber\\
&\times&2\pi\delta(m_n-E_e-E_{\bar\nu}-
\sqrt{(\vec{k}+\vec{k'})^2+m^2_p}\;)
\end{eqnarray}
Note that for the sake of brevity we denote the once integrated
differential decay rate by the same symbol as the original
quantity. In (\ref{eq1.47}) we have already neglected  the proton
momentum (setting $E_p\doteq m_p$) in the normalization
coefficient at the corresponding phase-space factor. We will
neglect quantities of the order of $\Delta m/M$ in the
course of our calculation, whenever such an approximation is
safely under control and does not lead to an inconsistency e.g.
in energy-momentum balance. The next step is an integration over
$d^3k'$. For a fixed direction of the electron momentum
$\vec{k}$ and fixed angle $\vartheta$ between $\vec{k}'$ and
$\vec{k}$, we will integrate first over the modulus
$|\vec{k}'| =E_{\bar\nu}$. Let us denote $x=|\vec{k}'|$
for brevity; the energy-conservation delta function in
(\ref{eq1.47}) can be written as $\delta(f(x))$, with
\begin{equation}
\label{eq1.48}
f(x)=m_n-E_e-x-\sqrt{|\vec{k}|^2+2x|\vec{k}|
\cos \vartheta+x^2+m^2_p}
\end{equation}
The energy-conservation condition $f(x_0)=0$ implies
\begin{equation}
\label{eq1.49}
x_0=E_{\bar\nu}=\frac{m^2_n-m^2_p+m^2_e-2m_n E_e}
{2m_n-2E_e+2|\vec{k}| \cos \vartheta}
\end{equation}
(note that from (\ref{eq1.49}) one can recover the kinematical
upper bound (\ref{eq1.6}) for electron energy). To carry out the
$x$-integration one can now use the well-known relation
\begin{equation}
\label{eq1.50}
\delta(f(x))=\frac{1}{| f'(x_0)|}\delta(x-x_0)
\end{equation}
The evaluation of the $f'(x_0)$ is straightforward;
differentiating plainly the expression (\ref{eq1.48}) and
utilizing the condition $f(x_0)=0$, i.e.
\begin{center}
$\sqrt{|\vec{k}|^2+2x_0|\vec{k}| \cos
\vartheta+x^2_0+m^2_p}=m_n-E_e-x_0$
\end{center}
\noindent
one gets readily
\begin{equation}
\label{eq1.51}
| f'(x_0)|=\frac{m_n-E_e+|\vec{k}|\cos \vartheta}
{m_n-E_e-x_0}
\end{equation}
Now we can make our usual approximations, neglecting the terms of
relative order $\Delta m/M$. The expression (\ref{eq1.49}) then
becomes
\begin{eqnarray}
\label{eq1.52}
x_0=E_{\bar\nu}&\doteq &\frac{1}{2m_n}(m^2_n-m^2_p+m^2_e-2m_nE_e)
\nonumber\\
&=&E^{max}_e-E_e
\end{eqnarray}
where we have taken into account (\ref{eq1.6}). We thus see that
for a given value of $E_e$, the antineutrino energy can
approximately be written as
\begin{equation}
\label{eq1.53}
E_{\bar\nu}\doteq E^{max}_e-E_e\doteq\Delta m-E_e
\end{equation}
(of course, this is an expected result -- it follows simply from
energy conservation if the proton motion is neglected). Similarly,
from (\ref{eq1.51}) we obtain
\begin{equation}
\label{eq1.54}
| f'(x_0)|\doteq 1
\end{equation}
The integration of (\ref{eq1.47}) over the $|\vec{k}'|$ can
now be done easily -- it essentially consists in dropping the delta
function and replacing everywhere the antineutrino energy by its
approximate physical value (\ref{eq1.53}). Since the $d^3 k'$ can be
written in spherical coordinates as $x^2 dxd\Omega_{\bar\nu}$
(where $d\Omega_{\bar\nu}$ is an element of a corresponding solid
angle), the result of the $x$ -integration can be written as
\begin{equation}
\label{eq1.55}
dw\doteq\frac{1}{8M^2}\overline{|{\cal M}|^2}E_{\bar\nu}
\frac{d^3\vec{k}}{(2\pi)^3 2E_e}\frac{d\Omega_{\bar\nu}}{4\pi^2}
\end{equation}
where we have also replaced $m_n$ and $m_p$ by the average nucleon
mass $M$. Note that $d\Omega_{\bar\nu}=\sin \vartheta d \vartheta
d\varphi$ with $\vartheta$ being the angle between the directions
of $\vec{k}'$ and $\vec{k}$. The decay rate (\ref{eq1.55}) thus
describes the angular correlation\index{angular correlation}
between electron and antineutrino -- we will discuss this
experimentally interesting quantity in more detail in the next
section.

To arrive at the electron energy spectrum, we have to integrate
over the angular variables in (\ref{eq1.55}). For the moment, let
us consider e.g. only the Fermi part of the $\overline{|{\cal M}
|^2}$, i.e. the expression (\ref{eq1.45}). Obviously, the terms
proportional to $\cos\vartheta$ vanish upon integration over
$d\vartheta$ (and the constant terms are simply multiplied by
$4\pi$). Further, $d^3k=|\vec{k}|^2d|\vec{k}| d\Omega_e$, where
$d\Omega_e$ is the  element of solid angle corresponding to the
electron momentum direction, related to an arbitrarily
(conventionally) chosen coordinate system. Of course, there is no
non-trivial angular dependence that would survive after the
preceding integration over $d\Omega_{\bar\nu}$ (as there is no
natural preferred spatial direction in the considered problem), so
the remaining angular integration over $d\Omega_e$ is trivial --
it amounts to a multiplication by $4\pi$. As the last step, one
should pass from the differential $d|\vec{k}|$ to $dE_e$; this is
done easily, since the relation $E_e=(|\vec{k}|^2+m^2_e)^{1/2}$
implies immediately $|\vec{k}|\ d|\vec{k}|=E_edE_e$. The form of
the electron energy spectrum is thus given by
\begin{eqnarray}
\label{eq1.56}
dw(E_e)&\doteq&\frac{1}{2\pi^3}[C^2_S(1+\alpha^2_S)+C^2_V(1+
\alpha^2_V)+2C_SC_V(1-\alpha_S\alpha_V)\frac{m_e}{E_e}]\times
\nonumber\\
&\times&|\vec{k}| E_e(\Delta-E_e)^2 dE_e
\end{eqnarray}
where we have employed the explicit expression for
$\overline{|{\cal M}_F|^2}$ given by (\ref{eq1.45}) and we
have also introduced the usual shorthand notation $\Delta$ for
the endpoint of the spectrum ($\Delta=E^{max}_e\doteq\Delta m$).

Let us now discuss the obtained result. We have performed the
calculation for a free neutron, but it is not difficult to realize
that the energy-dependence shown in (\ref{eq1.56}) should be valid
in the case of an allowed nuclear beta decay as well. Indeed, when
dealing with atomic nuclei, the usual quasi-static approximation
for nucleons can be employed; the relevant nuclear matrix element
is then of course independent of the electron energy and the form
of $dw(E_e)$ comes out to be the same as in (\ref{eq1.56}). In
other words, for an allowed nuclear beta decay only a constant
factor in (\ref{eq1.56}) may get modified, but not the functional
dependence on the $E_e$. The result (\ref{eq1.56}) is thus
appropriate for description of the electron energy spectrum
corresponding to a pure F transition. Similarly, one could use the
form (\ref{eq1.46}) for the spin-averaged matrix element squared
and obtain thus a straightforward analogy of the relation
(\ref{eq1.56}) for pure GT transitions (of course, for a free
neutron decay one has to include both types of matrix elements).
The contribution proportional to $m_e/E_e$ in the generic formula
(\ref{eq1.56}) is called, for historical reasons, a Fierz
interference term\index{Fierz interference term|(}. The existing
experimental data show that the value of the corresponding
coefficient of such a term is consistent with zero for both F and
GT transitions (for typical numbers, see e.g. \cite{CoB} or
\cite{Ren}). Obviously, this empirical fact represents a certain
constraint on the parameters $C_j$ and $\alpha_j$. If taken at
face value (i.e. assuming that the Fierz interference term is
exactly zero), this would mean that for F transitions one has
\begin{equation}
\label{eq1.57}
C_SC_V(1-\alpha_S\alpha_V)=0
\end{equation}
and this in turn implies that either $C_S=0$, or $C_V=0$, or
$1-\alpha_S\alpha_V=0$. We will obtain further constraints on the
parameters later on (by utilizing other relevant experimental data)
and for the time being we will simply keep in mind the condition
(\ref{eq1.57}). Similarly, we will interpret the empirical evidence
for the absence of Fierz interference terms in GT transitions as a
constraint
\begin{equation}
\label{eq1.58}
C_AC_T(1-\alpha_A\alpha_T)=0
\end{equation}
(cf. (\ref{eq1.46})). Let us emphasize that we do not attempt to
accomplish a \qq{best fit} of the free parameters of the general
four-fermion Lagrangian (\ref{eq1.27}) to the available
experimental data -- rather we will try to show that the wealth of
empirical data clearly point towards a very particular and
remarkably simple theoretical scheme for weak interactions.

\begin{figure}[h]
\centering
\s{\includegraphics{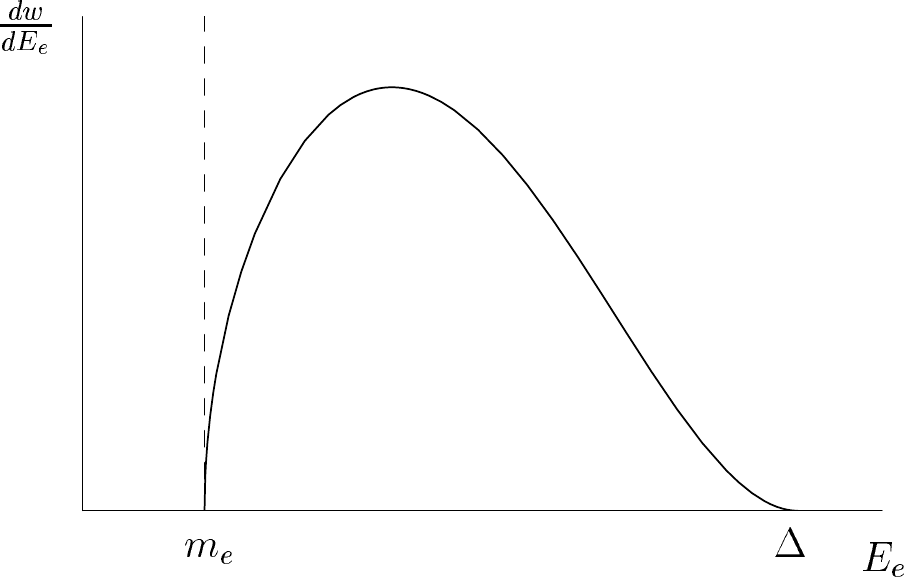}}
\caption{The typical form of the electron energy spectrum for
an allowed beta decay.}
\label{fig2}
\end{figure}

From the preceding considerations it is clear that -- in the
absence of the Fierz interference terms -- the electron energy
spectra do not provide any further information about the
properties of the weak interaction (for example, on the basis of the
spectrum alone one cannot distinguish between $S$ and $V$ couplings
in Fermi transitions, etc.). The characteristic functional
dependence
\begin{equation}
\label{eq1.59}
\frac{dw(E_e)}{dE_e}=const.\times\sqrt{E^2_e-m^2_e}\;E_e(\Delta-E_e)^2
\end{equation}
is sometimes called the \qq{statistical form} of the energy
spectrum, since this is essentially determined by the phase-space
factors (it is instructive to trace the origin of the individual
factors in (\ref{eq1.59}) back to our starting point
(\ref{eq1.35}) and to the normalization of the matrix element
$\cal{M}$). The function (\ref{eq1.59}) is depicted in
Fig.\,\ref{fig2}.

\begin{figure}[h]
\centering
\s{\includegraphics{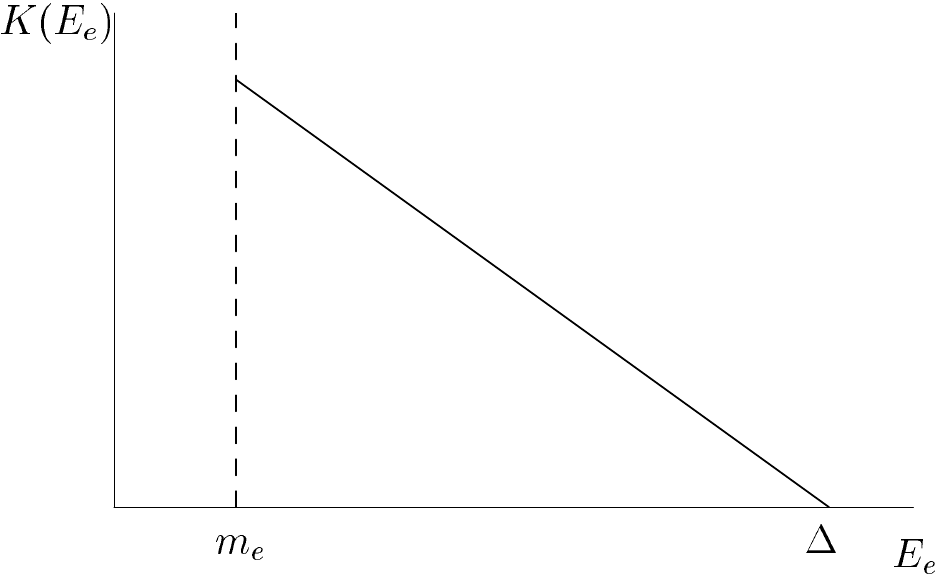}}
\caption{The straight-line Kurie plot for a beta decay
spectrum, which exhibits the absence of a Fierz interference
term.}
\label{fig3}
\end{figure}

Note that beta-decay spectra are usually represented in the form
of the so-called Kurie (or Fermi--Kurie\index{Fermi--Kurie
plot|see{Kurie plot}}\index{Kurie plot}) plot, which displays the
energy dependence of the quantity
\begin{equation}
\label{eq1.60}
K(E_e)=\left(\frac{1}{|\vec{k}| E_e}\frac{dw}{dE_e}\right)
^{1/2}
\end{equation}
For $dw(E_e)$ given by (\ref{eq1.59}) the Kurie plot is a falling
straight line, $K(E_e)=const.\times(\Delta-E_e)$, as shown in
Fig.\,\ref{fig3}.
It is obvious that the presence of a Fierz interference term would
manifest itself as a distortion of the straight-line Kurie plot,
that would be most pronounced in the low-energy part of the
spectrum, i.e. near its beginning at $E_e=m_e$.

Last but not least, one should note that the linear dependence
(\ref{eq1.60}) also relies on the assumption of vanishing rest
mass of the neutrino\index{neutrino!mass}. Indeed, from the
antineutrino phase-space volume element in (\ref{eq1.35}) one
gets, in general, a factor of $|\vec{k}'| E_{\bar\nu}$ in
(\ref{eq1.59}) (which of course coincides with $E^2_{\bar\nu}$ for
$m_\nu=0$), with
$|\vec{k}'|=\sqrt{E^2_{\bar\nu}-m^2_\nu}<E_{\bar\nu}$ for
$m_\nu\neq 0$. This could then also yield a deviation from the
straight-line Kurie plot (namely a downward deflection), in
particular near its endpoint: if the electron energy is close to
its maximum value, the antineutrino energy is small and a relative
difference between $|\vec{k}'|$ and $E_{\bar\nu}$ due to
$m_\nu\neq 0$ then becomes largest. This simple observation in
fact provides a conceptual basis for several experiments that play
an important role in the present-day quest for a neutrino mass.
For details, the reader is referred e.g. to the monographs
\cite{Vog}, \cite{Kay} or \cite{CoB} (see also the textbook
\cite{Gre})\index{energy spectrum of the electron|)}.

%\input{kniha15}  %            1.5
%%%%%%%%%%%%%%%%%%%%%%%%%%%%%%%%%%%%%%%%%%%%%%%%%%%%%%%%%%%%%%%%%%%
%%%%%%%%%%%%%%%%%%%%%%%%%%%%%%%%%%%%%%%%%%%%%%%%%%%%%%%%%%%%%%%%%%%%%%%%%%%%%%%%%%%%%%%%%%%%%%%%%%%%%%%%%%%%%%%%%%%%%%%%%%%%%%%%%%%%%%%%
\section[The $e-\bar\nu$ angular correlation]{The $\boldsymbol{e-\bar{\nu}}$ angular correlation: dominance of $V$ and
$A$ couplings}\markright{\thesection\quad THE
$\boldsymbol{e-\bar{\nu}}$ ANGULAR CORRELATION} \index{angular
correlation|ff}\label{sec1.5}

  We are now going to examine the angular distribution of leptons
produced in a beta-decay process. Such an observable quantity is
rather sensitive to the type of coupling responsible for a given
process and its analysis hence provides some powerful constraints
on the parameters of the effective Lagrangian (\ref{eq1.27}). Before
discussing this issue in detail, let us derive -- as promised in
the preceding section -- the formula (\ref{eq1.46}) for the spin-averaged
GT matrix element squared.

Looking back at (\ref{eq1.33}), we have
\begin{eqnarray}
{\cal M}_{GT} &=& {\cal M}_{A} + {\cal M}_{T}
\nonumber\\
&=& C_A ( U_p^\dagger \sigma_j U_n ) [\bar u_e
(1+\alpha_A\gamma_5) \gamma_5 \gamma^j v_\nu]
\nonumber\\
&& \hspace{-1.3em} + 2 C_T ( U_p^\dagger \sigma_j U_n ) [\bar u_e
(1+\alpha_T\gamma_5) \Sigma^j v_\nu]
\label{eq1.61}
\end{eqnarray}
Squaring (\ref{eq1.61}), using the familiar trace techniques and summing
over spins, one gets first
\begin{eqnarray}
\overline{|{\cal M}_{GT}|^2} &=& \frac12 \sum_{spins} |{\cal M}_{GT}|^2
\nonumber\\
&=& 2 C_A^2 M^2 \Tr(\sigma_j \sigma_k)
\Tr[(\slashed{k}+m_e)(1+\alpha_A\gamma_5) \gamma_5 \gamma^j
\slashed{k}'(1+\alpha_A\gamma_5) \gamma_5 \gamma^k]
\nonumber\\
&+& 8 C_T^2 M^2 \Tr(\sigma_j \sigma_k)
\Tr[(\slashed{k}+m_e)(1+\alpha_T\gamma_5) \Sigma^j
\slashed{k}'(1-\alpha_T\gamma_5) \Sigma^k]
\nonumber\\
&+& 4 C_A C_T M^2 \Tr(\sigma_j \sigma_k)
\Tr[(\slashed{k}+m_e)(1+\alpha_A\gamma_5) \gamma_5 \gamma^j
\slashed{k}'(1-\alpha_T\gamma_5) \Sigma^k]
\nonumber\\
&+& {\rm c. c.} \label{eq1.62}
\end{eqnarray}
(where we have employed, among other things, the completeness relation
(\ref{eq1.39}) for the two-component nucleon spinors); the complex conjugation
of course refers only to the last line in (\ref{eq1.62}). To work out the traces
involving the spin matrix $\Sigma^j$ (see (\ref{eq1.34})) it is useful to
remember the identity
\begin{equation}\label{eq1.63}
\Sigma^j = \gamma_5 \alpha^j = \gamma_5 \gamma_0 \gamma^j
\end{equation}
(cf. (\ref{eqA.71})). The expression (\ref{eq1.62}) then becomes,
after some manipulations
\begin{eqnarray}
\overline{|{\cal M}_{GT}|^2} &=& 2 C_A^2 M^2 \Tr(\sigma_j
\sigma_k) \Tr[\slashed{k} \gamma^j \slashed{k}'\gamma^k
(1+\alpha_A^2 - 2\alpha_A \gamma_5)]
\nonumber\\
&& + 8 C_T^2 M^2 \Tr(\sigma_j \sigma_k)
\Tr[\slashed{k}\gamma^j\gamma_0 \slashed{k}'\gamma_0 \gamma^k
(1+\alpha_T^2 - 2\alpha_T \gamma_5)]
\nonumber\\
&& + 4 C_A C_T M^2 m_e \Tr(\sigma_j \sigma_k) \Tr[\gamma^j
\slashed{k}'\gamma_0 \gamma^k (1-\alpha_A \alpha_T + (\alpha_A -
\alpha_T)\gamma_5)]
\nonumber\\
&& + {\rm c. c.}
\label{eq1.64}
\end{eqnarray}
The number of gamma matrices under the second trace can be easily
reduced to four by using the standard anticommutation relations --
one has $\gamma_0 \gamma^j \gamma_0 = - \gamma^j$ and hence $
\gamma_0 \slashed{k} \gamma_0 = \tilde{\slashed{k}}$ with $\tilde
k = (k_0, -\vec{k})$. Further, we use the identity $\Tr(\sigma_j
\sigma_k)=2\delta_{jk}$ for the Pauli matrices and some well-known
trace properties of Dirac matrices; in particular, one may observe
that traces of the type
$\Tr(\slashed{a}\gamma^j\slashed{b}\gamma^j\gamma_5)$ vanish
identically because of antisymmetry of the Levi-Civita tensor
$\epsilon_{\alpha\beta\gamma\delta}$. We are thus left with
\begin{eqnarray}
\overline{|{\cal M}_{GT}|^2} &=& 4 C_A^2 (1+\alpha_A^2) M^2
\Tr(\slashed{k} \gamma^j \slashed{k}'\gamma^j)
\nonumber\\
&& + 16 C_T^2 (1+\alpha_T^2) M^2 \Tr(\tilde{\slashed{k}}\gamma^j
\slashed{k}'\gamma^j)
\nonumber\\
&& + 8 C_A C_T (1-\alpha_A \alpha_T) M^2 m_e \Tr(\gamma^j
\slashed{k}'\gamma_0 \gamma^j)
\nonumber\\
&& + {\rm c. c.}
\label{eq1.65}
\end{eqnarray}
The evaluation of the remaining traces is then straightforward and one
obtains
\begin{eqnarray}
\overline{|{\cal M}_{GT}|^2} &=& 16 C_A^2 (1+\alpha_A^2) M^2
(2\vec{k}\cdot\vec{k}'-g^{jj}k\cdot k')
\nonumber\\
&& + 64 C_T^2 (1+\alpha_T^2) M^2
(-2\vec{k}\cdot\vec{k}'-g^{jj}\tilde k\cdot k')
\nonumber\\
&& + 64 C_A C_T (1-\alpha_A \alpha_T) M^2 m_e
k_0' g^{jj}
\label{eq1.66}
\end{eqnarray}
However, $g^{jj}=-3$ and the scalar products of the
four-momenta can be written as
$k\cdot k'= E_e E_{\bar\nu}(1-\beta_e \cos\vartheta)$
and
$\tilde k\cdot k'= E_e E_{\bar\nu}(1+\beta_e \cos\vartheta)$.
One thus finally gets,
after some simple manipulations
\begin{eqnarray}\label{eq1.67}
\overline{|{\cal M}_{GT}|^2} &=& 16 M^2 E_e E_{\bar\nu} \Bigl[3
C_A^2 (1+\alpha_A^2)(1-\frac13 \beta_e\cos\vartheta)\\
&& + 12 C_T^2 (1+\alpha_T^2)(1+\frac13 \beta_e\cos\vartheta)
         - 12 C_A C_T
         (1-\alpha_A\alpha_T)\frac{m_e}{E_e}\Bigr]\nonumber
\end{eqnarray}
and (\ref{eq1.46}) is thereby proved.

     Let us now examine the angular correlation of the electron and
antineutrino. As we have noted before, this is described in terms of
a differential decay rate of the type (\ref{eq1.55}). The angular dependence
is contained in the leptonic part of the relevant matrix element
squared. The results (\ref{eq1.45}) and (\ref{eq1.67}) are applicable
to the allowed
nuclear F and GT transitions resp. -- the nuclear wave functions can
only contribute an overall constant factor which of course does not
influence the lepton angular distribution in question. We may now also
use the conditions (\ref{eq1.57}) and (\ref{eq1.58}) which express
the absence of the
Fierz interference. For the F transitions we thus have
\begin{equation}\label{eq1.68}
\frac{dw_F}{d\Omega_{\bar\nu}} = {\rm const.} \times (1+a_F
\beta_e \cos\vartheta)
\end{equation}
with the correlation coefficient
\begin{equation}\label{eq1.69}
a_F=\frac{C_V^2(1+\alpha_V^2)-C_S^2(1+\alpha_S^2)}
         {C_V^2(1+\alpha_V^2)+C_S^2(1+\alpha_S^2)}
\end{equation}
and for GT transitions similarly
\begin{equation}\label{eq1.70}
\frac{dw_{GT}}{d\Omega_{\bar\nu}} = {\rm const.} \times (1+a_{GT}
\beta_e \cos\vartheta)
\end{equation}
with
\begin{equation}\label{eq1.71}
a_{GT}=-\frac 13 \frac{C_A^2(1+\alpha_A^2)-4 C_T^2(1+\alpha_T^2)}
         {C_A^2(1+\alpha_A^2)+4 C_T^2(1+\alpha_T^2)}
\end{equation}
It should be stressed that the angular correlations (\ref{eq1.68})
or (\ref{eq1.70}) resp. have nothing to do with a possible parity
violation: the $\cos\vartheta$ in the considered case is
determined by a scalar product of the particle {\it momenta},
which of course is a ${\cal P}$-even quantity (it is also seen
that the correlation coefficients do not vanish for $\alpha_j =
0$). The relations (\ref{eq1.69}) and (\ref{eq1.71}) make it
obvious that a pure $S$ coupling in F transitions would lead to
the angular correlation coefficient $a_S=-1$, while the $V$
coupling gives $a_V=+1$. Similarly, for the GT transitions, a pure
$A$ coupling would produce $a_A=-\frac 13$, while the $T$ coupling
yields $a_T=+\frac 13$.

   The available experimental data show that $a_F \doteq 1$ while
$a_{GT} \doteq -\frac 13$, within some $1 \%$ -- $10 \%$ accuracy
(examples of typical numbers can be found e.g. in \cite{Ren} or
\cite{CoB}). One may interpret this as an indication that the
underlying theory of weak interactions yields $a_F = 1$ and
$a_{GT} = -\frac 13$ {\it exactly\/}. Assuming this, the relations
(\ref{eq1.69}) and (\ref{eq1.71}) then immediately imply
\begin{equation}\label{eq1.72}
C_S = 0, \qquad C_T = 0
\end{equation}
Note that (\ref{eq1.72}) also automatically satisfies the conditions
(\ref{eq1.57}), (\ref{eq1.58}); the absence of $S$ and $T$ couplings
actually provides a simple and natural explanation for vanishing of the
Fierz interference terms.

Of course, at this stage there are other possible interpretations
of the existing data as well: for example, the experimental result
$a_F \doteq 1$ (along with the condition (\ref{eq1.57})) can also
be reproduced if the $C_S$ and $C_V$ are approximately equal, but
$\alpha_S \alpha_V = 1$ with
$\alpha_S \ll \alpha_V$  %%%%!!!
(and similarly for GT transitions). In fact, such a scenario is
excluded by further empirical data to be discussed in the next
section. For the time being, we adopt -- at least tentatively --
the straightforward conclusion (\ref{eq1.72}), which means that a
theoretical description of beta-decay processes can be formulated
in terms of the {\it $V$ and $A$ couplings
alone}.\footnote{Looking back in history, it is amusing to notice
that in 1950s a prevailing opinion was just opposite: the data
available then (mostly before the recognition of parity violation)
seemed to favour $S$ and $T$ couplings. In fact, there was a
period of confusion and the experimental situation was only
clarified in the late 1950s and early 1960s, in a remarkable
interplay with some successful theoretical conjectures formulated
at that time -- this theme we shall discuss later on.} Let us
emphasize once again that our -- rather dramatic -- conclusion
(\ref{eq1.72}) is based on an idealization of the existing
empirical data; we simply interpret the real data as a {\it strong
evidence\/} in favour of an effective theory of weak interactions
dominated by $V$ and $A$ couplings, without seeking an optimum fit
for all possible parameters in (\ref{eq1.27}).

     Now it remains to determine the parameters $\alpha_V$ and $\alpha_A$,
which characterize the non-invariance of our effective Lagrangian
under space reflection\index{space reflection}. To this end, one
has to examine phenomena which manifest directly the parity
violation in beta-decay processes.

%\input{kniha16}  %            1.6
%%%%%%%%%%%%%%%%%%%%%%%%%%%%%%%%%%%%%%%%%%%%%%%%%%%%%%%%%%%%%%%%%%%
%%%%%%%%%%%%%%%%%%%%%%%%%%%%%%%%%%%%%%%%%%%%%%%%%%%%%%%%%%%%%%%%%%%%%%%%%%%%%%%%%%%%%%%%%%%%%%%%%%%%%%%%%%%%%%%%%%%%%%%%%%%%%%%%%%%%%%%%
%\documentstyle[12pt]{report}
%\begin{document}
%\newcommand{\detr}{\mathop{\rm det'}\nolimits}
%\newcommand{\Tr}{\mathop{\rm Tr}\nolimits}
%\newcommand{\asym}{\mathop{\rm asym}\nolimits}
%\newcommand{\Pexp}{\mathop{\rm P exp}\nolimits}
%\newcommand{\antiPexp}{\mathop{\rm \bar P exp}\nolimits}
%\newcommand{\Ker}{\mathop{\rm Ker}\nolimits}
%\newcommand{\re}{\mathop{\rm Re}\nolimits}
%\newcommand{\im}{\mathop{\rm Im}\nolimits}
%\newcommand{\asympt}{\mathop{\sim}}
%\newcommand{\leftpartial}{\mathop{\stackrel{\leftarrow}{\partial}}\nolimits}
%\newcommand{\rightpartial}{\mathop{\stackrel{\rightarrow}{\partial}}\nolimits}
%\newcommand{\leftD}{\mathop{\stackrel{\leftarrow}{D}}\nolimits}
%\newcommand{\st}{\mathop{\rm st}\nolimits}

%\let\eps = \varepsilon

\section{Longitudinal polarization of electrons}\label{sec1.6}

\index{polarization!of the electron|(}

  There are several observable quantities that may reveal parity
violation in weak interactions;  a detailed account of the
relevant experiments can be found e.g. in \cite{Adv}.
Historically, the first example was provided by the celebrated
experiment of C.~S.~Wu et al. \cite{ref4} who measured the angular
correlation between electron momentum and nuclear spin in the beta
decay of the polarized nucleus of ${\rm Co}^{60}$ (recall that the
scalar product of a spin and a momentum is certainly a ${\cal
P}$-odd quantity). We will discuss this type of angular
correlation (for the free neutron) later on and now let us examine
another parity-violating observable, which can provide the desired
information about the parameters $\alpha_j$ in a very
straightforward and efficient way. The quantity we have in mind is
the {\it degree of polarization of the electrons\/} (or positrons)
produced in the beta decay of an {\it unpolarized\/} nucleon
system. In particular, one may consider longitudinal polarization
(helicity)\index{helicity}\index{polarization!longitudinal} and
study a relative difference between the rates of emission of a
right-handed  and a left-handed electron. To put it explicitly,
the degree of longitudinal polarization is defined
as\index{polarization!degree of}
\begin{equation}\label{eq1.73}
P_e = \frac{N_R - N_L}{N_R + N_L}
\end{equation}
where the $N_R$ and $N_L$ denote the number of emitted electrons with
positive and negative helicity resp. For a given energy $E_e$ this can
be calculated in terms of the corresponding differential decay rates
\begin{equation}\label{eq1.74}
P_e = \frac{dw_R(E_e) - dw_L(E_e)}{dw_R(E_e) + dw_L(E_e)}
\end{equation}
Intuitively, it should be clear that such a quantity, if non-zero,
is a direct manifestation of parity violation in weak
interactions. First of all, one should recall that the space
inversion transforms a right-handed electron into a left-handed
one. An asymmetry between different spin states of the final
particles could in fact occur simply as a consequence of the
angular momentum conservation if the initial nucleon system is
polarized (i.e. if it has a well-defined spin projection). If an
asymmetry between $N_R$ and $N_L$ appears in the case of
unpolarized initial nucleons, it can be only accounted for by an
intrinsic parity violation (the corresponding interaction is able
to distinguish between \qq{right} and \qq{left}).

Let us now calculate the quantity (\ref{eq1.74}), starting from
the matrix elements (\ref{eq1.33}); in view of our preceding
results we will consider now only the vector ($V$) and
axial-vector ($A$) couplings\index{axial vector!coupling}. We can
perform the calculation for the $V$ and $A$ terms separately,
since we know that in the corresponding decay rate there is no
interference between the Fermi and Gamow--Teller parts of the
amplitude in case of unpolarized nucleons.

Let us start with the $V$ term. The matrix element corresponding to the
emission of a right-handed electron is written as
\begin{equation}\label{eq1.75}
{\cal M}_V^R = C_V ( U_p^\dagger U_n)
[\bar u_{eR} (1+\alpha_V \gamma_5) \gamma_0 v_\nu]
\end{equation}
where the right-handed Dirac spinor $u_{eR}$ satisfies
\begin{equation}\label{eq1.76}
u_{eR}(k) \bar u_{eR}(k) = (\slashed{k} + m_e) \frac{1+\gamma_5
\slashed{s}_R}{2}
\end{equation}
with $s_R$ being the longitudinal spin four-vector
\begin{equation}\label{eq1.77}
s_R^\mu(k) = \left( \frac{|\vec{k}|}{m_e},
\frac{E_e}{m_e} \frac{\vec{k}}{|\vec{k}|} \right)
\end{equation}
(see (\ref{eqA.67})). The corresponding matrix element for a
left-handed electron is obtained by replacing the $s_R$ by
\begin{equation}\label{eq1.78}
s_L = - s_R
\end{equation}
Looking back at our calculation of the electron energy spectrum
carried out in Section~\ref{sec1.4}, it is easy to realize that the ratio
of differential decay rates shown in (\ref{eq1.74}) is in fact
equal to
\begin{equation}\label{eq1.79}
P_e^{(V)} =\frac{\int \limits_{-1}^1 d(\cos\vartheta) \left(
\overline{|{\cal M}_V^R|^2} - \overline{|{\cal M}_V^L|^2}
\right)}{\int \limits_{-1}^1 d(\cos\vartheta) \left(
\overline{|{\cal M}_V^R|^2} + \overline{|{\cal M}_V^L|^2}
\right)}
\end{equation}
where the bar over a $|{\cal M}|^2$ now indicates summing over
the $n, p, \bar\nu$ spins and averaging over the initial neutron spin;
$\vartheta$ denotes, as usual, an angle between the electron and antineutrino
directions. Employing the standard trace techniques (including in
particular the relations (\ref{eq1.76}) and (\ref{eq1.78})), the integrand
of the numerator in (\ref{eq1.79}) becomes, after some algebra
\begin{eqnarray}\label{eq1.80}
X_V &\equiv& \overline{|{\cal M}_V^R|^2} - \overline{|{\cal
M}_V^L|^2}
\nonumber\\
&=& 4 M^2 C_V^2 \Tr[(\slashed{k}+m_e)\gamma_5\slashed{s}_R
(1+\alpha_V \gamma_5) \gamma_0 \slashed{k}' (1+\alpha_V \gamma_5)
\gamma_0 ]
\end{eqnarray}
while the denominator has in fact been calculated before -- it is
precisely the $V$ part of the result (\ref{eq1.45}), namely
\begin{equation}\label{eq1.81}
\overline{|{\cal M}_V^R|^2} + \overline{|{\cal M}_V^L|^2}
= 16 M^2 E_e E_{\bar\nu} C_V^2 (1+\alpha_V^2)
(1+\beta_e\cos\vartheta)
\end{equation}
Using the well-known properties of Dirac matrices, the expression
(\ref{eq1.80}) can be further simplified to
\begin{eqnarray}\label{eq1.82}
X_V &=&  4 M^2 C_V^2 (-2\alpha_V) m_e \Tr(\slashed{s}_R \gamma_0
\slashed{k}'\gamma_0)
\nonumber\\
&=&  16 M^2 C_V^2 (-2\alpha_V) m_e
(2s_R^0 E_{\bar\nu} - s_R \cdot k')
\end{eqnarray}
Using now the explicit expression for the $s_R$ (see (\ref{eq1.77})),
one gets readily
\begin{eqnarray}\label{eq1.83}
X_V &=&  16 M^2 C_V^2 (-2\alpha_V) m_e
\left( \beta_e \frac{1}{m_e} E_e E_{\bar\nu} +
\frac{E_e}{m_e} E_{\bar\nu} \cos\vartheta \right)
\nonumber\\
&=&  16 M^2 E_e E_{\bar\nu} C_V^2 (-2\alpha_V)
( \beta_e + \cos\vartheta )
\end{eqnarray}
The terms in (\ref{eq1.81}) and (\ref{eq1.83}), proportional to
$\cos\vartheta$, obviously vanish upon the angular integration indicated in
(\ref{eq1.79}) and we thus finally obtain
\begin{equation}\label{eq1.84}
P_e^{(V)} = -\frac{2 \alpha_V}{1+\alpha_V^2} \beta_e
\end{equation}

For the axial-vector coupling\index{axial vector!coupling} one can
proceed in a similar way. After some simple manipulations one gets
first
\begin{eqnarray}\label{eq1.85}
X_A &\equiv&
\overline{|{\cal M}_A^R|^2} - \overline{|{\cal M}_A^L|^2}\\
&=& 2 M^2 C_A^2 \Tr(\sigma_j\sigma_k)
\Tr[(\slashed{k}+m_e)\gamma_5\slashed{s}_R \gamma^j \slashed{k}'
\gamma^k (1+\alpha_A^2-2\alpha_A\gamma_5)] \nonumber
\end{eqnarray}
and using the familiar trace identities this is simplified to
\begin{equation}\label{eq1.86}
X_A = 4 M^2 C_A^2 (-2\alpha_A) m_e \Tr(\slashed{s}_R \gamma^j
\slashed{k}'\gamma^j)
\end{equation}
Working out the last trace, one obtains
\begin{eqnarray}\label{eq1.87}
X_A &=&  16 M^2 C_A^2 (-2\alpha_A) m_e
(2\vec{s}_R\cdot\vec{k}' - g^{jj} s_R\cdot k')
\nonumber\\
&=&  16 M^2 E_e E_{\bar\nu} C_A^2 (-2\alpha_A)
( 3\beta_e + \cos\vartheta )
\end{eqnarray}
On the other hand, a corresponding result for the sum over electron
helicities can be retrieved from (\ref{eq1.46}); it reads
\begin{equation}\label{eq1.88}
\overline{|{\cal M}_A^R|^2} + \overline{|{\cal M}_A^L|^2}
= 16 M^2 E_e E_{\bar\nu} C_A^2 (1+\alpha_A^2)
(3-\beta_e\cos\vartheta)
\end{equation}
Now, the degree of electron polarization can again be evaluated as
\begin{equation}\label{eq1.89}
P_e^{(A)} =\frac{\int \limits_{-1}^1 d(\cos\vartheta) \left(
\overline{|{\cal M}_A^R|^2} - \overline{|{\cal M}_A^L|^2}
\right)}{\int \limits_{-1}^1 d(\cos\vartheta) \left(
\overline{|{\cal M}_A^R|^2} + \overline{|{\cal M}_A^L|^2}
\right)}
\end{equation}
Thus, inserting into (\ref{eq1.89}) the expressions (\ref{eq1.87}) and
(\ref{eq1.88}), we get immediately
\begin{equation}\label{eq1.90}
P_e^{(A)} = -\frac{2 \alpha_A}{1+\alpha_A^2} \beta_e
\end{equation}
i.e. a result completely analogous to that obtained for the vector
coupling.

Formally, we have performed our calculation for a free neutron, but the
results (\ref{eq1.84}) and (\ref{eq1.90}) are in fact valid also for
pure F and GT allowed nuclear beta decays resp. -- the energy-dependence
of the quantity in question is determined solely by the leptonic factor
of  the matrix element and a constant factor coming from nucleons drops
out from the ratio (\ref{eq1.79}) or (\ref{eq1.89}) resp. Note also that
the full answer in the free-neutron case obviously reads
\begin{equation}\label{eq1.91}
P_e^{(V,A)} = -\beta_e \frac{2\alpha_V C_V^2 + 2\alpha_A \,3C_A^2}
{(1+\alpha_V^2) C_V^2+(1+\alpha_A^2) \,3C_A^2}
\end{equation}
since the neutron decay is a mixed transition\index{mixed
transition} and, as noted before, there is no F-GT interference
for the considered observable.

Now we are in a position to confront our theoretical results with empirical
data. Various measurements of beta-electron helicities (for both the F
and GT nuclear transitions) show that -- over a wide energy range and
with a rather high accuracy -- the degree of electron longitudinal
polarization is simply related to its velocity:
\begin{equation}\label{eq1.92}
P_e^{(exp)} = -\beta_e
\end{equation}
(for an overview of the data we refer the reader e.g. to
\cite{CoB}). The remarkable result (\ref{eq1.92}) means, among
other things, that highly relativistic beta electrons are almost
completely polarized, being predominantly {\it left-handed\/}.
Comparing (\ref{eq1.92}) with the formulae (\ref{eq1.84}),
(\ref{eq1.90}) we may then conclude that
\begin{equation}\label{eq1.93}
\alpha_V = 1, \quad \alpha_A =1
\end{equation}
We thus see that the parity-violating effects due to weak
interactions are substantial; the considered quantity in fact
reaches its maximum possible value (obviously, the function
$2\alpha/(1+\alpha^2)$ has a maximum for $\alpha=1$). For this
reason, it is usually said that weak interactions exhibit {\it
maximal parity violation}; note that (\ref{eq1.93}) also means
that the ${\cal P}$-even terms and their ${\cal P}$-odd
counterparts contained in the Lagrangian (\ref{eq1.27}) have an
equal strength.

When an analogous calculation is carried out for positrons (i.e. starting
from the h.c. part of (\ref{eq1.27})), one finds that the overall sign
in the relevant results is reversed (a verification of this statement
is recommended to the reader as an instructive exercise). The experimental
data, though less ample and less accurate than those for electrons, show
that indeed
\begin{equation}\label{eq1.94}
P_{positron} = +\beta_{positron}
\end{equation}
(cf. \cite{CoB}), i.e. positrons emitted in beta-decay processes
are mostly right-handed at relativistic velocities. The Lagrangian
(\ref{eq1.27}) with $\alpha_V = \alpha_A = 1$  thus provides a
very good description of the data from longitudinal polarization
measurements for both electrons and positrons.

The spectacular result (\ref{eq1.93}) provides a very important
piece of information in our search for a realistic effective
theory of beta decay. At present we are left with two parameters
(coupling constants) $C_V$ and $C_A$\index{coupling
constants!weak|(} that remain to be determined from some further
empirical data. We will complete this task later on -- an
impatient reader may pass immediately to Section~\ref{sec1.8}. However, now
we would like to pause for a moment, and mention a possible
modification of the procedure that has led us to our present
position. In particular, we might interchange the last two steps:
instead of eliminating plainly the $S$ and $T$ couplings on the
basis of the $e - \bar\nu$ angular correlation data, we might
examine the longitudinal polarization first, keeping for the
moment all the parameters $C_S,C_V,C_A,C_T$ in the game. Of
course, in any case we have to account for the observed absence of
the Fierz interference; this can be simply achieved by assuming
\begin{equation}\label{eq1.95}
1-\alpha_S\alpha_V = 0, \quad 1-\alpha_A\alpha_T = 0
\end{equation}
(cf. (\ref{eq1.45}), (\ref{eq1.46}))\index{Fierz interference
term|)}. Let us consider neutron decay, where all types of
couplings may contribute. The evaluation of the degree of electron
polarization starts from the full matrix element ${\cal M}_S+{\cal
M}_V+{\cal M}_A+{\cal M}_T$ (see (\ref{eq1.33})) and proceeds
along similar lines as before. As we already know, for the
considered quantity one need not worry about an F-GT interference;
moreover, if one makes use of the conditions (\ref{eq1.95}), the
$S$-$V$ and $A$-$T$ interference terms turn out to vanish
completely as well. The final result reads
\begin{equation}\label{eq1.96}
P_e = -\beta_e \frac{2\alpha_S C_S^2 + 2\alpha_V C_V^2 +
2\alpha_A \,3C_A^2  + 2\alpha_T \,12C_T^2}
{(1+\alpha_S^2) C_S^2+(1+\alpha_V^2) C_V^2+(1+\alpha_A^2) \,3C_A^2
+(1+\alpha_T^2) \,12C_T^2}
\end{equation}
A detailed derivation of the last expression is left to an interested
reader as an instructive (though somewhat tedious) exercise. From
(\ref{eq1.96}) it is also easy to guess the corresponding answers for
pure F and GT nuclear transitions. Comparing now our theoretical formula
(\ref{eq1.96}) with the experimental observation (\ref{eq1.92}), one gets
readily the condition
\begin{equation}\label{eq1.97}
C_S^2 (1-\alpha_S)^2 +C_V^2 (1-\alpha_V)^2 + 3C_A^2 (1-\alpha_A)^2
+12C_T^2 (1-\alpha_T)^2 = 0
\end{equation}
Obviously, if one wants to keep momentarily all the $C_j$ non-zero,
the last relation can only be satisfied if
\begin{equation}\label{eq1.98}
\alpha_S = \alpha_V = \alpha_A =  \alpha_T =1
\end{equation}
(note also that the conditions (\ref{eq1.95}) are then fulfilled
\qq{trivially}). Equipped with this knowledge, we may reconsider
the $e-\bar\nu$ angular correlations. Remembering the formulae
(\ref{eq1.69}) and (\ref{eq1.71}), it is obvious that the relevant
experimental data along with the values of the parameters
$\alpha_j$ shown in (\ref{eq1.98}) force us to set
\begin{equation}\label{eq1.99}
C_S=0, \quad C_T= 0
\end{equation}
in accordance with the option chosen tentatively in Section~\ref{sec1.5}.
In particular, as a by-product of our analysis one can see -- as
we have promised before -- that e.g. a pattern with $C_S \doteq
C_V$, $\alpha_S \ll \alpha_V$ is clearly excluded by the available
empirical data on the electron longitudinal
polarization\index{polarization!of the electron|)}.

%\end{document}

%\input{kniha17}  %            1.7
%%%%%%%%%%%%%%%%%%%%%%%%%%%%%%%%%%%%%%%%%%%%%%%%%%%%%%%%%%%%%%%%%%%
%%%%%%%%%%%%%%%%%%%%%%%%%%%%%%%%%%%%%%%%%%%%%%%%%%%%%%%%%%%%%%%%%%%%%%%%%%%%%%%%%%%%%%%%%%%%%%%%%%%%%%%%%%%%%%%%%%%%%%%%%%%%%%%%%%%%%%%%
\section{Neutrino helicity}\index{helicity!of neutrino|ff}\label{sec1.7}

In view of the preceding arguments, the original Lagrangian
(\ref{eq1.27}) is now effectively reduced to
\begin{eqnarray}
\label{eq1.100}
\lagr^{(\beta)}_{int}&=&C_V(\bar\psi_p\gamma_\mu\psi_n
[\bar\psi_e(1+\gamma_5)\gamma^\mu\psi_\nu]\nonumber\\
&+&C_A(\bar\psi_p\gamma_5\gamma_\mu\psi_n)[\bar\psi_e
(1+\gamma_5)\gamma^\mu\psi_\nu]
\end{eqnarray}
(notice that one $\gamma_5$ factor in the leptonic part of the
axial-vector term has been absorbed into the $1+\gamma_5$
because of $\gamma^2_5=1$). The Hermitean conjugate of
(\ref{eq1.100}) reads
\begin{eqnarray}
\label{eq1.101}
\lagr^{(\beta)\dagger}_{int}&=&C_V(\bar\psi_n\gamma_\mu\psi_p)
[\bar\psi_\nu(1+\gamma_5)\gamma^\mu\psi_e]\nonumber\\
&+&C_A(\bar\psi_n\gamma_5\gamma_\mu\psi_p)
[\bar\psi_\nu(1+\gamma_5)\gamma^\mu\psi_e]
\end{eqnarray}
It is easy to see that the form (\ref{eq1.100}) or (\ref{eq1.101})
resp. gives a definite {\it prediction\/} for the helicity of the
antineutrino or neutrino resp. Indeed, making use of the
$\gamma_5$ anticommutativity, the matrix element for $n\rightarrow
p+e^-+\bar\nu$ corresponding to (\ref{eq1.100}) can obviously be
written as\footnote{Throughout this section, we don't need to use
the non-relativistic approximation for nucleons.}
\begin{eqnarray}
\label{eq1.102}
{\cal M}^{(\bar\nu)}_{fi}&=&C_V(\bar u_p\gamma_\mu u_n)[\bar
u_e\gamma^\mu(1-\gamma_5)v_\nu]\nonumber\\
&+&C_A(\bar u_p\gamma_5\gamma_\mu u_n)[\bar u_e\gamma^\mu
(1-\gamma_5)v_\nu]
\end{eqnarray}
and, similarly, for an inverse process $p\rightarrow n+e^++\nu$
one gets from (\ref{eq1.101})
\begin{eqnarray}
\label{eq1.103}
{\cal M}^{(\nu)}_{fi}&=&C_V(\bar u_n\gamma_\mu u_p)[\bar
u_\nu(1+\gamma_5)\gamma^\mu v_e]\nonumber\\
&+&C_A(\bar u_n\gamma_5\gamma_\mu u_p)[\bar
u_\nu(1+\gamma_5)\gamma^\mu v_e]
\end{eqnarray}
Now it is obvious that only right-handed antineutrino can be
emitted: indeed, the $v_L$ satisfies
$v_L=\frac{1}{2}(1+\gamma_5)v_L$ in the massless case, so that the
factor $1-\gamma_5$ contained in (\ref{eq1.102}) makes it vanish.
On the other hand,  the $v_R$, satisfying
$v_R=\frac{1}{2}(1-\gamma_5)v_R$ clearly survives in
(\ref{eq1.102}). In other words, the $(V,A)$ structure of the
interaction and the presence of the factor $1+\gamma_5$ in
(\ref{eq1.100}) (enforced by the empirical data on the {\it
electron\/} helicity) together lead to a definite prediction for
the value of antineutrino helicity. In a similar way, from
(\ref{eq1.103}) it is seen that the neutrino should always be
produced as left-handed: indeed, one has
\begin{eqnarray}
\label{eq1.104}
u_L&=&\frac{1-\gamma_5}{2}u_L\;\;\Rightarrow\;\;\bar u_L=\bar u_L
\frac{1+\gamma_5}{2}\nonumber\\
u_R&=&\frac{1+\gamma_5}{2}u_R\;\;\Rightarrow\;\;\bar u_R=\bar u_R
\frac{1-\gamma_5}{2}
\end{eqnarray}
and hence only $u_L$ can survive in (\ref{eq1.103}).

To verify the above predictions experimentally is an extremely
difficult task, since the neutrino has no electromagnetic
interactions and hence its helicity cannot be measured directly
(as e.g. that of an electron or photon). Nevertheless, one
(indirect) measurement does exist -- it has been accomplished in
an ingenious experiment by M.~Goldhaber et al. \cite{ref6}. The
process investigated in \cite{ref6} was essentially
$e^-+p\rightarrow n+\nu$. In particular, Goldhaber et al. studied
the capture of an electron from an inner atomic orbit in
Eu$^{152}(0^-)$, which produces an excited state Sm$^{152*}(1^-)$
and a neutrino is emitted (this particular reaction was chosen
because of some exceptionally favourable properties of the nuclei
involved). The neutrino helicity can then be deduced from the spin
and momentum of the daughter nucleus; this is accomplished through
a measurement of the circular polarization of the
photon\index{polarization!of the photon} emitted (in deexcitation
of the samarium nucleus) along the direction of flight of the
Sm$^{152*}$. More details of this unique experiment are described
in many places; see e.g. \cite{CaG}, \cite{Gre} and, in
particular, \cite{Tel}. Goldhaber et al. found that the neutrino
was always emitted with negative helicity, i.e. left-handed, which
confirms the prediction given above.

Such an independent check of our effective beta-decay theory is
gratifying, but we should perhaps add one more remark concerning
the importance of the measurement of neutrino helicity. Imagine
that we have already exploited the data concerning electron
helicity (longitudinal polarization), but all the couplings
$S,V,A,T$ are still preserved in the effective Lagrangian -- in
other words, we set $\alpha_S=\alpha_V=\alpha_A=\alpha_T=1$, but
ignore temporarily the available data on the $e-\bar\nu$ angular
correlations. The process studied by Goldhaber et al. \cite{ref6}
is a pure GT transition (notice the spin assignments of the parent
and daughter nuclei), so that both $A$ and $T$ couplings can
contribute to the relevant matrix element. It is easy to see that
a measurement of the neutrino helicity provides, in fact, a
clear-cut test of the type of the coupling responsible for the
beta transition in question. Indeed, using the Hermitean conjugate
term in (\ref{eq1.27}) with the particular values of the
parameters, a general GT matrix element for $e^- +p\rightarrow
n+\nu$ can formally be written as
\begin{eqnarray}
\label{eq1.105}
{\cal M}_{GT}&=&C_A(\bar u_n\gamma_5\gamma_\mu u_p)[\bar u_\nu(1+\gamma_5)
\gamma^\mu u_e]\nonumber\\
&+&C_T(\bar u_n\sigma_{\mu\nu}u_p)[\bar
u_\nu(1-\gamma_5)\sigma^{\mu\nu}u_e]
\end{eqnarray}
Of course, the appearance of the $1+\gamma_5$ and $1-\gamma_5$ in
the $A$ and $T$ terms resp. is due to the different commutation
properties of the Dirac matrices involved: the $\gamma^\mu$
anticommutes with $\gamma_5$ while the $\sigma^{\mu\nu}$ commutes.
Now, taking into account (\ref{eq1.104}), it is clear that
neutrino helicity clearly distinguishes between the $A$ and $T$
couplings: neutrinos produced through the $A$ coupling are purely
left-handed (the observed case), while the $T$ coupling would
yield right-handed ones. These considerations can be easily
generalized to the Fermi transitions -- the $V$ coupling, as noted
before, can only produce left-handed neutrinos while those due to
an $S$ coupling would be right-handed. Again, such a
\qq{dichotomy} is simply related to the commutation properties of
the corresponding matrix structures. However, one should keep in
mind that for Fermi transitions, there is no corresponding
measurement of the neutrino helicity.

It is important to realize that the above conclusions concerning
neutrino helicity and the possible algebraic types of the relevant
couplings are intimately related to the empirical data on electron
helicity, which tell us that relativistic beta-electrons are
left-handed (precisely this fact has led us to set
$\alpha_S=\alpha_V=\alpha_A=\alpha_T=1$). Clearly, a pattern which
thus emerges is the following. The presence of $e_L$ and $\nu_L$
reveals a $(V,A)$ structure of the underlying effective theory,
while the combination of $e_L$ and $\nu_R$ would correspond to an
$(S,T)$ model; other equivalent variants are obvious.

Historically, the measurement of neutrino helicity played a very
important role in determining the right form of the beta-decay
effective Lagrangian (at least for its GT part). Before the advent
of parity violation, there were some controversial results
concerning the $e-\bar\nu$ angular correlation in GT transitions,
which preferred the $T$, rather than $A$ coupling (in this context
see, in particular, the paper by B. M. Rustad and S. L. Ruby,
Phys.\ Rev.\ 97 (1955) 991, dealing with the decay of
He$^6$\index{decay!of He$^6$}). The helicity measurements for
electron and neutrino, which followed the discovery of parity
violation, provided a powerful argument in favour of the $A$
coupling. In any case -- in view of the absence of a measurement
of neutrino helicity in Fermi (or mixed) transitions -- it is
gratifying that the relevant data on the $e-\bar\nu$ angular
correlation for both F and GT transitions now support the $(V,A)$
effective theory.

%\input{kniha18}  %            1.8
%%%%%%%%%%%%%%%%%%%%%%%%%%%%%%%%%%%%%%%%%%%%%%%%%%%%%%%%%%%%%%%%%%%
%%%%%%%%%%%%%%%%%%%%%%%%%%%%%%%%%%%%%%%%%%%%%%%%%%%%%%%%%%%%%%%%%%%%%%%%%%%%%%%%%%%%%%%%%%%%%%%%%%%%%%%%%%%%%%%%%%%%%%%%%%%%%%%%%%%%%%%%
\section{The $V$ and $A$ coupling constants}\label{sec1.8}

Let us now show how the remaining free parameters in the
Lagrangian (\ref{eq1.100}), namely the coupling constants $C_V$ and
$C_A$, can be determined. Our earlier results (\ref{eq1.45}) and
(\ref{eq1.67}) imply that within the effective theory described by
(\ref{eq1.100}) (and within the usual non-relativistic approximation),
the spin-averaged squared matrix element for neutron decay
becomes
\begin{eqnarray}
\label{eq1.106} \overline{|{\cal M}|^2}=\overline{|{\cal M}_V|^2}+
\overline{|{\cal M}_A|^2}=32M^2E_eE_{\bar\nu}[C^2_V+
3C^2_A+(C^2_V-C^2_A)\beta_e\cos\vartheta]\nonumber\\
\end{eqnarray}
The last expression clearly indicates that a measurement of the
$e-\bar\nu$ angular correlation in the free neutron decay could
fix at least the ratio of the coupling constants squared. Indeed,
denoting
\begin{equation}
\label{eq1.107}
f=C_A/C_V
\end{equation}
the angular distribution corresponding to (\ref{eq1.106}) can obviously
be written as
\begin{equation}
\label{eq1.108}
\frac{dw}{d(\cos\vartheta)}=\text{const.}\times(1+a_n\beta_e\cos
\vartheta)
\end{equation}
with
\begin{equation}
\label{eq1.109}
a_n=\frac{1-f^2}{1+3f^2}
\end{equation}
The experimental value of the correlation coefficient is
$a_n=-0.1049\pm0.0013$ (the weighted world average \cite{ref5}).
Using this in (\ref{eq1.109}), we get roughly
\begin{equation}
\label{eq1.110} | f|\doteq1.27
\end{equation}
i.e. the $V$ and $A$ couplings turn out to be of comparable, yet
unequal, strength. Note that such a closeness of the $C_V$ and
$C_A$ is essentially accidental -- we will comment on this
point in the next chapter.

Of course, the data on the particular angular correlation
considered so far can only provide information on the absolute
value of the ratio $f$, since (\ref{eq1.106}) does not involve any
interference between the $V$ and $A$ couplings -- as we know, this
is a general feature of the observable quantities calculated for
unpolarized nucleons. Thus, in order to find the sign of the $f$,
one obviously has to exploit an observable related to {\it
polarized\/} nucleons. In particular, a suitable experimentally
accessible quantity is the angular correlation between electron
momentum and neutron spin in the decay of a polarized neutron. As
we have noted at the beginning of Section~\ref{sec1.6}, such an angular
correlation represents a parity-violating effect, so it would
perhaps be also instructive to demonstrate this aspect explicitly
in the result of our calculation. For this purpose, let us restore
temporarily arbitrary parameters $\alpha_V$ and $\alpha_A$ in our
effective Lagrangian; it means that we start the calculation from
the matrix element
\begin{eqnarray}
\label{eq1.111}
{\cal M}&=&C_V(U^\dagger_pU_n)[\bar u_e(1+\alpha_V\gamma_5)
\gamma_0v_\nu]\nonumber\\
&+&C_A(U^\dagger_p\sigma_j U_n)[\bar u_e(1+\alpha_A\gamma_5)
\gamma_5\gamma^jv_\nu]
\end{eqnarray}
The coordinate system can be conventionally chosen so that the
initial neutron spin is directed along the third axis. For a
practical calculation it then implies that
\begin{equation}
\label{eq1.112}
U_nU^\dagger_n=2M\frac{1+\sigma_3}{2}
\end{equation}
The evaluation of the matrix element squared is somewhat tedious
and we have therefore relegated the technical details to the
Appendix~\ref{appenC}. Here let us quote only the result; it reads
\begin{eqnarray}
\label{eq1.113}
\int\frac{d\Omega_{\bar\nu}}{4\pi}\sum_{spin\;p,e,
\bar\nu}|{\cal M}|^2
&=&16M^2E_eE_{\bar\nu}\bigl[C^2_V(1+\alpha^2_V)+3C^2_A(1+\alpha^2_A)
\nonumber\\
&+&\bigl(2C_VC_A(\alpha_V+\alpha_A)-4\alpha_AC^2_A\bigr)\beta_e\cos
\theta_e\bigr]
\end{eqnarray}
where $\theta_e$ denotes the angle between the electron momentum
and neutron spin (i.e. the polar angle for the electron direction,
in our coordinate frame). Now the parity-violating nature of the
considered angular dependence should be obvious -- as expected,
the term involving the $\cal{P}$-odd $\cos\theta_e$ is
proportional to the parameters $\alpha_V,\alpha_A$ and thereby it
is trivial for $\alpha_V=\alpha_A=0$. Another remarkable feature
of the result (\ref{eq1.113}) is that the coefficient at
$\cos\theta_e$ also vanishes for $C_A=0$ (for arbitrary values of
$\alpha_V,\alpha_A$); in other words, the effect would be trivial
for a pure F transition (recall that C.~S.~Wu et al. in their
celebrated experiment \cite{ref4} measured the angular
distribution of the considered type for a pure GT transition
Co$^{60}\rightarrow$Ni$^{60}$).

Let us now proceed to determine the ratio $f=C_A/C_V$, as
indicated above. Returning to the known values $\alpha_V=\alpha_A=1$,
the expression (\ref{eq1.113}) becomes
\begin{eqnarray}
\label{eq1.114}
\lefteqn{\int\frac{d\Omega_{\bar\nu}}{4\pi}\sum_{spin\;p,e,
\bar\nu}|{\cal M}|^2=}\nonumber\\
&&=32M^2E_eE_{\bar\nu}C^2_V[1+3f^2+2(f-f^2)\beta_e\cos\theta_e]
\end{eqnarray}
The corresponding angular distribution then obviously can be
written as
\begin{equation}
\label{eq1.115}
\frac{dw}{d(\cos\theta_e)}=\text{const.}\times(1+A_n\beta_e\cos
\theta_e)
\end{equation}
with the coefficient $A_n$ given by
\begin{equation}
\label{eq1.116}
A_n=2\frac{f-f^2}{1+3f^2}
\end{equation}
The experimental value of the \qq{$\beta$ asymmetry parameter}
$A_n$ is $=-0.11958\pm0.00021$ (the rounded world average
\cite{ref5}). Using this in the equation (\ref{eq1.116}), one
obtains two solutions for the $f$, namely $f^{(1)}\doteq1.27$ and
$f^{(2)}\doteq-0.06$. Obviously, the latter possibility is not
compatible with our preceding result for the $| f|$ (see
(\ref{eq1.110})). Thus, one may conclude that
\begin{equation}
\label{eq1.117}
f\doteq1.27
\end{equation}
i.e. the coupling constants $C_V$ and $C_A$ have the same sign,
within our system of definitions (the reader should be warned,
however, that a definition of the axial-vector coupling
constant\index{axial vector!coupling} with opposite sign occurs
rather frequently in the literature -- cf. e.g. \cite{ref5}).
Looking now back at the formula (\ref{eq1.113}) (with
$\alpha_V=\alpha_A=1$), it is clear that the F-GT interference
acts \qq{destructively} on the magnitude of the correlation
coefficient in question -- this is one more reason why Wu et al.
\cite{ref4} have chosen a pure GT transition for their
investigation of parity violation. In any case, the value of the
correlation coefficient $A_n$ is negative (similarly to the case
considered in \cite{ref4}), which means that the electrons are
emitted preferentially in the direction opposite to the neutron
spin.

At this place, it is worth noting that the calculation leading to
(\ref{eq1.116}) (see Appendix~\ref{appenC}) can be easily modified to yield
an analogous result for the coefficient of the correlation of
neutron spin and antineutrino momentum. This \qq{$\bar\nu$
asymmetry parameter} comes out to be
\begin{equation}
\label{eq1.118}
B_n=2\frac{f+f^2}{1+3f^2}
\end{equation}
With the known value of the $f$ (fixed by other experiments), the
last result represents a {\it prediction\/} of our effective
beta-decay theory. For $f\doteq 1.27$, one gets from
(\ref{eq1.118}) $B_n\doteq 0.988$, to be compared with the
experimental value $B_n=0.981\pm0.003$ (the weighted world average
\cite{ref5}).

For a complete knowledge of the coupling constants $C_V$ and $C_A$
it is now sufficient to fix the absolute value of one of them by
means of a suitable experiment. Obviously, an appropriate
observable quantity would be a fully integrated decay rate (the
decay width), which determines the mean lifetime $\tau$ of the
neutron\index{lifetime!of the neutron} or of a beta-radioactive
nucleus. Such a decay width is obtained by integrating the
electron energy distribution function over the whole kinematical
range and it obviously comes out to be a linear combination of the
$C^2_V$ and $C^2_A$ with calculable coefficients. (Needless to
say, we have in mind the first order of perturbation theory. In
the case of a nuclear beta transition, the practical calculability
is of course limited by our knowledge of the wave functions of the
nuclei involved.) An elementary example of such an integration is
given in the next section. Thus, any measured lifetime would do,
provided that we are able to carry out a reasonably accurate
theoretical calculation indeed. This is possible e.g. in the case
of a free neutron decay, but in fact the most favourite and
practical method consists in exploiting the pure F transition
O$^{14}\rightarrow$N$^{14*}+e^++\nu$, which occurs within an
isospin multiplet (isotriplet).\footnote{Other examples of this
kind are C$^{10}\rightarrow$ B$^{10}$, Co$^{54}\rightarrow$
Fe$^{54}$ etc. (see \cite{CoB} and \cite{Gre}). Such transitions
are sometimes called super-allowed\index{super-allowed
transition}. The point is that in such a case the nuclear matrix
element is easily calculable -- it is determined by the isospin
lowering or raising operator since the internal structure of the
parent and daughter nuclei is essentially identical, up to small
electromagnetic corrections. An instructive and rather detailed
discussion of the O$^{14}$ decay can be found in \cite{HaM},
Section 12.3.} The lifetime of the O$^{14}$\index{lifetime!of the
O$^{14}$} is known with a very good accuracy (note that the
half-life $T_{1/2}=\tau\ln2$ is about 71s). One thus gets directly
the absolute value of the $C_V$; by convention, the $C_V$ is
expressed in terms of a \qq{beta-decay Fermi constant} $G_\beta$
as
\begin{equation}
\label{eq1.119}
C_V=-\frac{G_\beta}{\sqrt{2}}
\end{equation}
The $G_\beta$ is taken to be positive and the data then yield
\begin{equation}
\label{eq1.120} G_\beta=(1.136\pm0.001)\times10^{-5}\ \GeV^{-2}
\end{equation}
Note that the minus sign in the definition (\ref{eq1.119}) is pure
convention at the present level, but we shall see that it becomes
very natural in the context of weak interaction theory involving
an intermediate vector boson. The factor of $\sqrt{2}$ is of
historical origin -- it serves to reproduce the value of the
coupling constant $G$ appearing in the old parity-conserving
Fermi theory (cf. (\ref{eq1.14})).

Thus, we have got through the determination of the form of an
effective beta-decay Lagrangian. Having fixed the values of all
relevant free parameters, let us now return to the original
relativistic form (\ref{eq1.27}) (with only $V$ and $A$ terms
preserved). Making use of anticommutativity of $\gamma_5$ and the
notation (\ref{eq1.107}), (\ref{eq1.119}), it is easy to see that
the $\lagr^{(\beta)}_{int}$ can now be written as
\begin{equation}
\label{eq1.121}
\lagr^{(\beta)}_{int}=-\frac{G_\beta}{\sqrt{2}}[\bar\psi_p
\gamma_\mu(1-f\gamma_5)\psi_n][\bar\psi_e\gamma^\mu(1-\gamma_5)
\psi_\nu]\;+\text{h.c.}
\end{equation}
Paraphrasing the famous H. Andersen's work \cite{ref7}, one might
say that the original \qq{ugly-duckling form}\index{ugly duckling}
(\ref{eq1.27}) has now matured, through some stringent
experimental tests, to a \qq{swan-like} appearance
(\ref{eq1.121}). In fact, the realistic effective Lagrangian now
in a way resembles the old Fermi model: the vectorial currents of
the Fermi theory are replaced by linear combinations of the $V$
and $A$ currents; in particular, the leptonic part has a pure
structure $V-A$. This remarkable feature of the weak interaction
Lagrangian will be discussed in detail in the next chapter.

%\input{kniha19}  %            1.9
%%%%%%%%%%%%%%%%%%%%%%%%%%%%%%%%%%%%%%%%%%%%%%%%%%%%%%%%%%%%%%%%%%%
%%%%%%%%%%%%%%%%%%%%%%%%%%%%%%%%%%%%%%%%%%%%%%%%%%%%%%%%%%%%%%%%%%%%%%%%%%%%%%%%%%%%%%%%%%%%%%%%%%%%%%%%%%%%%%%%%%%%%%%%%%%%%%%%%%%%%%%%
%\documentstyle[12pt]{report}
%\begin{document}
%\newcommand{\detr}{\mathop{\rm det'}\nolimits}
%\newcommand{\Tr}{\mathop{\rm Tr}\nolimits}
%\newcommand{\asym}{\mathop{\rm asym}\nolimits}
%\newcommand{\Pexp}{\mathop{\rm P exp}\nolimits}
%\newcommand{\antiPexp}{\mathop{\rm \bar P exp}\nolimits}
%\newcommand{\Ker}{\mathop{\rm Ker}\nolimits}
%\newcommand{\re}{\mathop{\rm Re}\nolimits}
%\newcommand{\im}{\mathop{\rm Im}\nolimits}
%\newcommand{\asympt}{\mathop{\sim}}
%\newcommand{\leftpartial}{\mathop{\stackrel{\leftarrow}{\partial}}\nolimits}
%\newcommand{\rightpartial}{\mathop{\stackrel{\rightarrow}{\partial}}\nolimits}
%\newcommand{\leftD}{\mathop{\stackrel{\leftarrow}{D}}\nolimits}
%\newcommand{\st}{\mathop{\rm st}\nolimits}

%\let\eps = \varepsilon

\section{Mean lifetime of the neutron}
\index{lifetime!of the neutron|(}

With the effective Lagrangian (\ref{eq1.121}) at hand, we may now
make a {\it prediction\/} for another physical observable quantity
not exploited within our parameter-fixing procedure. In
particular, we can calculate the total decay rate (decay width)
for the free neutron, which in turn determines the mean lifetime
of such an unstable particle. The decay width is obtained by
integrating the original differential rate (\ref{eq1.35}) over all
kinematical variables of the final-state particles. We have
implemented some of the relevant integration steps in Section~\ref{sec1.4}
when deriving the form of the electron energy
spectrum\index{energy spectrum of the electron}. To apply our
previous results in the case of a free neutron, we may start with
the intermediate result (\ref{eq1.55}) and employ the expression
(\ref{eq1.106}) for the matrix element squared. The integration
over the angular variables is essentially trivial and one thus
arrives at the electron energy spectrum
\begin{equation}\label{eq1.122}
\frac{dw(E_e)}{dE_e} =
\frac{1}{\pi^3}(C_V^2+3C_A^2)\sqrt{E_e^2-m_e^2} E_e
(\Delta-E_e)^2
\end{equation}
which agrees, as expected, with the generic form (\ref{eq1.59}).
The decay width $\Gamma$ is then obtained by means of an
integration over the whole range of electron energies, i.e.
\begin{displaymath}
\Gamma = \int_{m_e}^\Delta \frac{dw(E_e)}{dE_e} dE_e
\end{displaymath}
Using (\ref{eq1.122}), the last expression becomes
\begin{equation}\label{eq1.123}
\Gamma= \frac{1}{\pi^3}(C_V^2+3C_A^2) I_F
\end{equation}
where the symbol $I_F$ stands for the so-called Fermi
integral\index{Fermi!integral}\footnote{ Note that for nuclear
beta transitions the Fermi integral includes also a coulombic
correction factor $F(Z,E_e)$, which may be important especially
for higher atomic numbers $Z$. For more details, see e.g.
\cite{CoB}}
\begin{equation}\label{eq1.124}
I_F = \int_{m_e}^\Delta (\Delta-E)^2 \sqrt{E^2-m_e^2} E dE
\end{equation}
The evaluation of the integral (\ref{eq1.124}) is straightforward and
the result can be written as
\begin{equation}\label{eq1.125}
I_F = \frac{1}{30} \Delta^5 \left(\beta_{max}^5
-\frac 52 \frac{m_e^2}{\Delta^2} \beta_{max}^3
-\frac{15}{2} \frac{m_e^4}{\Delta^4} \beta_{max}
+\frac{15}{2} \frac{m_e^4}{\Delta^4}
\ln\frac{\Delta+\sqrt{\Delta^2-m_e^2}}{m_e}
\right)
\end{equation}
where the $\beta_{max}$ denotes the maximum electron velocity, i.e.
$\beta_{max} = (1-m_e^2/\Delta^2)^{1/2}$
(cf.(\ref{eq1.11})). Numerically, (\ref{eq1.125}) means that
\begin{equation}\label{eq1.126}
I_F = \frac{1}{30} \Delta^5 K
\end{equation}
with $K \doteq 0.46$. Thus, within the effective theory
(\ref{eq1.121}), the decay width of a free neutron is given by a formula
\begin{equation}\label{eq1.127}
\Gamma(n \to p+e^- + \bar\nu) =
K \frac{G_\beta^2 \Delta^5}{60\pi^3} (1+3f^2)
\end{equation}
Putting in numbers, one gets $\Gamma\doteq 6.77\times 10^{-25} \
\MeV$. The mean lifetime is the reciprocal value of the $\Gamma$,
so that $\tau = \Gamma^{-1} \doteq 1.48\times 10^{24}\ \MeV^{-1}$.
Converting this to ordinary units (using $\hbar=6.58\times
10^{-22}\ \MeV\,\text{s}$) one gets finally
\begin{equation}\label{eq1.128}
\tau_n \doteq 974 \ \text{s}
\end{equation}
The experimental value quoted in \cite{ref5} is $(878.4 \pm 0.5)
\,\text{s}$. In order to get from (\ref{eq1.128}) closer to the
experimental result, one should include some additional minor
effects (coulombic and radiative corrections\index{radiative
corrections} in particular), but this would go beyond the scope of
this introductory treatment. Anyway, the agreement between our
simple theoretical prediction and the empirical value (within
about $10 \%$) is quite satisfactory as it stands.

The formula (\ref{eq1.127}) is an example of a rather general
rule\index{rule!G@$G^2 \Delta^5$|(}
\begin{equation}\label{eq1.129}
\Gamma \propto G^2 \Delta^5
\end{equation}
which is highly useful for making the order-of-magnitude estimates
of the decay rates of allowed beta transitions (and of many other
semileptonic decays as well). Let us explain briefly the origin of
such a rule. The characteristic form of the electron energy spectrum
(\ref{eq1.59}) clearly suggests that, at least for
$\Delta \gg m_e$, a dominant contribution to the Fermi integral
(\ref{eq1.124}) amounts to $\Delta^5$ (up to a pure numerical factor).
Indeed, neglecting the $m_e$ in (\ref{eq1.124}), one gets
\begin{eqnarray}\label{eq1.130}
I_F &\doteq& \int_0^\Delta (\Delta-E)^2 E^2 dE
= \Delta^5 \int_0^1 (1-x)^2 x^2 dx
\nonumber\\
&=& \frac{1}{30} \Delta^5
\end{eqnarray}
and the effects of $m_e \ne 0$ are expected to be of a relative order
$O(m_e^2/\Delta^2)$ (cf. (\ref{eq1.125})). On the other hand, the
decay rate must include a factor of $G^2$ (with  $G$ being a pertinent
Fermi-type coupling constant, $G=G_\beta$ for nuclear beta decays), as
the corresponding matrix element is proportional to $G$ when calculated
in the first order of perturbation theory. The product
$G^2\Delta^5$ already has right dimension of a decay width, so any
other factor on the right-hand side of (\ref{eq1.129}) can only be a
dimensionless number.

Of course, the condition $\Delta \gg m_e$ is not always satisfied
sufficiently well (e.g. $m_e/\Delta \doteq 0.4$ for neutron decay)
and there may be some particular extra factors present (as e.g.
the $1+3f^2$ in (\ref{eq1.127})), but for a wide variety of
beta-decay processes the \qq{rule $G^2
\Delta^5$}\index{rule!G@$G^2 \Delta^5$|)} does provide quite
reasonable order-of-magnitude estimates of the lifetimes -- the
point is that the usual corrections to the leading behaviour
(\ref{eq1.129}) do not influence the result dramatically. In fact,
one only has to be careful to take into account properly such
ubiquitous numerical factors as e.g. the $1/(60\pi^3)$ in
(\ref{eq1.127}), since these typically change a naive guess for a
$\Gamma$ by three orders of magnitude. The safest way of including
these numerical effects is to relate the estimated decay rate to
some \qq{reference value} (for which one may take e.g. the neutron
lifetime); the large universal factors cancel when a ratio of
decay rates is taken and one should thus expect a realistic
result, within one order of magnitude or so. To put it in explicit
terms, let us denote quantities referring to an atomic nucleus and
neutron by indices $A$ and $n$ respectively. For the ratio of the
decay rates we have
\begin{equation}\label{eq1.131}
\Gamma_A / \Gamma_n \doteq (\Delta_A / \Delta_n)^5
\end{equation}
(the coupling constants squared are cancelled in the ratio as
well). A mean lifetime $\tau$ is equal to $\Gamma^{-1}$, and
(\ref{eq1.131}) thus implies
\begin{equation}\label{eq1.132}
\tau_A \doteq \tau_n \left(\frac{\Delta_n}{\Delta_A}\right)^5
\end{equation}
Let us now illustrate by some numerical examples how our rule of
thumb (\ref{eq1.132}) works in practice. We will consider two
processes mentioned before, namely the pure GT transition ${\rm
He}^6 \to {\rm Li}^6 + e^- + \bar\nu$ and the pure F transition
${\rm O}^{14} \to {\rm N}^{14^*} + e^+ + \nu$. In the first case
one has $\Delta_A \doteq 2.3 \ \MeV$ (for $\tau_n$ we take
approximately $900 \,{\rm s}$ and $\Delta_n \doteq 1.3 \ \MeV$).
From (\ref{eq1.132}) we then get $\tau_{\rm He^6} \doteq 3.3
\,{\rm s}$; for the corresponding half-life $\tau_{1/2} = \tau \ln
2$ this yields the value of about $2.25 \,{\rm s}$ which is
reasonably close to the value $0.81 \,{\rm s}$ found in tables of
isotopes (see in particular \cite{ref8}). For the ${\rm O}^{14}$
decay one has $\Delta_A \doteq 4 \ \MeV$ and (\ref{eq1.132}) then
yields an estimate $\tau_{1/2} \doteq 36 \,{\rm s}$ which agrees,
as to the order of magnitude (actually within a factor of $2$),
with the measured value $71 \,{\rm s}$. The approximate relation
(\ref{eq1.132}) is thus seen to be quite reliable and we will
appreciate the efficiency of such a rule again in the next
chapter, in connection with semileptonic decays of
baryons\index{baryon} (other than nucleons)\index{coupling
constants!weak|)} and mesons\index{meson}.

In concluding this chapter, let us add a remark on the role that
weak interactions play in our universe in a somewhat broader
context. It is well known that apart from being responsible for
the beta radioactivity of atomic nuclei, the weak interaction of
nucleons and leptons is also crucial for starting up the
thermonuclear reactions occurring in visible stars. In particular,
the \qq{proton burning} process\index{proton burning process}
$p+p\rightarrow p+n+e^++ \nu\rightarrow D+e^++\nu$ (where $D$
denotes the deuteron) constitutes the beginning of a chain of
reactions producing most of the energy radiated by the Sun (see
e.g.\cite{CoB}). Thus, one should bear in mind that the weak
interaction is in fact of immense {\it practical\/} importance --
without it, life on the Earth could not exist in its present form.
In this connection, one may also say that the character of our
environment depends rather dramatically on the weak interaction
strength: the magnitude of the weak coupling
constant\index{coupling constants!weak} determines the rate of
solar energy production and this in turn influences the
temperature of the Earth's atmosphere, the intensity of
ultraviolet radiation etc. For more details, see
\cite{Cah}\index{lifetime!of the neutron|)}.

%\end{document}

%\input{problems1}
%%%%%%%%%%%%%%%%%%%%%%%%%%%%%%%%%%%%%%%%%%%%%%%%%%%%%%%%%%%%%%%%%%%
%%%%%%%%%%%%%%%%%%%%%%%%%%%%%%%%%%%%%%%%%%%%%%%%%%%%%%%%%%%%%%%%%%%%%%%%%%%%%%%%%%%%%%%%%%%%%%%%%%%%%%%%%%%%%%%%%%%%%%%%%%%%%%%%%%%%%%%%
\begin{priklady}{11}
\item  Derive the formula (\ref{eq1.96}).
\item  Derive the formula (\ref{eq1.118}).
\item Using the beta-decay matrix element following directly from
(\ref{eq1.121}) (without making the quasi-static approximation for
proton) one can calculate the proton energy spectrum. Perform such
a calculation and show that the distribution function
$dw(E_p)/dE_p$ vanishes at both ends of the spectrum, i.e. both
for $E_p^\ti{min.}$ and for $E_p^\ti{max.}$.\\{\it Hint}: For the
phase-space integration over the $e$ and $\bar{\nu}$ momenta one
can employ the formulae (\ref{eq2.36}), (\ref{eq2.37}) quoted in
Chapter~\ref{chap2}.
\item Calculate longitudinal polarization of the proton produced in the
decay of a free neutron at rest (employing the same matrix element
as in the preceding problem). The degree of longitudinal
polarization ($P$) is defined in analogy with (\ref{eq1.74}). In
particular, consider the value of the $P(E_p)$ at the endpoint of
the spectrum, $E_p=E_p^\ti{max.}$. Show that for $m_e=0$ the
result is simplified to
$$
P(E^\ti{max.}_p)\Bigl|_{m_e=0} = -\frac{2f}{1+f^2}
$$
\item Compute the cross section of the process $\bar{\nu}_e + p
\rightarrow n + e^+$ for low energies of the incident antineutrino
(typically, $1\ \MeV \lesssim E_{\bar{\nu}} \lesssim 10\ \MeV$).
\end{priklady}

%\input{kniha21}  %kapitola 2  2.1 2.2
%%%%%%%%%%%%%%%%%%%%%%%%%%%%%%%%%%%%%%%%%%%%%%%%%%%%%%%%%%%%%%%%%%%
%%%%%%%%%%%%%%%%%%%%%%%%%%%%%%%%%%%%%%%%%%%%%%%%%%%%%%%%%%%%%%%%%%%%%%%%%%%%%%%%%%%%%%%%%%%%%%%%%%%%%%%%%%%%%%%%%%%%%%%%%%%%%%%%%%%%%%%%
\chapter{Universal $V-A$ theory}\label{chap2}
\section{Two-component neutrino}\index{V-A
theory@$V-A$ theory|(}

In the preceding chapter we have arrived at a remarkably simple
form of the effective Lagrangian for beta decay. The result
(\ref{eq1.121}) is written as a product of two \qq{currents} --
linear combinations of Lorentz vectors and axial
vectors\index{axial vector} (pseudovectors) and, in particular,
the leptonic current\index{leptonic current} has a pure $V-A$
structure. The currents are composed of fermionic fields differing
by one unit of electric charge and this is why such objects are
usually called \qq{weak charged currents}, or simply \qq{charged
currents}\index{charged current}\index{weak!charged
current|see{charged current}}. The $V-A$ form of the leptonic
current -- deduced from empirical data within our approach -- is a
rather striking feature of the effective Lagrangian
(\ref{eq1.121}), and it certainly calls for a theoretical
interpretation. Of course, such a problem is intimately related to
the remarkable phenomenon of maximal parity
violation\index{parity!violation|(}, revealed e.g. by the data on
the electron longitudinal polarization (see Section~\ref{sec1.6}).
Historically, a first attempt to formulate a \qq{theory} of parity
violation in weak interactions appeared almost simultaneously with
its experimental discovery (see \cite{ref9}, \cite{ref10}). It
relied on a revival of the two-component relativistic equation for
a massless spin-$\frac{1} {2}$ particle (written first by H.~Weyl
in 1929) and it has become known as the \qq{two-component neutrino
theory} (more concisely, \qq{the theory of two-component
neutrino}). We are now going to summarize briefly this simple
idea.

To begin with, let us remember the ordinary Dirac equation for a
massive spin-$\frac{1}{2}$ particle. This can be written
as\index{Dirac!equation}
\begin{equation}
\label{eq2.1}
i\frac{\partial\psi}{\partial t}=(-i\vec{\alpha}\cdot\vec{\nabla}
+\beta m)\psi
\end{equation}
where the $\vec{\nabla}$ stands for $\partial/\partial x^j$,
$j=1,2,3$. The matrices $\vec\alpha$ (i.e. $\alpha^j,\; j=1,2,3$)
and $\beta$ must satisfy
\begin{equation}
\label{eq2.2}
\{\alpha^j,\alpha^k\}=2\delta^{jk}\,,\quad\{\beta,\alpha^j\}=0\,,\quad \beta^2=1
\end{equation}
in order to reproduce correctly the standard relation between the
particle energy and momentum known in special relativity. It is
well known that the algebraic conditions (\ref{eq2.2}) can only be
satisfied by matrices of dimension four (or higher). For $m$
= 0 one is left with an equation
\begin{equation}
\label{eq2.3}
i\frac{\partial\psi}{\partial t}=-i\vec\alpha\cdot\vec\nabla\psi
\end{equation}
where the matrices $\alpha^j$ satisfy the anticommutation
relations shown in (\ref{eq2.2}), i.e.
\begin{equation}
\label{eq2.4}
\{\alpha^j,\alpha^k\}=2\delta^{jk}
\end{equation}
but now there is no $\beta$. The relations (\ref{eq2.4}) alone can be
satisfied by 2 $\times$ 2 matrices; in fact, there are two
inequivalent options, namely
\begin{equation}
\label{eq2.5}
\alpha^j=\sigma_j
\end{equation}
and
\begin{equation}
\label{eq2.6}
\alpha^j=-\sigma_j
\end{equation}
with $\sigma_j$ being the standard Pauli matrices. (Of course, it
is just the need for a fourth matrix $\beta$ that forces one to
work with 4 $\times$ 4 matrices in the massive case -- there is no
non-trivial 2 $\times$ 2 matrix anticommuting with all Pauli
matrices.) Note that the non-equivalence of the sets (\ref{eq2.5})
and (\ref{eq2.6}) is obvious for the same technical reason: there
is no regular matrix that would implement a similarity
transformation between the two sets, since the transformation
matrix would have to anticommute with the $\sigma_j$ for any $j$=
1,2,3. On the other hand, one has infinitely many equivalent
representations of the $\alpha^j$, obtained from (\ref{eq2.5}) or
(\ref{eq2.6}) resp. by means of arbitrary similarity
transformations. The two basic options (\ref{eq2.5}) and
(\ref{eq2.6}) define two possible types of two-component Weyl
equations, namely\index{Weyl!equation|(}
\begin{equation}
\label{eq2.7}
i\frac{\partial\psi}{\partial t}=-i\vec\sigma\cdot\vec\nabla\psi
\end{equation}
and
\begin{equation}
\label{eq2.8}
i\frac{\partial\psi}{\partial t}=+i\vec\sigma\cdot\vec\nabla\psi
\end{equation}
An experienced reader may notice that the last two equations are
relativistically invariant and correspond to the spinor
representations of Lorentz group\index{Lorentz!group} denoted
usually as ($\frac{1}{2}$, 0) and (0, $\frac{1}{2}$) resp., or, in
an alternative terminology, to dotted and undotted (Weyl)
spinors\index{Weyl!spinor}.

Let us now examine the plane-wave solutions of these equations,
corresponding to a positive energy $E=|\vec p|$, with $\vec p$
being the particle momentum. Such a plane wave can be written as
\begin{equation}
\label{eq2.9} \psi_+=N(p)u(p) \text{e}^{-ipx}
\end{equation}
where the $N(p)$ stands for an appropriate normalization factor
and $px= |\vec p| t-\vec p\cdot\vec x$. Inserting now
(\ref{eq2.9}) into equation (\ref{eq2.7}) one gets
\begin{equation}
\label{eq2.10} (\vec\sigma\cdot\vec p)u=|\vec p| u
\end{equation}
This is a remarkable result, as it obviously means that a solution
of the Weyl equation of the type (\ref{eq2.7}) with positive
energy automatically has positive helicity\index{helicity} (for a
negative-energy plane wave we would get negative helicity). In a
similar way, for the Weyl equation of the type (\ref{eq2.8}) one
finds that positive-energy solutions have negative helicity. Of
course, such a strict correspondence between energy and helicity
is a specific feature of the two-component equations -- if we use
a four-component Dirac equation, we always have both helicities
for a given energy, even in the massless case.

Thus, if one assumes that neutrino is strictly massless, it seems
to be natural to describe it by means of a two-component Weyl
equation (since it is then the most economical choice). To decide
which variant is relevant in nature is essentially an experimental
problem -- one has to determine the neutrino helicity. Here we may
refer to the famous experimental result \cite{ref6} quoted in the
preceding chapter (cf. Section~\ref{sec1.7}) which states that the neutrino
produced in beta decay is left-handed. This suggests that the
relevant Weyl equation is that given by (\ref{eq2.8}). It is easy
to see that the Weyl equations are not invariant under space
inversion -- technically, it is again due to the algebraic fact
that there is no 2 $\times$ 2 matrix anticommuting with Pauli
matrices (remember that for the four-component Dirac equation, the
parity transformation is implemented through the matrix $\beta$,
which is missing in the two-component case).

The parity non-invariance of the Weyl equation was precisely the
reason for its rejection in 1929, but it has become a blessing
after 1956 when parity violation turned out to be an experimental
reality. If a two-component field for negative-helicity neutrino
is to be incorporated into an interaction Lagrangian involving
four-component Dirac fields of other fermions (electron, proton,
etc.), one has to find an equivalent four-dimensional description
of  Weyl neutrino\index{Weyl!neutrino}. This can be achieved by
making use of the left-handed part of a four-component neutrino
field, which of course is obtained by applying the projector
$\frac{1}{2}(1-\gamma_5)$. In other words, a two-component
neutrino with negative helicity is taken into account
automatically if the corresponding field operator occurs in the
form $\psi_{\nu L}=\frac{1}{2}(1-\gamma_5)\psi_\nu$ (note that the
$\psi_{\nu L}$ then describes left-handed neutrinos and
right-handed antineutrinos). When one adopts such a principle, a
general parity-violating Lagrangian for beta decay can be written
in a straightforward way as
\begin{equation}
\label{eq2.11}
\lagr^{(\beta)}_{int}=\sum_{j=S,V,A,T,P}C_j(\bar\psi_p
\Gamma_j\psi_n)[\bar\psi_e\Gamma^j(1-\gamma_5)\psi_\nu] +
\text{h.c.}
\end{equation}
where the $C_j$ are arbitrary Fermi-type constants. Thus we see
that the idea of a two-component massless neutrino automatically
yields maximal parity violation in weak interactions (i.e. the
parity violation is simply due to left-handed Weyl neutrino), but
otherwise any algebraic type of coupling is possible. Obviously,
to restrict further the relevant couplings, one needs data (or an
educated guess) concerning the electron
helicity\index{Weyl!equation|)}.

To conclude this section, one should stress that from today's
point of view the theory of two-component neutrino can hardly be
taken seriously as an explanation of parity violation in weak
interactions, since it is well known by now that maximal parity
violation is observed even for interactions of massive particles
(e.g. quarks). Moreover, there are hints from various experiments
that neutrinos have non-zero (though tiny) masses. Parity
violation\index{parity!violation|)} thus seems to be simply an
inherent property of the interaction itself and, in general, has
nothing to do with massless neutrinos. It is perhaps fair to say
that its deeper origin still remains rather mysterious -- an
explanation will hopefully be provided by a future more
fundamental theory (note that the present-day standard model of
electroweak interactions in fact does not shed much light on this
issue). Nevertheless, the idea of a two-component left-handed
neutrino played an important heuristic role in the history of weak
interactions as it stimulated significantly the development of
relevant theory.

\section[Left-handed chiral leptons]{Left-handed chiral leptons: elimination of the
$S,P,T$ couplings}

Motivated by the two-component neutrino theory, R.~Feynman and
M.~Gell-Mann \cite{ref11} (and independently R.Marshak and
E.~Sudarshan \cite{ref12}) set forth the idea that, in general,
any elementary fermion (regardless of its mass) can participate in
weak interactions only through the left-handed chiral component of
the corresponding spinor field, i.e. through
$\psi_L=\frac{1}{2}(1-\gamma_5)\psi$.\footnote{Note that the
adjective \qq{chiral} used here thus means \qq{with a definite
chirality} -- e.g. $\gamma_5\psi_L=-\psi_L$\index{chirality}.} It
is not difficult to find that such a simple assumption leads to a
radical simplification of the Lagrangian (\ref{eq2.11}) -- in
fact, only the $V$ and $A$ terms then survive. To see this, let us
assume that, in addition to the left-handed massless neutrino, the
(massive) electron field also appears in the form $\psi_{eL}$.
Instead of (\ref{eq2.11}), one can then write a general beta-decay
Lagrangian as
\begin{equation}
\label{eq2.12}
\lagr^{(\beta)}_{int.}=\sum_{j=S,V,A,T,P}\widetilde{C}_j(\bar\psi_p
\Gamma_j\psi_n)(\bar\psi_{eL}\Gamma^j\psi_{\nu L}) + \text{h.c.}
\end{equation}
with $\widetilde{C}_j$ being some Fermi-type coupling
constants\index{coupling constants!weak|ff}. Taking into account
that $\bar\psi_L=\frac{1}{2}\bar\psi(1+\gamma_5)$, it becomes
clear that the leptonic factors appearing in (\ref{eq2.12})
contain the matrix products
\begin{equation}
\label{eq2.13}
(1+\gamma_5)\Gamma^j(1-\gamma_5)
\end{equation}
However, the well-known (anti)commutation properties of the Dirac
matrices now make it clear that the expression (\ref{eq2.13}) vanishes
identically for $j=S,P,T$ (remember that the $\Gamma_S,\Gamma_P$ and
$\Gamma_T$ commute with the $\gamma_5$ -- cf. (\ref{eq1.28})). Thus,
we are indeed left with only $V$ and $A$ terms in (\ref{eq2.12}), as
stated above.

The lesson to be learnt from this simple exercise is that the
\qq{law of left-handed chiral leptons} obviously represents an
extremely efficient organizational principle in weak interaction
theory: such a theoretical {\it tour de force\/} yields
immediately the right structure of the beta-decay effective
Lagrangian, which in the preceding chapter was obtained via a
rather lengthy systematic investigation of the empirical data. On
the other hand, if the ($V,A$) structure is deduced from the
Feynman--Gell-Mann (or Marshak--Sudarshan) conjecture, it must be
verified experimentally anyway, so that the work we have done in
Chapter~\ref{chap1} was certainly not in vain. In any case, one should bear
in mind that such a simple theoretical rule is not substantiated
(at least at the present level of understanding) by any deeper
physical principle and may be perceived as a fortunate educated
guess of an effective theory (which may be a manifestation of a
more fundamental underlying theory).

Nevertheless, it is quite remarkable that the theoretical
construction \cite{ref11}, \cite{ref12} was proposed at a time,
when some respected experimental data preferred the $T$ coupling
for Gamow--Teller beta transitions\index{Gamow--Teller
transitions}, instead of the $A$ coupling predicted by the simple
theory. Feynman and Gell-Mann \cite{ref11} went so far as to
suggest that these data might be wrong -- a guess that turned out
to be right somewhat later, when the controversial experiments
were repeated independently by other groups. In the meantime,
measurements of the electron and neutrino helicities were carried
out, with results confirming the $V-A$ theory. One can thus say
that the ultimate triumph of the ($V,A$) scheme for weak
interactions in the early 1960s resulted from an interplay between
the simple theoretical ideas \cite{ref11}, \cite{ref12} and a
careful analysis of the available experimental data.

Of course, if one adopts the principle of negative chirality for
nucleons as well, one gets a pure $V-A$ nucleon current in the
beta-decay Lagrangian (i.e.  $f$ = 1 in (\ref{eq1.121})). As we
know now from experiments, the $f$ is definitely different from 1
(which was not quite clear in the late 1950s, when the papers
\cite{ref11}, \cite{ref12} were published). It seems to suggest
that the rule of negative chirality can be reasonably used only
for elementary fermions (leptons and quarks). We will discuss the
quark interactions and related problems later on, and next we are
going to analyze a \qq{canonical} purely leptonic process -- the
muon decay\index{muon|ff}, which played a crucial role in
establishing the concept of weak interaction as a universal force,
not necessarily associated with nuclear beta decay.

%\input{kniha23}  %            2.3
%%%%%%%%%%%%%%%%%%%%%%%%%%%%%%%%%%%%%%%%%%%%%%%%%%%%%%%%%%%%%%%%%%%
%%%%%%%%%%%%%%%%%%%%%%%%%%%%%%%%%%%%%%%%%%%%%%%%%%%%%%%%%%%%%%%%%%%%%%%%%%%%%%%%%%%%%%%%%%%%%%%%%%%%%%%%%%%%%%%%%%%%%%%%%%%%%%%%%%%%%%%%

%\documentstyle[12pt]{report}
%\begin{document}
%\newcommand{\detr}{\mathop{\rm det'}\nolimits}
%\newcommand{\Tr}{\mathop{\rm Tr}\nolimits}
%\newcommand{\asym}{\mathop{\rm asym}\nolimits}
%\newcommand{\Pexp}{\mathop{\rm P exp}\nolimits}
%\newcommand{\antiPexp}{\mathop{\rm \bar P exp}\nolimits}
%\newcommand{\Ker}{\mathop{\rm Ker}\nolimits}
%\newcommand{\re}{\mathop{\rm Re}\nolimits}
%\newcommand{\im}{\mathop{\rm Im}\nolimits}
%\newcommand{\asympt}{\mathop{\sim}}
%\newcommand{\leftpartial}{\mathop{\stackrel{\leftarrow}{\partial}}\nolimits}
%\newcommand{\rightpartial}{\mathop{\stackrel{\rightarrow}{\partial}}\nolimits}
%\newcommand{\leftD}{\mathop{\stackrel{\leftarrow}{D}}\nolimits}
%\newcommand{\st}{\mathop{\rm st}\nolimits}

%\let\eps = \varepsilon

\section{Muon decay}\label{sec2.3}

\index{decay!of the muon|(}By now it is well known that muon disintegrates into an electron and
two neutrinos according to
\begin{equation}\label{eq2.14}
\mu^- \to e^- + \nu_\mu + \bar\nu_e
\end{equation}
In (\ref{eq2.14}) we have marked explicitly two different neutrino
species;  in particular, the $\nu_\mu$ carries a muonic lepton
number equal to that of the initial muon. We are not going to
review here the historical development of muon physics, but a few
remarks concerning (\ref{eq2.14}) are in order. The fact that the
muon (discovered in 1937) decays into more than two particles was
recognized around 1949, simply on the basis of the continuous
energy spectrum of the final electron. It was also immediately
obvious that the remaining decay products are electrically neutral
and rather light -- information about masses is contained e.g.
in the maximum electron energy that can be calculated along the
same lines as in the case of beta decay. If one assumes that the
decay products other than electron are massless, one gets
\begin{equation}\label{eq2.15}
E_e^{max} = \frac{m_\mu^2 + m_e^2}{2m_\mu}
\end{equation}
in good agreement with observed data (note that the current upper
bound \cite{ref5} is $m_{\nu_\mu} < 0.19$\ \MeV). The idea of the
muon decay scheme of the type (\ref{eq2.14}) seems to have been
accepted in the late 1940s, but the non-trivial question whether
$\nu_\mu \ne \nu_e$ has been answered directly only in the early
1960s (see \cite{ref13} and also e.g. \cite{CaG}). In this
context, one should also note that muon decays provide impressive
evidence in favour of separate conservation of muonic and
electronic lepton numbers -- let us quote e.g. the bounds for
unseen processes like $\mu^- \to e^- \gamma$ or $\mu^- \to e^- e^+
e^-$,  with branching ratios less than $1.2 \times 10^{-11}$ and
$1.0 \times 10^{-12}$ respectively \cite{ref5}\index{branching
ratios!of $\mu$-decay}.

  Let us now try to describe the decay process (\ref{eq2.14}) in
quantitative terms. If one takes for granted the theory of
left-handed chiral leptons \cite{ref11,ref12} described in the
preceding section, one can write immediately the corresponding
effective Lagrangian as
\begin{equation}\label{eq2.16}
\lagr_{int}^{(\mu)} = -\frac{G_\mu}{\sqrt{2}}
[\bar\psi_{(\nu_\mu)}\gamma_\rho (1-\gamma_5) \psi_{(\mu)}]
[\bar\psi_{(e)}\gamma^\rho (1-\gamma_5) \psi_{(\nu_e)}]
\end{equation}
where the $G_\mu$ is an appropriate Fermi-type coupling constant.
An important goal of our subsequent analysis will be the
determination of the relevant coupling strength -- this can be
done by comparing the calculated muon lifetime\index{lifetime!of
the muon|ff} with its measured value. Apart from this task, it
would also be interesting to test the (postulated) $V-A$ structure
of the currents in (\ref{eq2.16}). As a simple example of such a
check, we will temporarily modify (\ref{eq2.16}) to
\begin{equation}\label{eq2.17}
\lagr_{int} = -\frac{G_\mu}{\sqrt{2}}
[\bar\psi_{(\nu_\mu)}\gamma_\rho (1-\lambda\gamma_5) \psi_{(\mu)}]
[\bar\psi_{(e)}\gamma^\rho (1-\gamma_5) \psi_{(\nu_e)}]
\end{equation}
with $\lambda$ being an arbitrary real parameter, and show that
the observed shape of the electron energy spectrum clearly favours
the value $\lambda = 1$ corresponding to the $V-A$ theory (note
that here we essentially follow the treatment of \cite{BjD},
Chapter 10).\footnote{Notice that we are not trying to perform
here a general analysis of the muon-decay effective Lagrangian
that would be analogous to the procedure applied to neutron decay
in the preceding chapter. Such an analysis is in a sense more
difficult for muon decay, as there is only one charged particle in
the final state and e.g. the simple electron-antineutrino angular
correlations\index{angular correlation} cannot be studied
experimentally. For a detailed discussion of muon physics from
this point of view, see \cite{ref14}.}

  Thus, we start our calculation with the lowest-order matrix element
corresponding to (\ref{eq2.17}), i.e.
\begin{equation}\label{eq2.18}
{\cal M} = -\frac{G_\mu}{\sqrt{2}}
[\bar u(k) \gamma_\rho (1-\lambda\gamma_5) u(P)]
[\bar u(p) \gamma^\rho (1-\gamma_5) v(k')]
\end{equation}
where the four-momenta of the $\mu, e, \nu_\mu, \bar\nu_e$ are denoted by
$P, p, k, k'$ respectively. The spin-averaged matrix element squared then
becomes, after some simple algebraic manipulations
\begin{eqnarray}\label{eq2.19}
\overline{|{\cal M}|^2} &=& \frac 12 \sum_{spins} |{\cal M}|^2 =\\
&=& \frac{1}{2} G_\mu^2 \Tr[\slashed{k} \gamma_\rho
(\slashed{P}+m_\mu) \gamma_\sigma (1+\lambda^2-2\lambda\gamma_5)]
\Tr[(\slashed{p}+m_e) \gamma^\rho \slashed{k'} \gamma^\sigma
(1-\gamma_5)] \nonumber
\end{eqnarray}
Obviously, the mass terms appearing in (\ref{eq2.19}) in fact do not
contribute and the product of traces can then be easily evaluated with
the help of the formulae
\begin{eqnarray}\label{eq2.20}
\Tr(\slashed{a} \gamma_\rho \slashed{b} \gamma_\sigma)
\Tr(\slashed{c} \gamma^\rho \slashed{d} \gamma^\sigma)
&=& 32 [(a\cdot c)(b\cdot d) + (a\cdot d)(b\cdot c)]
\nonumber\\
\Tr(\slashed{a} \gamma_\rho \slashed{b} \gamma_\sigma \gamma_5)
\Tr(\slashed{c} \gamma^\rho \slashed{d} \gamma^\sigma \gamma_5)
&=& 32 [(a\cdot c)(b\cdot d) - (a\cdot d)(b\cdot c)]
\nonumber\\
\Tr(\slashed{a} \gamma_\rho \slashed{b} \gamma_\sigma)
\Tr(\slashed{c} \gamma^\rho \slashed{d} \gamma^\sigma \gamma_5)
&=& 0
\end{eqnarray}
(see (\ref{eqA.50})). We thus get finally
\begin{equation}\label{eq2.21}
\overline{|{\cal M}|^2} =
16 G_\mu^2 \left[ (1+\lambda)^2 (k\cdot p)(k'\cdot P) +
(1-\lambda)^2 (k\cdot k')(p\cdot P) \right]
\end{equation}

\begin{figure}[h]
\centering \s{\includegraphics{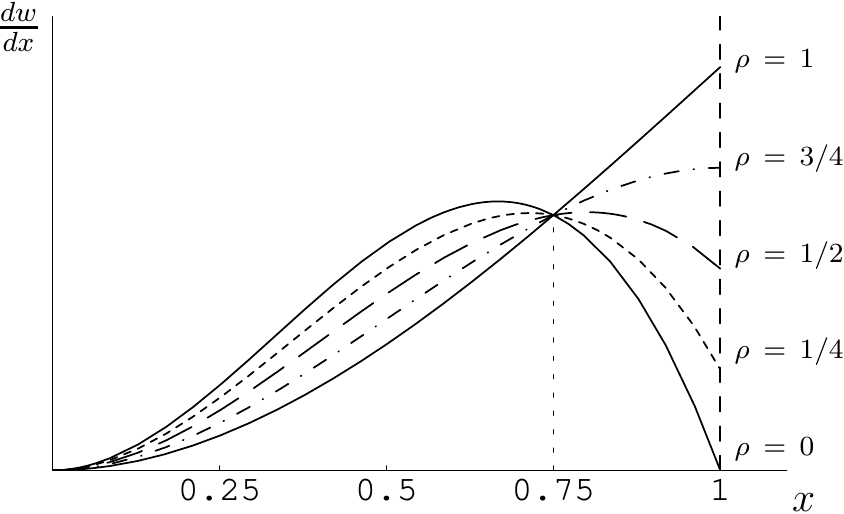}}
\caption{Variation of the shape of muon decay spectrum with
respect to the value of the Michel parameter $\rho$, as described
by the approximate formula (\ref{eq2.31}).} \label{fig4}
\end{figure}

  The differential decay rate is given by the standard formula
\begin{equation}\label{eq2.22}
dw = \frac{1}{2m_\mu} \overline{|{\cal M}|^2}
\frac{d^3p}{2E(p)(2\pi)^3} \frac{d^3k}{2E(k)(2\pi)^3}
\frac{d^3k'}{2E(k')(2\pi)^3} (2\pi)^4 \delta^4(P-p-k-k')
\end{equation}
where of course $E(p) = \sqrt{\vec{p}{\,}^2+m_e^2}, E(k) =
|\vec{k}|$ and $E(k') = |\vec{k'}|$. To obtain the electron energy
spectrum, the expression (\ref{eq2.22}) could be integrated in a
similar fashion as in the case of neutron beta decay, but now we
are not allowed to make the simplifying kinematical approximations
used before -- in muon decay all final-state particles may be
relativistic, so that no momentum can be neglected. A
straightforward integration of (\ref{eq2.22}) is left to the
reader as a useful (though somewhat tedious) exercise; here we
offer an alternative method, which may be of a more general
interest. From the structure of the expression (\ref{eq2.21}) it
is clear that one needs the integral
\begin{equation}\label{eq2.23}
I_{\alpha\beta}(Q) = \int \frac{d^3k}{2E(k)} \frac{d^3k'}{2E(k')}
k_\alpha k_\beta' \delta^4(Q-k-k')
\end{equation}
where we have denoted $Q=P-p$. Now, the crucial observation is
that the $I_{\alpha\beta}$ is a 2nd rank tensor under Lorentz
transformations\index{Lorentz!transformation}. Indeed, using some
simple tricks for the integration involving the delta functions,
the expression (\ref{eq2.23}) can be recast as
\begin{equation}\label{eq2.24}
I_{\alpha\beta}(Q) = \int d^4k \,\theta(k_0) \delta(k^2)
\delta\left((Q-k)^2\right) k_\alpha (Q-k)_\beta
\end{equation}
which makes the tensor character of the $I_{\alpha\beta}$ rather
obvious. The most general 2nd rank tensor $I_{\alpha\beta}(Q)$ has
the form
\begin{equation}\label{eq2.25}
I_{\alpha\beta}(Q) = A g_{\alpha\beta} + B Q_\alpha Q_\beta
\end{equation}
with $A$ and $B$ being arbitrary functions of $Q^2$. To determine these
coefficients, it suffices to evaluate two independent components of
the tensor (\ref{eq2.23}) in an arbitrary reference frame. 
A most convenient choice is the c.m. system of the two neutrinos, where
the $Q$ has components $Q=(Q_0,\vec{0})$, with
$Q_0=2k_0=2|\vec{k}|$. Let us consider e.g. the tensor components
$I_{00}$ and $I_{33}$. According to (\ref{eq2.25}), these are related
to $A$ and $B$ by
\begin{eqnarray}\label{eq2.26}
I_{00} &=& A + B Q_0^2
\nonumber\\
I_{33} &=& -A
\end{eqnarray}
and a direct integration of the original form (\ref{eq2.23}) in the c.m.
system gives
\begin{equation}\label{eq2.27}
I_{00} = \frac{\pi}{8} Q_0^2, \quad I_{33} = -\frac{\pi}{24} Q_0^2
\end{equation}
This result, together with (\ref{eq2.26}), then yields
\begin{equation}\label{eq2.28}
I_{\alpha\beta}(Q) = \frac{\pi}{24}
( Q^2 g_{\alpha\beta} + 2 Q_\alpha Q_\beta)
\end{equation}
With (\ref{eq2.21}) and (\ref{eq2.28}) at hand, the evaluation of
the electron energy spectrum is reduced to purely algebraic
manipulations. These are elementary but somewhat lengthy, so we
give only the final answer\index{energy spectrum of the electron}
\begin{equation}\label{eq2.29}
\frac{dw(E_e)}{dE_e} = \frac{1}{3\pi^3} G_\mu^2
\frac{1+\lambda^2}{2} m_\mu |\vec{p}| E_e \Bigl[ 3(W-E_e) + \frac
14 \frac{(1+\lambda)^2}{1+\lambda^2}
\Bigl(4E_e-3W-\frac{m_e^2}{E_e} \Bigr) \Bigr]
\end{equation}
where we have used the symbol $W$ for the maximum electron energy
(see (\ref{eq2.15})). Note that for some historical reasons, it has
become customary to denote
\begin{equation}\label{eq2.30}
\frac 14 \frac{(1+\lambda)^2}{1+\lambda^2} = \frac 23 \rho
\end{equation}
where $\rho$ is the so-called Michel\index{Michel parameter|(}
parameter.\footnote{The $\rho$ is in fact one of the four or five
parameters used for the description of muon decay in a general
case when one takes into account also particle polarizations. In
the unpolarized case, two parameters are usually introduced --
apart from the $\rho$ there is another one denoted by $\eta$,
which characterizes the shape of the low-energy end of the
electron spectrum. We will be mostly interested in the upper
endpoint of the spectrum, so that only the $\rho$ is relevant for
our further considerations. The parameters are named after
L.~Michel, who in the 1950s performed a comprehensive analysis of the
muon decay within the framework of a general Fermi-type model
involving all the $S, V, A, T, P$ couplings \cite{ref14}. For
details, see also \cite{CoB} and \cite{Gre}.} For energies $E_e
\gg m_e$ one can neglect the term $m_e^2 / E_e$ in (\ref{eq2.29}).
In terms of the dimensionless variable $x=E_e/W$ the high-energy
part of the spectrum (\ref{eq2.29}) can then be approximately
written as
\begin{equation}\label{eq2.31}
\frac{dw}{dx} = \frac{1}{48\pi^3} G_\mu^2  m_\mu^5
\frac{1+\lambda^2}{2} x^2
\left[ 3(1-x) + \frac 23 \rho (4x-3) \right]
\end{equation}
(note that in the last expression we have also set
$W \doteq \frac 12 m_\mu$). Obviously, the value of the Michel parameter
\begin{equation}\label{eq2.32}
\rho = \frac 38 \frac{(1+\lambda)^2}{1+\lambda^2}
\end{equation}
determines the shape of the energy spectrum near its endpoint.
From (\ref{eq2.32}) it is clearly seen that for $\rho = 0$ the
distribution function $dw/dx$ would vanish at $x=1$, but in
general it is non-zero at the endpoint (in contrast to the case of
neutron beta decay). Some illustrative examples are shown in
Fig.\,\ref{fig4}
(note that $0\leq\rho\leq 3/4$ for any $\lambda$ in
(\ref{eq2.32})). The different character of  muon decay spectrum
in comparison with the beta decay is due to the different
kinematical conditions in both processes; we will comment on this
point later on. The current experimental value is $\rho = 0.74979
\pm 0.00026$ (the world average according to \cite{ref5}), i.e.
$\rho \doteq 3/4$ with high accuracy. In view of (\ref{eq2.32})
this immediately implies $\lambda = 1$, which confirms the
anticipated validity of the $V-A$ theory for muon decay.

  To determine the coupling constant $G_\mu$, let us calculate the full
decay width. For simplicity, we will employ the approximate expression
(\ref{eq2.31}) over the whole electron energy range, as one may thus
presumably lose only small correction terms of the relative order
$O(m_e^2/m_\mu^2)$. Setting $\lambda = 1$ in (\ref{eq2.31}) one has
\begin{equation}\label{eq2.33}
\frac{dw}{dx} = \frac{1}{48\pi^3} G_\mu^2  m_\mu^5
x^2 \left(\frac 32 -x \right)
\end{equation}
and the decay width is then obtained by integrating (\ref{eq2.33}) over
the $x$ from 0 to 1. One thus gets readily the result
\begin{equation}\label{eq2.34}
\Gamma = \frac{G_\mu^2  m_\mu^5}{192\pi^3}
\end{equation}
Let us remark that the characteristic dependence on the $m_\mu^5$
can be easily understood on dimensional grounds: the decay width
must have dimension of a mass, the coupling constant squared
supplies a mass to minus four and for $m_e= 0$ the $m_\mu$ is the
only mass scale left in the game. Since there are essentially no
other relevant decay modes for muon, the inverse of the width
(\ref{eq2.34}) gives the muon lifetime. This is measured with a
rather high accuracy, $\tau_\mu=(2.1969811 \pm 0.0000022)\times
10^{-6}$ s. The experimental number is to be compared with the
result of our calculation, corrected for the electron mass effects
and for QED\index{quantum!electrodynamics (QED)} effects
(radiative corrections\index{radiative corrections}). Such a
detailed calculation goes beyond the scope of our treatment, so
let us only quote the result of such an analysis. The value of the
coupling constant $G_\mu$ corresponding to the measured muon
lifetime is\index{coupling constants!Fermi}
\begin{equation}\label{eq2.35}
G_\mu=(1.1663787 \pm 0.0000006)\times  10^{-5} \ \GeV^{-2}
\end{equation}
and is identified with the \qq{canonical} Fermi constant $G_F$
recorded in the Review of Particle Physics \cite{ref5}\index{Fermi
constant}.

  The value (\ref{eq2.35}) almost coincides with the beta-decay
constant $G_\beta$ (cf. (\ref{eq1.120})) and this clearly
indicates that muon decay is a manifestation of essentially the
same basic force that is responsible for the nuclear beta decay.
In other words, the results of the analysis of muon decay strongly
support the idea that the original \qq{weak nuclear
force}\index{weak!nuclear force} discovered in connection with
beta decay represents in fact only one aspect of a universal weak
interaction acting on widely different subatomic particles. On the
other hand, although the $G_\mu$ is very close to the $G_\beta$,
these two coupling parameters clearly differ by many standard
deviations, so one obviously needs an additional small parameter
to characterize the universality of weak interactions
properly\index{universality of weak interaction}. As we shall see
in subsequent sections, it makes sense to describe the difference
between $G_\mu$ and $G_\beta$ by means of the so-called Cabibbo
angle\index{Cabibbo angle} -- this observation lies at the basis
of the \qq{Cabibbo universality} formulated in the early 1960s. In
any case, it should be stressed that through our analysis we have
arrived at a simple explanation of the widely different muon and
neutron lifetimes (about $10^{-6}$ seconds for muon and 15 minutes
for neutron): such a difference of many orders of magnitude is
entirely due to the phase space\index{phase space|ff} factors
being proportional to the fifth power of the relevant energy
scales -- the coupling strengths (i.e. the basic dynamics) are
essentially the same.

  Before closing this section, let us return briefly to the problem of
the shape of electron energy spectrum. As we have seen in
(\ref{eq2.29}), the distribution function $dw(E_e)/dE_e$ in
general does not vanish at the endpoint $E_e=W$, unless $\rho=0$.
It is not difficult to realize that such a peculiar dissimilarity
to the case of neutron beta decay is due to the assumption that
both neutrinos produced in muon decay are massless. Indeed, when
the electron energy reaches its maximum value $W$, the neutrinos
carry off the remaining part $m_\mu-W$. In the massless case,
there are infinitely many ways how to divide it between $\nu_\mu$
and $\bar\nu_e$: any kinematical configuration such that both
neutrinos are emitted in the direction opposite to the electron
momentum, with otherwise arbitrary energies satisfying $E(k)+E(k')
= m_\mu-W$, fulfills the required simultaneous conservation of
energy and momentum. Thus, the endpoint of the electron energy
spectrum corresponds to infinitely many degenerate states and,
consequently, the volume of the phase space at $E_e=W$ can be
non-zero. On the other hand, if at least one of the neutrinos, say
$\nu_\mu$, is massive, there is only one possible kinematical
configuration corresponding to $E_e=W$ and the phase-space volume
then vanishes at the endpoint of the spectrum. This can be nicely
illustrated if one calculates explicitly the integral
(\ref{eq2.23}) for a massive $\nu_\mu$ (leaving the $\nu_e$
massless for simplicity). Denoting the $\nu_\mu$ mass by $m$, one
gets
\begin{equation}\label{eq2.36}
I_{\alpha\beta}(Q;m) = A(Q^2;m) g_{\alpha\beta} + B(Q^2;m) Q_\alpha Q_\beta
\end{equation}
with
\begin{eqnarray}\label{eq2.37}
A(Q^2;m) &=& \frac{\pi}{24} \frac{1}{(Q^2)^2} (Q^2-m^2)^3
\nonumber\\
B(Q^2;m) &=& \frac{\pi}{12} \frac{1}{(Q^2)^3} (Q^2-m^2)^2 (Q^2+2m^2)
\end{eqnarray}
(a derivation of (\ref{eq2.37}) is left to the reader as an
exercise). Notice that the expressions (\ref{eq2.37})  reduce to
(\ref{eq2.28}) for $m = 0$. Now, the maximum electron energy
corresponds to $Q^2=m^2$, where the formfactors $A$ and $B$ are
seen to vanish and this confirms our previous considerations
concerning the phase space at the endpoint of the electron
spectrum. Of course, if the neutrino mass\index{neutrino!mass} is
very small, one cannot practically distinguish the shape of a
spectrum falling steeply to zero near the endpoint from the case
where the energy distribution function is truly non-vanishing for
$E_e = W$\index{Michel parameter|)}.

%\end{document}

%\input{kniha24}  %            2.4
%%%%%%%%%%%%%%%%%%%%%%%%%%%%%%%%%%%%%%%%%%%%%%%%%%%%%%%%%%%%%%%%%%%
%%%%%%%%%%%%%%%%%%%%%%%%%%%%%%%%%%%%%%%%%%%%%%%%%%%%%%%%%%%%%%%%%%%%%%%%%%%%%%%%%%%%%%%%%%%%%%%%%%%%%%%%%%%%%%%%%%%%%%%%%%%%%%%%%%%%%%%%
\section{Universal interaction of $V-A$ currents}

In the preceding discussion we have seen two examples of physical
processes -- the nuclear beta decay and the decay of muon -- that
involve quite different particles and also have widely different
lifetimes, yet they turn out to be governed by essentially the
same force. We have found that neutron beta decay (and associated
processes) can be successfully described by an effective
Lagrangian of the form
\begin{equation}
\label{eq2.38} \lagr^{(\beta)}_{int}=-\frac{G_\beta}{\sqrt
2}[\bar\psi_p
\gamma^\rho(1-f\gamma_5)\psi_n][\bar\psi_e\gamma_\rho(1-\gamma_5)
\psi_\nu]+\text{h.c.}
\end{equation}
while the muon decay corresponds to
\begin{equation}
\label{eq2.39} \lagr^{(\mu)}_{int}=-\frac{G_\mu}{\sqrt
2}[\bar\psi_{\nu_\mu}
\gamma^\rho(1-\gamma_5)\psi_\mu][\bar\psi_e\gamma_\rho(1-\gamma_5)
\psi_{\nu_e}]+\text{h.c.}
\end{equation}
and the coupling constants $G_\beta$ and $G_\mu$ nearly
coincide\index{decay!of the muon|)}.

Another important reaction, studied experimentally since the 1940s, is
the so-called \qq{weak muon capture}
\begin{equation}
\label{eq2.40}
\mu^-+p\rightarrow n+\nu_\mu
\end{equation}
Processes of the type (\ref{eq2.40}) occur (at the level of atomic
nuclei) with both $\mu^-$ and $\mu^+$; as for an effective
interaction, the experimental data show that the $V-A$ leptonic
current\index{leptonic current}
$\bar\psi_{\nu_\mu}\gamma^\rho(1-\gamma_5)\psi_\mu$ must be
involved and an overall strength of the coupling of nucleons to
muon-type leptons practically coincides with $G_\beta$ (for
details, see e.g. \cite{CoB}).\footnote{Of course, it took some
time to establish the $V-A$ nature of the relevant interaction,
but the observation that an overall coupling strength is of the
order of the beta-decay Fermi constant has already been made in
the late 1940s -- cf. e.g. \cite{ref16} and \cite{Jac}. Note,
however, that a simple effective Lagrangian analogous to
(\ref{eq2.38}) would provide only a rough description of the
nucleonic part of a corresponding scattering matrix element, since
the momentum transfer in the muon capture processes can be
relatively high in comparison with the neutron beta decay. We will
discuss the general structure of nucleonic matrix elements later
on.}

Observations of such diverse processes that occur among different
particles, yet with an essentially equal coupling strength, led
soon to the idea of a universal weak interaction connecting
leptons with nucleons and leptons with themselves. In those early
days, such a universal coupling scheme was symbolized by the
so-called \qq{Tiomno--Wheeler triangle}\index{Tiomno--Wheeler
triangle} (see \cite{ref16} and \cite{Jac}) -- an equilateral
triangle with pairs ($n,p$), ($e,\nu$) and ($\mu,\nu$) at
vertices. In the 1950s it gradually became clear that the decays of
charged pions\index{decay!of the pion} and of the newly discovered
strange mesons\index{meson} (to say nothing of new
baryons\index{baryon}) could also be accounted for by a force of
the \qq{Fermi strength}. Thus it appeared desirable to incorporate
mesons (i.e. spin-zero particles) into the weak interaction scheme
as well, although this was originally conceived as a model of
direct four-fermion coupling\index{four-fermion interaction}.
Anyway, with the number of observed decay processes proliferating
rapidly and with the accumulating evidence for a universal
magnitude of the corresponding couplings, there was obviously
growing need for a coherent unified picture of weak interaction
phenomenology that would involve all known leptons and
hadrons\index{hadrons|ff}. In this context, one might perhaps use
the following pictorial description of the status of the
provisional weak interaction models discussed so far: the simple
effective Lagrangians, directly applicable in particular cases
mentioned before, look rather like individual pieces of a
\qq{jigsaw puzzle}, presumably with some missing parts to be added
in order to get a complete pattern.

Feynman and Gell-Mann \cite{ref11} made an important step forward
by postulating a universal current-current form of weak
interaction, namely\index{Fermi-type interaction}
\begin{equation}
\label{eq2.41} \lagr^{(w)}_{int}=-\frac{G_F}{\sqrt 2}J^\rho
J^\dagger_\rho
\end{equation}
where the \qq{universal Fermi constant} $G_F$\index{Fermi
constant} is to be identified with the muon decay constant $G_\mu$
(cf.(\ref{eq2.35})) and the weak current $J^\rho$ consists of
leptonic and hadronic parts
\begin{equation}
\label{eq2.42}
J^\rho=\bar\psi_{\nu_e}\gamma^\rho(1-\gamma_5)\psi_e+\bar\psi_
{\nu_\mu}
\gamma^\rho(1-\gamma_5)\psi_\mu+J^\rho_{(hadron)}
\end{equation}
The hadronic part is assumed to have the structure $V-A$ with
respect to Lorentz transformations\index{Lorentz!transformation},
but it need not be expressed explicitly in terms of the field
operators of physical hadrons. One should only require that the
$J^\rho_{(hadron)}$ is an operator having non-trivial matrix
elements between physical hadronic states and, eventually, between
a meson state and vacuum\index{vacuum} (as we shall see later,
this last property is an important prerequisite for describing
e.g. the leptonic decays of charged pions). Such a
phenomenological device, which bypasses the field-theoretic
treatment of the hadronic sector, seems to be quite reasonable in
a situation when one faces the rich spectrum of hadrons with
different spins. Needless to say, a simple organizing principle
for hadronic world emerged somewhat later with the advent of the
quark model -- more about this later. By combining various pieces
of the two currents in (\ref{eq2.42}), one is obviously able to
reproduce the weak processes discussed before (i.e. beta decay,
muon decay, weak muon capture, etc.). Moreover, the form
(\ref{eq2.42}) also predicts some new (\qq{diagonal}) processes,
in particular the elastic\index{neutrino-electron scattering}
(anti)neutrino-electron scattering.\footnote{Note that the elastic
scattering processes $\bar\nu_e e\rightarrow\bar\nu_e e$ and
$\nu_e e\rightarrow\nu_e e$ were in fact first observed
experimentally only many years after their prediction by
Feynman--Gell-Mann theory -- see \cite{ref17}, \cite{ref18}.} It
is perhaps in order to remark here that by considering in
(\ref{eq2.41}) the interaction of the weak current (\ref{eq2.42})
{\it with itself}, Feynman and Gell-Mann clearly envisaged a
possible alternative description of weak interactions in terms of
an exchange of a massive charged \qq{intermediate vector
boson}\index{intermediate vector boson} -- a scheme that leads to
essentially identical predictions as the current-current form at
sufficiently low energies (this is a point to be discussed in
detail in Chapter~\ref{chap3}).

In any case, one important aspect of the weak hadronic
current\index{hadronic current} was missing in the pioneering
treatment \cite{ref11}. Owing to the lack of sufficiently accurate
measurements at that time, Feynman and Gell-Mann were not aware of
the subtle difference between the muon decay constant $G_\mu=G_F$
and the $G_\beta$ that we know now (cf. (\ref{eq2.35}) and
(\ref{eq1.120})) and it was also not clear what is precisely the
relative strength of the strangeness-changing weak transitions in
comparison with the ordinary beta decay. In this respect, the
Feynman--Gell-Mann model \cite{ref11} was significantly improved
by N.~Cabibbo \cite{ref19}.

%\input{kniha25}  %            2.5
%%%%%%%%%%%%%%%%%%%%%%%%%%%%%%%%%%%%%%%%%%%%%%%%%%%%%%%%%%%%%%%%%%%
%%%%%%%%%%%%%%%%%%%%%%%%%%%%%%%%%%%%%%%%%%%%%%%%%%%%%%%%%%%%%%%%%%%%%%%%%%%%%%%%%%%%%%%%%%%%%%%%%%%%%%%%%%%%%%%%%%%%%%%%%%%%%%%%%%%%%%%%
\section[Cabibbo angle and selection rules for strangeness]{Cabibbo angle and selection rules\\ for strangeness}\label{sec2.5}

\index{strangeness|(}\index{Cabibbo angle|(} In order to explain
the development that led to the notion of \qq{Cabibbo
universality}, let us focus on the difference between the $G_F$
and $G_\beta$. As we noted before, $G_F$ ( =$G_\mu$) $\doteq 1.166
\times 10^{-5}\ \GeV^{-2}$ with high accuracy, whereas
$G_\beta\doteq 1.136\times 10^{-5}\ \GeV^{-2}$ with about one per
mille accuracy. Thus, since $G_\beta<G_F$, one may introduce a
parametrization
\begin{equation}
\label{eq2.43} G_\beta/G_F=\cos \theta_C
\end{equation}
with $\theta_C$ being the so-called \qq{Cabibbo angle}. Putting in
numbers, one has $\cos\theta_C\doteq0.974$, so that
\begin{equation}
\label{eq2.44}
\theta_C\doteq 13^\circ
\end{equation}
At this stage, such a parametrization may seem somewhat
artificial, as it is not clear why a particular $angle$ should be
appropriate for describing the simple fact that $G_\beta<G_F$. The
true relevance of the parameter $\theta_C$ can be appreciated only
when one considers more fancy weak processes, namely the hadron
decays in which strangeness is not conserved. As an instructive
example, let us consider the semileptonic decay\footnote{It should
be noted that the considered process is one of the relatively rare
decay modes of the $\Sigma^-$, as its branching ratio is about
10$^{-3}$ (see \cite{ref5})\index{branching ratios!of
$\Sigma$-decay}. The $\Sigma^-$ decays predominantly via the
non-leptonic mode $\Sigma^-\rightarrow n+\pi^-$, with the
corresponding branching ratio being 99.85$\%$. Nevertheless, the
relevant characteristics of decay (\ref{eq2.45}) are measured with
rather good accuracy and the data constitute a valuable source of
information on the weak interactions of strange particles. For
completeness, let us recall that the $\Sigma^-$ mass is $1197\
\MeV$ and the mean lifetime\index{lifetime!of the $\Sigma$}
$\tau_{\Sigma^-}\doteq 1.48\times 10^{-10}$s.}
\begin{equation}
\label{eq2.45}
\Sigma^-\rightarrow n+e^-+\bar\nu_e
\end{equation}
The baryon $\Sigma^-$ carries the strangeness $S = -1$, while for
neutron one has, of course, $S$ = 0. The process (\ref{eq2.45})
may be viewed as a \qq{strangeness-changing beta decay} of the
$\Sigma^-$\index{decay!of the hyperon $\Sigma$|(}. Measurements
analogous to those performed for ordinary neutron beta decay lead
to the conclusion that a corresponding decay matrix element can
approximately be written as
\begin{equation}
\label{eq2.46} {\cal M}_{fi}\doteq-\frac{G_F}{\sqrt 2}\sin
\theta_C[\bar u_n\gamma_\rho(1-\tilde{f}\gamma_5)u_{\Sigma^-}]
[\bar u_e\gamma^\rho(1-\gamma_5)v_\nu]
\end{equation}
with $\tilde{f}\doteq - 0.34$ (cf. \cite{ref5}); the meaning of
the other symbols is obvious. Let us stress that in writing
(\ref{eq2.46}) we have neglected effects associated with the
corresponding momentum transfer (i.e. effects of the order of
$(m_{\Sigma^-}-m_n)/ m_{\Sigma^-})$; of course, such an
approximation is of poorer quality than in the case of neutron
beta decay. From (\ref{eq2.46}) one may infer readily a
corresponding effective Lagrangian of the form analogous to
(\ref{eq1.121}). The crucial (experimental) result embodied in
(\ref{eq2.46}) is that, instead of the $G_\beta=G_F\cos \theta_C$
appearing in (\ref{eq1.121}), the relevant Fermi coupling constant
is now $G_F\sin\theta_C$! (In view of (\ref{eq2.44}) it
practically means that the coupling strength now constitutes only
about $23\%$ of the $G_\beta$.) This is the essence of the crucial
observation made by Cabibbo \cite{ref19} (who analyzed primarily
the strangeness-changing kaon decays) -- now it is at least clear
that it makes sense to parametrize the subtle difference between
the $G_\beta$ and $G_F$ in terms of an angle. Nevertheless, the
origin of such an empirical angle remained entirely obscure within
the framework of the weak interaction theory in the 1960s. Looking
ahead, let us remark at this place that the situation is slightly
better now, since within the present-day electroweak Standard
Model the appearance of an angle like the $\theta_C$ is quite
natural; it turns out to be intimately related to the mechanism of
fermion mass generation. While the reader may be pleased by this
encouraging news, it is fair to admit, on the other hand, that the
numerical value (\ref{eq2.44}) remains mysterious even within SM
(in fact, a prediction of the $\theta_C$ value constitutes one of
the major challenges for theories attempting to go beyond SM).

Up to now, we have only compared the strangeness-changing process
(\ref{eq2.45}) with the ordinary beta decay, where the strangeness
does not play any role. In fact, it turns out that while all
strangeness-changing decays proceed with strength
$G_F\sin\theta_C$, the relevant Fermi constant for any
strangeness-conserving decay is $G_F\cos\theta_C$ (irrespective of
whether the hadrons involved are strange or not). For example, if
the decay\index{decay!of the hyperon $\Sigma$|)}
$\Sigma^-\rightarrow\Lambda e^-\bar\nu_e$ is analyzed, one finds
that the corresponding coupling strength is $G_F\cos\theta_C$ as
for the ordinary beta decay, although both the $\Sigma^-$ and
$\Lambda$ have strangeness $S = -1$. Since the $\theta_C$ is
numerically small, one may thus conclude that, at a
phenomenological level, the role of the Cabibbo angle consists in
suppressing the strangeness-changing weak processes relatively to
the strangeness-conserving ones.

The previous considerations may be summarized by writing the
hadronic weak current schematically as
\begin{equation}
\label{eq2.47}
J^{(hadron)}_\rho=\cos\theta_C J^{(\Delta S=0)}_\rho +\sin\theta_C
J^{(\Delta S\neq 0)}_\rho
\end{equation}
where the operators $J^{(\Delta S=0)}_\rho$ and $J^{(\Delta S\neq
0)}_\rho$ have the form $V-A$ and it is tacitly assumed that they
do not incorporate any other suppression factors related to
strangeness. To get a deeper insight into the structure of the
operators appearing in (\ref{eq2.47}), we have to consider some further
empirical selection rules that hold for weak transitions. There
are essentially two such rules that should be taken into account:

First, it is a well-established fact that processes in which the
strangeness is changed by more than one unit are very strongly
suppressed\index{rule!Delta S@$\lvert\Delta S\rvert\leq 1$}. As an
example, one may consider the process $\Xi^-\rightarrow n+\pi^-$,
where $S(\Xi^-)-S(n)= - 2$. The experimental upper bound for the
corresponding branching ratio is about 1.9 $\times 10^{-5}$,
though the available phase space for the decay products is
certainly more favourable than in the case of the dominant mode
$\Xi^-\rightarrow \Lambda+\pi^-$ (which has the branching ratio of
99.88$\%$)\index{branching ratios!of $\Xi$-decay}\index{decay!of
the hyperon $\Xi$}. There are other similar examples for the
decays of the $\Xi^0$ and $\Omega^-$ ($S = -3$) as well.

The second non-trivial phenomenological constraint is provided by
the well-known \qq{rule $\Delta S = \Delta
Q$}\index{rule!Delta@$\Delta S = \Delta Q$}. This holds for
strangeness-changing semileptonic decays of hadrons (both
mesons\index{meson} and baryons\index{baryon}) and can be
formulated concisely as follows. Let us denote by $h_i$ and $h_f$
the initial and final hadron respectively. For the process
\begin{equation}
\label{eq2.48} h_i\rightarrow h_f  + \text{lepton pair}
\end{equation}
define $\Delta S=S(h_f)-S(h_i)$ and $\Delta Q=Q(h_f)-Q(h_i)$ (with
$Q$ being, as usual, a charge expressed in units of $e$). Then, if
$\Delta S\neq 0$, an allowed transition\index{allowed transitions}
satisfies
\begin{equation}
\label{eq2.49}
\Delta S=\Delta Q
\end{equation}
whereas the processes with $\Delta S\neq\Delta Q$ are strongly
suppressed.

As an illustrative example of validity of this second rule one may
consider e.g. the decay $\Sigma^-\rightarrow ne^-\bar\nu_e$
discussed earlier in this section. For this rare, yet
well-established process one has $\Delta S$ = $\Delta Q$ = 1. On
the other hand, its counterpart $\Sigma^+\rightarrow ne^+\nu_e$
(which naively would be conceivable) has not been observed; there
is an upper bound for its decay width that can be expressed as
\begin{equation}
\label{eq2.50}
\Gamma(\Sigma^+\rightarrow ne^+\nu_e)/\Gamma(\Sigma^-\rightarrow
ne^-\bar\nu_e)<0.04
\end{equation}
Another example that should be quoted here is the spectacular
suppression of weak decays with $\Delta S\neq 0$ and $\Delta Q=0$
(\qq{weak neutral-current processes}). In particular, the decay
$K^+\rightarrow\pi^+ e^+ e^-$ (for which $\Delta S=-1$ and $\Delta
Q=0$) has been measured to have the branching ratio of about
$3\times 10^{-7}$, while the branching ratio of its natural
counterpart $K^+\rightarrow\pi^0 e^+\nu_e$ (called
$K_{e3}$\index{Ke3@$K_{e3}$ decay}) is roughly
$5\%$\index{branching ratios!of $K$-decay}\index{decay!of the
kaon}. There are many other examples of unseen processes with
$\Delta S\neq 0$, $\Delta Q=0$, for which rather stringent upper
bounds are available.\footnote{Note that two other weak neutral
strangeness-changing processes have been observed recently, namely
$K^+\rightarrow\pi^+\nu\bar\nu$ and
$K^+\rightarrow\pi^+\mu^+\mu^-$ (see \cite{ref20}), with branching
ratios of about $10^{-10}$ and $10^{-8}$ resp.}

Of course, in a model involving only charged
currents\index{charged current|(}, there is no place for $\Delta
Q=0$ weak transitions in the lowest order (though they are
conceivable as higher-order effects, e.g. at the level of one-loop
Feynman graphs). The original assumption of Feynman and Gell-Mann
\cite{ref11} (adopted by Cabibbo \cite{ref19} as well) actually
was that in the weak interaction Lagrangian there were $no$
neutral currents\index{neutral current} at all, since at that time
there had been no phenomenological need for them. Such an
assumption is obviously correct as far as the strangeness-changing
neutral currents\index{strangeness-changing neutral current} are
concerned; however, as we know now, {\it strangeness-conserving\/}
weak neutral currents do play an important role in the standard
model of electroweak unification -- this issue will be discussed
in detail in Chapter~\ref{chap7}.

Now we are going to focus on the structure of hadronic weak
charged currents complying with the above-mentioned
phenomenological constraints. In fact, it is quite easy to
construct currents satisfying the empirical rules $|\Delta S|\leq
1$ and $\Delta S=\Delta Q$, if one adopts the quark model
\cite{ref21}. Within such a framework it is natural to view a
hadronic weak transition as a process involving a pair of quarks
(or antiquarks, or a quark-antiquark pair), possibly with some
other ones playing the role of  \qq{spectators}. One is then led
to write the operators in (\ref{eq2.47}) in terms of the Dirac
spinor fields of quarks $u$ (up), $d$ (down) and $s$ (strange)
simply
as\index{u-quark@$u$-quark}\index{dquark@$d$-quark|(}\index{s-quark@$s$-quark|(}
\begin{eqnarray}
\label{eq2.51}
J_\rho^{(\Delta S=0)}&=&\bar\psi_u\gamma_\rho(1-\gamma_5)\psi_d
\nonumber\\
J_\rho^{(\Delta S\neq 0)}&=&\bar\psi_u\gamma_\rho(1-\gamma_5)\psi_s
\end{eqnarray}
Taking into account the charge assignments $Q_u\;=\;2/3$ and
$Q_d\;=\;Q_s\;=-1/3$, as well as the strangeness of the $s$-quark
being $-1$ (the $u$ and $d$ quarks of course have zero
strangeness), it is easy to see that weak hadronic transitions
mediated by the currents (\ref{eq2.51}) indeed obey automatically
the necessary empirical rules: at the quark level, the basic
transitions are $d\rightarrow u$ and $s\rightarrow u$ (or
conjugated processes) so that one has clearly $\Delta S$ = 0 or
$\Delta S$ = 1 and for $s\rightarrow u$ the relation $\Delta S$ =
$\Delta Q$ holds obviously.

An important aspect of the representation of weak hadronic
currents\index{hadronic current|(} in terms of quark fields is
that the original concept of a universal {\it four-fermion\/}
interaction\index{four-fermion interaction} is thereby restored --
as we shall see later, this plays a crucial role in the
formulation of the electroweak SM. Within the context of the
provisional phenomenological weak interaction theory, it is
certainly gratifying that the expressions (\ref{eq2.51}) reproduce
automatically the empirically established selection rules, but
they in fact represent more than mere mnemonics. The point is that
the quark currents (\ref{eq2.51}) can be conveniently recast in a
form exhibiting their transformation properties under the
approximate \qq{flavour $SU(3)$} symmetry\index{flavour!$SU(3)$
symmetry|(}\index{SU(3) group@$SU(3)$ group|ff} (actually this was
originally done by Cabibbo \cite{ref19} before the emergence of
the quark model). In particular, one has
\begin{equation}
\label{eq2.52}
\bar\psi_u\gamma_\rho(1-\gamma_5)\psi_d=\bar\psi_q\gamma_\rho
(1-\gamma_5)\frac{\lambda^1+i\lambda^2}{2}\psi_q
\end{equation}
and
\begin{equation}
\label{eq2.53}
\bar\psi_u\gamma_\rho(1-\gamma_5)\psi_s=\bar\psi_q\gamma_\rho(1-\gamma_5)
\frac{\lambda^4+i\lambda^5}{2}\psi_q
\end{equation}
where $\psi_q$ denotes the triplet
\begin{equation}
\label{eq2.54} \psi_q=\bm{
\psi_u\\
\psi_d\\
\psi_s}
\end{equation}
(belonging to the fundamental representation of the
$SU(3)_\ti{flavour}$) and the $\lambda^a$,~$a=1,2,4,5$ are
Gell-Mann matrices\index{Gell-Mann matrices}
\begin{alignat}{2}
\label{eq2.55} \lambda^1&=\left(
\begin{array}{ccc}0&1&0\\1&0&0\\0&0&0\end{array}
\right) &\hspace{1cm}&\lambda^2=\left(
\begin{array}{ccc}0&-i&0\\i&0&0\\0&0&0\end{array}
\right)\notag\\[0.2cm]
%\nonumber\\
%\notag\\
\lambda^4&=\left(
\begin{array}{ccc}0&0&1\\0&0&0\\1&0&0\end{array}
\right) &\hspace{1cm}&\lambda^5=\left(
\begin{array}{ccc}0&0&-i\\0&0&0\\i&0&0\end{array} \right)
\end{alignat}
The verification of the identities (\ref{eq2.52}), (\ref{eq2.53})
is a trivial algebraic exercise. Thus, the weak hadronic current
operator can be written as
\begin{equation}
\label{eq2.56}
J^{(hadron)}_\rho=\cos\theta_C J^{(\Delta S=0)}_\rho+\sin\theta_C
J^{(\Delta S=1)}_\rho
\end{equation}
where
\begin{equation}
\label{eq2.57}
J^{(\Delta S=0)}_\rho=V^{1+i2}_\rho-A^{1+i2}_\rho\;,\;
J^{(\Delta S=1)}_\rho=V^{4+i5}_\rho-A^{4+i5}_\rho
\end{equation}
(the superscripts in (\ref{eq2.57}) represent an obvious shorthand
notation for the combinations of Gell-Mann matrices appearing in
(\ref{eq2.52}), (\ref{eq2.53})). The utility of such an algebraic
form becomes clear when one takes into account the familiar
classification of known hadrons within (approximate) $SU(3)$
flavour multiplets (the famous \qq{eightfold way}\index{eightfold
way} \cite{GeN}): in the limit of exact $SU(3)$ flavour symmetry
one can calculate matrix elements of weak currents between
hadronic states by means of the general Wigner--Eckart theorem; it
turns out that e.g. all relevant matrix elements between baryon
octet states can be expressed in terms of only two independent
\qq{reduced matrix elements}\index{reduced matrix element}, which
must eventually be measured (these phenomenological parameters are
directly related to the coefficients $f$ and $\tilde{f}$
describing neutron beta decay and the process $\Sigma^-\rightarrow
ne^-\bar\nu_e$). In this way, one is able to get some non-trivial
predictions for semileptonic baryon decays. Of course, the $SU(3)$
flavour symmetry is in fact broken, so that such a calculational
scheme provides only an approximate description of the weak decays
of real hadrons. Nevertheless, the results agree reasonably well
with empirical data, so one can say that the weak interaction
theory formulated in terms of the quark fields does have some
predictive power at the level of physical hadrons. More details
concerning the Cabibbo theory of semileptonic baryon decays can be
found e.g. in \cite{CoB} or \cite{Geo}.

In any case, it should be stressed that for weak processes of the
beta-decay type (i.e. for semileptonic decays of mesons and
baryons) one may use again the \qq{rule
$G^2\Delta^5$}\index{rule!G@$G^2 \Delta^5$} to estimate
approximately the corresponding partial decay rates; one only has
to include correctly the relevant coupling strength, namely
$G_F\cos\theta_C$ or $G_F\sin\theta_C$ for strangeness-conserving
or strangeness-changing decays resp. and relate the estimated
quantity to that of an appropriate \qq{reference} process (e.g.
neutron beta decay) -- cf. the discussion around the formula
(\ref{eq1.131}).

Let us now summarize the model of universal weak interaction,
established in the 1960s and generally accepted before the advent of
the modern gauge theories. The interaction Lagrangian (due to
Feynman, Gell-Mann, Cabibbo, etc.) can be written as
\begin{equation}
\label{eq2.58} \lagr^{(w)}_{int}=-\frac{G_F}{\sqrt 2}J^\rho
J^\dagger_\rho
\end{equation}
with the charged current
\begin{equation}
\label{eq2.59}
J_\rho=\bar\nu_e\gamma_\rho(1-\gamma_5)e+\bar\nu_\mu\gamma_\rho
(1-\gamma_5)\mu+\bar u\gamma_\rho(1-\gamma_5)(d\cos\theta_C+s\sin
\theta_C)
\end{equation}
\index{charged current|)}(for the sake of brevity we have denoted
here all fermion fields by means of the corresponding particle
labels -- we will stick to this shorthand notation henceforth).
The last expression indicates that we actually have to do with a
rotation in the space of quark fields. One can also say that the
Cabibbo angle describes a mixing between the quarks
$d$\index{dquark@$d$-quark|)} and $s$\index{s-quark@$s$-quark|)},
that carry the same electric charge but differ in flavour. As we
shall see in Chapter~\ref{chap7}, these hints become transparent within the
framework of the electroweak SM.

The reader should bear in mind that the simple and elegant form
(\ref{eq2.58}), (\ref{eq2.59}) resulted primarily from intricate
confrontation of the earlier provisional Fermi-type models with
experimental data; nevertheless, at some stages of the
development, brilliant insight and intuition of theorists played
an important role as well. The current-current four-fermion
Lagrangian (\ref{eq2.58}) will serve later on as an appropriate
starting point of our path towards the electroweak unification,
but now we are going to examine further phenomenological
applications of this low-energy effective theory of weak
interactions\index{Cabibbo angle|)}\index{strangeness|)}.

%\input{kniha26}  %            2.6
%%%%%%%%%%%%%%%%%%%%%%%%%%%%%%%%%%%%%%%%%%%%%%%%%%%%%%%%%%%%%%%%%%%
%%%%%%%%%%%%%%%%%%%%%%%%%%%%%%%%%%%%%%%%%%%%%%%%%%%%%%%%%%%%%%%%%%%%%%%%%%%%%%%%%%%%%%%%%%%%%%%%%%%%%%%%%%%%%%%%%%%%%%%%%%%%%%%%%%%%%%%%
\section{Pion decays into leptons}

\index{decay!of the pion|(}A familiar weak process that can be
calculated rather easily is the decay of a charged pion into a
pair of leptons, i.e.
\begin{equation}
\label{eq2.60} \pi^-\rightarrow \ell^-+\bar\nu_\ell
\end{equation}
(or $\pi^+\rightarrow \ell^+ +\nu_\ell$), where $\ell = e$ or
$\mu$. For the sake of brevity, we will denote such a decay
process as $\pi_{\ell 2}$\index{pil2@$\pi_{\ell 2}$ decay|ff}. The
evaluation of the lowest-order matrix element for (\ref{eq2.60})
may proceed as follows. The final state in (\ref{eq2.60}) can be
obtained by applying the appropriate creation operators to
vacuum\index{vacuum|ff}, namely $| f\rangle = b^+(p)d^+(k)|
0\rangle$, where the $k$ and $p$ denote the antineutrino and
charged lepton four-momenta respectively. As for the initial
state, we will write this rather symbolically as $| i \rangle =
|\pi^-(q)\rangle$ with $q=k+p$; without introducing an effective
pion field one can hardly do more. The first-order $S$-matrix
element is expressed through $\langle f|\lagr_{int}| i \rangle$,
which involves the conjugated (bra) vector $\langle f| =\langle 0|
d(k)b(p)$. Thus it becomes clear that the part of the weak
interaction Lagrangian responsible for (\ref{eq2.60}) should
certainly contain a piece $\bar
\ell\gamma_\rho(1-\gamma_5)\nu_\ell$ (descending from the
Hermitean conjugate current $J^\dagger_\rho$ in (\ref{eq2.58})).
On the other hand, (\ref{eq2.60}) is clearly a
strangeness-conserving process -- the $\pi^-$ can be considered as
a $\bar u d$ state within the quark model. Hence, the relevant
interaction Lagrangian describing the decay (\ref{eq2.60}) is
\begin{equation}
\label{eq2.61} \lagr^{(\pi_{\ell 2})}_{int}=-\frac{G_F}{\sqrt
2}\cos\theta_C[\bar \ell\gamma_\rho(1-\gamma_5)\nu_\ell][\bar
u\gamma^\rho(1-\gamma_5)d]
\end{equation}
Of course, we are not able to take straightforwardly a matrix
element of the operator (\ref{eq2.61}) between the initial pion
state and the final leptonic state -- the hadronic current is
expressed in terms of quark fields while the pion\index{pion} is a
composite state involving strong-interaction dynamics that cannot
be treated by means of perturbative methods. Nevertheless, we may
at least try to write the hadronic part of the matrix element on
general grounds; as we shall see, even so one is able to make some
interesting predictions for the corresponding decay rates. To this
end, we will naturally assume that the quark current connects only
hadronic states (including vacuum) and, similarly, that the lepton
current has no non-trivial matrix elements between leptonic and
hadronic states (with the only possible exception of hadronic
vacuum). Now, when evaluating the matrix element of the
interaction Lagrangian (\ref{eq2.61}) in question, let us imagine
inserting a complete set of states between the quark and lepton
currents. Taking into account the above-mentioned assumptions, one
is then obviously led to the conclusion that out of the whole
infinite sum of such intermediate states, the only non-trivial
contribution is provided by the vacuum insertion, i.e., one may
write
\begin{eqnarray}
\label{eq2.62} \lefteqn{\langle
\ell^-(p)\bar\nu_\ell(k)|\lagr_{int}^{(\pi_{\ell
2})}(x)|\pi^-(q)\rangle=
-\frac{G_F}{\sqrt 2}\cos\theta_C\times}\\
& &\langle \ell^-(p)\bar\nu_\ell(k)|\bar \ell
(x)\gamma_\rho(1-\gamma_5)\nu_\ell(x)| 0 \rangle \times \langle
0|\bar u(x)\gamma^\rho(1-\gamma_5)d(x)|\pi^-(q) \rangle \nonumber
\end{eqnarray}
Proceeding from (\ref{eq2.62}) to the lowest-order $S$-matrix
element $S_{fi}$ and subsequently to the usual relativistically
invariant matrix element ${\cal M}_{fi}$, it is easy to see that
the leptonic part will contribute  a factor of the form $\bar
u(p)\gamma_\rho(1-\gamma_5)v(k)$ to the $\cal M$, with $u$ and $v$
being the corresponding Dirac spinors for the $\ell^-$ and
$\bar\nu_\ell$ respectively. The hadronic part of (\ref{eq2.62})
must then supply the necessary further factors making the $\cal M$
Lorentz-invariant -- in other words, it must be a four-vector. Of
course, this can only depend on the pion four-momentum $q$, so
that the most general covariant hadronic contribution entering the
decay matrix element $\cal M$ can be written as $F(q^2)q^\rho$
with the \qq{formfactor} $F$ being an essentially arbitrary
function. However, we consider the decay of a physical (on-shell)
pion, i.e. $q^2$ = $m^2_\pi$ and therefore $F(q^2=m^2_\pi)$ is
simply a constant, which we denote as $F_\pi$. As a result of
these considerations, the matrix element for the decay
$\pi^-(q)\rightarrow \ell^-(p)+\bar\nu_\ell(k)$ can be written as
\begin{equation}
\label{eq2.63} {\cal M}_{\pi_{\ell 2}}=-\frac{G_F}{\sqrt
2}\cos\theta_C F_\pi q^\rho \bar u(p)\gamma_\rho(1-\gamma_5)v(k)
\end{equation}
The last expression suggests a convenient change of notation,
namely
\begin{equation}
\label{eq2.64}
F_\pi=f_\pi\sqrt 2
\end{equation}
that we will use in the sequel. The $f_\pi$, called \qq{pion decay
constant}\index{pion!decay constant}, is the only free parameter
entering our description of the $\pi_{\ell 2}$ decay processes and
its value must be determined experimentally (from the measured
lifetime of the charged pion\index{lifetime!of the pion|ff}).

  The main message of the preceding simple considerations should
be that the relevant matrix element for a $\pi_{\ell 2}$ decay can
in fact be written almost by heart. Nevertheless, before
proceeding further, a brief commentary on the definition of the
$f_\pi$ may be useful. In formal terms, this is actually defined
as follows. First, using translational covariance of the field
operators, the $x$-dependence of the matrix element of the $V-A$
hadronic current in (\ref{eq2.62}) is easily factorized as
\begin{eqnarray}
\langle 0| V^{1+i2}_\rho (x)-A^{1+i2}_\rho(x)|\pi^-(q)\rangle=
\text{e}^{-iqx} \langle 0| V^{1+i2}_\rho
(0)-A^{1+i2}_\rho(0)|\pi^-(q)\rangle \nonumber
\end{eqnarray}
Now, since the pion is a pseudoscalar meson\index{pseudoscalar
meson}, only the axial-vector current in the last expression can
give a non-zero contribution and this will be a true Lorentz
vector (apart from the conventional normalization factor for the
one-pion state) depending on the pion four-momentum $q$ only. Note
that a corresponding matrix element of the vector current would
have to be a pseudovector, but obviously there is no such thing
that could be written in terms of a single four-momentum $q$. For
the on-shell pion one may thus write finally
\begin{eqnarray}
\langle 0| A^{1+i2}_\rho (0)|\pi^-(q)\rangle=-N(q)\sqrt{2}f_\pi
q_\rho \nonumber
\end{eqnarray}
where the normalization factor $N(q)$ is taken to be
$(2\pi)^{-3/2}(2E(q))^{-1/2}$ in accordance with our conventions
(cf. Appendix~\ref{appenB}). This relation can eventually be used in the
calculation of the lowest-order $S$-matrix element and the result
(\ref{eq2.63}) for the $\cal M$ is thus reproduced. Let us remark
that our definition of the constant $f_\pi$ differs from the
convention used in the literature by an inessential phase factor
of $-i$. Finally, one should also note that the $f_\pi$ is in fact
a fundamental parameter describing the spontaneous breakdown of
chiral symmetry\index{chiral symmetry} in the theory of strong
interactions\index{strong interaction} (see e.g. \cite{Geo}), but
we will not elaborate here on this profound aspect of the pion
decay constant.

Let us now proceed to calculate the decay width corresponding to
(\ref{eq2.63}). To this end, it is useful first to simplify the
matrix element by means of the equations of motion. In particular,
setting in (\ref{eq2.63}) $q=k+p$, one may utilize the Dirac
equations $\bar u(p)\slashed{p}=m_\ell\bar u(p)$ and
$\slashed{k}v(k)=0$ (as usual, neutrino is taken to be massless
for simplicity) and the $\pi_{\ell 2}$ matrix element thus becomes
\begin{equation}
\label{eq2.65} {\cal M}_{\pi_{\ell 2}}=-G_F\cos\theta_C f_\pi
m_\ell\bar u(p)(1-\gamma_ 5)v(k)
\end{equation}
Squaring the last expression, summing over the lepton spins and
employing the usual trace techniques, one gets
\begin{eqnarray}
\label{eq2.66} \overline{|{\cal M}|^2}&=&G^2_F\cos^2\theta_C
f^2_\pi m^2_\ell
\Tr[(\slashed{p}+m_\ell)(1-\gamma_5)\slashed{k}(1+\gamma_5)]
\nonumber\\
&=&2G^2_F\cos^2\theta_C f^2_\pi m^2_\ell
\Tr(\slashed{k}\slashed{p})
\nonumber\\
&=&8G^2_F\cos^2\theta_C f^2_\pi m^2_\ell (k\cdot p)
\end{eqnarray}
and using $2k\cdot p = (k+p)^2-k^2-p^2 = m^2_\pi -m^2_\ell$ this
is recast as
\begin{equation}
\label{eq2.67} \overline{|{\cal M}_{\pi_{\ell
2}}|^2}=4G^2_F\cos^2\theta_C f^2 _\pi m^2_\ell(m^2_\pi -m^2_\ell)
\end{equation}
To get the decay rate we also need the two-body phase space for
the final-state leptons. According to the general formula shown in
the Appendix~\ref{appenB} this is $\LIPS_2 = (4\pi m_\pi)^{-1}|{\vec p} |$,
where $\vec p$ is a lepton momentum in the rest system of the
decaying pion. The energy conservation $\sqrt{|\vec p
|^2+m^2_\ell}+|\vec p|=m_\pi$ yields the solution $|\vec
p|=(m^2_\pi -m^2_\ell)/(2m_\pi)$, so that
\begin{equation}
\label{eq2.68} \LIPS_2=\frac{1}{8\pi}\Bigl(1-\frac{m^2_\ell}{m^2_\pi}\Bigr)
\end{equation}
Putting all the necessary factors together, the result for the
decay width can be written as
\begin{equation}
\label{eq2.69} \Gamma(\pi^-\rightarrow
\ell^-+\bar\nu_\ell)=\frac{G^2_F}{4\pi}\cos^2 \theta_C f^2_\pi
m^2_\ell m_\pi\Bigl(1-\frac{m^2_\ell}{m^2_\pi}\Bigr)^2
\end{equation}
It is easy to see that the same result holds for a process
$\pi^+\rightarrow \ell^+\nu_\ell$ as well. Of course, the formula
(\ref{eq2.69}) does not represent a pure prediction, as it
involves the hitherto arbitrary pion decay constant $f_\pi$.
Rather it can be used for a determination of the $f_\pi$ by
comparing the calculated pion decay rate with its measured
lifetime. Using the result (\ref{eq2.69}) for $\ell = e,\mu$ and
the experimental value $\tau_{\pi^\pm}\doteq 2.6\times 10^{-8}$\
s, one gets
\begin{equation}
\label{eq2.70} f_\pi\doteq 93\ \MeV
\end{equation}
(needless to say, for an accurate measurement of the $f_\pi$ one
should take into account also the electromagnetic radiative
corrections\index{radiative corrections} etc., but in fact the
approximate value (\ref{eq2.70}) already represents a right number
to be remembered for further applications). On the other hand, the
result (\ref{eq2.69}) does entail a clear-cut prediction for the
ratio of the decay rates corresponding to the electronic and
muonic modes. Indeed, taking
\begin{equation}
\label{eq2.71}
R_{e/\mu}=\frac{\Gamma(\pi^-\rightarrow e^-+\bar\nu_e)}
{\Gamma(\pi^-\rightarrow \mu^-+\bar\nu_\mu)}
\end{equation}
the $f_\pi$ drops out and one gets readily
\begin{equation}
\label{eq2.72}
R_{e/\mu}=\frac{m^2_e}{m^2_\mu}\frac{(m^2_\pi-m^2_e)^2}
{(m^2_\pi-m^2_\mu)^2}
\end{equation}
Putting in numbers, namely $m_\pi\doteq 139.6\ \MeV$, $m_\mu\doteq
105.6\ \MeV$ and $m_e\doteq 0.5\ \MeV$, the relation
(\ref{eq2.72}) yields $R_{e/\mu}\doteq1.28\times 10^{-4}$, which
agrees very well with the experimental result
$R^{exp.}_{e/\mu}=(1.230\pm 0.004) \times 10^{-4}$ (note that the
branching ratio for the muonic decay mode thus constitutes about
99.99\%).

Such a dramatic suppression of the electronic decay mode
relatively to the muonic mode may be somewhat surprising at first
sight, since the phase-space volume obviously prefers the
$\pi_{e2}$ mode:
\begin{equation}
\label{eq2.73} \frac{\LIPS_2(\pi_{e2})}{\LIPS_2(\pi_{\mu2})}=
\frac{m^2_\pi-m^2_e}{m^2_\pi-m^2_\mu}\doteq2.34
\end{equation}
Of course, the suppression of the $\pi_{e2}$ decay mode is due to
the factorization of the squared lepton mass in the formula
(\ref{eq2.69}) and this in turn can be easily traced back to the
factor of $q^\rho$ in the basic matrix element (\ref{eq2.63}).
This factor clearly reflects the (pseudo)vector character of the
weak current; one may thus say that the observed suppression of
the $\pi_{e2}$ decay mode provides a stringent test of the  nature
of weak interactions.\footnote{The beautiful prediction
(\ref{eq2.72}) was made first by Feynman and Gell-Mann in their
fundamental paper \cite{ref11}. It is amusing to notice that they
considered the predicted numerical value of $R_{e/\mu}$ a serious
problem for their theory, since no $\pi_{e2}$ decay was observed
experimentally at that time and the estimated upper bound for the
$\pi_{e2}$ branching ratio was consequently much too low to fit in
(\ref{eq2.72}). Feynman and Gell-Mann also remarked that they
\qq{had no idea on how such a discrepancy could be resolved}.
Needless to say, the problem was solved by the experimentalists
later on and the result (\ref{eq2.72}) became a triumph for the
theory of weak interactions based on vector and axial-vector
currents.} Indeed, if the weak interaction had e.g. \qq{scalar
character} (i.e. the weak currents were a combination of Lorentz
scalars and pseudoscalars), the pion-to-vacuum matrix element of
the (pseudoscalar) hadronic current\index{hadronic current|)}
would essentially reduce to a constant $f_\pi$ alone and thus one
would be led to a prediction
\begin{equation}
\label{eq2.74}
R^{(scalar)}_{e/\mu}=\frac{(m^2_\pi-m^2_e)^2}
{(m^2_\pi-m^2_\mu)^2}\doteq 5.5
\end{equation}
On the other hand, the reader should realize that
%the successful result (\ref{eq2.72})
in fact one does not need precisely the $V-A$ currents to achieve
(\ref{eq2.72}); from our derivation it should be obvious that any
combination of $V$ and $A$ currents would give the same result,
provided that the $A$ component is non-trivial (let us recall once
again that $A$ is needed because the intrinsic parity of pion is
$-1$).

Two remarks are in order here. First, the proportionality of the
$\pi_{\ell 2}$ matrix element to $m_\ell$ and the ensuing
suppression of the electronic decay mode, shown to be
characteristic feature of a weak interaction model of the ($V,A$)
type, can also be easily understood on the basis of chirality and
angular momentum conservation. \index{chirality}Indeed, let us
consider the limit $m_\ell=0$. Both $V$ and $A$ interactions
preserve chirality (formally, this is due to $\gamma_ \alpha$ and
$\gamma_\alpha\gamma_5$ anticommuting with $\gamma_5$). For
massless leptons, this entails a simple helicity\index{helicity}
selection rule: the $\ell^-$ and $\bar\nu_\ell$ can only be
produced with opposite helicities (i.e. a left-handed $\ell$ must
be accompanied by right-handed $\bar\nu_\ell$ and vice versa). On
the other hand, since the pion spin is zero, angular momentum
conservation obviously requires that $\ell^-$ and $\bar\nu_\ell$
helicities be equal in their c.m. system. Thus, the $\pi_{\ell 2}$
decay is forbidden for $m_\ell=0$ within a ($V,A$) weak
interaction theory. For $m_\ell\neq 0$, a helicity flip is
possible (though the chirality selection rules remain valid) and
the matrix element corresponding to the $\ell^-$ and
$\bar\nu_\ell$ with like helicities is then naturally proportional
to $m_\ell$. In particular, within the $V-A$ theory, the
(massless) $\bar\nu_\ell$ is of course produced as right-handed in
$\pi^-\rightarrow \ell^-\bar\nu_\ell$ and hence the (massive)
$\ell^-$ must also be emitted with positive helicity, owing to the
angular momentum conservation (the reader is recommended to
verify, by means of an explicit calculation, that the probability
of an emission of left-handed electron or negative muon indeed
vanishes).

Second, looking towards the present-day Standard Model, it is
important to emphasize that  experimental confirmation of the
remarkable result (\ref{eq2.72}) implies a definite message for
weak interaction models involving an intermediate
boson\index{intermediate boson} (which were \qq{on the market}
since the early days of weak interaction theory): {\it vector\/}
$W$ boson\index{W boson@$W$ boson} (i.e. that of spin 1) coupled
to $V$ and $A$ currents is thereby clearly favoured over the other
possibilities (e.g. $W$ of spin 0 or 2 corresponding to scalar or
tensor couplings).

The simple calculational techniques discussed above can also be
successfully applied to other processes, closely related to the
$\pi_{\ell 2}$ decays. There are at least two \qq{classic}
applications that should be mentioned before closing this section,
namely the leptonic decays of a charged kaon\index{decay!of the
kaon} (the $K_{\ell 2}$ decays\index{Kl2@$K_{\ell 2}$ decay} with
$\ell$ = $e$ or $\mu$) and the decay process $\tau^-\rightarrow
\pi^-+\nu_\tau$ (or $\tau^-\rightarrow K^-+\nu_\tau$ resp.). Let
us start with the $K_{\ell 2}$ decays. In such a case one may
repeat essentially all the steps that led us previously to the
formula (\ref{eq2.63}), except that now one has to replace
$\cos\theta_C$ by $\sin\theta_C$ (we are dealing with a
strangeness-changing process) and the corresponding \qq{kaon decay
constant}\index{kaon decay constant} $f_K$ may in general be
different from the $f_\pi$. It is useful to remember that under an
assumption of exact $SU(3)$ flavour symmetry, the $f_K$ and
$f_\pi$ would be equal  -- although such a statement may not be
immediately obvious, it is not difficult to realize that the kaon
carries the same quantum numbers as the $\Delta S$ = 1 weak
current (cf. (\ref{eq2.53})). Thus, knowing the $f_\pi$ value from
$\pi_{\ell 2}$ decays, one can make quite reasonable predictions
for the $K_{\ell 2}$ decay rates at least in the $SU(3)$ flavour
symmetry limit. Of course, in the real world the
$SU(3)_\ti{flavour}$ is broken and such a symmetry prediction may
be reliable only with an accuracy of about 20\%. One may best assess
this accuracy by comparing a measured value of the $f_K$ with the
$f_\pi$ found before. The $f_K$ can be determined by matching the
formula for the $K_{\ell 2}$ decay rate
\begin{equation}
\label{eq2.75} \Gamma(K^-\rightarrow
\ell^-+\bar\nu_\ell)=\frac{G^2_F}{4\pi}\sin^2 \theta_C f^2_K
m^2_\ell m_K(1-m^2_\ell/m^2_K)^2
\end{equation}
(cf. (\ref{eq2.69})) with relevant experimental data. In
particular, one may use the branching ratio
$\BR(K^-\rightarrow\mu^-+\bar\nu_\mu) \doteq
63.5\%$\index{branching ratios!of $K$-decay} together with the
kaon mean lifetime $\tau_{K^\pm} \doteq 1.24\times 10^{-8}$s (for
completeness, let us also recall that $m_{K^\pm}\doteq 494\
\MeV$). One thus gets $f_K\doteq 111\ \MeV$, i.e. $f_K/f_\pi\doteq
1.2$\index{flavour!$SU(3)$ symmetry|)}.

To describe the process $\tau^-\rightarrow
\pi^-+\nu_\tau$\index{decay!of the tau@of the $\tau$ lepton}, one
adds a term $\bar\nu_\tau\gamma_\rho(1-\gamma_5)\tau$ to the
leptonic current\index{leptonic current} in (\ref{eq2.59}). The
corresponding matrix element can then be written on similar
grounds as that obtained earlier for the $\pi_{\ell 2}$ decays.
Armed with our previous experience, we may guess the relevant
result rather easily; this obviously reads
\begin{equation}
\label{eq2.76} {\cal M}(\tau^-\rightarrow
\pi^-+\nu_\tau)=-G_F\cos\theta_C f_\pi \bar
u(k)\slashed{p}(1-\gamma_5)u(q)
\end{equation}
where we have denoted the $\tau$, $\pi$ and $\nu_\tau$
four-momenta by $q$, $p$ and $k$ resp. (the reader is
recommended to recover the formal steps leading to the last
expression). For the corresponding decay rate we then have
\begin{equation}
\label{eq2.77}
\Gamma(\tau^-\rightarrow\pi^-+\nu_\tau)=\frac{G^2_F}{8\pi}\cos^2
\theta_C f^2_\pi m^3_\tau(1-\frac{m^2_\pi}{m^2_\tau})^2
\end{equation}
Having fixed the $f_\pi$ value through the $\pi_{\ell 2}$ decays,
the last result now represents a definite prediction for the
partial decay width in question. Putting in numbers (in particular
$m_\tau= 1.777\ \GeV$)\index{tau lepton@$\tau$ lepton} and taking
into account that the $\tau$ lifetime is about $2.9\times
10^{-13}$s, the formula (\ref{eq2.77}) yields the branching ratio
$\BR(\tau^-\rightarrow\pi^-+\nu_\tau)\doteq
10.8\%$\index{branching ratios!of $\tau$-decay}, in good agreement
with the experimental value, which is $10.82\pm0.05\%$ according
to \cite{ref5}. In a similar way, we can calculate the branching
ratio for the mode $\tau^-\rightarrow K^-+\nu_\tau$, or,
alternatively, the ratio
\begin{equation}
\label{eq2.78}
\frac{\Gamma(\tau^-\rightarrow K^-+\nu_\tau)}{\Gamma(\tau^-\rightarrow
\pi^-+\nu_\tau)}=\tan^2\theta_C\frac{f^2_K}{f^2_\pi}
\frac{(1-m^2_K/m^2_\tau)^2}{(1-m^2_\pi/m^2_\tau)^2}
\end{equation}
Numerically, the last expression gives approximately 0.06, which
agrees well with experimental data (note that
$\BR(\tau^-\rightarrow K^-\nu_\tau)\doteq 7\times 10^{-3})$.

Thus, we have seen that the meson decay constants $f_\pi$ and
$f_K$ measured in $\pi_{\ell 2}$ and $K_{\ell 2}$ decays can be
used in other places as well, enabling one to make useful physical
predictions e.g. for $\tau$ decays. The argument could be reversed
-- one may e.g. imagine determining the $f_\pi$ from
$\tau^-\rightarrow\pi^-\nu_\tau$ and employing it to predict the
$\pi^\pm$ lifetime. In other words, although the pion lifetime
cannot be simply calculated from the first principles (i.e. by
using the Lagrangian (\ref{eq2.58}) only), one additional
measurement (of another process) is sufficient for accomplishing
such a prediction.

In this context, one last remark should be added. The
phenomenological parameter $f_\pi$ also plays an important role in
the decay of the neutral pion. It is well known that the $\pi^0$
decays predominantly into a pair of photons, i.e. through an
electromagnetic interaction (note that the branching ratio for
$\pi^0\rightarrow\gamma\gamma$ constitutes about 98.8$\%$ and the
mean lifetime $\tau_{\pi^0}\doteq 8.4\times 10^{-17}$s).
\index{branching ratios!of
$\pi$-decay}\index{electromagnetic!interaction}The matrix element
for the $\pi^0\rightarrow\gamma\gamma$ decay can be written with
good accuracy as
\begin{equation}
\label{eq2.79} {\cal
M}(\pi^0\rightarrow\gamma\gamma)=-\frac{1}{f_\pi}\frac
{\alpha}{\pi}\epsilon_{\mu\nu\rho\tau}k^\rho p^\tau
\varepsilon^{\ast\mu}(k)\varepsilon^{\ast\nu}(p)
\end{equation}
where the $\varepsilon^\mu(k),\varepsilon^\nu(p)$ denote the
polarization\index{polarization!of the
photon}\index{polarization!vector} vectors of the final-state
photons with four-momenta $k,p$ and $\alpha$ is the
electromagnetic fine-structure constant, $\alpha\doteq$ 1/137. The
uninitiated reader should be warned that the remarkable formula
(\ref{eq2.79}) is by no means obvious -- a comprehensive treatment
of this subject can be found e.g. in \cite{ChL}, \cite{Geo},
\cite{Don}. Here let us only add that (\ref{eq2.79}) holds in the
limit of zero pion mass (the \qq{soft pion limit}) and eventual
corrections due to finite $m_\pi$ may be of the order of one per
cent. The decay rate corresponding to (\ref{eq2.79}) is then
\begin{equation}
\label{eq2.80}
\Gamma(\pi^0\rightarrow\gamma\gamma)=\frac{1}{64\pi}m^3_\pi(\frac
{\alpha}{\pi})^2\frac{1}{f^2_\pi}
\end{equation}
Numerically, the formula (\ref{eq2.80}) yields the prediction
$7.65\ \eV$ for the decay width in question, in good agreement
with the experimental value. In view of the preceding
considerations one may now say that, alternatively, the
experimental value of the $\pi^0$ lifetime could provide the
necessary input for making a prediction for the charged pion
lifetime. In principle it is true, yet one should bear in mind
that the $\pi^\pm$ lifetime is measured with much better accuracy
than the $\pi^0$ lifetime ($\tau_{\pi^\pm}$ = (2.6033 $\pm$
0.0005) $\times 10^{-8}$s, to be compared with $\tau_{\pi^0}$ =
(8.43 $\pm$ 0.13)$\times 10^{-17}$s), i.e. the charged pion decays
provide the most accurate data for determination of the $f_\pi$
value.

%\input{kniha27}  %            2.7
%%%%%%%%%%%%%%%%%%%%%%%%%%%%%%%%%%%%%%%%%%%%%%%%%%%%%%%%%%%%%%%%%%%
%%%%%%%%%%%%%%%%%%%%%%%%%%%%%%%%%%%%%%%%%%%%%%%%%%%%%%%%%%%%%%%%%%%%%%%%%%%%%%%%%%%%%%%%%%%%%%%%%%%%%%%%%%%%%%%%%%%%%%%%%%%%%%%%%%%%%%%%
\section{Beta decay of charged pion}\label{sec2.7}
%section 2.7
\index{beta decay!of charged pion|ff} There is another possible
decay channel for the charged pion, namely
\begin{equation}
\label{eq2.81}
\pi^+\rightarrow\pi^0+e^++\nu_e
\end{equation}
(or, equivalently, $\pi^-\rightarrow\pi^0 e^-\bar\nu_e$). This may
be naturally called \qq{pion beta decay} and is sometimes denoted
as $\pi_{e3}$\index{Pie3@$\pi_{e3}$ decay|ff}. Its muonic analogue
is obviously precluded by energy conservation. The decay mode
(\ref{eq2.81}) is in fact very rare -- its branching ratio
constitutes only about $10^{-8}$, as one can easily guess on the
basis of the \qq{rule $G^2\Delta^5$}\index{rule!G@$G^2 \Delta^5$}
(cf. Section~\ref{sec2.5}); the essential point is that masses of the
charged and neutral pion are rather close
($m_\pi^+-m_{\pi^0}\doteq 4.6\ \MeV$) so that the available phase
space for the decay products is small. Nevertheless, despite being
so rare, the pion beta decay is extremely interesting, since it
serves as a \qq{test bench} for some basic ideas of the theory of
hadronic weak interactions. In particular, it provides an
important check on the properties of the vector part of the weak
current. As we shall see, the $\pi_{e3}$ decay width is calculable
in a theoretically clean way, without introducing further
phenomenological parameters (in contrast to the $\pi_{\ell 2}$
decays discussed in the preceding section), i.e. one essentially
gets a pure prediction based on the Lagrangian (\ref{eq2.58}).
Such a theoretical result then can be compared with the
corresponding experimental value that has been measured with good
accuracy; according to \cite{ref5} one has
\begin{equation}
\label{eq2.82} \text{BR}_{exp}(\pi^+\rightarrow\pi^0
e^+\nu_e)=(1.036\pm 0.006)\times 10^{-8}
\end{equation}
In the rest of this section we will explain how the calculation
can be done. From the lepton content of the final state in
(\ref{eq2.81}) it is easy to guess that the relevant part of the weak
interaction Lagrangian (\ref{eq2.58}) is
\begin{equation}
\label{eq2.83}
\lagr^{(\pi_{e3})}_{int}=-\frac{G_F}{\sqrt 2}\cos\theta_C[\bar d
\gamma_\mu(1-\gamma_5)u][\bar\nu_e\gamma^\mu(1-\gamma_5)e]
\end{equation}
In analogy with the discussion of preceding section, it should be
clear that the $S$-matrix element in the first order of
perturbation theory will be factorized into the hadronic and
leptonic parts. Thus, one needs to know matrix elements of the
currents appearing in (\ref{eq2.83}), namely
\begin{equation}
\label{eq2.84} \langle \pi^0(p)|\bar
d(x)\gamma_\mu(1-\gamma_5)u(x)|\pi^+(P) \rangle
\end{equation}
 and
\begin{equation}
\label{eq2.85} \langle
e^+(k)\nu_e(l)|\bar\nu_e(x)\gamma^\mu(1-\gamma_5)e(x)|0 \rangle
\end{equation}
Note that in (\ref{eq2.84}) and (\ref{eq2.85}) we have marked
explicitly the corresponding particle momenta. Of course, the
leptonic term can be evaluated in a straightforward way, so let us
focus first on the hadronic matrix element. Before we proceed to
work it out, the reader should be warned that we are going to
employ some tricks characteristic for the so-called \qq{current
algebra}\index{current algebra} -- a basic and powerful technique
in hadron physics developed mostly in the 1960s, which, however, is
not commonly used in the bulk of this text. A brief review of the
current-algebra ideas can be found e.g. in \cite{ChL}.

  For our purpose, the basic observations are as follows. First
of all, one should realize that both $\pi^+$ and $\pi^0$ are
pseudoscalar mesons\index{pseudoscalar meson} and therefore only
the $vector$ part of the weak quark current in (\ref{eq2.84}) can
contribute to the matrix element in question.\footnote{The matrix
element of the axial-vector part would be a pseudovector in the
considered case, but a pseudovector obviously cannot be
constructed from two independent four-vectors (the four-momenta
$p$ and $P$). Thus, in a sense, we now have a situation opposite
to that encountered previously in the case of the $\pi_{\ell 2}$
decays.} Next, one may recall that the algebraic structure of the
quark current corresponds to
\begin{equation}
\label{eq2.86}
\bar d\gamma_\mu u=V^1_\mu-iV^2_\mu
\end{equation}
where the $V^1_\mu$ and $V^2_\mu$ are defined in accordance with
Section~\ref{sec2.5} (cf. the discussion around the relation
(\ref{eq2.57})). In the present context, one actually does not have
to use the Gell-Mann matrices -- the Pauli matrices will do. The
last relation thus reads, explicitly
\begin{eqnarray}
\label{eq2.87} \bar d\gamma_\mu u&=&\bm{\bar u\;, &\!\!\bar
d}\gamma_\mu\left [\frac{1}{2} \bm{0 & 1\\ 1 &
0}-i\frac{1}{2}\bm{0& -i\\ i & 0}
\right ]\bm{u\\ d}\nonumber\\
&=&\bar\psi_q\gamma_\mu\frac{\tau^1}{2}\psi_q-i\bar\psi_q
\gamma_\mu\frac{\tau^2}{2}\psi_q
\end{eqnarray}
(the notation should be self-explanatory). The meaning of the
symbolic identity (\ref{eq2.86}) consists in exhibiting the
isospin properties of the strangeness-conserving weak current: it
is seen that we are working with components of an isospin triplet
$V^a_\mu=\bar\psi_q\gamma_\mu\frac{\tau^a}{2}\psi_q,a=1,2,3$, in
particular with the \qq{isospin-lowering} combination $V^1_\mu-i
V^2_\mu$. Now we come to a point that is crucial for our
calculation. {\bf It turns out that the weak current in
(\ref{eq2.86}) can be expressed as a commutator of the
electromagnetic current and a pertinent combination of the isospin
charges,} namely\index{electromagnetic!current|ff}
\begin{equation}
\label{eq2.88}
V^1_\mu-iV^2_\mu=[Q^1-iQ^2,J^{(em)}_\mu]
\end{equation}
where $J^{(em)}_\mu=\frac{2}{3}\bar u\gamma_\mu u-\frac{1}{3}\bar
d\gamma_\mu d$ and
\begin{equation}
\label{eq2.89}
Q^a=\int V^a_0(\vec x,x_0)d^3 x
\end{equation}
Some technical details of the derivation of the important relation
(\ref{eq2.88}) can be found in \cite{Bai}. At this place let us
only remark that such a relation is in fact quite natural: it is
easy to realize that the electromagnetic current of the quarks $u$
and $d$ can be expressed in terms of the third component of the
isotopic triplet and an isosinglet
as\index{dquark@$d$-quark}\index{u-quark@$u$-quark}
\begin{eqnarray}
\label{eq2.90}
J^{(em)}_\mu&=&\frac{2}{3}\bar u\gamma_\mu
u-\frac{1}{3}\bar
d\gamma_\mu d\nonumber\\
&=&\bar\psi_q\gamma_\mu\frac{\tau^3}{2}\psi_q+\frac{1}{6}\bar
\psi_q \gamma_\mu \J \psi_q\nonumber\\
&=&V^3_\mu+\frac{1}{6}V^0_\mu
\end{eqnarray}
The commutators are trivial for the singlet term (since this
involves the unit matrix) and the identity (\ref{eq2.88}) thus
essentially corresponds to the familiar algebraic relations among
the isospin $SU(2)$ generators\index{SU(2) group@$SU(2)$ group}.
Last but not least, let us emphasize that the isospin currents may
be considered, with a rather good accuracy, as $conserved$
quantities (isospin is a good approximate symmetry of strong
interactions\index{strong interaction}); consequently, the charges
(\ref{eq2.89}) can be taken (approximately) as time-independent
generators of a corresponding $SU(2)$ algebra.

  The above considerations form a basis of what has been called,
historically, the \qq{CVC hypothesis} -- the acronym stands for
\qq{conserved vector current}\index{conserved vector current
(CVC)}. Within our quark picture this emerges quite naturally and
almost automatically, but in the early days of the weak
interaction theory it was a rather non-trivial assumption (an
essential \qq{leap of faith} was precisely placing the vector part
of a weak transition operator into the same multiplet with the
relevant part of the electromagnetic current -- this undoubtedly
also represents a major step towards a conceptual unification of
both forces).\footnote{There is another important conceptual
aspect of the above discussion that should perhaps be mentioned
here. The currents entering the commutators of the current algebra
originate in strong-interaction symmetries (such as the isospin)
and these \qq{symmetry currents} are subsequently identified with
physical currents participating in weak interaction dynamics.
Before the advent of quark-model Lagrangians (that gradually led
to the present-day Standard Model) such an identification was by
no means obvious.} The concept of CVC is originally due to
Gershtein and Zeldovich \cite{ref22} and Feynman and Gell-Mann
\cite{ref11}; for a rather detailed discussion see e.g. the books
\cite{BjD}, \cite{MRR}.

  Let us now show how the matrix element (\ref{eq2.84}) can be
evaluated. In view of the preceding discussion this is equal to
\begin{eqnarray}
\label{eq2.91}
&&\langle\pi^0(p)| V^1_\mu-iV^2_\mu|\pi^+(P)\rangle\nonumber\\
&=&\langle\pi^0(p)| [Q^1-iQ^2,J^{(em)}_\mu]|\pi^+(P)\rangle
\end{eqnarray}
Note that here and in what follows we may take the current
operators at the point $x$ = 0; as ever, the coordinate dependence
of the matrix element in question is essentially trivial -- it is
carried by a usual exponential factor obtained through an
appropriate space-time translation. To work out the last
expression, one has to realize that the pion states are (with good
accuracy) isospin eigenstates, while the combination $Q^1-iQ^2$ is
an isospin-lowering operator. One may then utilize the familiar
relations known e.g. from the quantum-mechanical theory of angular
momentum\index{angular momentum} (remember that algebraic
properties of the isospin and ordinary spin are formally the
same); we thus get, in particular\footnote{For reader's
convenience, let us remark that in order to arrive at
(\ref{eq2.92}) one employs the relation
\begin{center}
$Q_\pm| T,T_3\rangle=\sqrt{T(T+1)-T_3(T_3\pm 1)}| T,T_3\pm
1\rangle$
\end{center}
with $Q_\pm=Q^1\pm iQ^2$, $T$ = 1 for pion isotriplet, $T_3$ = 0
and 1 for $\pi^0$ and $\pi^+$ respectively.}
\begin{eqnarray}
\label{eq2.92}
(Q^1-iQ^2)|\pi^+(P)\rangle&=&\sqrt{2}|\pi^0(P)\rangle\nonumber\\
\langle\pi^0(p)| (Q^1-iQ^2)&=&\sqrt{2}\langle\pi^+(p)|
\end{eqnarray}
Using this, the last expression in (\ref{eq2.91}) becomes
\begin{equation}
\label{eq2.93} \sqrt{2}\bigl(\langle\pi^+(p)|
J^{(em)}_\mu|\pi^+(P)\rangle-\langle\pi^0(p) |
J^{(em)}_\mu|\pi^0(P)\rangle\bigr)
\end{equation}
but the last term vanishes identically (to see this formally, one
should realize that the electromagnetic current changes its sign
under charge conjugation $\cal C$ while the neutral pion is a
$\cal C$ eigenstate). We thus arrive at the result
\begin{eqnarray}
\label{eq2.94}
&&\langle\pi^0(p)| V^1_\mu(0)-iV^2_\mu(0)|\pi^+(P)\rangle=\nonumber\\
&=&\sqrt{2}\langle\pi^+(p)| J^{(em)}_\mu(0)|\pi^+(P)\rangle
\end{eqnarray}
which embodies, technically, the essence of the \qq{CVC relation}
relevant for the considered process. Up to conventional
normalization factors associated with the one-particle states, the
quantity (\ref{eq2.94}) is a Lorentz vector that must be made of
two independent four-momenta. Clearly, the most general form of
such a vector can be described as
\begin{equation}
\label{eq2.95} \langle\pi^+(p)|
J^{(em)}_\mu(0)|\pi^+(P)\rangle=F_+(q^2)(P+p)_\mu+F_-
(q^2)(P-p)_\mu
\end{equation}
where $q = P -p$ and the coefficients $F_\pm(q^2)$ are essentially
arbitrary functions (formfactors). Note that we consider the
on-mass-shell pions, i.e. $p^2=m^2_{\pi^0}$ and $P^2=m^2_{\pi^+}$.
However, one should not forget that the assumption of exact
isospin symmetry actually means that $m_{\pi^0}=m_{\pi^+}$, so one
has to set $p^2=P^2$ whenever the current conservation is used
explicitly. In terms of the parametrization (\ref{eq2.95}), the
current conservation is tantamount to
\begin{equation}
\label{eq2.96}
0=q^\mu[F_+(q^2)(P+p)_\mu+F_-(q^2)(P-p)_\mu]
\end{equation}
Setting there $p^2=P^2$, the first term in (\ref{eq2.96}) drops out
automatically and one is left with the condition $q^2F_-(q^2)$ = 0, i.e.
the formfactor $F_-$ has to vanish (in the considered symmetry limit).
Thus, the matrix element of the electromagnetic current is given by
\begin{equation}
\label{eq2.97} \langle\pi^+(p)|
J^{(em)}_\mu(0)|\pi^+(P)\rangle=F_\pi(q^2)(P+p)_\mu
\end{equation}
where we have denoted $F_+=F_\pi$ so as to introduce a standard
symbol for the pion electromagnetic
formfactor\index{electromagnetic!formfactors}. For small $q^2$
(which is our case) the $F_\pi(q^2)$ is a slowly varying function
and the value $F_\pi(0)$ is determined by the pion charge, i.e.
\begin{equation}
\label{eq2.98}
F_\pi(0)=1
\end{equation}
Thus, we may now state our main result as follows. With a rather
good accuracy, the matrix element (\ref{eq2.84}) is given by
\begin{eqnarray}
\label{eq2.99} &&\langle\pi^0(p)|\bar
d\gamma_\mu(1-\gamma_5)u|\pi^+(P)\rangle=
\nonumber\\
&=&\langle\pi^0(p)| V^1_\mu-iV^2_\mu|\pi^+(P)\rangle\doteq\sqrt{2}
(p+P)_\mu
\end{eqnarray}

This result is indeed remarkable, since -- as we indicated earlier
in this section -- one needs no extra phenomenological parameter
to describe the hadronic matrix element in question. Obviously, an
essential point  was that we were able to recast the whole problem
in terms of the electromagnetic current (see (\ref{eq2.94})) whose
properties are well known. It is also useful to realize that,
since only the vector part of the weak current contributes in the
considered case, the pion beta decay is in fact an example of a
pure Fermi transition within the domain of  particle physics.
  With the result (\ref{eq2.99}) at hand, we are in a position to
write down the complete matrix element for the pion beta decay,
corresponding to the Lagrangian (\ref{eq2.83}) in the lowest order. The
evaluation of the leptonic factor descending from (\ref{eq2.85}) is
essentially trivial and one thus gets readily
\begin{equation}
\label{eq2.100}
{\cal M}_{\pi_{e3}}=-G_F\cos\theta_C(p+P)^\mu[\bar
u(l)\gamma_\mu(1-\gamma_5)v(k)]
\end{equation}
The calculation of the corresponding decay rate is then routine.
For the squared matrix element summed over the lepton spins one
gets, after some algebra
\begin{eqnarray}
\label{eq2.101} {\overline{|{\cal
M}_{\pi_{e3}}|^2}}&=&2G^2_F\cos^2\theta_C
\Tr[(\slashed{k}-m_e)\slashed{Q}\slashed{l}\slashed{Q}(1-\gamma_5)]
\nonumber\\
&=&2G^2_F\cos^2\theta_C\Tr(\slashed{k}\slashed{Q}\slashed{l}
\slashed{Q})
\end{eqnarray}
where we have denoted $Q$ = $P$ + $p$ for brevity. In what
follows, we shall work in the rest frame of the decaying $\pi^+$.
Kinematically, the considered process is similar to the neutron
beta decay, as one of the decay products (the $\pi^0$) has a mass
that is very close to $m_{\pi^+}$. The maximum positron energy is
$\Delta=(m^2_{\pi^+}-m^2_{\pi^0}+m^2_e)/(2m_{\pi^+}) = 4.52\ \MeV$
($\Delta\doteq m_{\pi^+}-m_{\pi^0}$), which means that the
positron can be highly relativistic near the endpoint of its
spectrum ($\beta^{max}_e\doteq$ 0.99). On the other hand, the
recoil $\pi^0$ is safely non-relativistic over the whole
kinematical range -- it is easy to check that the maximum $\pi^0$
momentum is of the order of the mass difference
$m_{\pi^+}-m_{\pi^0}$ and constitutes thus only about 3$\%$ of its
rest mass. Using the static approximation for the $\pi^0$ (i.e.
setting $Q\doteq(m_{\pi^+}+m_{\pi^0},{\vec 0})$), one gets from
(\ref{eq2.101}), after some simple manipulations
\begin{equation}
\label{eq2.102} {\overline{|{\cal M}_{\pi_{e3}}|^2}}\doteq
32G^2_F\cos^2 \theta_C m^2_\pi E_e E_\nu(1+\beta_e\cos\vartheta)
\end{equation}
where we have also introduced (in the spirit of our kinematical
approximation) an average pion mass $m_\pi=\frac{1}{2}(m_{\pi^+}
+m_{\pi^0})$. The expression (\ref{eq2.102}) illustrates
explicitly the pure Fermi character of the considered transition
-- the coefficient of the $e^+ - \nu$ angular
correlation\index{angular correlation} is seen to be equal $+1$
(for a comparison with nuclear beta decay see e.g.
(\ref{eq1.45})).

  To obtain the $\pi_{e3}$ decay rate, one can now proceed in full
analogy with the case of neutron decay described in detail in
Section~\ref{sec1.4}. Within our kinematical approximation, the positron
energy spectrum has the familiar \qq{statistical} form
\begin{equation}
\label{eq2.103}
\frac{dw_{\pi_{e3}}(E)}{dE}\doteq\frac{1}{\pi^3}G^2_F\cos^2
\theta_C|\vec k| E(\Delta-E)^2
\end{equation}
and this can be integrated over the $E$ from $m_e$ to
$\Delta$. The $m_e$ can be neglected with reasonable accuracy
(note that such an approximation is better here than for
neutron decay, since in the present case $m_e/\Delta\doteq$ 0.1).
One thus has
\begin{eqnarray}
\label{eq2.104}
\Gamma(\pi^+\rightarrow\pi^0 e^+\nu_e)&=&\int^\Delta_{m_e}\frac
{dw_{\pi_{e3}}(E)}{dE}dE\doteq\nonumber\\
&\doteq&\frac{1}{\pi^3}G^2_F\cos^2\theta_C\int^\Delta_0 E^2
(\Delta-E)^2 dE
\end{eqnarray}
so that our final answer reads
\begin{equation}
\label{eq2.105}
\Gamma(\pi^+\rightarrow\pi^0 e^+\nu_e)\doteq\frac{G^2_F\cos^2
\theta_C}{30\pi^3}\Delta^5
\end{equation}
Putting in numbers (in particular, $\Delta = 4.52\ \MeV$), the
last expression yields
\begin{equation}
\label{eq2.106} \Gamma_{theor}(\pi^+\rightarrow\pi^0
e^+\nu_e)\doteq 2.62\times 10^{-22}\ \MeV
\end{equation}
Taking into account the measured $\pi^+$ lifetime, which
corresponds to the total width of about 2.53 $\times 10^{-14}\
\MeV$, one gets from (\ref{eq2.106}) a prediction for the
branching ratio\index{branching ratios!of $\pi$-decay}
\begin{equation}
\label{eq2.107}
\text{BR}_{theor}(\pi^+\rightarrow\pi^0e^+\nu_e)\doteq1.03\times
10^{-8}
\end{equation}
in agreement with the experimental value (\ref{eq2.82}).
  It should be stressed that, in view of the approximations made
in the course of our calculation, one should in general only
expect an accuracy at the level of several per cent (say, up to
10$\%$) in our theoretical prediction. Indeed, for possible
corrections to the basic approximation (\ref{eq2.105}) one would
have to take into account the isospin violation effects, the
recoil $\pi^0$ motion and also the $m_e\neq 0$ effects neglected
in the evaluation of the Fermi integral\index{Fermi!integral} in
(\ref{eq2.104}); each of these corrections may typically represent
several per cent for the calculated decay width. The possible
corrections to (\ref{eq2.105}) seem to be well under control and
the agreement between theory and experiment is very good; for more
details, see e.g. \cite{Bai}, \cite{MRR}. In any case, the
successful theoretical prediction (\ref{eq2.105}) certainly
represents a remarkable test of our ideas about the structure of
weak currents and the rare process $\pi_{e3}$ therefore occupies a
very important niche in theoretical particle
physics\index{decay!of the pion|)}.

%\input{kniha28}  %            2.8%
%%%%%%%%%%%%%%%%%%%%%%%%%%%%%%%%%%%%%%%%%%%%%%%%%%%%%%%%%%%%%%%%%%%
%%%%%%%%%%%%%%%%%%%%%%%%%%%%%%%%%%%%%%%%%%%%%%%%%%%%%%%%%%%%%%%%%%%%%%%%%%%%%%%%%%%%%%%%%%%%%%%%%%%%%%%%%%%%%%%%%%%%%%%%%%%%%%%%%%%%%%%%
\section{Nucleon matrix elements of the weak current}
%section 2.8
Let us now return to the process that has been a starting point of
our discussion in Chapter~\ref{chap1} -- the neutron beta decay. When this
is to be calculated within the theory described by the Lagrangian
(\ref{eq2.58}), one has to know the relevant matrix elements of
the weak current (\ref{eq2.59}). The evaluation of the leptonic
part is straightforward as ever, but a matrix element of the
hadronic current\index{hadronic current} between nucleon states is
not directly calculable. The reason is clear: in contrast to the
phenomenological approach adopted in Chapter~\ref{chap1}, the currents
appearing now in our deeper theory are not expressed explicitly in
terms of nucleon fields and, at the same time, we do not know a
precise quantitative connection between the quark fields and the
nucleon states.\footnote{The nucleon is a  composite state made of
confined quarks and its description would involve complicated
strong-interaction bound-state dynamics. A corresponding {\it ab
initio\/} calculation is thus beyond the reach of the present-day
techniques of quantum field theory.} One might say that this is
the price we have to pay for a more elegant formulation of the
weak interaction theory in terms of fundamental degrees of
freedom. Nevertheless, it should be clear that from the point of
view of practical phenomenology we in fact do not lose anything.
Indeed, for the purpose of a practical beta-decay calculation, we
may resort to the method used in preceding sections in connection
with pion decays. In particular, we can write down a most general
form of the nucleon matrix element in question, compatible with
some obvious requirements such as Lorentz covariance etc.; as
usual, the corresponding expression then involves a few
phenomenological coefficients that have to be measured anyway. As
we shall see, the resulting picture represents a natural
generalization of our old treatment (that was based on an
effective Lagrangian involving local nucleon fields) and
incorporates also some new subtle phenomena, absent within the old
provisional framework; the description developed in Chapter~\ref{chap1} is
only recovered in the limit of vanishing nucleon momentum
transfer.

Thus, how can one write the desired matrix element on general
grounds? For definiteness, let us start with the corresponding
vector part.\footnote{The discussion given here essentially
follows the lecture notes by C.~Jarlskog \cite{ref23}.} Up to the
normalization factors associated with one-particle nucleon states,
the matrix element must transform as a Lorentz four-vector and
this should be constructed in terms of the relevant Dirac spinors
for nucleons and the corresponding four-momenta (we will restrict
ourselves to the spacetime point $x$ = 0, since an appropriate
shift can be performed trivially). Denoting the neutron and proton
four-momenta as $p_1$ and $p_2$ respectively, it is not difficult
to realize that the matrix element $\langle p| V_\mu| n\rangle$
can then be written in a most general way as
\begin{multline}
\label{eq2.108}
\langle p(p_2)| V_\mu (0)| n(p_1)\rangle
=N_p N_n\bar u_p(p_2)(A\gamma_\mu +\gamma_\mu B+C\gamma_\mu
D+Ep_{1\mu}+Fp_{2\mu})u_n(p_1)
\end{multline}
where the coefficients $E$ and $F$ are Lorentz scalars (they may
depend on $p^2_1$, $p^2_2$ and $p_1\cdot p_2$ only) while the
$A,B,C$ and $D$ are to be understood as matrices made of products
of $\slashed{p}_1$ and $\slashed{p}_2$ (consequently, they do not
necessarily commute with $\gamma_\mu$). Having in mind the
identity $\slashed{a}\slashed{b} + \slashed{b}\slashed{a} =
2a\cdot b$, it is easy to see that without loss of generality one
may take
\begin{equation}
\label{eq2.109} A=a_0 +
a_1\slashed{p}_1+a_2\slashed{p}_2+a_3\slashed{p}_1\slashed{p}_2
\end{equation}
with $a_j,j=0,...,3$ being some scalar coefficients (and similarly
for $B, C$ and $D$). By using (\ref{eq2.109}) and the Dirac equations
for the $\bar u_p$ and $u_n$ one can then get rid of the matrix
coefficients in (\ref{eq2.108}). Indeed, one has e.g.
\begin{eqnarray}
\nonumber \bar u_p \slashed{p}_1 \gamma_\mu u_n&=&2p_{1\mu}\bar
u_p u_n -\bar u_p
\gamma_\mu\slashed{p}_1 u_n\\
\nonumber
&=&2p_{1\mu}\bar u_p u_n -m_n\bar u_p\gamma_\mu u_n
\end{eqnarray}
and other relevant relations of such a type can be obtained
easily. Needless to say, the anticommutation relation
$\{\gamma_\mu,\gamma_\nu\}=2g_{\mu\nu}$ is amply used throughout
all these calculations. Thus, one arrives at the form
\begin{equation}
\label{eq2.110} \langle p| V_\mu| n\rangle=\text{N.f}.\times\bar
u_p[K_0(t)\gamma_\mu+ K_1(t)p_{1\mu}+K_2(t)p_{2\mu}]u_n
\end{equation}
where the symbol N.f. stands for the normalization factors $N_p N_n$
and the coefficients $K_j,j=0,1,2$ are scalars. Since the nucleons
are taken on the mass shell, the $K_j$ in fact depend only on the
kinematical variable $t=(p_2-p_1)^2$ and we may call them formfactors.
In (\ref{eq2.110}) one can replace the variables $p_1,p_2$ by
\begin{eqnarray}
\nonumber
P=p_1+p_2\quad,\quad q=p_2-p_1
\end{eqnarray}
and employ the well-known Gordon identity\index{Gordon identity},
which in the present case reads
\begin{equation}
\label{eq2.111} \bar u(p_2)P_\mu u(p_1)=(m_p+m_n)\bar
u(p_2)\gamma_\mu u(p_1)- i\bar u(p_2)\sigma_{\mu\nu}q^\nu u(p_1)
\end{equation}
(cf. (\ref{eqA.79})). One thus immediately gets the standard
representation
\begin{equation}
\label{eq2.112} \langle p| V_\mu| n\rangle=\text{N.f.}\times\bar
u_p[\gamma_\mu f^{pn}_1 (t)-i\frac{1}{2M}\sigma_{\mu\nu}q^\nu
f^{pn}_2(t)+\frac{1}{2M} q_\mu f^{pn}_3(t)]u_n
\end{equation}
where $M$ denotes the average nucleon mass, introduced on
dimensional grounds (the correspondence between the old
formfactors $K_j$ and the new ones is straightforward). In a
similar way, we would find that a general form of the
axial-vector matrix element can be written as
\begin{equation}
\label{eq2.113} \langle p| A_\mu| n\rangle=\text{N.f.}\times\bar
u_p[\gamma_\mu g^{pn}_1 (t)-i\frac{1}{2M}\sigma_{\mu\nu} q^\nu
g^{pn}_2(t)+\frac{1}{2M} q_\mu g^{pn}_3(t)]\gamma_5 u_n
\end{equation}
The terminology used in the literature for the formfactors
appearing in (\ref{eq2.112}) and (\ref{eq2.113}) is as
follows\index{weak!magnetism}\index{weak!formfactors}:

$\hspace{3.3cm} f^{pn}_1(t)$ = vector formfactor,

$\hspace{3.3cm}f^{pn}_2(t)$ = weak magnetism,

$\hspace{3.3cm}f^{pn}_3(t)$ =  induced scalar,

$\hspace{3.3cm}g^{pn}_1(t)$ = axial vector formfactor\index{axial
vector}

$\hspace{3.3cm}g^{pn}_2(t)$ = pseudotensor formfactor

$\hspace{3.3cm}g^{pn}_3(t)$ = induced pseudoscalar

Thus, the nucleon matrix element of the weak current has, in
general, a more complicated structure than that appearing within
the effective Lagrangian approach of Chapter~\ref{chap1}
(cf.(\ref{eq1.111})). However, for nucleon beta decay one has the
kinematical limits
\begin{equation}
\label{eq2.114} (0.51\ \MeV)^2=m^2_e\leq t\leq(m_n-m_p)^2=(1.29\
\MeV)^2
\end{equation}
and thus obviously $| q_\mu|/2M\ll 1$. Then it is natural to
expect that the formfactors $f_2,f_3,g_2$ and $g_3$ do not give
sizable contributions to observable quantities (unless they are
anomalously large near zero -- but this does not seem to be the
case). Of course, their contributions would be of the same order
of magnitude as the proton recoil effects ignored throughout our
previous discussion in Chapter~\ref{chap1}. Such effects can also be
neglected for most of nuclear beta decays, with only a few
exceptions.\footnote{In some cases the relevant momentum transfer
(the energy release) may be large enough so that e.g. the weak
magnetism does give a measurable effect. In particular, such
effects were studied for the processes $B^{12}\rightarrow
C^{12}+e^-+\bar\nu_e$ and $N^{12}\rightarrow C^{12}+e^++\nu_e$,
where the energy release can be as large as 13 and $16\ \MeV$
respectively. M.~Gell-Mann \cite{ref24} was the first who
calculated the effect of the weak magnetism in the electron (or
positron) energy spectra for these beta decays and subsequent
experiments \cite{ref25} confirmed the theoretical results. For
further details, the interested reader is referred e.g. to
\cite{CoB}.} For the free neutron decay, one may thus safely
assume that the only essential contributions are due to the vector
and axial vector formfactors $f_1$ and $g_1$. Further, according
to (\ref{eq2.114}), the variable $t$ is restricted to be near zero
and one can thus presumably neglect the $t$-dependence of these
formfactors altogether (barring some unexpectedly wild behaviour),
setting simply $f_1(t)\doteq f_1(0)$ and $g_1(t)\doteq g_1(0)$.
Thus, in the limit of zero momentum transfer we arrive at
\begin{equation}
\label{eq2.115} \langle p| V_\mu-A_\mu|
n\rangle=\text{N.f.}\times\bar u_p\gamma_\mu[f_1
(0)-g_1(0)\gamma_5]u_n
\end{equation}
which corresponds formally to the approximate description employed
in Chapter~\ref{chap1}.

Returning to the interaction Lagrangian (\ref{eq2.58}) with the
current (\ref{eq2.59}) and using (\ref{eq2.115}), it is clear that
the beta-decay matrix element can now be approximately written as
\begin{equation}
\label{eq2.116} {\cal M}^{(\beta)}=-\frac{G_F}{\sqrt
2}\cos\vartheta_C[\bar u_p
\gamma_\mu(f_1(0)-g_1(0)\gamma_5)u_n][\bar
u_e\gamma^\mu(1-\gamma_5) v_\nu]
\end{equation}
The measurements of the effective beta-decay constants that we
have discussed earlier (see (\ref{eq1.117}), (\ref{eq1.120}) and
Section~\ref{sec2.5}) now clearly show that
\begin{equation}
\label{eq2.117}
f_1(0)\doteq 1
\end{equation}
with high accuracy, while $g_1(0)\doteq$ 1.27. It should be
emphasized that the result (\ref{eq2.117}) does not represent
merely a convenient normalization of the weak hadronic matrix
element. In fact, it is a rather non-trivial statement: the
coupling constant for the vector part of the weak current
appearing in the Lagrangian at the quark level is essentially left
unchanged when one passes to physical nucleons.\footnote{Note that
the weak coupling strength appearing in the quark-level
interaction Lagrangian manifests itself directly e.g. in
high-energy processes of deep-inelastic neutrino scattering, where
we can \qq{look inside} the target nucleons.} In other words, the
strong interactions\index{strong interaction} binding quarks in
nucleons do not renormalize the vectorial weak coupling. On the
other hand, the $g_1(0)$ is seen to differ significantly from the
axial-vector coupling appearing in the Lagrangian. The remarkable
relation (\ref{eq2.117}) is due to the conserving nature of the
vectorial weak current (which lies in a common isospin multiplet
with the electromagnetic current) and to the fact that neutron and
proton are classified as members of an isospin doublet. Thus, we
have another example of a \qq{CVC relation}\index{conserved vector
current (CVC)|ff} -- it is basically of the same origin as those
discussed in the preceding section in connection with pion beta
decay. It should also be stressed that the result (\ref{eq2.117})
is analogous to the equality of electric charges of e.g. electron
and proton: strong interactions forming the proton do not
renormalize the \qq{bare charge}, i.e. the parameter appearing in
the electromagnetic Lagrangian and associated with conserved
current written in terms of elementary fields (charged leptons and
quarks).

Although this topic goes slightly beyond the basic framework of
our treatment, let us summarize here, for completeness and for
reader's convenience, a set of CVC relations valid for the vector
weak formfactors appearing in (\ref{eq2.112}). For this purpose,
let us define first the electromagnetic
formfactors\index{electromagnetic!formfactors} of a nucleon (e.g.
proton):
\begin{equation}
\label{eq2.118} \langle p(p')| J_\mu^{em}(0)|
p(p)\rangle=\text{N.f.}\times\bar u (p')[\gamma_\mu
F_1^p(t)-i\frac{1}{2M}\sigma_{\mu\nu} q^\nu F_2^p(t)]u(p)
\end{equation}
Note that the general form (\ref{eq2.118}) follows from
considerations analogous to those employed for weak current; we
have discarded the third formfactor in order to satisfy the
current conservation. The relevant CVC relations can be written as
\begin{eqnarray}
\label{eq2.119}
f_1^{pn}(t)&=&F_1^p(t)-F_1^n(t)\nonumber\\
f_2^{pn}(t)&=&F_2^p(t)-F_2^n(t)\nonumber\\
f_3^{pn}(t)&=&0
\end{eqnarray}
where the electromagnetic formfactors of neutron are defined in
complete analogy with the proton case. A derivation of these
relations is outlined e.g. in \cite{FaR}. The normalization of
nucleon electromagnetic formfactors is
\begin{equation}
\label{eq2.120}
F_1^p(0)=1,\qquad F_2^p(0)=\mu_p
\end{equation}
for the proton and
\begin{equation}
\label{eq2.121}
F_1^n(0)=0,\qquad F_2^n(0)=\mu_n
\end{equation}
for the neutron, where $\mu_{p,n}$ are the corresponding magnetic
moments (given in units of nuclear magneton, i.e. $\mu_p=2.79$,
$\mu_n=-1.91$). From (\ref{eq2.119}) one thus gets first
$f^{pn}_1(0)$ = 1 (cf. (\ref{eq2.117})) and
\begin{equation}
\label{eq2.122}
f_2^{pn}(0)=\mu_p-\mu_n
\end{equation}
The remarkable prediction (\ref{eq2.122}) has been confirmed
experimentally in the beta decays of $B^{12}$ and $N^{12}$
that we have mentioned earlier.

In closing this section, we should perhaps recapitulate briefly
the circular path we have gone through in our theoretical
description of the free neutron beta decay. In Chapter~\ref{chap1} we have
started with an effective Lagrangian, written directly in terms of
nucleon fields and involving some unknown constants that have to
be determined experimentally. The universal weak interaction
theory of the 1960s, which crystallized from the wealth of
empirical data as a masterpiece of theoretical insight, is
certainly more elegant and includes only the Fermi constant
(inferred from muon lifetime) and the Cabibbo angle\index{Cabibbo
angle}. However, the other phenomenological parameters enter
through the back door when physical matrix elements are
considered: instead of constant parameters of the effective
Lagrangian, we get first a set of formfactors and the expression
for the decay amplitude acquires a more general structure that
can, in principle at least, be tested experimentally. At low
energies, the formfactors can be replaced by constants, the novel
effects (as the weak magnetism etc.) may be safely neglected and
the result stemming originally from the nucleonic effective
Lagrangian is thus recovered. Of course, some phenomenological
parameters must be fixed experimentally anyway, so that for most
practical purposes both approaches are essentially equivalent.

%\input{kniha29}  %            2.9
%%%%%%%%%%%%%%%%%%%%%%%%%%%%%%%%%%%%%%%%%%%%%%%%%%%%%%%%%%%%%%%%%%%
%%%%%%%%%%%%%%%%%%%%%%%%%%%%%%%%%%%%%%%%%%%%%%%%%%%%%%%%%%%%%%%%%%%%%%%%%%%%%%%%%%%%%%%%%%%%%%%%%%%%%%%%%%%%%%%%%%%%%%%%%%%%%%%%%%%%%%%%
\section{$\cal{C,P}$ and $\cal{CP}$}\label{sec2.9}
%section 2.9
\index{CP symmetry@${\cal CP}$ symmetry|ff}\index{CP
violation@${\cal CP}$ violation|ff}\index{charge conjugation|(}
The last topic to be discussed in this chapter concerns discrete
symmetries -- the space reflection\index{space reflection}
(parity\index{parity}) $\cal {P}$, charge conjugation $\cal {C}$
and their combination $\cal {CP}$. We will show that the
provisional weak interaction Lagrangian (\ref{eq2.58}) is
invariant under the \qq{combined parity} transformation $\cal
{CP}$, though both the $\cal {P}$ and the $\cal {C}$ are violated
maximally. On the other hand, there is a long-standing and
well-established experimental evidence for tiny $\cal
{CP}$-violating effects in the neutral kaon system \cite{ref26}.
The standard model (SM) of electroweak interactions that we will
discuss in detail later on, can incorporate $\cal {CP}$ violation
in a rather natural way -- it is quite remarkable that the
occurrence of $\cal {CP}$-violating terms in the SM interaction
Lagrangian is due to the existence of the third generation of
quarks ($b$ and
$t$)\index{bquark@$b$-quark}\index{t-quark@$t$-quark}. Relevant
experimental data are accumulating from measurements on $B$-mesons
(i.e. those containing $b$-quarks) and the underlying theoretical
picture should consequently be further clarified. All this makes
the $\cal {CP}$ violation one of the most prominent open problems
in modern particle physics. The present section serves as a
prelude to our later discussion of this issue within the standard
electroweak model.

The non-invariance of weak interactions under the space
reflection\index{space reflection} $\cal {P}$ (parity
violation\index{parity!violation}) has already been described
earlier in this text. As we know, the chiral $V-A$ structure of
the current (\ref{eq2.59}) corresponds to maximum parity violation
-- the $\cal {P}$-odd terms of the type $VA$ descending from the
product of currents enter the interaction Lagrangian with the same
strength as the $\cal {P}$-even contributions of the type $VV$ and
$AA$. It is quite remarkable that the chiral structure of weak
currents also leads to non-invariance of the Lagrangian under a
discrete internal symmetry -- the charge conjugation $\cal {C}$.
To see this, let us work out the corresponding transformations of
vector and axial-vector fermionic currents explicitly. Throughout
our calculation we will use the standard representation of gamma
matrices. A Dirac spinor transforms under the charge conjugation
as
\begin{equation}
\label{eq2.123}
\psi_c=C\bar\psi^T
\end{equation}
where the superscript $T$ denotes transposed matrix and the $C$ is
defined by
\begin{equation}
\label{eq2.124}
C^{-1}\gamma_\mu C=-\gamma^T_\mu
\end{equation}
(let us recall that the free-field Dirac equation is then
invariant under (\ref{eq2.123})). It is well known that within the
standard representation the matrix $C$ can be written as
\begin{equation}
\label{eq2.125}
C=i\gamma^2\gamma^0
\end{equation}
From the last expression some useful relations follow
immediately, in particular
\begin{eqnarray}
\label{eq2.126}
\lefteqn{C^{-1}=C^\dagger=C^T=-C,}\nonumber\\
&&\{C,\gamma_0\}=0,\quad [C,\gamma_5]=0
\end{eqnarray}
Now it is easy to evaluate the Dirac conjugate of the $\psi_c$.
One obtains first
\begin{displaymath}
%\nonumber
\bar\psi_c=(C\bar\psi^T)^\dagger\gamma_0=(C\gamma_0\psi^\ast)^
\dagger\gamma_0=\psi^T\gamma_0C^\dagger\gamma_0
\end{displaymath}
and with the help of (\ref{eq2.126}) the last result is easily
recast as
\begin{equation}
\label{eq2.127}
\bar\psi_c=\psi^T C
\end{equation}
From (\ref{eq2.123}) and (\ref{eq2.127}) one obtains readily the
transformation of a fermionic current. For a vector (made generally
of two different Dirac fields) one has
\begin{eqnarray}
\label{eq2.128}
\lefteqn{\bar\psi_{1c}\gamma_\mu\psi_{2c}=\psi^T_1C\gamma_\mu
C\bar\psi_2^T=}\nonumber\\
&&=-\psi^T_1C^{-1}\gamma_\mu C\bar\psi^T_2=\psi^T_1\gamma^T_\mu
\bar\psi^T_2
\end{eqnarray}
For classical fields, one may write
\begin{equation}
\label{eq2.129}
\psi^T_1\gamma^T_\mu\bar\psi^T_2=(\bar\psi_2\gamma_\mu\psi_1)^T
\end{equation}
and the last expression is, of course, equal to
$\bar\psi_2\gamma_\mu\psi_1$. We thus arrive at the result
\begin{equation}
\label{eq2.130} \text{for classical
fields:}\quad\bar\psi_1\gamma_\mu \psi_2 \stackrel{\cal
C}{\longrightarrow} \bar\psi_2\gamma_\mu\psi_1
\end{equation}
For quantized Dirac fields one has to take into account their
anticommuting nature (if $\psi_1$ = $\psi_2$, we assume that
the current is normal-ordered) and the relation (\ref{eq2.129})
then obviously acquires a negative sign. Thus we have
\begin{equation}
\label{eq2.131} \text{for quantum
fields:}\quad\bar\psi_1\gamma_\mu\psi_2\stackrel
{\cal{C}}{\longrightarrow}-\bar\psi_2\gamma_\mu\psi_1
\end{equation}
(in particular, for $\psi_1$ = $\psi_2$ this leads to the
desirable result that electromagnetic current changes its sign
upon charge conjugation). Similarly, for the axial-vector current one
gets first\index{electromagnetic!current}
\begin{eqnarray}
\label{eq2.132}
\lefteqn{\bar\psi_{1c}\gamma_\mu\gamma_5\psi_{2c}=\psi^T_1C
\gamma_\mu\gamma_5C\bar\psi^T_2=}\nonumber\\
&&=-\psi^T_1 C^{-1}\gamma_\mu CC^{-1}\gamma_5C\bar\psi^T_2=
\psi^T_1\gamma^T_\mu C^{-1}\gamma_5 C\bar\psi^T_2
\end{eqnarray}
Using (\ref{eq2.124}) it is easy to find that
$C^{-1}\gamma_5C=\gamma^T_5$ (note also that $\gamma^T_5$ =
$\gamma_5$ in the standard representation) and we thus have the
result
\begin{equation}
\label{eq2.133} \text{for classical fields:}\quad
\bar\psi_1\gamma_\mu\gamma_5
\psi_2\stackrel{\cal{C}}{\longrightarrow}-\bar\psi_2\gamma_\mu\gamma_5
\psi_1
\end{equation}
For quantum fields there is an extra minus sign due to
anticommutators, so that one has
\begin{equation}
\label{eq2.134} \text{for quantum fields:}\quad
\bar\psi_1\gamma_\mu\gamma_5
\psi_2\stackrel{\cal{C}}{\longrightarrow}\bar\psi_2\gamma_\mu\gamma_5
\psi_1
\end{equation}
Thus, we see that the vector and axial-vector current have an
opposite $\cal{C}$-parity\index{C-parity@$\mathcal{C}$-parity} and
this in turn means that the ${\it VA}$ term in the interaction
Lagrangian is $\cal{C}$-odd while the $\it{AA}$ or $\it{VV}$ terms
are $\cal{C}$-even. Since the overall strength of all these terms
is the same, one can say that the Lagrangian (\ref{eq2.58})
exhibits maximum $\cal{C}$-violation (similarly to $\cal{P}$);
\index{C-parity@$\mathcal{C}$-parity!violation} moreover, one may
also observe that the $\cal{C}$ and $\cal{P}$ violation have the
same algebraic origin in the chiral structure of the weak current.

Our next goal is finding the $\cal{CP}$ transformation law for
currents\index{currents}. For this purpose and for reader's
convenience let us first summarize here the relevant formulae for
the space inversion $\cal{P}$. Starting with the well-known
transformation law for Dirac spinors
\begin{equation}
\label{eq2.135}
\psi_P(x)=\gamma_0\psi(\tilde{x})
\end{equation}
where $\tilde{x}=(x^0,-\vec{x})$, it is straightforward to obtain
\begin{equation}
\label{eq2.136}
\bar\psi_1(x)\gamma_\mu\psi_2(x)\stackrel{\cal{P}}{\longrightarrow}
\bar\psi_1(\tilde{x})\gamma^\mu\psi_2(\tilde{x})
\end{equation}
for the vector current and
\begin{equation}
\label{eq2.137}
\bar\psi_1(x)\gamma_\mu\gamma_5\psi_2(x)\stackrel{\cal{P}}
{\longrightarrow}-\bar\psi_1(\tilde{x})\gamma^\mu\gamma_5\psi_2
(\tilde{x})
\end{equation}
for the axial-vector current. Note that these results follow from
simple properties of the Dirac gamma matrices and hold for both
classical and quantum fields.

Putting now together (\ref{eq2.130}), (\ref{eq2.133}),
(\ref{eq2.136}) and (\ref{eq2.137}), it is clear that for the combined
transformation $\cal{CP}$ one has
\begin{equation}
\text{for classical fields:}\quad
\begin{cases}
\bar\psi_1(x)\gamma_\mu\psi_2(x)\stackrel{\cal{CP}}{\longrightarrow}
\bar\psi_2(\tilde{x})\gamma^\mu\psi_1(\tilde{x})\\[0.4cm]
%\end{displaymath}
%\begin{equation}
\label{eq2.138}
\bar\psi_1(x)\gamma_\mu\gamma_5\psi_2(x)\stackrel{\cal{CP}}
{\longrightarrow}
\bar\psi_2(\tilde{x})\gamma^\mu\gamma_5\psi_1(\tilde{x})
\end{cases}
\end{equation}
For quantum fields there is an extra overall minus sign
descending from the $\cal{C}$ transformation, so that
\begin{equation}
\text{for quantum fields:}\quad
\begin{cases}
\bar\psi_1(x)\gamma_\mu\psi_2(x)\stackrel{\cal{CP}}
{\longrightarrow}-\bar\psi_2(\tilde{x})\gamma^\mu\psi_1(\tilde{x})\\[0.4cm]
\label{eq2.139}
\bar\psi_1(x)\gamma_\mu\gamma_5\psi_2(x)\stackrel{\cal{CP}}
{\longrightarrow}
-\bar\psi_2(\tilde{x})\gamma^\mu\gamma_5\psi_1(\tilde{x})
\end{cases}
\end{equation}
The $\cal{CP}$-invariance of the weak interaction Lagrangian
(\ref{eq2.58}) should now be clear. Indeed, using e.g.
(\ref{eq2.138}), the transformation of the current (\ref{eq2.59})
can be written as
\begin{equation}
\label{eq2.140}
J^\mu(x)\stackrel{\cal{CP}}{\longrightarrow}J^\dagger_\mu(\tilde{x})
\end{equation}
since $(\bar\psi_1\Gamma_\mu\psi_2)^\dagger=\bar\psi_2\Gamma_\mu
\psi_1$ for $\Gamma_\mu$ equal to $\gamma_\mu$ or
$\gamma_\mu\gamma_5$. However, the Lagrangian density is given by
the product $J^\mu J^\dagger_\mu$, where, of course, raising of
the Lorentz index in (\ref{eq2.140}) becomes irrelevant; the
$\lagr_{int}$ is thus scalar under $\cal{CP}$, i.e.
\begin{equation}
\label{eq2.141}
\lagr_{int}(x)\stackrel{\cal{CP}}{\longrightarrow}\lagr_{int}
(\tilde{x})
\end{equation}
Note that the restriction to classical fields in the present
context has not been important  -- for quantum fields there is a
twofold sign change in the transformed currents and
(\ref{eq2.141}) is recovered anyway. Another remark is in order
here. It is useful to realize that for the relation
(\ref{eq2.140}) to be valid it is essential that the current
$J_\mu$ involves, apart from Dirac matrices, only real
coefficients like $\cos\theta_C$ and $\sin\theta_C$. This,
however, need not be the case in the world built upon three
generations of quarks. It turns out that the simple Cabibbo mixing
(\qq{rotation}) is then naturally replaced by elements of a 3
$\times$ 3 unitary matrix that may be imaginary and give rise to
$\cal{CP}$-violating terms in the interaction Lagrangian. As we
noticed earlier in this section, a natural framework for such a
discussion is provided by the standard electroweak model and we
will have more to say about this in Chapter~\ref{chap7}\index{charge
conjugation|)}.

%\input{problems2}
%%%%%%%%%%%%%%%%%%%%%%%%%%%%%%%%%%%%%%%%%%%%%%%%%%%%%%%%%%%%%%%%%%%
%%%%%%%%%%%%%%%%%%%%%%%%%%%%%%%%%%%%%%%%%%%%%%%%%%%%%%%%%%%%%%%%%%%%%%%%%%%%%%%%%%%%%%%%%%%%%%%%%%%%%%%%%%%%%%%%%%%%%%%%%%%%%%%%%%%%%%%%
\begin{priklady}{11}

\item  Derive the result (\ref{eq2.31}) by means of a straightforward
integration of the differential decay rate (\ref{eq2.22}) (i.e.
without using the \qq{tensor trick} employed in Section~\ref{sec2.3}).

\item Within the $V-A$ theory of weak interactions calculate the
degree of polarization of electrons\index{polarization!of the
electron} in the decay of an unpolarized muon at rest.

\item  Calculate the angular distribution of electrons in the decay
of a polarized muon at rest. For simplicity, neglect the electron
mass throughout the calculation (for an instructive discussion of
this problem see \cite{BjD}, chapter 10 therein).

\item Calculate the probability of production of left-handed
electron in the pion decay $\pi^- \rightarrow e^- \bar{\nu}_e$
within $V-A$ theory, assuming that the neutrino is massless. How
the result would change if the lepton weak current had the form
$V-\lambda A$, with $\lambda$ being an arbitrary real parameter?
How is the result obtained within the $V-A$ theory changed, when
the neutrino has a non-zero mass?

\item  Consider scattering processes $\bar{\nu}_e + e^- \rightarrow \bar{\nu}_\mu + \mu^-$ and
$\nu_\mu + e^- \rightarrow \mu^- + \nu_e$ in a high-energy domain,
i.e. for $E_{c.m.}\gg m_\mu$ (thus, lepton masses can be
neglected). Suppose that the weak charged current has the structure
$V-aA$ for electron-type leptons and $V-bA$ for muon-type leptons,
where $a$ and $b$ are essentially arbitrary real parameters. Show
that the ratio
$$
R = \frac{\sigma(\nu_\mu e^- \rightarrow \mu^-
\nu_e)}{\sigma(\bar{\nu}_e e^- \rightarrow \bar{\nu}_\mu \mu^-)}
$$
satisfies inequality $1\leq R \leq 3$.

\item  Within the $V-A$ theory of weak interactions calculate the cross section of the
process $e^+e^- \rightarrow \nu_e \bar{\nu}_e$. Calculate also (at
the tree level) the QED cross section $\sigma(e^+e^- \rightarrow
\mu^+ \mu^-)$. In both cases assume that the collision energy is
sufficiently large ($E_{c.m.} \gg m_\mu$) and neglect lepton
masses. Evaluate the ratio $\sigma(e^+e^- \rightarrow \nu_e
\bar{\nu}_e)/\sigma(e^+e^- \rightarrow \mu^+ \mu^-)$ as a function
of energy in the considered domain. For which energy the two cross
sections become comparable?

\item \label{pro2.7}  Employing the \qq{$G^2 \Delta^5$ rule}, estimate the branching
ratios for
\begin{align*}
\Sigma^- &\rightarrow \; n + e^- + \bar{\nu}_e\\
K^- &\rightarrow \; \pi^0+ e^- +\bar{\nu}_e\\
\Sigma^- &\rightarrow \; \Lambda+ e^- +\bar{\nu}_e\\
\Xi^- &\rightarrow \;\Sigma^0 +e^- +\bar{\nu}_e\\
\Omega^- &\rightarrow \;\Xi^0 +e^- +\bar{\nu}_e.
\end{align*}

\item The decay amplitude for $\Sigma^- \rightarrow \Lambda e^-
\bar{\nu}_e$ can be written approximately as
$$
\mathcal{M}_{fi} = \frac{G_F}{\sqrt{2}} \cos\theta_C
\sqrt{\frac{2}{3}}\ a \bigl[\bar{u}(p)\gamma_\mu \gamma_5
u(P)\bigr]\bigl[\bar{u}(k)\gamma^\mu (1-\gamma_5)v(k')\bigr]
$$
where $P,p,k,k'$ denote consecutively the four-momenta of
$\Sigma^-, \Lambda, e^-, \bar{\nu}_e$ and the constant $a$
reflects non-perturbative nature of the hadronic matrix element
(numerically, $a\doteq 0.81$); the remaining symbols have a
standard meaning. Calculate the decay width for the considered
process as a function of the maximum electron energy $\Delta$ and
of the other parameters. Compute also the corresponding branching
ratio and compare the result with the estimate obtained in solving
the Problem \ref{pro2.7}. Throughout the calculation, neglect the $\Lambda$
momentum wherever it is possible and set also
$m_e=0$ (why is such an approximation good?).\\
{\it Remark:} It is amusing to notice that the above amplitude
corresponds to a pure Gamow--Teller transition\index{Gamow--Teller
transitions} for {\it hyperons\/} (only the axial-vector term
contributes to the hadronic matrix element). Further details
concerning Cabibbo theory of semileptonic decays of baryons can be
found in the monograph \cite{CoB}.

\item Consider the beta decay of charged kaon, $K^- \rightarrow
\pi^0 e^- \bar{\nu}_e$. Compute the electron energy spectrum and
partial decay width.\\
{\it Hint:} For the necessary current-algebra background, see
\cite{FaR}

\item Consider the decay $\pi^- \to e^- + \bar{\nu}\;$ involving the neutrino with a non-zero mass. 
Let us denote as $|{\cal M}_L|^2$ the squared matrix element for the production of left-handed (i.e. negative-helicity) electron, with the neutrino spin states summed over. The analogous quantity for the production of right-handed electron is denoted as $|{\cal M}_R|^2$ and for the full decay matrix element squared the usual symbol $\overline{|{\cal M}|^2}$ is employed. Show that
$$
\frac{|{\cal M}_L|^2}{\overline{|{\cal M}|^2}} = \frac{ (a-b)^2}{2 (a^2 +b^2)}\,,\qquad
\frac{|{\cal M}_R|^2}{\overline{|{\cal M}|^2}} = \frac{ (a+b)^2}{2 (a^2 +b^2)}
$$
where
\begin{align*}
&a  = (m_e - m_\nu)\sqrt{m_\pi^2 - (m_e +m_\nu)^2}\\
&b  = (m_e + m_\nu)\sqrt{m_\pi^2 - (m_e -m_\nu)^2}
\end{align*}
For a consistency check, one may notice immediately that $a = b$ if $m_\nu = 0$; the result $|{\cal M}_L|^2 = 0$, anticipated a priori in the case of massless neutrino, is thus recovered. Further, as a simple algebraic exercise, show that expanding the above result for $|{\cal M}_L|^2$ in powers of $m_\nu$ one obtains
$$
\frac{|{\cal M}_L|^2}{\overline{|{\cal M}|^2}} = \frac{m_\nu^2}{m_e^2}\Bigl(\frac{m_\pi^2}{m_\pi^2-m_e^2}\Bigr)^2 (1 + {\cal O}(m_\nu))
$$
Such a result demonstrates clearly that the possibility of producing a left-handed electron in the considered $\pi_{e2}$ decay process is due to the distinction between helicity and chirality for massive neutrino.
\index{V-A theory@$V-A$
theory|)}\index{weak!interaction|)}
\end{priklady}

%\input{kniha31}
%%%%%%%%%%%%%%%%%%%%%%%%%%%%%%%%%%%%%%%%%%%%%%%%%%%%%%%%%%%%%%%%%%%
%%%%%%%%%%%%%%%%%%%%%%%%%%%%%%%%%%%%%%%%%%%%%%%%%%%%%%%%%%%%%%%%%%%%%%%%%%%%%%%%%%%%%%%%%%%%%%%%%%%%%%%%%%%%%%%%%%%%%%%%%%%%%%%%%%%%%%%%
\chapter {Intermediate vector boson $W$}\label{chap3}
\section{Difficulties of Fermi-type theory}\index{W
boson@$W$ boson|(}\label{sec3.1}

The weak interaction theory built according to Fermi's paradigm,
discussed at length in the preceding two chapters, was certainly
one of the highlights of the particle physics in 1960s. The simple
and elegant Feynman--Gell-Mann interaction Lagrangian
(\ref{eq2.58}) was capable to describe a lot of experimental data
available then and -- as we have demonstrated in several examples
-- it also had a considerable predictive power. Since its early
days, the theory was successfully tested for a variety of decay
processes and, with less accuracy, also for some particular
scattering processes at low energies (the famous reaction $\bar\nu
+p\rightarrow n+e^+$ used for the first direct neutrino detection
\cite{ref27} can serve as one such example). In any case, the
relevant theoretical predictions were verified at that time within
a rather limited kinematical region, corresponding to low energy
and low momentum transfer -- certainly less than $1\ \GeV$ or so.
Having established an effective weak interaction theory,
phenomenologically successful at low energies, we should
scrutinize its behaviour at higher energies as well. To this end,
one must naturally consider scattering processes, as these can be
studied (at least in principle) at an arbitrarily high collision
energy. We shall see below that the usual Feynman-diagram methods
become rather problematic for sufficiently high energies and it
will also be immediately clear that such difficulties are common
to all Fermi-type models, i.e. to those involving a direct
interaction of four fermionic fields.

\begin{figure}[h]
\centering \s{\includegraphics{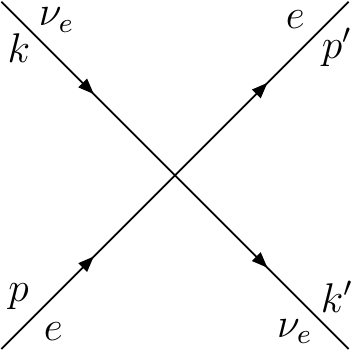}}
\caption{Tree-level diagram for $\nu_e-e$ elastic scattering
within a Fermi-type weak interaction theory.}
\label{fig5}\index{Feynman diagrams!for $\nu_e-e$ scattering}
\end{figure}

The problem we have in mind can be demonstrated on an example of
any binary reaction proceeding in lowest order through the
interaction Lagrangian (\ref{eq2.58}). For definiteness, we may
consider e.g. the neutrino-electron elastic
scattering\index{neutrino-electron scattering}. In the lowest
order of perturbation expansion this process is described by the
simple Feynman diagram shown in Fig.\,\ref{fig5}.

%\noindent
Before carrying out a technical calculation of the corresponding
cross section, it will be instructive to estimate its high-energy
behaviour on dimensional grounds. According to the arguments given
in Chapter~\ref{chap1}, the Fermi constant $G_F$ has dimension of inverse
squared mass (see (\ref{eq1.15})). In the first order of
perturbation expansion, the relevant matrix element is
proportional to $G_F$; consequently, the corresponding cross
section must contain the factor $G^2_F$, which is of dimension
$M^{-4}$. On the other hand, the dimension of a cross section is
(length)$^2$, i.e. $M^{-2}$. Thus, one needs an additional factor
of energy (mass) squared to balance the dimensionality of $G^2_F$
and get the quantity with right dimension of a cross section. At
high energies ($E\gg m_e$), i.e. in the ultrarelativistic limit,
the effects of electron mass can be neglected (such a guess is
indeed confirmed by an explicit calculation -- see below) and the
only quantity with dimension of mass that remains in the game is
the collision energy. Since the cross section is Lorentz
invariant, it can only depend on the Mandelstam invariant
$s=E^2_{c.m.}$, where $E_{c.m.}$ is the full centre-of-mass
energy\index{centre-of-mass energy|see{Mandelstam invariants}}.
Thus, we arrive at the following \qq{rule of thumb} for the
high-energy behaviour of the neutrino-electron cross section:
\begin{equation}
\label{eq3.1}
\sigma\simeq G^2_F E^2_{c.m.}
\end{equation}
Similarly, one can estimate the behaviour of the scattering
amplitude. Within our normalization conventions the matrix element
${\cal M}_{fi}$ is dimensionless for any binary process (see
Appendix~\ref{appenB}) and in the first order of perturbation theory it is
proportional to $G_F$. Thus, barring the irrelevant mass
dependence, one can expect that
\begin{equation}
\label{eq3.2}
{\cal M}_{fi}\simeq G_F E^2_{c.m.}f(\Omega)
\end{equation}
in the high-energy limit, with $f(\Omega)$ being a (dimensionless)
function of scattering angles. Obviously, it is the dimensionality
of the Fermi constant that plays a crucial role in preceding
considerations. Consequently, the \qq{scaling laws} (\ref{eq3.1}) and
(\ref{eq3.2}) should be valid for any binary reaction involving
four elementary fermions and within any particular model of the
Fermi type. We will explain shortly what is wrong with such a
high-energy behaviour, but now let us verify -- just to be
sure -- the results of our simple dimensional analysis by means
of an explicit calculation.

The scattering amplitude corresponding to Fig.\,\ref{fig5} can be
written as
\begin{equation}
\label{eq3.3}
{\cal M}_{fi}=-\frac{G_F}{\sqrt 2}[\bar u (p')\gamma_\mu
(1-\gamma_5)u(k)][\bar u(k')\gamma^\mu(1-\gamma_5)u(p)]
\end{equation}
where we have suppressed, for the sake of brevity, the spin
labels of the Dirac spinors. For the spin-averaged matrix element
squared one then gets, by means of the usual trace techniques
\begin{eqnarray}
\label{eq3.4} \lefteqn{\overline{|{\cal M}_{fi}| ^2}=\frac{1}{2}
\sum_{spins}|{\cal M}_{fi}| ^2=}
\nonumber\\
&&=G^2_F \Tr[\slashed{p}'\gamma^\rho\slashed{k}
\gamma^\sigma(1-\gamma _5)]\cdot \Tr[\slashed{k}' \gamma_\rho
\slashed{p}\gamma_\sigma(1-\gamma_5)]
\end{eqnarray}
where we have used $(1-\gamma_5)^2$ = $2(1-\gamma_5)$ and other
familiar properties of the gamma matrices (needless to say, we
have set $m_\nu$ = 0 from the very beginning). The spinor traces
in (\ref{eq3.4}) can be evaluated most economically with the help
of the formulae (\ref{eqA.50}). One thus gets immediately
\begin{equation}
\label{eq3.5} \overline{|{\cal M}_{fi}| ^2}=64G^2_F(k\cdot
p)(k'\cdot p')
\end{equation}
and this can be further recast in terms of the Mandelstam
variable $s = (k+p)^2$ as
\begin{equation}
\label{eq3.6} \overline{|{\cal M}_{fi}| ^2}=16G^2_F(s-m^2_e)^2
\end{equation}
For the differential cross section (angular distribution in the
c.m. system) one then has
\begin{equation}
\label{eq3.7}
\frac{d\sigma^{(\nu e)}}{d\Omega_{c.m.}}=\frac{G^2_F}{4\pi^2}
\frac{(s-m^2_e)^2}{s}
\end{equation}
The angular integration of the last expression is trivial and
yields the result
\begin{equation}
\label{eq3.8}
\sigma^{(\nu e)}=\frac{G^2_F}{\pi}\frac{(s-m^2_e)^2}{s}
\end{equation}
which makes it clear that the effect of electron mass can indeed
be neglected in the high-energy limit. For $s\gg m^2_e$
thus (\ref{eq3.8}) becomes simply
\begin{equation}
\label{eq3.9} \sigma^{(\nu e)}|_{s\gg
m^2_e}\approx\frac{G^2_F}{\pi}s
\end{equation}
which confirms our previous estimate (\ref{eq3.1}) made on
simple dimensional grounds.\footnote{The reader may wonder why we
emphasize the verification of an intuitively plausible claim that
the electron mass effects in the considered cross section can be
neglected at high energy. The point is that in some other cases
(within other field-theory models) one can get results that
appear, in this sense, rather counter-intuitive. In particular, as
we shall see later in this chapter, for processes involving a
physical massive charged particle with spin 1 (the vector boson
$W$) the corresponding mass effects persist even at high energies
-- the limit of taking the vector boson mass to zero becomes
singular.} Other examples of this kind (such as the $\bar\nu -e$
scattering and the annihilation process
$e^+e^-\rightarrow\nu\bar\nu$) can be worked out easily, but they
will not be immediately necessary here and the corresponding
calculation can be left as an instructive exercise to the
interested reader (some technical details can also be found in the
book \cite{Hor}).

Now, let us explain what is wrong with the high-energy behaviour
shown in (\ref{eq3.1}) or (\ref{eq3.2}) resp. To put it briefly,
such a power-like growth of a scattering amplitude leads to rapid
violation of the $S$-matrix unitarity\index{S matrix
unitarity@$S$-matrix unitarity}. This statement may be understood
rather easily at an intuitive level: the absolute value of an
element of a unitary matrix is bounded from above (it must be less
than unity) and one thus naturally expects that the scattering
amplitude ${\cal M}_{fi}$ should not rise indefinitely with
energy. For an explicit discussion of such a \qq{unitarity
bound}\index{unitarity bound|(}, one has to invoke the technique
of partial-wave expansion (some basic formulae can be found in
Appendix~\ref{appenB}). Using (\ref{eq3.6}) we notice that in the considered
case the relevant scattering amplitude ${\cal M}_{fi}$ does not
depend on the scattering angle\footnote {Obviously, only the
negative-helicity\index{helicity} states of electron and neutrino
contribute in (\ref{eq3.6}) in the high-energy limit and the
considered scattering amplitudes should consequently be labelled
e.g. as ${\cal M}_{----}$. In what follows, we will usually
suppress the helicity indices for the sake of brevity.} and this
in turn means that the whole partial-wave expansion is reduced to
the lowest term carrying the angular momentum $j$ =
$0$\index{angular momentum}. The amplitude of the partial
wave\index{partial-wave expansion} with $j = 0$ can then be easily
inferred from (\ref{eq3.6}); for its absolute value one gets
\begin{equation}
\label{eq3.10} |{\cal M}^{(0)}(s)|=\frac{1}{2\pi\sqrt 2}G_F s
\end{equation}
In the high-energy limit, an ${\cal M}^{(j)}$ can be written as
${\cal M}^{(j)}=(S^{(j)} -1)/2i$, where the $S^{(j)}$ is an
element of a finite-dimensional unitary matrix (the $S$-matrix
restricted to the subspace characterized by a given value of $j$)
and this obviously yields the bound
\begin{equation}
\label{eq3.11} |{\cal M}^{(j)}(s)|\leq 1
\end{equation}
Applying now the constraint (\ref{eq3.11}) to our result (\ref{eq3.10}),
it becomes clear that only for energies within the range
$s\leq2\pi\sqrt 2G^{-1}_F$, i.e. for
\begin{equation}
\label{eq3.12} E_{c.m.}\leq\left (\frac{2\pi\sqrt 2}{G_F}\right
)^{\frac{1}{2}} \doteq 870\ \GeV
\end{equation}
one avoids a manifest violation of unitarity -- in other words,
outside the domain (\ref{eq3.12}) our calculation cannot be
reliable. A restriction of the type (\ref{eq3.12}) is usually
called \qq{unitarity bound}. The unitarity violation at high
energy within weak interaction theory of Fermi type has been first
emphasized in the early 1960s (see \cite{ref28}). The power-law
rise of the cross section (\ref{eq3.9}) has often been referred to
as the weak interaction \qq{becoming strong} at high energies.

Several remarks are in order here. The numerical value of the
\qq{critical energy} shown in (\ref{eq3.12}) is in fact rather
high; one can hardly expect that the $\nu -e$ collisions would be
studied experimentally at such energies in foreseeable future
(note that $E_{c.m.} = 870\ \GeV$ corresponds to the incident
neutrino energy of about 7.5 $\times 10^5\ \TeV$ in the electron
rest system!). However, it is not the particular value of the
unitarity bound that really matters. The important point is that
the considered scattering amplitude {\it grows as a positive power
of energy\/} -- this is precisely what we have in mind when saying
that there is a \qq{rapid violation} of unitarity. This feature
distinguishes the Fermi-type theory of weak interactions from e.g.
quantum electrodynamics\index{quantum!electrodynamics (QED)} of
electrons and photons (spinor QED), where the high-energy
behaviour of lowest-order Feynman diagrams is much softer (so that
a possible conflict with unitarity is deferred to the realm of
astronomically high energies). Thus, at least from a technical
point of view, the weak interaction theory of Fermi type seems to
be inferior to some other field theory models that work
successfully in different areas of particle physics. On the other
hand, one may object that the violation of unitarity discussed
here is not of fundamental nature: since the interaction
Hamiltonian\index{Hamiltonian density} is hermitean, the {\it
exact\/} $S$-matrix must be unitary and the offending behaviour
(\ref{eq3.10}) is just an artefact of the lowest-order
perturbation theory. This argument is perfectly true, but rather
academic. Indeed, nobody can solve exactly a quantum field theory
model of considered type to see that the full scattering amplitude
behaves decently. One may e.g. try to calculate higher orders of
perturbation theory, but then one obviously runs into more severe
difficulties than in the basic approximation -- as there are
higher powers of the Fermi constant $G_F$, there must also be
higher powers of the energy in order to get a dimensionless
scattering amplitude. Thus, although there is no fundamental
inconsistency in the Fermi-type models, their practical
applicability is limited, as it is notoriously difficult to go
beyond the framework of perturbation theory\index{unitarity
bound|)}.

\begin{figure}[h]\centering
\s{\includegraphics{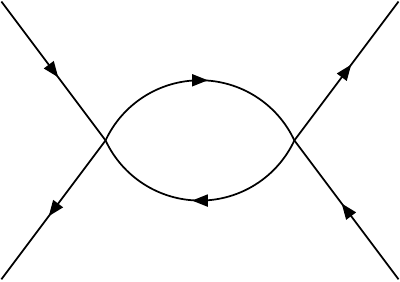}}\hspace{2cm}\s{\includegraphics{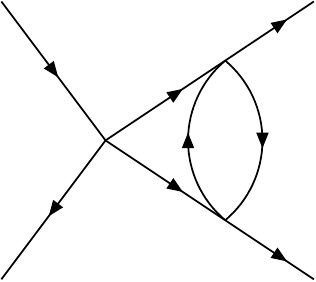}}
\caption{Examples of closed-loop Feynman graphs representing
higher-order contributions to the $\nu-e$ scattering.}
\label{fig6}\index{Feynman diagrams!for $\nu_e-e$ scattering}
\end{figure}

There is another important aspect of perturbation expansion that
should be mentioned separately. In our discussion of the
neutrino-electron elastic scattering we have used the lowest-order
approximation, which corresponds to the simple Feynman graph shown
in Fig.\,\ref{fig5}. In the standard terminology of perturbative
quantum field theory, the diagrams of such a type are called {\bf
tree diagrams}\index{tree diagrams} as they do not contain closed
loops of internal lines. In higher orders of perturbation
expansion, closed-loop diagrams\index{loop diagrams} necessarily
appear; some examples are depicted in Fig.\,\ref{fig6}.
Contributions of the closed loops are expressed in terms of
integrals over the four-momenta of \qq{virtual particles}
associated with the internal lines. Unfortunately, such integrals
usually diverge in the ultraviolet region (i.e. in the
neighbourhood of infinity) -- this phenomenon is in fact typical
for most of the quantum field theory models. These ultraviolet
(UV) divergences\index{ultraviolet divergences} can be tamed
successfully within some QFT models by means of the
renormalization\index{renormalization} procedure, which
essentially consists in a redefinition of a certain (finite)
number of parameters of the model. Such a procedure has been first
formulated in the late 1940s for quantum electrodynamics, where
one is then able to calculate explicitly some finite higher-order
\qq{radiative} corrections\index{radiative corrections} to
observable quantities (which are tiny but measurable). A
discussion of the renormalization techniques can be found in any
textbook on quantum field theory (for a concise summary, see e.g.
the book \cite{ChL}). As for the QED, this became one of the most
precise physical theories ever conceived -- for a relatively
recent overview of the successes of QED, see e.g. \cite{ref29}.
However, the Fermi-type weak interaction theory is {\it not\/}
renormalizable\index{renormalizable theory|(} in such a manner. A
detailed analysis shows that there are infinitely many types of UV
divergences that would require introducing an infinite number of
parameters in the interaction Lagrangian -- needless to say, the
theory thus loses considerably its predictive power. The point is
that according to a standard \qq{power counting} for Feynman
graphs, a QFT model can only be renormalizable if its interaction
Lagrangian incorporates terms with dimension less than or equal to
four\footnote{The dimension we have in mind here does not include
the corresponding coupling constant and is to be understood as the
pertinent power of a mass; thus, fermionic and bosonic fields have
dimensions 3/2 and 1 resp. and a derivative carries dimension 1
(inverse length has a dimension of mass in the natural system of
units). The dimension of a given term in (polynomial) interaction
Lagrangian is then the sum of dimensions of all fermion and boson
fields and derivatives occurring therein.}. The dimension of the
four-fermion interaction\index{four-fermion interaction} is
obviously equal to {\it six\/} and this becomes fatal for
renormalizability of any Fermi-type theory of weak interactions.
We will not go into further technical details here and rather
refer the interested reader to standard textbooks on quantum field
theory (see e.g. \cite{ItZ}).

Finally, let us emphasize what is perhaps the most interesting
moment of the considered situation. It turns out that -- for a
general QFT model -- the
non-renormalizability\index{non-renormalizability} of UV
divergences in higher orders of perturbation expansion is closely
connected with the character of high-energy behaviour of
scattering amplitudes at lowest order (i.e. at the tree level):
{\bf the power-like growth of a tree-level scattering amplitude
implies non-renormalizability in higher orders.} This statement is
perhaps more useful in the reverse direction: {\bf absence of a
power-like growth of tree-level scattering amplitudes with energy
is a necessary condition for perturbative
renormalizability\index{perturbative renormalizability} at higher
orders of perturbation expansion.} In view of its relation to
unitarity, the absence of a power-like high-energy rise of
scattering amplitudes is usually termed technically as \qq{tree
unitarity}\index{tree unitarity|ff} (cf. e.g. \cite{ref30}). Let
us note that this remarkable connection of two different aspects
of perturbative QFT -- the tree unitarity and UV renormalizability
-- has never been proved quite rigorously, but still it seems to
be valid beyond any reasonable doubt. The point is that there is
no known exception from this rule and, beside that, there is a
rather plausible hand-waving argument in its favour, based on the
technique of dispersion relations\index{dispersion relations} (for
a more detailed discussion, the interested reader is referred e.g.
to \cite{Hor} and to the relevant literature quoted therein, in
particular \cite{ref30}). Throughout this text we will adopt the
criterion of \qq{tree-level unitarity} as a simple and practical
necessary condition for perturbative renormalizability and it will
often serve as a subsidiary guiding principle in our road toward
the unified theory of weak and electromagnetic interactions.

%\input{kniha32}
%%%%%%%%%%%%%%%%%%%%%%%%%%%%%%%%%%%%%%%%%%%%%%%%%%%%%%%%%%%%%%%%%%%
%%%%%%%%%%%%%%%%%%%%%%%%%%%%%%%%%%%%%%%%%%%%%%%%%%%%%%%%%%%%%%%%%%%%%%%%%%%%%%%%%%%%%%%%%%%%%%%%%%%%%%%%%%%%%%%%%%%%%%%%%%%%%%%%%%%%%%%%
\section{The case for intermediate vector
boson}\index{intermediate vector boson|ff}

Having described the \qq{splendeurs et mis\`{e}res} of the
Fermi-type weak interaction theory, one should now seek a viable
alternative, which would lead to a more satisfactory high-energy
behaviour of scattering amplitudes already in lowest
approximation. To this end, it is important to realize that the
source of all difficulties arising within a Fermi-type model is
the {\it dimensionality\/} of the relevant coupling constant
$G_F$. Indeed, as we have seen, this leads to the quadratic growth
of tree-level scattering amplitudes with c.m. energy and is also
obviously related to the fact that the dimension of any
four-fermion interaction is equal to six. Formally, one can get
rid of the dimensionful coupling constant if the original
\qq{current $\times$ current} interaction is replaced by a
coupling of the weak current (\ref{eq2.59}) to a vector field
\begin{equation}
\label{eq3.13}
\lagr^{(w)}_{int}=\frac{g}{2\sqrt 2}(J^\mu W^+_\mu+
J^{\mu\dag}W^-_\mu)
\end{equation}
Obviously, the new coupling constant $g$ is dimensionless, in
analogy with spinor electrodynamics (note that dim $J^\mu$ = 3 and
dim $W_\mu$ = 1). The numerical factor $1/2\sqrt 2$ in
(\ref{eq3.13}) is purely conventional; its origin will become
clear in the context of the gauge theory of weak interactions. The
field $W^+_\mu$ must be complex (non-hermitean) as it is coupled
to the charged current\index{charged current}; of course, we use a
natural notation $W^-_\mu$ = $(W^+_\mu)^{\dag}$. The corresponding
quanta (vector bosons $W^\pm$) are spin-1 particles carrying
electric charge $\pm 1$. Taking into account the structure of the
current $J^\mu$, it is not difficult to realize that the field
$W^+_\mu$ must contain annihilation operator for the $W^+$ boson
and creation operator for the $W^-$, if the charge conservation is
to be maintained in (\ref{eq3.13}). In the theory described by
(\ref{eq3.13}), the vector field $W_\mu$ mediates weak
interactions of fermions and the particle $W^+$ or $W^-$ is
therefore usually called {\bf intermediate vector boson} (IVB).
Note that in the first order of perturbation expansion, the
Lagrangian (\ref{eq3.13}) gives rise to the two-fermion decays of
the $W$ boson (for example, the first term produces
$W^+\rightarrow e^+ +\nu_e$ while its hermitean conjugate leads to
$W^-\rightarrow e^-+\bar\nu_e$)\index{decay!of the W@of the $W$
boson}.

  Some additional remarks are perhaps in order here. It should be
clear that the intermediary of an interaction between fermion
pairs must be a boson -- this is an obvious general consequence of
angular momentum (spin) conservation. Historically, theorists
contemplated the idea of an intermediate weak boson (in analogy
with the description of strong and electromagnetic interactions)
since the late 1930s, i.e. long before the generic technical flaws
of four-fermion models have been appreciated. As we know now, it
took more than 20 years to clarify that such a hypothetical
particle must carry {\it spin one\/} (this, of course, was
tantamount to establishing the dominance of {\it vector and
axial-vector currents\/} in weak interactions). Since the early
1960s, the IVB concept has been taken quite seriously and over the
years, it was discussed in numerous theoretical papers. At the
same time, direct experimental searches have shown soon that if a
$W$ boson exists, its mass must be larger than e.g. $1\ \GeV$, a
typical hadronic mass. Still further 20 years were then necessary
to prove its real existence (for a rather detailed survey of the
IVB history and discovery see e.g. the book \cite{Wat}).

\begin{figure}[h]\centering
\begin{tabular}{cc}
\subfigure[]{\s{\includegraphics{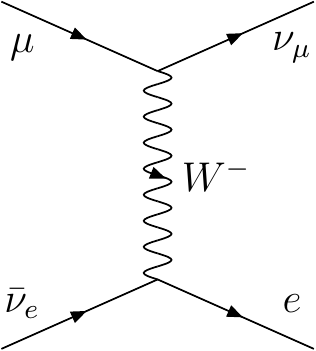}}}&\hspace{1.5cm}\subfigure[]{\s{\includegraphics{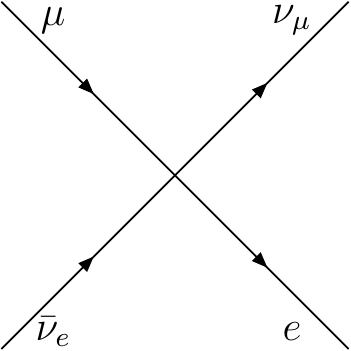}}}
\vspace{-0.2cm}
\end{tabular}
\caption{Tree-level diagrams for the muon decay (a) within the W
boson model (b) within a Fermi-type model.}
\label{fig6n}\index{Feynman diagrams!for the muon decay}
\end{figure}

   Coming back to the IVB interaction Lagrangian (\ref{eq3.13}), we
now have to find out whether such a model can indeed reproduce the
successes of the Fermi-type theory at low energies and whether it
is able to remedy its maladies in the high-energy limit. In order
to examine the correspondence between the two versions of weak
interaction theory in the low-energy limit, let us consider a
particular decay process involving four fermions, e.g. muon
decay\index{decay!of the muon}. Within the IVB model
(\ref{eq3.13}), one needs at least one $W$ exchange to connect the
lepton pairs of muon and electron type -- in other words, the
lowest approximation in which such a process can appear is the
second order of perturbation expansion. The corresponding tree
diagram is shown in Fig.\,\ref{fig6n}, along with its counterpart
arising within the Fermi-type theory.
Note that one must assume a priori that the $W$ boson is massive,
since the weak interaction is known to be of a (very) short range.
Thus, the $W$ exchange in Fig.\,\ref{fig6n}a is described by the
vector boson propagator\index{propagator!of massive vector
boson|(} with $m_W\neq 0$ and the decay matrix element can then be
written as
\begin{eqnarray}
\label{eq3.14}
i{\cal M}_a&=&i^3\left(\frac{g}{2\sqrt 2}\right)^2[\bar u(k)\gamma_\rho
(1-\gamma_5)u(P)][\bar u(p)\gamma_\sigma(1-\gamma_5)v
(k')]\times\nonumber\\
& &\times\frac{-g^{\rho\sigma}+m^{-2}_W q^\rho q^\sigma}{q^2-m^2_W}
\end{eqnarray}
(for a concise summary of basic properties of the massive vector
field see Appendix~\ref{appenD}). On the other hand, for Fig.\,\ref{fig6n}b
we have
\begin{equation}
\label{eq3.15} {\cal M}_b=-\frac{G_F}{\sqrt 2}[\bar
u(k)\gamma_\rho (1-\gamma_5)u(P)][\bar
u(p)\gamma^\rho(1-\gamma_5)v (k')]
\end{equation}
Now, how can one reduce -- at least approximately -- the form
(\ref{eq3.14}) to (\ref{eq3.15})? In fact, this can be done quite
easily. First of all, one has to realize that the kinematical limits
for the four-momentum of the virtual $W$ boson are given by
\begin{equation}
\label{eq3.16}
m^2_e\leq q^2\leq m^2_\mu
\end{equation}
(proving (\ref{eq3.16}) is left to the reader as a simple
exercise). Then, taking $m^2_W\gg m^2_\mu$ (as we have noted
earlier, such a bound for the $W$ mass has been established long
before its discovery), one can safely neglect the $q$-dependence
in the denominator of the propagator in (\ref{eq3.14}). Further,
it is easy to see that the second term in the numerator becomes in
fact proportional to $m_e m_\mu/m^2_W$; indeed, using the
four-momentum conservation and equations of motion for the $u$ and
$v$ spinors, one gets readily
\begin{eqnarray}
\label{eq3.17}
\lefteqn{q^\rho q^\sigma[\bar u(k)\gamma_\rho(1-\gamma_5)u(P)]
[\bar u(p)\gamma_\sigma(1-\gamma_5)v(k')]=}\nonumber\\
&=&[\bar u(k)(\slashed{P}-\slashed{k})(1-\gamma_5)u(P)]
[\bar u(p)(\slashed{p}+\slashed{k}')(1-\gamma_5)
v(k')]\nonumber\\
&=&m_e m_\mu[\bar u(k)(1+\gamma_5)u(P)]
[\bar u(p)(1-\gamma_5)v(k')]
\end{eqnarray}
Thus, the effect of the $q^\rho q^\sigma$ term in the $W$ propagator
can be reliably neglected as well. Putting all this together, we see
that the matrix element (\ref{eq3.14}) is approximately equal to
\begin{equation}
\label{eq3.18}
{\cal M}_a\approx -\frac{g^2}{8m^2_W}[\bar u(k)\gamma_\rho
(1-\gamma_5)u(P)][\bar u(p)\gamma^\rho(1-\gamma_5)v(k')]
\end{equation}
which is indeed of the form (\ref{eq3.15}). Matching the two
expressions, one gets a condition for the parameters of the IVB
model, namely
\begin{equation}
\label{eq3.19}
\frac{G_F}{\sqrt 2}=\frac{g^2}{8m^2_W}
\end{equation}
One should also notice that the origin of the minus sign in the
Fermi-type Lagrangian (\ref{eq2.58}) becomes transparent through
our calculation: such a convention is necessary for the
correspondence relation (\ref{eq3.19}) to be valid with a positive
value of the $G_F$.

  Thus, we have shown that for the considered process the IVB model
(\ref{eq3.13}) leads to the same result as the original Fermi-type
theory (up to corrections of the relative order $O(m^2_\mu/m^2_W)$
or less), provided that parameters of the IVB Lagrangian satisfy
the relation (\ref{eq3.19}). In fact, our reasoning makes it clear
that such an equivalence should hold for any process involving
four light fermions, whenever the relevant momentum transfer
(energy) is small in comparison with the $W$ boson mass. In any
particular example of that kind, the steps described above can be
repeated and one may consequently ignore all momentum-dependence
in the IVB propagator, which is thereby effectively reduced to a
constant with the dimension of (mass)$^{-2}$; a Fermi-type matrix
element thus emerges as a low-energy approximation to the original
expression. A generic structure of the relation (\ref{eq3.19}) is
also transparent: the IVB model at second order ($g^2$) and at low
energy ($m^{-2}_W$ replacing the propagator) is equivalent to a
corresponding Fermi-type model in the first order ($G_F$). Our
preceding considerations can now be concisely summarized as
follows. If (\ref{eq3.19}) is valid, then predictions of the IVB
model (\ref{eq3.13}) and those of the current-current model
(\ref{eq2.58}) are practically indistinguishable for energies and
momentum transfers much smaller than the $W$ boson mass; the
four-fermion Lagrangian (\ref{eq2.58}) thus represents a
low-energy effective theory corresponding to the underlying IVB
model (\ref{eq3.13}).

\begin{figure}[h]\centering
\s{\includegraphics{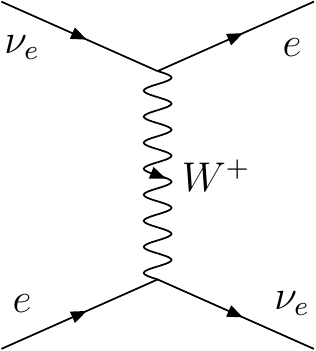}} \caption{Tree-level
$W$-exchange diagram for the $\nu_e - e$ elastic scattering.}
\label{fig7}\index{Feynman diagrams!for $\nu_e-e$ scattering}
\end{figure}

  Now that we have made sure of the right low-energy properties
of the IVB model, let us investigate its behaviour in the
high-energy limit. To this end, we will consider again the
neutrino-electron scattering\index{neutrino-electron scattering},
now described by the tree-level (i.e. second-order) Feynman
diagram shown in Fig.\,\ref{fig7}.

\begin{figure}[h]\centering
\s{\includegraphics{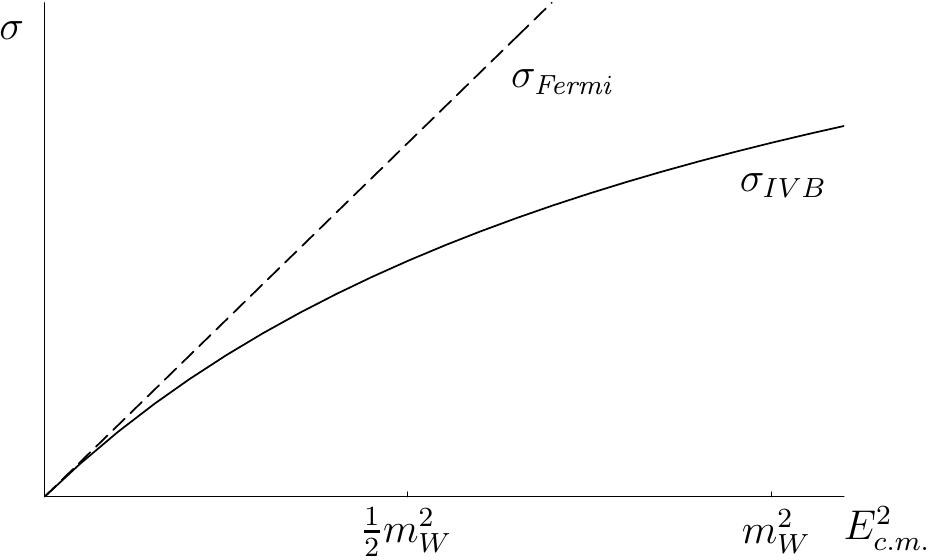}}
\caption{Energy dependence of the $\nu -e$ elastic scattering
cross section within a Fermi-type model (dashed line) and the IVB
model (solid line).}
\label{fig8}
\end{figure}

\noindent
The corresponding matrix element can be written as
\begin{eqnarray}
\label{eq3.20}
i{\cal M}^{(\nu e)}_{IVB}&=&i^3\left(\frac{g}{2\sqrt 2}\right)^2
[\bar u(p')\gamma_\rho(1-\gamma_5)u(k)]
[\bar u(k')\gamma_\sigma(1-\gamma_5)u(p)]\times\nonumber\\
&\times&\frac{-g^{\rho\sigma}+m^{-2}_W q^\rho q^\sigma}
{q^2-m^2_W}
\end{eqnarray}
At first glance, one might worry that we have actually won nothing
in comparison with Fermi theory: the $W$ boson propagator contains
a piece proportional to $m^{-2}_W$ and one could thus expect, on
dimensional grounds, a quadratic growth of the (dimensionless)
matrix element (\ref{eq3.20}) for $E\rightarrow\infty$. However, a
closer look reveals that it is not so. As in the previous example,
one may employ the equations of motion to factorize $m^2_e$ from
the potentially dangerous $q^\rho q^\sigma$ term. Thus, it becomes
in fact strongly suppressed (by the factor of $m^2_e/m^2_W$) in
comparison with the $g^{\rho\sigma}$ term and we will drop it in
subsequent manipulations. Using the standard trace techniques, the
spin-averaged square of the matrix element (\ref{eq3.20}) then
comes out to be
\begin{equation}
\label{eq3.21}
\overline{|{\cal M}^{(\nu e)}_{IVB}|^2}=\frac{1}{2}g^4
\frac{(s-m^2_e)^2}{(u-m^2_W)^2}
\end{equation}
where we have denoted, as usual, $s=(k+p)^2$ and $u=(k-p')^2$
(an astute reader may notice that (\ref{eq3.21}) can in fact be
obtained essentially without any calculation, by utilizing our previous
result (\ref{eq3.6})). With the high-energy limit in mind, we will
of course ignore the effects of electron mass. Then, when recast in
terms of the c.m. scattering angle, the expression (\ref{eq3.21}) becomes
\begin{equation}
\label{eq3.22}
\overline{|{\cal M}^{(\nu e)}_{IVB}|^2}=2g^4\frac{1}
{(1+\cos\vartheta_{c.m.}+2m^2_W/s)^2}
\end{equation}
The corresponding differential cross section is then integrated
easily; one obtains
\begin{eqnarray}
\label{eq3.23} \sigma^{(\nu
e)}_{IVB}&=&\frac{g^4}{16\pi}\frac{1}{s}\int \limits
^1_{-1}\frac{1}{(1+\cos\vartheta_{c.m.}+2m^2_W/s)^2}
d(\cos\vartheta_{c.m.})\nonumber\\
&=&\frac{G^2_F}{\pi}m^2_W\frac{s}{s+m^2_W}
\end{eqnarray}
where we have used the relation (\ref{eq3.19}) in the last step.
From (\ref{eq3.23}) it is obvious that the cross section tends
asymptotically (i.e. for $s\gg m^2_W$) to a constant:
\begin{equation}
\label{eq3.24}
\lim_{s\rightarrow\infty}\sigma^{(\nu e)}_{IVB}(s)=
\frac{G^2_F}{\pi}m^2_W
\end{equation}
Thus we see that the $W$ boson exchange indeed ameliorates the
high-energy behaviour observed earlier within the theory of Fermi
type -- instead of rising rapidly, the considered cross section is
now asymptotically flat. The difference between the two theories
is schematically depicted in Fig.\,\ref{fig8} -- the $W$ boson
mass obviously plays the role of a natural high-energy
\qq{cut-off}.
(Note that for the antineutrino-electron scattering the effect of
the $W$ exchange leads to a cross section that vanishes for
$s\rightarrow\infty$; proving this is left to the reader as an
instructive exercise.) Of course, the suppression of a power-like
growth of the $\nu -e$ cross section becomes clear immediately,
when one observes the effective elimination of the $m^{-2}_W
q^\rho q^\sigma$ term from the $W$ boson propagator: as there
remain no uncompensated constant factors with the dimension of a
negative power of mass, the relevant scattering matrix element can
behave at most as $O(1)$ for $E\rightarrow\infty$ (at a fixed
scattering angle)\footnote{More precisely, the result
(\ref{eq3.22}) makes it clear that our scattering amplitude is
asymptotically flat for any $\theta_{c.m.}\neq 180^\circ$; on the
other hand, for $\theta_{c.m.}=180^\circ$ (i.e. for backward
scattering) it rises as $s/m^2_W$. It is not difficult to realize
that such an isolated singularity is in fact responsible for the
non-zero limit in (\ref{eq3.24}); if  the matrix element ${\cal
M}^{(\nu e)}_{IVB}$ were bounded uniformly, the angular
integration in (\ref{eq3.23}) would obviously yield a cross
section decreasing as $1/s$ for $s\rightarrow\infty$.}  and the
formula for the cross section includes an additional factor of
$1/s$.

Let us now discuss the problem of unitarity bound\index{unitarity
bound}. We will only summarize here briefly the main results; more
technical details can be found e.g. in \cite{Hor}. The relevant
high-energy scattering amplitude (corresponding to
negative-helicity leptons) that can be guessed from (\ref{eq3.22})
has the form
\begin{equation}
\label{eq3.25} {\cal M}^{(\nu
e)}_{IVB}=2g^2\frac{1}{ 1 + \cos \vartheta_{c.m.}+2m^2_W/s}
\end{equation}
The non-trivial angular dependence in the denominator (which of
course is due to the $W$ propagator) implies that now there is an
infinite number of partial waves contributing to the expansion of
(\ref{eq3.25}) (one expands in Legendre polynomials\index{Legendre
polynomials} in the considered case). Up to a normalization
factor, the amplitude of the lowest ($j= 0$) partial
wave\index{partial-wave expansion} is obtained by integrating
(\ref{eq3.25}) over the $\cos\vartheta_{c.m.}$ from $-1$ to $1$;
the result is
\begin{equation}
\label{eq3.26}
{\cal M}^{(0)}_{IVB}(s)=\frac{g^2}{16\pi}\ln(\frac{s}{m^2_W}+1)
\end{equation}
Obviously, such a logarithmic dependence on $s/m_W^2$ is due to the
singularity occurring in (\ref{eq3.25}) at $\vartheta_{c.m.}= 180^\circ$
for $s\rightarrow\infty$ (or, equivalently, for $m_W=0$). Note that
an analogous result holds for higher partial waves as well. Taking
into account that the dimensionless coupling constant $g$ is rather
small (say, of the order of electromagnetic coupling constant $e$),
the slow logarithmic rise of a partial-wave amplitude with energy means
that a conflict with unitarity may only occur at an astronomically
high energy. Indeed, the critical value $s^\star$ for which
(\ref{eq3.26}) saturates the bound (\ref{eq3.11}) is (with a very
good accuracy) given by
\begin{equation}
\label{eq3.27}
s^\star=m^2_W\,\exp\,\left(\frac{16\pi}{g^2}\right)
\end{equation}
which amounts to $E^\star_{c.m.}\approx 10^{29}\ \GeV$ if one
employs the present-day values of the relevant parameters,
$m_W\doteq 80\ \GeV$ and $g\doteq 0.6$. For other processes the
situation may be even better -- in particular, for the $\bar\nu
-e$ scattering there is no logarithmic term in the relevant
partial-wave amplitude and the bound (\ref{eq3.11}) is not
violated even at $s\rightarrow\infty$. In any case, from the above
discussion it should be clear that the $W$-exchange mechanism
suppressing a rapid (power-like) violation of unitarity is rather
general, in the sense that it must work for any fermion-fermion
scattering process.

   To summarize the results obtained so far, one may say that we
have demonstrated explicitly how the IVB model alleviates the
unitarity violation problem encountered earlier within the
Fermi-type weak interaction theory. In fact, the logarithmic rise
of a partial-wave scattering amplitude with energy (at a fixed
order of perturbation expansion) cannot be avoided even within a
renormalizable field theory -- in this sense the \qq{logarithmic
violation of unitarity} exhibited in (\ref{eq3.26}) is the best
high-energy behaviour attainable within a variety of perturbative
QFT models. Let us stress that according to our criterion
formulated in Section~\ref{sec3.1}, within a renormalizable QFT model one
can have {\it at worst\/} a logarithmic violation of perturbative
unitarity since there can be no scattering amplitude rising
asymptotically as a positive power of energy\index{propagator!of
massive vector boson|)}.

%\input{kniha33}
%%%%%%%%%%%%%%%%%%%%%%%%%%%%%%%%%%%%%%%%%%%%%%%%%%%%%%%%%%%%%%%%%%%
%%%%%%%%%%%%%%%%%%%%%%%%%%%%%%%%%%%%%%%%%%%%%%%%%%%%%%%%%%%%%%%%%%%%%%%%%%%%%%%%%%%%%%%%%%%%%%%%%%%%%%%%%%%%%%%%%%%%%%%%%%%%%%%%%%%%%%%%
\section{Difficulties of the simple IVB model}\label{sec3.3}
%3.3
The progress we have achieved so far is not the whole story of the
IVB model. Apart from the four-fermion scattering processes
discussed previously, the interaction Lagrangian (\ref{eq3.13})
describes, at second order of perturbation expansion, also the
production of $W^+ W^-$  pairs in fermion-antifermion
annihilation. As we shall see, the corresponding tree-level
amplitudes can lead, for certain combinations of $W$ boson
helicities, to a rapid (power-like) violation of unitarity at high
energies. In other words, the history repeats itself: the
difficulties characteristic of the Fermi-type theory occur here
just for another class of physical processes.

\begin{figure}[h]\centering
\s{\includegraphics{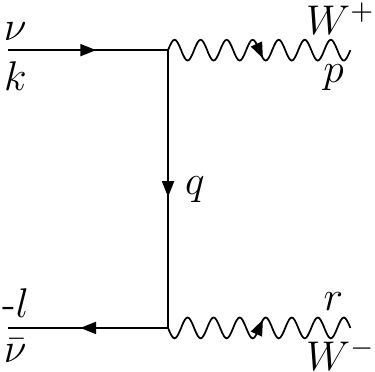}} \caption{Tree-level graph
describing the process $\nu\bar\nu \rightarrow W^+W^-$ within the
IVB model (\ref{eq3.13}).} \label{fig9}\index{Feynman diagrams!for
$\nu\bar\nu\rightarrow W^+W^-$}
\end{figure}

For an explicit illustration of the problems we have in mind, let
us start with neutrino-antineutrino annihilation process
$\nu\bar\nu\rightarrow W^-W^+$. Historically, the earliest known
reference to this example is probably the paper \cite{ref31}
(published two years after the formulation of the electroweak
standard model!). In the lowest non-trivial order, it is described
by the Feynman diagram shown in Fig.\,\ref{fig9}.

The corresponding matrix element can be written as
\begin{equation}
\label{eq3.28}
i{\cal M}^{(e)}_{\nu\bar\nu}=i^3\left(\frac{g}{2\sqrt 2}\right)^2
\bar v(l)\gamma_\mu(1-\gamma_5)\frac{1}{\slashed{q}
-m_e}\gamma_\nu(1-\gamma_5)u(k)\varepsilon^{\ast\mu}(r)
\varepsilon^{\ast\nu}(p)
\end{equation}
(note that we have labelled the $\cal M$ with regard to our later
calculations within electroweak standard model). The polarization
vectors $\varepsilon$ appearing in (\ref{eq3.28}) characterize the
spin (helicity)\index{helicity}\index{polarization!vector} states
of the final-state vector bosons. Their properties are summarized
in Appendix~\ref{appenD}. There are two possible transverse
polarizations\index{transverse polarization}, corresponding to
helicities $\pm 1$ and the longitudinal polarization corresponding
to helicity 0. The existence of the zero-helicity state is a
characteristic distinguishing feature of a massive vector
boson\index{massive vector boson|ff} -- there is no such thing for
massless photon. In fact, it is precisely the $W$ boson
longitudinal polarization vector $\varepsilon_L$ that will play a
crucial role in our subsequent considerations. The important
property of the $\varepsilon_L$ is that its components grow
indefinitely in the high-energy limit as the corresponding
four-momentum itself:
\begin{equation}
\label{eq3.29}
\varepsilon^\mu_L(p)=\frac{1}{m_W}p^\mu+O\left(\frac{m_W}{p_0}
\right)
\end{equation}
The last expression makes it clear, on simple dimensional grounds,
why one should worry about the high-energy behaviour of a
scattering amplitude involving longitudinally
polarized\index{longitudinally polarized vector boson|ff} $W$
bosons: the leading contribution from each polarization vector
$\varepsilon_L$ introduces a factor of inverse mass and one thus
expects that a corresponding positive power of energy will be
needed to get a dimensionless matrix element (for a binary
process). In particular, for (\ref{eq3.28}) one expects a
quadratic growth with energy when both final-state $W$'s are
longitudinally polarized. To make this claim more transparent, let
us now work out the corresponding leading asymptotic term (in a
form that will be useful also in our later calculations within the
standard electroweak model). First of all, from the decomposition
(\ref{eq3.29}) one can infer easily that
\begin{equation}
\label{eq3.30}
\varepsilon^{\ast\mu}_L(r)\varepsilon^{\ast\nu}_L(p)=\frac{r^\mu}
{m_W} \frac{p^\nu}{m_W}+O(1)
\end{equation}
Consequently, (\ref{eq3.28}) can be rewritten as
\begin{equation}
\label{eq3.31}
{\cal M}^{(e)}_{\nu\bar\nu}=-\frac{g^2}{8m^2_W}\bar v(l)
\slashed{r}(1-\gamma_5)\frac{1}{\slashed{q}-m_e}\slashed{p}
(1-\gamma_5)u(k)+O(1)
\end{equation}
Further, one can employ momentum conservation, equations of
motion and some simple algebraic manipulations to cancel
partially the denominator in (\ref{eq3.31}); one thus gets
\begin{eqnarray}
\label{eq3.32} {\cal
M}^{(e)}_{\nu\bar\nu}&=&-\frac{g^2}{4m^2_W}\bar v(l)
\slashed{p}(1-\gamma_5)u(k)\nonumber\\
&&-\frac{g^2}{8m^2_W}m_e\bar
v(l)(1+\gamma_5)\frac{\slashed{q}+m_e}
{q^2-m^2_e}\slashed{p}(1-\gamma_5)u(k)\nonumber\\
&&+O(1)
\end{eqnarray}
Clearly, owing to the presence of an otherwise uncompensated
factor of $m^{-2}_W$, the first term in (\ref{eq3.32}) embodies
the quadratic high-energy divergence.\footnote{We are not going to
work it out as an explicit function of energy, but it is useful to
observe that the $u$ and $v$ spinors behave (within our
normalization convention) as $E^{1/2}$ in the high-energy limit;
together with the factor of $\slashed{p}$, this then makes up the
quadratic rise with energy anticipated on dimensional grounds.} In
the second term, only a contribution proportional to $m^2_e$
survives, which obviously can be absorbed into the asymptotically
flat $O(1)$ remainder in (\ref{eq3.32}). Thus, the tree-level
matrix element for $\nu\bar\nu\rightarrow W_L^+W^-_L$ can be
decomposed as
\begin{equation}
\label{eq3.33}
{\cal M}^{(e)}_{\nu\bar\nu}=-\frac{g^2}{4m^2_W}\bar v(l)
\slashed{p}(1-\gamma_5)u(k)+O(1)
\end{equation}
In this result, the leading asymptotic $O(E^2)$ part of the
considered matrix element is singled out in a rather simple form.
As we shall see later, such a form is in general well suited for a
discussion of divergence cancellations among different diagrams
contributing within the standard electroweak model. The
high-energy divergence in (\ref{eq3.33}) cannot vanish
identically, for an arbitrary scattering angle (unless $g$ = 0).
Thus, in the corresponding partial-wave expansion one must
necessarily run into the problem with rapid violation of
unitarity, completely analogous to that encountered within the old
Fermi-type theory. We will not calculate here explicitly the
relevant partial-wave amplitudes (the interested reader is
referred to the original paper \cite{ref31}). Instead, it may be
instructive to see what is the high-energy behaviour of the
corresponding cross section. Using (\ref{eq3.33}) and the usual
trace techniques, it is straightforward to show that
\begin{eqnarray}
\label{eq3.34} \sigma^{(e)}(\nu\bar\nu\rightarrow W^+_L W^-_L)
&\approx &
\frac{1}{64\pi^2}\frac{1}{s}\left(\frac{g^2}{4m^2_W}\right)^2 \int
\Tr[\slashed{l}\slashed{p} (1-\gamma_5)\slashed{k}
\slashed{p}(1-\gamma_5)] d\Omega_{c.m.}\nonumber\\
&\approx&
\frac{g^4}{512\pi}\frac{s}{m^4_W}\int\limits ^1_{-1}(1-\cos^2
\vartheta_{c.m.})d(\cos\vartheta_{c.m.})\nonumber\\
&=& \frac{G^2_F}{12\pi}s
\end{eqnarray}
for $s\gg m^2_W$. Note that in the last step we have reintroduced
the Fermi constant $G_F$ through the relation (\ref{eq3.19}), in
order to stress the close analogy of the considered case with our
earlier results. Of course, one would get the same asymptotic
behaviour for the unpolarized $W$ boson cross section, calculated
directly from (\ref{eq3.28}) with the help of the standard formula
for the polarization sum\index{polarization!sum}
\begin{equation}
\label{eq3.35}
\sum^3_{\lambda=1}\varepsilon_\mu(k,\lambda)\varepsilon^\ast_\nu
(k,\lambda)=-g_{\mu\nu}+\frac{1}{m^2_W}k_\mu k_\nu
\end{equation}
(the point is that the longitudinally polarized $W$ bosons give
dominant contribution in the high-energy limit). There is an
important general feature of the above results that should be
noticed here. While fermion masses become irrelevant in the
high-energy limit, the mass of a physical $W$ boson cannot be
generally neglected for $E\rightarrow\infty$ simply because this
appears in a negative power in the expressions like
(\ref{eq3.34}). Of course, the source of such an anomalous
behaviour is the longitudinal polarization vector (\ref{eq3.29}),
which is also responsible for the factor of $m^{-2}_W$ in the
polarization sum (\ref{eq3.35}).\footnote{An astute reader might
object that the different character of the mass dependence
manifested in the spin sums for fermions and vector bosons is due
merely to our normalization conventions: we take $\bar
u(k)u(k)=2m$ for Dirac spinors (with the corresponding spin sum
being $\slashed{k}+m$), while
$\varepsilon(k)\cdot\varepsilon^\ast(k)=-1$ for vector boson
polarization vectors (leading to (\ref{eq3.35})). In fact, these
normalization conventions do match each other, for simple
dimensional reasons -- since the vector and Dirac fields have
dimensions of $M$ and $M^{3/2}$ resp., the one-particle states
thus become normalized in the same way for both cases and this in
turn fits into the general cross-section formula given in Appendix~\ref{appenB}.}

The rapid violation of tree-level unitarity, observed here for the
process $\nu\bar\nu\rightarrow W^+_L W^-_L$, also indicates --
according to the criterion formulated at the end of Section~\ref{sec3.1} --
that the model based on the Lagrangian (\ref{eq3.13}) is not
renormalizable in higher orders of perturbation expansion. Such a
claim was indeed proved in 1960s (see \cite{ref32}). Thus, the
considered IVB model of weak interactions constitutes in fact only
a partial improvement of the Fermi-type theory: while some old
problems (concerning four-fermion processes) are solved, new
difficulties show up, due to longitudinally polarized physical $W$
bosons. Such a flaw obviously cannot be removed within the simple
model (\ref{eq3.13}) itself and thus it is clear that a further
amelioration of technical properties of the weak interaction
theory may only be achieved within a broader theory.

%\input{kniha34}
%%%%%%%%%%%%%%%%%%%%%%%%%%%%%%%%%%%%%%%%%%%%%%%%%%%%%%%%%%%%%%%%%%%
%%%%%%%%%%%%%%%%%%%%%%%%%%%%%%%%%%%%%%%%%%%%%%%%%%%%%%%%%%%%%%%%%%%%%%%%%%%%%%%%%%%%%%%%%%%%%%%%%%%%%%%%%%%%%%%%%%%%%%%%%%%%%%%%%%%%%%%%

\section{Electromagnetic interactions of $W$ bosons}

\index{electrodynamics of vector bosons|(}
\index{electromagnetic!interaction|ff}One possible extension of
the IVB theory is immediately clear. Since the $W$ boson carries
electric charge, one should also consider its electromagnetic
interactions. This subject has been discussed in considerable
detail in \cite{Hor}, so we are going to give here only a concise
summary of the most important results.

Let us start with the free-field Lagrangian for $W^\pm_\mu$
(understood here as the \qq{matter fields}). It can be written as
\begin{equation}
\label{eq3.36}
\lagr_0=-\frac{1}{2}W^-_{\mu\nu}W^{+\mu\nu}+m^2_W W^-_\mu
W^{+\mu}
\end{equation}
where we have denoted $W^\pm_{\mu\nu}=\partial_\mu W^\pm_\nu
-\partial_\nu W^\pm_\mu$. A usual way of introducing the
electromagnetic interaction consists in the \qq{minimal
substitution} for the derivatives in the corresponding kinetic
term. In the present case this means that (\ref{eq3.36}) is
replaced by
\begin{eqnarray}
\label{eq3.37}
\lagr^{(min.)}_{em}&=&-\frac{1}{2}(D_\mu
W^-_\nu-D_\nu W^-_\mu)
(D^{\mu\ast}W^{+\nu}-D^{\nu\ast}W^{+\mu})\nonumber\\
&&+m^2_W W^-_\mu W^{+\mu}
\end{eqnarray}
where $D_\mu=\partial_\mu +ie A_\mu$ and
$D^\ast_\mu=\partial_\mu -ieA_\mu$, with $A_\mu$ being the
electromagnetic four-potential and $e$ denoting the relevant
coupling constant ($e > 0$ and $e^2/4\pi =\alpha$ is the
fine-structure constant, $\alpha\doteq$ 1/137). Note that
(\ref{eq3.37}) is built in a straightforward analogy with the
familiar electrodynamics of charged Dirac field (which the reader
is supposed to know from an introductory field-theory course). One
may observe that the Lagrangian (\ref{eq3.37}) is invariant under
the gradient transformations
\begin{equation}
\label{eq3.38}
A'_\mu(x)=A_\mu(x)+\frac{1}{e}\partial_\mu\omega(x)
\end{equation}
accompanied with the corresponding local phase transformations of
the charged fields $W^\pm_\mu$
\begin{eqnarray}
\label{eq3.39}
W^{-'}_\mu(x)&=&\text{e}^{-i\omega(x)}W^-_\mu(x)\nonumber\\
W^{+'}_\mu(x)&=&\text{e}^{i\omega(x)}W^+_\mu(x)
\end{eqnarray}
Using a standard terminology, (\ref{eq3.38}) and (\ref{eq3.39})
represent local gauge transformations or, simply, gauge
transformations. We will discuss the concept of gauge
invariance\index{gauge invariance} more thoroughly in the next
chapter -- here we only stress again the similarity with
electrodynamics of a charged Dirac field. The interaction terms
descending from (\ref{eq3.37}) are
\begin{eqnarray}
\label{eq3.40} \lagr^{(min.)}_{int}&=& - ie[(A^\mu W^{-\nu}-A^\nu
W^{-\mu})
\partial_\mu W^+_\nu -(A^\mu W^{+\nu}-A^\nu W^{+\mu})\partial_\mu
W^-_\nu]\nonumber\\
&& - e^2(A_\mu A^\mu W^-_\nu W^{+\nu}-A^\mu A^\nu W^-_\mu W^+_\nu)
\end{eqnarray}
In fact, one may consider more general gauge-invariant interaction
terms than those contained in (\ref{eq3.37}). We restrict ourselves
{\it a priori\/} to interaction Lagrangians with dimension not greater
than four, in order to avoid a coupling constant with dimension of a
negative power of mass (that would lead automatically to the by now
familiar difficulties in high-energy limit) and for simplicity we will
also assume the parity invariance. Then there is only one possible
addition to (\ref{eq3.40}), namely
\begin{equation}
\label{eq3.41}
\lagr'_{int}=-i\kappa eW^-_\mu W^+_\nu F^{\mu\nu}
\end{equation}
where, of course, $F_{\mu\nu}=\partial_\mu A_\nu-\partial_\nu
A_\mu$ and $\kappa$ is an arbitrary real parameter (this
determines the value of the $W$ boson magnetic moment and electric
quadrupole moment -- see e.g. \cite{ref32}, \cite{ref33} and also
the book \cite{Tay}).\footnote {Note that within electrodynamics
of spin-$\frac{1}{2}$ fermions a term analogous to (\ref{eq3.41})
would have the form $\bar\psi\sigma_{\mu\nu}\psi F^{\mu\nu}$.
However, in contrast to (\ref{eq3.41}), this is of dimension 5 and
spoils perturbative renormalizability\index{perturbative
renormalizability}.} Thus, a general electromagnetic interaction
of $W$ bosons can be written as
\begin{eqnarray}
\label{eq3.42} \lagr^{(em)}_{int}&=&-ie[A^\mu(W^{-\nu}\partial_\mu
W^+_\nu
-\partial_\mu W^-_\nu W^{+\nu})\nonumber\\
&&+W^{-\mu}(\kappa W^{+\nu}\partial_\mu A_\nu -\partial_\mu
W^{+\nu}A_\nu)\nonumber\\
&&+W^{+\mu}(A^\nu\partial_\mu W^-_\nu -\kappa\partial_\mu
A^\nu W^-_\nu)]\nonumber\\
&&-e^2(A_\mu A^\mu W^-_\nu W^{+\nu}-A^\mu A^\nu W^-_\mu W^+_\nu)=
\nonumber\\
&=&\lagr^{(\kappa)}_{WW\gamma}+\lagr_{WW\gamma\gamma}
\end{eqnarray}
where we have marked explicitly the trilinear ($WW\gamma$) and
quadrilinear ($WW\gamma\gamma$) parts respectively (we prefer to
label the interaction Lagrangians in terms of the corresponding
particle symbols, i.e. $WW\gamma$ instead of $WWA$ etc.). Notice
that the $WW\gamma\gamma$ part is independent of $\kappa$. The
interaction terms appearing in (\ref{eq3.42}) are of
renormalizable type as they have dimension four (consequently, the
corresponding coupling constants are dimensionless). On the other
hand, we are already well aware of the difficulties associated
with zero-helicity\index{helicity} states of the physical $W$
bosons. Therefore one might worry that the rapid violation of
unitarity (and the ensuing loss of perturbative renormalizability)
could show up for $W$ boson electromagnetic interactions as well,
in analogy with the weak interaction case discussed previously.
The problem is analyzed in detail in \cite{Hor} and it turns out
that such expectations are indeed fulfilled. We will summarize
here the salient points of such an analysis.

\begin{figure}[h]\centering
\begin{tabular}{cc}
\subfigure[]{\s{\includegraphics{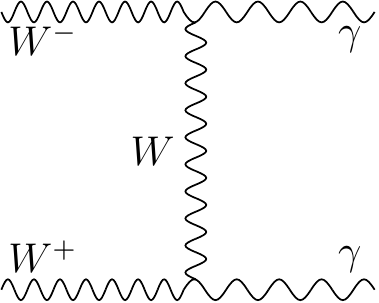}}}&\hspace{1.5cm}\subfigure[]{\s{\includegraphics{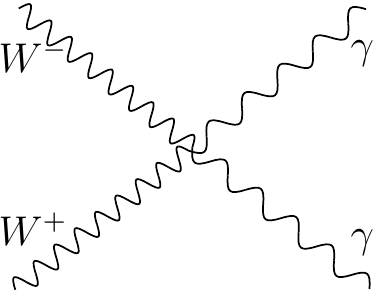}}}
\end{tabular}\vspace{-0.2cm}
\caption{Tree-level diagrams contributing to the electromagnetic
process $W^+ W^-\rightarrow\gamma\gamma$}
\label{fig10}\index{Feynman diagrams!for
$W^+W^-\rightarrow\gamma\gamma$}
\end{figure}

One may start with a particular tree-level electromagnetic
process, e.g. with the two-photon annihilation of the $W^+W^-$
pair. The relevant Feynman diagrams of the order $O(e^2)$ in the
electromagnetic coupling are depicted in Fig.\,\ref{fig10}.
Note that the considered process is particularly interesting for
our purpose since the corresponding diagrams involve both external
and internal $W$ lines and both potential sources of a \qq{bad}
high-energy behaviour (longitudinal polarization vectors and the
$W$ propagator) thus occur here. Since the interaction Lagrangian
contains an arbitrary parameter $\kappa$, one may also wonder how
its value can influence the high-energy asymptotics of the
diagrams in question. The corresponding calculations are somewhat
tedious, but the conclusion that emerges is rather remarkable.
{\bf The tree-level $\boldsymbol{W^+ W^- \rightarrow\gamma\gamma}$
amplitude is asymptotically flat (i.e. free of power-like
divergences) for any combination of external $W$ boson helicities
if and only if $\boldsymbol{\kappa = 1}$.} The lesson to be learnt
from this example is that the only possible candidate for a
renormalizable electrodynamics of $W$ bosons is the model with
$\kappa$ = 1. Indeed, our little theorem claims that if
$\kappa\neq 1$, the tree-level unitarity (which is a necessary
condition for renormalizability) would be violated for a
particular combination of $W$ boson polarizations in the amplitude
of $W^+W^-\rightarrow\gamma\gamma$.

\begin{figure}[h]\centering
\s{\includegraphics{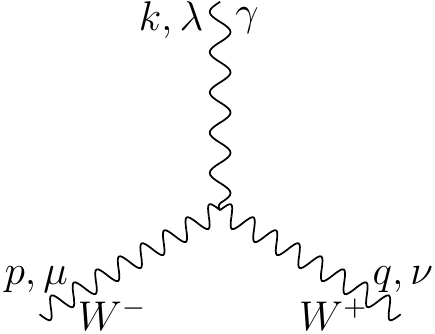}} \caption{An example of the
labelling of the $WW\gamma$ vertex. All four-momenta are taken as
outgoing.} \label{fig11}\index{Feynman diagrams!for vertex
$WW\gamma$}
\end{figure}

Thus, let us adopt the interaction Lagrangian (\ref{eq3.42}) with
$\kappa$ = 1. Its $WW\gamma$\index{WWgamma@$WW\gamma$ interaction}
part can then be written as\footnote{For reasons that will become
clear in the following chapters, the interaction (\ref{eq3.43})
may be called the $WW\gamma$ coupling of Yang--Mills type.}
\begin{equation}
\label{eq3.43} \lagr^{(\kappa=1)}_{WW\gamma}=-ie(A^\mu
W^{-\nu}\dmd W^+_\nu+W^{-\mu}W^{+\nu} \dmd A_\nu +W^{+\mu}
A^\nu\dmd W^-_\nu)
\end{equation}
where the symbol $\partialvob$ is defined by $f \dmd
g=f(\partial_\mu g)-(\partial_\mu f)g$. This in turn leads to the
following momentum-space Feynman rule: when each line involved in
the corresponding vertex is labelled by a corresponding
four-momentum and a Lorentz index as shown in Fig.\,\ref{fig11},
then the contribution of the $WW\gamma$ vertex is given by the
function\index{trilinear (triple) coupling of vector bosons}
\begin{equation}
\label{eq3.44}
V_{\lambda\mu\nu}(k,p,q)=g_{\lambda\mu}(k-p)_\nu +g_{\mu\nu}(p-q)
_\lambda +g_{\lambda\nu}(q-k)_\mu
\end{equation}
multiplied by the coupling constant $e$. For a vertex involving an
incoming line, the corresponding four-momentum in (\ref{eq3.44}) is
taken with opposite sign. Note also that an incoming $W^\pm$ is
equivalent to an outgoing $W^\mp$.

\begin{figure}[h]\centering
\begin{tabular}{cc}
\subfigure[]{\s{\includegraphics{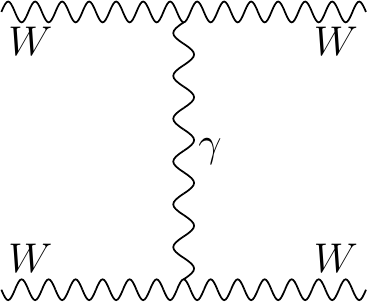}}}&\hspace{1.5cm}\subfigure[]{\s{\includegraphics{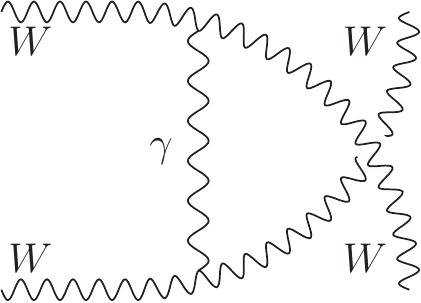}}}
\end{tabular}
\vspace{-0.2cm} \caption{Tree-level QED diagrams for the process
$WW\rightarrow WW$.} \label{fig12}\index{Feynman diagrams!for
$WW\rightarrow WW$}
\end{figure}

In view of our previous observations, the form (\ref{eq3.43}) (or,
equivalently, (\ref{eq3.44})) represents, in a sense, an
\qq{optimal choice} for the QED of $W$ bosons. However, it is not
difficult to demonstrate that even such an option for the
$WW\gamma$ vertex is not able to tame the bad high-energy
behaviour for all possible electromagnetic processes. As a
pertinent example illustrating this one may choose the $WW$
elastic scattering. The corresponding lowest-order Feynman graphs
are shown in Fig.\,\ref{fig12}.
Using (\ref{eq3.44}), the contribution of Fig.\,\ref{fig12}a can
be written as
\begin{equation}
\label{eq3.45} i{\cal M}_a=i^3
e^2\varepsilon^{\ast\mu}(p)\varepsilon^\nu(k)V_
{\mu\nu\rho}(p,-k,q)\frac{-g^{\rho\sigma}}{q^2}V_{\sigma\alpha\beta}
(-q,r,-l)\varepsilon^{\ast\alpha}(r)\varepsilon^\beta(l)
\end{equation}
where we have also employed the standard form of the photon
propagator\index{photon propagator} in Feynman gauge. Of course,
the corresponding expression for Fig.\,\ref{fig12}b is obtained
from (\ref{eq3.45}) by interchanging $p$ and $r$. Now, the worst
high-energy behaviour can be expected in the case when all the
external $W$ bosons are longitudinally polarized. Taking into
account (\ref{eq3.29}) one may guess, on simple dimensional
grounds, that the leading asymptotic term in (\ref{eq3.45})
behaves as $O(E^4/m^4_W)$. This is indeed confirmed by a direct
calculation. Substituting into (\ref{eq3.45}) the decomposition
(\ref{eq3.29}) for each polarization vector, one obviously gets an
expansion
\begin{eqnarray}
\label{eq3.46} {\cal M}_a&=&-e^2\frac{p^\mu}{m_W}\frac{k^\nu}{m_W}
V_{\mu\nu\rho}
(p,-k,q)\frac{-g^{\rho\sigma}}{q^2}V_{\sigma\alpha\beta}
(-q,r,-l)\frac{r^\alpha}{m_W}\frac{l^\beta}{m_W}\nonumber\\
&&+O(\frac{E^2}{m^2_W})+O(1)
\end{eqnarray}
To work out the first (leading) term explicitly, one can employ
the\index{t Hooft identity@'t Hooft identity}
identity\footnote{Note that the relation (\ref{eq3.47}) is
sometimes called 't Hooft identity, since it has been probably
given first in the paper \cite{ref34}.}
\begin{equation}
\label{eq3.47}
p^\mu V_{\lambda\mu\nu}(k,p,q)=(k^2 g_{\lambda\nu}-k_\lambda
k_\nu)-(q^2g_{\lambda\nu}-q_\lambda q_\nu)
\end{equation}
(A practically important feature of this formula is that its
right-hand side is a difference of two transverse expressions. It
should be stressed that the last identity is valid for arbitrary
four-momenta satisfying $k + p + q = 0$. A proof of (\ref{eq3.47})
is left to the reader as an easy exercise.) After some simple
manipulations and taking into account that $k^2=l^2=p^2=r^2=m^2_W$,
the expression (\ref{eq3.46}) can then be recast as
\begin{equation}
\label{eq3.48}
{\cal M}_a=\frac{e^2}{4m^4_W}(t^2+2ts)+O(\frac{E^2}{m^2_W})
+O(1)
\end{equation}
where we have used the standard notation $s=(k+l)^2$ and
$t=(k-p)^2$. The contribution of Fig.\,\ref{fig12}b is then
obtained from (\ref{eq3.48}) by the replacement $t\rightarrow u$,
with $u=(k-r)^2$. Adding the two contributions and using the
kinematical identity $s+t+u=4m^2_W$, the $W_L W_L$ scattering
amplitude thus finally becomes
\begin{equation}
\label{eq3.49}
{\cal M}^{(\gamma)}_{WW}=\frac{e^2}{4m^4_W}(t^2+u^2-2s^2)+O
(\frac{E^2}{m^2_W})+O(1)
\end{equation}
This result exhibits clearly the quartic growth of the considered
amplitude with energy. The leading $O(E^4)$ term depends on the
scattering angle through the Mandelstam variables $t$ and $u$, but
obviously it cannot vanish identically (unless $e$ = 0). The
next-to-leading term $O(E^2)$ has a rather complicated form, but
we will not need it now.

Thus, we may conclude that there is no choice of the parameter
$\kappa$ in (\ref{eq3.42}), which would eliminate all potential
high-energy divergences\index{high-energy divergences} in the
tree-level scattering amplitudes. Consequently, {\bf the quantum
electrodynamics of $W$ bosons cannot be
renormalizable}\index{quantum!electrodynamics (QED)}, in contrast
to the \qq{textbook} case of the spinor QED. In any case, the
$WW\gamma$ interaction of Yang--Mills type, corresponding to
$\kappa$ = 1, seems to be the \qq{best} choice for QED of $W$
bosons and we will use it in what follows as an appropriate
reference model.

%\input{kniha35}
%%%%%%%%%%%%%%%%%%%%%%%%%%%%%%%%%%%%%%%%%%%%%%%%%%%%%%%%%%%%%%%%%%%
%%%%%%%%%%%%%%%%%%%%%%%%%%%%%%%%%%%%%%%%%%%%%%%%%%%%%%%%%%%%%%%%%%%%%%%%%%%%%%%%%%%%%%%%%%%%%%%%%%%%%%%%%%%%%%%%%%%%%%%%%%%%%%%%%%%%%%%%
\section{The case for electroweak unification}

\begin{figure}[h]\centering
\begin{tabular}{cc}
\subfigure[]{\s{\includegraphics{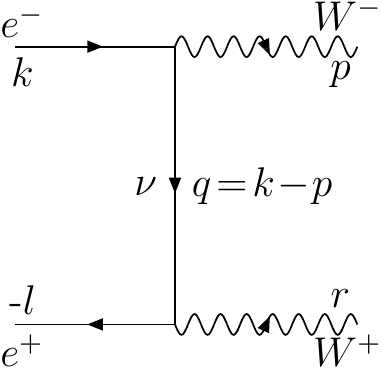}}}&\hspace{1.5cm}\subfigure[]{\s{\includegraphics{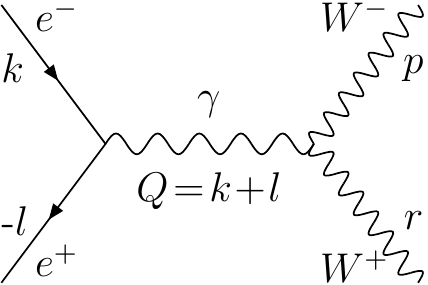}}}
\end{tabular}\vspace{-0.3cm}
\caption{Weak and electromagnetic second-order contributions to
$e^+e^-\rightarrow W^+W^-$.} \label{fig13}\index{Feynman
diagrams!for $e^+e^-\rightarrow W^+W^-$}
\end{figure}

\index{electroweak unification|ff} One can find other examples
showing that the IVB model of weak interactions and
electrodynamics of $W$ bosons suffer from the same technical
difficulties. In particular, there are processes that receive both
weak and electromagnetic contributions at the level of tree
diagrams\index{tree diagrams}. One such example is the process
$e^+e^-\rightarrow W^+W^-$. The relevant lowest-order Feynman
diagrams are depicted in Fig.\,\ref{fig13}.
Let us examine the weak and electromagnetic contribution
separately. For Fig.\,\ref{fig13}a one has
\begin{equation}
\label{eq3.50} i{\cal M}^{(\nu)}_{e^+e^-}=i^3\left(\frac{g}{2\sqrt
2} \right)^2\bar v(l)
\gamma_\mu(1-\gamma_5)\frac{1}{\slashed{q}}\gamma_\nu(1-\gamma_5)
u(k)\varepsilon^{\ast\mu}(r)\varepsilon^{\ast\nu}(p)
\end{equation}
Invoking the usual dimensional arguments, one may guess easily
that such a matrix element grows quadratically with energy when
both final-state $W$ bosons are longitudinally polarized. To
evaluate the leading $O(E^2)$ divergence, one can proceed in
analogy with the process $\nu\bar\nu\rightarrow W^+_L W^-_L$
discussed in Section~\ref{sec3.3}. Substituting
$\varepsilon^\mu_L(r)\varepsilon^\nu_L(p)$ in the general
expression (\ref{eq3.50}) and using the decomposition
(\ref{eq3.29}) one gets, after some manipulations, the result
\begin{equation}
\label{eq3.51} {\cal M}^{(\nu)}_{e^+e^-}=-\frac{g^2}{4m^2_W}\bar
v(l)\slashed{p}(1-\gamma_5)u(k)+O(\frac{m_e}{m^2_W}E)+O(1)
\end{equation}
Notice that in contrast to the $\nu\bar\nu$ annihilation case,
here one also gets a linearly divergent term.

The contribution of Fig.\,\ref{fig13}b is given by
\begin{equation}
\label{eq3.52} i{\cal M}^{(\gamma)}_{e^+e^-}=-i^3e^2\bar v(l)
\gamma_\alpha u(k)\frac{-g^{\alpha\nu}}{Q^2}V_{\lambda\mu\nu}
(p,r,-Q) \varepsilon^{\ast\lambda}(p)\varepsilon^{\ast\mu}(r)
\end{equation}
(note that in writing (\ref{eq3.52}) we have taken into account
that the coupling factor for the $e^+e^-\gamma$ vertex is ($-e$)).
Again, for longitudinally polarized $W^\pm$ one can use
(\ref{eq3.29}) and the identity (\ref{eq3.47}). One then gets,
after some algebra
\begin{equation}
\label{eq3.53} {\cal M}^{(\gamma)}_{e^+e^-}=\frac{e^2}{m^2_W}\bar
v(l)\slashed{p} u(k)+O(1)
\end{equation}
Although the considered two diagrams look rather different
(Fig.\,\ref{fig13}a corresponds to a $t$-channel fermion exchange,
while Fig.\,\ref{fig13}b represents a bosonic exchange in the
$s$-channel), the leading divergent\index{leading divergences}
terms in (\ref{eq3.51}) and (\ref{eq3.53}) come out in a similar
form. Such a similarity raises a hope that the high-energy
divergences of weak and electromagnetic origin might cancel within
a broader unified theory if e.g. the ratio of $e$ and $g$ is
chosen appropriately. In fact, if one simply adds (\ref{eq3.51})
and (\ref{eq3.53}), a complete cancellation of the $O(E^2)$ terms
obviously cannot be achieved since the matrix factor of
$1-\gamma_5$ occurs in (\ref{eq3.51}), while (\ref{eq3.53}) can be
split into two equal parts involving $1-\gamma_5$ and
$1+\gamma_5$. In other words, weak interactions violate parity
maximally\index{parity!violation}\index{parity!conservation},
while the electromagnetic interactions are parity-conserving --
such a deep difference cannot be simply compensated in the two
diagrams themselves. Moreover, (\ref{eq3.51}) includes another
term that diverges linearly for $E\rightarrow\infty$, but this is
absent in (\ref{eq3.53}). Thus, a new particle exchange would be
clearly needed to cancel the divergence in the sum of
Figs.\ref{fig13}a and \ref{fig13}b.

\begin{figure}[h]\centering
\begin{tabular}{cc}
\subfigure[]{\s{\includegraphics{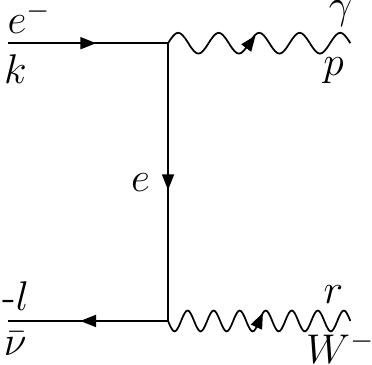}}}&\hspace{1.5cm}\subfigure[]{\s{\includegraphics{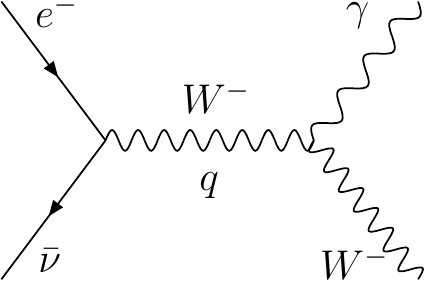}}}
\end{tabular}\vspace{-0.3cm}
\caption{Tree-level graphs for the electro-weak process $\bar\nu_e
e^-\rightarrow W^-\gamma$. These two graphs actually represent the
full second-order contribution of the current standard model.}
\label{fig14}\index{Feynman diagrams!for $\bar\nu_e e^-\rightarrow
W^-\gamma$}
\end{figure}

The above example -- as well as those discussed in preceding
sections -- make it obvious that a simple addition of the weak and
electromagnetic interaction Lagrangians cannot remedy, in general,
the technical flaws inherent in these models. Nevertheless, it is
in order to remark that there is at least one type of an
\qq{electro-weak} process, for which the interactions considered
so far do produce a well-behaved scattering amplitude. The
simplest example is provided by the reaction $\bar\nu e\rightarrow
W\gamma$, described by the diagrams shown in Fig.\,\ref{fig14}
(a variant of such a process, which is far more realistic from the
point of view of present-day experiments, is $\bar u d\rightarrow
W\gamma$, where the $d$\index{dquark@$d$-quark} and
$u$\index{u-quark@$u$-quark} are quarks with charges $-1/3$ and
$2/3$ resp.). We are not going to perform the corresponding
calculation in detail, but one salient point should perhaps be
emphasized here. Remembering the usual power-counting dimensional
analysis, one might worry that the graph in Fig.\,\ref{fig14}b
diverges faster than Fig.\,\ref{fig14}a in the high-energy limit,
since the longitudinal term in the $W$ boson propagator introduces
an extra factor of $m^{-2}_W$. In fact, it is easy to show that
such a term leads to a contribution to the scattering matrix
element that is always asymptotically flat (even for
longitudinally polarized external $W$ boson!). To see this, let us
consider the expression
\begin{equation}
\label{eq3.54}
\bar v(l)\gamma_\rho(1-\gamma_5)u(k)q^\rho q^\nu V_{\lambda\mu\nu}
(p,r,-q)\varepsilon^{\ast\lambda}(p)\varepsilon^{\ast\mu}(r)
\end{equation}
which constitutes the potentially dangerous part of the
contribution of Fig.\,\ref{fig14}b. Using the obvious cyclicity
property of the $WW\gamma$ vertex function (\ref{eq3.44}) (i.e.
$V_{\lambda\mu\nu}(p,r,-q) = V_{\mu\nu\lambda}(r,-q,p)$ etc.), the
't Hooft identity (\ref{eq3.47}) and equations of motion (i.e.
$p\cdot\varepsilon^\ast(p)=0$, $r\cdot\varepsilon^\ast(r)=0$,
$p^2=0$, $r^2=m^2_W$ and $\bar v(l)\slashed{l}=0$, $\slashed{k}
u(k)=m_e u(k)$), it becomes
\begin{equation}
\label{eq3.55}
-m_em^2_W\bar v(l)(1+\gamma_5)u(k)\varepsilon^\ast(p)\cdot
\varepsilon^\ast(r)
\end{equation}
Thus, the expression (\ref{eq3.54}) is proportional to a factor of
(mass)$^3$ and this is sufficient to swamp completely any negative
power of $m_W$ that would arise from the $W$ propagator and a
polarization vector\index{polarization!vector} of the external $W$
boson (this can be $m^{-3}_W$ at worst). The diagram in
Fig.\,\ref{fig14}b can therefore only produce a linear high-energy
divergence (in case of a longitudinally polarized $W$ boson) and
this is exactly compensated by the contribution of
Fig.\,\ref{fig14}a.

The lesson to be learnt from the examples given in this chapter is
that there are certainly some {\it technical\/} reasons for a
non-trivial unification of weak and electromagnetic forces: when
the weak interaction theory and the electrodynamics of $W$ bosons
are taken separately, one encounters rapid violation of
perturbative unitarity at various places and, consequently, the
renormalizability is lost. Thus, if one wishes to cancel somehow
the high-energy divergences in both theories, the only logical
possibility apparently consists in unifying the two interactions.
However, as we have seen, the simple addition
$\lagr^{(w)}_{int}+\lagr^{(em)}_{int}$ is not sufficient for such
a purpose. Therefore, one obviously has to envisage a broader
unification framework, including additional particles and
interactions. Since these additional interactions are designed to
compensate the high-energy divergences of both weak and
electromagnetic origin, they should in a sense \qq{interpolate}
between the original two forces, i.e. they must necessarily mix
the weak and electromagnetic couplings (in other words, one should
expect that the coupling strengths of the additional
\qq{compensating} interactions are non-trivial combinations of the
$e$ and $g$). Such a theoretical scheme can then be naturally
called {\bf electroweak unification}.

Taking into account our previous knowledge, one may envisage
the corresponding interaction Lagrangian in a form
\begin{equation}
\label{eq3.56}
\lagr^{(ew)}_{int}=\lagr_{CC}+\lagr^{(em)}_{fermion}+\lagr_
{WW\gamma}+\lagr_{WW\gamma\gamma}+...
\end{equation}
where the first term represents the charged-current weak
interaction (\ref{eq3.13}) and the remaining ones stand for the
electromagnetic interactions of fermions (quarks and leptons) and
$W$ bosons (note that the Yang--Mills form (\ref{eq3.43}) is
assumed tacitly for the $\lagr_{WW\gamma}$). The ellipsis
symbolizes the \qq{missing links} of the electroweak unification
that should presumably make the theory well-behaved in the
high-energy limit. One can indeed construct a solution to this
problem by adding in (\ref{eq3.56}) new interaction terms so as to
cancel systematically the high-energy divergences arising within
the provisional model. In fact, there are infinitely many
solutions that may be obtained in this way, but in turns out that
the minimal\footnote{The adjective \qq{minimal} refers to the
particle content of a considered model.} electroweak theory
satisfying the criterion of tree unitarity\index{tree unitarity}
is just the present-day Standard Model. The construction of
renormalizable models of weak and electromagnetic interactions
from the high-energy constraints on tree-level Feynman diagrams
has been first implemented in the papers \cite{ref30},
\cite{ref35}, \cite{ref36} and for SM it is also described in
detail in \cite{Hor} (for another pedagogical exposition see the
lecture notes \cite{ref37}). Such a derivation of  the electroweak
standard model \qq{from scratch} is quite remarkable not only
technically, but also conceptually: it shows that the whole
structure of SM (which admittedly may seem rather complicated to
an uninitiated person) is in fact inevitable, if one insists on
perturbative renormalizability\index{perturbative
renormalizability}.

However, the right solution to the electroweak unification problem
has originally been found in a completely different way
\cite{ref38}, \cite{ref39}, \cite{ref40}. Instead of going through
a tedious diagram analysis, the inventors of the electroweak SM
simply had a right inspiration: they employed a rather abstract
principle of (broken) gauge symmetry, which in fact has become
subsequently the theoretical backbone of the whole modern particle
physics. This is precisely the path we are going to follow in the
subsequent chapters. The desired cancellations of high-energy
divergences must then be verified {\it a posteriori}, but such a
\qq{symmetry construction} does have certain advantage over the
aforementioned \qq{engineering approach} -- a specific formulation
of the scheme of broken gauge symmetry within SM provides a deeper
insight into the meaning of the cancellation mechanism.

As a prelude to the discussion of this fundamental method, one may
observe that there is in fact a simple a priori {\it aesthetic\/}
argument in favour of a unified treatment of weak and
electromagnetic forces: both interactions are of {\it vectorial
nature\/} (Lorentz vector or pseudovector currents are involved in
both cases) and they are {\it universal} -- that is, they act
between widely different particles (such as quarks and leptons)
with equal strength. Moreover, the pure vector part of the weak
current belongs to the same isospin multiplet (isotriplet) as the
electromagnetic current (cf. the discussion of CVC\index{conserved
vector current (CVC)} in Section~\ref{sec2.7}). The vectorial character of
the two forces means that the corresponding mediators (photon and
$W^\pm$) have spin 1 and one thus may imagine placing them -- at
least formally -- into a common symmetry multiplet. It turns out
that the concept of non-Abelian gauge
symmetry\index{non-Abelian!symmetry} \cite{ref41} (discovered
originally without any direct motivation from the side of weak
interaction theory) fits, in fact, precisely to this situation.

Thus, as we have seen, {\bf there are both technical and
\qq{moral} (aesthetic) arguments in favour of a unification of
weak and electromagnetic interactions}. In the following chapters
we will develop the ideas and techniques of broken gauge symmetry
that are crucial for the construction of a technically successful
(renormalizable) electroweak theory\index{space
reflection|)}\index{renormalizable theory|)}\index{electrodynamics
of vector bosons|)}.

%\input{problems3}
%%%%%%%%%%%%%%%%%%%%%%%%%%%%%%%%%%%%%%%%%%%%%%%%%%%%%%%%%%%%%%%%%%%
%%%%%%%%%%%%%%%%%%%%%%%%%%%%%%%%%%%%%%%%%%%%%%%%%%%%%%%%%%%%%%%%%%%%%%%%%%%%%%%%%%%%%%%%%%%%%%%%%%%%%%%%%%%%%%%%%%%%%%%%%%%%%%%%%%%%%%%%
\begin{priklady}{11}
\item Calculate the decay width for $W^- \rightarrow \ell^- + \bar{\nu}_\ell$
for unpolarized particles. Neglect $m_{\bar{\nu}}$, but keep
$m_\ell\neq 0$.

\item Calculate longitudinal polarization of charged leptons produced
in the decay of an unpolarized $W$ boson. Check correctness of the
obtained result by setting there $m_\ell = 0$.

\item  Calculate the angular distribution of charged leptons
produced in the decay of a polarized $W$ boson at rest.

\item Determine asymptotic behaviour of the photon-exchange
contribution to the tree-level amplitude for $e^+e^- \rightarrow
W^+ W^-$, assuming that the vertex $WW\gamma$ corresponds to the
{\it minimal\/} electromagnetic interaction (\ref{eq3.40}).
\\ {\it Hint:} As a relevant Feynman rule, take the expression (4.13) in
\cite{Hor} with $\kappa = 0$.

\item  Consider the process $e^+e^-\rightarrow \gamma_L\gamma_L$ within spinor QED
with a {\it massive\/} photon. Show that the corresponding
tree-level amplitude behaves as $O(1)$ in high-energy limit (i.e.
for $E_{c.m.}\gg m_\gamma$).

\item  Consider the quark-antiquark annihilation process
$d + \bar{u} \rightarrow W^- + \gamma$ within the provisional
electro-weak theory described by the first four terms in the
interaction Lagrangian (\ref{eq3.56}). Show that the corresponding
tree-level amplitude behaves asymptotically as $O(1)$ for any
polarization of the $W$\index{W boson@$W$ boson|)}.

%\item  Evaluate amplitudes of the first three partial waves ($j = 0,
%1, 2$) corresponding to the tree-level matrix element for $\nu e
%\rightarrow \nu e$ (see (\ref{eq3.25})). What is the leading
%asymptotic behaviour of the partial-wave amplitude for a general
%$j$?

\end{priklady}

%\input{skript2}  %kapitola   4.1
%%%%%%%%%%%%%%%%%%%%%%%%%%%%%%%%%%%%%%%%%%%%%%%%%%%%%%%%%%%%%%%%%%%
%%%%%%%%%%%%%%%%%%%%%%%%%%%%%%%%%%%%%%%%%%%%%%%%%%%%%%%%%%%%%%%%%%%%%%%%%%%%%%%%%%%%%%%%%%%%%%%%%%%%%%%%%%%%%%%%%%%%%%%%%%%%%%%%%%%%%%%%
\chapter{Gauge invariance and Yang--Mills field}\label{chap4}\index{gauge
invariance|ff}
\section{Abelian gauge invariance}\index{Yang--Mills field|(}

  Let us consider e.g. the Lagrangian of a free classical Dirac field
\begin{equation}
\label{eq4.1}
\lagr_0 = i \bar{\psi} \gamma^\mu \partial_\mu \psi - m \bar{\psi}
\psi
\end{equation}
\noindent where $\psi$ denotes the corresponding bispinor field
variable. It is easy to verify that the expression (\ref{eq4.1})
is invariant under {\it global\/} phase transformations
\begin{eqnarray}
\label{eq4.2}
\psi'(x)& = & \text{e}^{i\omega}\psi(x) \\
\label{eq4.3} \bar{\psi}'(x) &= &\text{e}^{-i\omega} \bar{\psi}(x)
\end{eqnarray}
\noindent where the $\omega$ is a constant independent of
coordinates (the adjective \qq{global} refers to the
$x$-independence of the transformation parameter). The $\omega$
can otherwise take on an arbitrary real value and the unitary
transformations (\ref{eq4.2}), (\ref{eq4.3}) thus form an Abelian
(i.e. commutative) group\index{Abelian group} called
$U(1)$\index{U(1) group@$U(1)$ group}. Let us recall that such a
continuous one-parameter symmetry leads in general to a conserved
Noether current\index{Noether current}, which in the present case
has the familiar form
\begin{eqnarray}
\label{eq4.4}
J_\mu=\bar{\psi}\gamma_\mu\psi
\end{eqnarray}
\noindent One may now ask what happens if we let the parameter
$\omega$ depend on $x$, i.e. if we consider {\it local\/}
transformations
\begin{eqnarray}
\label{eq4.5}
\psi'(x)& = & \text{e}^{i\omega(x)}\psi(x) \\
\label{eq4.6} \bar{\psi}'(x) &= &\text{e}^{-i\omega(x)}
\bar{\psi}(x)
\end{eqnarray}
\noindent
When the Lagrangian (\ref{eq4.1}) is transformed according to
(\ref{eq4.5}), (\ref{eq4.6}) one obtains
\begin{align}
\lagr'_0&=i\bar\psi'\gamma^\mu\partial_\mu\psi' -m\bar\psi'\psi'=i
\text{e}^{-i\omega}\bar\psi \gamma^\mu(i\partial_\mu\omega{e}^
{i\omega}\psi+\text{e}^{i\omega}\partial_\mu\psi)-m\bar\psi\psi\notag\\
&=-\bar\psi \gamma^\mu\psi \partial_\mu\omega +
i\bar\psi\gamma^\mu\partial_\mu\psi-m\bar\psi\psi
=-\bar\psi\gamma^\mu\psi\partial_\mu\omega+\lagr_0\label{eq4.7}
\end{align}
Thus, the $\lagr_0$ is {\it not\/} invariant under local phase
transformations and its non-invariance (which is obviously due to
the derivative involved in the kinetic term) can be represented as
a coupling of the gradient of the local phase parameter to the
vector current (\ref{eq4.4}). Now one can make a simple
observation, which will be of crucial importance for our later
considerations. The contribution proportional to
$\partial_\mu\omega$, which has shown up in the last expression,
can be cancelled by adding to the original free Lagrangian an {\it
interaction term\/} involving a new vector field (coupled to the
current (\ref{eq4.4})), endowed with appropriate transformation
properties. In particular, the term to be added may be written as
\begin{equation}
\label{eq4.8}
\lagr_{int}=g\bar\psi\gamma^\mu\psi A_\mu
\end{equation}
\noindent
where $g$  denotes a coupling constant, and the vector field
$A_\mu$ is required  to transform according to
\begin{equation}
\label{eq4.9}
A'_\mu(x)=A_\mu(x)+\frac{1}{g}\partial_\mu\omega(x)
\end{equation}
\noindent
The extended Lagrangian
\begin{equation}
\label{eq4.10}
\lagr=\lagr_0+g\bar\psi\gamma^\mu\psi A_\mu
\end{equation}
\noindent
is then invariant under the transformations (\ref{eq4.5}),
(\ref{eq4.6}), (\ref{eq4.9}),
as now we have
\begin{eqnarray}
\label{eq4.11} \lagr'&=&\lagr'_0+g\bar\psi'\gamma^\mu\psi' A'_\mu
=\lagr_0-\bar\psi \gamma^\mu\psi \partial_\mu \omega +g\bar\psi
\gamma^\mu\psi (A_\mu+\frac{1}{g}
\partial_\mu\omega)\nonumber\\
&=&\lagr
\end{eqnarray}
In  the standard terminology, the relations (\ref{eq4.5}),
(\ref{eq4.6}), (\ref{eq4.9}) represent the (Abelian) {\bf gauge
transformations} and the vector field $A_\mu$ is called
accordingly the Abelian {\bf gauge field}. The Lagrangian
(\ref{eq4.10}) can be recast as
\begin{eqnarray}
\label{eq4.12}
\lagr&=&i\bar\psi\gamma^\mu\partial_\mu\psi+g\bar\psi\gamma^\mu
\psi A_\mu-m\bar\psi\psi\nonumber\\
&=&i\bar\psi\gamma^\mu(\partial_\mu-igA_\mu)\psi
-m\bar\psi\psi=i\bar\psi\slashed{D}\psi-m\bar\psi\psi
\end{eqnarray}
where we have introduced a usual symbol $D_\mu$ denoting the {\bf
covariant derivative}\index{covariant derivative|ff}
\begin{eqnarray}
\label{eq4.13}
D_\mu=\partial_\mu-ig A_\mu
\end{eqnarray}
\noindent
(this is another piece of the standard gauge-theory vocabulary).

Let us now pause here  to discuss briefly the meaning of the
preceding manipulations. Of course, in (\ref{eq4.12}) one may
easily recognize the \qq{minimal electromagnetic coupling} well
known from classical electrodynamics, and the electromagnetic
vector potential thus can serve as  an obvious example of an
Abelian gauge field\index{Abelian gauge field}. Historically, the
classical Maxwell electrodynamics\index{Maxwell electrodynamics}
has been deduced from the wealth of known experimental data, and
its gauge (or \qq{gradient}) invariance shows up as an additional
mathematical property of the relevant system of equations (this
should presumably be familiar to everybody who followed a
corresponding introductory course). Here, however, we have
proceeded in a reverse direction: starting with a free field
Lagrangian (which violates the local gauge symmetry) we have
subsequently extended it by including a particular interaction (of
an \qq{electromagnetic} type) to meet the requirement of local
gauge invariance. This is actually the most important lesson to be
learnt from the preceding discussion, so let us formulate it once
again in a more concise form. {\bf Promoting the global
phase\index{global symmetry} invariance of a free matter-field
Lagrangian to the local gauge symmetry\index{local symmetry|ff},
one is forced to introduce an interaction involving a vector
(gauge) field with rather specific properties.} Such a simple
observation is in fact the core of all modern gauge theories,
which are based on the non-Abelian generalization of the concept
of local symmetry (to be discussed in the next section). The
$a\;priori$ requirement of local gauge symmetry, though in a sense
natural, is a rather abstract mathematical principle and its
physical meaning is not immediately obvious. Nevertheless, it has
proved to be an immensely successful heuristic principle in modern
particle theory, as it led to the formulation of the present-day
standard model of fundamental interactions (incorporating the
quantum chromodynamics (QCD)\index{quantum!chromodynamics (QCD)}
for strong interactions\index{strong interaction} and the
Glashow--Weinberg--Salam theory of electroweak unification).

Let us now return to the technical development of  the gauge
theory ideas. First, let us observe that the gradient
transformation of the gauge field (\ref{eq4.9}) corresponds to the
transformation of the covariant derivative
\begin{equation}
\label{eq4.14} D'_\mu=\text{e}^{i\omega}D_\mu \text{e}^{-i\omega}
\end{equation}
(which in fact justifies the adjective \qq{covariant}). The last
relation is easy to prove; letting act the relevant differential
operator on an arbitrary test function $f$, one gets, on the one
hand
\begin{eqnarray}
\label{eq4.15}
(\partial_\mu-igA'_\mu)f&=&\partial_\mu
f-ig(A_\mu+\frac{1}{g}\partial_\mu
\omega)f\nonumber\\
&=&\partial_\mu f-igA_\mu f-i\partial_\mu
\omega f
\end{eqnarray}
\noindent
On the other hand,
\begin{eqnarray}
\label{eq4.16}
\text{e}^{i\omega}(\partial_\mu-igA_\mu)\text{e}^{-i\omega}f&=&\text{e}^{i\omega}
\partial_\mu(\text{e}^{-i\omega}f)-igA_\mu f\nonumber\\
&=&\text{e}^{i\omega}(-i\partial_\mu\omega \text{e}^{-i\omega}f+
\text{e}^{-i\omega}\partial_\mu f)-igA_\mu f\nonumber\\
&=&-i\partial_\mu\omega f+\partial_\mu f
-igA_\mu f
\end{eqnarray}
\noindent so comparing the results (\ref{eq4.15}) and
(\ref{eq4.16}), the identity (\ref{eq4.14}) is seen to be valid.
Notice also that the transformation property (\ref{eq4.14}) now
makes the gauge invariance of the  Lagrangian (\ref{eq4.12})
transparent.

  Once we have introduced a new field $A_\mu$, we should add a
corresponding kinetic term (i.e. a term involving the $A_\mu$
derivatives) as well, in order to arrive at non-trivial
Euler-Lagrange equations of motion for the $A_\mu$. If one wants
to maintain gauge invariance, one has to invoke the familiar
antisymmetric electromagnetic field
tensor\index{electromagnetic!field tensor}
\begin{eqnarray}
\label{eq4.17}
F_{\mu\nu}=\partial_\mu A_\nu-\partial_\nu A_\mu
\end{eqnarray}
\noindent
which is manifestly invariant under (\ref{eq4.9}). The Lagrangian
(\ref{eq4.10}) may now be completed by adding a term quadratic in
the $F_{\mu\nu}$ to get finally

\begin{equation}
\label{eq4.18}
\lagr_{g.inv.}=-\frac{1}{4}F_{\mu\nu}F^{\mu\nu}+i\bar\psi
\slashed{D}\psi-m\bar\psi\psi
\end{equation}
\noindent
where the relevant coefficient has been fixed so as to reproduce
correctly the standard Maxwell-Dirac equations.

  When elaborating on the gauge theory formalism, it is important
to realize that the gauge field tensor $F_{\mu\nu}$\index{field
tensor|seealso{electromagnetic field tensor}}\index{field tensor}
can in fact be expressed in terms of the commutator of covariant
derivatives, namely
\begin{eqnarray}
\label{eq4.19}
-igF_{\mu\nu}=[D_{\mu}, D_\nu]
\end{eqnarray}

\noindent
Indeed, let the commutator act on an arbitrary test function; one
gets readily
\begin{eqnarray}
\label{eq4.20}
[D_\mu, D_\nu]f&=&(\partial_\mu-ig A_\mu)(\partial_\nu-ig A_\nu)f-
(\mu\leftrightarrow\nu)\nonumber\\
&=&\partial_\mu\partial_\nu f-ig\partial
_\mu A_\nu f-ig A_\nu\partial_\mu f-igA_\mu\partial_\nu f-g^2
A_\mu A_\nu f   \nonumber\\
&-&(\mu\leftrightarrow\nu)=-ig(\partial_\mu A_\nu-\partial_\nu
A_\mu)f=-igF_{\mu\nu}f
\end{eqnarray}

\noindent Note that the identity (\ref{eq4.19}) makes the gauge
invariance of the $F_{\mu\nu}$ obvious (taking into account the
transformation properties of the covariant derivative shown in
(\ref{eq4.14})). Of course, in the Abelian case we know a right
form of  the gauge field kinetic term anyway, so the identity
(\ref{eq4.19}) is actually not of vital importance here (and
essentially the same can be said about the transformation law of
the covariant derivative (\ref{eq4.14})). However, the knowledge
of such identities (which in the Abelian case can be viewed merely
as an elegant reformulation of some familiar elementary relations)
is extremely useful for a successful generalization of the gauge
theory concepts to the non-Abelian case. This crucial development
is the subject of the next section.

%\input{skript3}  %kapitola 4   4.2
%%%%%%%%%%%%%%%%%%%%%%%%%%%%%%%%%%%%%%%%%%%%%%%%%%%%%%%%%%%%%%%%%%%
%%%%%%%%%%%%%%%%%%%%%%%%%%%%%%%%%%%%%%%%%%%%%%%%%%%%%%%%%%%%%%%%%%%%%%%%%%%%%%%%%%%%%%%%%%%%%%%%%%%%%%%%%%%%%%%%%%%%%%%%%%%%%%%%%%%%%%%%
%\documentstyle[12pt]{article}
%\newcommand\skrt[1]{#1\!\!\!\! / }  % pro psan¡ ¨krtlch p¡smenek,
                                  % nap©. \skrt{p}
%\newcommand\lagr{{\cal L}}

%\begin{document}

\section{Non-Abelian gauge invariance}

The ideas and techniques of the preceding section can be extended
in a non-trivial way to the field theory models involving
non-Abelian (i.e. non-commut\-ative) internal symmetries, such as
isospin etc. This extension is due to C.~N.~Yang and R.~Mills
\cite{ref41} and it has become a true conceptual foundation of
modern particle theory, fully recognized since the early 1970s.
The famous Yang--Mills construction can be described in the
following way. Let us consider again a free-field Lagrangian
\begin{equation}
\label{eq4.21}
\lagr_0=i\bar\Psi\gamma^\mu\partial_\mu\Psi-m\bar\Psi\Psi
\end{equation}
where the $\Psi$  now means a doublet of Dirac spinors
\begin{equation}
\label{eq4.22}
\Psi =\begin{pmatrix}\psi_1 \\ \psi_2\end{pmatrix}
\end{equation}
(the $\Psi$ is thus in fact an eight-component object). The
individual spinor fields $\psi_1$ and $\psi_2$ may be viewed as
corresponding e.g. to proton and neutron, or neutrino and electron
(or any other natural \qq{isotopic doublet} in a generalized
sense), but our considerations in this section will be in fact
purely methodical and stay on a rather abstract level. Note also
that in (\ref{eq4.21}) one obviously has
\begin{equation}
\label{eq4.23}
m\bar{\Psi}\Psi=m(\bar\psi_1\psi_1+\bar\psi_2\psi_2)
\end{equation}
\noindent
so that the components of the doublet are degenerate in mass. It
is easy to realize that the Lagrangian (\ref{eq4.21}) is invariant under
matrix transformations
\begin{eqnarray}
\label{eq4.24}
\Psi'(x)&=&U\Psi(x)\nonumber\\
\bar\Psi'(x)&=&\bar\Psi(x)U^\dagger
\end{eqnarray}
where the $U$ is a 2 $\times$ 2 unitary matrix (i.e. $U^\dagger=
U^ {-1}$) with constant elements. The $U$ can otherwise be
arbitrary, so the transformations (\ref{eq4.24}) constitute the
group $U(2)$\index{U(2) group@$U(2)$ group}. In what follows, we
shall restrict ourselves to matrices with unit determinant, i.e.
we consider only the special unitary group $SU(2)$\index{SU(2)
group@$SU(2)$ group}, which is the lowest-dimensional non-Abelian
group suitable for our discussion. When imposing such a
restriction, one actually does not lose any essential feature
connected with the non-Abelian nature of the general
transformations (\ref{eq4.24}), since any $U(2)$ matrix can be
written as a $SU(2)$ matrix multiplied by a $U(1)$ phase factor.
In other words, the $U(2)$\index{U(2) group@$U(2)$ group} group is
actually factorized as $SU(2)\times U(1)$ and the Abelian factor
$U(1)$ can be treated separately, in the manner already described
in the preceding section. Thus, we will examine symmetry
properties of the Lagrangian (\ref{eq4.21}) with respect to the
transformations
\begin{eqnarray}
\label{eq4.25}
\Psi'&=&S\Psi\nonumber\\
\bar\Psi'&=&\bar\Psi S^{-1}
\end{eqnarray}
with $S^{-1}=S^\dagger$,  $\det S = 1$. Any $SU(2)$ matrix can be described
in terms of three independent real parameters; in particular,
the $S$ may be conveniently written in exponential form as
\begin{equation}
\label{eq4.26}
S=\exp(i\omega^a T^a)
\end{equation}
with $T^a=\frac{1}{2}\tau^a$, where the $\tau^a$, $a = 1,2,3$
denote the Pauli matrices\index{Pauli matrices}, and $\omega^a$
are the relevant parameters. The $SU(2)$ matrix (\ref{eq4.26})
represents a rotation in an abstract internal-symmetry (isospin)
space. Note that in a more general context the exponential form
(\ref{eq4.26}) reflects the fact that $SU(2)$ is a particular
example of a Lie group, with generators $T^a$ satisfying
commutation relations of the corresponding Lie algebra
\begin{equation}
\label{eq4.27}
[T^a,T^b]=if^{abc}T^c
\end{equation}
where the symbol $f^{abc}$ denotes generally the relevant
structure constants; in the particular $SU(2)$\index{SU(2)
group@$SU(2)$ group} case, $f^{abc}=\epsilon^{abc}$ with
$\epsilon^{abc}$ being the totally antisymmetric
(three-dimensional) Levi-Civita symbol.

In analogy with the previously discussed Abelian case, let us now
consider {\it local\/} $SU(2)$ transformations, i.e. let the
parameters $\omega^a$ in (\ref{eq4.26}) depend on $x$. As before,
the derivative kinetic term in the Lagrangian (\ref{eq4.21})
obviously violates such a local symmetry, and the corresponding
non-invariance can now be expressed in terms of gradients of the
three parameters $\omega^a$. Invoking the ideas developed in the
preceding section, one may therefore try to compensate the
\qq{local isospin} non-invariance by introducing an appropriate
number of vector fields (three in the present case) endowed with
suitable transformation properties. With the identity
(\ref{eq4.14}) in mind, it is not difficult to guess how such a
procedure can be implemented technically: one can introduce the
relevant compensation term by means of a covariant derivative (in
analogy with (\ref{eq4.12}), (\ref{eq4.13})), required to obey a
transformation law  which would represent a straightforward
generalization of (\ref{eq4.14}). The relevant transformation
properties of the vector fields can then be deduced from the rule
for the covariant derivative. Thus, we will introduce a triplet of
vector fields $A^a_\mu$, $a = 1,2,3$ (corresponding to the three
\qq{phases} $\omega^a(x)$) which can equivalently be described in
terms of the matrix
\begin{equation}
\label{eq4.28}
A_\mu (x)=A^a_\mu (x)T^a
\end{equation}
In the original free-field Lagrangian (\ref{eq4.21}) we replace
the ordinary derivative by the covariant one, i.e. extend
(\ref{eq4.21}) to the form
\begin{eqnarray}
\label{eq4.29}
\lagr&=&i\bar\Psi\gamma^\mu D_\mu\Psi-m\bar\Psi\Psi\nonumber\\
&=&i\bar\Psi\gamma^\mu(\partial_\mu-igA_\mu)\Psi-m\bar\Psi\Psi
\end{eqnarray}
As we have stated above, the $D_\mu$ should  transform
\qq{covariantly} under the local $SU(2)$, i.e.
\begin{equation}
\label{eq4.30}
D'_\mu=SD_\mu S^{-1}
\end{equation}
where $D'_\mu=\partial_\mu-igA'_\mu$. From (\ref{eq4.30}) the
corresponding transformation law for the matrix field $A_\mu$ (see
(\ref{eq4.28})) can be deduced easily. Indeed, using
(\ref{eq4.30}) for an arbitrary test function (two-component
column vector) $f$, one gets
\begin{eqnarray}
\label{eq4.31}
(\partial_\mu-igA'_\mu)f&=&S(\partial_\mu-igA_\mu)S^{-1}f
\nonumber\\
&=&S(\partial_\mu S^{-1}f+S^{-1}\partial_\mu f-
igA_\mu S^{-1}f) \nonumber\\
&=&S\partial_\mu S^{-1}f+\partial_\mu f-igSA_\mu S^{-1}f
\end{eqnarray}
and from (\ref{eq4.31}) then immediately follows
\begin{equation}
\label{eq4.32}
A'_\mu=SA_\mu S^{-1}+\frac{i}{g}S\partial_\mu S^{-1}
\end{equation}
We should now make sure that the local $SU(2)$ transformation
(\ref{eq4.32})
is compatible with the structure (\ref{eq4.28}), namely that the
transformed matrix field $A'_\mu$ can be decomposed in terms of the
$SU(2)$ generators in accordance with (\ref{eq4.28}) (in other
words, the $A'_\mu$ should also be equivalent to a triplet of
components $A'^a_\mu$). Having in mind that a basis in the space
of 2 $\times$ 2
matrices can be taken as consisting of the three Pauli matrices
(which are traceless) and the unit matrix, it is clear that the
problem  reduces to showing that Tr $A'_\mu=0$. The first term on
the right-hand side of (\ref{eq4.32}) is manifestly traceless as a
consequence of Tr $A_\mu=0$. As for the second term, vanishing of
its trace is not immediately obvious, but the proof can be
accomplished in an elementary way. Indeed, for any matrix
$M(x) =\exp\Omega (x)$ one can show that
\begin{equation}
\label{eq4.33} \Tr(M^{-1}\partial_\mu
M)=\Tr(\partial_\mu\Omega)\;\; (=\Tr(\partial_\mu\ln M))
\end{equation}
(this can be done by means of a straightforward power-series
expansion of the relevant exponentials, and by employing the
cyclic property of the trace -- of course, the trace symbol in
(\ref{eq4.33}) is absolutely essential for the validity of such an
identity). From (\ref{eq4.33}) the desired result
\begin{equation}
\label{eq4.34}
\Tr(S\partial_\mu S^{-1})=0
\end{equation}
follows immediately, if one takes into account (\ref{eq4.26}). Let
us remark that an alternative proof of (\ref{eq4.34}) can be
accomplished by invoking an elegant general formula for
differentiating a matrix exponential, namely
\begin{equation}
\label{eq4.35}
\partial_\mu \text{e}^{\Omega(x)}=\int\limits^1_0 dt \text{e}^{t\Omega(x)}
\partial_\mu\Omega(x)\text{e}^{(1-t)\Omega(x)}
\end{equation}
It is clear that the knowledge of the last identity already makes
the proof of (\ref{eq4.34}) trivial. The formula (\ref{eq4.35}) is
also highly useful in other field-theory applications; we leave
its proof to the interested reader as an instructive exercise.

  The matrix field $A_\mu$ (or an individual component $A^a_\mu$
of the corresponding \qq{isomultiplet}) obeying the local
transformation law (\ref{eq4.32}) is called the {\bf non-Abelian
gauge field}\index{non-Abelian!gauge field} or {\bf Yang--Mills
field} corresponding to the gauge group $SU(2)$. Of course, the
preceding construction can be generalized in a straightforward way
e.g. to any unitary group $SU(n)$. There we would have a traceless
$n \times n$ matrix field, equivalent to a multiplet of $n^2-1$
Yang--Mills components; in particular, for $n=3$ a relevant set of
generators is represented by the well-known Gell-Mann
matrices\index{Gell-Mann matrices}. The rule (\ref{eq4.32})
represents a non-trivial generalization of the original gradient
transformation (\ref{eq4.9}); it is easy to check that in the
Abelian case, i.e. when the $S$ is taken simply as exp
$(i\omega)$, the form (\ref{eq4.32}) is indeed reduced to
(\ref{eq4.9}):
\begin{equation}
\label{eq4.36} A'_\mu=\text{e}^{i\omega}A_\mu
\text{e}^{-i\omega}+\frac{i}{g}\text{e}^{i\omega}
(-i\partial_\mu\omega)\text{e}^{-i\omega}=A_\mu+\frac{1}{g}\partial_\mu
\omega
\end{equation}
There is still one point concerning the non-Abelian transformation
(\ref{eq4.32}) that should be clarified here. Once we have shown
that the $A'_\mu$ can be written as
\begin{equation}
\label{eq4.37} A'_\mu=A^{a \prime}_\mu T^a
\end{equation}
one may wonder whether the relation (\ref{eq4.32}) could be recast
in terms of the isotriplet components. For finite gauge
transformations, it is not possible to obtain a transformation
relation for the Yang--Mills components in a closed form
(technically, this is precluded by complications stemming from the
multiplication of matrix exponentials). However, for {\it
infinitesimal\/} gauge transformations one can get a  simple and
intuitively transparent result, which we are going to derive now.
To this end, let us write the transformation matrices $S$ and
$S^{-1}$ in the form
\begin{align}
S(x)&=1+i\epsilon^a(x)T^a\notag\\
S^{-1}(x)&=1-i\epsilon^a(x)T^a\label{eq4.38}
\end{align}
where the $\epsilon^a(x)$ is an infinitesimal local parameter.
Substituting (\ref{eq4.38}) into (\ref{eq4.32}), neglecting
systematically terms of the order $O(\epsilon^2)$ and employing
the commutation relation (\ref{eq4.27}) one gets the desired
result
\begin{equation}
\label{eq4.39} A^{a \prime}_\mu=A^a_\mu-f^{abc}\epsilon^b
A^c_\mu+\frac{1}{g}\partial _\mu\epsilon^a
\end{equation}
which is a standard (and in fact most frequently used) form of the
gauge transformation of a Yang--Mills field (the rule
(\ref{eq4.39}) is of course quite general, not restricted to the
$SU(2)$ case we have started with). Note that the second term on
the right-hand side of (\ref{eq4.39}) clearly reflects the
non-Abelian nature of the considered transformation, and it is
non-vanishing even for the parameters $\epsilon^a$ independent of
the space-time coordinates (i.e. for {\it global\/}
transformations), while the last term is simply an infinitesimal
gradient transformation analogous to the Abelian case.

Now, in analogy with the Abelian case, we should look for an
appropriate kinetic term for the Yang--Mills field. If one
attempts to introduce simply a term quadratic in the first
derivatives of the $A^a_\mu$, one finds that in the non-Abelian
case there is no straightforward way of doing it in a gauge
invariant way. In particular, the simplest expression that would
come first to one's mind, namely
\begin{equation}
\label{eq4.40}
\lagr_{kin}=-\frac{1}{4}A^a_{\mu\nu}A^{a\mu\nu}
\end{equation}
with $A^a_{\mu\nu}=\partial_\mu A^a_\nu-\partial_\nu A^a_\mu$ is
not gauge invariant (the reader is recommended to check this
statement explicitly, using the transformation rule
(\ref{eq4.39})). At this point one may invoke  the identity
(\ref{eq4.19}), which provides a crucial inspiration. Indeed, let
us define the quantity $F_{\mu\nu}$ in terms of the matrix
covariant derivatives as
\begin{equation}
\label{eq4.41}
-igF_{\mu\nu}=[D_\mu ,D_\nu]
\end{equation}
Then one has
\begin{gather}\label{eq4.42}
[D_\mu ,D_\nu]f=(\partial_\mu-igA_\mu)(\partial_\nu-igA_\nu)f-
(\mu\leftrightarrow\nu)\\
=\partial_\mu\partial_\nu f-ig\partial_\mu A_\nu f-
igA_\nu\partial_\mu f-igA_\mu\partial_\nu f-
g^2 A_\mu A_\nu f-(\mu\leftrightarrow\nu)\notag\\
=-ig(\partial_\mu A_\nu-\partial_\nu A_\mu)f-g^2[A_\mu, A_\nu]
f=-ig(\partial_\mu A_\nu -\partial_\nu A_\mu- ig[A_\mu ,A_\nu
])f\notag
\end{gather}
since, in contrast with the Abelian case, the commutator
[$A_\mu ,A_\nu$] is now non-zero. The $F_{\mu\nu}$ defined by
(\ref{eq4.41}) thus becomes
\begin {equation}
\label{eq4.43}
F_{\mu\nu}=\partial_\mu A_\nu-\partial_\nu
A_\mu-ig[A_\mu,A_\nu]
\end {equation}
For components defined by $F_{\mu\nu}= F^a_{\mu\nu} T^a$ one then
gets, using the commutation relation (\ref{eq4.27})
\begin{equation}
\label{eq4.44}
F^a_{\mu\nu}=\partial_\mu A^a_\nu -\partial_\nu A^a_\mu
+gf^{abc}A^b_\mu A^c_\nu
\end{equation}
Obviously, the meaning of the construction (\ref{eq4.41}) is that the
$F_{\mu\nu}$ now transforms covariantly under (\ref{eq4.32}), i.e. in the
same way as the covariant derivative:
\begin{equation}
\label{eq4.45}
F'_{\mu\nu}=SF_{\mu\nu}S^{-1}
\end{equation}
Notice that (\ref{eq4.45}) also means that the change of the
$F_{\mu\nu}$ under local and global transformations is the same.
For infinitesimal transformations one gets
\begin{equation}
\label{eq4.46} F^{a \prime
}_{\mu\nu}=F^a_{\mu\nu}-f^{abc}\epsilon^bF^c_{\mu\nu}
\end{equation}
Thus, while the $F_{\mu\nu}$ is $not$ gauge invariant (in contrast
to the Abelian case), it is gauge {\it covariant} and therefore it can
be used to construct readily a quadratic invariant, namely
\begin{equation}
\label{eq4.47} \lagr^{(2)}=c \Tr(F_{\mu\nu} F^{\mu\nu})
\end{equation}
where $c$ is an arbitrary constant. Taking into account that the
generators $T^a$ are normalized as
\begin{equation}
\label{eq4.48} \Tr (T^a T^b) =\frac{1}{2}\delta^{ab}
\end{equation}
we can recast (\ref{eq4.47}) as
\begin{equation}
\label{eq4.49}
\lagr^{(2)}=\frac{1}{2}cF^a_{\mu\nu}F^{a\mu\nu}
\end{equation}
For convenience, we will fix the overall coefficient in
(\ref{eq4.49}) in analogy with the Abelian (Maxwell) case (cf.
(\ref{eq4.18})); then, putting together the gauge invariant pieces
(\ref{eq4.29}) and (\ref{eq4.49}), the full Yang--Mills Lagrangian
can be written as
\begin{equation}
\label{eq4.50}
\lagr_{YM}=-\frac{1}{4}F^a_{\mu\nu}F^{a\mu\nu}+i\bar\Psi
\slashed{D}\Psi-m\bar\Psi\Psi
\end{equation}
The most remarkable contribution is contained in the first term,
made entirely of the gauge fields. Let us denote it as
$\lagr_{gauge}$; using for the $F^a_{\mu\nu}$  the expression
(\ref{eq4.44}) one gets, after a simple manipulation
\begin{eqnarray}
\lagr_{gauge}&=&-\frac{1}{4}F^a_{\mu\nu}F^{a\mu\nu}
=-\frac{1}{4}A^a_{\mu\nu}A^{a\mu\nu}
-\frac{1}{2}gf^{abc}(\partial_\mu A^a_\nu -\partial_\nu A^a_\mu)
A^{b \mu} A^{c \nu}
\nonumber\\&&\phantom{-\frac{1}{4}F^a_{\mu\nu}F^{a\mu\nu}
=}-\frac{1}{4}g^2f^{abc}f^{ajk}A^b_\mu A^c_\nu A^{j\mu}A^{k\nu}
\label{eq4.51}
\end{eqnarray}
\noindent Thus, the $\lagr_{gauge}$ is seen to contain the desired
kinetic terms, but in addition we have  earned some new cubic and
quartic terms, i.e. contributions corresponding to {\it
self-interactions\/} of the Yang--Mills fields, which have no
analogue in the Abelian case (note that a term quartic in the
electromagnetic field would describe classical light-by-light
scattering, which of course does not exist in Maxwell
electrodynamics). Obviously, the form of the Yang--Mills
interaction terms is severely constrained by the gauge
symmetry\index{gauge invariance} -- notice e.g. that the coupling
constant at the quartic term is the square of a relevant factor
corresponding to the triple gauge field interaction, and the
polynomial structure of both terms is also determined completely.
Let us emphasize, however, that we have restricted ourselves to
the terms with lowest dimension (equal to four). In principle, we
could introduce higher powers of the $F_{\mu\nu}$  as well;
however, at the quantum level, such contributions would spoil
perturbative renormalizability\index{perturbative
renormalizability}, which is a desired technical aspect of the
theory of electroweak interactions, and in fact has been a primary
goal of the inventors of the Standard Model in the late 1960s.

  One more remark is perhaps in order here, concerning the gauge
coupling constant appearing in the covariant derivative. Although
we have been working with a multiplet ($SU(2)$ triplet) of vector
Yang--Mills fields, we are allowed to introduce only a single
coupling constant $g$. This is due to the fact that the gauge
group under consideration (let us say $SU(n)$, in general) is
simple (mathematically, this means that the corresponding Lie
algebra does not contain any non-trivial invariant subalgebra); in
other words, introducing more coupling constants in the covariant
derivative would not be compatible with the commutation relations
(\ref{eq4.27}). The standard model of electroweak interactions is
based on the gauge group $SU(2)\times U(1) = U(2)$ (which, from
the mathematical point of view, is not even semi-simple because of
the Abelian factor $U(1)$) and one then has to introduce, in
general, two independent coupling constants $g$ and $g'$. Some
models of the so-called grand unification\index{grand unification}
(unifying all forces except gravity) are based on simple groups
(e.g. $SU(5)$\index{SU(5) group@$SU(5)$ group} or $SO(10)$) and
this leads to an interrelation between the coupling strengths of
the electromagnetic, weak and strong
interactions\index{electromagnetic!interaction}\index{strong
interaction}.

In the preceding discussion, we have not included any mass term
for the Yang--Mills fields. The reason why we did not do so is
that a mass term of the vector fields violates gauge invariance.
In fact, from our pragmatic point of view it is the
renormalizability which is of interest to us, rather than a
symmetry. As we shall see later, the bad news is that a naive mass
term for the Yang--Mills field would in general spoil the
renormalizability as well. Massless gauge fields do describe at
least a part of our physical world: the modern theory of strong
interaction (quantum chromodynamics)\index{quantum!chromodynamics
(QCD)} is based on the idea of exact gauge invariance under
$SU(3)_\ti{colour}$\index{colour} and the corresponding (eight)
gauge fields represent massless gluons\index{gluons} interacting
with coloured quarks (and with themselves). On the other hand, for
the weak interaction theory we need (a multiplet of) massive
vector bosons to reproduce the familiar phenomenology of the beta
decay, muon decay,~ etc. We will defer the subtle issue of mass
generation in gauge theories to the Chapter~\ref{chap6}. In the next chapter
we will show how far one can get if the concept of Yang--Mills
field is applied to the unification of weak and electromagnetic
interactions, assuming for the moment that the relevant mass terms
are added simply by hand. In other words, we will first discuss
the theory proposed by S.Glashow in 1961, which now constitutes a
well-established part of the present-day standard electroweak
model and leads by itself to remarkable valid predictions.
%\end{document}

%\input{problems4}
%%%%%%%%%%%%%%%%%%%%%%%%%%%%%%%%%%%%%%%%%%%%%%%%%%%%%%%%%%%%%%%%%%%
%%%%%%%%%%%%%%%%%%%%%%%%%%%%%%%%%%%%%%%%%%%%%%%%%%%%%%%%%%%%%%%%%%%%%%%%%%%%%%%%%%%%%%%%%%%%%%%%%%%%%%%%%%%%%%%%%%%%%%%%%%%%%%%%%%%%%%%%
\begin{priklady}{2.2}
\item  Prove the identity (\ref{eq4.35}).

\item  We have seen that the term $\Tr (F_{\mu\nu}F^{\mu\nu})$ is gauge invariant by
construction and has dimension {\it four}. Is there any other
gauge invariant expression made of Yang-Mills fields $A_\mu$ only
and carrying the same dimension?

\item Consider the $SU(2)$ gauge theory involving Yang--Mills
fields coupled to a doublet of fermions. Write down the relevant
equations of motion. What are the conserved Noether
currents\index{Noether current} corresponding to the global
$SU(2)$ symmetry\index{Yang--Mills field|)}?

\item  Find an appropriate set of generators for the group
$SU(2)\times SU(2)$.

\item  What are generators of the group $SU(3)$, satisfying the
normalization condition $\Tr(T^a T^b) = \frac{1}{2}\delta^{ab}$
(c.f. (\ref{eq4.48}))? Find such a set of generators for
$SU(4)$\index{SU(4) group@$SU(4)$ group} and $SU(5)$\index{SU(5)
group@$SU(5)$ group}.
\end{priklady}

%\input{skript31} %kapitola 5   5.1
%%%%%%%%%%%%%%%%%%%%%%%%%%%%%%%%%%%%%%%%%%%%%%%%%%%%%%%%%%%%%%%%%%%
%%%%%%%%%%%%%%%%%%%%%%%%%%%%%%%%%%%%%%%%%%%%%%%%%%%%%%%%%%%%%%%%%%%%%%%%%%%%%%%%%%%%%%%%%%%%%%%%%%%%%%%%%%%%%%%%%%%%%%%%%%%%%%%%%%%%%%%%
\chapter[Electroweak unification and gauge symmetry]{Electroweak unification\\ and gauge symmetry}\index{non-Abelian!symmetry|ff}\label{chap5}

\section{$SU(2)\times U(1)$ gauge theory for
leptons}\index{electroweak interactions|ff}\label{sec5.1}

In Chapter~\ref{chap3} we have emphasized that new particles and new
interactions must be added to the old theory of weak and
electromagnetic forces if one wants to tame divergent  high-energy
behaviour\index{tree unitarity} of the tree-level $S$-matrix
elements, and we have argued that accomplishing this goal would
also bear on the issue of perturbative renormalizability at higher
orders\index{perturbative renormalizability}. From a purely
theoretical point of view, any such scenario should introduce
either a new (neutral) massive vector boson or \qq{exotic}
fermions (such as heavy leptons\index{heavy leptons}). Of course,
a combination of both schemes would be possible as well. At the
same time, looking back in history, the technical experience
gained by various people from the early studies of Yang--Mills
theories suggested that non-Abelian gauge symmetry might control
at least a part of the desirable divergence cancellations (one of
the pioneering personalities in this direction was M. Veltman). A
rigorous proof of perturbative renormalizability of a broad class
of non-Abelian gauge models (incorporating the by now famous Higgs
mechanism\index{Higgs mechanism} for the mass generation) was
finally invented by G. 't~Hooft in 1971 and this triggered the
boom of \qq{gauge model building} in the early 1970s. Various
options were discussed, but the crucial moment was the
experimental discovery of weak neutral currents\index{neutral
current} in 1973 that pointed rather clearly towards a model
involving a neutral vector boson, as a viable candidate for
realistic description of the physical world. This development
ultimately led to recognizing a \qq{minimal} gauge theory model,
proposed by Weinberg and Salam in the late 1960s (who followed, in
a sense, the earlier Glashow's
attempt\index{Glashow--Weinberg--Salam theory|ff}\index{standard
model of electroweak interactions|ff}), as the \qq{standard model}
of electroweak interactions. The theory passed many stringent
experimental tests in subsequent years and represents today one of
the most successful physical theories of the 20th century. The
only essential missing link of the standard model is represented
by the Higgs scalar boson, which emerges in the theory as a
leftover of the electroweak symmetry breaking mechanism.

Thus, from now on we will follow a path leading to the standard
electroweak model. In this chapter, we will discuss the gauge
structure of the model, leaving aside, for the moment, the subtle
issue of the mass generation via Higgs mechanism. We will restrict
ourselves to the leptonic sector of the elementary fermion
spectrum, since most of the important aspects of the electroweak
gauge symmetry can be displayed within such a reduced framework.
Thus, in the present chapter we will stay essentially within the
1961 Glashow model, which constitutes a part of the present-day
Standard Model.

The idea of unifying weak and electromagnetic interactions on the
basis of a non-Abelian gauge symmetry is in fact rather appealing
{\it a priori\/}. Both forces are universal and involve vector (or
axial-vector) currents that can be coupled naturally to vector
fields, and these may constitute a Yang--Mills multiplet (note
that a pioneering work in this direction is due to J. Schwinger
(1957)). Using then our previous considerations as a technical
guide, one may guess that in a \qq{minimal variant} of the
electroweak unification, {\it four\/} gauge fields are actually
needed, corresponding to the $W^+, W^-$, photon and a new neutral
vector boson. Thus, an appropriate gauge group is $SU(2) \times
U(1)$\index{SU(2) times U(1) group@$SU(2)\times U(1)$ group|ff}.
Later on we shall see that a fourth vector boson is indeed
necessary for a successful electroweak gauge unification
(involving only conventional leptons) even for purely
\qq{algebraic} reasons, i.e. without making any further reference
to the high-energy behaviour of Feynman diagrams. Obviously, an
important conceptual problem is how to accommodate in the
envisaged unified theory both the vectorial
(parity-conserving)\index{parity!conservation} electromagnetic
current and the left-handed weak charged current\index{charged
current} manifesting maximum parity
violation\index{parity!violation}. At first sight, this dramatic
difference between the two forces might seem to be a major
obstacle to their unification, but as we shall see, the distinct
chiral structure of the relevant currents can in fact be
incorporated quite easily. The right idea is to consider the
chiral components of the fermion fields as independent
\qq{building blocks}, and assign different transformation
properties to the left-handed and right-handed components when
writing down the $SU(2)\times U(1)$ gauge invariant Lagrangian. In
particular, the left-handed fermion fields are placed in $SU(2)$
doublets, while the right-handed fermions are taken to be singlets
-- as we shall see, this is precisely the choice leading
automatically to the desired $V-A$ structure of the charged weak
current.

Thus, let us now proceed to construct the relevant Lagrangian
invariant under the local $SU(2) \times U(1)$. Needless to say, as
in the preceding chapter we start at the level of classical field
theory; we will comment on the quantization later on. To begin
with, we are going to consider leptons of the electron type. The
left-handed components $\nu_L =\frac{1}{2}(1-\gamma_5)\nu$ and
$e_L=\frac{1}{2}(1- \gamma_5)e$  form an $SU(2)$ doublet
\begin{equation}
\label{eq5.1}
L^{(e)} =\begin{pmatrix}
\nu_L \\ e_L \end{pmatrix}
\end{equation}
(as usual, we denote the individual fields by letters labelling
normally the corresponding particles, and write for the moment
$\nu$ instead of $\nu_e$; in what follows, we will also drop the
superscript on the $L$ for brevity). The right-handed fields $e_R
=\frac{1}{2}(1+\gamma_5)e$ and $\nu_R=\frac{1}{2}(1+\gamma_5) \nu$
are $SU(2)$ singlets. Note that we have included the right-handed
component of the neutrino field in addition to the mandatory
$\nu_L$ (by doing it, we keep an open mind about a possibility of
non-vanishing neutrino mass\index{neutrino!mass}). We should also
specify the transformation properties of lepton fields under the
Abelian subgroup $U(1)$\index{U(1) group@$U(1)$ group} and clarify
the form of the relevant covariant derivatives acting on the
lepton fields. To this end, the following simple observation will
be helpful: if an Abelian gauge field $B_\mu$  transforms
as\index{Abelian gauge field}
\begin{equation}
\label{eq5.2}
B'_\mu=B_\mu+\frac{1}{g}\partial_\mu\omega
\end{equation}
and a Dirac field $\Psi$  is transformed according to
\begin{equation}
\label{eq5.3} \Psi'=\text{e}^{iY\omega}\Psi
\end{equation}
with $Y$ being a real number (note that the $\Psi$ may in general mean a
multiplet of fields), then the expression
\begin{equation}
\label{eq5.4}
\bar\Psi\gamma^\mu D_\mu \Psi
\end{equation}
is invariant under the local $U(1)$ if the covariant derivative
$D_\mu$ has the form\index{covariant derivative}
\begin{equation}
\label{eq5.5}
D_\mu=\partial_\mu-igYB_\mu
\end{equation}
(the proof of this statement is left to the reader as a
trivial exercise).

The meaning of (\ref{eq5.3}) consists in pointing out the
existence of infinitely many (inequivalent) representations of the
Abelian group $U(1)$\index{Abelian group}, labelled here by an
arbitrary real parameter $Y$. We may now use  this  freedom to
assign different (in general arbitrary) values of the real
parameter -- called usually \qq{weak
hypercharge}\index{weak!hypercharge} -- to the doublet $L$ and
singlets $e_R,\;\nu_R$, to characterize their transformation
properties with respect to the Abelian factor $U(1)$ of the
considered gauge group. The $SU(2) \times U(1)$ gauge invariant
Lagrangian involving lepton interactions thus can be written as
\begin{eqnarray}
\label{eq5.6}
\lagr_{lepton}=i\bar
L\gamma^\mu(\partial_\mu-igA^a_\mu
\frac{\tau^a}{2}-ig'Y_LB_\mu)L\nonumber\\
+i\bar e_R\gamma^\mu(\partial_\mu-ig'Y^{(e)}_RB_\mu)e_R
+i\bar\nu_R\gamma^\mu(\partial_\mu-ig'Y^{(\nu)}_R B_\mu)\nu_R
\end{eqnarray}
where the $A^a_\mu$, $a = 1,2,3$ denote the triplet of Yang--Mills
fields corresponding to the \qq{weak isospin}\index{weak!isospin}
subgroup $SU(2)$, and the $B_\mu$ is the gauge field associated
with the weak hypercharge subgroup $U(1)$. Of course, in the
covariant derivatives acting on the $e_R$ and $\nu_R$, the
non-Abelian part is absent since the $SU(2)$ generators are
trivial in the singlet representation.

In (\ref{eq5.6}) we have introduced arbitrary values of the weak
hypercharges for the $L, e_R$ and $\nu_R$. It is important to
employ such a general parametrization at this initial stage; we
shall see later that the relevant values of weak hypercharge are
constrained non-trivially by the requirement of recovering --
within the considered unified theory -- a standard electromagnetic
interaction of leptons carrying the usual charges. Looking ahead,
let us state already here that a rule providing automatically the
right values of $Y$ reads
\begin{equation}
\label{eq5.7}
Q = T_3 +  Y
\end {equation}
where $Q$ is the relevant charge (in units of the positron charge,
so e.g. $Q_e = -1$ etc.) and $T_3$ is the value of weak isospin, defined
as the eigenvalue of the corresponding $SU(2)$ generator; in particular,
for the considered $SU(2)$ doublet and singlets resp. one has
\begin{equation}
\label{eq5.8}
T_3(\nu_L)=+\frac{1}{2},\;\;\;T_3(e_L)=-\frac{1}{2},\;\;\;
T_3(\nu_R)=0,\;\;\;T_3(e_R)=0
\end{equation}
From (\ref{eq5.7}) and (\ref{eq5.8}) one then gets
\begin{equation}
\label{eq5.9}
Y_L=-\frac{1}{2},\;\;\;\;Y^{(e)}_R=-1,\;\;\;\;Y^{(\nu)}_R=0
\end{equation}
The textbook expositions of the standard electroweak model usually
start immediately with the relation (\ref{eq5.7}) yielding the
\qq{physical values} (\ref{eq5.9}). Here we will keep the general
parametrization (\ref{eq5.6}) and derive the rule (\ref{eq5.7})
yielding the values (\ref{eq5.9}) in Section~\ref{sec5.3}, where the
electromagnetic interaction will be discussed in detail.

  Notice that in (\ref{eq5.6}) we have included two independent
coupling constants $g$ and $g'$. This, of course, is related to
the fact that the considered gauge group is not simple (cf. the
discussion at the end of the previous chapter). Such a dichotomy
represents an obvious aesthetic flaw of the envisaged electroweak
unification -- one would certainly prefer a unified picture of the
two interactions based on a single common coupling constant.
However, one cannot arbitrarily set $g = g'$, since such a
relation would be violated by
renormalization\index{renormalization} effects at the quantum
level (and, as we know today, it would also contradict
experimental facts). Nevertheless, in the course of the subsequent
discussion it will become clear that the term \qq{electroweak
unification} does match with the $SU(2) \times U(1)$ model (the
discussion of the weak neutral currents in Section~\ref{sec5.4} is
particularly instructive in this respect). On the other hand, it
may well be that a simple group of \qq{grand
unification}\index{grand unification} of the electroweak and
strong interactions\index{strong interaction} lies ahead in our
future. The goal of the \qq{incomplete} electroweak model is much
more modest: it unifies the electrodynamics with the low-energy
$V-A$ theory\index{V-A theory@$V-A$ theory} of weak interactions,
and does provide a realistic and highly accurate description of
the electroweak forces at the currently accessible energies. In
this sense, the standard electroweak model can be viewed as an
effective approximation (at relatively low energies) of a deeper
theory whose contours we may now only guess.

  Coming back to the structure of the Lagrangian (\ref{eq5.6}), it
should be stressed that the gauge fields $A^a_\mu$ and $B_\mu$
need not (and in fact do not) have any direct physical meaning.
The physical vector fields will emerge as their linear
combinations, displaying thus a characteristic feature of the
electroweak unification. The physical contents of the leptonic
Lagrangian will be discussed in subsequent sections, but before
proceeding to this fundamental task, we should add to
(\ref{eq5.6}) a gauge invariant contribution involving the kinetic
term of the vector fields. In the spirit of the general
Yang--Mills construction described in the preceding chapter we may
write
\begin{equation}
\label{eq5.10}
\lagr_{gauge}=-\frac{1}{4}F^a_{\mu\nu}F^{a\mu\nu}-\frac{1}{4}
B_{\mu\nu}B^{\mu\nu}
\end{equation}
where
\begin{equation}
\label{eq5.11}
F^a_{\mu\nu}=\partial_\mu A^a_\nu-\partial_\nu
A^a_\mu +g\epsilon^{abc}A^b_\mu A^c_\nu
\end{equation}
and
\begin{equation}
\label{eq5.12}
B_{\mu\nu}=\partial_\mu B_\nu-\partial_\nu B_\mu
\end{equation}
In the rest of this chapter we will thus analyze the form
\begin{equation}
\label{eq5.13}
\lagr_1=\lagr_{gauge}+\lagr_{lepton}
\end{equation}
which in fact constitutes the first part of gauge invariant
Glashow--Weinberg--Salam (GWS) electroweak Lagrangian.

%\input{skript32} %             5.2
%%%%%%%%%%%%%%%%%%%%%%%%%%%%%%%%%%%%%%%%%%%%%%%%%%%%%%%%%%%%%%%%%%%
%%%%%%%%%%%%%%%%%%%%%%%%%%%%%%%%%%%%%%%%%%%%%%%%%%%%%%%%%%%%%%%%%%%%%%%%%%%%%%%%%%%%%%%%%%%%%%%%%%%%%%%%%%%%%%%%%%%%%%%%%%%%%%%%%%%%%%%%
%\documentstyle[12pt]{article}
%\newcommand\skrt[1]{#1\!\!\!\! / }  % pro psan¡ ¨krtlch p¡smenek,
%                                  % nap©. \skrt{p}
%\newcommand\lagr{{\cal L}}

%\begin{document}
\section{Charged current weak interaction}\label{sec5.2}

Let us consider the lepton Lagrangian (\ref{eq5.6}). We should
check that it contains, among other things, the conventional weak
interaction of the left-handed (i.e. $V-A$) charged current\index{charged current|ff}
(made of the neutrino and electron fields) with a charged
intermediate vector boson. It is not difficult to guess that such
a term could originate from the non-Abelian part of the covariant
derivative in (\ref{eq5.6}), in particular from the two terms
involving the anti-diagonal Pauli matrices $\tau^1$ and $\tau^2$.
Indeed, the interaction part of (\ref{eq5.6}) reads
\begin {align}
\lagr^{(int.)}_{lepton}&= g \bar
L\gamma^\mu\frac{\tau^a}{2}LA^a_\mu+
g'Y_L\bar L\gamma^\mu LB_\mu\nonumber\notag\\
& + g'Y^{(e)}_R\bar e_R\gamma^\mu e_R B_\mu+g'Y^{(\nu)}_R\bar\nu_R
\gamma^\mu\nu_R B_\mu\label{eq5.14}
\end{align}
This can be recast as
\begin{align}
\label{eq5.15}
\lagr^{(int.)}_{lepton}&=g(\bar
L\gamma^\mu\frac{\tau^+} {2}LA^-_\mu +\bar
L\gamma^\mu\frac{\tau^-}{2}LA^+_\mu+
\bar L\gamma^\mu\frac{\tau^3}{2}LA^3_\mu)\notag\\
&+g'Y_L\bar L\gamma^\mu LB_\mu+g'Y^{(e)}_R\bar e_R\gamma^\mu
e_R B_\mu+g'Y^{(\nu)}_R\bar\nu_R\gamma^\mu\nu_R B_\mu
\end{align}
where $\tau^\pm=\frac{1}{\sqrt 2}(\tau^1\pm i\tau^2)$  and
$A^\pm_\mu=\frac{1}{\sqrt 2} (A^1_\mu\pm iA^2_\mu)$, that is
\begin{eqnarray}
\label{eq5.16}
\frac{1}{2}\tau^+ =\frac{1}{\sqrt 2}
\begin{pmatrix}0&1\\0&0
\end{pmatrix},
\quad \frac{1}{2}\tau^-=\frac{1}{\sqrt 2}
\begin{pmatrix}0&0\\1&0
\end{pmatrix}
\end{eqnarray}
Using (\ref{eq5.16}) in (\ref{eq5.15}) and working out the simple
matrix products in the first two terms one gets readily
\begin{equation}
\label{eq5.17} \lagr^{(int.)}_{lepton}=\frac{g}{\sqrt
2}(\bar\nu_L\gamma^\mu e_L W^+_\mu+\bar e_L\gamma^\mu\nu_L
W^-_\mu)+\lagr_{diag.}
\end{equation}
where we have denoted $W^\pm_\mu =A^\mp_\mu$ and under the symbol
$\lagr_{diag.}$ we have collected all the remaining terms from
(\ref{eq5.15}), i.e. those involving diagonal $2 \times 2$
matrices $(\tau^3\;\text{or}\;\J)$. We have passed from $A^\pm_\mu$
to $W^\mp_\mu$ so as to recover precisely the structure of the old
charged current weak interaction discussed in the previous
chapters. Indeed, the first two terms in (\ref{eq5.17}) can be
obviously written in the form
\begin{equation}
\label{eq5.18} \lagr_{CC}=\frac{g}{2\sqrt
2}\bar\nu\gamma^\mu(1-\gamma_5) eW^+_\mu+\text{h.c.}
\end{equation}
which is seen to coincide with the electron part of the weak
interaction Lagrangian (\ref{eq3.13}), and explains also the convention
used for the definition of the weak coupling constant in the old
theory.

Of course, recovering the charged current leptonic weak
interaction within the framework of the considered gauge theory
should not come as a surprise -- we have actually \qq{ordered} this
result by imposing different $SU(2)$\index{SU(2)
group@$SU(2)$ group} transformation properties for
the left-handed and right-handed fields resp. Now it is also clear
that introducing a doublet of right-handed leptons along with the
$L$ in (\ref{eq5.6}) would produce a purely vector-like charged weak
current (which would be a phenomenological disaster). Thus, the
assignment of the $SU(2)$ transformation properties to the chiral
components of lepton fields that we have chosen here simply means
that we have \qq{translated} the requirement of the $V-A$
structure of weak charged currents into the gauge theory language
-- by specifying the representation contents of the matter
(lepton) fields.
%\end{document}
%\left(\matrix{10\cr 0-1}\right)

%\input{skript33} %             5.3
%%%%%%%%%%%%%%%%%%%%%%%%%%%%%%%%%%%%%%%%%%%%%%%%%%%%%%%%%%%%%%%%%%%
%%%%%%%%%%%%%%%%%%%%%%%%%%%%%%%%%%%%%%%%%%%%%%%%%%%%%%%%%%%%%%%%%%%%%%%%%%%%%%%%%%%%%%%%%%%%%%%%%%%%%%%%%%%%%%%%%%%%%%%%%%%%%%%%%%%%%%%%
%\documentstyle[12pt]{article}
%\newcommand\skrt[1]{#1\!\!\!\! / }  % pro psan¡ çkrtlìch p¡smenek,
%                                  % napý. \skrt{p}
%\newcommand\lagr{{\cal L}}

%\begin{document}
\section{Electromagnetic interaction}\label{sec5.3}

\index{electromagnetic!interaction|ff}Let us now proceed to
analyze the \qq{diagonal} part of the leptonic interaction
Lagrangian (\ref{eq5.17}), i.e. the contribution
\begin{align}
\lagr_{diag.}&=\frac{1}{2}g\bar L\gamma^\mu\tau^3 LA^3_\mu+
g'Y_L\bar L\gamma^\mu LB_\mu\notag\\
& + g'Y^{(e)}_R\bar e_R\gamma^\mu e_R B_\mu+g'Y^{(\nu)}_R
\bar\nu_R\gamma^\mu\nu_R B_\mu \label{eq5.19}
\end{align}
Taking into account that
\begin{equation}
\label{eq5.20} \tau^3 =\begin{pmatrix}1&0\\0&-1\end{pmatrix}
\end{equation}
and working out the matrix multiplication in (\ref{eq5.19}), one
gets readily
\begin{align}
\label{eq5.21} \lagr_{diag.}&=\frac{1}{2}g\bar\nu_L\gamma^\mu\nu_L
A^3_\mu- \frac{1}{2}g\bar e_L\gamma^\mu e_L A^3_\mu+
g'Y_L\bar\nu_L\gamma^\mu\nu_L B_\mu\notag\\
&+g'Y_L\bar e_L\gamma^\mu e_L B_\mu+ g'Y^{(e)}_R\bar e_R\gamma^\mu
e_R B_\mu+ g'Y^{(\nu)}_R\bar\nu_R\gamma^\mu\nu_R B_\mu
\end{align}
From the last expression it is obvious that neither the $A^3_\mu$
nor $B_\mu$ can be identified directly with the electromagnetic
field; more precisely, there is no choice of the weak
hypercharges\index{weak!hypercharge|(} that would enable one to
make such an identification -- both these fields are generally
coupled to the neutrino and none of them has a purely vectorial
coupling to the electron. Now it is also clear why an electroweak
unification based on the simple group $SU(2)$\index{SU(2)
group@$SU(2)$ group} (and involving only the ordinary leptons)
would not work: discarding the field $B_\mu$, one is left with the
$A^3_\mu$ couplings only, which certainly are not of an
electromagnetic type. In other words, the $SU(2)$ gauge theory
would not be able to accommodate both the left-handed weak charged
currents and the vectorial electromagnetic current made of the
conventional leptons. We thus arrive at an independent, \qq{purely
algebraic} argument in favour of the $SU(2)\times
U(1)$\index{SU(2) times U(1) group@$SU(2)\times U(1)$ group}
electroweak unification, without invoking an analysis of the
high-energy behaviour\index{tree unitarity} of Feynman diagrams.
It should be stressed, however, that an $SU(2)$ unification does
work if one introduces some extra leptons of the electron type;
such a theoretical scenario was developed by H. Georgi and S.
Glashow in 1972, but no exotic (heavy) leptons demanded by that
theory have been observed so far.

Although the gauge fields $A^3_\mu$ and $B_\mu$ have no direct
physical interpretation, one may try to produce physical fields by
making appropriate linear combinations of the former. In
particular, we are going to consider an orthogonal
transformation\index{orthogonal transformation|ff}
\begin{alignat}{3}
\label{eq5.22} A^3_\mu&=&&\cos \theta_W Z_\mu&+&\,\sin \theta_W A_
\mu \notag\\
B_\mu&=-&&\sin \theta_W Z_\mu&+&\,\cos \theta_W A_ \mu
\end{alignat}
where the $A_\mu$ will be required to have properties of the
electromagnetic field and the $Z_\mu$ represents a new neutral
vector field. The $\theta_W$ is an arbitrary angle at the present
moment, but it will be expressed through the other parameters of
the theory after imposing the necessary physical requirements. It
is usually called the \qq{Weinberg angle} or \qq{weak mixing
angle}\index{Weinberg angle|see{weak mixing
angle}}\index{weak!mixing angle|ff}. We should perhaps explain
here the reason why we have chosen an {\it orthogonal\/}
transformation. The orthogonality is in fact necessary for
preserving the diagonal structure of kinetic terms of the vector
fields, as one can see easily. For the original gauge fields
$A^a_\mu$ and $B_\mu$ one has
\begin{equation}
\label{eq5.23}
\lagr^{(kin.)}_{gauge}=-\frac{1}{4}A^a_{\mu\nu}A^{a\mu\nu}-
\frac{1}{4}B_{\mu\nu}B^{\mu\nu}
\end{equation}
where $A^a_{\mu\nu}=\partial_\mu A^a_\nu-\partial_\nu A^a_\mu$ and
$B_{\mu\nu}=\partial_\mu B_\nu-\partial_\nu B_\mu$ (cf.
(\ref{eq5.10})). In the preceding section we have already passed
from the $A^1_\mu,~A^2_\mu$ to the charged vector fields
$W^{\pm}_\mu$  through another (complex) orthogonal
transformation; the expression (\ref{eq5.23}) can thus be recast
as
\begin{equation}
\label{eq5.24} \lagr^{(kin.)}_{gauge}=-\frac{1}{2}W^-_{\mu\nu}
W^{+\mu\nu}-
\frac{1}{4}A^3_{\mu\nu}A^{3\mu\nu}-\frac{1}{4}B_{\mu\nu}B^{\mu\nu}
\end{equation}
Using (\ref{eq5.22}) in (\ref{eq5.24}) one gets finally
\begin{equation}
\label{eq5.25}
\lagr^{(kin.)}_{gauge}=-\frac{1}{2}W^-_{\mu\nu}W^{+\mu\nu}
-\frac{1}{4}A_{\mu\nu}A^{\mu\nu}-\frac{1}{4}Z_{\mu\nu}Z^{\mu\nu}
\end{equation}
The orthogonality of (\ref{eq5.22}) thus prevents any $A-Z$ mixing
terms from appearing in (\ref{eq5.25}).

After these explanatory remarks, let us now substitute the
transformation (\ref{eq5.22}) into (\ref{eq5.21}). One gets
\begin{equation}
\label{eq5.26}
\lagr_{diag.}=\lagr^{(A)}_{diag.}+\lagr^{(Z)}_{diag.}
\end{equation}
where
\begin{multline}
\label{eq5.27} \lagr^{(A)}_{diag.}=\Bigl[\frac{1}{2}g\sin
\theta_W\bar\nu_L \gamma^ \mu\nu_L+Y_Lg'\cos
\theta_W\bar\nu_L\gamma^\mu\nu_L+ Y^{(\nu)}_R g'\cos
\theta_W\bar\nu_R\gamma^\mu\nu_R
\\
-\frac{1}{2}g\sin \theta_W \bar e_L\gamma^\mu e_L+ Y_L g'\cos
\theta_W\bar e_L\gamma^\mu e_L+ Y^{(e)}_R g'\cos \theta_W\bar
e_R\gamma^\mu e_R\Bigr]A_\mu
\end{multline}
and the $\lagr^{(Z)}_{diag.}$ denotes the part involving the $Z$
field\index{Z boson@$Z$ boson|ff}; this term will be discussed in
detail later on. We would like to interpret (\ref{eq5.27}) as the
standard electromagnetic interaction of leptons, so it should have
the corresponding familiar properties: in particular, the neutrino
fields should be absent from (\ref{eq5.27}) and the right-handed
and left-handed components of the electron field should interact
with the $A_\mu$ with an equal strength (in other words, the
electromagnetic current must be pure vector). Of course, to meet
these requirements we can use the freedom we still have in the
assignments of the weak hypercharge values. The first requirement
thus leads to the conditions
\begin{equation}
\label{eq5.28} Y^{(\nu)}_R=0
\end{equation}
and
\begin{equation}
\label{eq5.29} \frac{1}{2}g\sin \theta_W+Y_Lg'\cos \theta_W=0
\end{equation}
Similarly, the requirement of vectorial (i.e.
parity-conserving)\index{parity!conservation} nature of the
electromagnetic current leads to
\begin{equation}
\label{eq5.30} -\frac{1}{2}g\sin \theta_W+Y_Lg'\cos \theta_W=Y^
{(e)}_R g'\cos \theta_W
\end{equation}
Combining (\ref{eq5.29}) with (\ref{eq5.30}) we obtain immediately
\begin{equation}
\label{eq5.31} Y^{(e)}_R=2Y_L
\end{equation}
and from (\ref{eq5.29}) the weak mixing angle can be expressed as
\begin{equation}
\label{eq5.32} \tan\theta_W=-2Y_L\frac{g'}{g}
\end{equation}
Writing now the electromagnetic interaction conventionally as
\begin{equation}
\label{eq5.33} \lagr^{(EM)}_{lepton}=-e~\bar e\gamma^\mu e A_\mu
\end{equation}
the relevant coupling constant $e$ is given by one of the
equivalent expressions in (\ref{eq5.30}) with the negative sign;
using (\ref{eq5.31}) and (\ref{eq5.32}) one then obtains a
remarkably simple relation
\begin{equation}
\label{eq5.34} e=g\sin \theta_W
\end{equation}
or, in terms of the $g$, $g'$ and $Y_L$
\begin{equation}
\label{eq5.35} e=-2Y_L\frac{gg'}{\sqrt{g^2+4Y_L^2g'^2}}
\end{equation}
Note that (\ref{eq5.34}) means
\begin{equation}
\label{eq5.36} e<g
\end{equation}
From the previous discussion it is clear that the {\it strict
inequality\/} must hold indeed, as the $\cos \theta_W$ must not be
zero. The relation (\ref{eq5.36}) (or (\ref{eq5.34}) resp.) is
usually called the {\bf unification condition}\index{unification
condition} as it relates the coupling strengths of the old weak
interaction and electromagnetism, unified within the $SU(2)\times
U(1)$ gauge theory; note that before the unification, the ratio of
$e$ and $g$ was completely unconstrained. We will discuss an
important physical consequence of the unification condition in the
next section.

We have seen that the weak hypercharge values are essentially
fixed by the requirement of internal consistency of the
electroweak unification, up to the $Y_L$, which can be arbitrary
(but non-zero). We believe that it may be instructive for the
reader, in particular for a beginner in the field, to fully
realize such a freedom of parametrization before adopting the
conventional values of $Y$ mentioned in Section~\ref{sec3.1} (cf.
(\ref{eq5.7}) through (\ref{eq5.9})) -- this is why we have
devoted a relatively large space to this general discussion.

In fact, it is natural to expect that the electric charge should
be a linear combination of the weak isospin\index{weak!isospin}
and weak hypercharge, simply because the generators $T^3$ and $Y$
are both represented by diagonal matrices. Summarizing now our
previous knowledge, one may  notice that the $Y$  values indeed
satisfy a relation
\begin{equation}
\label{eq5.37}
Q=T_3+cY
\end{equation}
where $c$ is a real coefficient (its value being fixed e.g. by an
arbitrarily chosen $Y_L$). The conventional choice corresponds to
$c = 1$: this leads to $Y_L = - 1/2$ and the formula
(\ref{eq5.32}) for the weak mixing angle is thus simplified to
\begin{equation}
\label{eq5.38} \tan\theta_W=\frac{g'}{g}
\end{equation}
i.e. the $\sin\theta_W$ and $\cos\theta_W$ are expressed by the
aesthetically pleasing formulae
\begin{equation}
\label{eq5.39} \cos \theta_W=\frac{g}{\sqrt{g^2+g'^2}}\;\;,\;\;
\sin \theta_W=\frac{g'}{\sqrt{g^2+g'^2}}
\end{equation}

The upshot of all this is that {\bf in further study of the
standard electroweak model the reader can, for convenience, use
the weak hypercharge values determined by the rule
\begin{equation} \label{eq5.40} \boldsymbol{Q=T_3+Y}
\end{equation} \index{weak!hypercharge|)}which automatically lead to the correct structure of the
electromagnetic current and to a simple relation for the mixing
angle $\boldsymbol{\theta_W}$}. Nevertheless, we will come back to
the general parametrization in Section~\ref{sec5.5} (and also in the next
chapter, in connection with a mass formula for the $W$ and $Z$
fields) to show that physical results do not depend on the choice
of the non-zero value of the $Y_L$.
%\end{document}

%\input{skript34} %             5.4
%%%%%%%%%%%%%%%%%%%%%%%%%%%%%%%%%%%%%%%%%%%%%%%%%%%%%%%%%%%%%%%%%%%
%%%%%%%%%%%%%%%%%%%%%%%%%%%%%%%%%%%%%%%%%%%%%%%%%%%%%%%%%%%%%%%%%%%%%%%%%%%%%%%%%%%%%%%%%%%%%%%%%%%%%%%%%%%%%%%%%%%%%%%%%%%%%%%%%%%%%%%%
%\documentstyle[12pt]{article}
%\newcommand\skrt[1]{#1\!\!\!\! / }  % pro psan¡ ¨krtlch p¡smenek,
%                                  % nap©. \skrt{p}
%\newcommand\lagr{{\cal L}}

%\begin{document}
\section{Unification condition and $W$ boson mass}\index{W
boson@$W$ boson|ff}\label{sec5.4}

Let us now return to the unification condition (\ref{eq5.36}). It
gives a simple lower bound for the weak coupling constant $g$ (at
least at the level of the classical Lagrangian, i.e. at the tree
level within quantum theory) and one might therefore employ it to
obtain useful constraints on the physical quantities, expressed in
terms of the $g$. In particular, in this section we will discuss a
lower bound for the $W$ boson mass which follows from
(\ref{eq5.36}). We have not introduced any mass terms for the
vector fields so far, but the good old $W$ boson model should of
course be fully reproduced within the $SU(2) \times U(1)$
unification scheme. Thus, at the present stage we may simply put a
mass term for the $W^\pm$ fields by hand (the $Z$ boson should
also become massive, in order to avoid a new long-range force
different from the electromagnetism) and the photon  will remain
massless. We may then adopt, for the moment, a standard
(perturbative) canonical quantization procedure and consider the
corresponding Feynman graphs -- such a program can be successfully
carried out at least at the tree level. As we have seen in Chapter~\ref{chap3}, in a model involving charged intermediate vector boson there is
a relation between the weak coupling constant $g$, the $W$ boson
mass and the Fermi coupling $G_F$\index{unification condition}
\begin{equation}
\label{eq5.41}
\frac{G_F}{\sqrt{2}}=\frac{g^2}{8m^2_W}
\end{equation}
which tells us that in the low-energy limit a Fermi-type model
represents a good effective weak interaction theory. The relation
(\ref{eq5.41}) must be then also valid within the $SU(2) \times
U(1)$ unified theory and using the unification condition
(\ref{eq5.34}) in (\ref{eq5.41}) one gets a formula for the $W$
boson mass
\begin{equation}
\label{eq5.42} m_W=\left( \frac{\pi\alpha}{G_F\sqrt{2}}
\right)^{1/2} \frac{1}{\sin \theta_W}
\end{equation}
where we have introduced the fine structure constant
\index{alpha@$\alpha$, the fine structure constant}
\index{coupling constants|seealso{$\alpha$, $G_F$}}
$\alpha=e^2/(4\pi)$. The weak mixing angle $\theta_W$ is a free
parameter of the considered model of electroweak unification,
which must be measured independently (the $\theta_W$ can be traded
for other physical parameters, but its numerical value is not
predicted by the standard electroweak model -- such a prediction
can only be accomplished within an appropriate grand
unification\index{grand unification} scheme). The $\theta_W$ can
be measured e.g. in neutrino scattering processes (see Section~\ref{sec5.6}
for an explicit example) so that the formula (\ref{eq5.42})
(corrected by including higher-order quantum effects within the
full standard model) did provide a {\it prediction\/} for the $W$
boson mass before its actual discovery in 1983. Using in
(\ref{eq5.42}) the current experimental value $\sin
^2\theta_W\doteq$ 0.23, taking $\alpha\doteq$ 1/137 and
$G_F\doteq1.166\times 10^{-5}\ \GeV^{-2}$, one gets $m_W\doteq
77.7\ \GeV$. For the corrected value one then obtains roughly $80\
\GeV$ (the relevant corrections were calculated first by M.
Veltman in 1980), which is in agreement with the current
experimental value $m_W =(80.377~\pm~ 0.012)\ \GeV$ (cf.
\cite{ref5}). Note that the main effect of these higher-order
corrections on the $W$ mass can be reproduced by replacing the
traditional low-energy value of the fine structure constant by the
\qq{running electromagnetic coupling} at the $W$ mass scale, i.e.
$\alpha(m^2_W)\doteq$ 1/128.

In any case, (\ref{eq5.42}) obviously implies a {\it lower
bound\/} for the $W$ mass, namely
\begin{equation}
\label{eq5.43} m_W>\left( \frac{\pi\alpha}{G_F\sqrt{2}}
\right)^{1/2}
\end{equation}
(of course, the bound (\ref{eq5.43}) is an immediate consequence
of (\ref{eq5.41}) and the unification condition written as the
inequality $g > e$). For $\alpha$ = 1/137  one thus gets roughly
\begin{equation}
\label{eq5.44} m_W>37\ \GeV
\end{equation}
Let us also remark that at this stage the $Z$ boson mass can be
entirely arbitrary; we will touch the problem of the $Z$ mass
determination from the low-energy scattering experiments in
Section~\ref{sec5.6}. To get a prediction for the $Z$ mass, one has to
settle the subtle issue of mass generation in gauge theories
(which we have trivialized for the moment). This will be a subject
of the next chapter.

Finally, let us stress that the lower bound for the $W$ boson mass
(\ref{eq5.43}) is not a universal feature of any electroweak
unification -- the condition (\ref{eq5.36}) is indeed intimately
connected with the particular $SU(2) \times U(1)$ unification
scheme. For example, in the $SU(2)$ (or $O(3)$\index{O(3)
group@$O(3)$ group}) model of Georgi and
Glashow\index{Georgi--Glashow model} \cite{ref62} involving heavy
leptons\index{heavy leptons} the unification condition reads
\begin{equation}
\label{eq5.45}
g\leq e\sqrt{2}
\end{equation}
which implies an {\it upper\/} bound for the $W$ mass, namely
\begin{equation}
\label{eq5.46} m_W\leq
\left(\frac{\pi\alpha\sqrt{2}}{G_F}\right)^{1/2} \doteq 53\ \GeV
\end{equation}
The currently known experimental value of the $m_W$  thus certainly excludes
the minimal scenario based on heavy leptons, but this still remains to be
of methodical interest as a construction of electroweak unification
alternative to the standard model. For details of the heavy lepton
scheme the interested reader is referred to the literature.
%\end{document}

%\input{skript35} %             5.5
%%%%%%%%%%%%%%%%%%%%%%%%%%%%%%%%%%%%%%%%%%%%%%%%%%%%%%%%%%%%%%%%%%%
%%%%%%%%%%%%%%%%%%%%%%%%%%%%%%%%%%%%%%%%%%%%%%%%%%%%%%%%%%%%%%%%%%%%%%%%%%%%%%%%%%%%%%%%%%%%%%%%%%%%%%%%%%%%%%%%%%%%%%%%%%%%%%%%%%%%%%%%
%\documentstyle[12pt]{article}
%\newcommand\skrt[1]{#1\!\!\!\! / }  % pro psan¡ ¨krtlch p¡smenek,
%                                  % nap©. \skrt{p}
%\newcommand\lagr{{\cal L}}

%\begin{document}
\section{Weak neutral currents}\label{sec5.5}

\index{weak!neutral current|see{neutral current}}\index{neutral current|ff}
Let us now examine the interactions of leptons with the $Z$ boson,
i.e. the term denoted by $\lagr^{(Z)}_{diag.}$ in (\ref{eq5.26}).
To keep the discussion as general as possible, we will maintain an
arbitrary value of the weak hypercharge $Y_L$\index{weak!hypercharge|ff} -- of course, at the
same time we will utilize the relations $Y^{(e)}_R = 2Y_L$ and
$Y^{(\nu)}_R = 0$ established in Section~\ref{sec5.3}. We will show that
physical results do not depend on the $Y_L$. According to
(\ref{eq5.21}) (where the substitution (\ref{eq5.22}) is
performed) we have, grouping together the interactions of the
individual chiral components of lepton fields
\begin{eqnarray}
\label{eq5.47}
\lagr^{(Z)}_{diag.}&=&(\frac{1}{2}g\cos \theta_W- Y_L
g'\sin \theta_W)\bar\nu_L\gamma^\mu\nu_L
Z_\mu\nonumber\\
&+&(-\frac{1}{2}g\cos \theta_W-Y_L g'\sin \theta_W)
\bar e_L\gamma^\mu e_L Z_\mu\nonumber\\
&-&2Y_L g'\sin \theta_W\bar e_R\gamma^\mu e_R Z_\mu
\end{eqnarray}
The last expression can be conveniently recast as
\begin{eqnarray}
\label{eq5.48}
\lagr^{(Z)}_{diag.}&=&\frac{g}{\cos \theta_W}
(\frac{1}{2}\cos ^2\theta_W-Y_L\frac{g'}{g}
\sin \theta_W\cos \theta_W)\bar\nu_L
\gamma^\mu\nu_L Z_\mu\nonumber\\
&+&\frac{g}{\cos \theta_W}(-\frac{1}{2}\cos ^2
\theta_W-Y_L\frac{g'}{g}\sin \theta_W\cos
\theta_W)\bar e_L\gamma^\mu e_L Z_\mu\nonumber\\
&+&\frac{g}{\cos \theta_W}(-2Y_L\frac{g'}{g}\sin
\theta_W\cos \theta_W)\bar e_R\gamma^\mu e_R Z_\mu
\end{eqnarray}
and employing the relation  $\tan \theta_W=-2Y_L g'/g$ (see
(\ref{eq5.32})) one gets, after a simple manipulation
\begin{equation}
\label{eq5.49}
\lagr^{(Z)}_{diag}=\frac{g}{\cos \theta_W}[\frac{1}{2}
\bar\nu_L\gamma^\mu\nu_L+(-\frac{1}{2}+\sin ^2\theta_W) \bar
e_L\gamma^\mu e_L+\sin ^2\theta_W\bar e_R\gamma^\mu
e_R]Z_\mu\qquad
\end{equation}
The form (\ref{eq5.49}) represents an interaction of the $Z$ boson
field with \qq{weak neutral leptonic currents}\index{leptonic current}. Note that the
adjective \qq{neutral} in the present context means that the
corresponding current is composed of fermion fields carrying the
same charge -- in this sense, the electromagnetic current is
neutral as well. We see that any possible dependence on the $Y_L$
drops out, and the neutral current (NC) interaction is fully
parametrized in terms of the CC coupling strength $g$ and the weak
mixing angle $\theta_W$. Let us stress that the form of the NC
interaction (\ref{eq5.49}) is a non-trivial {\it prediction\/} of
the considered electroweak unification -- having fixed the values
of the free parameters (weak hypercharges) so as to recover the
standard electromagnetic interaction, the weak NC interactions are
fully determined. From now on, we will denote the term
(\ref{eq5.49}) by the symbol $\lagr^{(e)}_{NC}$ (referring
explicitly to leptons of electron type) and introduce a frequently
used notation for the relevant coupling strengths by writing
\begin{equation}
\label{eq5.50}
\lagr^{(e)}_{NC}=\frac{g}{\cos \theta_W}\sum_{f=\nu ,e}
(\varepsilon_L^{(f)}\bar f_L\gamma^\mu f_L+\varepsilon^{(f)}_R
\bar f_R\gamma^\mu f_R)Z_\mu
\end{equation}
where
\begin{equation}
\label{eq5.51}
\varepsilon^{(\nu)}_L=\frac{1}{2}~,~\varepsilon^{(\nu)}_R=0~,~
\varepsilon^{(e)}_L=-\frac{1}{2}+\sin ^2\theta_W~,~
\varepsilon^{(e)}_R=\sin ^2\theta_W
\end{equation}
This description of the neutral current structure in terms of the
parameter $\sin ^2\theta_W$  exhibits a famous rule
characteristic of the $SU(2)\times U(1)$ standard electroweak
model, namely
\begin{equation}
\label{eq5.52}
\varepsilon^{(f)}_{L,R}=T^{(f)}_{3L,R}-Q^{(f)}\sin ^2\theta_W
\end{equation}
The reader can easily verify that (\ref{eq5.51}) is indeed
reproduced when one uses in (\ref{eq5.52}) the relevant values of
the electric charge and weak isospin\index{weak!isospin} for the chiral components of
lepton fields (cf. (\ref{eq5.8})).

We could also express the coupling constants for NC interactions
in terms of the $g$ and $e$ -- in other words, in terms of the
parameters of the \qq{old physics} (weak and electromagnetic
interactions before the gauge unification). Introducing an
alternative notation for the NC couplings (\ref{eq5.50})
\begin{equation}
\label{eq5.53}
\lagr^{(e)}_{NC}=g^{(\nu)}_L\bar\nu_L\gamma^\mu\nu_L
Z_\mu+ g^{(e)}_L\bar e_L\gamma^\mu e_L Z_\mu+ g^{(e)}_R\bar
e_R\gamma^\mu e_R Z_\mu
\end{equation}
and using the relation
\begin{equation}
\label{eq5.54}
\sin \theta_W=\frac{e}{g}
\end{equation}
(which is valid independently of the $Y_L$ value -- cf.
(\ref{eq5.34})), one obtains
\begin{equation}
\label{eq5.55}
g^{(\nu)}_L=\frac{g^2}{2\sqrt{g^2-e^2}}~,~
g^{(e)}_L=\frac{-\frac{1}{2}g^2+e^2}{\sqrt{g^2-e^2}}~,~
g^{(e)}_R=\frac{e^2}{\sqrt{g^2-e^2}}
\end{equation}
From (\ref{eq5.55}) it is particularly clear that the term
\qq{electroweak unification} is indeed justified in connection with
the considered $SU(2)\times U(1)$ gauge model: the NC couplings
are non-trivial functions of the $e$ and $g$ and \qq{interpolate}
thus between the electromagnetic and weak interactions. We will
see more examples of such a functional dependence of the
electroweak couplings (and mass ratios) in other sectors of the
standard model -- in particular, in the sector of vector bosons
discussed in detail in Section~\ref{sec5.7}. In any case, the weak mixing
angle, which can be expressed in terms of the ratio $e/g$, is an
arbitrary parameter of the electroweak unification and must be
measured independently. In the next section we will show in an
example how the parameter $\sin ^2\theta_W$ can be determined
from the low energy neutrino scattering processes mediated by the
weak neutral currents.
%\end{document}

%\input{skript36} %             5.6
%%%%%%%%%%%%%%%%%%%%%%%%%%%%%%%%%%%%%%%%%%%%%%%%%%%%%%%%%%%%%%%%%%%
%%%%%%%%%%%%%%%%%%%%%%%%%%%%%%%%%%%%%%%%%%%%%%%%%%%%%%%%%%%%%%%%%%%%%%%%%%%%%%%%%%%%%%%%%%%%%%%%%%%%%%%%%%%%%%%%%%%%%%%%%%%%%%%%%%%%%%%%

%\documentstyle[12pt]{article}
%\newcommand\skrt[1]{#1\!\!\!\! / }  % pro psan¡ ¨krtlch p¡smenek,
                                  % nap©. \skrt{p}
%\newcommand\lagr{{\cal L}}

%\begin{document}
\section{Low energy neutrino-electron scattering}\index{neutrino-electron scattering|ff}\label{sec5.6}

Scattering of muon neutrino or antineutrino on the electron is a
typical process which goes via neutral currents; in the old
Feynman--Gell-Mann theory it can only occur at one-loop (and
higher) level (the reader is recommended to draw a one-loop
diagram\index{loop diagrams} describing this process in the old
weak interaction theory). Before proceeding to a detailed
discussion of the
$\stackrel{\scriptscriptstyle(-)}{\nu_\mu}\!\!-e$ scattering
within the $SU(2)\times U(1)$ electroweak model, we have to
incorporate leptons of muon type into this framework. In fact,
this can be done in an almost trivial way. In complete analogy
with the scheme explained in Section~\ref{sec5.1}, one introduces
left-handed doublet and right-handed singlets for the second
(muonic) generation
\begin{equation}
\label{eq5.56} L^{(\mu)}=\begin{pmatrix}\nu_{\mu L}\\ \mu_L
\end{pmatrix},\;\mu_R,\; {\nu_{\mu R}}
\end{equation}
with weak hypercharges following the pattern of the electron-type
leptons. Then the  muonic contributions to the weak charged
current and to electromagnetic current have the right form and the
corresponding weak neutral currents obviously repeat precisely the
structure shown in (\ref{eq5.49}). Let us stress that in writing
(\ref{eq5.56}) along with (\ref{eq5.1}) we neglect {\it a
priori\/} a possible mixing between the two lepton generations
(and exclude thus phenomena like the neutrino
oscillations\index{neutrino!oscillations}) -- we will comment on
this issue later on, in the context of the full standard
electroweak model including the mechanism for generating masses.
It is also clear that the $SU(2)\times U(1)$ electroweak model can
be extended in this way to an arbitrary number of lepton
generations; as we know now, in our physical world there are
precisely three generations of leptons (labelled as
$e,\mu,\tau$)\index{tau lepton@$\tau$ lepton} involving light
neutrinos.

\begin{figure}[h]\centering
\begin{tabular}{cc}
\subfigure[]{\s{\includegraphics{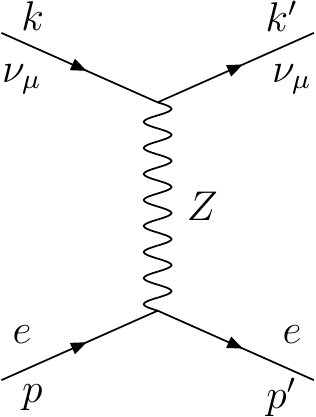}}}&\hspace{1.5cm}\subfigure[]{\s{\includegraphics{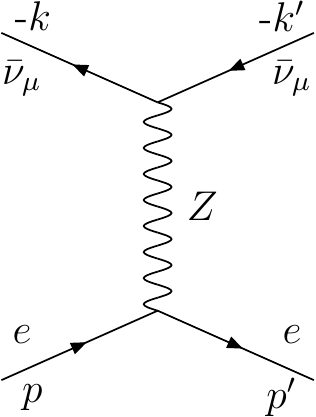}}}
\end{tabular}
\caption{Tree-level Feynman graphs for the $\nu_\mu -e$ scattering
(a) and $\bar\nu_\mu -e$ scattering (b).}
\label{fig15}\index{Feynman diagrams!for $\nu_\mu-e$ and
$\bar\nu_\mu-e$ scattering}
\end{figure}

With the above remarks in mind, we are ready to write down the
part of the neutral-current interaction Lagrangian relevant for
the description of the
$\stackrel{\scriptscriptstyle(-)}{\nu_\mu}\!\!-e$ scattering
processes. This can be written as
\begin{equation}
\label{eq5.57}
\lagr^{(\nu_\mu
e)}_{NC}=\frac{g}{2\cos \theta_W}
[\frac{1}{2}\bar\nu_\mu\gamma^\alpha(1-\gamma_5)\nu_\mu+ \bar
e\gamma^\alpha(v-a\gamma_5)e]Z_\alpha
\end{equation}
where the axial-vector and vector NC couplings\index{axial
vector!coupling} for the electron are, according to the results of
the preceding section
\begin{eqnarray}
\label{eq5.58}
v&=&\varepsilon_L+\varepsilon_R=-\frac{1}{2}+2\sin ^2
\theta_W\nonumber\\
a&=&\varepsilon_L-\varepsilon_R=-\frac{1}{2}
\end{eqnarray}
(cf. (\ref{eq5.50}), (\ref{eq5.51})). The lowest-order Feynman
graphs for the considered processes are shown in
Fig.\,\ref{fig15}.
Let us start with the neutrino process. The Lorentz invariant
matrix element corresponding to Fig.\,\ref{fig15}a is, following
(\ref{eq5.57})
\begin{eqnarray}
\label{eq5.59}
{\cal M}_a&=&-\frac{g^2}{8\cos^2\theta_W}[\bar u(k') \gamma_\alpha(1-\gamma_5)u(k)][\bar
u(p')\gamma_\beta
(v-a\gamma_5)u(p)]\times\nonumber\\
~&~&\times\frac{-g^{\alpha\beta}+m^{-2}_Z
q^\alpha q^\beta}{q^2-m^2_Z}
\end{eqnarray}
The kinematical conditions are assumed to be such that
\begin{eqnarray}
\label{eq5.60}
m^2_e\ll s\ll m^2_Z
\end{eqnarray}
where $s=(k+p)^2$, so we will neglect the electron mass in what
follows, and the $q^2$ in the $Z$ boson
propagator\index{propagator!of massive vector boson} can be
neglected as well. Of course, the contribution of the longitudinal
term in the numerator of the $Z$ propagator is also strongly
suppressed (it vanishes exactly for a massless neutrino). Thus,
(\ref{eq5.59}) is approximately equal to
\begin{equation}
\label{eq5.61}
{\cal M}_a\doteq
-\frac{g^2}{8\cos ^2\theta_W}\frac{1}{m^2_Z} [\bar
u(k')\gamma_\alpha(1-\gamma_5)u(k)][\bar u(p')\gamma^\alpha
(v-a\gamma_5)u(p)]
\end{equation}
The last expression may be conveniently recast as
\begin{equation}
\label{eq5.62}
{\cal M}_a=-\frac{G_F}{\sqrt 2}\rho[\bar
u(k')\gamma_\alpha(1-\gamma_5) u(k)][\bar
u(p')\gamma^\alpha(v-a\gamma_5)u(p)]
\end{equation}
where we have introduced the Fermi coupling constant through the
relation $G_F/\sqrt 2=g^2/8m^2_W$ and $\rho$ denotes the ratio
\begin{equation}
\label{eq5.63}
\rho=\frac{m^2_W}{m^2_Z\cos ^2\theta_W}\qquad
\end{equation}
As we shall see in the next chapter, the Weinberg--Salam standard
model predicts classically (i.e. at the tree level) the value
$\rho =1$ as a consequence of the specific realization of the
Higgs mechanism\index{Higgs mechanism} generating the vector boson
masses -- a prediction that has indeed turned out to be
phenomenologically successful. At the present stage of our
discussion the $\rho$ value is essentially arbitrary, but we will
see shortly that it may  be determined experimentally (along with
the weak mixing angle) when both neutrino and antineutrino low
energy cross sections are measured.

\begin{figure}[h]
\centering \s{\includegraphics{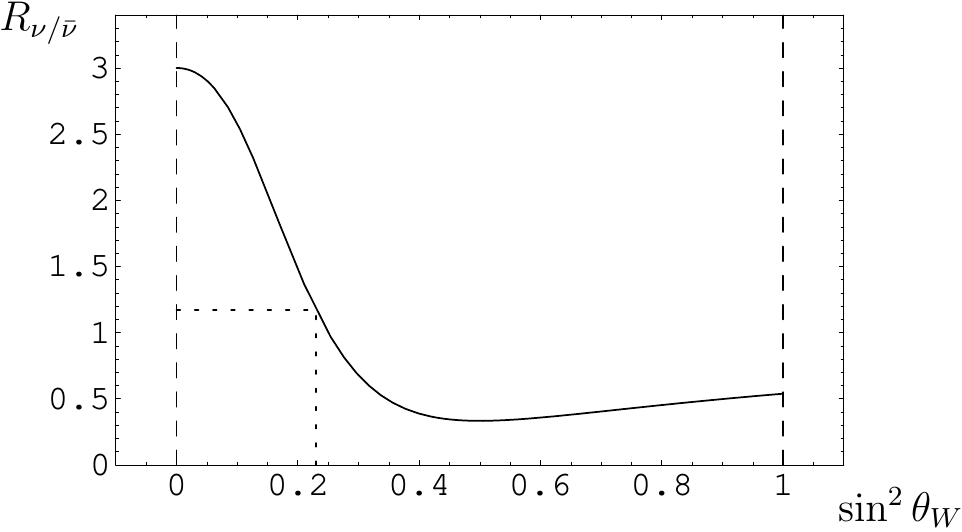}} \caption{The
dependence of the ratio of neutrino and antineutrino cross
sections on the parameter $\sin^2\theta_W$.} \label{fig16}
\end{figure}

In calculation of the cross section for the neutrino process we
will assume that the electron is unpolarized; the spin-averaged
square of the matrix element (\ref{eq5.62}) then becomes
\begin{eqnarray}
\label{eq5.64} \lefteqn{\overline{|{\cal
M}_a|^2}=\frac{1}{2}\sum_{pol.}
|{\cal M}_a|^2=}\\
& &=\frac{1}{4}G^2_F\rho^2\Tr[\slashed{k}'\gamma_\alpha
(1-\gamma_5)
\slashed{k}\gamma_\beta(1-\gamma_5)]\cdot\Tr[\slashed{p}'\gamma^\alpha
(v-a\gamma_5)\slashed{p}\gamma^\beta(v-a\gamma_5)]\nonumber
\end{eqnarray}
After some simple manipulations and using the formulae
(\ref{eqA.50}) one gets from (\ref{eq5.64})
\begin{align}
\label{eq5.65} \overline{|{\cal M}_a|^2}&=\frac{1}{2}G^2_F\rho^2
[(v^2+a^2)\Tr(\slashed{k}'\gamma_\alpha\slashed{k}\gamma_\beta)\cdot
\Tr(\slashed{p}'\gamma^\alpha\slashed{p}\gamma^\beta)
\notag\\
&\phantom{=}+2va\Tr(\slashed{k}'\gamma_\alpha\slashed{k}\gamma_\beta\gamma_5)\cdot
\Tr(\slashed{p}'\gamma^\alpha\slashed{p}\gamma^\beta\gamma_5)]
\notag\\
&=16G^2_F\rho^2[(v^2+a^2)\left((k\cdot p)(k'\cdot p')+(k\cdot
p')(k'\cdot p)\right)
\notag\\
&\phantom{=}+2va\left((k\cdot p)(k'\cdot p')-(k\cdot p')(k'\cdot p)\right)]\notag\\
&=16G^2_F\rho^2[(v+a)^2(k\cdot p)(k'\cdot p')+(v-a)^2(k\cdot
p')(k'\cdot p)]
\end{align}
The last expression can be rewritten in terms of the Mandelstam
invariants $s = (k + p)^2$  and $u = (k - p')^2$ as
\begin{equation}
\label{eq5.66} \overline{|{\cal M}_a|^2}=4G^2_F\rho^2[(v+a)^2 s^2+
(v-a)^2 u^2]
\end{equation}
(let us stress again that we neglect systematically the electron
mass). Alternatively, one may introduce the dimensionless
invariant $y = p\cdot q / p\cdot k$ which in the considered
massless case satisfies a simple relation $u= - s (1 - y)$  (cf.
Appendix~\ref{appenB}); one thus obtains
\begin{equation}
\label{eq5.67} \overline{|{\cal M}_a|^2}=4G^2_F\rho^2 s^2[(v+a)^2
+ (v-a)^2 (1-y)^2]
\end{equation}
Similarly, one can calculate a corresponding quantity for the
antineutrino process described by the graph in Fig.\,\ref{fig15}b.
One gets
\begin{equation}
\label{eq5.68} \overline{|{\cal M}_b|^2}=4G^2_F\rho^2
s^2[(v+a)^2(1-y)^2 + (v-a)^2]
\end{equation}
(of course, the result (\ref{eq5.68}) can also be obtained
directly from (\ref{eq5.66}) by employing the crossing symmetry,
i.e. interchanging the $s$ and $u$ variables). Using now a
standard formula for the differential cross section (cf. Appendix~\ref{appenB}) one gets
\begin{eqnarray}
\label{eq5.69}
\frac{d\sigma}{dy}(\nu_\mu e\rightarrow\nu_\mu
e)&=& \frac{G^2_F
s}{\pi}\rho^2[\varepsilon^2_L+\varepsilon^2_R(1-y)^2]
\nonumber\\
\frac{d\sigma}{dy}(\bar\nu_\mu e\rightarrow\bar\nu_\mu e)&=&
\frac{G^2_F s}{\pi}\rho^2[\varepsilon^2_L(1-y)^2+\varepsilon^2_R]
\end{eqnarray}
where we have retrieved the original \qq{chiral} parameters for the
electron neutral current (cf. (\ref{eq5.58})). Integrating the
expressions (\ref{eq5.69}) over the $y$ from 0 to 1, one obtains
the total cross sections
\begin{eqnarray}
\label{eq5.70}
\sigma(\nu_\mu e\rightarrow\nu_\mu e)&=&
\frac{G^2_F s}{\pi}\rho^2(\varepsilon^2_L+\frac{1}{3}
\varepsilon^2_R)\nonumber\\
\sigma(\bar\nu_\mu e\rightarrow\bar\nu_\mu e)&=&
\frac{G^2_F s}{\pi}\rho^2(\frac{1}{3}\varepsilon^2_L+
\varepsilon^2_R)
\end{eqnarray}
Now it is clear that a measurement of the cross sections
(\ref{eq5.69}) or (\ref{eq5.70}) leads to a determination of both
the weak mixing angle and the parameter~$\rho$. In particular,
taking the ratio of the neutrino and antineutrino total cross
sections (\ref{eq5.70}) and using $\varepsilon_L
=-\frac{1}{2}+\sin ^2\theta_W,\varepsilon_R=\sin ^2 \theta_W$ one
gets
\begin{eqnarray}
\label{eq5.71}
R_{\nu/\bar\nu}=\frac{\sigma(\nu_\mu
e\rightarrow\nu_\mu e)} {\sigma(\bar\nu_\mu
e\rightarrow\bar\nu_\mu e)}=\frac
{3-12\sin ^2\theta_W+16\sin ^4\theta_W}
{1-4\sin ^2\theta_W+16\sin ^4\theta_W}
\end{eqnarray}
The functional dependence (\ref{eq5.71}) is graphically depicted
in Fig.\,\ref{fig16}.
The relevant experimental value is roughly around $R=1.2$, where
the slope of the curve in Fig.\,\ref{fig16} is rather favourable
for a reasonably accurate determination of the $\sin^2\theta_W$.
From an absolute value of one of the cross sections (\ref{eq5.70})
one can then determine the parameter $\rho$. Currently the best
data for the considered scattering processes are provided by the
collaboration CHARM II. An analysis of the data accumulated till
1991 led to the results
\begin{eqnarray}
\label{eq5.72}
\sin ^2\theta_W&=&0.232~\pm~0.008\nonumber\\
\rho&=&1.006~\pm~0.047
\end{eqnarray}
(see P. Vilain et al.: Phys. Lett. B335 (1994) 246). Of course,
the knowledge of the parameters $\rho$ and $\sin ^2\theta_W$
together with the result for the $W$ mass (see (\ref{eq5.42}))
enables one to determine the $Z$ mass. In particular, taking
$\rho=1$, $\sin ^2\theta_W=0.23$ and $m_W=80\ \GeV$, one obtains
$m_Z\doteq 90\ \GeV$ (the current experimental value is $m_Z\doteq
91.19\ \GeV$). Let us stress again that the $Z$ boson mass --
obtained here from an experimental value of the parameter $\rho$
-- is in fact a successful prediction of the full GWS standard
electroweak model; this will be made clear in the next chapter.
The cross sections (\ref{eq5.70}) are very small; using the values
(\ref{eq5.72}), one has roughly
\begin{eqnarray}
\label{eq5.73} \sigma(\nu_\mu e\rightarrow\nu_\mu
e)&\doteq&1.5\,E_{lab.}
(\GeV)\times10^{-42}\text{cm}^2\nonumber\\
\sigma(\bar\nu_\mu e\rightarrow\bar\nu_\mu e)&\doteq&1.3\,E_{lab.}
(\GeV)\times10^{-42}\text{cm}^2
\end{eqnarray}
The corresponding experimental measurement therefore represents a
formid\-able task -- on the other hand, these purely leptonic
processes are theoretically clean and provide a simple and
instructive example of a calculation involving the neutral current
interactions. Needless to say, {\bf there are other more accurate
determinations of the relevant NC parameters from processes with
higher statistics} (such as lepton-nucleon scattering,
electron-positron annihilation, etc.). Nevertheless, from the
historical point of view, the $\nu_\mu - e$ scattering was
actually the first neutral current process observed (in 1973) and
provided thus a decisive experimental support to the $SU(2)\times
U(1)$ gauge theory of weak and electromagnetic interactions.
%\end{document}

%\input{skript37} %             5.7
%%%%%%%%%%%%%%%%%%%%%%%%%%%%%%%%%%%%%%%%%%%%%%%%%%%%%%%%%%%%%%%%%%%
%%%%%%%%%%%%%%%%%%%%%%%%%%%%%%%%%%%%%%%%%%%%%%%%%%%%%%%%%%%%%%%%%%%%%%%%%%%%%%%%%%%%%%%%%%%%%%%%%%%%%%%%%%%%%%%%%%%%%%%%%%%%%%%%%%%%%%%%
%\documentstyle[12pt]{article}
%\newcommand\skrt[1]{#1\!\!\!\! / }  % pro psan¡ ¨krtlch p¡smenek,
                                  % nap©. \skrt{p}
%\newcommand\lagr{{\cal L}}

%\begin{document}
\section{Interactions of vector bosons}\label{sec5.7}
Let us now turn to the investigation of the term $\lagr_{gauge}$
in the Lagrangian (\ref{eq5.13}), which contains the interactions
of the gauge fields with themselves. According to (\ref{eq4.51})
and (\ref{eq5.10}) through (\ref{eq5.12}), the relevant
interaction term can be written as
\begin{align}
\label{eq5.74} \lagr^{int.}_{gauge}=\ &-\frac{1}{2}g\epsilon^{abc}
(\partial_\mu A^a_\nu-\partial_\nu A^a_\mu)A^{b\mu}A^{c\nu}
\notag\\
&-\frac{1}{4}g^2\epsilon^{abc}\epsilon^{ajk} A^b_\mu A^c_\nu
A^{j\mu}A^{k\nu}
\end{align}
Working out explicitly the first term in (\ref{eq5.74}) and
employing the identity $\epsilon^{abc}\epsilon^{ajk}=
\delta^{bj}\delta^{ck}-\delta^{bk}\delta^{cj}$ in the second term,
one gets first
\begin{align}
\label{eq5.75} \lagr^{int.}_{gauge}=\ &-g[(\partial_\mu
A^1_\nu-\partial_\nu A^1_\mu)
A^{2\mu} A^{3\nu}+\ \text{cycl. perm.}(123)]\notag\\
&-\frac{1}{4}g^2[(A^a_\mu A^{a\mu})(A^b_\nu A^{b\nu})- (A^a_\mu
A^a_\nu)(A^{b\mu}A^{b\nu})]
\end{align}
which can be further recast as
\begin{align}
\label{eq5.76} \lagr^{(int.)}_{gauge}=\ &-g(A^1_\mu A^2_\nu\dmm
A^{3\nu}+A^2_\mu A^3_\nu\dmm A^{1\nu}+A^3_\mu
A^1_\nu\dmm A^{2\nu})\notag\\
&-\frac{1}{4}g^2[(\vec{A}_\mu \cdot \vec{A}^\mu)^2
-(\vec{A}_\mu\cdot\vec{A}_\nu)(\vec{A}^\mu\cdot\vec{A}^\nu)]
\end{align}
where the symbol $\partialvob$ is defined by
$f\partialvob g=f(\partial g)-(\partial f)g$
and we have used the standard shorthand notation,
$\vec{A}_\mu\cdot\vec{A}^\mu=A^a_\mu A^{a\mu}$
etc. Replacing the Yang--Mills fields $A^1_\mu$ and $A^2_\mu$ by the
physical charged vector fields $W^\pm_\mu$  according to
\begin{center}
$A^1_\mu=\frac{1}{\sqrt 2}(W^+ +W^-)~~,~~
A^2_\mu=\frac{i}{\sqrt 2}(W^+_\mu -W^-_\mu)$\\
\end{center}
(cf. (\ref{eq5.17})), the form (\ref{eq5.76}) becomes
\begin{align}
\label{eq5.77} \lagr^{(int.)}_{gauge}=&-ig(W^0_\mu W^-_\nu \dmm
W^{+\nu}+W^-_\mu W^+_\nu \dmm W^{0\nu}+W^+_\mu W^0_\nu
\dmm W^{-\nu})\nonumber\\
&-g^2[\frac{1}{2}(W^+_\mu W^{-\mu})^2-\frac{1}{2}(W^+_\mu
W^{+\mu})(W^-_\nu W^{-\nu})+(W^0_\mu W^{0\mu})(W^+_\nu W^{-\nu})
\nonumber\\
&-(W^-_\mu W^+_\nu)(W^{0\mu}W^{0\nu})]
\end{align}
where the $W^0_\mu$ denotes the linear combination
\begin{equation}
\label{eq5.78}
W^0_\mu=\cos \theta_W Z_\mu+\sin \theta_W
A_\mu
\end{equation}
($W^0_\mu$ is simply a different name for the original Yang--Mills
field $A^3_\mu$ -- cf. (\ref{eq5.22})).

The expression (\ref{eq5.77}) is seen to contain trilinear and
quadrilinear interactions\index{quadrilinear vertex} of the vector
fields; when it is worked out by employing (\ref{eq5.78}), one can
identify two trilinear and four quadrilinear couplings of the
$W,~Z$ and photon, namely
\begin{eqnarray}
\lagr_{WW\gamma}&=&-ie(A_\mu W^-_\nu\dmm W^{+\nu}+W^-_\mu
W^+_\nu\dmm A^\nu \nonumber
\\
\label{eq5.79}
~&~&+W^+_\mu
A_\nu\dmm W^{-\nu})\qquad\\
%\end{equation}
%\begin{equation}
\label{eq5.80} \lagr_{WWZ}&=&-ig\cos \theta_W(Z_\mu W^-_\nu \dmm
W^{+\nu} +W^-_\mu
W^+_\nu \dmm Z^\nu\nonumber\\
~&~&+W^+_\mu Z_\nu\dmm W^{-\nu})\\
%\end{eqnarray}
%\begin{equation}
\label{eq5.81}
\lagr_{WW\gamma\gamma}&=&-e^2(W^-_\mu W^{+\mu}A_\nu
A^\nu-W^-_\mu
A^\mu W^+_\nu A^\nu)\\
%\end{equation}
%\begin{equation}
\label{eq5.82}
\lagr_{WWWW}&=&\frac{1}{2}g^2(W^-_\mu
W^{-\mu}W^+_\nu W^{+\nu}-
W^-_\mu W^{+\mu}W^-_\nu W^{+\nu})\\
%\end{equation}
%\begin{equation}
\label{eq5.83}
\lagr_{WWZZ}&=&-g^2\cos ^2\theta_W(W^-_\mu
W^{+\mu}Z_\nu
Z^\nu-W^-_\mu Z^\mu W^+_\nu Z^\nu)\\
%\end{equation}
%\begin{equation}
\label{eq5.84}
\lagr_{WWZ\gamma}&=&g^2\sin \theta_W\cos \theta_W
(-2W^-_\mu W^{+\mu}A_\nu Z^\nu+W^-_\mu Z^\mu W^+_\nu A^\nu
\nonumber\\
~&~&+W^-_\mu A^\mu W^+_\nu Z^\nu)
\end{eqnarray}
Note that in the electromagnetic interactions of $W$ bosons, i.e.
in (\ref{eq5.79}) and (\ref{eq5.81}), we have used the unification
condition $e=g\sin\theta_W$ (see (\ref{eq5.34})). One may observe
that the triple coupling (\ref{eq5.79}) corresponds to the value
$\kappa$ = 1 in the Lagrangian (\ref{eq3.42}) discussed in Chapter~\ref{chap3}. In other words, the $SU(2)\times U(1)$ electroweak gauge model
predicts automatically a very specific {\it non-minimal\/}
electromagnetic interaction of the $W$ bosons.

In general, the coupling constants of vector boson interactions\index{coupling
constants!of vector bosons}
appearing in (\ref{eq5.79}) through (\ref{eq5.84}) can obviously
be expressed in terms of the $e$ and $g$ when the unification
condition is employed. For example,
\begin{equation}
\label{eq5.85}
g_{WWZ}=g\cos \theta_W=\sqrt{g^2-e^2}
\end{equation}
etc. This is again an explicit illustration of the characteristic
feature of the $SU(2)\times U(1)$ electroweak unification,
mentioned earlier in this chapter: the new interactions stemming
from the Yang--Mills construction involve coupling constants that
are non-trivial functions of the parameters of the old theory of
weak interactions and electromagnetism.

\begin{figure}[h]
\centering
\s{\includegraphics{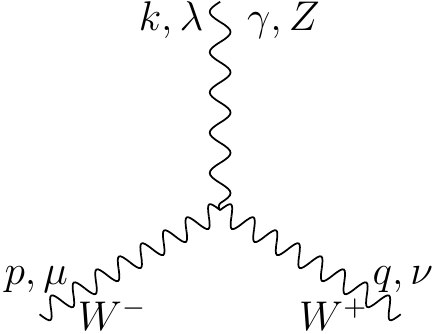}}
\caption{Feynman-graph vertex for the triple vector boson
interactions $WW\gamma$ or $WWZ$.}
\label{fig17}\index{Feynman diagrams!for
vertex $WW\gamma$}\index{Feynman diagrams!for
vertex $WWZ$}
\end{figure}

In quantum theory (i.e. at the level of Feynman diagrams) the
interaction Lagrangians (\ref{eq5.79}) or (\ref{eq5.80}) resp.
lead to the vertex shown in Fig.\,\ref{fig17}.
The relevant Feynman rule is given by the function\index{trilinear
(triple) coupling of vector bosons}\index{WWgamma@$WW\gamma$
interaction}\index{WWZ@$WWZ$ interaction}
\begin{equation}
\label{eq5.86}
V_{\lambda\mu\nu}(k,p,q)=(k-p)_\nu
g_{\lambda\mu}+(p-q)_\lambda g_{\mu\nu}+(q-k)_\mu g_{\lambda\nu}
\end{equation}
(multiplied by an appropriate coupling constant). The function
(\ref{eq5.86}) is obviously invariant under cyclic permutations
\begin{equation}
\label{eq5.87}
V_{\lambda\mu\nu}(k,p,q)=V_{\mu\nu\lambda}(p,q,k)=
V_{\nu\lambda\mu}(q,k,p)
\end{equation}
and satisfies also a highly useful relation
\begin{equation}
\label{eq5.88}
p^\mu
V_{\lambda\mu\nu}(k,p,q)=(k^2g_{\lambda\nu}-k_\lambda k_\nu)-(q^2
g_{\lambda\nu}-q_\lambda q_\nu)
\end{equation}
(called the 't Hooft identity\index{t Hooft identity@'t Hooft
identity}, cf. (\ref{eq3.47})). A proof of the relation
(\ref{eq5.88}) (which is valid for any four-momenta satisfying $k
+ p + q = 0$) is left to the reader as an easy exercise. The form
(\ref{eq5.88}) can be used in Feynman diagrams for different
configurations of the outgoing and incoming particles -- one has
to remember that an incoming $W^\pm$ line is equivalent to the
outgoing $W^\mp$ line carrying opposite momentum. As for the
Feynman rules for the quartic interactions\index{quartic
coupling}, these can be read off rather easily from the
Lagrangians (\ref{eq5.81}) -- (\ref{eq5.84}); some illustrations
will be provided in the subsequent calculations.

The Yang--Mills structure of the vector boson interactions and the
non-trivial relations among the relevant coupling parameters have
dramatic consequences for the cancellation of  high-energy
divergences\index{high-energy divergences|ff} in tree-level scattering amplitudes for various
electroweak processes. Some examples of such \qq{gauge cancellations}
are discussed in the next section.
%\end{document}

%\input{skript38} %             5.8
%%%%%%%%%%%%%%%%%%%%%%%%%%%%%%%%%%%%%%%%%%%%%%%%%%%%%%%%%%%%%%%%%%%
%%%%%%%%%%%%%%%%%%%%%%%%%%%%%%%%%%%%%%%%%%%%%%%%%%%%%%%%%%%%%%%%%%%%%%%%%%%%%%%%%%%%%%%%%%%%%%%%%%%%%%%%%%%%%%%%%%%%%%%%%%%%%%%%%%%%%%%%
%\documentstyle[12pt]{article}
%\newcommand\skrt[1]{#1\!\!\!\! / }  % pro psan¡ ¨krtlch p¡smenek,
                                  % nap©. \skrt{p}
%\newcommand\lagr{{\cal L}}

%\begin{document}
\section{Cancellation of leading divergences}\index{leading divergences|(}\label{sec5.8}

\begin{figure}[h]\centering
\begin{tabular}{cc}
\subfigure[]{\s{\includegraphics{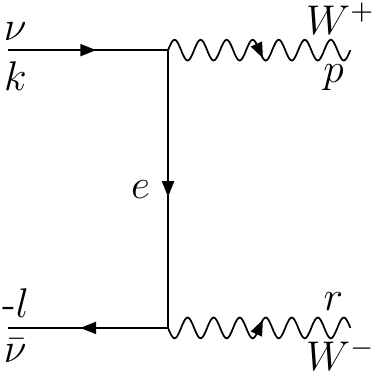}}}&\hspace{1.5cm}\subfigure[]{\s{\includegraphics{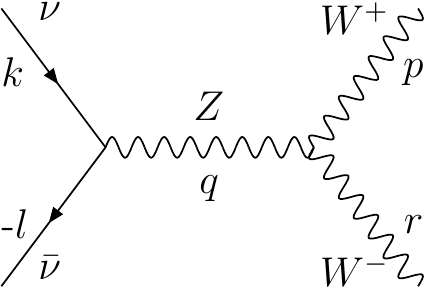}}}
\end{tabular}
\caption{Tree-level diagrams contributing to the process
$\nu\bar\nu\rightarrow W^+ W^-$ within the  $SU(2)\times U(1)$
gauge model of electroweak interactions. No further graphs arise
in the  full standard electroweak model if the neutrino mass is neglected.}
\label{fig18}\index{Feynman diagrams!for
$\nu\bar\nu\rightarrow W^+W^-$}
\end{figure}

To begin with, let us consider the process $\nu\bar\nu\rightarrow W^+
W^-$, which we have already mentioned briefly in Chapter~\ref{chap3}. Within
our $SU(2)\times U(1)$ electroweak model this is described, at the
tree level, by the two Feynman graphs shown in Fig.\,\ref{fig18}.
We will examine
the case of longitudinally polarized $W$ bosons, where one can expect
the worst high-energy behaviour\index{tree unitarity}. The matrix element corresponding to
the diagram (a) can then be written, according to the results of
Chapter~\ref{chap3}, as
\begin{equation}
\label{eq5.89}
{\cal M}^{(e)}_{\nu\bar\nu}=-\frac{g^2}{4m^2_W}\bar
v(l)\slashed{p} (1-\gamma_5)u(k)+O(1)
\end{equation}
The quadratic divergence occurring in (\ref{eq5.89}) for
$E\rightarrow\infty$ is embodied in the first term. Now we are
going to show that this divergence is cancelled by a corresponding
contribution coming from the diagram (b).

According to the standard Feynman rules the matrix element for
the graph (b) can be written (for arbitrary polarizations of
the external $W$ bosons) as

\begin{align}
\label{eq5.90}
{\cal M}^{(Z)}_{\nu\bar\nu}=\ &-\frac{1}{4}\frac{g}
{\cos \theta_W} \cdot g \cos\theta_W\bar v(l)\gamma_\rho
(1-\gamma_5)u(k)\notag\\
&\times\frac{-g^{\rho\nu}+m^{-2}_Zq^\rho q^\nu}{q^2-m^2_Z}
V_{\nu\mu\lambda}(q,r,p)\varepsilon^{\ast\mu}(r)
\varepsilon^{\ast\lambda}(p)
\end{align}
where we have employed the earlier results for the neutral-current
$\nu\nu Z$ vertex (see (\ref{eq5.49})) and the Yang--Mills $WWZ$
vertex (see (\ref{eq5.80}) and (\ref{eq5.86})). By naive
power counting one might expect that the leading high-energy
divergence associated with this graph could be more severe than
that occurring in (a), owing to the extra factor $m^{-2}_Z$ in
the longitudinal part of the $Z$ boson propagator. However, using
the cyclicity property (\ref{eq5.87}) and the 't Hooft identity
(\ref{eq5.88}) for the Yang--Mills vertex, along with the familiar
properties of the polarization vectors\index{polarization!vector}, it is not difficult to
show that the potentially dangerous contribution proportional to
$m^{-2}_Z$ vanishes exactly, for any combination of the external
polarizations (the proof is left to the reader as a simple
exercise). For longitudinally polarized external $W$ bosons one
then gets, using the usual high-energy decomposition of the
polarization vectors,
\begin{equation}
\label{eq5.91}
{\cal M}^{(Z)}_{\nu\bar\nu}=\frac{1}{4}g^2\bar
v(l)\gamma^\nu
(1-\gamma_5)u(k)\frac{1}{q^2-m^2_Z}V_{\nu\mu\lambda}(q,r,p)
\frac{r^\mu}{m_W}\frac{p^\lambda}{m_W}+O(1)
\end{equation}
where we have singled out explicitly the diverging part of the
whole contribution -- now it is obvious that we are left with only
a quadratic divergence, similarly to the graph (a). Employing
once more the 't Hooft identity, as well as the equations of
motion for the Dirac spinors, the expression (\ref{eq5.91}) can be
finally recast, after some simple algebraic manipulations, as
\begin{equation}
\label{eq5.92}
{\cal M}^{(Z)}_{\nu\bar\nu}=\frac{g^2}{4m^2_W}\bar
v(l)\slashed{p} (1-\gamma_5)u(k)+O(1)
\end{equation}
Comparing this with (\ref{eq5.89}), one can see that for the
considered process we have indeed achieved the desired
compensation of the original high-energy divergence lurking in the
old $W$ boson weak interaction model -- it was the gauge structure
of the electroweak $SU(2)\times U(1)$ theory, manifested in the
combination of the two relevant graphs, which played an important
role in the cancellation mechanism. Let us add that the above
treatment remains unaltered even within the full GWS standard
electroweak model, at least if the neutrino is taken to be
massless. In fact, there are several other processes of similar
type (involving a fermion pair along with a pair of vector
bosons), where complete cancellation of the high-energy
divergences is achieved already at the level of the
$SU(2) \times U(1)$ gauge theory (i.e.without invoking the Higgs
mechanism\index{Higgs mechanism} of the full GWS standard model); finding some relevant
examples is left as a challenge for the reader.

\begin{figure}[h]\centering
\begin{tabular}{ccc}
\subfigure[]{\s{\includegraphics{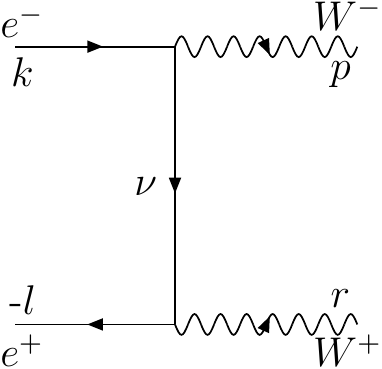}}}&\hspace{0.2cm}\subfigure[]{\s{\includegraphics{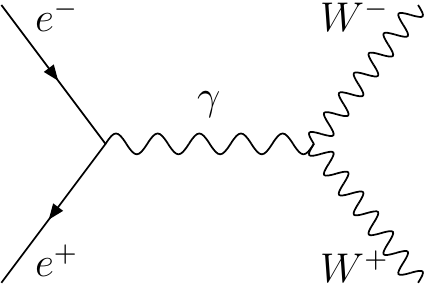}}}&\hspace{0.2cm}\subfigure[]{\s{\includegraphics{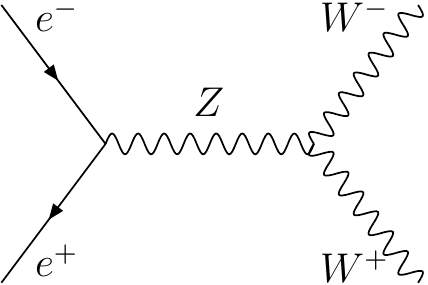}}}
\end{tabular}
\caption{Tree-level diagrams for the process
$e^+e^-\rightarrow W^+ W^-$ in the $SU(2)\times U(1)$ electroweak
theory, including the t-channel neutrino exchange and s-channel
photon or $Z$ boson exchange.}
\label{fig19}\index{Feynman diagrams!for
$e^+e^-\rightarrow W^+W^-$}
\end{figure}

Next, let us turn to the process $e^+ e^-\rightarrow W^+W^-$ for the
longitudinally polarized external $W$ bosons. Within our
$SU(2) \times U(1)$ electroweak theory there are now three
Feynman graphs that contribute at the tree level, namely those
depicted in Fig.\,\ref{fig19}.
The graphs (a) and (b) were discussed in
Chapter~\ref{chap3} and the results can be written as
\begin{equation}
\label{eq5.93}
{\cal M}^{(\nu)}_{e^+e^-}=-\frac{g^2}{4m^2_W}\bar
v(l)\slashed{p} (1-\gamma_5)u(k)+O(\frac{m_e}{m^2_W}E)+O(1)
\end{equation}
for the weak contribution (i.e. the  neutrino exchange (a)) and
\begin{equation}
\label{eq5.94}
{\cal M}^{(\gamma)}_{e^+e^-}=\frac{e^2}{m^2_W}\bar
v(l)\slashed{p} u(k)+O(1)
\end{equation}
for the electromagnetic contribution (b) -- let us stress that
the last expression corresponds to the $WW\gamma$ vertex of the
Yang--Mills type (cf. (\ref{eq5.79})). The calculation of the
$Z$-exchange graph (c) proceeds essentially along the same lines
as in the neutrino-antineutrino case described earlier. Again, the
longitudinal part of the $Z$ propagator does not contribute, and
the final result can be written as
\begin{align}
{\cal
M}^{(Z)}_{e^+e^-}=&-\frac{1}{2m^2_W}g\cos
\theta_W\frac{g}{\cos \theta_W}(-\frac{1}{2}
+\sin^2\theta_W)\bar v(l)\slashed{p}(1-\gamma_5)u(k)\notag\\
&-\frac{1}{2m^2_W}g\cos \theta_W\frac{g}{\cos \theta_W}
\sin ^2\theta_W\bar v(l)\slashed{p}(1+\gamma_5)u(k)\notag\\
&+O(\frac{m_e}{m^2_W} E)+O(1)\label{eq5.95}
\end{align}
where we have singled out a term involving the leading (quadratic)
high-energy divergences and used the neutral current
parametrization (\ref{eq5.49}). An explicit form of the
non-leading (linear) divergences can be found in \cite{Hor}. We
will ignore these terms for the moment, but we will return to them
in the next chapter. Adding now the expressions (\ref{eq5.93}),
(\ref{eq5.94}) and (\ref{eq5.95}) and using the familiar relation
$\sin\theta_W= e / g$, it is seen that the quadratic divergences
indeed cancel, but -- in contrast to the preceding example -- a
residual linear divergence still persists in the sum of the graphs
(a), (b) and (c). Thus, in the present case the electroweak
gauge structure alone cannot ensure a complete cancellation of the
high-energy divergences -- one may observe that, technically, this
is related to the non-vanishing electron mass. Nevertheless, as
before, the gauge couplings do control the $leading$ divergences
-- this is the most important lesson to be learnt from the present
example.

\begin{figure}[h]\centering
\begin{tabular}{ccc}
\subfigure[]{\s{\includegraphics{figs/fig12a}}}&\hspace{0.7cm}\subfigure[]{\s{\includegraphics{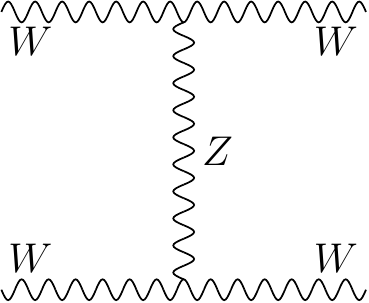}}}&\hspace{0.7cm}\subfigure[]{\s{\includegraphics{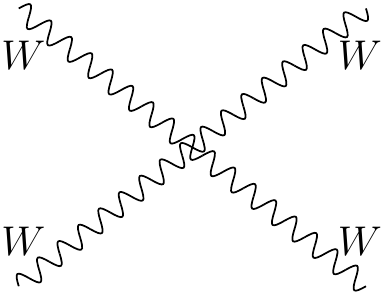}}}
\end{tabular}
\caption{Tree-level Feynman graphs for the process
$WW\rightarrow WW$, including the photon and Z boson exchange and
the direct coupling of four $W$ bosons.}
\label{fig20}\index{Feynman diagrams!for $WW\rightarrow WW$}
\end{figure}

Last but not least, let us reconsider the process of $WW$ scattering,
which we have been able to describe only via photon exchange in the
naive $W$ boson model. Within the $SU(2)\times U(1)$ electroweak
gauge theory there are three types of tree-level diagrams representing
such a process, shown in Fig.\,\ref{fig20}
(of course, for graphs (a) and (b) one must also take into account the relevant crossing of the
external lines).
We will consider again the case of longitudinally polarized external
$W$ bosons. For the photon exchange contribution we have obtained
earlier (cf. Chapter~\ref{chap3}) the result
\begin{equation}
\label{eq5.96} {\cal
M}^{(\gamma)}_{WW}=\frac{e^2}{4m^4_W}(t^2+u^2-2s^2)+ O(E^2)+O(1)
\end{equation}
(where $s,~t,~u$ are the standard Mandelstam variables), which
exhibits the quartic high-energy divergence expected on dimensional
grounds. As for the $Z$ boson exchange, the longitudinal part of the
propagator does not contribute, owing to the Yang--Mills structure
of the $WWZ$ vertex. This in turn means that the worst divergent
behaviour to be expected for this graph is the same as for the
photon exchange. Further, the $Z$ mass cannot play any role in the
coefficient of the leading divergence, so one may readily write
for the sum of the graphs (a) and (b)
\begin{equation}
\label{eq5.97} {\cal
M}^{(Z,\gamma)}_{WW}=(e^2+g^2_{WWZ})\frac{1}{4m^4_W}
(t^2+u^2-2s^2)+O(E^2)+O(1)
\end{equation}
Using the relations $e=g\sin \theta_W$ and $g_{WWZ}=
g\cos\theta_W$ (see (\ref{eq5.34}), (\ref{eq5.80})), the coupling
factor in (\ref{eq5.97}) is reduced to
\begin{equation}
\label{eq5.98} e^2+g^2_{WWZ}=g^2
\end{equation}
The contribution of the diagram (c) can be calculated in a
straightforward way from the interaction term (\ref{eq5.82}); for
the leading divergences one gets, after some simple algebraic
manipulations
\begin{equation}
\label{eq5.99} {\cal
M}^{(direct)}_{WW}=-g^2\frac{1}{4m^4_W}(t^2+u^2)+g^2
\frac{1}{2m^4_W}s^2+O(E^2)+O(1)
\end{equation}
Let us remark that the non-leading terms (quadratic in energy)
have, in general, rather complicated form for the individual
diagrams; similarly to the
preceding example, we relegate their treatment to the next
chapter. From (\ref{eq5.97}), (\ref{eq5.98}) and (\ref{eq5.99}) it
is clear that the quartic divergences are cancelled in the sum of
the three considered graphs -- again, the Yang--Mills structure of
the vector boson sector (manifested in the interplay of the
three-boson and four-boson couplings) is responsible for a
\qq{miraculous} cancellation of the leading high-energy divergences.

The examples discussed in this section illustrate nicely some
remarkable technical consequences of the non-Abelian gauge
invariance in the theory of electroweak unification. It turns out
that within such a theory the scattering amplitudes involving
massive vector bosons have much softer high-energy behaviour than
one might naively guess on simple dimensional grounds; in particular,
the leading high-energy divergences are cancelled owing to the gauge
structure of the relevant interactions. After such cancellations,
there are still some residual non-leading divergences and their
ultimate elimination within the full GWS standard model is related
to the subtle issue of  the mass generation in gauge theories,
which is a subject of the next chapter\index{leading divergences|)}.
%\end{document}

%\input{problems5}
%%%%%%%%%%%%%%%%%%%%%%%%%%%%%%%%%%%%%%%%%%%%%%%%%%%%%%%%%%%%%%%%%%%
%%%%%%%%%%%%%%%%%%%%%%%%%%%%%%%%%%%%%%%%%%%%%%%%%%%%%%%%%%%%%%%%%%%%%%%%%%%%%%%%%%%%%%%%%%%%%%%%%%%%%%%%%%%%%%%%%%%%%%%%%%%%%%%%%%%%%%%%
\begin{priklady}{11}

\item  What is the effective four-fermion Lagrangian describing low-energy neutral current interactions?

\item \label{pro52} Calculate the partial decay width $\Gamma(Z\rightarrow \ell^+ \ell^-)$ for
unpolarized $Z$ and $\ell^\pm$. First, set $m_\ell=0$ for the sake
of simplicity (of course, this is expected to be a very good
approximation, since $m_\ell^2\ll m_Z^2$ for any
$\ell=e,\mu,\tau$). How is the result changed (numerically) when
the effects of $m_\ell \neq 0$ are taken into account? In
particular, make such a comparison for the heaviest known lepton,
the $\tau$\index{tau lepton@$\tau$ lepton}. Next, calculate the
decay width $\Gamma(Z\rightarrow \nu_\ell
\bar{\nu}_\ell)$\index{decay!of the Z@of the $Z$ boson}.

\item Evaluate longitudinal polarization\index{polarization!of the electron} of a lepton $\ell$ produced in the decay of an unpolarized $Z$ boson at rest. The degree of polarization in question is defined as
$$
P = \frac{w_R - w_L}{w_R+w_L}
$$
with $w_L$ and $w_R$ denoting the probability of the production of left-handed and right-handed lepton respectively. First of all, set $m_\ell = 0$ for simplicity. Actually, in such a massless case, the outcome can be guessed quite easily. An astute expert should then anticipate the result (to be verified by an explicit calculation)
$$
P = \frac{\varepsilon_R^2 - \varepsilon_L^2}{\varepsilon_R^2 + \varepsilon_L^2} = - \frac{2 v a}{v^2 + a^2}
$$
where the lepton coupling factors are given by (\ref{eq5.58}) (these are independent of the lepton species).
For $m_\ell \neq 0$, the calculation is algebraically more complicated (and its result cannot be guessed so easily). Anyway, any hard-working reader is encouraged to derive the formula
$$
P = - \frac{2 v a\sqrt{1- \frac{4 m_\ell^2}{m_Z^2}}}{(v^2+a^2)\bigl(1-\frac{m_\ell^2}{m_Z^2}\bigr) + 3 (v^2-a^2)\frac{m_\ell^2}{m_Z^2}}
$$

\item  \label{pro5.4}Calculate the angular distribution of electrons produced in
decays of a polarized Z boson at rest (again, work in the
approximation $m_\ell=0$). As a follow-up, evaluate the up-down
asymmetry for electrons produced in decays of a $Z$ with spin
\qq{up} (i.e. directed along positive $z$ axis). The asymmetry in
question is defined as $(\Gamma_+-\Gamma_-)/(\Gamma_+ +
\Gamma_-)$, where the $\Gamma_+$ denotes the angular distribution
integrated over the upper hemisphere (with the azimuthal angle
$\vartheta$ lying between $0$ and $\frac{\pi}{2}$), and the
$\Gamma_-$ has an analogous meaning with respect to the lower
hemisphere ($\vartheta \in (0,\, \frac{\pi}{2})$).

\item  Consider the process $e^+e^- \rightarrow \mu^+ \mu^-$ in the c.m. system and
at a sufficiently high energy ($E_{c.m.}\gg m_\mu$), so that the
lepton masses can be safely neglected. In lowest order, the
relevant matrix element can be written as
$\mathcal{M}_\gamma+\mathcal{M}_Z$, with $\mathcal{M}_\gamma$ and
$\mathcal{M}_Z$ corresponding to the exchange of photon and $Z$
boson respectively. Let us denote the cross sections obtained from
$|\mathcal{M}_\gamma|^2$ and $|\mathcal{M}_Z|^2$ as
$\sigma_\gamma$ and $\sigma_Z$. What is the numerical value of the
ratio $\sigma_Z/\sigma_\gamma$ for $E_{c.m.}= 1\ \GeV,\ 20\ \GeV,\
200\ \GeV$? Next, calculate the full cross section involving both
$\gamma$ and $Z$ exchange. For the above-mentioned energies,
determine a relative magnitude of the interference term
$\sigma_{\gamma Z}$, descending from
$|\mathcal{M}_\gamma+\mathcal{M}_Z|^2 = |\mathcal{M}_\gamma|^2 +
|\mathcal{M}_Z|^2 + \mathcal{M}_\gamma \mathcal{M}_Z^* +
\mathcal{M}^*_\gamma \mathcal{M}_Z$.

\item  \label{pro5.6}For the process $e^+e^-\rightarrow \mu^+\mu^-$ evaluate also the
forward-backward\index{forward-backward asymmetry} (or front-back)
asymmetry $A_{FB}$ of the cross section. The $A_{FB}$ is defined
in close analogy with the quantity considered in the Problem \ref{pro5.4},
namely
$$
A_{FB}=\frac{\sigma_F - \sigma_B}{\sigma_F + \sigma_B}
$$
where the $\sigma_F$ and $\sigma_B$ are cross sections obtained by
integrating over the front and back hemisphere respectively, i.e.
over the azimuthal angle $\vartheta$ lying in the interval
$(0,\,\frac{\pi}{2})$ and $(\frac{\pi}{2},\,\pi)$ resp. (note that
the $\vartheta$ is conventionally chosen as the angle between the
momentum of $\mu^-$ and that of the incident electron). Is a
non-vanishing value of the $A_{FB}$ related to the $\mathcal{C}$
or $\mathcal{P}$ violation in the weak neutral current
interaction?

\item  Show that the tree-level amplitude for $e^+ e^- \rightarrow
Z_L \gamma$ behaves as $O(1)$ in the high-energy limit. What kind
of asymptotic behaviour one gets for the process $e^+e^-
\rightarrow Z_L Z_L$?

\item  Compute the limiting value of the cross section
$\sigma(\nu_\mu e \rightarrow \nu_\mu e)$ for $s\rightarrow
\infty$. Neglect lepton masses throughout the calculation.

\item Consider the annihilation process $e^- e^+ \to \nu_\mu \bar{\nu}_\mu$. At the tree level, evaluate its cross section as a function of the collision energy $E_{c.m.} = s^{1/2}$, in the kinematic region $m_e \ll E_{c.m.} \ll m_Z$. Throughout the calculation, neglect everything that may be safely neglected. Further, find the limit of the relevant cross section for $s \to \infty$.

\item Calculate cross sections for processes $\nu_e e \rightarrow
\nu_e e$ and $\bar{\nu}_e e \rightarrow \bar{\nu}_e e$ in a low
energy domain $m_e^2 \ll s \ll m_W^2$, taking into account both CC
and NC contributions. Compare the results with those obtained
within the old IVB model involving the $W^\pm$ only. Examine these
processes also in high energy region, where $s \gg m_Z^2$. What is
the asymptotic value of the cross section ratio
$\sigma(\bar{\nu}_e e \rightarrow \bar{\nu}_e e)/\sigma(\nu_e e
\rightarrow \nu_e e)$?

\item An instructive illustration of the ``miraculous'' cancellations of high-energy divergences within SM is provided by the process $W^- W^+ \to Z \gamma$. Show that its amplitude satisfies the condition of tree-level unitarity. 
\\{\it Hint:} Any attentive reader may guess immediately that the cancellation in question is due to the interplay of the $WW\gamma$, $WWZ$ and $WWZ\gamma$ Yang-Mills couplings shown in (\ref{eq5.79}), (\ref{eq5.80}) and (\ref{eq5.84}).

\item As another example of the divergence cancellation mechanism due to the Yang-Mills structure of the vector boson sector of SM, one may consider the reaction $W_L W_L \to Z_L Z_L$. Demonstrate a compensation of the leading divergences occurring in contributions of the relevant tree-level diagrams. Clearly, for this purpose one has to invoke just the couplings $WWZ$ and $WWZZ$ shown in (\ref{eq5.80}) and (\ref{eq5.83}). Note that in this case some residual quadratic divergences persist, which are ultimately eliminated by means of the exchange of the Higgs scalar boson (to be introduced in the next chapter).

\end{priklady}

%\input{skript41} %kapitola 6
%%%%%%%%%%%%%%%%%%%%%%%%%%%%%%%%%%%%%%%%%%%%%%%%%%%%%%%%%%%%%%%%%%%
%%%%%%%%%%%%%%%%%%%%%%%%%%%%%%%%%%%%%%%%%%%%%%%%%%%%%%%%%%%%%%%%%%%%%%%%%%%%%%%%%%%%%%%%%%%%%%%%%%%%%%%%%%%%%%%%%%%%%%%%%%%%%%%%%%%%%%%%
%\documentstyle[12pt]{article}
%\newcommand\skrt[1]{#1\!\!\!\! / }  % pro psan¡ ¨krtlch p¡smenek,
                                  % nap©. \skrt{p}
%\newcommand\lagr{{\cal L}}

%\begin{document}
\chapter{Higgs mechanism for masses}\label{chap6}\index{Higgs mechanism|(}
%\section{4 Higgs mechanism for masses}
\section{Residual divergences: need for scalar
bosons}\index{residual divergences|(}\label{sec6.1}
%\subsection{4.1 Motivation for scalar bosons}

The bulk of the present chapter is devoted to the so-called Higgs
mechanism. This is a tool for generating particle masses in gauge
theories through specific interactions involving scalar fields,
without spoiling perturbative renormalizability\index{perturbative
renormalizability}. Historically, the first field theory models
exhibiting such a mechanism \cite{ref46} were developed
independently of the program of electroweak unification -- only a
few years after its discovery, the magic Higgs trick has been
applied successfully by Weinberg and Salam to the Glashow's
$SU(2)\times U(1)$ gauge model of weak and electromagnetic
interactions. In most of the current textbooks on particle theory,
the Higgs mechanism is usually introduced immediately when
formulating the Standard Model. However, the corresponding
construction might seem, at first sight, somewhat bizarre to an
uninitiated reader and, subsequently, the beginner in the field
could wonder whether the electroweak SM {\it must\/} indeed be built
precisely as it is -- in particular, whether the Higgs scalars are
necessary or not. Of course, since 2012 we know that a spin-0 particle, which resembles closely the Higgs scalar, indeed exists (though we still cannot be quite sure that the observed scalar is just {\it the\/} Higgs boson of SM), and this may dispel possible doubts of a skeptical reader. Nevertheless, it may be instructive to explain the role of a scalar boson in the electroweak theory independently of the current experimental data. So, we will describe the idea of the Higgs construction (and its realization within the GWS standard model) later in this chapter, and in this section we will start by
showing first a rather straightforward motivation for a scalar
boson in the electroweak gauge theory, in connection with the
issue of divergence cancellations investigated in the preceding
chapter. Of course, such a preliminary discussion cannot provide
us with a detailed insight into the subtle aspects of the Higgs
mechanism, but will at least indicate that a scalar boson is a
{\it necessary\/} ingredient for achieving the tree-level
unitarity (which in turn is necessary for renormalizability) in a
gauge model incorporating mass terms for vector bosons and
fermions. Moreover, it will become clear that the relevant scalar
boson couplings must be intimately related to the particle masses.

\begin{figure}[h]\centering
\begin{tabular}{cc}
\subfigure[]{\s{\includegraphics{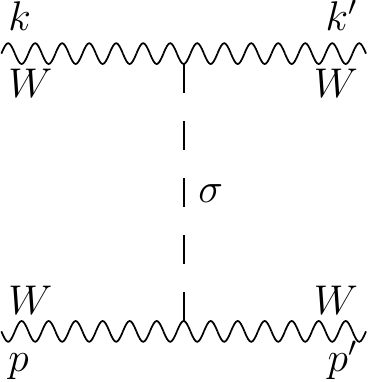}}}&\hspace{1.5cm}\subfigure[]{\s{\includegraphics{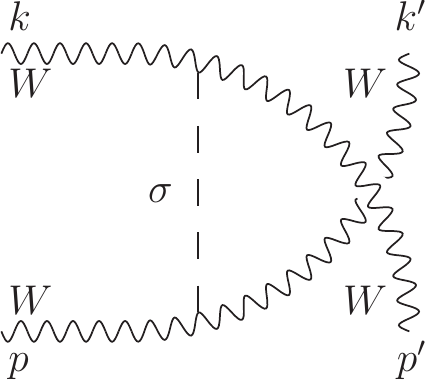}}}
\end{tabular}
\caption{Neutral scalar exchange graphs for the $WW$ scattering.}
\label{fig21}\index{Feynman diagrams!for $WW\rightarrow
WW$!neutral scalar exchange}
\end{figure}

First, let us come back to the $W_L W_L$ scattering process. In
Section~\ref{sec5.8} we have found that the leading (quartic) high-energy
divergences cancel in the sum of the diagrams shown in
Fig.\,\ref{fig20}. A detailed (somewhat tedious) calculation
reveals that the remaining quadratically divergent contribution
has a remarkably simple form
\begin{equation}
\label{eq6.1}
{\cal M}^{(\gamma)}_{WW}+{\cal M}^{(Z)}_{WW}+{\cal
M}^{(direct)} _{WW}=-g^2\frac{s}{4m^2_W}+O(1)
\end{equation}
({\it A technical remark:\/} Along with the quartic divergences,
some ugly-looking $O(E^2)$ terms from the individual graphs are
cancelled as well and one is thus happily left with the result
(\ref{eq6.1}); details of the calculation can be found in the
Appendix J of the book \cite{Hor}.) Since the coupling factor
occurring in this expression is definitely non-zero, there is
obviously no way how the divergent term in (\ref{eq6.1}) could be
eliminated without introducing a new particle and a corresponding
new interaction. The crucial observation is that the quadratic
divergence in (\ref{eq6.1}) can be cancelled in a most natural way
by means of an additional diagram involving the exchange of a
scalar boson (in fact one can hardly imagine any other option that
would be feasible). An interaction of a pair of the $W$'s\index{W
boson@$W$ boson} with a single neutral scalar field $\sigma$ has
an essentially unique form if it is required to be of
renormalizable\index{renormalizable theory} type (i.e. have
dimension not greater than four), namely
\begin{equation}
\label{eq6.2}
\lagr_{WW\sigma}=g_{WW\sigma}W^-_\mu W^{+\mu}\sigma
\end{equation}
The relevant coupling constant\index{coupling constants!of Higgs
sector} then obviously has a dimension of mass. The
$\sigma$-exchange diagrams contributing to the $WW$ scattering in
the lowest order are shown in Fig.\,\ref{fig21}.
The corresponding matrix element can be written as
\begin{eqnarray}
\label{eq6.3}
i{\cal M}^{(\sigma)}_{WW}&=&i^3
g^2_{WW\sigma}\varepsilon^\alpha(k)\varepsilon^\ast_\alpha(k')
\frac{1}{(k-k')^2-m^2_\sigma}\varepsilon^\beta (p)\varepsilon
^\ast_\beta(p')\nonumber\\
&+&\{k'\leftrightarrow p'\}
\end{eqnarray}
The leading divergence\index{leading divergences} associated with
the graphs of Fig.\,\ref{fig21}  for longitudinally polarized
$W$'s is obtained easily from the last expression by replacing the
$\varepsilon^\alpha_L(k)$ by $k^\alpha/m_W$, etc. After some
simple algebraic manipulations, one thus gets
\begin{equation}
\label{eq6.4}
{\cal
M}^{(\sigma)}_{WW}=-g^2_{WW\sigma}\frac{1}{4 m^4_W}(t+u)+O(1)
\end{equation}
As we could have anticipated on simple dimensional grounds
(keeping in mind the dimensionality of the coupling constant
$g_{WW\sigma}$), the high-energy divergence embodied in
(\ref{eq6.3}) is indeed quadratic. Using now the kinematical
identity $s+t+u=4m^2_W$, the result (\ref{eq6.3}) can be recast as
\begin{equation}
\label{eq6.5}
{\cal
M}^{(\sigma)}_{WW}=g^2_{WW\sigma}\frac{s}{4m^4_W}+O(1)
\end{equation}
It is obvious that the divergent terms in (\ref{eq6.1}) and
(\ref{eq6.5}) cancel each other if and only if
\begin{equation}
\label{eq6.6}
g_{WW\sigma}=gm_W
\end{equation}
We thus see that the extra interaction of $W$ bosons with a
neutral scalar field $\sigma$, introduced in a rather {\it ad
hoc\/} way in the context of the $WW$ scattering process, does
provide a remedy for the residual non-leading divergence
(\ref{eq6.1}). At the same time, the result (\ref{eq6.6}) reveals
a remarkable connection of such a \qq{compensating} coupling with
the $W$ boson mass.

\begin{figure}[h]\centering
\s{\includegraphics{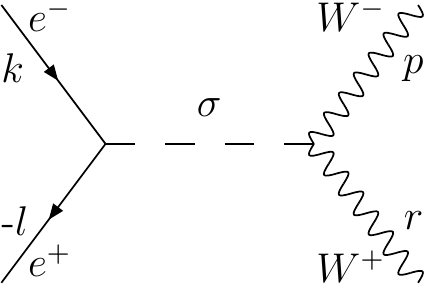}} \caption{Scalar boson
exchange contribution to the process $e^+e^-\rightarrow W^+W^-$.}
\label{fig22}\index{Feynman diagrams!for $e^+e^-\rightarrow
W^+W^-$!neutral scalar exchange}
\end{figure}

Let us work out one more example displaying the above-mentioned
feature, i.e. a link between a primordial mass term and a scalar
field coupling necessary for achieving the tree-level unitarity
within the electroweak gauge theory. The case we have in mind is
the process $e^+e^-\rightarrow W^+W^-$. In Section~\ref{sec5.8} we have
observed that  the sum of the three diagrams in Fig.\,\ref{fig19}
is already free of the leading quadratic divergences, but a linear
divergence may still persist. Indeed, an explicit calculation
yields the result
\begin{equation}
\label{eq6.7} {\cal M}^{(\nu)}_{e^+e^-}+{\cal
M}^{(\gamma)}_{e^+e^-}+ {\cal
M}^{(Z)}_{e^+e^-}=-\frac{g^2}{4m^2_W}m_e\bar v(l)u(k)+O(1)
\end{equation}
The formula (\ref{eq6.7}) exhibits the residual (linear)
high-energy divergence, which obviously cannot be eliminated
without an additional diagram. As before, the scalar boson
exchange offers a possible way out. Of course, we will try to
utilize the $\sigma$ field introduced in the previous example. We
already know the precise form of the $WW\sigma$ coupling, so one
only has to add an interaction of the scalar with leptons. The
simple matrix structure of the linearly divergent term in
(\ref{eq6.7}) makes it obvious that one has to postulate a Yukawa
coupling\index{Yukawa coupling}
\begin{equation}
\label{eq6.8}
\lagr_{ee\sigma}=g_{ee\sigma}\bar e e\sigma
\end{equation}
The $\sigma$-exchange graph designed to cancel the divergent
behaviour of (\ref{eq6.7}) is shown in Fig.\,\ref{fig22}.
The matrix element corresponding to Fig.\,\ref{fig22}  reads
\begin{equation}
\label{eq6.9} i{\cal M}^{(\sigma)}_{e^+ e^-}=i^3gm_W
g_{ee\sigma}\bar v(l)u(k)
\frac{1}{q^2-m^2_\sigma}\varepsilon^\ast_\mu(p)\varepsilon^
{\ast\mu}(r)
\end{equation}
(note that we have already taken into account the result
(\ref{eq6.6})). On simple dimensional grounds, one may guess that
the matrix element (\ref{eq6.9}) will be at most linearly
divergent for longitudinal external $W$'s. An explicit expression
is easily calculated; proceeding in the usual way, one gets
\begin{equation}
\label{eq6.10} {\cal M}^{(\sigma)}_{e^+
e^-}=-\frac{1}{2m^2_W}g_{ee\sigma}gm_W \bar v(l)u(k)+O(1)
\end{equation}
Comparing now (\ref{eq6.10}) with (\ref{eq6.7}), it is obvious
that the divergent parts are cancelled if and only if
\begin{equation}
\label{eq6.11}
g_{ee\sigma}=-\frac{g}{2}\frac{m_e}{m_W}
\end{equation}
Thus, we have identified another $\sigma$ boson coupling that
implements successfully the desired divergence cancellation;
similarly to the previous case, the corresponding coupling
constant is proportional to a bare mass. Recall that masses
introduced into the electroweak Lagrangian simply by hand break
the $SU(2)\times U(1)$ gauge symmetry, so one may also say that
the interactions of the scalar $\sigma$ field compensate the
effects of the symmetry-breaking terms in the electroweak
Lagrangian.

The examples discussed above show that the mass terms incorporated
in the electroweak Lagrangian must be tightly correlated with
couplings of a newly postulated neutral scalar boson, if one wants
to accomplish the delicate divergence cancellations necessary for
perturbative renormalizability. In fact, this heuristic discussion
seems to offer an important clue for building renormalizable
electroweak\index{renormalizable theory} models: instead of
introducing the phenomenologically needed mass terms directly, one
should perhaps try {\it to generate masses through appropriate
interactions involving scalar fields\/}. Such a vague statement
can indeed be given a more precise meaning within some particular
field theory models, developed (though in a slightly different
context) in the early 1960s. These field-theoretic constructions
will be described in the following sections and finally employed
in completing the construction of the full standard model of
electroweak interactions\index{residual divergences|)}.
%\end{document}

%\input{kniha62}
%%%%%%%%%%%%%%%%%%%%%%%%%%%%%%%%%%%%%%%%%%%%%%%%%%%%%%%%%%%%%%%%%%%
%%%%%%%%%%%%%%%%%%%%%%%%%%%%%%%%%%%%%%%%%%%%%%%%%%%%%%%%%%%%%%%%%%%%%%%%%%%%%%%%%%%%%%%%%%%%%%%%%%%%%%%%%%%%%%%%%%%%%%%%%%%%%%%%%%%%%%%%
\section{Goldstone model}\index{Goldstone!model|ff}\label{sec6.2}

\begin{figure}[h]\centering
\s{\includegraphics{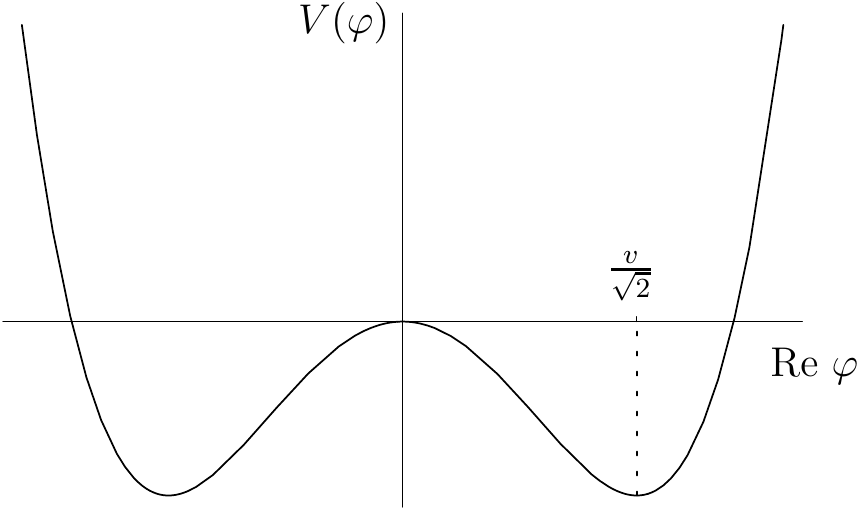}}
\caption{A visualization of the Goldstone potential\index{Goldstone!potential} given by
(\ref{eq6.13}). In fact, the full picture would consist of a surface
formed by rotating this curve around the ordinate axis.}
\label{fig23}
\end{figure}

A basic ingredient of the Higgs mechanism is the so-called
\qq{Goldstone phenomenon}, associated with \qq{spontaneous
symmetry breakdown}\index{spontaneous symmetry breakdown|(}. We
shall start our discussion with a simple model of classical scalar
field theory (invented originally by J.~Goldstone \cite{ref42})
that illustrates these concepts. The model we have in mind is
described by the Lagrangian density of the type
\begin{equation}
\label{eq6.12}
\lagr=\partial_\mu\varphi\partial^\mu\varphi^\ast-V(\varphi)
\end{equation}
with
\begin{equation}
\label{eq6.13}
V(\varphi)=-\mu^2\varphi\varphi^\ast+\lambda(\varphi\varphi^\ast)
^2
\end{equation}
where $\varphi$ is a complex scalar field, the $\mu$ is a real
parameter with dimension of mass and $\lambda$ is a
(dimensionless) coupling constant.\footnote{We assume $\lambda >
0$ in order that the energy density corresponding to
(\ref{eq6.12}) be bounded from below.} In what follows, the
function (\ref{eq6.13}) will sometimes be called the
\qq{potential} (though, of course, it has nothing to do with the
potential of a classical force). It is depicted schematically in
Fig.~\ref{fig23}.
The essential feature of the considered Lagrangian is the
\qq{wrong sign} of the mass term in (\ref{eq6.13}). Indeed,
discarding temporarily the $\lambda\varphi^4$ term, one is left
with a quadratic form that leads to the equation of motion $(\Box
-\mu^2)\varphi=0$, i.e. to the Klein--Gordon
equation\index{Klein--Gordon equation} with reversed sign of mass
squared. Thus, it is not immediately clear how the classical model
described by (\ref{eq6.12}), (\ref{eq6.13}) should be quantized --
the quadratic part of the Lagrangian cannot be simply interpreted
in terms of free particles and a straightforward perturbative
treatment thus becomes inapplicable.

In order to guess a plausible interpretation of the considered
model, it is instructive to calculate the corresponding energy
(Hamiltonian) density\index{Hamiltonian density|(}. This is given
by the component ${\mathscr T}_{00}$ of the canonical
energy-momentum tensor\index{energy-momentum tensor}, namely
\begin{equation}
\label{eq6.14} {\mathscr H}={\mathscr
T}_{00}=\frac{\partial\lagr}{\partial(\partial_0
\varphi)}\partial_0\varphi+\frac{\partial\lagr}{\partial(\partial
_0\varphi^\ast)}\partial_0\varphi^\ast-\lagr
\end{equation}
(of course, we take $\varphi$ and $\varphi^\ast$ as independent
dynamical variables). Using (\ref{eq6.12}) in (\ref{eq6.14}) one
thus gets, after some simple manipulations
\begin{equation}
\label{eq6.15} {\mathscr
H}=\partial_0\varphi\partial_0\varphi^\ast+\vec{\nabla}
\varphi\vec{\nabla}\varphi^\ast+V(\varphi)
\end{equation}
One may now ask what is the field configuration $\varphi(x)$
corresponding to a {\it minimum\/} of the energy density.
Obviously, the derivative terms in (\ref{eq6.15}) always give a
positive contribution for a $\varphi$ that is not a space-time
constant. Thus, one should consider a constant $\varphi$ and find
a minimum of the potential~$V$. This can be done easily. The $V$
in fact depends only on one real variable $\rho$ defined as
$\rho^2=\varphi\varphi^\ast$ so that instead of (\ref{eq6.13}) one
could write
\begin{equation}
\label{eq6.16}
V(\rho)=-\mu^2 \rho^2+\lambda \rho^4
\end{equation}
The first derivative $V'(\rho)$ vanishes for $\rho$ = 0
and for $\rho^2=\mu^2/2\lambda$. The value $\rho= 0$
corresponds to a local maximum, while for
$\rho=\mu/\sqrt {2\lambda}$ one has an absolute minimum of the
$V$. In terms of the original variable $\varphi$ it means that the
minimum of the energy density corresponds to a one-parametric set
of constant values
\begin{equation}
\label{eq6.17} \varphi_0=\frac{v}{\sqrt 2}\text{e}^{i\alpha}
\end{equation}
\noindent where $\alpha$ is an arbitrary real number and we have
denoted\footnote{The symbol $v$ introduced in (\ref{eq6.18})
stands for \qq{vacuum}\index{vacuum|(} -- this refers to the fact
that the value $|\varphi|=v/\sqrt 2$ corresponds to the ground
state of the considered field system. The term \qq{vacuum} for the
ground state would be more appropriate at quantum level, but such
a loose terminology is quite customary even in the context of
classical field theory. In a systematic quantum treatment of this
problem one is led to the notion of \qq{effective potential} (see
e.g. \cite{Hua}) and the $v$ then represents a non-zero vacuum
expectation value\index{vacuum expectation value} of the quantum
scalar field. The relation (\ref{eq6.18}) fixes a notation that
has become standard for electroweak theories involving Higgs
mechanism.}
\begin{equation}
\label{eq6.18}
v=\frac{\mu}{\sqrt\lambda}
\end{equation}
In other words, the $\varphi_0$ values that minimize the energy
density lie on a circle in the complex plane with radius $v/\sqrt
2$ and the energy minimum is thus infinitely (continuously)
degenerate. Such a finding, namely the observation that the ground
state of the considered system is described by a non-zero constant
field, leads to the following simple idea: instead of the
$\varphi$, one should perhaps use its deviation from the
\qq{vacuum value} (\ref{eq6.17}) as a true dynamical variable.
Indeed, it seems to be more promising to study small oscillations
around a stable ground state with $|\varphi|=v/\sqrt 2$, rather
than take as a reference point the value $\varphi$ = 0
corresponding to an unstable state. This idea can be implemented
mathematically in a rather elegant way if the original Lagrangian
(\ref{eq6.12}) is first rewritten in terms of radial and angular
field variables defined by
\begin{equation}
\label{eq6.19}
\varphi(x)=\rho(x)\exp\left(i\frac{\pi(x)}{v}\right)
\end{equation}
(note that we have introduced the factor of $1/v$ in the exponent
in order to get the angular field $\pi(x)$ with the right dimension
of mass). Using (\ref{eq6.19}) in (\ref{eq6.12}) one gets easily
\begin{equation}
\label{eq6.20}
\lagr=\partial_\mu\rho\partial^\mu\rho+\frac{1}{v^2}\rho^2\partial
_\mu\pi\partial^\mu\pi-V(\rho)
\end{equation}
For further discussion it is now also useful to recast the
potential in a slightly different form; in particular, from
(\ref{eq6.16}) one gets immediately
\begin{eqnarray}
\label{eq6.21}
V(\rho)&=&\lambda\Bigl[(\rho^2-\frac{\mu^2}{2\lambda})^2-(\frac
{\mu^2}{2\lambda})^2\Bigr]\nonumber\\
&=&\lambda(\rho^2-\frac{v^2}{2})^2-\frac{1}{4}\lambda v^4
\end{eqnarray}
Obviously, the additive constant appearing in the last line can be
dropped without changing anything essential -- only the energy
density thus becomes automatically non-negative. In what follows
we shall therefore replace the Lagrangian (\ref{eq6.20}) by the
equivalent form
\begin{equation}
\label{eq6.22}
\lagr=\partial_\mu\rho\partial^\mu\rho+\frac{1}{v^2}\rho^2\partial_\mu
\pi\partial^\mu\pi-\lambda(\rho^2-\frac{v^2}{2})^2
\end{equation}
(of course, we could have started with such a positively definite
potential from the very beginning, but we find the expression
(\ref{eq6.13}) to be a more natural starting point since the
\qq{wrong-sign mass term} is explicitly singled out there). Let us
now perform the shift of the field variable suggested above. The
$\rho$ may be written as
\begin{equation}
\label{eq6.23}
\rho=\frac{1}{\sqrt 2}(\sigma+v)
\end{equation}
where the variable $\sigma$ involves the rescaling factor of
$1/\sqrt 2$ so as to get a proper normalization of its kinetic
term. Using (\ref{eq6.23}) in (\ref{eq6.22}), one gets easily
\begin{eqnarray}
\label{eq6.24}
\lagr&=&\frac{1}{2}\partial_\mu\sigma\partial^\mu\sigma+
\frac{1}{2}\partial_\mu\pi\partial^\mu\pi-\frac{1}{4}\lambda
(\sigma^2+2v\sigma)^2\nonumber\\
&+&\frac{1}{2v^2}\sigma^2\partial_\mu\pi\partial^\mu\pi+
\frac{1}{v}\sigma\partial_\mu\pi\partial^\mu\pi
\end{eqnarray}
that is
\begin{equation}
\label{eq6.25}
\lagr=\frac{1}{2}\partial_\mu\sigma\partial^\mu\sigma+
\frac{1}{2}\partial_\mu\pi\partial^\mu\pi
-\lambda v^2\sigma^2+\text{interactions}
\end{equation}
where all terms higher than quadratic have been generically
denoted as \qq{interactions}. The important point is that the
$\sigma$ field now has a mass term with the {\it right sign},
while the $\pi$ came out to be {\it massless}. In particular, the
$\sigma$ mass value that can be read off from (\ref{eq6.25}) is
given by $\frac{1}{2}m^2_\sigma = \lambda v^2$, i.e. $m^2_\sigma =
2\mu^2$ in view of (\ref{eq6.18}). In fact, the appearance of a
mass term with correct sign should not be surprising. Our
redefinition (shift) of the radial field variable actually means
that we perform Taylor expansion around a local minimum of the
potential, where its second derivative is of course positive.
However, this second derivative determines the coefficient of the
term quadratic in the relevant field, which is precisely the mass
term in the Lagrangian.

Thus, in the above simple exercise we have seen that the model
(\ref{eq6.12}) describes in fact two real scalar fields $\sigma$ and
$\pi$, where
\begin{equation}
\label{eq6.26}
m_\sigma=\mu\sqrt 2\;\;\;,\;\;\;m_\pi=0
\end{equation}
although such an interpretation is completely hidden in the
original form of the Lagrangian written in terms of the variables
$\varphi$ and $\varphi^\ast$. A most remarkable feature of the
considered model is the appearance of the massless field $\pi$,
 since this provides an illustration of the so-called Goldstone
phenomenon alluded to earlier in this section. To explain this
point, we have to make a brief digression here and recall first
some important general concepts concerning the problem of
symmetry breaking in field-theory models of particle physics.

A familiar manifestation of an (approximate) internal symmetry of
such a model is the existence of multiplets of particles with
(nearly) degenerate masses. The multiplets correspond to
irreducible representations of the relevant symmetry group and
become truly degenerate in the limit of exact symmetry, while the
observed deviations from degeneracy within multiplets are
attributed to small symmetry-violating terms in the Hamiltonian --
in this context, the term \qq{explicit symmetry breaking} is used
(a good example of such an approximate symmetry is the isospin in
strong interaction\index{strong interaction} physics). This
pattern corresponds to what is usually called the {\bf
Wigner--Weyl realization}\index{Wigner--Weyl realization} of
symmetry (cf. e.g. \cite{Mar}); another well-known aspect of a
symmetry realized in the Wigner--Weyl mode is the existence of
certain selection rules for transition matrix elements with
respect to the relevant quantum numbers (an illustration of this
is in fact provided by the calculation of pion beta decay in
Chapter~\ref{chap2}). To put it briefly, in the Wigner--Weyl mode the
physical states transform according to the symmetry group
representations; in particular, the vacuum can be taken as
invariant.

On the other hand, there is a radically different possibility for
a symmetry realization, which corresponds to the case of an
invariant Hamiltonian or Lagrangian possessing non-invariant
ground state (vacuum). Such a mode is indeed relevant for a wide
variety of systems with infinite number of degrees of freedom,
both relativistic and non-relativistic.\footnote{Of course, for a
system with finite number of degrees of freedom one can also have
the ground state that does not share a symmetry of the
corresponding Hamiltonian, but in such a case this has no further
dramatic consequences.} This scheme means that the symmetry is not
realized on physical states in the usual way and the structure of
quasi-degenerate multiplets as well as the selection rules typical
for the Wigner--Weyl realization are completely lost. In current
parlance, the term {\bf spontaneous symmetry breaking}
\cite{ref43} is usually used for such a situation (\qq{symmetry
breaking} because the symmetry is no longer manifest in the
physical spectrum and \qq{spontaneous} because one may imagine
that the system occupies spontaneously a non-invariant
lowest-energy state, e.g. under the influence of an arbitrarily
small asymmetric perturbation that picks a particular ground
state). One may note that such a term is slightly deceptive as the
symmetry is in fact only {\it hidden} -- it is still present at
the level of the Hamiltonian or Lagrangian (cf. e.g. \cite{Col}).
The most important aspect of spontaneous symmetry breakdown is
that it has a generic signature described by the celebrated {\bf
Goldstone theorem}\index{Goldstone!theorem} \cite{ref44} (see also
e.g. the textbook \cite{Wei}): If the symmetry of the considered
Hamiltonian\index{Hamiltonian density|)} or Lagrangian is
continuous, the non-invariance of its ground state (which is then
necessarily continuously degenerate) implies the existence of a
{\it massless bosonic excitation} (Goldstone
boson\index{Goldstone!boson|(}) in the physical spectrum of the
system. In particular, in the context of relativistic quantum
field theory, the Goldstone boson is a spin-zero massless particle
(its spinless nature is related to the requirement of Lorentz
invariance\index{Lorentz!invariance} of the vacuum state, but it
can be both scalar and pseudoscalar). A familiar example of an
(approximate) Goldstone boson in particle physics is the
pion\index{pion}: many experimental facts in low-energy
hadron\index{hadrons} phenomenology are naturally explained in
terms of an effective theory where pions $\pi^\pm,\pi^0$ are
massless in the limit of exact chiral symmetry\index{chiral
symmetry} $SU(2)_L\times SU(2)_R$ of the strong interaction
Lagrangian; their masses as observed in the real world are assumed
to be due to an additional explicit symmetry breaking.
Historically, it was probably Y.~Nambu \cite{ref45} who came up
first with this idea (for more details, see also e.g. \cite{Wei},
\cite{ChL} or \cite{Geo}). To close this general digression, the
last terminological remark is perhaps in order here. For reasons
that should be obvious from the above discussion, the term {\bf
Nambu--Goldstone realization}\index{Nambu--Goldstone realization}
of a symmetry (or simply Goldstone realization) is also frequently
used instead of \qq{spontaneous symmetry breakdown} (in fact, it
is even more appropriate), but the latter name has certainly
become more popular in the present-day particle physics.

Now it is easy to see how the scalar field model discussed before
illustrates the Goldstone phenomenon associated with spontaneous
symmetry breakdown. The Lagrangian (\ref{eq6.12}) is invariant
under global phase transformations
\begin{eqnarray}
\label{eq6.27}
\varphi'(x)&=&\text{e}^{i\omega}\varphi(x)\nonumber\\
\varphi^{*\prime}(x)&=&\text{e}^{-i\omega}\varphi^\ast(x)
\end{eqnarray}
where $\omega$ is a constant parameter (an arbitrary real number,
independent of $x$). In other words, the symmetry group of our
model is $U(1)$\index{U(1) group@$U(1)$ group} (which is
isomorphic to $O(2)$\index{O(2) group@$O(2)$ group} -- the
rotation group of two-dimensional plane). The ground state
$\varphi_0$ shown in (\ref{eq6.17}) is obviously not invariant
under such transformations (by applying (\ref{eq6.27}) one moves
around the circle in the complex plane corresponding to
(\ref{eq6.17})) and the ground-state energy is thus continuously
degenerate. The massless field $\pi$ then may be understood as
corresponding to a Goldstone boson (note, however, that we are
staying at the {\it classical\/} level!). It would be a highly
non-trivial task to reformulate this simple Goldstone model for
quantum fields, but the manipulations that led to (\ref{eq6.24})
and (\ref{eq6.26}) are nevertheless quite instructive -- a
difficult part of the discussion has been done for classical
fields, with a result that is in accordance with the general
Goldstone theorem. The Lagrangian (\ref{eq6.24}) can then be
quantized in the usual perturbative way. Notice that this of
course retains the original symmetry of (\ref{eq6.12}), but in
terms of the variables $\rho$ and $\pi$ the transformation law
(\ref{eq6.27}) is recast as
\begin{eqnarray}
\label{eq6.28}
\sigma'(x)&=&\sigma(x)\nonumber\\
\pi'(x)&=&\pi(x)+v\omega
\end{eqnarray}
It is also easy to guess how one can get, in the present context,
a classical picture of a Wigner--Weyl
realization\index{Wigner--Weyl realization} of symmetry. Clearly,
this would correspond to the Lagrangian of the type
(\ref{eq6.12}), with the opposite sign of $\mu^2$ in the
potential, i.e. with $V(\varphi)$ =
$\mu^2\varphi\varphi^\ast+\lambda(\varphi \varphi^\ast)^2$ instead
of (\ref{eq6.13}). In such a case, the ground state is unique and
corresponds to $\varphi$ = 0. The model then can be interpreted
e.g. as a system of two real scalar fields $\varphi_1$ and
$\varphi_2$ (with $\varphi=\varphi_1+i\varphi_2)$ corresponding to
particles with equal mass $\mu$, i.e. with the $O(2)$ symmetry
manifested directly in the particle spectrum.

As we noted before, the discussion of the simple Goldstone model
carried out here is only a necessary prerequisite for the
formulation of the mass-generation mechanism in the electroweak
theory. Actually, the physics of massless scalar bosons is not of
primary interest to us. The truly important thing, from our point
of view, happens when the interaction with an Abelian gauge
field\index{Abelian gauge field} is switched on in the Lagrangian
(\ref{eq6.12}). As we shall see in the next section, the magic
Higgs trick then works, which means that the Goldstone boson
becomes unphysical and one gets a mass term for the vector field.

%\input{kniha63}
%%%%%%%%%%%%%%%%%%%%%%%%%%%%%%%%%%%%%%%%%%%%%%%%%%%%%%%%%%%%%%%%%%%
%%%%%%%%%%%%%%%%%%%%%%%%%%%%%%%%%%%%%%%%%%%%%%%%%%%%%%%%%%%%%%%%%%%%%%%%%%%%%%%%%%%%%%%%%%%%%%%%%%%%%%%%%%%%%%%%%%%%%%%%%%%%%%%%%%%%%%%%
\section{Abelian Higgs model}\label{sec6.3}
\index{Abelian Higgs model|(} Let us now introduce, following P.
Higgs \cite{ref46}, the interaction with an Abelian gauge
field\index{Abelian gauge field} into the Goldstone model
considered in preceding section. As we know from Chapter~\ref{chap4}, there
is a standard way of doing that: ordinary derivatives in the
kinetic term in (\ref{eq6.12}) are replaced by the covariant ones
and the usual kinetic term for the gauge field is added. One thus
gets, formally, the scalar electrodynamics incorporating also
quartic self-coupling\index{quartic coupling} of the complex
scalar field and its mass term with the wrong sign. The
corresponding Lagrangian can be written as
\begin{align}
\lagr_{Higgs}= & -\frac{1}{4}F_{\mu\nu}F^{\mu\nu}+(\partial_\mu
-igA_\mu)\varphi(\partial^\mu+igA^\mu)\varphi^\ast
\notag\\
& -\lambda(\varphi\varphi^\ast-\frac{v^2}{2})^2 \label{eq6.29}
\end{align}
where, of course, $F_{\mu\nu}=\partial_\mu A_\nu -\partial_\nu
A_\mu$ and $g$ denotes the gauge coupling constant. Note that in
(\ref{eq6.29}) we have used the form (\ref{eq6.22}) for the
Goldstone potential\index{Goldstone!potential}, with
$v=\mu/\sqrt\lambda$ (see (\ref{eq6.18})). By construction, the
Lagrangian (\ref{eq6.29}) is invariant under local gauge
transformations\index{local symmetry}
\begin{eqnarray}
\label{eq6.30}
\varphi '(x)&=&\text{e}^{i\omega(x)}\varphi(x)\nonumber\\
\varphi^{\ast\prime} (x)&=&\text{e}^{-i\omega(x)}\varphi^\ast(x)\nonumber\\
A'_\mu(x)&=&A_\mu(x)+\frac{1}{g}\partial_\mu\omega(x)
\end{eqnarray}
In analogy with the discussion of previous section, one may now
trade the $\varphi$ and $\varphi^\ast$ for the corresponding radial
and angular variables, and shift the radial field according to
(\ref{eq6.23}); in other words, the complex field $\varphi$ is
reparametrized as
\begin{equation}
\label{eq6.31}
\varphi(x)=\rho(x)\exp\left(i\frac{\pi(x)}{v}\right)
\end{equation}
In terms of the variables $\rho$ and $\pi$ the gauge transformations
(\ref{eq6.30}) are recast as
\begin{eqnarray}
\label{eq6.32}
\rho'(x)&=&\rho(x)\nonumber\\
\pi' (x)&=&\pi(x)+v\omega(x)\nonumber\\
A'_\mu(x)&=&A_\mu(x)+\frac{1}{g}\partial_\mu\omega(x)
\end{eqnarray}
The gauge invariance means that a field configuration described by
some functions $\rho(x) , \pi(x)$ and $A_\mu(x)$ (solutions of the
corresponding equations of motion) is equivalent to the set
$\rho'(x) , \pi'(x) , A'_\mu(x)$ obtained by the transformation
(\ref{eq6.32}) (the equivalence is to be understood in the sense
that any physical quantity can be calculated either from $\rho,
\pi , A_\mu$ or from $\rho' , \pi' , A'_\mu$, with the same
result). In particular, for a given set $\rho , \pi, A_\mu$ one
can choose $\omega = -\pi/v$ in (\ref{eq6.32}) and eliminate thus
completely the angular field variable $\pi$; in other words, the
original field configuration is equivalent, up to a gauge
transformation, to that described by
\begin{eqnarray}
\label{eq6.33}
\rho'(x)&=&\rho(x)\nonumber\\
\pi'(x)&=&0\nonumber\\
A'_\mu(x)&=&A_\mu(x)-\frac{1}{gv}\partial_\mu\pi(x)
\end{eqnarray}
One can thus also say that -- owing to the local gauge invariance
-- the angular field $\pi$ (i.e. the erstwhile Goldstone boson)
becomes unphysical within the Higgs model, since it can be
eliminated by an appropriate choice of gauge. The gauge fixed by
the condition (\ref{eq6.33}), i.e. by the requirement $\pi\equiv
0$, is usually called unitary gauge
($U$-gauge\index{U-gauge@$U$-gauge|(}).\footnote{The adjective
\qq{unitary} may seem totally obscure at the present moment, but
this in fact refers to the envisaged quantum version of the
considered model: it is well known that, in general, the
$S$-matrix unitarity\index{S matrix unitarity@$S$-matrix
unitarity} becomes transparent in a theory that does not involve
any auxiliary unphysical fields. The label \qq{physical gauge}
(which would be perhaps most appropriate in the present context)
is also sometimes used, but the term \qq{unitary gauge} has become
standard in modern electroweak theories.} Now it is clear that the
equations of motion for the $U$-gauge dynamical variables can be
obtained directly from the Lagrangian (\ref{eq6.29}) where one
fixes the gauge by setting simply $\pi = 0$, i.e. $\varphi =
\varphi^\ast = \rho$. (The reason is obvious: the constraint $\pi
= 0$ is implemented via a special gauge transformation and the
original Lagrangian (\ref{eq6.29}) is gauge invariant.) Further,
in full analogy with our previous analysis of the Goldstone model,
the radial field $\rho$ should be shifted as
\begin{equation}
\label{eq6.34}
\rho=\frac{1}{\sqrt 2}(\sigma+v)
\end{equation}
(see (\ref{eq6.23})), in order to get rid of the wrong-sign scalar
mass term (obviously, the resulting mass term for $\sigma$ must be
the same as in the case of the Goldstone model since it is fully
determined by the scalar field potential $V$). Thus, the $U$-gauge
Higgs Lagrangian (\ref{eq6.29}) can be written as
\begin{align}
\lagr^{(U)}_{Higgs}=& -\frac{1}{4}G_{\mu\nu}G^{\mu\nu}+\frac{1}{2}
(\partial_\mu-igB_\mu)(\sigma+v)(\partial^\mu+igB^\mu)(\sigma+v)
\notag\\
& -\frac{1}{4}\lambda[(\sigma+v)^2-v^2]^2 \label{eq6.35}
\end{align}
where we have introduced, for definiteness, an extra symbol
$B_\mu$ for the $U$-gauge value of the vector field (cf.
(\ref{eq6.33})), and $G_{\mu\nu} = \partial_\mu B_\nu
-\partial_\nu B_\mu$. The form (\ref{eq6.35}) can be easily worked
out as
\begin{align}
\lagr^{(U)}_{Higgs}=& -\frac{1}{4}G_{\mu\nu}G^{\mu\nu}-\frac{1}{4}
\lambda(\sigma^2+2v\sigma)^2\notag\\
& +\frac{1}{2}(\partial_\mu\sigma-ig\sigma B_\mu -igvB_\mu)
(\partial^\mu\sigma+ig\sigma B^\mu+igvB^\mu)\notag\\
=& -\frac{1}{4}G_{\mu\nu}G^{\mu\nu}+\frac{1}{2}\partial_\mu
\sigma\partial^\mu\sigma-\frac{1}{4}\lambda(\sigma^2+2v\sigma)^2
\notag\\
& +\frac{1}{2}g^2(\sigma+v)^2B_\mu B^\mu \label{eq6.36}
\end{align}
Separating now in the last expression its quadratic part and the
interaction terms, one has
\begin{align}
\lagr^{(U)}_{Higgs}&= \frac{1}{2}\partial_\mu\sigma\partial^\mu
\sigma-\lambda v^2\sigma^2-\frac{1}{4}G_{\mu\nu}G^{\mu\nu}+
\frac{1}{2}g^2v^2B_\mu B^\mu\notag\\
&+\,g^2v\sigma B_\mu B^\mu+\frac{1}{2}g^2\sigma^2B_\mu B^\mu-
\lambda v\sigma^3-\frac{1}{4}\lambda\sigma^4 \label{eq6.37}
\end{align}
As expected, there is a mass term of the field $\sigma$ that
coincides with (\ref{eq6.25}), but the truly remarkable feature of
the expression (\ref{eq6.37}) is the presence of a {\bf mass term
for the vector field $\boldsymbol{B_\mu}$}. Although there was no
such thing in the original form (\ref{eq6.29}), eventually it has
shown up as a consequence of the scalar field shift
(\ref{eq6.34}). This, in fact, is the essence of the famous
\qq{Higgs mechanism} or \qq{Higgs trick}, demonstrated here (at
the classical level) within the simplest Abelian theory. {\bf When
the spontaneously broken symmetry of a scalar-field model is
gauged, the original Goldstone boson disappears from physical
spectrum and the gauge field acquires a mass.} (In a common
physical \qq{folklore} this situation is sometimes characterized
by saying that the would-be Goldstone boson is eaten by the gauge
boson, which becomes heavy.) At the same time, a massive scalar
field survives in the physical spectrum (we shall call $\sigma$
the Higgs field). Obviously, the whole mechanism is triggered by
the wrong-sign scalar mass term in the original symmetric
Lagrangian (\ref{eq6.29}). The $B_\mu$ mass can be easily read off
from (\ref{eq6.37}); this is
\begin{equation}
\label{eq6.38}
m_B=gv
\end{equation}
The reader should notice the natural and easy-to-remember
structure of the last formula: the induced vector-field mass is a
product of the generic mass scale $v$ (the scalar field vacuum
value characteristic for spontaneous symmetry breaking) and the
gauge interaction strength $g$.
 As for the interaction part of (\ref{eq6.37}), it basically exhibits
a pattern that will be recovered later on within the electroweak
standard model. In particular, using (\ref{eq6.38}), one can see
that the strength of the trilinear coupling $\sigma BB$ is
proportional to the $B_\mu$ mass, namely
\begin{equation}
\label{eq6.39}
g_{\sigma BB}=gm_B
\end{equation}
\index{coupling constants!of Higgs sector}For the quadrilinear
coupling\index{quadrilinear vertex} $\sigma\sigma BB$ one has
$g_{\sigma \sigma BB}=\frac{1}{2}g^2$ and the coupling constants
for the cubic and quartic self-interactions of the Higgs field
$\sigma$ can be easily expressed e.g. in terms of the $m_\sigma ,
m_B$ and $g$ (needless to say, values of the $\sigma$
self-couplings are the same as in the Goldstone model).

  The $U$-gauge Higgs model can be quantized in a straightforward way
and the $B_\mu$ propagator\index{propagator!in the $U$- and
$R$-gauge} then has the canonical form
\begin{equation}
\label{eq6.40}
D^{(U)}_{\mu\nu}(k)=\frac{-g_{\mu\nu}+m^{-2}_Bk_\mu k_\nu}
{k^2-m^2_B}
\end{equation}
This, in combination with the well-known behaviour of the
longitudinal polarization vector\index{polarization!vector} for a
massive spin-1 particle (cf. (\ref{eq3.29})) may lead to
power-like growth of some tree-level Feynman diagrams in the
high-energy limit. However, while such divergences indeed occur
for individual diagrams, they get cancelled when all relevant
contributions to a given physical process are summed. In fact,
what one observes here is a simplified variant of the mechanism
suggested in Section~\ref{sec6.1}. As an instructive exercise, the reader
is recommended to verify such a divergence cancellation e.g. for
the process $BB\rightarrow\sigma\sigma$. ({\it Hint\/}: both
$\sigma BB$ and $\sigma\sigma BB$ couplings enter the game in this
case.) As we know, the soft high-energy behaviour at the tree
level (\qq{tree unitarity}\index{tree unitarity}) suggests that
the theory might be renormalizable\index{renormalizable
theory|ff}. In the present case it is indeed so; quite generally,
renormalizability of a gauge theory with the Higgs
mechanism\footnote{The alternative term \qq{spontaneously broken
gauge theory} is also frequently used in this context.} was proved
first by G. 't Hooft and M.~Veltman \cite{ref47} and nowadays this
topic is covered by most of the modern textbooks on quantum field
theory.
  For a general proof of renormalizability of a spontaneously broken
gauge theory a different quantization procedure is used, namely
the so-called $R$-gauge\index{R-gauge@$R$-gauge|(} formulation
(invented originally by 't Hooft \cite{ref48}). In contrast to the
$U$-gauge, the $R$-gauge propagator of a massive vector
boson\index{massive vector boson|ff} behaves for
$k^2\rightarrow\infty$ as $1/k^2$ (i.e. in the same way as in the
massless case). This in turn means that the convergence properties
of higher-order (closed-loop) Feynman diagrams become much better
than in the $U$-gauge and the usual power-counting analysis
indicates immediately a renormalizable behaviour (see e.g.
\cite{ChL} or the Appendix G in \cite{Hor}). The price to be paid
for that is the presence of the unphysical would-be Goldstone
boson, which is not eliminated explicitly by means of the gauge
choice -- instead, it is preserved as an auxiliary field variable.
Within such a quantization scheme, a proof of the $S$-matrix
unitarity is consequently more complicated as one has to prove
that the additional contributions of unphysical particles are
irrelevant. Moreover, since there is in fact a whole class of the
$R$-gauges, one has to demonstrate the gauge-independence of the
physical $S$-matrix (in particular, one has to prove an
equivalence of the $U$-gauge with any of the $R$-gauges). All this
has by now become \qq{common wisdom} in modern field theory and
the internal consistency of different formulations of a gauge
theory with the Higgs mechanism has been firmly established. We
are not going into further details here; some technicalities
concerning the $R$-gauges will be described in later sections,
within the framework of the full standard electroweak model. In
any case, one should bear in mind that the basic idea behind the
question of renormalizability of a gauge theory with Higgs
mechanism is in fact extremely simple: in such a theory one starts
with a massless gauge field, which {\it a priori\/} cannot produce
any non-renormalizable behaviour of Feynman graphs. The physically
interpretable Lagrangian is then obtained by means of a mere
redefinition of the relevant dynamical variables (along with an
appropriate gauge fixing) and one thus expects, intuitively, that
the convergence properties of the $S$-matrix remain basically the
same.

   However, it is also fair to stress the following point. Within
the Abelian model considered here it is actually not necessary to
invoke the Higgs mechanism for obtaining a renormalizable theory
with a massive vector boson -- one could as well introduce the
corresponding mass term into the scalar
QED\index{quantum!electrodynamics (QED)} Lagrangian simply by hand
without spoiling renormalizability (this is analogous to the case
of spinor QED with massive photon; an essential point is that in
both cases the Abelian gauge field is coupled to a conserved
current). The example of the Abelian Higgs model can serve as a
prototype for more complicated situations (involving non-Abelian
gauge symmetry) encountered within electroweak theory, where the
vector field mass term cannot be put in by hand with impunity, and
an appropriate variant of the Higgs trick becomes necessary.

  One more remark is perhaps in order at this place. The basic
feature of the Higgs model, namely the appearance of an
\qq{induced} vector-boson mass term and simultaneous disappearance
of a Goldstone boson, may intuitively be understood as a
transformation of the would-be Goldstone boson into the
zero-helicity\index{helicity} state of the vector boson (i.e. the
state corresponding to longitudinal
polarization\index{polarization!longitudinal}) -- such a state is
of course absent in the massless case. In this spirit, one can
say, somewhat loosely, that the total number of \qq{degrees of
freedom} is preserved throughout the Higgs mechanism: at the
beginning, there are two real scalar fields and two (transverse)
polarizations\index{transverse polarization} of a massless gauge
field, and we end up with one physical scalar and three
polarization states of a massive vector boson. It is interesting
that such a vague connection between the unphysical Goldstone
boson and physical longitudinal vector boson can be given a more
precise meaning within the $R$-gauge formulation of the Higgs
model. This is described by the famous \qq{equivalence
theorem}\index{equivalence theorem} \cite{ref49}, stating roughly
that in the high-energy limit an $S$-matrix element for
longitudinal vector bosons is asymptotically equal (possibly up to
a constant factor) to its unphysical counterpart involving the
corresponding would-be Goldstone scalars (the asymptotic region
here corresponds to energies much larger than the vector boson
mass). We will discuss this remarkable statement in more detail in
the context of the electroweak standard model.

  In closing this section, let us add a brief historical commentary.
In fact, the first hint of the Higgs mechanism appeared in the
context of non-relativistic condensed-matter physics \cite{ref50}.
Then it was discussed independently by several authors \cite{ref51,ref52,ref53}
within the framework of relativistic quantum field theory, without
resorting to explicit models of the type described above (a good
review of the non-perturbative QFT aspects of spontaneous symmetry
breaking and the Higgs phenomenon can be found e.g. in the article
\cite{Brn}). The explicit model \cite{ref46} was originally
conceived as a mere illustration of the rather abstract
field-theory concepts involved, but as we know today it had an
immense heuristic value. In particular, its straightforward
non-Abelian generalization \cite{ref54} was subsequently utilized
by S.~Weinberg \cite{ref39} and A.~Salam \cite{ref40} for building
the first potentially renormalizable unified theory of weak and
electromagnetic interactions. In the early 1970s the principles of
gauge symmetry\index{gauge invariance} and the Higgs mechanism
were widely accepted by particle theorists and this led to an
explosion of \qq{model building} following the Weinberg--Salam
paradigm (cf. e.g.
\cite{AbL})\index{R-gauge@$R$-gauge|)}\index{Abelian Higgs
model|)}.
Nevertheless, despite the technical attractiveness of the Higgs mechanism, many theorists were reluctant to accept the real existence of a physical elementary scalar boson as an ingredient of the electroweak gauge theory. In particular, the main controversial point consisted in distinguishing a more general „Higgs mechanism“ for generating gauge boson masses\footnote{For obvious reasons, the extended label Brout-Englert-Higgs (BEH) mechanism is quite frequently used in such a context. The point is that Brout and Englert in their celebrated original paper apparently did not care about the possible existence of a physical scalar boson that may (but need not) occur as a „by-product“ of the mass-generation mechanism for gauge fields.}) and the „Higgs boson“ that emerges within a specific model like \cite{ref46}, which involves  an elementary scalar field (for a detailed discussion of the subtle issue in question, see e.g. the nice instructive essay \cite{refWel}). To put it briefly, a standard statement reads that the Higgs mechanism and the Higgs boson are two different things. This long-standing dilemma was apparently resolved in 2012, when the observation of a Higgs-like boson was announced by two independent experimental collaborations (ATLAS and CMS) working at the Large Hadron Collider (LHC) at CERN (see \cite{ATLAS} and \cite{CMS} for the original discovery papers and \cite{ref5} for a comprehensive review of current data). So, we may continue with confidence towards constructing the full edifice of the electroweak SM.

%\input{kniha64}
%%%%%%%%%%%%%%%%%%%%%%%%%%%%%%%%%%%%%%%%%%%%%%%%%%%%%%%%%%%%%%%%%%%
%%%%%%%%%%%%%%%%%%%%%%%%%%%%%%%%%%%%%%%%%%%%%%%%%%%%%%%%%%%%%%%%%%%%%%%%%%%%%%%%%%%%%%%%%%%%%%%%%%%%%%%%%%%%%%%%%%%%%%%%%%%%%%%%%%%%%%%%
\section{Higgs mechanism for $SU(2)\times U(1)$ gauge theory}\label{sec6.4}

Before proceeding to the formulation of the Higgs mechanism that
operates within the standard electroweak theory, we will describe
some characteristic general features of non-Abelian extensions of
the field-theory models considered in preceding two sections. The
general statements we are going to specify below will not be
proved here as the corresponding proofs can be found in many other
books (see e.g. \cite{ChL}, \cite{Hua}, \cite{Rai}); rather we
will utilize the available general knowledge for motivating the
choice of the SM Higgs sector.

Let us start with the Goldstone-type models. In the example
discussed in Section~\ref{sec6.2}, a particular ground state (\qq{vacuum})
belonging to the set (\ref{eq6.17}) does not share the one-parametric
$U(1)$ symmetry of the Lagrangian (\ref{eq6.12}) and, as a result, one
Goldstone boson appears. It turns out (see \cite{ref54} for an
original paper) that such a pattern can be generalized as follows.
One may consider a model involving a multiplet of scalar fields,
with dynamics described by means of a Lagrangian of the type
(\ref{eq6.12}) possessing a continuous $n$-parametric internal symmetry
group $G$ and with a ground state (defined as a minimum of the
corresponding \qq{potential} $V(\varphi )$) that is less symmetric
than the Lagrangian. In particular, let us assume that the vacuum
state remains invariant under an $r$-parametric ($r<n$) subgroup
$H \subset G$. Then there are $n-r$ massless Goldstone bosons; in
other words, the number of Goldstone bosons is equal to the number
of broken symmetry generators. Needless to say, some massive
scalar bosons always appear as well -- their number depends on the
dimension of the original multiplet.

Next, let us see what happens when (a part of) the global
symmetry\index{global symmetry} of a general Goldstone-type model
with the symmetry-breaking pattern indicated above is made local,
that is, when a subgroup of the $G$ is gauged by introducing a set
of Yang--Mills fields\index{Yang--Mills field|ff} associated with
the corresponding generators. Let the total number of gauge fields
be $m$ ($m\leq n$) and suppose that $k$ of them are coupled to
broken symmetry generators, i.e. to those connected with Goldstone
bosons (of course, $k$ $\leq n-r$). Then it turns out \cite{ref54}
that upon shifting scalar fields by the relevant vacuum values one
gets $k$ massive vector fields and $k$ Goldstone bosons become
unphysical -- they can be eliminated by an appropriate choice of gauge (the $%
U$-gauge, analogous to that discussed earlier in the Abelian
case). The other $m-k$ gauge fields (coupled to unbroken
generators) remain massless. Note that the existence of the
physical $U$-gauge in a general case was proved in \cite{ref55}.

Thus, a general scheme of the non-Abelian generalization of the
Higgs mechanism that emerges from the preceding discussion is
quite elegant and easy to remember: within a gauged Goldstone-type
model, the Yang--Mills fields coupled to \qq{spontaneously broken}
symmetry generators give rise to massive vector bosons and the
associated scalar Goldstone bosons disappear from physical
spectrum. Such a result actually implies an important rule for
building models of electroweak interactions: {\bf for each vector
boson mass, which is to be generated via Higgs mechanism, one
needs a (would-be) Goldstone boson in the scalar sector.} Needless
to say, this also represents a certain constraint on the contents
of scalar multiplets involved in the theory.

One may employ the above general observations to make a right
guess for the Higgs--Goldstone sector of the standard electroweak
theory\index{Glashow--Weinberg--Salam theory|ff}\index{standard
model of electroweak interactions|ff}. Since we want to get three
massive vector bosons, we must have three Goldstone bosons in the
underlying scalar field model. Further, it is also known that at
least one physical scalar boson (the Higgs boson) survives the
Higgs mechanism. Thus, it is clear that one has to start with at
least four real scalar fields. In order to get the desired
spectrum of vector boson masses, the scalars must be coupled in a
non-trivial way to the $SU(2)$ gauge fields; it means that the two
complex scalars should constitute a doublet representation of the
$SU(2)$. The upshot of these considerations is that the minimal
Higgs--Goldstone scalar sector for the $SU(2) \times
U(1)$\index{SU(2) times U(1) group@$SU(2)\times U(1)$ group}
electroweak theory consists of one complex (weak
isospin\index{weak!isospin}) $SU(2)$ doublet; this must also be
endowed with some specific transformation properties under the
$U(1)$\index{U(1) group@$U(1)$ group} (weak
hypercharge\index{weak!hypercharge|ff}) subgroup, as we shall
discuss in the sequel. The weak isodoublet can be written as
\begin{equation}
\label{eq6.41}
\Phi =\left(
\begin{array}{c}
\varphi ^{+} \\
\varphi ^0
\end{array}
\right)
\end{equation}
where the two complex components $\varphi ^{+}$ and $\varphi ^0$ are of
course equivalent to four real fields, e.g. through a straightforward
parametrization
\begin{equation}\label{eq6.42}
\Phi =\left(
\begin{array}{c}
\varphi _1+i\varphi _2 \\
\varphi _3+i\varphi _4
\end{array}
\right)
\end{equation}
The superscripts of the components of (\ref{eq6.41}) indicate that the $\varphi ^{+}$
and $\varphi ^0$ should represent fields carrying charges +1 and 0
respectively (this becomes clear when one specifies the interaction terms
involving the scalar doublet and other fields with definite charge
assignments).

Now we are in a position to discuss the Higgs mechanism within the
$SU(2)\times U(1)$ gauge theory in explicit terms. The starting
point of our discussion will be the underlying Goldstone-type
model. Using (\ref{eq6.41}) as a basic building block it is easy to
construct the corresponding Lagrangian possessing the necessary
symmetry. In analogy with (\ref{eq6.12}) this can be written as
\begin{equation}\label{eq6.43}
\lagr_{Goldstone}=(\partial_\mu \Phi^{\dagger})(\partial^\mu \Phi)
-V(\Phi)
\end{equation}
with the potential $V$ given by
\begin{equation}\label{eq6.44}
V(\Phi) =-\mu ^2\Phi^{\dagger}\Phi +\lambda (\Phi^{\dagger}\Phi)^2
\end{equation}
It is interesting to notice that such a Lagrangian has, in fact,
an \qq{accidental} symmetry larger than the originally required
$SU(2) \times U(1)$. Indeed, using the parametrization (\ref{eq6.42}), one
sees that
\begin{equation}\label{eq6.45}
\Phi ^{\dagger }\Phi =\varphi _1^2+\varphi _2^2+\varphi _3^2+\varphi _4^2
\end{equation}
which means that the full symmetry of the $V$ is $O(4)$\index{O(4)
group@$O(4)$ group} (of course, the same is true for the kinetic
term in (\ref{eq6.43})). One may observe immediately that this accidental
symmetry is due precisely to the doublet character of the basic
Higgs--Goldstone field -- when starting from (\ref{eq6.42}), one must
necessarily employ the form (\ref{eq6.45}) in order to construct an
$SU(2)$ invariant Lagrangian. We shall discuss these deeper
symmetry aspects of the standard electroweak theory later in this
chapter (see Section~\ref{sec6.8}).

It is not difficult to see that the Lagrangian (\ref{eq6.43}) describes three
massless Goldstone bosons and one massive scalar. Indeed, it can be recast as
\begin{equation}\label{eq6.46}
\lagr_\ti{Goldstone}=\text{derivative terms}+\mu^2\rho^2-\lambda
\rho^4
\end{equation}
where we have denoted
\begin{equation}
\rho ^2=\Phi ^{\dagger }\Phi
\end{equation}
Similarly to the Abelian case, one may argue that the minimum of energy
density occurs for space-time constant field configurations $\Phi _0$ such
that
\begin{equation}\label{eq6.48}
\Phi _0^{\dagger }\Phi _0=\frac{v^2}2
\end{equation}
where
\begin{equation}\label{eq6.49}
v=\frac \mu {\sqrt{\lambda }}
\end{equation}
(cf. (\ref{eq6.18})). Subtracting then from the field variable $\rho $ its
\qq{vacuum value} mentioned above, one gets rid of the wrong-sign
mass term in (\ref{eq6.46}) and the shifted field acquires an ordinary
mass in the by now familiar way. The other three real fields that
parametrize our complex doublet remain massless as they enter only
the derivative terms in (\ref{eq6.46}). For an explicit description of the
Goldstone bosons and the massive (Higgs) scalar it is again useful
to introduce an exponential parametrization of (\ref{eq6.41}) analogous to
the relation (\ref{eq6.19}) employed in the Abelian case. Now we can write
\begin{equation}\label{eq6.50}
\Phi \left( x\right) =\exp \Bigl( \frac{i}v\pi ^a(x) \tau ^a\Bigr)
\left(
\begin{array}{c}
0 \\
\frac 1{\sqrt{2}}\left( v+H\left( x\right) \right)
\end{array}
\right)
\end{equation}
where we have already marked explicitly the shift of the
\qq{radial} variable $\rho $, defining thus the Higgs field $H.$
The \qq{angular} fields $\pi^a$, $a$ = 1, 2, 3 represent the
Goldstone bosons (the $\tau ^a$ denote, as usual, the Pauli
matrices). Using (\ref{eq6.50}) in (\ref{eq6.43}) it is then elementary to find
the Higgs field mass; this is
\begin{equation}
\label{eq6.51}
m_H^2=2\lambda v^2
\end{equation}
(which becomes $m_H=\mu \sqrt{2}$ when one takes into account (\ref{eq6.49})). We
should note that a particular vacuum field configuration $\Phi _0$ is, for
example
\begin{equation}\label{eq6.52}
\Phi _0^{\left( 0\right) }=\frac 1{\sqrt{2}}\left(
\begin{array}{c}
0 \\
v
\end{array}
\right)
\end{equation}
Of course, any $\Phi _0$ obtained from (\ref{eq6.52}) by means of a global
$SU(2)$ transformation can represent the ground state as well,
since the potential minimum is determined by the $\Phi _0^{\dagger
}\Phi _0$ value only. In other words, there is a three-parametric
degenerate set of vacua associated with the potential (\ref{eq6.44}).
Thus, in the considered classical field theory model we can indeed
recognize characteristic features of spontaneous symmetry
breakdown: the Lagrangian (\ref{eq6.43}) is invariant under $SU(2)$ while
the ground state (represented e.g. by (\ref{eq6.52})) is not. As a result,
three Goldstone bosons appear, corresponding to the three
generators of the $SU(2)$ broken by the vacuum state.

The passage from (\ref{eq6.43}) to a Higgs-type Lagrangian with local
$SU(2) \times U(1)$ symmetry is accomplished in a similar manner
as in the Abelian model discussed in preceding section. From a
purely technical point of view, the covariant
derivative\index{covariant derivative} acting on the scalar
doublet $\Phi$ can be written in a straightforward analogy with
the case of lepton sector described in Chapter~\ref{chap5}. The $\Phi$,
apart from being a doublet under the $SU(2)$, carries also a weak
hypercharge $Y_\Phi $ associated with the $U(1)$ subgroup (we
shall denote it simply as $Y$ in what follows). The gauge
invariant Lagrangian can then be written as
\begin{eqnarray}
\label{eq6.53}
\lagr_{Higgs} &=&\Phi ^{\dagger }\left( \partialvb_\mu
+igA_\mu ^a\frac{\tau ^a}2+ig'YB_\mu \right) \left(
\partialv^\mu -igA^{b\mu }\frac{\tau ^b}2-ig'YB^\mu \right) \Phi
\nonumber \\
&&-\lambda \left( \Phi ^{\dagger }\Phi -\frac{v^2}2\right) ^2
\end{eqnarray}
where the $A_\mu ^a$, $a$ = 1, 2, 3 and $B_\mu $ are Yang--Mills
fields corresponding to $SU(2)$ and $U(1)$ resp. and $g$, $g'$ are
the associated coupling constants. For the sake of brevity, we
have not included here the kinetic terms of gauge fields and their
pure self-interactions that have been discussed in detail earlier;
these can be retrieved from Chapter~\ref{chap5} whenever necessary. As usual
(for later convenience) we have also shifted the bottom of the
scalar-field potential (\ref{eq6.44}) to zero by adding an otherwise
inessential constant. The exponential parametrization (\ref{eq6.50}) can
be used for fixing the physical $U$-gauge in a similar manner as
in the Abelian case discussed in preceding section. Such a gauge
fixing is formally equivalent to a local $SU(2)$ transformation
that removes the angular fields from the $\Phi$ (this indicates
the unphysical nature of the would-be Goldstone bosons in the
present context). When this is done, one is left with
\begin{equation}
\label{eq6.54}
\Phi _U\left( x\right) =\left(
\begin{array}{c}
0 \\
\frac 1{\sqrt{2}}\left( v+H\left( x\right) \right)
\end{array}
\right)
\end{equation}
The Lagrangian (\ref{eq6.53}) in the $U$-gauge can then be written in terms of
(\ref{eq6.54}) and correspondingly transformed gauge fields, without changing its
original form. It reads
\begin{eqnarray}
\lagr_{Higgs}^{\left( U\right) } &=&\Phi _U^{\dagger }\left(
\partialvb_\mu +igA_\mu ^a\frac{\tau ^a}2+ig'YB_\mu
\right) \left( \partialv^\mu -igA^{b\mu }\frac{\tau ^b}%
2-ig'YB^\mu \right) \Phi _U  \nonumber \\
&&-\lambda \left( \Phi _U^{\dagger }\Phi _U-\frac{v^2}2\right) ^2
\label{eq6.55}\end{eqnarray}
where we have used, for notational simplicity, the same symbols for the
transformed gauge fields as for the old ones. Writing (\ref{eq6.54}) as
\begin{equation}
\Phi _U\left( x\right) =\frac 1{\sqrt{2}}\left( v+H\left( x\right)
\right) \xi,\qquad \qquad \xi =\left(
\begin{array}{c}
0 \\
1
\end{array}
\right)
\end{equation}
the $U$-gauge Lagrangian (\ref{eq6.55}) can be easily worked out as
\begin{eqnarray}
\lagr_{Higgs}^{\left( U\right) } &=&\frac 12\partial _\mu H\partial ^\mu
H-\frac 14\lambda \left[ \left( v+H\right) ^2-v^2\right] ^2  \\
&&+\frac 12\left( v+H\right) ^2\xi ^{\dagger }\left( gA_\mu ^a\frac{\tau ^a}%
2+g'YB_\mu \right) \left( gA^{b\mu }\frac{\tau ^b}2+g'YB^\mu
\right) \xi  \nonumber
\end{eqnarray}
The last expression can be further simplified by means of the relations
\begin{eqnarray}
\tau ^a\tau ^b+\tau ^b\tau ^a &=&2\delta ^{ab}\J  \nonumber
\\
\xi ^{\dagger }\tau ^a\xi &=&-\delta ^{3a}  \nonumber \\
\xi ^{\dagger }\xi &=&1
\end{eqnarray}
and one thus gets finally
\begin{equation}\label{eq6.59}
\begin{split}
\lagr_{Higgs}^{\left( U\right) } = \phantom{+}&\frac 12\partial
_\mu H\partial ^\mu
H-\lambda v^2H^2-\lambda vH^3-\frac 14\lambda H^4  \\
+&\frac 18\left( v+H\right) ^2\left( g^2A_\mu ^aA^{a\mu }-4Ygg'
A_\mu ^3B^\mu +4Y^2g'^2 B_\mu B^\mu \right)
\end{split}
\end{equation}
Obviously, the Higgs field mass is the same as before (cf.
(\ref{eq6.51})). The part of the Lagrangian (\ref{eq6.59}) quadratic in gauge
fields is diagonalized immediately and mass terms of intermediate
vector bosons can be thus identified easily. For the relevant
quadratic form one obtains from (\ref{eq6.59})
\begin{eqnarray}
\lagr_{mass}^{\left( IVB\right) } &=&\frac 18v^2\left[ g^2\left( \left(
A_\mu ^1\right) ^2+\left( A_\mu ^2\right) ^2\right) +\left( gA_\mu
^3-2g'YB_\mu \right) ^2\right]  \nonumber \\
&=&\frac 18\left( g^2+4Y^2g'^2\right) v^2\left( \frac g{\sqrt{%
g^2+4Y_{}^2g'^2}}A_\mu ^3-\frac{2Yg'}{\sqrt{%
g^2+4Y_{}^2g'^2}}B_\mu \right) ^2  \nonumber \\
&&+\frac 14g^2v^2W_\mu ^{-}W^{+\mu }\label{eq6.60}
\end{eqnarray}
where the $W_\mu ^{\pm }$ stand for combinations $\frac
1{\sqrt{2}}\left( A_\mu ^1\mp i A_\mu ^2\right) $ familiar from
our previous analysis of charged current interaction (cf. Section~\ref{sec5.2}) and we have also introduced a \qq{normalized} linear
combination of the $A_\mu ^3$ and $B_\mu $, which should
presumably be identical with the $Z$ boson field\index{Z boson@$Z$
boson|ff} discussed in Section~\ref{sec5.3}. Let us recall that for the $Z$
field coupled to weak neutral currents\index{neutral current} we
had
\begin{equation}\label{eq6.61}
Z_\mu =\cos \theta _WA_\mu ^3-\sin \theta _WB_\mu
\end{equation}
(cf. (\ref{eq5.22})), where the mixing angle $\theta
_W$\index{weak!mixing angle} is in general given by $\tan \theta
_W=-2Y_Lg'/g$ (see (\ref{eq5.32})) with $Y_L$ being the weak
hypercharge of the left-handed leptonic doublet. Comparing this
result with (\ref{eq6.60}), it is clear that we must set (returning to the
notation $Y_\Phi =Y$ for a moment)
\begin{equation}
Y_\Phi =-Y_L
\end{equation}
if the $Z$ coming from the mass matrix diagonalization is to be the same as
that in (\ref{eq6.61}) -- that is, if we want the two ends of electroweak theory to
be mutually consistent.

The reader may remember that we have eventually set $Y_L=-1/2$
(cf. (\ref{eq5.38})) for the sake of simplicity of the resulting
formulae. This implies the conventional choice
\begin{equation}
Y_\Phi =+\frac 12
\end{equation}
which means that one can always use the rule
\begin{equation}\label{eq6.64}
Q=T_3+Y
\end{equation}
(cf. eq. (\ref{eq5.40})). Notice that for the scalar doublet
(\ref{eq6.41}) this reflects the fact that the upper component (with
$T_3=+\frac{1}{2}$) has $Q = + 1$. Actually, in most textbooks the
convention (\ref{eq6.64}) is usually adopted automatically when the Higgs
sector of the standard electroweak theory is described. We have
discussed here the general case at some length for completeness; a
pragmatically minded reader might omit the analysis involving an arbitrary $%
Y $ value and use immediately the law (\ref{eq6.64}) from the very start.

Thus, we have seen that there is just one massive combination of $A_\mu ^3$
and $B_\mu $; upon setting $Y=1/2$ in (\ref{eq6.60}) this becomes
\begin{equation}\label{eq6.65}
Z_\mu =\frac 1{\sqrt{g^2+g'^2}}\left( gA_\mu ^3-g'B_\mu
\right)
\end{equation}
and coincides with (\ref{eq6.61}). Since the total number of gauge fields is four,
we will introduce also a combination \qq{orthogonal} to (\ref{eq6.65}), namely
\begin{equation}
A_\mu =\frac 1{\sqrt{g^2+g'^2}}\left( g'A_\mu ^3+gB_\mu
\right)
\end{equation}
which is obviously massless (simply because there is no such mass
term in (\ref{eq6.60})) and coincides with the electromagnetic
field\index{electromagnetic!field} appearing in (\ref{eq5.22}).
Let us recall that the orthogonality is imposed so as to preserve
the diagonal structure of the kinetic term for vector fields (cf.
(\ref{eq5.25})). Taking into account the normalization of the
kinetic term, the non-zero masses can now be read off directly
from (\ref{eq6.60}). For $Y$ $=+1/2$ one has
\begin{equation}\label{eq6.67}
\lagr_{mass}^{\left( IVB\right) }=\frac 14g^2v^2W_\mu ^{-}W^{+\mu }+\frac
18\left( g^2+g'^2\right) v^2Z_\mu Z^\mu
\end{equation}
which yields
\begin{align}
m_W^2&=\frac 14g^2v^2 \notag \\ \frac 12m_Z^2&=\frac 18\bigl(
g^2+g'^2\bigr) v^2
\label{eq6.68}
\end{align}
Thus, as a result of the Higgs mechanism described above we have
the mass formulae\index{mass formula for $W$ and $Z$}
\begin{align}
m_W&=\frac 12gv \notag \\ m_Z&=\frac 12 (g^2+g'^2)^{1/2} v
\label{eq6.69}
\end{align}
derived first by S. Weinberg in his celebrated paper \cite{ref39}.
The relations (\ref{eq6.69}) imply, in particular
\begin{equation}\label{eq6.70}
\frac{m_W}{m_Z}=\cos \theta _W
\end{equation}
or, in other words
\begin{equation}\label{eq6.71}
\frac{m_W^2}{m_Z^2}=1-\frac{e^2}{g^2}
\end{equation}
if one uses the relation $e=g\sin \theta _W$ for the electromagnetic
coupling constant (see (\ref{eq5.34})). It is easy to realize that the relation
(\ref{eq6.70}) (or (\ref{eq6.71}) resp.) holds for a general value of the weak hypercharge
of the scalar isodoublet $\Phi$.

The formula (\ref{eq6.69}) for $m_W$ has a rather remarkable consequence
that should be emphasized here. When the expression $m_W=\frac
12gv$ is inserted into the familiar relation for the Fermi
constant $G_F/\sqrt{2}=g^2/\left( 8m_W^2\right) $ (see
(\ref{eq3.19})), one gets immediately
\begin{equation}\label{eq6.72}
v=\left( G_F\sqrt{2}\right) ^{-1/2}\doteq 246\ \GeV
\end{equation}
Thus, the vacuum value of the Higgs scalar field turns out to be directly
related to the Fermi constant $-$ the parameter of the old weak interaction
physics. This may be somewhat surprising at first sight, since the $v$ has
originally been expressed (see (\ref{eq6.49})) in terms of the $\mu $ and $\lambda $%
, the totally unknown parameters of the \qq{Goldstone
potential}\index{Goldstone!potential} $V(\Phi)$. On the other
hand, the $v$ is obviously the only relevant mass scale that
enters the Higgs mechanism and the $G_F$ can be considered as the
only dimensionful parameter describing weak interactions. Thus,
from this point of view the relation (\ref{eq6.72}) appears to be quite
natural\index{vacuum|)}.

One should notice that our specific example of the Higgs mechanism
confirms indeed the general statements formulated earlier in this
section. We have started with a model that exhibits three
Goldstone bosons associated with a spontaneously broken global
symmetry $SU(2)$. Within a corresponding Higgs-type model
invariant under local $SU(2)\times U(1)$ the erstwhile
Goldstone bosons become unphysical (they completely disappear in the $U$%
-gauge) and one gets three massive vector bosons $W^{+}$, $W^{-}$
and $Z$. As an additional bonus, one obtains an interesting
relation (\ref{eq6.70}) which also means that the parameter $\rho
=m_W^2/\left( m_Z^2\cos ^2\theta _W\right) $ (cf. (\ref{eq5.63}))
is equal to unity at the classical level -- we have mentioned this
remarkable fact already in Section~\ref{sec5.6}. As noted there, the
relation $\rho =1$ (which receives a small correction at the
quantum level) is indeed phenomenologically successful, i.e. it is
experimentally confirmed with good accuracy. Thus, one can say that long before the experimental discovery of the Higgs boson, there was a clear indirect argument in favour of the assumption that masses of $W$ and $Z$ are generated through the Higgs mechanism implemented by means of a complex scalar doublet.

In closing this section let us summarize, for reader's
convenience, formulae for the $W$ and $Z$ masses written in terms
of $\alpha $, $G_F$ and $\theta _W$. We have already found such a
formula for $m_W$ in Section~\ref{sec5.4} (see (\ref{eq5.42})) and now we
are able to add the corresponding expression for the $m_Z$, by
making use of (\ref{eq6.70}). Thus, we have\index{mass formula for $W$ and
$Z$}
\begin{eqnarray}
m_W &=&\left( \frac{\pi \alpha }{G_F\sqrt{2}}\right) ^{1/2}\frac 1{\sin
\theta _W}  \nonumber \\
m_Z &=&\left( \frac{\pi \alpha }{G_F\sqrt{2}}\right) ^{1/2}\frac 1{\sin
\theta _W\cos \theta _W}\label{eq6.73}
\end{eqnarray}
Since $\left( \pi \alpha /G_F\sqrt{2}\right) ^{1/2}\doteq  37\
\GeV$ and $\sin \theta _W\cos \theta _W=\frac 12\sin 2\theta _W$,
from (\ref{eq6.73}) it is obvious that in addition to the lower bound $m_W
\gtrsim  37\ \GeV$ derived earlier (cf. (\ref{eq5.44})) one also
has $m_Z \gtrsim 74\ \GeV$\index{spontaneous symmetry
breakdown|)}.

%\end{document}

%\input{kniha65}
%%%%%%%%%%%%%%%%%%%%%%%%%%%%%%%%%%%%%%%%%%%%%%%%%%%%%%%%%%%%%%%%%%%
%%%%%%%%%%%%%%%%%%%%%%%%%%%%%%%%%%%%%%%%%%%%%%%%%%%%%%%%%%%%%%%%%%%%%%%%%%%%%%%%%%%%%%%%%%%%%%%%%%%%%%%%%%%%%%%%%%%%%%%%%%%%%%%%%%%%%%%%

\section{Higgs boson interactions}

Having identified physical scalar and vector fields resulting from
the Higgs mechanism within the $SU(2)\times U(1)$ gauge theory, we
are now ready to describe their interactions. When the $U$-gauge
Lagrangian (\ref{eq6.59}) is recast in terms of the $W_\mu ^{\pm }$ and
$Z_\mu $ (cf. the discussion around (\ref{eq6.60}) and the relation
(\ref{eq6.67})), we have
\begin{eqnarray}
\lagr_{Higgs}^{\left( U\right) } &=&\frac 12\partial _\mu
H\partial ^\mu
H-\lambda v^2H^2-\lambda vH^3-\frac 14\lambda H^4  \nonumber \\
&&+\frac 18\left( v+H\right) ^2[2g^2W_\mu ^{-}W^{+\mu }+\left(
g^2+g'^2\right) Z_\mu Z^\mu ]\label{eq6.74}
\end{eqnarray}
so that the interaction Lagrangian reads
\begin{eqnarray}
\lagr_{Higgs}^{\left( int\right) } &=&\frac 18\left(
2vH+H^2\right) [2g^2W_\mu ^{-}W^{+\mu }+\left( g^2+g'^2\right)
Z_\mu Z^\mu ]
\nonumber \\
&&-\lambda vH^3-\frac 14\lambda H^4\label{eq6.75}
\end{eqnarray}
Let us now focus on the interactions of the $W$\index{W boson@$W$
boson} and $Z$ with the Higgs boson $H$. Similarly as in the
Abelian model of Section~\ref{sec6.3}, in (\ref{eq6.75}) one can recognize
essentially two types of couplings, namely the trilinear and
quadrilinear ones\index{quadrilinear vertex}. These are
\begin{eqnarray}
\lagr_{WWH} &=&gm_WW_\mu ^{-}W^{+\mu }H  \nonumber \\
\lagr_{ZZH} &=&\frac{gm_Z}{2\cos \theta _W}Z_\mu Z^\mu H
\label{eq6.76}
\end{eqnarray}
and
\begin{eqnarray}
\lagr_{WWHH} &=&\frac 14g^2W_\mu ^{-}W^{+\mu }H^2  \nonumber \\
\lagr_{ZZHH} &=&\frac 18\frac{g^2}{\cos ^2\theta _W}Z_\mu Z^\mu
H^2 \label{eq6.77}
\end{eqnarray}
Note that in writing (\ref{eq6.76}), (\ref{eq6.77}) we have eventually used the mass
relations (\ref{eq6.69}) as well as the familiar expression $\cos \theta _W=g/\sqrt{%
g^2+g'^2}$ for the Weinberg mixing
angle\index{U-gauge@$U$-gauge|)}\index{weak!mixing angle}.

It is quite remarkable that the form of the $WWH$ interaction
resulting from the Higgs mechanism coincides with the $WW\sigma $
coupling obtained in Section~\ref{sec6.1} through the analysis of residual
high-energy divergences of tree-level Feynman diagrams for
$W_LW_L\rightarrow W_LW_L$ (cf. (\ref{eq6.2}) and (\ref{eq6.6})). This indicates
that the Higgs mechanism within a gauge theory is essentially the
only means of saving the good asymptotic behaviour of scattering
amplitudes involving massive vector bosons and hence is of vital
importance for perturbative renormalizability\index{perturbative
renormalizability}. It is also not difficult to see that the
interactions (\ref{eq6.76}) and (\ref{eq6.77}) lead to the right high-energy
behaviour of the tree-level amplitudes for processes $WW\rightarrow HH$ and $%
ZZ\rightarrow HH$. In particular, one may observe that the
contribution of the direct $WWHH$ interaction (\ref{eq6.77}) compensates
the high-energy (quadratic) divergences\index{high-energy
divergences} produced by the second-order graph involving the $W$
exchange
and two $WWH$ vertices; an analogous mechanism operates in the $%
ZZ\rightarrow HH$ channel as well. The corresponding calculation
is left to the reader as an instructive exercise. Note that
converse is also true: the set of couplings (\ref{eq6.76}), (\ref{eq6.77}) is
fixed uniquely by the requirement of tree-level unitarity for the
relevant scattering amplitudes (for details, the reader is
referred to \cite{Hor}).

Finally, it should be noticed that the interaction Lagrangian
(\ref{eq6.75}) also contains cubic and quartic
self-couplings\index{quartic coupling} of the Higgs boson.
Denoting the corresponding coupling constants as $g_{HHH}$ and
$g_{HHHH}$ respectively, one has\index{coupling constants!of Higgs
sector|ff}
\begin{eqnarray}
g_{HHH} &=&-\lambda v  \nonumber \\
g_{HHHH} &=&-\frac 14\lambda
\label{eq6.78}
\end{eqnarray}
Using now the relations $m_H^2=2\lambda v^2$ (see (\ref{eq6.51})) and
$v=2m_W/g=(G_F\sqrt{2})^{-\frac{1}{2}}$ (see (\ref{eq6.72})), one can
recast (\ref{eq6.78}) as
\begin{eqnarray}
g_{HHH} &=&-\frac 14g\frac{m_H^2}{m_W}=-\left( \frac{G_F}{2\sqrt{2}}\right)
^{1/2}m_H^2  \nonumber \\
g_{HHHH} &=&-\frac 1{32}g^2\frac{m_H^2}{m_W^2}=-\frac{G_Fm_H^2}{4\sqrt{2}}
\label{eq6.79}
\end{eqnarray}
With current experimental data at hand (see \cite{ref5}), one may estimate the numerical value of the coupling constant $\lambda$: using (\ref{eq6.51}), (\ref{eq6.72}) and $m_H \doteq 125\,\text{GeV}$, one gets $\lambda \doteq 0.125$. Thus, one may conclude that also here one may rely on the implementation of  perturbation theory as in the other parts of the electroweak SM (note, however, that some relevant details of the Higgs boson interactions still require a thorough experimental study). Anyway, it may also be instructive to return briefly to the old times before the Higgs boson discovery, namely to some theoretical (technical) semi-quantitative constraints on the possible value of $m_H$. A basic hint is based on the simple-minded perturbativity argument: If the relevant Higgs-Goldstone Lagrangian in (\ref{eq6.74}) is to be used perturbatively, than any dimensionless coupling should not,
roughly speaking, exceed unity (otherwise the corresponding power
expansion would be doubtful {\it a priori\/}). In the considered
case the order-of-magnitude estimate $|g_{HHHH}|\lesssim 1$ yields
a simple upper bound for the Higgs mass, namely
\begin{equation}\label{eq6.80}
m_H\lesssim 2\sqrt{2}\left( G_F\sqrt{2}\right)
^{-1/2}=2\sqrt{2}\,v\doteq 700\ \GeV
\end{equation}
These considerations can be given more precise quantitative
meaning, if e.g. the unitarity condition for partial
waves\index{partial-wave expansion} is invoked for an appropriate
process at the tree level. Additional numerical factors then
modify slightly the straightforward bound (\ref{eq6.80}), but the overall
scale of the $m_H$ estimate remains the same. Moreover, such an
analysis can be further refined if one-loop diagrams\index{loop
diagrams} are taken into account. A more detailed discussion of
these issues would go beyond the scope of the present text and the
interested reader is therefore referred to the original literature
(see e.g. \cite{ref56}, \cite{ref57}).

Coming back to the present-day situation, the experimental value $m_H = 125.25 \pm 0.17\, \text{GeV}$ shown in \cite{ref5} certainly satisfies the perturbativity criterion, but the technical arguments outlined above are still useful in theoretical considerations concerning possible extensions SM, in particular when contemplating electroweak models involving several Higgs-like doublets (see, e.g. the book \cite{Gun}, the review \cite{Branco}, and the papers \cite{KKT, AAN, HK}.

One may also wonder whether the Higgs boson self-interactions
specified in (\ref{eq6.78}) or (\ref{eq6.79}) resp. play any role in the
high-energy divergence cancellations for some specific physical
processes. The answer is yes: it turns out that they are necessary
to ensure the tree-level unitarity for some $2\rightarrow 3$
reactions, such as e.g. $WW\rightarrow WWH$, $WW\rightarrow HHH$
etc. (note that the tree unitarity\index{tree unitarity} for
five-point amplitudes means that they decrease as $1/E$ in the
high-energy limit). For more details, see e.g. \cite{Hor} and the
references therein.

%\end{document}

%\input{kniha66}
%%%%%%%%%%%%%%%%%%%%%%%%%%%%%%%%%%%%%%%%%%%%%%%%%%%%%%%%%%%%%%%%%%%
%%%%%%%%%%%%%%%%%%%%%%%%%%%%%%%%%%%%%%%%%%%%%%%%%%%%%%%%%%%%%%%%%%%%%%%%%%%%%%%%%%%%%%%%%%%%%%%%%%%%%%%%%%%%%%%%%%%%%%%%%%%%%%%%%%%%%%%%
%\documentclass[12pt,report]{article}
%\input{tcilatex}
%\documentstyle{article}
%\input{tcilatex}
%\begin{document}
%\end{document}
%\binom 0{\frac{v+H}{\sqrt{2}}}
%\end{document}

\section{Yukawa couplings and lepton masses}
\index{Yukawa coupling|(}\label{sec6.6}

We will now show that lepton masses can also be generated through
appropriate interactions involving the Higgs doublet
$\Phi$\index{Higgs!doublet|ff} (the quark
sector will be discussed in the next chapter).\footnote{%
Let us recall that e.g. an electron mass term cannot be added to
our $SU(2) \times U(1)$ invariant Lagrangian simply by hand since
this would violate the required symmetry. Indeed, $m_e\bar{e}e=m_e
(\bar{e}_L e_R +\bar{e}_R e_L)$ and the chiral components $e_L$
and $e_R$ transform differently under the weak
isospin\index{weak!isospin|ff} $SU(2)$: the $e_L$ belongs to an
$SU(2)$ doublet while the $e_R$ is a singlet.} To this end, we are
going to employ a Yukawa-type coupling (that is, an interaction
bilinear in lepton fields and linear in $\Phi$). It is not
difficult to realize that such a (non-derivative) interaction term
is essentially the only renormalizable\index{renormalizable
theory} coupling that can still be added to the Lagrangian
considered so far. Following our symmetry principle, we should
construct it to be $SU(2)\times U(1)$ invariant. For the moment,
let us consider e.g. only leptons of the electron type. Of course,
as the basic building blocks we have to employ the left-handed
doublet
\begin{equation}\label{eq6.81}
L = \left(
\begin{array}{c}
\nu _L \\
e_L
\end{array}
\right)
\end{equation}
and the right-handed singlet $e_R$ (cf. Section~\ref{sec5.1}). From the doublets $L$
and $\Phi $, an $SU(2)$ singlet can be immediately formed as $\bar{L}%
\Phi $. Multiplying this by the $e_R$, one obtains an $SU(2)$ invariant
Yukawa interaction term
\begin{equation}
\label{eq6.82} \lagr_{Yukawa}=-h_e \widebar{L}\Phi e_R+\text{h.c.}
\end{equation}
where the $h_e$ is a (dimensionless) coupling constant; the minus sign has
been chosen for later convenience. Note that we suppress the lepton labels
whenever it does not lead to confusion. It is easy to see that the
Lagrangian (\ref{eq6.82}) is automatically invariant under the weak hypercharge $U(1)$
as well. Indeed, one has $Y_L$ $=-1/2$, which in turn means that the Dirac
conjugate $\widebar{L}$ carries $Y_{\widebar{L}}=+1/2$. Further, $%
Y_R^{(e)}=-1$ and $Y_\Phi =+1/2$. One thus gets
$Y_{\widebar{L}}+Y_\Phi +Y_R^{(e)}=0$, which proves our statement.
Now, fixing the unitary gauge, (\ref{eq6.82}) becomes
\begin{eqnarray}
\lagr_{Yukawa}^{(U)} &=&-h_e\bm{\bar{\nu}_L,\;\bar{e}_L}
\bm{0 \\ \frac 1{\sqrt{2}}(v+H)} e_R+\text{h.c.}  \nonumber \\
&=&-\frac 1{\sqrt{2}}h_e\left( v+H\right) \bar{e}_Le_R+\text{h.c.}
\nonumber \\
&=&-\frac 1{\sqrt{2}}h_e\left( v+H\right) \left( \bar{e}_Le_R+%
\bar{e}_Re_L\right)   \nonumber \\
&=&-\frac 1{\sqrt{2}}h_ev\bar{e}e-\frac 1{\sqrt{2}}h_e\bar{e}eH
\end{eqnarray}
The last expression contains an electron mass term with
\begin{equation}
\label{eq6.84}
m_e=\frac 1{\sqrt{2}}h_ev
\end{equation}
and a scalar Yukawa coupling
\begin{equation}
\lagr_{eeH}=g_{eeH}\bar{e}eH
\end{equation}
with $g_{eeH}=-\frac 1{\sqrt{2}}h_e$. Taking into account (\ref{eq6.84}), one then
has
\begin{equation}
g_{eeH}=-\frac{m_e}v
\end{equation}
that can be recast (by employing the familiar relation $v=2m_W/g$) as
\begin{equation}
g_{eeH}=-\frac g2\frac{m_e}{m_W}
\end{equation}
This is seen to coincide with the $g_{ee\sigma }$ coupling
obtained in Section~\ref{sec6.1} from the analysis of Feynman diagrams (cf.
(\ref{eq6.11})). Thus, similarly as in the previous section, one has
another indication that the mass generation through Higgs
mechanism is actually necessary for tree-level unitarity (and
thereby for perturbative renormalizability) of the electroweak
theory. An instructive exercise offered to the interested reader
is to check explicitly how the Higgs boson couplings derived here
and in the preceding section yield well-behaved tree-level
amplitudes e.g. for the processes $e^{+}e^{-}\rightarrow Z_LZ_L$
or $e^{+}e^{-}\rightarrow Z_LH$. Needless to say, mass terms for
muon or tau lepton\index{tau lepton@$\tau$ lepton} can be produced
in a completely analogous way -- one only needs different Yukawa
coupling constants to account for different lepton masses. Thus,
one has in general
\begin{equation}
\label{eq6.88} g_{\ell\ell H}=-\frac g2\frac{m_\ell}{m_W}
\end{equation}
for $\ell=e,\mu ,\tau $, which is characteristic for the standard
model Higgs boson. Obviously, the dependence of the interaction
strengths on the lepton type embodied in (\ref{eq6.88}) is rather
dramatic: it means that a reasonably accurate estimate for the
leptonic two-body rates would be
\begin{equation}
\label{eq6.89}
\Gamma \left( H\rightarrow e^{+}e^{-}\right):\Gamma (H\rightarrow \mu
^{+}\mu ^{-}):\Gamma (H\rightarrow \tau ^{+}\tau ^{-})
= m_e^2:m_\mu ^2:m_\tau ^2
\end{equation}
(when writing (\ref{eq6.89}) we have taken into account that $m_\ell^2\ll
m_H^2$ according to the current experimental bounds; the
lepton-mass dependence of the phase space\index{phase space}
volume etc. can then be essentially neglected).

Let us now consider the possibility of giving mass to a neutrino.
To begin with, we shall restrict ourselves to a single lepton
species (say, the electron type). As we have already noticed in
Chapter~\ref{chap5}, the right-handed component of neutrino field can be
introduced without violating any natural requirement of the
electroweak theory. Apart from being a weak isospin singlet, it
must then carry zero weak hypercharge (see (\ref{eq5.9})). Because
of the (mandatory) hypercharge assignments, one cannot construct an $SU(2)
\times U(1)$ invariant out of the doublets $L$ and $\Phi $ and the singlet $%
\nu _R$ (needless to say, any violation of the $U\left( 1\right)
_Y$ invariance would lead to the non-conservation of the electric
charge). However, it is possible to employ the following trick. It
can be shown that the quantity $\Phit$, defined in terms of the
original Higgs doublet $\Phi $ as
\begin{equation}
\label{eq6.90} \Phit=i\tau _2\Phi ^{*}
\end{equation}
with $\tau _2$ being the Pauli matrix
\begin{equation}
\tau _2=\left(
\begin{array}{cc}
0 & -i \\
i & 0
\end{array}
\right)
\end{equation}
transforms under $SU(2)$ in the same way as the $\Phi $, i.e. the
$\Phit$ is another scalar doublet. (Note that this observation is
also crucial for giving masses to all types of quarks and we will
utilize it in the next chapter as well.) We defer a formal proof
of the transformation properties of the $\Phit$ to the end of this
section and now let us proceed to see how it can be exploited for
our purpose. Since the definition (\ref{eq6.90}) involves complex
conjugation, it is clear that the $\Phit$ carries weak hypercharge
$Y_{\Phit}=-Y_\Phi =-\frac 12$ and one is then able to construct a
desired invariant form containing $\nu _R$. Indeed, one can write
\begin{equation}\label{eq6.92}
\widetilde{\lagr}_{Yukawa}=-h_\nu \widebar{L}\Phit\nu_R+
\text{h.c.}
\end{equation}
which is clearly $SU(2)\times U(1)$ invariant, as
$Y_{\widebar{L}}+Y_{\Phit}+Y_R^{(\nu)}=-Y_L-Y_\Phi
+Y_R^{(\nu)}=\frac 12-\frac 12+0=0$. In the unitary gauge, (\ref{eq6.92})
becomes
\begin{eqnarray}
\label{eq6.93} \widetilde{\lagr}_{Yukawa}^{(U) } &=&-h_\nu \bm{
\bar{\nu}_L,\; \bar{e}_L} \bm{ \frac 1{\sqrt{2}}(v+H)  \\ 0}
\nu _R+\text{h.c.}  \nonumber \\
&=&-\frac 1{\sqrt{2}}h_\nu ( \bar{\nu}_L\nu _R+\bar{\nu}_R\nu _L) (v+H)  \nonumber \\
&=&-\frac 1{\sqrt{2}}h_\nu v\bar{\nu} \nu -\frac 1{\sqrt{2}}h_\nu
\bar{\nu} \nu H
\end{eqnarray}
where we may identify immediately the neutrino
mass\index{neutrino!mass} term with $m_\nu =\frac 1{\sqrt{2}}h_\nu
v$ and a scalar Yukawa interaction
\begin{eqnarray}
\lagr_{\nu \nu H} &=&-\frac{m_\nu }v\bar{\nu}\nu H  \nonumber \\
&=&-\frac g2\frac{m_\nu }{m_W}\bar{\nu} \nu H
\label{eq6.94}
\end{eqnarray}

Now it is natural to ask, among other things, what is the impact
of such an additional neutrino interaction on the divergence
cancellations demonstrated earlier for various processes (see
Section~\ref{sec5.8}). In particular, we can reconsider the process
$\bar{\nu }\nu \rightarrow W_L^{-}W_L^{+}$. For massive neutrinos
in the initial state, one finds easily that the sum of the two
diagrams shown in Fig.\,\ref{fig18} (the exchange of the electron
and the $Z$)
still contains a residual $O\left( m_\nu E/m_W^2\right) $ divergence for $%
E\rightarrow \infty $. Once the coupling (\ref{eq6.94}) is present, there
is an additional graph involving $s$-channel Higgs boson exchange
and its contribution cancels exactly the linear divergence, in
close resemblance with the case of the $e^{+}e^{-}\rightarrow
W^{+}W^{-}$ process discussed earlier in this chapter. The reader
is recommended to verify this by means of an explicit calculation;
another instructive exercise would be to check that an analogous
mechanism also works for the process $\nu \bar{\nu} \rightarrow
Z_LZ_L$. In this context, it should be emphasized that
independently of its mass, the neutrino neutral
current\index{neutral current} remains purely left-handed -- the
$\nu _R$ remains uncoupled to the $Z$ boson.

It is obvious that the simple mechanism for generating neutrino
masses described above can be used for any lepton type. In fact,
it can be generalized in a more substantial way by producing also
possible mixings between different lepton species. We shall come
back to this issue later on, in connection with the discussion of
the quark sector of standard electroweak theory. To close this
section, let us now prove formally that the $\Phit$ defined in
(\ref{eq6.90}) is indeed an $SU(2)$ doublet. In particular, we are going
to prove that if
\begin{equation}
\Phi'=\text{e}^{i\omega _a\tau _a}\Phi
\end{equation}
(where the $\omega _a$ denote three arbitrary transformation parameters),
then
\begin{equation}\label{eq6.96}
%\stackrel{\thicksim }
\widetilde{\Phi}' =\text{e}^{i\omega _a\tau _a}\widetilde{\Phi }
\end{equation}
where the $\widetilde{\Phi}'$ is of course defined as $%
i\tau _2\Phi ^{\prime *}$. The crucial technical ingredient of the
proof is a simple identity for complex conjugation of the Pauli
matrices\index{Pauli matrices}, namely
\begin{equation}\label{eq6.97}
\tau _a^{*}=-\tau _2\tau _a\tau _2
\end{equation}
The verification of (\ref{eq6.97}) is straightforward and we leave it to the reader.
Now, according to our definitions, the left-hand side of (\ref{eq6.96}) can be
written as
\begin{equation}\label{eq6.98}
i\tau _2\left( \text{e}^{i\omega _a\tau _a}\Phi \right) ^{*}
\end{equation}
Expanding the exponential in (\ref{eq6.98}) in power series and employing (\ref{eq6.97}),
the $\widetilde{\Phi }'$ is worked out as
\begin{eqnarray}
\Phit' &=&i\tau _2\left( \J+\frac i{1!}\omega _a\tau
_a+\frac{i^2}{2!}\left( \omega _a\tau _a\right)
^2+...\right) ^{*}\Phi ^{*}=  \nonumber \\
&=&i\tau _2\left( \J+\frac{\left( -i\right) }{1!}\omega _a\tau
_a^{*}+\frac{\left( -i\right) ^2}{2!}\left( \omega _a\tau
_a^{*}\right) ^2+...\right) \Phi ^{*}  \nonumber \\
&=&i\tau _2 \cdot \tau _2\left( \J+\frac i{1!}\omega _a\tau _a+%
\frac{i^2}{2!}\left( \omega _a\tau _a\right) ^2+...\right) \tau _2\Phi ^{*}
\nonumber \\
&=&i\text{e}^{i\omega _a\tau _a}\tau _2\Phi ^{*}=\text{e}^{i\omega
_a\tau _a}\Phit
\end{eqnarray}
and (\ref{eq6.96}) is thus proved\index{Yukawa coupling|)}.

%\end{document}

%\end{document}

%\input{kniha67}
%%%%%%%%%%%%%%%%%%%%%%%%%%%%%%%%%%%%%%%%%%%%%%%%%%%%%%%%%%%%%%%%%%%
%%%%%%%%%%%%%%%%%%%%%%%%%%%%%%%%%%%%%%%%%%%%%%%%%%%%%%%%%%%%%%%%%%%%%%%%%%%%%%%%%%%%%%%%%%%%%%%%%%%%%%%%%%%%%%%%%%%%%%%%%%%%%%%%%%%%%%%%
%\input{tcilatex}
%\begin{document}
%\documentstyle{article}

%%%%%%%%%%%%%%%%%%%%%%%%%%%%%%%%%%%%%%%%%%%%%%%%%%%%%%%%%%%%%%%%%%%%%%%%%%%%%%%%%%%%%%%%%%%%%%%%%%%%%%%%%%%%%%%%%%%%%%%%%%%%
%TCIDATA{Created=Tue Sep 18 09:34:06 2001}
%TCIDATA{LastRevised=Wed Sep 19 15:19:31 2001}

%\begin{document}

\section{Higgs--Yukawa mechanism and parity violation}\index{parity!violation|ff}

The mechanism employed for generating masses within the standard GWS model
has another interesting aspect that deserves attention. In particular, it
turns out that the \qq{Higgs--Yukawa scheme} adopted here leads quite
naturally to the familiar parity-violating weak interactions as well as to
the parity-conserving electromagnetic current (the properties of the weak
neutral currents\index{neutral current} then follow automatically in the usual way). This
statement, that may seem somewhat surprising at first sight, will be
explained below (the argument is essentially due to M. Veltman \cite{ref58}).

For the sake of simplicity, let us restrict ourselves to the electron-type
leptons $\nu _e,$ $e$ -- in fact, adding further fermion species does not
bring anything new in the present context. As before, we shall assume that
the Higgs mechanism is realized via one complex doublet $\Phi $ (let us
recall that this is the minimum option giving the right values of vector
boson masses). Now, the Yukawa interaction that is supposed to produce the
electron mass must involve both $e_L$ and $e_R$ and it is also clear that
the two chiral components of the electron field must have different
transformation properties under the weak isospin $SU(2)$. Indeed, if they were
e.g. both singlets, then by coupling them to the doublet $\Phi $ one could
not get an $SU(2)$ singlet interaction term; a similar problem would occur if
both $e_L$ and $e_R$ belonged to doublets (note that one is certainly not
able to make a singlet out of three doublets -- mathematically, this would
be tantamount to adding three spins 1/2 to a resulting zero value). Thus, if
one considers only the lowest-dimensional representations of the $SU(2)$, the $%
e_L$ should belong to a doublet and $e_R$ to a singlet or the other way
round. Conventionally, we choose the first possibility, i.e. we place the $%
e_L$ into the usual doublet $L$ (cf. (\ref{eq6.81})) and the $e_R$ is taken to be
singlet under $SU(2)$; for simplicity we shall ignore here $\nu _R$.
Remembering now how the gauge interactions are constructed (see Chapter~\ref{chap5},
in particular the formulae (\ref{eq5.14}) and (\ref{eq5.17})), it becomes clear that weak
interactions necessarily exhibit maximum parity violation (the charged
currents are left-handed owing to our option). Of course, had we chosen the
other possibility (namely a doublet consisting of $\nu _R$ and $e_R$, with $%
e_L$ being an $SU(2)$ singlet), the charged weak currents would be purely
right-handed -- but this would mean a maximum parity violation anyway. Thus,
these simple considerations show that the familiar pattern of parity
violation in weak interactions emerges quite naturally: {\bf if one insists
on generating the lepton mass through a Yukawa coupling\index{Yukawa coupling|ff} involving the Higgs
doublet, different $\boldsymbol{SU(2)}$ transformation properties of the left- and
right-handed lepton fields are inevitable.}

Next, let us examine the consequences of the Higgs--Yukawa mechanism for
the parity properties of the electromagnetic interaction. We shall denote
the weak hypercharges of the $L$, $e_R$ and $\Phi $ as $Y_L$, $Y_R$ and $Y$.
Since the Yukawa interaction has the form $\bar{L}\Phi e_R$, the invariance
under the hypercharge $U(1)$ gauge subgroup requires that
\begin{equation}\label{eq6.100}
-Y_L+Y+Y_R=0
\end{equation}
We already know (cf. the discussion around the formula (\ref{eq6.60}))
that the diagonalization of the mass matrix for neutral vector
bosons leads to the massive field $Z_\mu $ and a massless $A_\mu
$, expressed in terms of the original gauge fields $A_\mu ^3$ and
$B_\mu $ as
\begin{eqnarray}
Z_\mu &=&cA_\mu ^3-sB_\mu  \nonumber \\
A_\mu &=&sA_\mu ^3+cB_\mu
\end{eqnarray}
where the $c$ and $s$ is a shorthand notation for the $\cos \theta _W$ and $%
\sin \theta _W$ resp.; one has
\begin{equation}\label{eq6.102}
c=\frac g{\sqrt{g^2+4Y^2 g'^2}},\qquad s=\frac{2Yg'}{\sqrt{g^2+4Y^2 g'^2}}
\end{equation}
To identify the interactions of the $A_\mu $ with leptons, one can employ
the formulae derived previously in Chapter~\ref{chap5} (see in particular (\ref{eq5.21}),
where we shall ignore the term involving $\nu _R$). Expressing the $A_\mu ^3$
and $B_\mu $ in terms of $Z_\mu $ and $A_\mu $, one obtains (cf. (\ref{eq5.27}))
\begin{eqnarray}
\lagr_{int}^{\left( A\right) } &=&\left( \frac 12gs+Y_Lg'c\right)
\bar{\nu}_L\gamma ^\mu \nu _LA_\mu  \nonumber \\
&&+\left( -\frac 12gs+Y_Lg'c\right) \bar{e}_L\gamma ^\mu e_LA_\mu
+Y_Rg'c\bar{e}_R\gamma ^\mu e_RA_\mu
\end{eqnarray}
The condition of vanishing neutrino charge reads
\begin{equation}\label{eq6.104}
\frac 12gs+Y_Lg'c=0
\end{equation}
and substituting into (\ref{eq6.104}) the expressions (\ref{eq6.102}) for $c$ and $s$, one
gets readily
\begin{equation}
Y_L=-Y
\end{equation}
This, in combination with (\ref{eq6.100}), yields
\begin{equation}
Y_R=2Y_L
\end{equation}
Using all the relations shown above, the interaction of the $A_\mu $ with
fermions can be worked out as
\begin{eqnarray}
\lagr_{int}^{\left( A\right) } &=&\left( -\frac 12gs-Yg'c\right)
\bar{e}_L\gamma ^\mu e_LA_\mu -2Yg'c\bar{e}_R\gamma ^\mu e_RA_\mu
\nonumber \\
&=&-\frac{2Ygg'}{\sqrt{g^2+4Y^2g'^2}}\left( \bar{e}%
_L\gamma ^\mu e_L+\bar{e}_R\gamma ^\mu e_R\right) A_\mu
\label{eq6.107}
\end{eqnarray}
and the last line of (\ref{eq6.107}) exhibits clearly the envisaged
parity-conserving\index{parity!conservation} nature of the $A_\mu $ (electromagnetic) interaction.

Thus, the preceding considerations can be summarized briefly as follows.

i) The maximum parity violation in charged-current weak interactions emerges
naturally within the GWS standard model, as a consequence of the Higgs--Yukawa
mechanism for generation of fermion masses. The essential point is
that the doublet character of the Higgs field enforces different $SU(2)$
transformation properties upon the left- and right-handed chiral components
of fermion fields.\footnote{%
Of course, we always assume tacitly that only the lowest-dimensional $SU(2)$
representations of fermion fields -- namely the singlets and doublets -- are
relevant.} In this sense, the parity violation in weak currents is
intimately connected with properties of the Higgs sector of the standard GWS
model; the connection is straightforward, though it may appear somewhat
surprising at first sight.

ii) The Higgs--Yukawa mechanism leads automatically to the
parity-con\-serving electromagnetic interaction. More precisely, one gets a
vector-like interaction of the massless physical gauge field emerging from
the standard Higgs mechanism, if one assumes that the Yukawa interaction
responsible for the lepton mass generation is $SU(2)\times U(1)$ invariant
and the neutrino charge is fixed to be zero.

%\input{kniha68}
%%%%%%%%%%%%%%%%%%%%%%%%%%%%%%%%%%%%%%%%%%%%%%%%%%%%%%%%%%%%%%%%%%%
%%%%%%%%%%%%%%%%%%%%%%%%%%%%%%%%%%%%%%%%%%%%%%%%%%%%%%%%%%%%%%%%%%%%%%%%%%%%%%%%%%%%%%%%%%%%%%%%%%%%%%%%%%%%%%%%%%%%%%%%%%%%%%%%%%%%%%%%
%\documentstyle{article}

%%%%%%%%%%%%%%%%%%%%%%%%%%%%%%%%%%%%%%%%%%%%%%%%%%%%%%%%%%%%%%%%%%%%%%%%%%%%%%%%%%%%%%%%%%%%%%%%%%%%%%%%%%%%%%%%%%%%%%%%%%%%
%TCIDATA{Created=Wed Sep 19 12:52:58 2001}
%TCIDATA{LastRevised=Thu Sep 20 09:25:41 2001}

%\input{tcilatex}
%\begin{document}

\section{Custodial symmetry}\index{custodial symmetry|(}\label{sec6.8}

Let us now turn to a discussion of some deeper symmetry aspects of
the standard \qq{minimal} Higgs system. As we have already noted
earlier (see the remarks around the relation (\ref{eq6.45})), the symmetry
of the Higgs--Goldstone potential\index{Goldstone!potential}
$V\left( \Phi \right) $ is in fact $O(4)$\index{O(4) group@$O(4)$
group|ff}, i.e. it is larger than the mandatory $SU(2)\times
U(1)$: the reason is simply that a $V$ with required properties
must inevitably depend on $\Phi ^{\dagger }\Phi =\varphi
_1^2+\varphi _2^2+\varphi _3^2+\varphi _4^2$, where $\varphi _1,$...,$%
\varphi _4$ are the four real scalar fields that parametrize the
complex doublet $\Phi $ according to (\ref{eq6.42}). For our present
purpose we shall employ such a four-dimensional real
parametrization explicitly, describing the Higgs multiplet as
\begin{equation}
\Phi =\left(
\begin{array}{r}
\varphi _1 \\
\varphi _2 \\
\varphi _3 \\
\varphi _4
\end{array}
\right)
\end{equation}
A \qq{vacuum configuration}\index{vacuum|ff} $\Phi _0$
corresponding to the Lagrangian (\ref{eq6.43}) is in general given by
\begin{equation}\label{eq6.109}
\Phi _0=\left(
\begin{array}{r}
v_1 \\
v_2 \\
v_3 \\
v_4
\end{array}
\right)
\end{equation}
with $v_1^2+v_2^2+v_3^2+v_4^2$=$v^2/2$ (cf. (\ref{eq6.48})). For convenience, we
will often use a particular representant of the manifold (\ref{eq6.109}), namely
\begin{equation}\label{eq6.110}
\Phi _0^{\left( 4\right) }=\left(
\begin{array}{c}
0 \\
0 \\
0 \\
v/\sqrt{2}
\end{array}
\right)
\end{equation}
(strictly speaking, this differs slightly from our previous conventional
choice (\ref{eq6.52}), but it does not matter).

The familiar Goldstone-type symmetry breakdown occurring in the considered
model can now be described in a concise way (in fact, the algebraic clarity
is the main virtue of the real-field formalism in the present context). Let
us start with the specification of the $O(4)$ symmetry generators. A
\qq{canonical} set is represented by the six real antisymmetric matrices
\begin{alignat}{2}
M_1&=\left(
\begin{array}{rrrr}
0 & 0 & 0 & 0 \\
0 & 0 & -1 & 0 \\
0 & 1 & 0 & 0 \\
0 & 0 & 0 & 0
\end{array}
\right),\qquad N_1&&=\left(
\begin{array}{rrrr}
0 & 0 & 0 & -1 \\
0 & 0 & 0 & 0 \\
0 & 0 & 0 & 0 \\
1 & 0 & 0 & 0
\end{array}
\right)
\notag\\
M_2&=\left(
\begin{array}{rrrr}
0 & 0 & 1 & 0 \\
0 & 0 & 0 & 0 \\
-1 & 0 & 0 & 0 \\
0 & 0 & 0 & 0
\end{array}
\right),\qquad N_2&&=\left(
\begin{array}{rrrr}
0 & 0 & 0 & 0 \\
0 & 0 & 0 & -1 \\
0 & 0 & 0 & 0 \\
0 & 1 & 0 & 0
\end{array}
\right)\notag\\
M_3 &=\left(
\begin{array}{rrrr}
0 & -1 & 0 & 0 \\
1 & 0 & 0 & 0 \\
0 & 0 & 0 & 0 \\
0 & 0 & 0 & 0
\end{array}
\right),\qquad N_3&&=\left(
\begin{array}{rrrr}
0 & 0 & 0 & 0 \\
0 & 0 & 0 & 0 \\
0 & 0 & 0 & -1 \\
0 & 0 & 1 & 0
\end{array}
\right)
\label{eq6.111}
\end{alignat}
satisfying the commutation relations
\begin{eqnarray}
\left[ M_j,M_k\right] &=&\epsilon _{jkl}M_l  \nonumber \\
\left[ M_j,N_k\right] &=&\epsilon _{jkl}N_l  \nonumber \\
\left[ N_j,N_k\right] &=&\epsilon _{jkl}M_l
\label{eq6.112}
\end{eqnarray}
Occasionally we will also use the shorthand notation $\vec{M},\ \vec{N}$
for the two triplets of $M_j$ and $N_j$.\footnote{%
Note that it is most natural to use real antisymmetric matrices as
the generators of real orthogonal transformations. Equivalently,
one could work with the hermitean generators $i\vec{M}$,
$i\vec{N}$; in fact, these would fit better into the gauge theory
formalism that we have developed so far. We will return to the
hermitean representation of the symmetry generators later in this
section, when we reconsider the Higgs mechanism.} To examine the
action of the generators (\ref{eq6.111}) on the vacuum state, one may
consider first e.g. the particular choice $\Phi _0^{\left(
4\right) }$ shown in (\ref{eq6.110}). It is immediately seen that the
three four-component vectors $\vec{N}\Phi _0^{\left( 4\right)
}$are non-zero and linearly independent, while the $\vec{M}\Phi
_0^{\left( 4\right) }$ all vanish. Thus, as expected, the $\Phi
_0^{\left( 4\right) }$ breaks the considered symmetry in three
directions, i.e. with respect to the three generators $\vec{N}$.
In fact, such a result can be easily generalized -- for an
arbitrary representant $\Phi _0$ of the vacuum manifold (as given
by (\ref{eq6.109})) one can show that the space made of
the linear combinations of the six real four-component vectors $%
\vec{M}\Phi _0$, $\vec{N}\Phi _0$ is $three$-$%
dimensional$ (the proof is left to the reader as an exercise in linear
algebra). Note that this is actually the precise contents of the statement
about the number of broken symmetry generators (associated with Goldstone
bosons), i.e. about the symmetry-breaking pattern occurring within our model.

Next one would like to choose an appropriate electroweak
$SU(2)\times U(1)$\index{SU(2) times U(1) group@$SU(2)\times U(1)$
group} subgroup of our $O(4)$; in other words, one has to identify
the generators corresponding to the weak isospin and hypercharge
(that are to be
gauged subsequently). To this end, it is convenient to pass from $%
\vec{M}$, $\vec{N}$ to another set (basis) of generators, defined
by
\begin{eqnarray}
\vec{L} &=&\frac 12\left( \vec{M}+\vec{N}%
\right)   \nonumber \\
\vec{R} &=&\frac 12\left( \vec{M}-\vec{N}%
\right)
\label{eq6.113}
\end{eqnarray}
Using (\ref{eq6.112}) one then gets easily
\begin{eqnarray}
\left[ L_j,L_k\right]  &=&\epsilon _{jkl}L_l  \nonumber \\
\left[ R_j,R_k\right]  &=&\epsilon _{jkl}R_l  \nonumber \\
\left[ L_j,R_k\right]  &=&0
\label{eq6.114}
\end{eqnarray}
which means that the sets $\vec{L}$, $\vec{R}$ generate two
independent (commuting) $O(3)$ subalgebras of the original
$O(4)$\index{O(3) group@$O(3)$ group}. Mathematically, this
remarkable fact corresponds to the known statement that the group
$O(4)$ is locally (i.e. at the level of Lie algebras) isomorphic
to the direct product $O(3)\times O(3)$; in the common symbolic
notation, $O(4)\simeq O(3)\times O(3)$. In the present context it
is important to recall that the $O(3)$ algebra is isomorphic to
that of $SU(2)$ (indeed, the hermitean matrices $i\vec{L}$ and
$i\vec{R}$ obviously satisfy the familiar $SU(2)$ commutation
relations) and the considered decomposition is therefore usually
also written as $O(4)\simeq SU(2)
\times SU(2)$. Thus, in view of (\ref{eq6.114}), one can take e.g. the $%
\vec{L}$ to be the isospin generators and the hypercharge (that
has to commute with isospin) is then selected among the matrices $%
\vec{R}$ (conventionally, the $R_3$ is chosen). For practical
purposes we shall employ the hermitean generators $\vec{T}$ and
$Y$ defined as
\begin{equation}\label{eq6.115}
\vec{T}=i\vec{L},\qquad Y=i R_3
\end{equation}
Note that their explicit matrix representation can then be written (using
(\ref{eq6.111}) and (\ref{eq6.113})) as
\begin{alignat}{2}
T_1&=\frac{i}{2}\left(
\begin{array}{rrrr}
0 & 0 & 0 & -1 \\
0 & 0 & -1 & 0 \\
0 & 1 & 0 & 0 \\
1 & 0 & 0 & 0
\end{array}
\right),\qquad T_2&&=\frac{i}{2}\left(
\begin{array}{rrrr}
0 & 0 & 1 & 0 \\
0 & 0 & 0 & -1 \\
-1 & 0 & 0 & 0 \\
0 & 1 & 0 & 0
\end{array}
\right)\notag\\
T_3 &=\frac{i}{2}\left(
\begin{array}{rrrr}
0 & -1 & 0 & 0 \\
1 & 0 & 0 & 0 \\
0 & 0 & 0 & -1 \\
0 & 0 & 1 & 0
\end{array}
\right),\qquad Y&&=\frac{i}2\left(
\begin{array}{rrrr}
0 & -1 & 0 & 0 \\
1 & 0 & 0 & 0 \\
0 & 0 & 0 & 1 \\
0 & 0 & -1 & 0
\end{array}
\right)\label{eq6.116}
\end{alignat}
Now, considering again the $\Phi _0^{\left( 4\right) }$ as a vacuum state,
one sees immediately that both $\vec{L}\Phi _0^{\left( 4\right) }$
and $\vec{R}\Phi _0^{\left( 4\right) }$ are different from zero.
However, as noted earlier, the sum $\vec{L}+\vec{R}=%
\vec{M}$ does annihilate the $\Phi _0^{\left( 4\right) }$ (let us
recall that there can be just three independent broken symmetry generators).
In particular, $L_3+R_3=M_3$ yields the electric charge, in correspondence
with the relation $T_3+Y=Q$. From now on, we shall use the hermitean
generators only, so in addition to (\ref{eq6.115}) let us also introduce the
notation
\begin{equation}
\vec{K}=i\vec{M}
\end{equation}
For the $\vec{K}$ and $\vec{T}$ we then have a set of
commutation relations
\begin{eqnarray}
\left[ K_j,K_k\right] &=&i\epsilon _{jkl}K_l  \nonumber \\
\left[ T_j,T_k\right] &=&i\epsilon _{jkl}T_l  \nonumber \\
\left[ T_j,K_k\right] &=&i\epsilon _{jkl}T_l
\end{eqnarray}
that follow immediately from (\ref{eq6.114}) and from the above definitions. These
relations mean, among other things, that the weak isospin generators $%
\vec{T}$ form a three-component vector under rotations generated
by the $\vec{K}$ (notice that the commutation relation between $%
\vec{T}$ and $\vec{K}$ are formally the same as e.g. those between
momentum and angular momentum\index{angular momentum} in ordinary
quantum mechanics). Passing from the algebra of commutators (i.e.
from infinitesimal transformations) to finite rotations, one can
write
\begin{equation}\label{eq6.119}
U^{\dagger }\left( \vec{\omega }\right) T_jU\left(
\vec{\omega }\right) =\mathscr{D}_{jk}\left( \vec{\omega }%
\right) T_k
\end{equation}
where the $U ( \vec{\omega }) =\exp( -i%
\vec{\omega}\cdot \vec{K})$ is unitary $4 \times 4$ matrix and the
$\mathscr{D}_{jk}\left( \vec{\omega }\right) $ represent a (real
orthogonal) matrix of three-dimensional rotation described
by the parameters $\vec{\omega }$.\footnote{%
Let us recall that the generators $\vec{K}$ are hermitean and
pure imaginary by construction, so the $U\left( \vec{\omega }%
\right) $ is in fact a real orthogonal matrix as well.} The relation (\ref{eq6.119})
is of crucial importance for our further considerations and we will return
to it shortly. At this point, let us summarize briefly the essential
features of the symmetry pattern discussed so far:

i) The conventionally chosen vacuum $\Phi _0^{\left( 4\right) }$
(denoted in what follows simply as $\Phi _0$) is invariant under
$SU(2)$ transformations generated by the matrices $\vec{K}$, i.e.
\begin{equation}\label{eq6.120}
U\left( \vec{\omega }\right) \Phi _0=\Phi _0
\end{equation}
for arbitrary $\vec{\omega}$ (of course, this is equivalent to $%
\vec{K}\Phi _0=0$).

ii) The weak isospin generators $\vec{T}$ (corresponding to the
broken symmetry) constitute a triplet with respect to the vacuum symmetry
subgroup -- in other words, they behave as a three-component vector under
rotations generated by the $\vec{K}.$

We are now in a position to reconsider the Higgs mechanism for $SU(2)\times
U(1)$ gauge bosons. The relevant Lagrangian can be written in analogy with
the formula (\ref{eq6.53}), replacing there $\vec{\tau }/2$ by the $%
\vec{T}$ and substituting the matrix
\begin{equation}\label{eq6.121}
Y=Q-T_3=K_3-T_3
\end{equation}
for the weak hypercharge. From the discussion following (\ref{eq6.53}) it is then
clear that the mass term for the vector bosons $A_\mu ^a$ and $B_\mu $ is
given by
\begin{equation}\label{eq6.122}
\lagr_{mass}^{\left( IVB\right) }=\Phi _0^{\dagger }\left( gA_\mu
^aT_a+g'B_\mu Y\right) \left( gA^{b\mu }T_b+g'B^\mu
Y\right) \Phi _0
\end{equation}
Taking into account (\ref{eq6.121}) and using the identities $Q\Phi _0=0$,
$\Phi _0^{\dagger }Q=0$, the expression (\ref{eq6.122}) can be worked out
as
\begin{eqnarray}
\lagr_{mass}^{\left( IVB\right) } &=&g^2\Phi _0^{\dagger }T_aT_b\Phi
_0A_\mu ^aA^{b\mu }-gg'\Phi _0^{\dagger }T_aT_3\Phi _0A_\mu ^aB^\mu
\nonumber \\
&&-gg'\Phi _0^{\dagger }T_3T_b\Phi _0A_\mu ^bB^\mu +g'^2\Phi _0^{\dagger }\left( T_3\right) ^2\Phi _0B_\mu B^\mu
\label{eq6.123}
\end{eqnarray}
All coefficients in (\ref{eq6.123}) are of the form $\Phi _0^{\dagger }T_aT_b\Phi _0$%
, so let us now examine the properties of such an algebraic expression.
Invoking the invariance of the $\Phi _0$ under the unitary transformations $%
U\left( \vec{\omega }\right) $ (see (\ref{eq6.120})) and making use of
the fundamental symmetry relation (\ref{eq6.119}), one thus obtains
\begin{eqnarray}
\Phi _0^{\dagger }T_aT_b\Phi _0 &=&\Phi _0^{\dagger }U^\dagger \left(
\vec{\omega }\right) T_aT_bU\left( \vec{\omega }%
\right) \Phi _0=  \nonumber \\
&=&\Phi _0^{\dagger }U^{\dagger }\left( \vec{\omega }\right)
T_aU\left( \vec{\omega }\right) U^{\dagger }\left(
\vec{\omega }\right) T_bU\left( \vec{\omega }\right)
\Phi _0=  \nonumber \\
&=&\mathscr{D}_{aj}\left( \vec{\omega }\right)
\mathscr{D}_{bk}\left( \vec{\omega }\right) \Phi _0^{\dagger
}T_jT_k\Phi _0
\end{eqnarray}
which means that the numerical coefficients $\Phi _0^{\dagger }T_aT_b\Phi _0$
behave as components of a 2nd rank tensor under three-dimensional rotations.
On the other hand, these coefficients are pure numbers (i.e. they are
obviously independent of the \qq{reference frame} characterized by the
transformation parameters $\vec{\omega }$); in other words, the
expression $\Phi _0^{\dagger }T_aT_b\Phi _0$ represents an \qq{isotropic
tensor} (i.e. such that its components are the same in all reference
frames). This, of course, is rather severe restriction and it is not
surprising that the 2nd rank tensor endowed with this property is
essentially unique: it is the Kronecker delta (up to a constant
multiplicative factor). Thus, solely on the basis of our symmetry arguments
we can write
\begin{equation}\label{eq6.125}
\Phi _0^{\dagger }T_aT_b\Phi _0=N\delta _{ab}
\end{equation}
where $N$ is a numerical (normalization) constant. The value of
the $N$ can be fixed by taking into account the explicit
representation of the generators $\vec{T}$ shown in (\ref{eq6.116}); one
thus finds easily that $N=v^2/8$. Substituting now the result
(\ref{eq6.125}) into (\ref{eq6.123}), one gets
\begin{equation}
\lagr_{mass}^{\left( IVB\right) }=g^2NA_\mu ^aA^{a\mu }-2gg'NA_\mu ^3B^\mu
+g'^2 NB_\mu B^\mu
\end{equation}
and the last expression is immediately diagonalized as
\begin{equation}
\lagr_{mass}^{\left( IVB\right) }=N\left[ g^2\left( A_\mu ^1A^{1\mu
}+A_\mu ^2A^{2\mu }\right) +\left( gA_\mu ^3-g'B_\mu \right)
^2\right]
\end{equation}
Identifying the physical vector fields $W_\mu ^{\pm },Z_\mu $ and their
masses in an analogous manner as in Section~\ref{sec6.4}, one has finally
\begin{equation}
\lagr_{mass}^{\left( IVB\right) }=m_W^2W_\mu ^{-}W^{+\mu }+\frac
12m_Z^2Z_\mu Z^\mu
\end{equation}
with
\begin{equation}\label{eq6.129}
m_W^2=2g^2N,\qquad m_Z^2=2\left( g^2+g'^2\right) N
\end{equation}
Notice that for the relevant value $N=v^2/8$ specified above one reproduces,
as expected, the standard formulae (\ref{eq6.69}). Even without using an explicit
value of the $N$, the result (\ref{eq6.129}) obviously yields the famous Weinberg
relation
\begin{equation}\label{eq6.130}
m_W^2/m_Z^2=g^2/\left( g^2+g'^2\right)
\end{equation}
In other words (taking into account that $%
g/(g^2+g'^2)^{1/2}=\cos \theta _W$), one recovers the value $\rho
=1$ for the parameter $\rho =m_W^2/( m_Z^2\cos ^2\theta _W)$
introduced earlier (cf. (\ref{eq5.63}) and the end of Section~\ref{sec6.4}).

It is important to realize that eq. (\ref{eq6.130}) has been derived here on the
symmetry grounds only\footnote{%
Remember that in the elementary treatment of Section~\ref{sec6.4} we had to
invoke some specific algebraic properties of the Pauli matrices
etc., to arrive at the same result.} and one thus gains a deeper
insight into the origin of the observed pattern of vector boson
masses.{\bf In particular, our argument relied substantially on
the fact that the vacuum symmetry is a global
$\boldsymbol{SU(2)}$, under which the (gauged) weak isospin
generators behave as a triplet.} In
this sense, the vacuum symmetry $SU(2)$ controls, or \qq{protects}, the value $%
\rho =1$ and therefore it is usually called the {\bf custodial
symmetry}. It should be stressed that the existence of such a
symmetry is not tied with a particular choice of the vacuum: for
the sake of technical simplicity we have chosen here the $\Phi
_0^{\left( 4\right) }$ shown in (\ref{eq6.110}), but in fact any $\Phi _0$
belonging to the set (\ref{eq6.109}) is invariant under an $SU(2)$,
generated by appropriately \qq{rotated} matrices $\vec{K}$ (the
gauged generators must then also be modified accordingly). Note
finally that the concept of custodial $SU(2)$ symmetry has
appeared for the first time in the paper \cite{ref59} and because
of its rather general nature one can utilize it also in some
schemes of electroweak symmetry breaking that go beyond the
standard model -- e.g. when one considers a generic model of
\qq{dynamical symmetry breaking} described by an effective
Lagrangian not involving elementary physical Higgs fields (see
e.g. \cite{ref60})\index{custodial symmetry|)}.

%\input{kniha69}
%%%%%%%%%%%%%%%%%%%%%%%%%%%%%%%%%%%%%%%%%%%%%%%%%%%%%%%%%%%%%%%%%%%
%%%%%%%%%%%%%%%%%%%%%%%%%%%%%%%%%%%%%%%%%%%%%%%%%%%%%%%%%%%%%%%%%%%%%%%%%%%%%%%%%%%%%%%%%%%%%%%%%%%%%%%%%%%%%%%%%%%%%%%%%%%%%%%%%%%%%%%%
%\documentstyle{article}

%%%%%%%%%%%%%%%%%%%%%%%%%%%%%%%%%%%%%%%%%%%%%%%%%%%%%%%%%%%%%%%%%%%%%%%%%%%%%%%%%%%%%%%%%%%%%%%%%%%%%%%%%%%%%%%%%%%%%%%%%%%%
%TCIDATA{Created=Mon Sep 17 14:09:13 2001}
%TCIDATA{LastRevised=Thu Sep 20 09:28:50 2001}

%%%%%%%%%%%%%%%%%%%%%%%%%%%%%%%%%%%%%%%%%%%%%%%%%%%%%%%%%%%%%%%%%%%%%%%%%%%%%%%%%%%%%%%%%%%%%%%%%%%%%%%%%%%%%%%%%%%%%%%%%%%%

\section{Non-standard Higgs multiplets}\index{Higgs!multiplets, non-standard|(}

Up to now we have focused our attention on the Higgs mechanism operating
within the standard electroweak theory. 
Despite current success of SM involving the solitary Higgs boson, it is not excluded that future experiments will reveal the existence of some siblings of $H$, i.e. some extra scalar bosons belonging to a broader Higgs-like family (in fact, many theorists and experimentalists do hope so). Thus, it may be instructive to discuss briefly extended Higgs-like scalar systems that go beyond SM, but still could basically fit into the overall picture of present-day phenomenology. 
In particular, it is useful
to know how the mass relation for the vector bosons $W$\index{W
boson@$W$ boson} and $Z$\index{Z boson@$Z$ boson|ff} is modified
in the presence of higher scalar $SU(2)$ multiplets -- in other words, how
such a relation depends on the values of weak isospin $T$ and hypercharge $Y$
labelling the Higgs multiplet in question.

To begin with, let us recall that the standard doublet carries $T=1/2$ and $%
Y=1/2$ and its electrically neutral component (which acquires a non-zero
vacuum value) has the third component of isospin $T_3=-Y=-1/2$, in
accordance with the relation $Q=T_3+Y$. For a general value of $T$ (integer
or half-integer) we have a multiplet consisting of $2T+1$ complex components%
\footnote{%
We return here to the complex Higgs fields since such a notation is most
compact and very convenient for our present purpose.} (corresponding to $%
T_3=-T,...,T$) that can be written as
\begin{equation}\label{eq6.131}
\Phi =\bm{\varphi _{T,T} \\ \varphi _{T,T-1} \\ \vdots\\ \varphi _{T,-T}}
\end{equation}
Let the weak hypercharge of this multiplet be $Y$. The neutral component (to
be shifted away from the vacuum value) then has the weak isospin projection $%
T_3=T_3^{\left( 0\right) }=-Y$ and the vacuum value of the multiplet (\ref{eq6.131})
(which minimizes the Higgs--Goldstone potential) is, in analogy with the
standard case
\begin{equation}
\Phi _0=\frac v{\sqrt{2}}\bm{0\\ \vdots \\ 1\\ \vdots\\ 0}
\end{equation}
i.e. the $\Phi _0$ is an eigenvector of the $SU(2)$ generator $T_3$
corresponding to the eigenvalue $T_3^{\left( 0\right) }\left( =-Y\right) $.
Let us now see what are the vector boson masses resulting from the Higgs
mechanism based on the scalar multiplet (\ref{eq6.131}). As before, from the
Higgs--Goldstone Lagrangian of the type (\ref{eq6.53}) one gets the quadratic mass
term
\begin{equation}\label{eq6.133}
\lagr_{mass}^{IVB}=\Phi _0^{\dagger }\left( gA_\mu ^aT_a+g'YB_\mu \right)
\left( gA^{b\mu }T_b+g'YB^\mu \right) \Phi _0
\end{equation}
where the $T_a$, $a=1,2,3$ are $\left( 2T+1\right) \times \left( 2T+1\right) $
matrices representing the $SU(2)$ generators, with $T_3$ taken to be diagonal
and the $Y$ is a multiple of unit matrix. Introducing the isospin raising
and lowering operators $T_{\pm }=T_1\pm $i$T_2$, as well as the $W^{\pm }$
fields (cf. the discussion around eq.\,(\ref{eq5.17})), the expression (\ref{eq6.133}) is recast
as
\begin{eqnarray*}
\lagr_{mass}^{IVB} &=&\Phi _0^{\dagger }\left\{ g\left[ \frac 1{\sqrt{2}%
}\left( T_{-}W_\mu ^{-}+T_{+}W_\mu ^{+}\right) +T_3A_\mu ^3\right]
+g'YB_\mu \right\} \\
&&\times \left\{ g\left[ \frac 1{\sqrt{2}}\left( T_{-}W^{-\mu }+T_{+}W^{+\mu
}\right) +T_3A^{3\mu }\right] +g'Y B^\mu \right\} \Phi _0
\end{eqnarray*}
and this becomes, after simple algebraic manipulations
\begin{equation}\label{eq6.134}
\lagr_{mass}^{IVB}=\Phi _0^{\dagger }\left[ \frac 12g^2\left(
T_{-}T_{+}+T_{+}T_{-}\right) W_{\mu \;}^{-}W^{+\mu }+Y^2\left( gA_\mu
^3-g'B_\mu \right) ^2\right] \Phi _0
\end{equation}
Note that in arriving at (\ref{eq6.134}) we have utilized the familiar properties of
the ladder operators $T_{\pm }$ and the fact that the $\Phi _0$ is an
eigenvector of the $T_3$ (with the eigenvalue $-Y$); one thus has, in
particular, $\Phi _0^{\dagger }\left( T_{\pm }\right) ^2\Phi _0=0$ and $\Phi
_0^{\dagger }T_{\pm }T_3\Phi _0=\Phi _0^{\dagger }T_3T_{\pm }\Phi _0=0$.
Finally, using the identities $T_{-}T_{+}+T_{+}T_{-}=2\left( \vec{%
T}^2-T_3^2\right) $ and $\vec{T}^2\Phi _0=T\left( T+1\right) \Phi
_0$, we get from (\ref{eq6.134})
\begin{equation}
\lagr_{mass}^{IVB}=\frac 12g^2v^2\left[ T\left( T+1\right) -Y^2\right]
W_\mu ^{-}W^{+\mu }+\frac 12\left( g^2+g'^2\right) v^2Y^2Z_\mu Z^\mu
\end{equation}
where $Z_\mu =\cos \theta _WA_\mu ^3-\sin \theta _WB_\mu $ with $\cos \theta
_W=g/\left( g^2+g'^2\right) ^{1/2}$.

Thus, the corresponding masses can be identified as follows\index{mass formula for $W$ and $Z$}
\begin{eqnarray}
m_W^2 &=&\frac 12g^2v^2\left[ T\left( T+1\right) -Y^2\right]  \nonumber \\
m_Z^2 &=&\left( g^2+g'^2\right) v^2Y^2
\label{eq6.136}
\end{eqnarray}
These formulae represent our desired goal. Let us now discuss the contents
of (\ref{eq6.136}) in more detail. Obviously, for the standard-model values $%
T=1/2,Y=1/2$ one recovers our previous result (cf. (\ref{eq6.68})). An interesting
feature of the formulae (\ref{eq6.136}) is the proportionality of the $m_Z$ to $Y$.
This means that e.g. for a Higgs triplet ($T=1$) with $Y=0$ (i.e. with
neutral middle component) one gets $m_Z=0$. In other words, a real scalar
triplet
\begin{equation}\label{eq6.137}
\Phi =\left(
\begin{array}{c}
\varphi ^{+} \\
\varphi ^0 \\
\varphi ^{-}
\end{array}
\right) ,\qquad\varphi ^{-}=\left( \varphi ^{+}\right) ^{*},\qquad \varphi
^0\quad \text{real}
\end{equation}
can only give mass to $W^{\pm }$ but not to the $Z$. Such an observation is
in fact quite instructive: it demonstrates explicitly that three real scalar
fields involved in a Goldstone-type potential are not enough for generating
realistic masses of the three vector bosons $W^{-}$, $W^{+}$ and $Z$
\footnote{%
Of course, this is equivalent to the fact that only two of the three real
scalar fields contained in (\ref{eq6.137}) can be identified as Goldstone bosons
when the corresponding potential $V\left( \Phi \right) $ is worked out.
Thus, in accordance with the general theorems, the Higgs mechanism results
in two massive vector bosons corresponding to the (unphysical) Goldstone
bosons and one real scalar acquires a mass and becomes physical. We shall
return to the example of the real Higgs triplet at the end of this section.}
-- as we have already noticed in Section~\ref{sec6.4}, one needs at least four real
scalars (e.g. those contained within the standard complex doublet).

Assuming generally that $Y\neq 0$, one obtains from (\ref{eq6.136}) a simple
formula for the ratio of the vector boson masses
\begin{equation}
\frac{m_W^2}{m_Z^2}=\frac{g^2}{g^2+g'^2}\frac{T\left( T+1\right) -Y^2%
}{2Y^2}
\end{equation}
In terms of the parameter $\rho =m_W^2/\left( m_Z^2\cos ^2\theta _W\right) $
it means that
\begin{equation}\label{eq6.139}
\rho =\frac{T\left( T+1\right) -Y^2}{2Y^2}
\end{equation}
As we have already noted, the value of the $\rho $ is very close to unity in
the real world, so it is desirable to have $\rho =1$ at the classical level.
From (\ref{eq6.139}) it is clear that such a relation is valid whenever the Higgs
multiplet is chosen so that the $T$ and $Y$ satisfy the equation
\begin{equation}\label{eq6.140}
T\left( T+1\right) -3Y^2=0
\end{equation}
(obviously, $T$ and $Y$ must be either both integer or both half-integer).
The first few solutions of eq. (\ref{eq6.140}) are
\begin{equation}
\left( T,Y\right) =\left( \frac 12,\frac 12\right) ,~\left( 3,2\right)
,~\left( \frac{25}2,\frac{15}2\right) ,...
\end{equation}
It is amusing to observe (see \cite{ref61}) that there are eleven solutions of
eq.\,(\ref{eq6.140}) less than $10^6$, the biggest one being $T=489060\frac
12,Y=282359\frac 12$.

The above result (\ref{eq6.139}) can be slightly generalized as follows. If there
are several Higgs scalar multiplets having in general different vacuum
values, one gets, instead of (\ref{eq6.139})
\begin{equation}
\rho =\frac{\sum_{T,Y}\left| v_{T,Y}\right| ^2\left[ T\left( T+1\right)
-Y^2\right] }{2\sum_{T,Y}\left| v_{T,Y}\right| ^2Y^2}
\end{equation}
From the last expression it is particularly clear that for a Higgs sector
consisting of doublets only, one always has $\rho =1,$ independently of the
values of $v_{T,Y}$ (note that for $T=\frac 12$, the only possible values of
$Y$ are $\pm \frac 12$).

In closing this section let us remark that we have not discussed the problem
of defining a $U$-gauge\index{U-gauge@$U$-gauge|ff} (i.e. that in which the would-be Goldstone bosons
are eliminated) for an extended Higgs sector considered here. As we have
noted before, there is a general proof that such a $U$-gauge always exists
(this can be found in ref. \cite{ref55}; see also the review article \cite{AbL}
and the book \cite{Hua}). Let us give at least an example of a non-standard
Higgs multiplet for which a $U$-gauge can be defined explicitly in a
straightforward manner, similarly to the case of the standard doublet (cf.
(\ref{eq6.50})). The example to be considered here is the real triplet (\ref{eq6.137}) (note
that such a Higgs sector was relevant e.g. in the old Georgi--Glashow $SU(2)$
(or $O(3)$) electroweak model\index{Georgi--Glashow model} \cite{ref62} that avoided a neutral vector boson $Z$ in
favour of heavy leptons\index{heavy leptons} -- see also \cite{Hor}). When working with
(\ref{eq6.137}), one has to choose a corresponding basis of the $SU(2)$ generators
carefully so that the $\Phi $ be transformed into the same form. It is easy
to find out that a suitable $SU(2)$ basis is
\[
T_1=\frac 1{\sqrt{2}}\left(
\begin{array}{rrr}
0 & -1 & 0 \\
-1 & 0 & 1 \\
0 & 1 & 0
\end{array}
\right),\qquad T_2=\frac 1{\sqrt{2}}\left(
\begin{array}{rrr}
0 & i & 0 \\
-i & 0 & -i \\
0 & i & 0
\end{array}
\right) ,
\]
\begin{equation}
T_3=\left(
\begin{array}{rrr}
1 & 0 & 0 \\
0 & 0 & 0 \\
0 & 0 & -1
\end{array}
\right)
\end{equation}
Then it is not difficult to show that a $\Phi $ defined by (\ref{eq6.137}) can be
written as
\begin{equation}
\Phi =\exp \left[i\left( \xi _{-}T_{+}+\xi _{+}T_{-}\right) \right]
\left(
\begin{array}{c}
0 \\
\eta \\
0
\end{array}
\right)
\end{equation}
where $T_{\pm }=T_1\pm \text{i} T_2$, $\xi _{-}=\xi _{+}^{*}$ and the $\eta $ is
real. Of course, $\xi _{\pm }$ then represent the would-be Goldstone bosons\index{Goldstone!boson|)}
if a scalar potential $V^{\pm }\left( \Phi \right) $ of the usual type is
considered and the $\eta $ (when shifted appropriately) becomes a physical
Higgs scalar boson\index{Higgs!multiplets, non-standard|)}.
%\end{document}

%\input{problems6}
%%%%%%%%%%%%%%%%%%%%%%%%%%%%%%%%%%%%%%%%%%%%%%%%%%%%%%%%%%%%%%%%%%%
%%%%%%%%%%%%%%%%%%%%%%%%%%%%%%%%%%%%%%%%%%%%%%%%%%%%%%%%%%%%%%%%%%%%%%%%%%%%%%%%%%%%%%%%%%%%%%%%%%%%%%%%%%%%%%%%%%%%%%%%%%%%%%%%%%%%%%%%
\begin{priklady}{11}

\item Show that the tree-level matrix element for the process
$W^+_L W^-_L \rightarrow HH$ behaves well in the high-energy limit
(i.e. for $s\gg m_W^2, m_H^2$).

\item Prove that an analogous statement holds also for the process
$e^+e^- \rightarrow Z_L H$. Keep $m_e \neq 0$, in order to
appreciate the mechanism of cancellations of high-energy
divergences arising from the individual diagrams.

\item Calculate the cross section $\sigma(e^+e^- \rightarrow ZH)$ as a function
of the c.m. energy and of the Higgs boson mass. For simplicity,
set $m_e=0$ throughout the calculation (obviously, such an
approximation is absolutely safe, because of the high threshold
energy for the considered process).

\item Show that the tree-level matrix element for the process $Z_L
Z_L \rightarrow Z_L Z_L$ is free of high-energy divergences.
Examine also the dependence of the scattering amplitude in
question on the Higgs boson mass.

\item For the SM Higgs boson with mass $m_H = 125\ \GeV$ make an order-of-magnitude estimate of the cross section for $e^+e^-
\rightarrow HH$ at the energy $E_{c.m.} = 500\ \GeV$ (at the tree level). Compare your
estimate with the cross section for $e^+e^- \rightarrow ZH$ and
also with the QED cross section for $\sigma(e^+e^-\rightarrow
\mu^+ \mu^-)$.

\item Write down a most general Higgs--Goldstone potential in the SM
extension involving two complex scalar doublets (this is currently
popular under the label THDM, an acronym for \qq{two-Higgs doublet
model}).\\ {\it Hint}: Consult the monograph \cite{Gun}.
\end{priklady}

%\input{kniha71}
%%%%%%%%%%%%%%%%%%%%%%%%%%%%%%%%%%%%%%%%%%%%%%%%%%%%%%%%%%%%%%%%%%%
%%%%%%%%%%%%%%%%%%%%%%%%%%%%%%%%%%%%%%%%%%%%%%%%%%%%%%%%%%%%%%%%%%%%%%%%%%%%%%%%%%%%%%%%%%%%%%%%%%%%%%%%%%%%%%%%%%%%%%%%%%%%%%%%%%%%%%%%
%\documentclass[12pt,tbtags]{report}
%\usepackage{amsmath,amssymb,epsfig}
%\newcommand{\qq}[1]{\textquotedblleft#1\/\textquotedblright}
%\newcommand{\J}{\mathbb{I}}
%\begin{document}
\chapter{Standard model of electroweak interactions}\label{chap7}

\section{Leptonic world -- brief recapitulation}

In previous chapters we have discussed in some detail the basic
principles upon which the GWS electroweak theory is built. As
regards the spectrum of elementary fermions, we have restricted
ourselves -- for simplicity of the exposition -- to its leptonic
part. In the following sections we will complete the edifice of
the standard electroweak model by incorporating its quark sector.
Before doing it, let us summarize here very briefly (mostly for
reference purposes and for reader's convenience) the relevant
results that we have achieved so far in our description of the
\qq{leptonic world}.

As we know, in building the GWS electroweak theory one relies on
two basic principles, namely\index{gauge invariance}

1) gauge symmetry $SU(2)\times U(1)$

2) Higgs mechanism\index{Higgs mechanism|)} realized via a complex
scalar doublet

\noindent For convenience, one may fix the physical
$U$-gauge\index{U-gauge@$U$-gauge}, which means that the Higgs
doublet\index{Higgs!doublet} $\Phi=\binom{\varphi^+}{\varphi^0}$
becomes
\begin{equation}
\label{eq7.1}
\Phi = \Phi_U =
\begin{pmatrix}
 0 \\ \frac{1}{\sqrt{2}}(v+H)
\end{pmatrix}
\end{equation}
(cf. (\ref{eq6.54})). Let us also recall that fixing the $U$-gauge is
formally equivalent to an $SU(2)$ gauge transformation and the
$U$-gauge GWS Lagrangian can thus be obtained from its gauge
invariant form simply by replacing the $\Phi$ with $\Phi_U$.

We have seen that the GWS Lagrangian of the leptonic world can be
written, schematically, as
\begin{equation}
\label{eq7.2} \lagr^{GWS} = \lagr_{gauge} +\lagr_{lepton}
+ \lagr_{Higgs} + \lagr_{Yukawa}
\end{equation}
and the individual terms appearing in (\ref{eq7.2}) were described
in great detail in preceding chapters. Now we are going to focus
our attention on the term $\lagr_{lepton}$ that describes the
interactions of leptons with vector bosons (it will serve as a
starting point for our preliminary discussion of quark sector in
the next section). The $\lagr_{lepton}$ for a particular lepton
species $\ell$ ($\ell = e, \mu$ or $\tau$\index{tau lepton@$\tau$
lepton}) can be written as
\begin{equation}
\label{eq7.3}
\begin{split}
\lagr_{lepton}^{(\ell)} = \phantom{+} &i\bar{L}^{(\ell)}\gamma^\mu
(\partial_\mu - i g A_\mu^a \frac{\tau^a}{2} - i g' Y_L^{(\ell)}
B_\mu) L^{(\ell)} \\ + \;& i \bar{R}^{(\ell)}\gamma^\mu
(\partial_\mu - i g' Y_R^{(\ell)} B_\mu) R^{(\ell)}
\end{split}
\end{equation}
where
\begin{equation}
\label{eq7.4} L^{(\ell)}=
\begin{pmatrix}
{\nu_\ell}_L \\ \ell_L
\end{pmatrix},
\qquad R^{(\ell)} = \ell_R
\end{equation}
(we ignore here momentarily the right-handed neutrino field
${\nu_\ell}_R$). The weak hypercharges\index{weak!hypercharge|(}
$Y_L^{(\ell)}, Y_R^{(\ell)}$ are fixed by the rule
\begin{equation}
\label{eq7.5}
Q = T_3 + Y
\end{equation}
(cf. (\ref{eq5.40})); in this way, one gets
$Y_L^{(\ell)}=-\frac{1}{2}$ and $Y_R^{(\ell)}=-1$. (Needless to
say, in a full Lagrangian incorporating all lepton species one has
to take the sum of the expressions (\ref{eq7.3}) over $\ell =
e,\mu,\tau$.) Working out (\ref{eq7.3}), one recovers easily the
interactions of familiar charged $V-A$ weak currents with vector
bosons $W^\pm$, namely
\begin{align}
\lagr_{CC}^{(\ell)} &= \frac{g}{\sqrt{2}} \bar{\nu}_{\ell L}
\gamma^\mu \ell_L W_\mu^+ + \text{h.c.} \notag \\
\label{eq7.6} &= \frac{g}{2\sqrt{2}} \bar{\nu}_\ell \gamma^\mu
(1-\gamma_5) \ell W_\mu^+ + \text{h.c.}
\end{align}
where
\begin{equation}
\label{eq7.7}
W_\mu^\pm = \frac{1}{\sqrt{2}} (A_\mu^1 \mp i A_\mu^2)
\end{equation}
Let us recall that (\ref{eq7.6}) descends from the part of the
covariant derivative in (\ref{eq7.3}) involving non-diagonal
matrices $\tau^1$ and $\tau^2$ (cf. (\ref{eq5.15}) through
(\ref{eq5.18})). Further, in the neutral sector of (\ref{eq7.3})
(containing diagonal matrices $\tau^3$ and $\J$) neither $A^3_\mu$
nor $B_\mu$ can be interpreted as the electromagnetic
field\index{electromagnetic!field}. Therefore, an orthogonal
transformation\index{orthogonal transformation}
\begin{align}
A^3_\mu &= \phantom{+}\cos{\theta_W} Z_\mu + \sin{\theta_W} A_\mu \notag\\
\label{eq7.8}
B_\mu &= -\sin{\theta_W} Z_\mu + \cos{\theta_W} A_\mu
\end{align}
introducing physical fields $A_\mu$ and $Z_\mu$ must be performed
and, if the $\theta_W$ is chosen so that
\begin{equation}
\cos{\theta_W} = \frac{g}{\sqrt{g^2+g'^2}}, \quad
\sin{\theta_W}=\frac{g'}{\sqrt{g^2 + g'^2}}
\end{equation}
the $A_\mu$ is coupled to the ordinary electromagnetic current.
The $Z_\mu$ interacts with the weak neutral current\index{neutral current|ff}; one gets
\begin{equation}
\label{eq7.10} \lagr_{NC}^{(\ell)} = \frac{g}{\cos{\theta_W}}
\sum_{f=\ell_L, \ell_R, \nu_{\ell L}} \varepsilon_f \bar{f}
\gamma_\mu f Z^\mu
\end{equation}
where
\begin{equation}
\label{eq7.11}
\varepsilon_f = T_{3f} - Q_f \sin^2{\theta_W}
\end{equation}
It is important to realize that (owing to the simplicity of
algebraic manipulations leading from (\ref{eq7.3}) to
(\ref{eq7.6})) there is a straightforward connection between the
contents of the leptonic doublet $L^{(\ell)}$ in (\ref{eq7.4}) and
the structure of the current: the charged current\index{charged
current|ff} is simply composed of the upper and lower component of
the $L^{(\ell)}$. In a similar way, the neutral current is made of
upper and lower components of $L^{(\ell)}$ separately and it also
gets a contribution from right-handed singlets.

Now one would like to generalize the GWS gauge theory construction
of lepton currents so as to reproduce the phenomenologically
successful Cabibbo form of the quark weak current. We already know
that recovering a desired form of the charged current is just a
matter of proper choice of the basic fermion building blocks
($SU(2)$ doublets and singlets) but, once such a choice is made, a
definite structure of the weak neutral current already follows as
a pure theoretical prediction. On the other hand, experimental
data put severe constraints on the phenomenology of hadronic
neutral current interactions. Thus, it is clear a priori that in
any attempt at extending the $SU(2)\times U(1)$ electroweak gauge
theory to the quark sector one has to deal seriously with the
issue of neutral currents.
%\end{document}

%\input{kniha72}
%%%%%%%%%%%%%%%%%%%%%%%%%%%%%%%%%%%%%%%%%%%%%%%%%%%%%%%%%%%%%%%%%%%
%%%%%%%%%%%%%%%%%%%%%%%%%%%%%%%%%%%%%%%%%%%%%%%%%%%%%%%%%%%%%%%%%%%%%%%%%%%%%%%%%%%%%%%%%%%%%%%%%%%%%%%%%%%%%%%%%%%%%%%%%%%%%%%%%%%%%%%%
%\documentclass[12pt,tbtags]{report}
%\usepackage{amsmath,amssymb,epsfig}
%\newcommand{\qq}[1]{\textquotedblleft#1\/\textquotedblright}
%\newcommand{\J}{\mathbb{I}}
%\begin{document}
\section{Difficulties with three quarks}\label{sec7.2}
In Chapter~\ref{chap2} we have written the hadronic part of the charged weak
current in terms of the quark fields $u,d,s$
as\index{dquark@$d$-quark|ff}\index{u-quark@$u$-quark|ff}\index{s-quark@$s$-quark|ff}
\begin{equation}
\bar{u} \gamma_\mu (1-\gamma_5) (d\cos{\theta_C} + s
\sin{\theta_C})
\end{equation}
(see (\ref{eq2.59})) where $\theta_C$ is the Cabibbo
angle\index{Cabibbo angle}. It means that the interaction of
quarks with charged vector bosons can be described by the
Lagrangian
\begin{align}
\lagr_{CC}^{(u,d,s)} &= \frac{g}{2\sqrt{2}} \bar{u} \gamma^\mu
(1-\gamma_5) (d \cos{\theta_C} + s \sin{\theta_C}) W_\mu^+ +
\text{h.c.} \notag \\ \label{eq7.13} &=
\frac{g}{\sqrt{2}}\bar{u}_L \gamma^\mu (d_L \cos{\theta_C} + s_L
\sin{\theta_C}) W_\mu^+ + \text{h.c.}
\end{align}
where the coupling constant $g$ is related to the Fermi constant
through $G_F/\sqrt{2}$ $= g^2/(8 m_W^2)$. Comparing the last line
of (\ref{eq7.13}) with the leptonic Lagrangian (\ref{eq7.6}), it
is obvious that the result (\ref{eq7.13}) is reproduced
automatically within the $SU(2)\times U(1)$ electroweak gauge
theory, if one chooses as one of the basic building blocks of the
quark sector an $SU(2)$ doublet of left-handed fields
\begin{equation}
\label{eq7.14}
U_L = \bm{u_L \\ d_L \cos{\theta_C} + s_L \sin{\theta_C}}
\end{equation}
Of course, right-handed components of quark fields
\begin{equation}
\label{eq7.15}
u_R, \; d_R, \; s_R
\end{equation}
are taken to be $SU(2)$ singlets. Weak hypercharges specifying the
$U(1)$ transformation properties of (\ref{eq7.14}) and
(\ref{eq7.15}) are given by the relation (\ref{eq7.5}). Taking
into account the charge assignments
\begin{equation}
Q_u = \frac{2}{3}, \quad Q_d=Q_s = -\frac{1}{3}
\end{equation}
one thus gets
\begin{equation}
Y_{U_L} = \frac{1}{6}, \quad Y_{u_R}=\frac{2}{3}, \quad
Y_{d_R} =Y_{s_R}=-\frac{1}{3}
\end{equation}
The $SU(2)\times U(1)$ invariant Lagrangian for quarks can be
written down in analogy with (\ref{eq7.3}) (using $U_L$ instead of
$L^{(\ell)}$ etc.) and physical vector fields are then introduced
according to (\ref{eq7.7}) and (\ref{eq7.8}). However, when the
interaction Lagrangian is worked out in detail, one finds out that
the result has a serious flaw: while the weak charged current
comes out right, the current coupled to the electromagnetic field
does not have the correct form; apart from the desired term
\begin{equation}
\label{eq7.18}
\frac{2}{3}\bar{u}\gamma_\mu u - \frac{1}{3}
\bar{d} \gamma_\mu d -\frac{1}{3} \bar{s}\gamma_\mu s
\end{equation}
it contains \qq{flavour non-diagonal} pieces of the type
$\bar{d}_L\gamma_\mu s_L$ (a verification of this statement is
left to the reader as an exercise). Of course, these non-diagonal
contributions are a direct consequence of the Cabibbo mixing
embodied in the doublet (\ref{eq7.14}). Such terms do conserve
electric charge, but they are in flagrant contradiction with the
empirical fact that electromagnetic interactions conserve
strangeness\index{strangeness}. To make things worse, a
contribution like $\bar{d}_L\gamma_\mu s_L$ would produce parity
violation\index{parity!violation} in the electromagnetic current.

In fact, these defects can be cured quite easily. It turns out
that if the naive model of electroweak quark interactions outlined
above is supplemented with an additional $SU(2)$ singlet
\begin{equation}
s'_L = -d_L \sin{\theta_C} + s_L \cos{\theta_C}
\end{equation}
(carrying the hypercharge $Y_{s'_L}=-\frac{1}{3}$), the unwanted
pieces of the electromagnetic current are cancelled and one ends
up with (\ref{eq7.18}) as it should be (again, an independent
verification of this result is left to the reader as a rewarding
exercise). However, a problem still persists. Working out the
interaction of the $Z_\mu$ with weak neutral current one gets,
after somewhat lengthy but elementary calculations,
\begin{equation}
\label{eq7.20}
\lagr_{NC}^{(u,d,s)} = \frac{g}{\cos{\theta_W}}
(\bar{U}_L\gamma_\mu \frac{\tau^3}{2} U_L - \sin^2{\theta_W}
J_\mu^{(em)}) Z^\mu
\end{equation}
where
\begin{equation}
J_\mu^{(em)} = \frac{2}{3} \bar{u}\gamma_\mu u - \frac{1}{3}
\bar{d}\gamma_\mu d -\frac{1}{3} \bar{s}\gamma_\mu s
\end{equation}
Let us now evaluate the first term of the neutral current in
(\ref{eq7.20}). This becomes
\begin{align}
\bar{U}_L \gamma_\mu \frac{\tau^3}{2} U_L &= \frac{1}{2} \bar{u}_L
\gamma_\mu u_L - \frac{1}{2}(\bar{d}_L \cos{\theta_C} + \bar{s}_L
\sin{\theta_C}) \gamma_\mu (d_L \cos{\theta_C} + s_L
\sin{\theta_C}) \notag\\ &= \frac{1}{2} \bar{u}_L \gamma_\mu u_L
-\frac{1}{2}\cos^2{\theta_C} \bar{d}_L \gamma_\mu d_L
-\frac{1}{2}\sin{\theta_C}\cos{\theta_C}\bar{d}_L\gamma_\mu s_L \notag\\
&\phantom{=}
-\frac{1}{2}\sin{\theta_C}\cos{\theta_C}\bar{s}_L\gamma_\mu d_L
-\frac{1}{2}\sin^2{\theta_C}\bar{s}_L \gamma_\mu s_L
\end{align}
It means that such a provisional model with three quarks $u,d,s$
leads inevitably to neutral-current interactions of the type
\begin{equation}
\label{eq7.23}
\lagr_{dsZ}=g_{dsZ} \bar{d}_L \gamma_\mu s_L
Z^\mu
\end{equation}
where the coupling strength is of the order
\begin{equation}
\label{eq7.24}
g_{dsZ}\simeq \frac{g}{\cos{\theta_W}}
\sin{\theta_C}\cos{\theta_C}
\end{equation}

The weak current appearing in (\ref{eq7.20}) is an example of
the so-called \qq{strangeness-changing neutral
current}\index{strangeness-changing neutral current|ff} (more
generally, \qq{flavour-changing neutral current}, usually referred
to by the acronym FCNC)\index{flavour-changing neutral current}.
The presence of a term like (\ref{eq7.23}) would be a
phenomenological disaster (remember the empirical selection rule
${\vartriangle}S = {\vartriangle}Q$ for semileptonic weak
decays!). For an instructive example let us recall the case of
kaon decays $K^+ \rightarrow \pi^0 e^+ \nu_e$ and $K^+ \rightarrow
\pi^+ e^+ e^-$ mentioned earlier (see Section~\ref{sec2.5})\index{decay!of
the kaon}. The former process, where ${\vartriangle}S =
{\vartriangle}Q=-1$, can be viewed at the quark level as
\begin{equation}
\label{eq7.25} \bar{s} \rightarrow \bar{u} + e^+ + \nu_e
\end{equation}
(since $K^+$ has the quark composition $u\bar{s}$ while the
$\pi^0$ is made of $u\bar{u}$ and $d\bar{d}$) and the latter, for
which ${\vartriangle}S=-1$ and ${\vartriangle}Q=0$, may be
represented as
\begin{equation}
\label{eq7.26}
\bar{s} \rightarrow \bar{d} + e^+ + e^-
\end{equation}
(since the quark contents of $\pi^+$ is $u\bar{d}$). Now,
(\ref{eq7.25}) proceeds at the tree level via $W^+$ exchange and
an overall coupling factor associated with such a diagram is of
the order
\begin{equation}
\label{eq7.27}
g^2 \sin{\theta_C}
\end{equation}
(cf. (\ref{eq7.6}) and (\ref{eq7.13})). If the FCNC interaction
(\ref{eq7.23}) were present, the process (\ref{eq7.26}) would
proceed at the tree level through the $Z^0$ exchange and the
overall coupling factor associated with the corresponding diagram
would be, in view of (\ref{eq7.10}) and (\ref{eq7.24}), of the
order
\begin{equation}
\Bigl(\frac{g}{\cos{\theta_W}}\Bigr)^2
\sin{\theta_C}\cos{\theta_C}
\end{equation}
which is numerically rather close to (\ref{eq7.27}). In other
words, the reactions (\ref{eq7.25}) and (\ref{eq7.26}) would occur
with roughly equal probability and this in turn means that one
would then expect comparable branching ratios for the two kaon
decays\index{branching ratios!of $K$-decay}. However, as we
already noted in Section~\ref{sec2.5}, $K^+ \rightarrow \pi^+ e^+ e^-$ is
in fact much less probable than $K^+ \rightarrow \pi^0 e^+ \nu_e$,
by about five orders of magnitude! Thus, the coupling
(\ref{eq7.23}) is clearly unacceptable from the phenomenological
point of view.

One final remark is in order here. We have seen that a
strangeness-changing weak neutral current necessarily appears
within the $SU(2)\times U(1)$ gauge theory of electroweak
interactions incorporating three quarks $u,d,s$. In fact, there is
another instructive argument showing that such an effect is
essentially unavoidable within an electroweak theory involving
$W^\pm, Z^0$ and three quarks with Cabibbo mixing. Requiring the
\qq{good high-energy behaviour}\index{tree unitarity} for all
tree-level scattering amplitudes (in the sense elucidated in
previous chapters) one may consider, in particular, the process
$d\bar{s} \rightarrow W^+W^-$. This certainly gets a contribution
from a $u$-quark exchange diagram (descending from charged-current
interactions), which produces a quadratic high-energy divergence
if both $W^+$ and $W^-$ are longitudinally polarized; in order to
compensate this divergence, one has to introduce a neutral-current
$dsZ$ coupling with the above-mentioned strength. More about this
line of argument can be found in \cite{Hor}.

Thus, the moral of this story is as follows. {\bf Cabibbo mixing
in a world built upon just three quarks is not compatible with the
$\boldsymbol{SU(2)\!\times U(1)}$ gauge symmetry of electroweak
interactions, because of the appearance of phenomenologically
unacceptable strangeness-changing neutral currents.} In the early
days of the GWS model this pathological feature was indeed a
mortal danger for the whole concept of gauge theories of
fundamental interactions. Fortunately, a simple and elegant
solution of the problem emerged in the early 1970s and this in
fact played a substantial role in the subsequent establishing the
GWS theory as a true \qq{standard model} of electroweak
interactions.

%\end{document}

%\input{kniha73}
%%%%%%%%%%%%%%%%%%%%%%%%%%%%%%%%%%%%%%%%%%%%%%%%%%%%%%%%%%%%%%%%%%%
%%%%%%%%%%%%%%%%%%%%%%%%%%%%%%%%%%%%%%%%%%%%%%%%%%%%%%%%%%%%%%%%%%%%%%%%%%%%%%%%%%%%%%%%%%%%%%%%%%%%%%%%%%%%%%%%%%%%%%%%%%%%%%%%%%%%%%%%
%\documentclass[12pt,tbtags]{report}
%\usepackage{amsmath,amssymb,epsfig}
%\newcommand{\qq}[1]{\textquotedblleft#1\/\textquotedblright}
%\newcommand{\J}{\mathbb{I}}
%\begin{document}
\section{Fourth quark and GIM construction}\index{GIM
construction|ff} 
\label{sec7.3}

The idea of how to get rid of the
strangeness-changing neutral currents within the GWS theory
originated from the work of S. Glashow, J. Iliopoulos and
L.~Maiani \cite{ref63}. They postulated a fourth quark, carrying
the charge $+2/3$ and labelled as $c$, which stands for
\qq{charm}\index{charm} as the new flavour was
named\index{cquark@$c$-quark|(}. Such a scheme has an obvious
aesthetic appeal because of a nice lepton--quark symmetry (four
quarks $u,d,c,s$ as counterparts of the four leptons
$\nu_e,e,\nu_\mu,\mu$ known then) but, what is more important, it
enables one to introduce a new doublet into the $SU(2)\times U(1)$
gauge theory of electroweak interactions. Thus, within the model
due to Glashow, Iliopoulos and Maiani (GIM) one may consider a set
of basic building blocks for the quark sector consisting of two
left-handed $SU(2)$ doublets
\begin{equation}
\label{eq7.29}
U_L = \bm{u_L\\ d_L \cos{\theta_C} + s_L \sin{\theta_C}},\quad
C_L = \bm{c_L\\ -d_L \sin{\theta_C} + s_L \cos{\theta_C}}
\end{equation}
and four right-handed singlets
\begin{equation}
\label{eq7.30}
u_R,\; c_R,\; d_R,\;s_R
\end{equation}
The choice of the \qq{orthogonal} combinations of $d_L$ and $s_L$
in the two doublets in (\ref{eq7.29}) is motivated by the presumed
cancellation of the unwanted strange\-ness-changing terms in the
electromagnetic and weak neutral currents. We will show now that
such a cancellation is indeed achieved.

To this end, let us start with the $SU(2)\times U(1)$ gauge
invariant Lagrangian made of the quark fields (\ref{eq7.29}) and
(\ref{eq7.30}). This can be written as
\begin{equation}
\label{eq7.31}
\begin{split}
\lagr^{(GIM)} = \phantom{+} &i \bar{U}_L \gamma^\mu (\partial_\mu
- i g A^a_\mu \frac{\tau^a}{2} - i g' Y_{U_L} B_\mu) U_L \\ +\ &i
\bar{C}_L \gamma^\mu (\partial_\mu - i g A^a_\mu \frac{\tau^a}{2} -
i g' Y_{C_L} B_\mu) C_L \\ +\ &\sum_{f=u,c,d,s} i \bar{f}_R
\gamma^\mu (\partial_\mu - i g' Y_{f_R} B_\mu) f_R
\end{split}
\end{equation}
with the weak hypercharges
\begin{equation}
\label{eq7.32} Y_{U_L} = Y_{C_L} = \frac{1}{6},\quad
Y_{u_R}=Y_{c_R}=\frac{2}{3}, \quad Y_{d_R}=Y_{s_R} = -\frac{1}{3}
\end{equation}
that follow from (\ref{eq7.5}) and from the quark charge
assignments. The evaluation of the relevant interaction Lagrangian
repeats essentially the steps that were already necessary in the
preceding section, but here we will be more explicit (for the
reader's convenience) as the envisaged result is rather important.
The interactions in the neutral current sector (i.e. those
involving $A^3_\mu$ and $B_\mu$) descending from (\ref{eq7.31})
are then
\begin{equation}
\begin{split}
\lagr^{(N)}_{int} = \phantom{+}&\bar{U}_L \gamma^\mu
(\frac{1}{2} g \tau^3 A_\mu^3 + \frac{1}{6} g' B_\mu) U_L +
\bar{C}_L \gamma^\mu
(\frac{1}{2}g \tau^3 A_\mu^3 + \frac{1}{6} g' B_\mu) C_L \\
+ &\frac{2}{3} g' \bar{u}_R \gamma^\mu u_R B_\mu + \frac{2}{3}g'
\bar{c}_R \gamma^\mu c_R B_\mu \\ -&\frac{1}{3} g' \bar{d}_R
\gamma^\mu d_R B_\mu - \frac{1}{3}g' \bar{s}_R \gamma^\mu s_R
B_\mu
\end{split}
\end{equation}
Introducing now the $A_\mu, Z_\mu$ according to (\ref{eq7.8}), the
interaction of quarks with the $A_\mu$ becomes
\begin{equation}
\lagr^{(em)}_{int} = e J_\mu^{(em)} A^\mu
\end{equation}
where $e=g\sin{\theta_W}$ and
\begin{equation}
\begin{split}
J_\mu^{(em)} = \phantom{+}&\frac{1}{2} \bar{U}_L \gamma_\mu \tau^3
U_L + \frac{1}{2} \bar{C}_L \gamma_\mu \tau^3 C_L +\frac{1}{6}
\bar{U}_L
\gamma_\mu U_L +\frac{1}{6} \bar{C}_L \gamma_\mu C_L \\
+&\frac{2}{3} \bar{u}_R \gamma_\mu u_R + \frac{2}{3}\bar{c}_R
\gamma_\mu c_R -\frac{1}{3} \bar{d}_R \gamma_\mu d_R
-\frac{1}{3}\bar{s}_R\gamma_\mu s_R
\end{split}
\end{equation}
The last expression is worked out as
\begin{equation}
\begin{split}
J_\mu^{(em)} = \phantom{+}&\frac{1}{2} \bar{u}_L \gamma_\mu
u_L -\frac{1}{2}(\bar{d}_L \cos{\theta_C} +\bar{s}_L
\sin{\theta_C})
\gamma_\mu(d_L \cos{\theta_C} + s_L \sin{\theta_C})\\
+&\frac{1}{2} \bar{c}_L \gamma_\mu c_L -\frac{1}{2}(-\bar{d}_L
\sin{\theta_C} +\bar{s}_L \cos{\theta_C})
\gamma_\mu(-d_L \sin{\theta_C} + s_L \cos{\theta_C})\\
+&\frac{1}{6} \bar{u}_L \gamma_\mu u_L +\frac{1}{6}(\bar{d}_L
\cos{\theta_C} +\bar{s}_L \sin{\theta_C})
\gamma_\mu(d_L \cos{\theta_C} + s_L \sin{\theta_C})\\
+&\frac{1}{6} \bar{c}_L \gamma_\mu c_L +\frac{1}{6}(-\bar{d}_L
\sin{\theta_C} +\bar{s}_L \cos{\theta_C})
\gamma_\mu(-d_L \sin{\theta_C} + s_L \cos{\theta_C})\\
+&\frac{2}{3}\bar{u}_R\gamma_\mu u_R
+\frac{2}{3}\bar{c}_R\gamma_\mu c_R -
\frac{1}{3}\bar{d}_R\gamma_\mu d_R -\frac{1}{3}\bar{s}_R\gamma_\mu
s_R
\end{split}
\end{equation}
which is readily simplified to
\begin{equation}
J_\mu^{(em)} = \frac{2}{3} \bar{u}\gamma_\mu u +
\frac{2}{3}\bar{c}\gamma_\mu c - \frac{1}{3}\bar{d}\gamma_\mu d
-\frac{1}{3}\bar{s}\gamma_\mu s
\end{equation}
Thus, the electromagnetic current has indeed the desired
flavour-diagonal form. For the $Z_\mu$ interaction one gets, after
some algebraic manipulations
\begin{equation}
\lagr_{NC}^{(GIM)} = \frac{g}{\cos{\theta_W}} J_\mu^{(GIM)}
Z^\mu
\end{equation}
where the current
\begin{equation}
\label{eq7.39} J_\mu^{({\it GIM})} = \frac{1}{2} \bar{U}_L
\gamma_\mu \tau^3 U_L +\frac{1}{2}\bar{C}_L\gamma_\mu \tau^3 C_L -
\sin^2{\theta_W} J_\mu^{(em)}
\end{equation}
has a form analogous to (\ref{eq7.20}), with the additional
contribution of the second (\qq{charmed}) doublet shown in
(\ref{eq7.29}). Of course, it is an expected result and one can
now show easily (in full analogy with what we have done for
electromagnetic current) that the $d-s$ mixing terms cancel as
needed. Indeed, one has
%\begin{equation}
\begin{multline}
\label{eq7.40} \frac{1}{2}\bar{U}_L \gamma_\mu \tau^3 U_L +
\frac{1}{2}\bar{C}_L
\gamma_\mu \tau^3 C_L =\\
\begin{aligned}
=\phantom{+}&\frac{1}{2} \bar{u}_L \gamma_\mu u_L
-\frac{1}{2}(\bar{d}_L \cos{\theta_C} +\bar{s}_L \sin{\theta_C})
\gamma_\mu(d_L \cos{\theta_C} + s_L \sin{\theta_C})\\
+&\frac{1}{2} \bar{c}_L \gamma_\mu c_L -\frac{1}{2}(-\bar{d}_L
\sin{\theta_C} +\bar{s}_L \cos{\theta_C}) \gamma_\mu(-d_L
\sin{\theta_C} + s_L \cos{\theta_C})
\end{aligned}\\
=\frac{1}{2}\bar{u}_L \gamma_\mu u_L + \frac{1}{2} \bar{c}_L
\gamma_\mu c_L - \frac{1}{2} \bar{d}_L \gamma_\mu d_L
-\frac{1}{2}\bar{s}_L \gamma_\mu s_L
\end{multline}
Thus, the form (\ref{eq7.40}) is diagonal in quark flavours and
its coefficients obviously coincide with the weak
isospin\index{weak!isospin} values for $u_L, c_L, d_L, s_L$. The
GIM neutral current (\ref{eq7.39}) can therefore be written as
\begin{equation}
\label{eq7.41} J_\mu^{(GIM)} = \sum_{f=u,c,d,s} \Bigl(
\varepsilon_L^{(f)} \bar{f}_L \gamma_\mu f_L + \varepsilon_R^{(f)}
\bar{f}_R \gamma_\mu f_R \Bigr)
\end{equation}
with
\begin{equation}
\varepsilon^{(f)} = T_3^{(f)} - Q^{(f)} \sin^2{\theta_W}
\end{equation}
In other words, weak neutral currents for quarks and leptons now
have the same structure (cf. (\ref{eq7.10})).

Let us also add that using (\ref{eq7.29}) and our previous
experience, the interactions of charged currents can be written
down almost immediately; obviously, one gets
\begin{align}
\lagr_{CC}^{(GIM)} =\frac{g}{\sqrt{2}} \bigl[
\,&\bar{u}_L\gamma^\mu(d_L \cos{\theta_C} + s_L \sin{\theta_C})\notag\\
+\,&\bar{c}_L \gamma^\mu (-d_L \sin{\theta_C} + s_L
\cos{\theta_C}) \bigr] W^+_\mu + \text{h.c.} \label{eq7.43}
\end{align}
Note that (\ref{eq7.43}) represents a specific prediction for
flavour-changing processes mediated by $W^\pm$: the $c
\rightarrow d$ transitions are Cabibbo-suppressed similarly to $u
\rightarrow s$ while $c \rightarrow s$ goes unsuppressed, in
analogy with $u \rightarrow d$.

Coming back to the crucial result (\ref{eq7.41}), one may say that
the GIM construction solved the problem of strangeness-changing
currents in the early 1970s and it saved, at least conceptually,
the idea of the GWS gauge electroweak theory. However, a
historical remark is in order here. When the fourth \qq{charmed}
quark has been postulated as a remedy for the difficulties
described above, there was absolutely no experimental sign of a
possible existence of such a particle. Fortunately enough, in 1974
a major discovery came. Two experimental teams, led by B. Richter
and S. Ting respectively, observed independently \cite{ref64} a
new meson\index{meson|ff}, denoted as $J/\psi$\index{J@$J/\Psi$
particle}, which found a natural interpretation as a bound state
$c\bar{c}$ (with $m_c\doteq 1.5\ \GeV$); for this reason, $J/\psi$
is also called charmonium (in analogy with positronium $e^+e^-$).
More precisely, the $J/\psi$ represents charmonium ground state
and soon after its discovery the Richter's group revealed the
existence of further mesons that could be interpreted as the
corresponding excited states. This spectacular result (which in
fact came not long after the discovery of weak neutral currents in
1973) was a real breakthrough, as it removed a major obstacle to
the recognition of the GWS gauge theory as a physically meaningful
model of electroweak interactions. In fact, the discovery of
hadrons\index{hadrons} containing the $c$-quark provided a great
support to the whole concept of gauge theories of fundamental
interactions, as well as to the quark model itself. A nice and
rather detailed description of the $J/\psi$ discovery and related
matters can be found e.g. in the book \cite{CaG}.

There is still one point to be mentioned here. Similarly as in the
case of leptons, quarks should acquire masses through Yukawa-type
interactions. However, a generalization of the procedure described
in Section~\ref{sec6.6} is not entirely straightforward in the quark case,
owing to the flavour mixing\index{flavour!mixing} embodied in the
basic left-handed doublets. From Yukawa interactions made of
(\ref{eq7.29}), (\ref{eq7.30}) and the Higgs field (\ref{eq7.1})
one gets $d-s$ mixing terms in the resulting quark mass matrix and
these have to be eliminated by imposing some appropriate relations
that should hold for the relevant coupling constants and masses.
This can be done, but we shall not proceed in this way. Instead,
we are going to put forward a slightly different formulation of
the GIM construction, based primarily on the discussion of general
quark mass matrices arising from Yukawa couplings {\it without\/}
assuming Cabibbo mixing {\it a priori\/}. A great virtue of such
an alternative approach is, among other things, that one thus
arrives at a rather natural understanding of the very existence of
the Cabibbo angle\index{cquark@$c$-quark|)}.
%\end{document}

%\input{kniha74}
%%%%%%%%%%%%%%%%%%%%%%%%%%%%%%%%%%%%%%%%%%%%%%%%%%%%%%%%%%%%%%%%%%%
%%%%%%%%%%%%%%%%%%%%%%%%%%%%%%%%%%%%%%%%%%%%%%%%%%%%%%%%%%%%%%%%%%%%%%%%%%%%%%%%%%%%%%%%%%%%%%%%%%%%%%%%%%%%%%%%%%%%%%%%%%%%%%%%%%%%%%%%
%\documentclass[12pt,tbtags]{report}
%\usepackage{amsmath,amssymb,epsfig,euscript,amsfonts,dsfont}
%\newcommand{\qq}[1]{\textquotedblleft#1\/\textquotedblright}
%\newcommand{\qs}[1]{\textquoteleft#1\/\textquoteright}
%\newcommand{\J}{\mathds{1}}
%\newcommand{\ti}[1]{\text{\it #1}}
%\newcommand{\U}{\mathcal{U}}
%\newcommand{\V}{\mathcal{V}}
%\newcommand{\M}{M}
%\newcommand{\Mp}{\mathfrak{M}}
%\newcommand{\Ut}{\widetilde{\mathcal{U}}}
%\newcommand{\Vt}{\widetilde{\mathcal{V}}}
%\begin{document}
\section[GIM construction]{GIM construction via diagonalization of quark mass matrices}\label{sec7.4}
To begin with, let us pinpoint some essential aspects of the GIM
construction as formulated in preceding section. It relies on the
empirical fact of the existence of Cabibbo angle\index{Cabibbo
angle|(}, which is then taken as an input parameter in the basic
$SU(2)$ doublets (\ref{eq7.29}). While the structure of the $U_L$
reflects the old phenomenology of 1960s (actually it defines the
$\theta_C$), the form of the $d-s$ mixing appearing in the $C_L$
is picked by hand so as to ensure the elimination of
strangeness-changing neutral currents. Quark fields entering the
relevant Lagrangian are supposed to be the physical ones (i.e.
corresponding to mass eigenstates). Within such a scheme, the
intriguing problem of a deeper origin of the Cabibbo angle remains
totally obscure and also the remarkable \qq{orthogonality} of the
lower components of $U_L$ and $C_L$ appears as a rather {\it ad
hoc\/} choice, enforced upon us by demands of hadron phenomenology
-- one might wonder rightfully whether it has a more profound
explanation.

As we have already noted, at least partial answer to these
questions can be found quite naturally, if the whole GIM
construction is formulated in a slightly different way. So, this
is what we are going to do now. The main idea is to start with
quark fields that need not, in general, coincide with the physical
ones; the latter will only emerge as a result of a mass matrix
diagonalization. Thus, let the basic building blocks for the
relevant quark Lagrangians be two left-handed $SU(2)$ doublets
\begin{equation}
\label{eq7.44} U_{0L} = \bm{u_{0L}\\d_{0L}}, \quad C_{0L} =
\bm{c_{0L}\\s_{0L}}
\end{equation}%
(corresponding to two \qq{generations} of quarks) and four
right-handed singlets
\begin{equation}
\label{eq7.45} u_{0R},\; d_{0R},\; s_{0R},\; c_{0R}
\end{equation}%
where the label \qs{$\scriptstyle 0$} indicates the presumed
unphysical nature of the fields in question. Again, the
corresponding weak hypercharges are determined according to
(\ref{eq7.5}), so that
\begin{equation}
\label{eq7.46} Y_{U_{0L}} = Y_{C_{0L}} = \frac{1}{6},\quad
Y_{u_{0R}} = Y_{c_{0R}} = \frac{2}{3}, \quad Y_{d_{0R}} =
Y_{s_{0R}} = -\frac{1}{3}
\end{equation}%
and the interactions of quarks with the $SU(2) \times U(1)$ gauge
fields thus become
\begin{equation}
\label{eq7.47}
\begin{split}
\lagr_\ti{int}^{(\ti{quark})} = \phantom{+}&\ \bar{U}_{0L}
\gamma^\mu (\frac{1}{2} g A^a_\mu \tau^a + \frac{1}{6}g'B_\mu )
U_{0L}\\ + &\ \bar{C}_{0L} \gamma^\mu (\frac{1}{2} g A^a_\mu
\tau^a + \frac{1}{6}g'B_\mu ) C_{0L} \\ +&\
\frac{2}{3}g'\bar{u}_{0R}\gamma^\mu u_{0R} B_\mu +
\frac{2}{3}g'\bar{c}_{0R} \gamma^\mu c_{0R} B_\mu
\\-&\ \frac{1}{3}g'\bar{d}_{0R} \gamma^\mu d_{0R} B_\mu
-\frac{1}{3} g'\bar{s}_{0R} \gamma^\mu s_{0R} B_\mu
\end{split}
\end{equation}
Now, a most general $SU(2) \times U(1)$ invariant Yukawa-type
interaction involving the quark fields (\ref{eq7.44}),
(\ref{eq7.45}) and the Higgs doublet\index{Higgs!doublet} $\Phi$
has the form
\begin{equation}
\label{eq7.48}
\begin{split}
\lagr_\ti{Yukawa}^{(\ti{d,s})} = &-h_{11} \bar{U}_{0L} \Phi
d_{0R} - h_{12} \bar{U}_{0L} \Phi s_{0R} \\
&-h_{21} \bar{C}_{0L} \Phi d_{0R} - h_{22} \bar{C}_{0L} \Phi
s_{0R} + \text{h.c.}
\end{split}
\end{equation}
where the $h_{ij},\ i,j=1,2$ are arbitrary (dimensionless)
coupling constants\index{coupling constants!of Higgs sector|ff};
for simplicity we may assume that they are real (we shall see
later on that such a restriction does not mean any loss of
generality). Note that the $SU(2)$ symmetry of the expression
(\ref{eq7.48}) is obvious and the hypercharge values shown in
(\ref{eq7.46}) guarantee its invariance under $U(1)$ (remember
that $\bar{U}_{0L}$ and $\bar{C}_{0L}$ carry $Y = -\frac{1}{6}$
!). In this context it is also important to realize that an
analogous coupling that would involve $u_{0R}$ and $c_{0R}$ is
forbidden precisely by the requirement of hypercharge $U(1)$
symmetry. Working out (\ref{eq7.48}) in the
$U$-gauge\index{U-gauge@$U$-gauge|(} one gets
\begin{equation}
\label{eq7.49}
\begin{split}
\lagr_\ti{Yukawa}^{(d,s)}=&-\frac{1}{\sqrt{2}}(v+H)
(h_{11}\bar{d}_{0L}d_{0R} + h_{12}\bar{d}_{0L}s_{0R}\\
&\phantom{-\sqrt{2}(v+H)(}+h_{21}\bar{s}_{0L}d_{0R} +
h_{22}\bar{s}_{0L}s_{0R}) +\text{h.c.}
\\
=&-\frac{1}{\sqrt{2}}(v+H)%
\bm{\bar{d}_{0L},\ \bar{s}_{0L}}
%      \negthickspace
\bm{h_{11}&h_{12}\\h_{21}&h_{22}}%
\bm{d_{0R}\\s_{0R}} + \text{h.c.}
\end{split}
\end{equation}
In (\ref{eq7.49}) one may identify both interactions and mass
terms. The latter can be collected in a compact form
\begin{equation}
\label{eq7.50}
\lagr^{(d,s)}_\ti{mass} = -\bm{\bar{d}_{0L},\ \bar{s}_{0L}}\M%
\bm{d_{0R}\\s_{0R}} + \text{h.c.}
\end{equation}
where
\begin{equation}
\label{eq7.51} \M = \frac{v}{\sqrt{2}}
\bm{h_{11}&h_{12}\\h_{21}&h_{22}}
\end{equation}
Thus, it becomes clear that only the \qq{down-type} quarks
$d,s$\index{s-quark@$s$-quark} can acquire masses through Yukawa
couplings involving the Higgs doublet $\Phi$. We shall explain a
bit later how the mass terms for $u$ and
$c$\index{u-quark@$u$-quark}\index{cquark@$c$-quark} are generated
and now let us discuss a diagonalization of the mass matrix
(\ref{eq7.51}) (this, of course, is necessary for a proper
identification of physical quark fields).

At first sight, this may seem rather problematic, since
(\ref{eq7.51}) is not, in general, hermitean and thus it cannot be
diagonalized simply by means of a unitary transformation matrix.
Fortunately, the job can be done with the help of a {\it
biunitary\/} transformation\index{biunitary transformation|ff}\footnote{An erudite reader may notice that the technique of biunitary transformations is in fact the construction that mathematicians call the ``singular value decomposition'' (SVD). Although in mathematics this has been known since the 19th century, particle physicists apparently developed it for their pragmatic needs independently, in the early 1970s. It turns out that mathematicians usually do not know that SVD has such an important application within SM, and particle physicists are rarely aware of the mathematical context and history of the currently popular algebraic tool of biunitary transformations.}. In
particular, one may rely on the following theorem:
\begin{flushright}
\begin{minipage}{0.95\textwidth}{\it
Any non-singular square complex matrix $\M$ can be decomposed as
\begin{equation}
\label{eq7.52} \M=\U^\dagger\Mp\V
\end{equation}
where $\U$, $\V$ are unitary matrices and the $\Mp$ is diagonal
and positive.\\ When $\M$ is real, the $\U$ and $\V$ are real
orthogonal matrices.}
\end{minipage}
\end{flushright}
A proof of this theorem is quite simple and we defer it to the end
of this section; now we are going to apply its statement to the
above quark mass term (we assume a priori that the matrix
(\ref{eq7.51}) is non-singular, in order to get non-zero quark
masses). Using (\ref{eq7.52}) in (\ref{eq7.50}), one has
\begin{equation}
\label{eq7.53} \lagr^{(d,s)}_\ti{mass}=-\bm{\bar{d}_{0L},\
\bar{s}_{0L}} \U^\dagger \Mp \V \bm{d_{0R}\\s_{0R}} + \text{h.c.}
\end{equation}
where the diagonal matrix $\Mp$ shall be written, for obvious
reasons, as
\begin{equation}
\label{eq7.54} \Mp=\bm{m_d&0\\0&m_s}
\end{equation}
Let us now define new quark fields $d_L, s_L$ and $d_R, s_R$
through unitary transformations
\begin{equation}
\label{eq7.55} \bm{d_L\\s_L}=\U \bm{d_{0L}\\s_{0L}},\quad
\bm{d_R\\s_R}=\V \bm{d_{0R}\\s_{0R}}
\end{equation}
Then (\ref{eq7.53}) becomes
\begin{equation}
\begin{split}
\lagr^{(d,s)}_\ti{mass} &= -\bm{\bar{d}_L,\ \bar{s}_L}\Mp
\bm{d_R\\s_R} + \text{h.c.}\\
&=-m_d \bar{d}_L d_R - m_s \bar{s}_L s_R + \text{h.c.}\\
&=-m_d (\bar{d}_L d_R + \bar{d}_R d_L) - m_s (\bar{s}_L s_R +
\bar{s}_R s_L)\\
&=-m_d \bar{d}d -m_s \bar{s}s
\end{split}
\end{equation}
It means that -- as anticipated in (\ref{eq7.54}) -- the $d,s$
correspond to quark mass eigenstates; in other words, they can be
identified with physical fields. In this context, it is also
important to realize that the kinetic terms remain diagonal: for
example, in terms of the original variables we had
\begin{equation}
\begin{split}
\lagr^{(d,s)}_\ti{kin} &= i \bar{d}_{0L} \gamma^\mu
\partial_\mu d_{0L} +i\bar{s}_{0L}\gamma^\mu\partial_\mu s_{0L}
+i \bar{d}_{0R} \gamma^\mu \partial_\mu d_{0R} + i \bar{s}_{0R}
\gamma^\mu \partial_\mu s_{0R} \\
&= i \bm{\bar{d}_{0L},\ \bar{s}_{0L}} \gamma^\mu
\partial_\mu \bm{d_{0L}\\s_{0L}} + i \bm{\bar{d}_{0R},\
\bar{s}_{0R}}\gamma^\mu
\partial_\mu \bm{d_{0R}\\s_{0R}}
\end{split}
\end{equation}
and this becomes
\begin{align}
\lagr^{(d,s)}_\ti{kin} &= i \bm{\bar{d}_L,\ \bar{s}_L}\gamma^\mu
\partial_\mu \bm{d_L\\s_L} + i \bm{\bar{d}_R,\ \bar{s}_R}\gamma^\mu
\partial_\mu \bm{d_R\\s_R} \notag\\
&= i \bar{d}\gamma^\mu \partial_\mu d + i\bar{s}\gamma^\mu
\partial_\mu s
\end{align}
when the transformation (\ref{eq7.55}) is implemented (simply
because $\U\U^\dagger = \V\V^\dagger = \J$).

Since the interaction terms descending from (\ref{eq7.49}) have an
algebraic structure completely analogous to that of the mass
terms, it is obvious that the Lagrangian describing the
interactions of quarks $d$ and $s$ with the physical Higgs boson
$H$ has the form
\begin{equation}
\label{eq7.59} \lagr^{(d,s)}_\ti{int} = -\frac{m_d}{v} \bar{d}d H
- \frac{m_s}{v}\bar{s}s H
\end{equation}
Thus, the relevant coupling constants, denoted in a
self-explanatory way as $g_{ddH}$ and $g_{ssH}$ respectively, are
given by
\begin{equation}
\label{eq7.60} g_{ddH} = -\frac{g}{2}\frac{m_d}{m_W}, \qquad
g_{ssH} = -\frac{g}{2}\frac{m_s}{m_W}
\end{equation}
in full analogy with the result valid for leptons (cf.
(\ref{eq6.88})).

As the next step, one should generate mass terms for the
\qq{up-type} quarks $u,c$. This is done by means of the trick that
we have already used for neutrinos in Chapter~\ref{chap6} (see the
discussion following after the formula (\ref{eq6.89})). Thus, let
us consider the conjugate doublet $\Phit$ defined in
(\ref{eq6.90}). It carries weak hypercharge $-\frac{1}{2}$ and
this in turn means that one can employ $\Phit$ for constructing an
$SU(2)\times U(1)$ invariant Yukawa interaction
\begin{equation}
\label{eq7.61}
\begin{split} \lagr^{(u,c)}_\ti{Yukawa} = &-\tilde{h}_{11} \bar{U}_{0L}
\Phit u_{0R} - \tilde{h}_{12} \bar{U}_{0L}
\Phit c_{0R}\\
&-\tilde{h}_{21}\bar{C}_{0L}\Phit u_{0R}
-\tilde{h}_{22}\bar{C}_{0L}\Phit c_{0R} + \text{h.c.}
\end{split}
\end{equation}
needed for our purpose (one may check readily that the relevant
weak hypercharge values fit precisely the invariance requirement
for (\ref{eq7.61})). Of course, the coupling constants
$\tilde{h}_{ij}$ are completely independent of the $h_{ij}$
appearing in (\ref{eq7.48}). For simplicity, we are again assuming
that all of the $\tilde{h}_{ij}$ are real. Now, in $U$-gauge the
$\Phit$ becomes
\begin{equation}
\label{eq7.62} \Phit_U = \bm{\frac{1}{\sqrt{2}}(v+H)\\0}
\end{equation}
(cf. (\ref{eq6.93})) and substituting (\ref{eq7.62}) into
(\ref{eq7.61}) one gets
\begin{equation}
\label{eq7.63} \lagr^{(u,c)}_\ti{Yukawa}= -\frac{1}{\sqrt{2}}(v+H)
\bm{\bar{u}_{0L},\ \bar{c}_{0L}}
\bm{\tilde{h}_{11}&\tilde{h}_{12}\\ \tilde{h}_{21}&\tilde{h}_{22}}
\bm{u_{0R}\\c_{0R}} + \text{h.c.}
\end{equation}
In particular, (\ref{eq7.63}) contains the expected mass term
\begin{equation}
\label{eq7.64} \lagr^{(u,c)}_\ti{mass} = -\bm{\bar{u}_{0L},\
\bar{c}_{0L}} \Mt \bm{u_{0R}\\c_{0R}} + \text{h.c.}
\end{equation}
where
\begin{equation}
\label{eq7.65} \Mt = \frac{v}{\sqrt{2}}
\bm{\tilde{h}_{11}&\tilde{h}_{12}\\ \tilde{h}_{21}&\tilde{h}_{22}}
\end{equation}
Again, (\ref{eq7.65}) can be diagonalized by means of a biunitary
(here in fact real biorthogonal) transformation, i.e., one can
write
\begin{equation}
\Mt = \Ut^\dagger \Mpt \Vt
\end{equation}
with
\begin{equation}
\Mpt = \bm{m_u&0\\0&m_c}
\end{equation}
and define new fields $u_{L,R}$, $c_{L,R}$ by rotating the
original ones according to
\begin{equation}
\label{eq7.68} \bm{u_L\\c_L} = \Ut \bm{u_{0L}\\c_{0L}},\quad
\bm{u_R\\c_R}=\Vt\bm{u_{0R}\\c_{0R}}
\end{equation}
Then the mass term (\ref{eq7.64}) is recast as
\begin{equation}
\lagr^{(u,c)}_\ti{mass} = -\bm{\bar{u}_L,\ \bar{c}_L} \Mpt
\bm{u_R\\c_R} + \text{h.c.} = -m_u \bar{u}u - m_c\bar{c}c
\end{equation}
so that the $u, c$ may be identified with physical fields, in
complete analogy with the preceding discussion of quarks $d,s$. Of
course, the corresponding interaction with the Higgs field becomes
\begin{equation}
\lagr^{(u,c)}_\ti{int} = -\frac{m_u}{v} \bar{u}u H
-\frac{m_c}{v}\bar{c}c H =-\frac{g}{2} \Bigl(
\frac{m_u}{m_W}\bar{u}u + \frac{m_c}{m_W}\bar{c}c\Bigr) H
\end{equation}
similarly to (\ref{eq7.59}), (\ref{eq7.60}).

Now it remains to be seen how the weak interactions of charged
currents\index{charged current} are expressed in terms of physical
quark fields and what happens in the sector of neutral currents.
First, from (\ref{eq7.47}) one gets readily the charged-current
interaction written in terms of the original fields
(\ref{eq7.44}):
\begin{equation}
\label{eq7.71}
\begin{split}
\lagr^{(\ti{quark})}_{CC} &= \frac{g}{\sqrt{2}}
(\bar{u}_{0L}\gamma^\mu d_{0L} + \bar{c}_{0L}\gamma^\mu s_{0L})
W^+_\mu + \text{h.c.}\\ &=\frac{g}{\sqrt{2}} \bm{\bar{u}_{0L},\
\bar{c}_{0L}} \gamma^\mu \bm{d_{0L}\\s_{0L}} W^+_\mu +\text{h.c.}
\end{split}
\end{equation}
When the physical quark fields are introduced through the rotations
\begin{equation}
\bm{d_{0L}\\s_{0L}} = \U^\dagger \bm{d_L\\s_L}, \qquad
\bm{u_{0L}\\c_{0L}} = \Ut^\dagger \bm{u_L\\c_L}
\end{equation}
(see (\ref{eq7.55}), (\ref{eq7.68})), the expression
(\ref{eq7.71}) is recast as
\begin{equation}
\label{eq7.73} \lagr^{(\ti{quark})}_{CC} = \frac{g}{\sqrt{2}}
\bm{\bar{u}_L,\ \bar{c}_L}\gamma^\mu\, \Ut\U^\dagger \bm{d_L\\s_L}
W^+_\mu +\text{h.c.}
\end{equation}
According to our conventions, the $\U$ and $\Ut$ are real
orthogonal matrices; thus, each of them is characterized by a
rotation angle. Denoting these angles as $\theta_1$ and $\theta_2$
respectively, the product $\Ut\U^\dagger$ is consequently
parametrized by the difference $\theta_2 - \theta_1$, which can
justly be called $\theta_C$. Indeed, it is seen immediately that
$\theta_2 - \theta_1$ plays the role of Cabibbo angle, because in
terms of $\theta_C=\theta_2 - \theta_1$, (\ref{eq7.73}) is written
as
\begin{equation}
\label{eq7.74}
\begin{split}
\lagr^{(\ti{quark})}_{CC} &= \frac{g}{\sqrt{2}} \bm{\bar{u}_L,\
\bar{c}_L}\gamma^\mu \bm{\cos{\theta_C}&\sin{\theta_C}\\
-\sin{\theta_C}&\cos{\theta_C}} \bm{d_L\\s_L} W^+_\mu +
\text{h.c.}\\
&= \frac{g}{\sqrt{2}} \bigr[ \bar{u}_L \gamma^\mu (d_L
\cos{\theta_C} + s_L \sin{\theta_C})\\ &\phantom{=\sqrt{2}\bigr[}+
\bar{c}_L \gamma^\mu (-d_L\sin{\theta_C}+s_L\cos{\theta_C})\bigl]
W^+_\mu + \text{h.c.}
\end{split}
\end{equation}
In other words, we have reproduced the GIM construction of the
charged-current interactions (cf. (\ref{eq7.43})). The crucial
aspect of our analysis is a natural appearance of the Cabibbo
angle\index{Cabibbo angle|)}, which originates in quark field
rotations necessary for diagonalization of their quark matrices;
more precisely, it is due to a mismatch between the rotations
performed on (left-handed) up-type and down-type quarks. At the
same time, the reason for \qq{orthogonality} of the $d_L, s_L$
combinations coupled to $u_L$ and $c_L$ in the original form
(\ref{eq7.43}) becomes manifest: the pattern of $d-s$ mixing is
determined by the orthogonal matrix $\Ut\U^\dagger$ shown
explicitly in the first line of (\ref{eq7.74}); for the purpose of
later references we will denote it as $U_\ti{GIM}$, i.e.
\begin{equation}
\label{eq7.75} U_\ti{GIM} =
\bm{\cos{\theta_C}&\sin{\theta_C}\\-\sin{\theta_C}&\cos{\theta_C}}
=\bm{U_{ud}&U_{us}\\U_{cd}&U_{cs}}
\end{equation}
It should be stressed that within such an approach, separate $d-s$
and $u-c$ mixings (characterized e.g. by the above-mentioned
angles $\theta_1, \theta_2$) would not make physical sense: only
the difference $\theta_2-\theta_1$ is physically relevant and this
is conventionally taken as the $d-s$ mixing angle.

Next, let us see what are the results for neutral currents.
Working out (\ref{eq7.47}), interaction terms involving the gauge
fields $A^3_\mu, B_\mu$ become
\begin{equation}
\begin{split}
\lagr_\ti{NC} &=\frac{1}{2}g (\bar{u}_{0L}\gamma^\mu u_{0L}
-\bar{d}_{0L}\gamma^\mu d_{0L} + \bar{c}_{0L}\gamma^\mu c_{0L}
-\bar{s}_{0L}\gamma^\mu s_{0L}) A^3_\mu\\
&+\frac{1}{6}g'(\bar{u}_{0L}\gamma^\mu u_{0L}
+\bar{d}_{0L}\gamma^\mu d_{0L} + \bar{c}_{0L}\gamma^\mu c_{0L}
+\bar{s}_{0L}\gamma^\mu s_{0L}) B_\mu\\
&+\frac{2}{3}g'(\bar{u}_{0R}\gamma^\mu u_{0R} +
\bar{c}_{0R}\gamma^\mu c_{0R})B_\mu
-\frac{1}{3}g'(\bar{d}_{0R}\gamma^\mu d_{0R} +
\bar{s}_{0R}\gamma^\mu s_{0R})B_\mu
\end{split}
\end{equation}
and this can be conveniently rewritten as
\begin{equation}
\label{eq7.77}
\begin{split}
\lagr_\ti{NC} &= \frac{1}{2}g\bigl[ \bm{\bar{u}_{0L},\
\bar{c}_{0L}}\gamma^\mu \bm{u_{0L}\\c_{0L}} -\bm{\bar{d}_{0L},\
\bar{s}_{0L}}\gamma^\mu \bm{d_{0L}\\s_{0L}}\bigr]A^3_\mu\\
&+\frac{1}{6}g'\bigl[ \bm{\bar{u}_{0L},\ \bar{c}_{0L}}\gamma^\mu
\bm{u_{0L}\\c_{0L}} +\bm{\bar{d}_{0L},\ \bar{s}_{0L}}\gamma^\mu
\bm{d_{0L}\\s_{0L}}\bigr]B_\mu\\
&+\frac{2}{3}g'\bm{\bar{u}_{0R},\ \bar{c}_{0R}} \gamma^\mu
\bm{u_{0R}\\c_{0R}}B_\mu -\frac{1}{3}g'\bm{\bar{d}_{0R},\
\bar{s}_{0R}}\gamma^\mu \bm{d_{0R}\\s_{0R}}B_\mu
\end{split}
\end{equation}
When one passes to the physical quark fields via (\ref{eq7.55})
and (\ref{eq7.68}), it is clear that only the products
$\U\U^\dagger$, $\V\V^\dagger$, $\Ut\Ut^\dagger$ and
$\Vt{\Vt}^\dagger$ can appear in the expression (\ref{eq7.77}).
However, any such product reduces to the unit matrix. Thus,
(\ref{eq7.77}) becomes immediately
\begin{equation}
\label{eq7.78}
\begin{split}\lagr_\ti{NC}&=\frac{1}{2}g(\bar{u}_L\gamma^\mu u_L
+\bar{c}_L \gamma^\mu c_L- \bar{d}_L \gamma^\mu d_L -\bar{s}_L
\gamma^\mu s_L)
A^3_\mu \\
&+\frac{1}{6}g'(\bar{u}_L\gamma^\mu u_L +\bar{c}_L \gamma^\mu c_L
+ \bar{d}_L \gamma^\mu d_L +\bar{s}_L \gamma^\mu s_L) B_\mu \\
&+\frac{2}{3}g'(\bar{u}_R\gamma^\mu u_R + \bar{c}_R \gamma^\mu
c_R)B_\mu -\frac{1}{3}g' (\bar{d}_R\gamma^\mu d_R + \bar{s}_R
\gamma^\mu s_R)B_\mu
\end{split}
\end{equation}
It is clear that by introducing $A_\mu, Z_\mu$ instead of
$A^3_\mu, B_\mu$ (through (\ref{eq7.8})) one cannot spoil the
flavour-diagonal structure of (\ref{eq7.78}); in fact, when this
is done, the desired expression for the electromagnetic current is
recovered, as well as the GIM result (\ref{eq7.39}) for the weak
neutral current.

Thus, we have arrived at a most transparent formulation of the GIM
mechanism that can be succinctly summarized as follows. Neutral
currents are manifestly (by construction) flavour-diagonal in the
basis of unphysical quark fields $u_0, d_0, c_0, s_0$ and quarks
of an equal charge are grouped in pairs that can be subsequently
transformed into the corresponding physical fields. The currents
remain diagonal under such transformations, because these are
implemented by means of unitary matrices and each transformation
matrix eventually gets multiplied by its inverse inside the
current (remember that in the case of charged currents we had
products like $\Ut\U^\dagger$, involving two {\it different\/}
matrices!).

In closing this section, we are going to prove the mathematical
theorem on biunitary transformations (expressed by the relation
(\ref{eq7.52})), which played a central role in our
considerations. Let $\M$ be an arbitrary non-singular complex
square $n\times n$ matrix. Then $\M\M^\dagger$ is hermitean and
positive and consequently it may be diagonalized by means of a
unitary transformation, i.e.
\begin{equation}
\M\M^\dagger = \U^\dagger \Mp^2 \U
\end{equation}
where $\U\U^+=\U^+\U = \J$ and $\Mp^2$ can be written as
\begin{equation}
\Mp^2 = \text{diag}(m_1^2,\ldots,m_n^2)
\end{equation}
with all $m_j^2,\ j=1,\ldots,n$ being positive numbers. Obviously,
for $\M$ real the $\U$ can be taken as a real orthogonal matrix.
Let us also define $\Mp$ as
\begin{equation}
\Mp = \sqrt{\Mp^2} = \text{diag}(|m_1|,\ldots,|m_n|)
\end{equation}
Now, with the relation (\ref{eq7.52}) in mind, one can define
\begin{equation}
\label{eq7.82} \V = \Mp^{-1}\U\M
\end{equation}
It is easy to show that $\V$ is unitary. Indeed,
\begin{equation}
\V\V^\dagger = \Mp^{-1}\U\M\M^\dagger\U^\dagger\Mp^{-1} =
\Mp^{-1}\cdot \Mp^2\cdot \Mp^{-1} = \J
\end{equation}
and $\V^\dagger\V=\J$ then follows automatically, since we are
dealing with finite-dimensional matrices (reality of the $\V$ for
a real $\M$ is also obvious from (\ref{eq7.82})). Thus, if for a
given $\M$ one chooses the matrices $\U$ and $\V$ as defined
above, (\ref{eq7.52}) is satisfied and our theorem is thereby
proved.
%\end{document}

%\input{kniha75}
%%%%%%%%%%%%%%%%%%%%%%%%%%%%%%%%%%%%%%%%%%%%%%%%%%%%%%%%%%%%%%%%%%%
%%%%%%%%%%%%%%%%%%%%%%%%%%%%%%%%%%%%%%%%%%%%%%%%%%%%%%%%%%%%%%%%%%%%%%%%%%%%%%%%%%%%%%%%%%%%%%%%%%%%%%%%%%%%%%%%%%%%%%%%%%%%%%%%%%%%%%%%
%\documentclass[12pt,tbtags]{report}
%\usepackage{amsmath,amssymb,epsfig,euscript,amsfonts,dsfont}
%\newcommand{\qq}[1]{\textquotedblleft#1\/\textquotedblright}
%\newcommand{\qs}[1]{\textquoteleft#1\/\textquoteright}
%\newcommand{\ti}[1]{\text{\it #1}}   % index v matematice
%\newcommand{\J}{\mathds{1}} %\newcommand{\ti}[1]{\text{\it #1}}
%\newcommand{\bm}[1]{\begin{pmatrix}#1\end{pmatrix}} % pro matice, vektory... napr.\bm{a&b\\c&d}
%\newcommand{\hv}{\tilde{h}}
%\newcommand{\Phit}{\widetilde{\Phi}}
%\newcommand{\U}{\mathcal{U}}
%\newcommand{\Ut}{\widetilde{\mathcal{U}}}
%\newcommand{\V}{\mathcal{V}}
%\newcommand{\Vt}{\widetilde{\mathcal{V}}}
%\newcommand{\M}{M}
%\newcommand{\Mp}{\mathfrak{M}}
%\begin{document}
\section{Kobayashi--Maskawa matrix}
So far we have considered electroweak interactions within a model
involving four quark flavours. However, as we know, the existence
of six quarks is now firmly established by experiments; more
precisely, the present-day picture of standard model of particle
physics incorporates six leptons and six quarks, i.e. three
\qq{generations} of elementary fermions. Therefore, we should
generalize our discussion so as to include two more quark types,
in addition to the four considered previously.

In fact, this can be done quite easily, if one adopts the strategy
described in the preceding section. It means that in building the
quark sector of the $SU(2)\times U(1)$ gauge theory of electroweak
forces one starts with three left-handed $SU(2)$ doublets
\begin{equation}
\label{eq7.84} U_{0L}=\bm{u_{0L}\\d_{0L}},\;
C_{0L}=\bm{c_{0L}\\s_{0L}},\; T_{0L}=\bm{t_{0L}\\b_{0L}}
\end{equation}
and six right-handed singlets
\begin{equation}
\label{eq7.85} u_{0R},\; d_{0R},\; c_{0R},\; s_{0R},\; t_{0R},\;
b_{0R}
\end{equation}
where $b$ stands for \qq{bottom} (or
\qq{beauty})\index{bottom}\index{beauty}\index{bquark@$b$-quark|(},
$t$ for \qq{top}\index{t-quark@$t$-quark}\index{top|ff} and the
other symbols have the by now familiar meaning. Quarks $b$ and $t$
carry electric charges $-\frac{1}{3}$ and $+\frac{2}{3}$
respectively and the weak hypercharges\index{weak!hypercharge|)}
of the fields (\ref{eq7.84}), (\ref{eq7.85}) are fixed by
(\ref{eq7.5}) as usual. Now we can proceed in full analogy with
the two-generation model. The relevant Yukawa interactions are
written as
%\begin{equation}
\begin{alignat}{3}
{\lagr}^{(d,s,b)}_\ti{Yukawa} = &-h_{11}\bar{U}_{0L}\Phi d_{0R}
&&-h_{12}\bar{U}_{0L}\Phi s_{0R} &&- h_{13}\bar{U}_{0L}\Phi b_{0R}\notag\\
&-h_{21}\bar{C}_{0L}\Phi d_{0R} &&-h_{22}\bar{C}_{0L}\Phi s_{0R}
&&-h_{23}\bar{C}_{0L}\Phi b_{0R}\notag\\ &-h_{31}\bar{T}_{0L}\Phi d_{0R}
&&-h_{32}\bar{T}_{0L}\Phi s_{0R} &&-h_{33}\bar{T}_{0L}\Phi b_{0R}
+\text{h.c.}\label{eq7.86}
\end{alignat}
%\end{equation}
and
\begin{alignat}{3}
{\lagr}^{(u,c,t)}_\ti{Yukawa} = &-\hv_{11}\bar{U}_{0L}\Phit u_{0R}
&&-\hv_{12}\bar{U}_{0L}\Phit c_{0R} &&- \hv_{13}\bar{U}_{0L}\Phit t_{0R}\notag\\
&-\hv_{21}\bar{C}_{0L}\Phit u_{0R} &&-\hv_{22}\bar{C}_{0L}\Phit c_{0R}
&&-\hv_{23}\bar{C}_{0L}\Phit t_{0R}\notag\\ &-\hv_{31}\bar{T}_{0L}\Phit u_{0R}
&&-\hv_{32}\bar{T}_{0L}\Phit c_{0R} &&-\hv_{33}\bar{T}_{0L}\Phit t_{0R}
+\text{h.c.}\label{eq7.87}
\end{alignat}
where the coupling constants $h_{ij}$ and $\hv_{ij}$ are, in
general, complex numbers. Substituting into (\ref{eq7.86}),
(\ref{eq7.87}) the $U$-gauge values for the $\Phi$ and $\Phit$,
one gets
\begin{equation}
\label{eq7.88} {\lagr}^{(d,s,b)}_\ti{Yukawa}
=-\frac{1}{\sqrt{2}}(v+H) \bm{\bar{d}_{0L},\ \bar{s}_{0L},\
\bar{b}_{0L}}
\bm{h_{11}&h_{12}&h_{13}\\h_{21}&h_{22}&h_{23}\\h_{31}&h_{32}&h_{33}}
\bm{d_{0R}\\s_{0R}\\b_{0R}} + \text{h.c.}
\end{equation}
and
\begin{equation}
\label{eq7.89}
{\lagr}^{(u,c,t)}_\ti{Yukawa}=-\frac{1}{\sqrt{2}}(v+H)
\bm{\bar{u}_{0L},\ \bar{c}_{0L},\ \bar{t}_{0L}}
\bm{\hv_{11}&\hv_{12}&\hv_{13}\\\hv_{21}&\hv_{22}&\hv_{23}\\\hv_{31}&\hv_{32}&\hv_{33}}
\bm{u_{0R}\\c_{0R}\\t_{0R}} + \text{h.c.}
\end{equation}
The mass terms for down- and up-type quarks contained in
(\ref{eq7.88}) and (\ref{eq7.89}) are diagonalized by means of
appropriate biunitary transformations\index{biunitary
transformation}. Denoting the relevant $3 \times 3$ transformation
matrices as $\U, \V$ (for down-type quarks) and $\Ut, \Vt$ (for
up-type quarks), one is thus led to redefine the quark fields as
\begin{equation}
\label{eq7.90} \bm{d_L\\s_L\\b_L} = \U
\bm{d_{0L}\\s_{0L}\\b_{0L}},\quad \bm{d_R\\s_R\\b_R} = \V
\bm{d_{0R}\\s_{0R}\\b_{0R}}
\end{equation}
and
\begin{equation}
\label{eq7.91} \bm{u_L\\c_L\\t_L} = \Ut
\bm{u_{0L}\\c_{0L}\\t_{0L}},\quad \bm{u_R\\c_R\\t_R} = \Vt
\bm{u_{0R}\\c_{0R}\\t_{0R}}
\end{equation}
The $d,s,b,u,c,t$ then represent a set of physical quark fields;
along with mass terms, also the $H$ interactions appearing in
(\ref{eq7.88}) and (\ref{eq7.89}) become flavour-diagonal, with
coupling constants obeying the law
\begin{equation}
g_{ffH}=-\frac{g}{2}\frac{m_f}{m_W}
\end{equation}
(where $f$ is a generic label for any flavour in question).

The charged-current interaction written in terms of the original
(unphysical) quark fields has the form
\begin{align}
{\lagr}^{(\ti{quark})}_{CC}&=\frac{g}{\sqrt{2}}(
\bar{u}_{0L}\gamma^\mu d_{0L} +\bar{c}_{0L}\gamma^\mu s_{0L}
+\bar{t}_{0L}\gamma^\mu b_{0L}) W^+_\mu + \text{h.c.} \notag\\
&=\frac{g}{\sqrt{2}}\bm{\bar{u}_{0L},\ \bar{c}_{0L},\
\bar{t}_{0L}} \gamma^\mu \bm{d_{0L}\\s_{0L}\\b_{0L}}W^+_\mu +
\text{h.c.}
\end{align}
and when one passes to the physical basis according to
(\ref{eq7.90}), (\ref{eq7.91}), this becomes
\begin{equation}
\label{eq7.94} {\lagr}^{(\ti{quark})}_{CC} =
\frac{g}{\sqrt{2}}\bm{\bar{u}_L,\ \bar{c}_L,\ \bar{t}_L}
\gamma^\mu\, \Ut \U^\dagger\bm{d_L\\s_L\\b_L}W^+_\mu + \text{h.c.}
\end{equation}
in full analogy with (\ref{eq7.73}). Thus, the flavour
mixing\index{flavour!mixing|ff} occurring in quark interactions
with $W$\index{W boson@$W$ boson|ff} bosons is now represented by
a $3 \times 3$ unitary matrix $\Ut\U^\dagger$ that has replaced
the $U_\ti{GIM}$ discussed previously.

The $U_\ti{GIM}$ was eventually described with the help of a
single real parameter -- the Cabibbo angle -- and one may wonder
what is a physically relevant parametrization of the $3 \times 3$
matrix $\Ut\U^\dagger$ in (\ref{eq7.94}). Let us start our
counting with a general unitary matrix $3 \times 3$. This has nine
complex (i.e. eighteen real) elements, which are constrained by
three real and three complex conditions (normalization of columns
to unit length and their mutual orthogonality). A complex
condition is equivalent to two real ones, so one can also say that
the elements of a matrix in question are subject to $3 + 2 \times
3 = 9$ real constraints. Thus, a unitary $3 \times 3$ matrix is
parametrized by means of $18 - 9 = 9$ real numbers. However, when
one has in mind the matrix $\Ut\U^\dagger$ entering the Lagrangian
(\ref{eq7.94}), the number of its independent parameters can be
further reduced: some phase factors become physically irrelevant,
as they can be absorbed into appropriate redefinitions of the
quark fields. We are now going to show explicitly, how this is
done. Denoting
\begin{equation}
\label{eq7.95} \Ut\U^\dagger = V =
\bm{V_{11}&V_{12}&V_{13}\\V_{21}&V_{22}&V_{23}\\V_{31}&V_{32}&V_{33}}
\end{equation}
the essential part of the expression (\ref{eq7.94}) reads
\begin{equation}
\label{eq7.96} \bm{\bar{u},\ \bar{c},\ \bar{t}}
\bm{V_{11}&V_{12}&V_{13}\\V_{21}&V_{22}&V_{23}\\V_{31}&V_{32}&V_{33}}
\bm{d\\s\\b}
\end{equation}
(for the moment, we may ignore Dirac gamma matrices within the
weak current, since in the present context these play the role of
an overall numerical factor). One may now factor out possible
complex phase factors from the first column of (\ref{eq7.95}) and,
having in mind a later redefinition of the quark fields $u, c, t$,
recast the $V$ identically as
\begin{equation}
V = \bm{{\rm e}^{i\delta_{11}}&0&0\\0&{\rm
e}^{i\delta_{21}}&0\\0&0&{\rm e}^{i\delta_{31}}} \bm{{\rm
e}^{-i\delta_{11}}&0&0\\0&{\rm e}^{-i\delta_{21}}&0\\0&0&{\rm
e}^{-i\delta_{31}}} \bm{R_{11}{\rm
e}^{i\delta_{11}}&V_{12}&V_{13}\\R_{21}{\rm
e}^{i\delta_{21}}&V_{22}&V_{23}\\R_{31}{\rm
e}^{i\delta_{31}}&V_{32}&V_{33}}
\end{equation}
where the $R_{11}, R_{21}, R_{31}$ are real numbers (as well as
the $\delta_{11}, \delta_{21}, \delta_{31}$). The matrix product
(\ref{eq7.96}) then becomes
\begin{multline}
\bm{\bar{u}{\rm e}^{i\delta_{11}},\ \bar{c}{\rm
e}^{i\delta_{21}},\ \bar{t}{\rm e}^{i\delta_{31}}}
\bm{R_{11}&V_{12}{\rm e}^{-i\delta_{11}}&V_{13}{\rm
e}^{-i\delta_{11}}\\R_{21}&V_{22}{\rm
e}^{-i\delta_{21}}&V_{23}{\rm
e}^{-i\delta_{21}}\\R_{31}&V_{32}{\rm
e}^{-i\delta_{31}}&V_{33}{\rm e}^{-i\delta_{31}}} \bm{d\\s\\b}\\ =
\bm{\bar{u}',\ \bar{c}',\ \bar{t}'}
\bm{R_{11}&V'_{12}&V'_{13}\\R_{21}&V'_{22}&V'_{23}\\R_{31}&V'_{32}&V'_{33}}
\bm{d\\s\\b} \label{eq7.98}
\end{multline} where
\begin{equation}
\label{eq7.99} u'={\rm e}^{-i\delta_{11}}u,\quad c'={\rm
e}^{-i\delta_{21}}c,\quad t'={\rm e}^{-i\delta_{31}}t
\end{equation}
and we have introduced an obvious shorthand notation for the
matrix elements in the last expression of (\ref{eq7.98}). Next, we
make a second step in this direction and write
\begin{multline}
\bm{R_{11}&V'_{12}&V'_{13}\\R_{21}&V'_{22}&V'_{23}\\R_{31}&V'_{32}&V'_{33}}=\\
= \bm{R_{11}&R_{12}{\rm e}^{i\delta'_{12}}&R_{13}{\rm
e}^{i\delta'_{13}}\\R_{21}&V'_{22}&V'_{23}\\R_{31}&V'_{32}&V'_{33}}
\bm{1&0&0\\0&{\rm e}^{-i\delta'_{12}}&0\\0&0&{\rm
e}^{-i\delta'_{13}}} \bm{1&0&0\\0&{\rm
e}^{i\delta'_{12}}&0\\0&0&{\rm e}^{i\delta'_{13}}}
\end{multline}
Using this, the expression (\ref{eq7.98}) can finally be recast as
\begin{equation}
\label{eq7.101} \bm{\bar{u}',\ \bar{c}',\ \bar{t}'}
\bm{R_{11}&R_{12}&R_{13}\\R_{21}&V''_{22}&V''_{23}\\R_{31}&V''_{32}&V''_{33}}
\bm{d\\s'\\b'}
\end{equation}
where
\begin{equation}
\label{eq7.102} s'={\rm e}^{i\delta'_{12}}s,\quad b'={\rm
e}^{i\delta'_{13}}b
\end{equation}
and the meaning of the other symbols should be clear.

Needless to say\index{Cabibbo--Kobayashi--Maskawa (CKM)!phase|(},
the quark field redefinitions (\ref{eq7.99}), (\ref{eq7.102}) have
no physical consequences and we will drop the primes in what
follows. What we have achieved is that we got rid of five complex
phase factors that could generally occur in the first column and
the first row of (\ref{eq7.95}). In other words, we have reduced
the number of physically relevant real parameters describing our
unitary matrix $V = \Ut\U^\dagger$ from nine to four. If the $V$
were purely real (i.e. real orthogonal), it would be parametrized
by just three rotation angles. Thus, the fourth real parameter
left over corresponds to the phase of a complex factor ${\rm
e}^{i\delta}$. Summing up these considerations, we see that the
six-flavour mixing matrix entering the interactions of quark
charged currents\index{charged current} (\ref{eq7.94}) can be
parametrized by means of three rotation angles $\theta_1,
\theta_2, \theta_3$ and one complex phase.

As regards the neutral currents, it is quite clear that within the
considered three-generation scheme they must exhibit essentially
the same properties as in the four-quark model discussed
previously. Indeed, one has a naturally diagonal structure in
terms of the unphysical quark fields, which now become grouped in
triplets $u_0, c_0, t_0$ and $d_0, s_0, b_0$ respectively. The
unitary transformations (\ref{eq7.90}), (\ref{eq7.91}) are then
implemented in a straightforward manner and, as we know, they
preserve automatically the diagonal character of the currents in
question -- the reason is that one encounters only products like
$\U\U^\dagger$ etc., equal to unit matrix. This is gratifying, as
all the available experimental data clearly show that processes in
which e.g. a $b$ quark would change into $s$ are strongly
suppressed, similarly as in the case of $d-s$ transitions
mentioned earlier.

The six-quark version of the $SU(2)\times U(1)$ electroweak theory
described above was formulated for the first time by M. Kobayashi
and T. Maskawa in their celebrated paper \cite{ref65}. This
remarkable work (especially its timing) certainly deserves an
additional historical commentary, but we postpone it to the end of
this section. Now let us come back to some important physical
aspects of the charged-current interactions
\begin{equation}
\label{eq7.103} {\lagr}^{(\ti{quark})}_{CC} = \frac{g}{2\sqrt{2}}
\bm{\bar{u},\ \bar{c},\ \bar{t}}\gamma^\mu(1-\gamma_5)
V\bm{d\\s\\b}W^+_\mu + \text{h.c.}
\end{equation}
involving the flavour-mixing matrix
\begin{equation}
\label{eq7.104}
V=\bm{V_{ud}&V_{us}&V_{ub}\\V_{cd}&V_{cs}&V_{cb}\\V_{td}&V_{ts}&V_{tb}}
\end{equation}
parametrized as indicated above (cf. (\ref{eq7.101})). This is
called {\bf Kobayashi--Maskawa} (or {\bf
Cabibbo--Kobayashi--Maskawa}) {\bf matrix}\footnote{In what
follows, we will usually employ the acronym \qq{CKM matrix} that
has become customary in the current literature.}
\index{Cabibbo--Kobayashi--Maskawa (CKM)!matrix|(}and its
essential feature is that it has, in general, a non-trivial
imaginary part -- as we have seen, it contains a complex phase
that cannot be removed by redefinitions of quark fields. The
original parametrization \cite{ref65} of the $V$ relies on a
simple generalization of Euler-type rotations;\index{Euler-type
rotation}\footnote{For other parametrizations that have become
more common in current literature see \cite{ref5}.} it is written
as
\begin{equation}
V=\bm{1&0&0\\0&c_2&s_2\\0&-s_2&c_2}\bm{c_1&s_1&0\\-s_1&c_1&0\\0&0&{\rm
e}^{i\delta}}\bm{1&0&0\\0&c_3&s_3\\0&-s_3&c_3}
\end{equation}
where $c_i = \cos{\theta_i}$, $s_i=\sin{\theta_i}$ for $i = 1, 2,
3$, so that the resulting form is
\begin{equation}
V = \bm{c_1&s_1 c_3&s_1 s_3\\-s_1 c_2& c_1 c_2 c_3 - s_2 s_3 {\rm
e}^{i\delta}&c_1 c_2 s_3 + s_2 c_3 {\rm e}^{i\delta}\\ s_1 s_2&
-c_1 s_2 c_3 - c_2 s_3 {\rm e}^{i\delta}&-c_1 s_2 s_3 + c_2 c_3
{\rm e}^{i\delta}}
\end{equation}
According to (\ref{eq7.103}), the pattern of $W$ boson couplings
to quarks is determined, up to an overall real factor, by the
elements of the CKM matrix. In this way, at least some coupling
constants in the charged-current sector can be imaginary and this
in turn has dramatic consequences for symmetry properties of the
relevant interaction Lagrangian: in general, it is no longer
invariant under ${\cal CP}$\index{CP violation@${\cal CP}$
violation|ff}, the combination of charge conjugation $C$ and space
inversion $\cal P$. In other words, apart from the separate
violation of $\cal C$ and $\cal P$
\index{C-parity@$\mathcal{C}$-parity!violation}(which is due to
the $V-A$ nature of charged currents), one can have a $\cal CP$
violation as well, if the phase $\delta$ is different from zero
(of course, the interaction Lagrangian is still invariant under
$\cal CPT$, so one can also say that $\delta\neq 0$ is tantamount
to a violation of $\cal T$, the time-reversal
invariance\index{time reversal}). We will now explain in more
detail, how the fact that a coupling constant is not real implies
the $\cal CP$ violation in considered interactions; to this end,
the results of Section~\ref{sec2.9} will be utilized in a substantial way
(it is sufficient to stay at the level of classical fields).

Let us consider a part of the interaction Lagrangian
(\ref{eq7.103}), written as
\begin{align}
{\lagr}^{(12)}_\ti{int} &= g_{12} \bar{\psi}_1 \gamma^\mu
(1-\gamma_5)\psi_2 W^+_\mu + \text{h.c.}\notag\\
&=g_{12}\bar{\psi}_1 \gamma^\mu (1-\gamma_5) \psi_2 W^+_\mu
+g^*_{12}\bar{\psi}_2\gamma^\mu (1-\gamma_5)\psi_1 W^-_\mu
\label{eq7.107}
\end{align}
where the fields $\psi_1, \psi_2$ represent two different quark
flavours and the coupling constant $g_{12}$ is, in general,
complex. According to (\ref{eq2.140}), a $V-A$\index{V-A
theory@$V-A$ theory} current transforms as
\begin{equation}
\bar{\psi}_1\gamma_\mu (1-\gamma_5)\psi_2(x) \xrightarrow{\cal CP}
\bar{\psi}_2(\tilde{x})\gamma^\mu(1-\gamma_5)\psi_1(\tilde{x})
\end{equation}
where $\tilde{x}= (x_0,\ -\vec{x})$ i.e. $\tilde{x}^\mu=x_\mu$.
The transformation law for the $W$ boson field should reflect its
four-vector character and the requirement that $W^+$ is changed
into $W^-$ under charge conjugation; thus, one has
\begin{equation}
W^\pm_\mu(x) \xrightarrow{\cal CP} W^{\mp\mu}(\tilde{x})
\end{equation}
(see e.g. \cite{Bra} for details). The resulting transformation of
the Lagrangian (\ref{eq7.107}) can therefore be written as
\begin{equation}
{\lagr}^{(12)}_\ti{int} \xrightarrow{\cal CP}
{\lagr}^{(12)'}_\ti{int} = g_{12}\bar{\psi}_2\gamma_\mu
(1-\gamma_5)\psi_1 W^{-\mu} + g^*_{12} \bar{\psi}_1 \gamma_\mu
(1-\gamma_5)\psi_2 W^{+\mu}
\end{equation}
(with the fields taken at the point $\tilde{x}$). Thus, it is seen
that for $g^*_{12}=g_{12}$ the original form of the Lagrangian
(\ref{eq7.107}) is not changed, but if $g^*_{12}\neq g_{12}$, the
$\cal CP$ invariance is lost. The possible non-invariance of the
Lagrangian (\ref{eq7.103}) under $\cal CP$, embodied in the CKM
matrix (\ref{eq7.104}), is very important from the
phenomenological point of view. $\cal CP$ violation in weak
interactions (in particular, in the system of neutral kaons) has
been an experimental fact for a long time \cite{ref26} and also
now it is a topic of paramount importance, in connection with
experimental studies of mesons containing the $b$ quark. We will
add more remarks on the history of the problem later in this
section and now, as a last technical point, let us generalize
slightly our previous discussion.

The generalization we have in mind is a model involving an
arbitrary number ($n$) generations of quarks, i.e. $n$ left-handed
doublets and $2n$ right-handed singlets, as a straightforward
extension of the pattern (\ref{eq7.84}), (\ref{eq7.85}).
Proceeding in analogy with the three-generation model described
above, one arrives at charged-current interaction involving an
$n\times n$ flavour mixing matrix, which is unitary by
construction. Obviously, one can also repeat the previous
considerations concerning the $V$ parametrization. As a unitary
$n\times n$ matrix, $V$ is in general described in terms of $n^2$
independent real parameters, since the $2n^2$ real numbers
representing its elements are subject to $n$ real and
$\binom{n}{2}=\frac{1}{2}n(n-1)$ complex constraints
(normalization and orthogonality of columns). Further, the first
column and row can be made real by appropriate redefinitions of
quark fields; in such a way, $n+(n-1)=2n-1$ parameters become
unphysical. Thus, an $n \times n$ generalization of the CKM matrix
involves $n^2-(2n-1)=(n-1)^2$ physically relevant real parameters.
Obviously, these comprise $\frac{1}{2}n(n-1)$ rotation angles
(that would provide complete description of a purely real $V$) and
the remaining $(n-1)^2-\frac{1}{2}n(n-1) = \frac{1}{2}(n-1)(n-2)$
parameters correspond to complex phases. Thus, one can conclude
that for $n$ generations of quarks (i.e. for $2n$ flavours) one
has
\begin{equation}
\label{eq7.111} \text{\# CKM phases} = \frac{1}{2}(n-1)(n-2)
\end{equation}

In particular, (\ref{eq7.111}) implies immediately that for $n =
2$ (i.e. for the four-quark model considered in preceding section)
there is no physically relevant complex
phase\index{Cabibbo--Kobayashi--Maskawa (CKM)!phase|)}; in other
words, the mixing matrix for four flavours can always be made
purely real and reduced thus to the GIM matrix (\ref{eq7.75}).
Thereby it is confirmed that our earlier result, obtained for the
GIM model in Section~\ref{sec7.4}, is in fact completely general (i.e., it
holds even in the case of complex Yukawa couplings\index{Yukawa
coupling}), though originally we have restricted ourselves to real
mass matrices. An important lesson to be learnt from the above
discussion is that a four-quark model cannot accommodate naturally
$\cal CP$ violation within the Lagrangian built according to the
principles of GWS theory.\footnote{It should be stressed that
another source of $\cal CP$ violation within a gauge theory of
electroweak interactions could be an extended Higgs sector
\cite{ref66}, but such a possibility has a highly speculative
status at present.}

Historically, the need for an incorporation of $\cal CP$ violation
into the $SU(2) \times U(1)$ gauge theory of electroweak
interactions was the prime motive that led Kobayashi and Maskawa
to consider a six-quark model as early as in 1973 -- at a time
when only three quarks $u, d, s$ were recognized \qq{officially}.
Of course, achieving right theoretical description of $\cal CP$
violation is an important goal. As we have already noted, $\cal
CP$ violating effects in weak interactions have been known since
1964; the crucial discovery was an observation of the decay of the
long-lived neutral kaon $K^0_L$ into two pions \cite{ref26}, a
process that would be strictly forbidden\index{forbidden
transition} if the $\cal CP$ symmetry were exact\index{decay!of
the kaon}. Following the GIM scheme, Kobayashi and Maskawa simply
noticed that the four-quark model is $\cal CP$ conserving, while
an extra generation of quarks would solve the problem quite
naturally. One should realize that this was a really bold
proposal, taking into account that even the fourth quark $c$ was
discovered only one year later, in 1974! Thus, it is not
surprising that the work of Kobayashi and Maskawa (KM) went almost
unnoticed when published, but it did gain some popularity after
the breakthrough discovery \cite{ref64} of the charmed quark
$c$\index{cquark@$c$-quark}\index{charm} (that we have already
mentioned at the end of Section~\ref{sec7.3}). Soon after that, the
\qq{heavy lepton} $\tau$\index{tau lepton@$\tau$ lepton} with mass
$m_\tau\doteq 1.8\ \GeV$ was observed \cite{ref67} (quite
unexpectedly at that time) and it has also become clear that
$\tau$ is accompanied by its own neutrino
$\nu_\tau$.\footnote{However, it should be noted that $\nu_\tau$
has been detected directly (see \cite{ref68}) only in 2000!} Then
in 1977 L. Lederman and collaborators \cite{ref69} found a new
resonance called $\Upsilon$\index{Ypsilon resonance@$\Upsilon$
(upsilon) resonance} that has been interpreted readily as a bound
state of a quark denoted as $b$ ($Q_b=-\frac{1}{3}$, $m_b\doteq
4.5\ \GeV$) and its antiquark; mesons carrying the $b$-flavour
were subsequently discovered during 1980s. Thus, the spectrum of
elementary fermions known since the late 1970s comprised six
leptons $\nu_e, e, \nu_\mu, \mu, \nu_\tau, \tau$ and five quarks
$u, d, s, c, b$. Of course, such a development provided much
support for the KM scheme, which has thus become a widely
recognized and trusted candidate for a realistic model of the
quark sector of electroweak theory. In 1980s, experimental data
for production of $b\bar{b}$ pairs in $e^+e^-$ annihilation
indicated clearly (though indirectly) that the $b$
quark\index{bquark@$b$-quark|)} must in fact belong to a weak
isospin\index{weak!isospin} $SU(2)$ doublet; moreover, decays
involving flavour-changing neutral currents\index{flavour-changing
neutral current} (i.e. transitions like $b \rightarrow s$ or $b
\rightarrow d$) were conspicuously absent. In a sense, the
situation of the early 1970s described in Section~\ref{sec7.2} thus
repeated itself: an {\it odd\/} number of quarks could not match
the demands of phenomenology and, at the same time, the elegant
and simple KM theory has already been at hand. The hunting for the
sixth quark $t$\index{t-quark@$t$-quark}, expected eagerly since
the late 1970s, was rather long and ended successfully in 1995
when its discovery was confirmed officially
\cite{ref70}.\footnote{Note that the top quark is much heavier
than intermediate vector bosons $W$ and $Z$: with its rest mass of
about $175\ \GeV$ it is as heavy as another $W$, the atom of
tungsten!} A remarkable quark-lepton symmetry thus has been
restored (three generations of quarks and leptons). Let us note
already here that apart from the absence of FCNC and an obvious
aesthetic appeal, such a scheme has another rather deep aspect:
equal number of quarks and leptons within each generation of
elementary fermions guarantees cancellation of the so-called ABJ
anomalies\index{Adler--Bell--Jackiw (ABJ) anomaly} and this in
turn is crucial for internal consistency of the considered
$SU(2)\times U(1)$ gauge theory. This technical aspect of the
electroweak standard model will be discussed in some detail in
Section~\ref{sec7.9}.

The KM model can serve as another example of a theoretical scheme
going far beyond the experimental knowledge of its time, yet
confirmed eventually in quite amazing way. The story of heavy
quark flavours described briefly in this chapter exhibits a
remarkable and typical feature of the electroweak standard model
--  an interplay of bold theoretical ideas and ingenious
experiments that resulted in a truly realistic and predictive
description of phenomena at a deep subnuclear level.

Finally, let us note that further study of $\cal CP$ violation is
a very important area of research; the present and forthcoming
experiments should determine in detail the elements of the flavour
mixing matrix and tell us whether the KM mechanism is indeed
sufficient for explaining all relevant phenomena. Of course,
sufficiently accurate data could also open up a window on a
possible new physics beyond the standard model. The literature
concerning the phenomenology of $\cal CP$ violation is enormous
and the subject is growing fast. As the present text is concerned
primarily with basic principles of the theory, the reader
interested in phenomenology (and/or in further technical details,
such as the various parametrizations of CKM matrix, etc.) is
referred e.g. to the monograph
\cite{Bra}\index{Cabibbo--Kobayashi--Maskawa (CKM)!matrix|)}.

%\end{document}

%\input{kniha76}
%%%%%%%%%%%%%%%%%%%%%%%%%%%%%%%%%%%%%%%%%%%%%%%%%%%%%%%%%%%%%%%%%%%
%%%%%%%%%%%%%%%%%%%%%%%%%%%%%%%%%%%%%%%%%%%%%%%%%%%%%%%%%%%%%%%%%%%%%%%%%%%%%%%%%%%%%%%%%%%%%%%%%%%%%%%%%%%%%%%%%%%%%%%%%%%%%%%%%%%%%%%%
%\documentclass[12pt,tbtags]{report}
%\usepackage{amsmath,amssymb,epsfig,euscript,amsfonts,dsfont}
%\newcommand{\qq}[1]{\textquotedblleft#1\/\textquotedblright} % to jsou uvozovky \qq{aaa} = "aaa"
%\newcommand{\qs}[1]{\textquoteleft#1\/\textquoteright} % apostrofy \qs{aaa}='aaa'
%\newcommand{\J}{\mathds{1}}     % jedn.matice
%\newcommand{\ti}[1]{\text{\it #1}}   % index v matematice
%\newcommand{\bm}[1]{\begin{pmatrix}#1\end{pmatrix}} % pro matice, vektory... napr.\bm{a&b\\c&d}
%\newcommand{\dmd}{\stackrel{\leftrightarrow}{\partial_\mu}\negthickspace} % pro oboustr.derivaci s index.\mu dole
%\newcommand{\dmn}{\stackrel{\leftrightarrow}{\partial^\mu}\negthickspace} % --""-- s indexem \mu nahore
%%%%%%%%%%%%%%%%%%%%%%%%%%%%%%%%%%%%%%%%%%%%%%%%%%%%%%%%%%%%%%%%%%%%%%%%%%%%%%%%%%%%%%%%%%%%%%%%%%%%%%%%%%%%%%%%%%%%%%%%%%%%
%\newcommand{\hv}{\tilde{h}}
%\newcommand{\Phit}{\widetilde{\Phi}}
%%%%%%%%%
%\newcommand{\lagr}{{\cal L}}
%\begin{document}
\section{$R$-gauges}\index{R-gauge@$R$-gauge|ff}\label{sec7.6}

As a last topic, we are going to discuss here and in the following
three sections some deeper gauge theory aspects of the GWS
standard model. From the technical point of view, our treatment
will be far from complete; nevertheless, it could serve,
hopefully, as a useful introduction to the concepts and techniques
involved. The main bonus to be gained is an additional insight
into some particular properties of the electroweak standard model,
encountered in previous chapters.

In our formulation of the SM we have employed so far the
$U$-gauge, in which the would-be Goldstone
bosons\index{Goldstone!boson} are eliminated explicitly and the
interaction Lagrangian is written solely in terms of the fields
corresponding to physical particles. However, it turns out that an
appropriate gauge can be fixed consistently in a completely
different way, such that the unphysical Goldstone boson fields are
kept in the Lagrangian and their effects only disappear at the
level of physical scattering amplitudes. As we have already noted
in Section~\ref{sec6.3}, the main advantage of such an approach (developed
originally by G. 't Hooft \cite{ref48}) is that the propagators of
massive vector bosons then exhibit the same asymptotic behaviour
as the photon propagator\index{photon propagator|ff} and this
makes the power-counting of the relevant ultraviolet
divergences\index{ultraviolet divergences} in higher-order Feynman
diagrams (related intimately to the problem of perturbative
renormalizability\index{perturbative renormalizability}) much more
transparent. Let us now describe such a gauge-fixing procedure in
detail. Since its formulation is essentially inspired by the
familiar case of quantum electrodynamics, we shall start with a
recapitulation of the covariant gauges in
QED\index{quantum!electrodynamics (QED)}.

The Lagrangian density for the free electromagnetic
(Maxwell)\index{Maxwell electrodynamics} field has the familiar
form\index{electromagnetic!field}
\begin{equation}
\label{eq7.112} \lagr_M = -\frac{1}{4}F_{\mu\nu}F^{\mu\nu}
\end{equation}
with
\begin{equation}
\label{eq7.113} F_{\mu\nu} = \partial_\mu A_\nu - \partial_\nu A_\mu
\end{equation}
For quantizing it, one has to cope with the problem of redundant
(unphysical) degrees of freedom involved in the four-potential
$A_\mu$ -- this, of course, is intimately related to the gauge
invariance of the Lagrangian (\ref{eq7.112}). In particular,
within a straightforward canonical approach based on
(\ref{eq7.112}) one cannot treat all components $A_\mu$,
$\mu=0,1,2,3$ on an equal footing, since the conjugate momentum
corresponding to $A_0$ vanishes identically. To see this, one
should first notice that
\begin{equation}
\frac{\delta\lagr_M}{\delta(\partial_\rho A_\sigma)} = -
F^{\rho\sigma}
\end{equation}
which means that the momentum $\pi_\mu$ associated with $A_\mu$ is
\begin{equation}
\pi_\mu = \frac{\delta\lagr_M}{\delta(\partial_0 A_\mu)} =
-F^{0\mu}
\end{equation}
and thus $\pi_0=-F^{00}=0$. One well-known way out is to quantize
only the physical degrees of freedom, but a manifest Lorentz
covariance is then lost. For accomplishing a covariant
quantization\index{quantization of!electromagnetic field|ff} one
has to keep all components $A_\mu$ in the game. This can be
consistently implemented by modifying the basic Lagrangian
(\ref{eq7.112}) (so as to break its gauge invariance) and imposing
subsequently an appropriate subsidiary condition on the physical
solutions.

In particular, instead of (\ref{eq7.112}) one can consider the
Lagrangian
\begin{equation}
\label{eq7.116} \widetilde{\lagr}_M = -\frac{1}{4}
F_{\mu\nu}F^{\mu\nu} - \frac{1}{2}(\partial\cdot A)^2
\end{equation}
where $\partial\cdot A$ is a shorthand notation for $\partial_\mu
A^\mu$. The equation of motion following from (\ref{eq7.116})
reads
\begin{equation}
\partial_\mu F^{\mu\nu} + \partial^\nu (\partial\cdot A) = 0
\end{equation}
which reduces to
\begin{equation}
\label{eq7.118} \Box A^\nu = 0
\end{equation}
when one takes into account (\ref{eq7.113}). Thus, one must add
the Lorenz condition\index{Lorenz!condition}
\begin{equation}
\label{eq7.119}
\partial\cdot A = 0
\end{equation}
to the d'Alembert equation (\ref{eq7.118}) if one wants to recover
the original Maxwell equations $\partial_\mu F^{\mu\nu}=0$. In
other words, the modified Lagrangian (\ref{eq7.116}) has to be
supplemented with the constraint (\ref{eq7.119}), if it is to
describe eventually the Maxwell field at the classical level. Note
also that the second term in (\ref{eq7.116}) is usually called the
\qq{gauge-fixing term}\index{gauge-fixing term|(} (and denoted
correspondingly
$\lagr_{g.f.}$) as it violates the gauge invariance.%
\footnote{Maxwell equations $\partial_\mu F^{\mu\nu}=0$ written in
terms of the $A_\rho$ read $\Box A^\nu -
\partial^\nu(\partial\cdot A) =0$ and when the gauge freedom is
constrained by $\partial\cdot A =0$, one is led to the d'Alembert
equation that follows directly from (\ref{eq7.116}). While this
observation may provide some additional justification of the term
\qq{gauge fixing} in the present context, one should keep in mind
that fixing a definite gauge for the electromagnetic
four-potential is in general rather subtle matter.}

In quantum theory, one can employ the Lagrangian (\ref{eq7.116})
and postulate a canonical commutation relation for any $A_\mu$,
$\mu=0,1,2,3$; note that the conjugate momenta are given by
\begin{equation}
\frac{\delta\widetilde{\lagr}_M}{\delta(\partial_0 A_\mu)} =
-F^{0\mu} - g^{0\mu}\partial\cdot A
\end{equation}
The Lorenz condition is then imposed, in an appropriate form, on
the physical states and the resulting (manifestly covariant)
theory is physically equivalent to a non-covariant formulation, in
which the unphysical degrees of freedom are eliminated from the
very beginning. Such a covariant procedure is originally due to S.
Gupta and K. Bleuler (see e.g. \cite{Ryd}); note also that for its
formulation, one has to introduce the state-vector space with
indefinite metric. The propagator of covariantly quantized field
$A_\mu$ can be calculated in a straightforward way and the result
(in the momentum space) reads
\begin{equation}
\label{eq7.121} D_{\mu\nu}(q) =
\frac{-g_{\mu\nu}}{q^2+i\varepsilon}
\end{equation}
The above procedure can be generalized in such a way that the
gauge-fixing term in (\ref{eq7.116}) is taken with an arbitrary
coefficient. Thus, the Lagrangian
\begin{equation}
\label{eq7.122} \lagr^{(\alpha)}_M = -\frac{1}{4}
F_{\mu\nu}F^{\mu\nu} - \frac{1}{2\alpha} (\partial\cdot A)^2
\end{equation}
is considered with $\alpha$ being a real \qq{gauge-fixing
parameter}; in order to make its physical contents equivalent to
the Maxwell field, an appropriate subsidiary condition has to be
added. Note that the Lagrangian (\ref{eq7.116}) considered
previously represents a particular case of (\ref{eq7.122}) with
$\alpha=1$. For a general $\alpha$ the canonical operator
quantization based on (\ref{eq7.122}) is more complicated than for
$\alpha=1$, because the equation of motion corresponding to
(\ref{eq7.122}) becomes
\begin{equation}
\partial_\lambda F^{\lambda\mu} + \frac{1}{\alpha} \partial^\mu
(\partial\cdot A) = 0
\end{equation}
or, equivalently
\begin{equation}
\label{eq7.124} \Box A^\mu + \Bigl(\frac{1}{\alpha}-1\Bigr)\partial^\mu
(\partial\cdot A) =0
\end{equation}
and this obviously does not coincide with the simple d'Alembert
equation for $\alpha\neq 1$. An appropriate generalization of the
Gupta--Bleuler method is the so-called Nakanishi--Lautrup
formalism (see e.g. \cite{Nak}), but we will not need such a
detailed treatment for our purpose.\footnote{Let us remark that
the modified Maxwell Lagrangian (\ref{eq7.122}) is also a
convenient starting point for a covariant quantization by means of
the path-integral method.} The object of our primary interest is
the propagator, which can be obtained as a Green's
function\index{Green's function} of the equation (\ref{eq7.124})
(in analogy with what we do e.g. for the massive vector field, cf.
Appendix~\ref{appenD}): to find it, one simply has to solve the equation
\begin{equation}
\label{eq7.125} \Box \mathcal{D}_\nu^\mu (x) +
\Bigl(\frac{1}{\alpha}-1\Bigr)\partial^\mu \bigl(\partial_\rho
\mathcal{D}^\rho_\nu (x) \bigr) = \delta^\mu_\nu \delta^4 (x)
\end{equation}
with $\mathcal{D}_\nu^\mu$ denoting the propagator in the
coordinate representation. Passing to the momentum space via
Fourier transformation, (\ref{eq7.125}) yields an algebraic
(matrix) equation
\begin{equation}
\label{eq7.126} L^\mu_\rho (q) D^\rho_\nu (q) = \delta^\mu_\nu
\end{equation}
%$D\mathcal{D}\mathfrak{D}\EuScript{D}$
where the $D^\rho_\nu$ stands for the Fourier transform of the
$\mathcal{D}^\rho_\nu$ and the coefficient matrix $L$ is given by
\begin{equation}
L^\mu_\rho(q)=-q^2 g^\mu_\rho + \Bigl(1-\frac{1}{\alpha}\Bigr)q^\mu q_\rho
\end{equation}
where we have taken into account that
$\delta^\mu_\rho=g^\mu_\rho$. The equation (\ref{eq7.126}) is
solved by inverting the matrix $L$ by means of the method
explained in Appendix~\ref{appenD} and one thus gets
\begin{equation}
\label{eq7.128} D_{\mu\nu}^{(\alpha)} =
\frac{1}{q^2+i\varepsilon}\Bigl[-g_{\mu\nu}+(1-\alpha)\frac{q_\mu
q_\nu}{q^2}\Bigr]
\end{equation}
where we have also introduced the usual $+i\varepsilon$
prescription for the Feynman propagator.\footnote{It is
instructive to note that for the original Maxwell Lagrangian
(\ref{eq7.112}), which corresponds formally to the limit $\alpha
\rightarrow \infty$ in (\ref{eq7.122}), the coefficient matrix $L$
in (\ref{eq7.126}) would become $-q^2g^\mu_\rho + q^\mu q_\rho$
and this is singular, i.e. has no inverse. Of course, such a
\qq{pathological} behaviour is due to the gauge invariance of the
$\lagr_M$. By adding the $\lagr_{g.f.}$ one breaks the original
gauge symmetry and the singularity is thus removed. From this
point of view, the $\lagr_{g.f.}$ can be understood as a simple
device for fixing a propagator of the electromagnetic field in a
consistent way.} In this way, we get a one-parametric set of
photon propagators that can be used in Feynman-diagram
calculations. Different choices of the parameter $\alpha$ are --
for reasons indicated above -- referred to as different
\qq{gauges}. In particular, the value $\alpha=1$ (for which one
recovers the previous result (\ref{eq7.121})) corresponds to the
{\bf Feynman gauge}\index{propagator!in the Feynman gauge}, as it
is called in common parlance. It is also interesting to note that
for $\alpha=0$ one gets a purely transverse propagator
\begin{equation}
\label{eq7.129} D_{\mu\nu}^{(\alpha=0)} (q) =
\frac{1}{q^2+i\varepsilon}\Bigl(-g_{\mu\nu} + \frac{q_\mu q_\nu}{q^2}\Bigr)
\end{equation}
that corresponds to the so-called {\bf Landau gauge}\index{Landau
gauge}\index{propagator!in the Landau gauge}. Obviously, such a
value of $\alpha$ cannot be accommodated directly in the
Lagrangian (\ref{eq7.122}); the result (\ref{eq7.129}) should be
understood as a limiting case of eq.\,(\ref{eq7.128}) within our
straightforward approach.

In calculations of physical scattering amplitudes, all gauges
should be equivalent, i.e. the results should be independent of
$\alpha$. For tree-level diagrams it is elementary to verify such
a statement explicitly; as an instructive example, one can
consider e.g. the process $e^+ e^- \rightarrow \mu^+\mu^-$ in the
lowest order of spinor QED and show that the $q_\mu q_\nu$ part of
the photon propagator (\ref{eq7.128}) does not contribute within
the relevant Feynman diagram (the reader is recommended to prove
it directly by using equations of motion for Dirac spinors in
external fermion lines). Of course, the crucial underlying fact is
that in a QED interaction vertex the electromagnetic
four-potential is coupled to conserved current.

\index{Z boson@$Z$ boson|(}After this somewhat long introduction
we are now in a position to examine an appropriate class of
covariant gauges for the GWS electroweak theory. The scheme
\cite{ref48}, \cite{ref71} we have in mind is conceptually rather
similar to that discussed above, but there is also a significant
difference in comparison with the QED case: the vector fields $W$
and $Z$ become massive through the Higgs mechanism\index{Higgs
mechanism} and thus they are mixed, in a sense, with the would-be
Goldstone bosons (that eventually disappear from the physical
spectrum). The key idea of the original papers \cite{ref48},
\cite{ref71} is to retain the unphysical Goldstone bosons as
auxiliary fields in the Lagrangian and add subsequently a suitable
gauge-fixing term to the original gauge invariant Lagrangian. For
implementing this, let us start with the Higgs--Goldstone scalar
doublet (\ref{eq6.41})
\begin{equation}
\label{eq7.130} \Phi = \bm{\varphi^+\\ \varphi^0} = \bm{\varphi_1 + i\varphi_2\\
\varphi_3 + i\varphi_4}
\end{equation}
and reparametrize it as
\begin{equation}
\label{eq7.131} \Phi = \bm{-iw^+\\ \frac{1}{\sqrt{2}}(v+H+iz)}
\end{equation}
where the $H$ stands for the physical Higgs boson and the constant
$v$ has the familiar meaning. Of course, the hermitean conjugate
$\Phi^\dagger$ is then written as
\begin{equation}
\label{eq7.132} \Phi^\dagger = \bm{iw^-,\
\frac{1}{\sqrt{2}}(v+H-iz)}
\end{equation}
The auxiliary fields $w^\pm$, $z$ correspond to the would-be
Goldstone bosons. Now, the Higgs part of the GWS Lagrangian (see
(\ref{eq6.53})) reads\footnote{Throughout this discussion we are
ignoring the fermionic sector of the GWS Lagrangian as this has no
impact on the problem of gauge fixing. We will retrieve the
interactions of fermions later on.}
\begin{equation}\begin{split}
\label{eq7.133} \lagr_\ti{Higgs} =&\phantom{-}
\Phi^\dagger(\partialvb_\mu + ig A_\mu^a \frac{\tau^a}{2} +
\frac{1}{2}ig'B_\mu)(\partialv^\mu - ig A^{b\mu}\frac{\tau^b}{2} -
\frac{1}{2}ig'B^\mu) \Phi \\&- \lambda(\Phi^\dagger \Phi -
\frac{v^2}{2})^2
\end{split}
\end{equation}
When the expressions (\ref{eq7.131}) and (\ref{eq7.132}) are
substituted into (\ref{eq7.133}), the vacuum\index{vacuum} shift
$v$ yields mass terms for vector bosons and the gauge fields
$A^a_\mu$, $B_\mu$ can be replaced by their physical combinations
$W^\pm_\mu$, $Z_\mu$ and $A_\mu$ in full analogy with what has
been done in Section~\ref{sec6.4}. Of course, the shift of the lower
component of the Higgs doublet\index{Higgs!doublet} $\Phi$ would
lead to mass terms {\it in any parametrization}, but an advantage
of the representation (\ref{eq7.131}) over (\ref{eq7.130}) is that
the $w^\pm$ and $z$ are, technically, direct counterparts of the
vector fields $W^\pm_\mu$ and $Z_\mu$. This becomes clear when one
considers the relevant mixing terms, i.e. terms bilinear in scalar
and vector fields that show up in the gauge invariant Lagrangian
(\ref{eq7.133}) upon substitutions (\ref{eq7.131}),
(\ref{eq7.132}). A straightforward (though somewhat tedious)
calculation yields the result
\begin{equation}
\label{eq7.134} (D_\mu\Phi)^\dagger(D^\mu\Phi) = m_W (\partial^\mu
w^- W_\mu^+ +
\partial^\mu w^+ W_\mu^-) + m_Z \partial^\mu z Z_\mu + \ldots
\end{equation}
where \qq{$\ldots$} denotes all remaining contributions to
(\ref{eq7.133}), including mass terms and true interactions. Now,
one has to add a gauge-fixing term. It is clearly desirable to get
rid of the mixing contributions (\ref{eq7.134}), since the
quadratic part of the Lagrangian should be diagonal in order to
define a conventional perturbation theory. An appropriate choice
of the $\lagr_{g.f.}$ can indeed do the job. To see this, let us
define
\begin{equation}
\label{eq7.135}
\begin{split} \lagr_{g.f.} = -&\frac{1}{2\xi} |\partial\cdot W^- - \xi m_W
w^-|^2
-\frac{1}{2\xi}|\partial\cdot W^+ - \xi m_W w^+|^2 \\
-&\frac{1}{2\eta}(\partial\cdot Z - \eta m_Z z)^2 -
\frac{1}{2\alpha} (\partial\cdot A)^2
\end{split}
\end{equation}
where the $\xi$, $\eta$ and $\alpha$ are arbitrary real
parameters. The expression (\ref{eq7.135}) can immediately be
recast as
\begin{equation}
\begin{split}
\lagr_{g.f.} = -&\frac{1}{\xi} (\partial\cdot W^- - \xi m_W
w^-)(\partial\cdot W^+ - \xi m_W w^+)\\ -&\frac{1}{2\eta}
(\partial\cdot Z - \eta m_Z z)^2 - \frac{1}{2\alpha}
(\partial\cdot A)^2
\end{split}
\end{equation}
which, in turn, is easily worked out as
\begin{equation}
\label{eq7.137}
\begin{split}
\lagr_{g.f.} = -&\frac{1}{\xi} (\partial\cdot W^-)(\partial\cdot
W^+) + m_W w^+ \partial\cdot W^- + m_W w^- \partial \cdot W^+\\
-&\xi m_W^2 w^-w^+ \\
-&\frac{1}{2\eta}(\partial\cdot Z)^2 + m_Z z\partial \cdot Z -
\frac{1}{2} \eta m_Z^2 z^2\\
-&\frac{1}{2\alpha} (\partial\cdot A)^2
\end{split}
\end{equation}
When this is combined with (\ref{eq7.134}), one gets
\begin{equation}
\begin{split}
\lagr_\ti{Higgs} + \lagr_{g.f.} = \phantom{+}&m_W w^+
\partial\cdot W^- + m_W
w^- \partial\cdot W^+ + m_Z z \partial\cdot Z \\
+\ &m_W W^{-\mu} \partial_\mu w^+ + m_W W^{+\mu} \partial_\mu w^-
+ m_Z Z^\mu \partial_\mu z + \ldots
\end{split}
\end{equation}
where we have singled out explicitly only the total contribution
to the scalar -- vector boson mixing, suppressing the other terms
for the moment. Obviously, the last expression can be recast as
\begin{equation}
\lagr_\ti{Higgs}+\lagr_{g.f.} = m_W \partial_\mu(w^+W^{-\mu}) +
m_W
\partial_\mu (w^-W^{+\mu}) + m_Z \partial_\mu(zZ^\mu) + \ldots
\end{equation}
Thus, the bilinear terms in question are combined into
four-divergences and therefore can be discarded from the
Lagrangian.\footnote{Let us remind the reader that a term of the
form $\partial_\mu X^\mu$ in a Lagrangian density does not
contribute to the action and thus does not influence the dynamical
contents (the equations of motion) of the theory.}

Let us now analyze the remaining quadratic terms in the considered
Lagrangian. First, for the scalar fields one gets
\begin{equation}
\label{eq7.140}
\begin{split}
\lagr_\ti{Higgs} + \lagr_{g.f.} = \phantom{-}&\partial_\mu w^-
\partial^\mu w^+ + \frac{1}{2} \partial_\mu z \partial^\mu z +
\frac{1}{2}
\partial_\mu H \partial^\mu H \\
-&\xi m_W^2 w^-w^+ - \frac{1}{2} \eta m_Z^2 z^2 - \lambda v^2 H^2
+\ldots
\end{split}
\end{equation}
Note that the kinetic terms for $w^\pm$, $z$ and $H$ (as well as
the $H$ mass term) descend from $\lagr_\ti{Higgs}$  (using
(\ref{eq7.133}) and (\ref{eq7.131}), (\ref{eq7.132})), while the
mass terms for $w^\pm$ and $z$ originate from the $\lagr_{g.f.}$
(see (\ref{eq7.137})). The result (\ref{eq7.140}) means that the
masses (or, more accurately, \qq{mass parameters}) of the $w^\pm$
and $z$ can be identified as
\begin{equation}
\label{eq7.141} m_{w^\pm}^2 = \xi m_W^2, \qquad m_z^2 = \eta m_Z^2
\end{equation}
while the $H$ mass is given by $m_H^2 = 2\lambda v^2$ as before
(see (\ref{eq6.51})). The dependence of the \qq{masses}
(\ref{eq7.141}) on the gauge parameters $\xi$, $\eta$ reflects
clearly the unphysical nature of the $w^\pm$ and $z$. Next,
collecting all quadratic terms for vector fields (including also
the kinetic term as given by eq. (\ref{eq5.25})), one has
\begin{equation}
\label{eq7.142}
\begin{split}
\lagr_\ti{gauge}^{(kin.)} + \lagr_\ti{Higgs} + \lagr_{g.f.} =
&-\frac{1}{2} W^-_{\mu\nu} W^{+\mu\nu} - \frac{1}{\xi}
(\partial\!\cdot\!W^-)(\partial\cdot W^+) + m_W^2 W^-_\mu W^{+\mu}\\
&-\frac{1}{4}Z_{\mu\nu}Z^{\mu\nu} - \frac{1}{2\eta} (\partial\cdot
Z)^2 + \frac{1}{2} m_Z^2 Z_\mu Z^\mu \\
&-\frac{1}{4} A_{\mu\nu}A^{\mu\nu} -\frac{1}{2\alpha}
(\partial\cdot A)^2 +\ldots
\end{split}
\end{equation}
Here, the mass terms are produced by $\lagr_\ti{Higgs}$ while the
additional derivative contributions dependent on $\xi$, $\eta$ and
$\alpha$ are obviously due to $\lagr_{g.f.}$. Of course, the part
corresponding to the electromagnetic field is the same as in pure
QED, as expected.

To sum up the preceding considerations, we have fixed the relevant
free-field Lagrangian, which is a prerequisite for defining the
perturbation expansion. From (\ref{eq7.142}) one can derive
equations of motion in a standard manner and, subsequently, the
propagators for $W^\pm_\mu$ and $Z_\mu$ are determined as the
corresponding Green's functions. To this end, we employ the same
technique as in the case of covariant photon propagator
(\ref{eq7.128}) (phrased in a common jargon, one has to \qq{invert
the quadratic part of the Lagrangian}). For the $W$ boson we thus
get\index{propagator!in the $U$- and $R$-gauge}
\begin{equation}
\label{eq7.143} D_{\mu\nu}^{(\xi)}(q; m_W) = \frac{1}{q^2 - m_W^2
+i\varepsilon} [-g_{\mu\nu} + (1-\xi) \frac{q_\mu q_\nu}{q^2 - \xi
m_W^2}]
\end{equation}
and, similarly, the $Z$ propagator becomes
\begin{equation}
\label{eq7.144} D_{\mu\nu}^{(\eta)} (q; m_Z) = \frac{1}{q^2
-m_Z^2+i\varepsilon} [-g_{\mu\nu} + (1-\eta) \frac{q_\mu
q_\nu}{q^2 - \eta m_Z^2}]
\end{equation}
The propagators of the unphysical Goldstone bosons follow
immediately from (\ref{eq7.140}); one has
\begin{eqnarray}
D_w^{(\xi)} &=& \frac{1}{q^2 - \xi m_W^2 + i\varepsilon}\nonumber\\
D_z^{(\eta)} &=& \frac{1}{q^2 - \eta m_Z^2 + i\varepsilon}
\label{eq7.145}
\end{eqnarray}
for the $w^\pm$ and $z$ respectively.

An astute reader may observe that in the limit $\xi, \eta
\rightarrow \infty$ one recovers the $U$-gauge results, namely
\begin{align}
\lim_{\xi\rightarrow \infty} D^{(\xi)}_{\mu\nu}(q;m_W) &=
\frac{1}{q^2 - m_W^2 + i\varepsilon} [ -g_{\mu\nu}
+\frac{1}{m_W^2} q_\mu q_\nu]\\ \intertext{and}
\lim_{\eta\rightarrow \infty} D^{(\eta)}_{\mu\nu}(q;m_Z) &=
\frac{1}{q^2 - m_Z^2 + i\varepsilon} [ -g_{\mu\nu}
+\frac{1}{m_Z^2} q_\mu q_\nu]
\end{align}
Further, from (\ref{eq7.145}) it is obvious that the propagators
of the $w^\pm$ and $z$ vanish identically for $\xi, \eta
\rightarrow \infty$. This is gratifying (as a consistency check of
our formalism), since the unphysical Goldstone bosons are absent
in the $U$-gauge by definition.

One should also note that the gauge fixing for the electromagnetic
field is obviously \qq{decoupled} from the procedure used for the
massive vector bosons $W$ and $Z$, as there is no unphysical
Goldstone boson associated with the massless photon. In
particular, the covariant term $\frac{1}{2\alpha}(\partial\cdot
A)^2$ can either be employed as a part of the scheme
(\ref{eq7.135}), or added directly to the $U$-gauge Lagrangian; of
course, one is free to use a non-covariant gauge for the photon as
well.

The above-described procedure based on adding the terms
(\ref{eq7.135}) to the GWS gauge invariant Lagrangian defines the
class of the so-called {\bf $\boldsymbol{R}$-gauges} (or
$R_\xi$-gauges), where the \qq{$R$} stands for
\qq{renormalizable}\index{renormalizable theory}. Such a label
refers to the fact that the massive vector boson propagators
(\ref{eq7.143}), (\ref{eq7.144}) fall off as $1/q^2$ for $q^2
\rightarrow \infty$, which in turn means that the theory is of
renormalizable type (this is indicated by the usual power-counting
analysis based on an evaluation of the index of divergence of a
general Feynman graph).\footnote{For the calculation of the
\qq{index} of a one-particle irreducible Feynman graph (called
also \qq{superficial degree of divergence}) within a general field
theory model see e.g. \cite{ItZ} or the Appendix G in \cite{Hor}.}
From the technical point of view, the formulation of the
$R$-gauges was a real breakthrough as it played a key role in the
proof of renormalizability of gauge theories with the Higgs
mechanism, accomplished first by G. 't Hooft and M. Veltman
\cite{ref47}. A more detailed discussion of this problem would go
far beyond the scope of this treatment, but at least one important
note is in order here. The $R$-gauges are instrumental in taming
the ultraviolet divergences of higher-order Feynman diagrams (that
are hard to control within the $U$-gauge formulation), but there
is a price to be paid for that. In particular, the presence of the
unphysical fields $w^\pm$ and $z$ requires special care when
proving unitarity of the $S$-matrix and its independence on the
gauge-fixing parameters; this, in fact, was the main issue of the
pioneering works \cite{ref47}.

In this context, it should be stressed that our discussion of the
$R$-gauges has been incomplete, in the technical sense, since
eventually one must also add another set of unphysical fields --
the so-called Faddeev--Popov (FP) ghosts\index{Faddeev--Popov
ghosts}. These fictitious particles enter only closed loops of
internal lines in Feynman diagrams and are necessary for
maintaining unitarity of the $S$-matrix in higher orders. Note
that they are absent in the $U$-gauge. In the present text, we
will not need FP ghosts in any of our calculations and therefore
we are not going to pursue this technical issue any further. A
detailed discussion of the FP term in the GWS electroweak
Lagrangian can be found e.g. in \cite{BaL}.

Finally, let us add a terminological remark. In his original paper
\cite{ref48}, 't~Hooft proposed the gauge corresponding to
$\xi=\eta=1$ in (\ref{eq7.143}), (\ref{eq7.144}). In a sense, this
is the simplest possible choice, since the $q_\mu q_\nu$ parts of
the $W$ and $Z$ propagators are then absent, in analogy with the
Feynman gauge for the photon propagator. Thus, such an option is
usually called the {\bf't Hooft--Feynman gauge}\index{t Hooft
Feynman@'t Hooft--Feynman gauge|ff}. Similarly, for $\xi=\eta=0$
one gets purely transverse vector boson propagators and this case
is therefore referred to as the {\bf't Hooft--Landau
gauge}\index{t Hooft Landau@'t Hooft--Landau gauge}. Note also
that in the 't Hooft--Feynman gauge the unphysical scalars $w^\pm$
and $z$\index{unphysical scalars} have the same masses as the
$W^\pm$ and $Z$ respectively, while in the 't Hooft--Landau gauge
they become massless. In the general case, there is a common
practice to take the same gauge for $W$ and $Z$, i.e. set
$\xi=\eta$.
%\end{document}

%\input{kniha77}
%%%%%%%%%%%%%%%%%%%%%%%%%%%%%%%%%%%%%%%%%%%%%%%%%%%%%%%%%%%%%%%%%%%
%%%%%%%%%%%%%%%%%%%%%%%%%%%%%%%%%%%%%%%%%%%%%%%%%%%%%%%%%%%%%%%%%%%%%%%%%%%%%%%%%%%%%%%%%%%%%%%%%%%%%%%%%%%%%%%%%%%%%%%%%%%%%%%%%%%%%%%%
%\documentclass[12pt,tbtags]{report}
%\usepackage{amsmath,amssymb,epsfig,euscript,amsfonts,dsfont,stmaryrd,subfigure,slashed}
%\newcommand{\qq}[1]{\textquotedblleft#1\/\textquotedblright} % to jsou uvozovky \qq{aaa} = "aaa"
%\newcommand{\qs}[1]{\textquoteleft#1\/\textquoteright} % apostrofy \qs{aaa}='aaa'
%\newcommand{\J}{\mathds{1}}     % jedn.matice
%\newcommand{\ti}[1]{\text{\it #1}}   % index v matematice
%\newcommand{\bm}[1]{\begin{pmatrix}#1\end{pmatrix}} % pro matice, vektory... napr.\bm{a&b\\c&d}
%\newcommand{\dmd}{\stackrel{\leftrightarrow}{\partial_\mu}\negthickspace} % pro oboustr.derivaci s index.\mu dole
%\newcommand{\dmn}{\stackrel{\leftrightarrow}{\partial^\mu}\negthickspace} % --""-- s indexem \mu nahore
%%%%%%%%%%%%%%%%%%%%%%%%%%%%%%%%%%%%%%%%%%%%%%%%%%%%%%%%%%%%%%%%%%%%%%%%%%%%%%%%%%%%%%%%%%%%%%%%%%%%%%%%%%%%%%%%%%%%%%%%%%%%
%\newcommand{\hv}{\tilde{h}}
%\newcommand{\Phit}{\widetilde{\Phi}}
%%%%%%%%%
%\newcommand{\lagr}{{\cal L}}
%%%%%%%%%%
%\newcommand{\partialv}{\partial\hspace{-4pt}\raisebox{9pt}[0pt]{$\scriptscriptstyle\shortrightarrow$}}
%\newcommand{\partialvb}{\partial\hspace{-5pt}\raisebox{9pt}[0pt]{$\scriptscriptstyle\shortleftarrow$}}
%\newcommand{\partialvob}{\partial\hspace{-6pt}\raisebox{9pt}[0pt]{$\scriptscriptstyle\leftrightarrow$}}
%\begin{document}
\section[Gauge independence of scattering amplitudes]{Gauge independence of scattering amplitudes:\\an example}
\label{sec7.7}

Having set up the basic framework of the $R$-gauge formulation of
the electroweak standard model, we should now demonstrate, at
least on an elementary example, that physical scattering
amplitudes are independent of the gauge-fixing parameters $\xi$,
$\eta$ and $\alpha$. To this end, we shall first collect the
relevant interaction terms contained in the $R$-gauge GWS
Lagrangian. It is clear a priori that the total number of the
$R$-gauge interaction vertices must be considerably larger than
within the $U$-gauge (just because the unphysical scalars $w^\pm$
and $z$\index{unphysical scalars} enter the game), so we shall
proceed step by step.

The Yang--Mills term $\lagr_\ti{gauge}$ (cf.(\ref{eq5.10}))
\begin{equation}
\lagr_\ti{gauge} = -\frac{1}{4} F^a_{\mu\nu}F^{a\mu\nu} -
\frac{1}{4}B_{\mu\nu}B^{\mu\nu}
\end{equation}
involves only the vector boson fields and, consequently, its form
in any of the $R$-gauges is the same as in the $U$-gauge. Next,
let us consider the interactions descending from
$\lagr_\ti{Higgs}$,
\begin{equation}
\label{eq7.149} \lagr_\ti{Higgs}=(D^\mu \Phi)^\dagger(D_\mu\Phi)
-\lambda(\Phi^\dagger\Phi - \frac{v^2}{2})^2
\end{equation}
(see (\ref{eq7.133})). When the covariant derivative $D_\mu$ is
expressed in terms of the physical vector fields $W^\pm_\mu$,
$Z_\mu$ and $A_\mu$ (see (\ref{eq7.7}), (\ref{eq7.8})), one
has\index{covariant derivative}
\begin{equation}\begin{split}
D_\mu \Phi = \Bigl[\partial_\mu - \frac{1}{2}igW^-_\mu \tau^-
&-\frac{1}{2}igW_\mu^+\tau^+ -\frac{1}{2}i g (\cos\theta_W Z_\mu +
\sin\theta_W A_\mu)\tau^3\\ &-\frac{1}{2}i g'(-\sin\theta_W Z_\mu
+ \cos\theta_W A_\mu)\cdot \J \Bigr]\Phi
\end{split}
\end{equation}
and
\begin{equation}\begin{split}
(D^\mu \Phi)^\dagger =
\Phi^\dagger\Bigl[\partialvb^\mu+ \frac{1}{2}igW^{+\mu} \tau^+
&+\frac{1}{2}igW^{-\mu}\tau^-
+\frac{1}{2}i g (\cos\theta_W Z^\mu + \sin\theta_W A^\mu)\tau^3\\
&+\frac{1}{2}i g'(-\sin\theta_W Z^\mu + \cos\theta_W A^\mu)\cdot
\J \Bigr]
\end{split}
\end{equation}
with $\tau^\pm=\frac{1}{\sqrt{2}}(\tau^1 \pm i \tau^2)$ (see
(\ref{eq5.16})). Substituting there the explicit matrix
representation
\begin{equation}
\tau^+ = \sqrt{2}\bm{0&1\\0&0},\quad
\tau^-=\sqrt{2}\bm{0&0\\1&0},\quad \tau^3=\bm{1&0\\0&-1}
\end{equation}
and using (\ref{eq7.131}), (\ref{eq7.132}) for the $\Phi$ and
$\Phi^\dagger$ respectively, the term
$(D^\mu\Phi)^\dagger(D_\mu\Phi)$ in (\ref{eq7.149}) yields a set
of interactions involving vector bosons and the scalars $w^\pm$,
$z$, $H$ (including, of course, the couplings $WWH$, $WWHH$, $ZZH$
and $ZZHH$ encountered earlier in the $U$-gauge). Further, from
the Goldstone potential in (\ref{eq7.149}) one obtains a set of
purely scalar interactions (including the $H$ self-couplings
occurring already in the $U$-gauge formulation). The scalar --
vector and/or pure scalar interactions obtained in this way are
trilinear or quadrilinear\index{quadrilinear vertex} in the fields
involved. Their complete list is rather long and can be found
(including explicit formulae for the corresponding Feynman rules)
e.g. in \cite{BaL}. Here we will only summarize, for the purpose
of an illustration, the types of trilinear interaction terms
appearing in an $R$-gauge (in addition to those already known from
the $U$-gauge treatment).

In the usual schematic notation, the couplings in question can be
grouped naturally in three subsets, namely
\begin{enumerate}
\item [(I)] $w^-W^+Z$, $w^+W^-Z$, $w^+w^-Z$, $w^-W^+z$, $w^+W^-z$
\item [(II)] $w^-W^+\gamma$, $w^+W^-\gamma$, $w^+w^-\gamma$
\item [(III)] $w^-W^+H$, $w^+W^-H$, $w^+w^-H$, $zZH$, $zzH$
\end{enumerate}
There is an obvious mnemonic rule for arriving at such a scheme.
One can take the trilinear interactions of vector bosons with
themselves or with the $H$, i.e. couplings $WWZ$, $WW\gamma$,
$WWH$ and $ZZH$ and replace consecutively the $W$ and $Z$ by $w$
and $z$. Note, however, that a coupling $W^+W^-z$ is missing in
the above catalogue (it is easy to realize that such terms exactly
cancel when working out the expression
$(D^\mu\Phi)^\dagger(D_\mu\Phi)$).

As an explicit example of a particular $R$-gauge interaction term
involving vector bosons let us consider the coupling of the
$w^+w^-$ pair to the photon or $Z$ boson. From the
$\lagr_\ti{Higgs}$ one gets, after some algebraic manipulations,
\begin{equation}
\label{eq7.153} \lagr_{w^+w^-\gamma} + \lagr_{w^+w^-Z} =
iew^-\partialvob^\mu w^+ A_\mu +
i\frac{g}{\cos{\theta_W}}(\tfrac{1}{2} - \sin^2\theta_W) w^-
\partialvob^\mu w^+ Z_\mu
\end{equation}
where we have utilized the familiar relations $g'=g\tan\theta_W$
and $e=g\sin\theta_W$.

An explicit evaluation of the purely scalar interactions is quite
straightforward. Using (\ref{eq7.131}), (\ref{eq7.132}) one gets
immediately
\begin{equation}
\Phi^\dagger\Phi = w^+w^- + \frac{1}{2}\bigl[(v+H)^2 + z^2 \bigr]
\end{equation}
and then
\begin{equation}
\lambda\bigl(\Phi^\dagger\Phi - \frac{v^2}{2}\bigr)^2 =
\lambda\bigl(w^+w^- + vH +\tfrac{1}{2}H^2 +
\tfrac{1}{2}z^2\bigr)^2
\end{equation}
From the last expression the individual scalar interactions can be
read off easily; in particular, the interaction $w^+w^-H$ is seen
to be described by the Lagrangian
\begin{equation}
\label{eq7.156} \lagr_{w^+w^-H} = -2\lambda v \ w^+w^-H
\end{equation}
Note that the coupling constant in the last expression can be
recast as
\begin{equation}
\label{eq7.157} g_{w^+w^-H} = -\frac{m_H^2}{v} = -\tfrac{1}{2}g
\frac{m_H^2}{m_W}
\end{equation}
when one makes use of the relations $m_H^2=2\lambda v^2$ and
$m_W=\frac{1}{2}gv$.

Let us now turn to the fermion sector of the GWS standard model.
The interactions of vector bosons with fermions obviously remain
intact when passing from the $U$-gauge to an $R$-gauge. On the
other hand, the Yukawa couplings\index{Yukawa coupling|ff} involve
the Higgs doublet\index{Higgs!doublet} $\Phi$ and thus can produce
interactions of fermions with unphysical Goldstone bosons. Below
we shall examine these new fermionic interaction terms in detail.
For simplicity, we restrict ourselves to leptons (in the quark
sector one would get similar results, but the flavour
mixing\index{flavour!mixing} must be properly taken into account).
Considering an arbitrary lepton type $\ell$ ($\ell=e,\mu,\tau$)
and ignoring the right-handed component of neutrino field, the
relevant Yukawa-type Lagrangian is written down as
\begin{equation}
\label{eq7.158} \lagr_\ti{Yukawa}^{(\ell)} = -h_\ell
\bar{L}^{(\ell)} \Phi \ell_R + \text{h.c.}
\end{equation}
(see (\ref{eq6.82})). Using the representation (\ref{eq7.131}) for
the $\Phi$, (\ref{eq7.158}) becomes
\begin{equation}
\label{eq7.159} \lagr^{(\ell)}_\ti{Yukawa} = i h_\ell
\bar{\nu}_{\ell L} \ell_R w^+ - \tfrac{i}{\sqrt{2}}h_\ell
\bar{\ell}_L \ell_R z -\tfrac{1}{\sqrt{2}}h_\ell \bar{\ell}_L
\ell_R H -\tfrac{1}{\sqrt{2}}h_\ell v \bar{\ell}_L\ell_R +
\text{h.c.}
\end{equation}
Now, making an obvious identification
$m_\ell=\frac{1}{\sqrt{2}}h_\ell v$ (cf. (\ref{eq6.84})) and using
the familiar relation $m_W=\frac{1}{2}gv$, the interaction part of
the expression (\ref{eq7.159}) is recast as
\begin{equation}\begin{split}
\lagr_{\nu\ell w} + \lagr_{\ell\ell z} + \lagr_{\ell\ell H} =
\phantom{+}&i \frac{g}{2\sqrt{2}}\frac{m_\ell}{m_W}\bar{\nu}_\ell
(1+\gamma_5)\ell w^+ - i\frac{g}{2\sqrt{2}}
\frac{m_\ell}{m_W}\bar{\ell}(1-\gamma_5)\nu_\ell w^-\\
-&i \frac{g}{2}\frac{m_\ell}{m_W}\bar{\ell}\gamma_5 \ell z -
\frac{g}{2}\frac{m_\ell}{m_W}\bar{\ell}\ell H
\end{split}
\label{eq7.160}
\end{equation}
Thus, we have arrived at a complete description of the anticipated
$R$-gauge couplings of the $w^\pm$ and $z$ to leptons. Notice that
we have also recovered (as expected) our earlier $U$-gauge result
for the coupling $\ell\ell H$.

\begin{figure}[h]\centering
\begin{tabular}{cc}
\subfigure[]{\s{\includegraphics{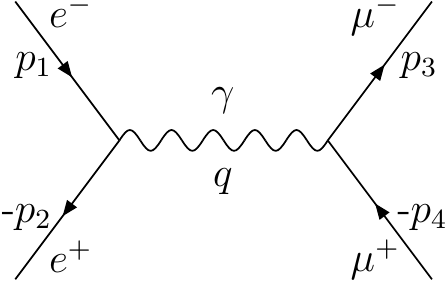}}}&\hspace{1cm}\subfigure[]{\s{\includegraphics{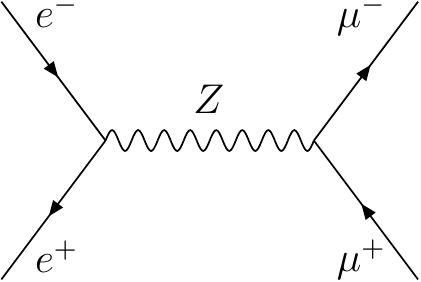}}}\\[-0.5cm]
\subfigure[]{\s{\includegraphics{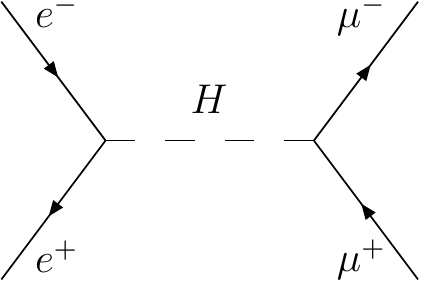}}}
\end{tabular}
\caption{Tree diagrams for the process $e^+e^-\rightarrow
\mu^+\mu^-$ in the $U$-gauge.} \label{fig24}\index{Feynman
diagrams!for $e^+e^-\rightarrow \mu^+\mu^-$!in the $U$-gauge}
\end{figure}

\begin{figure}[h]\centering
\begin{tabular}{cc}
\subfigure[]{\s{\includegraphics{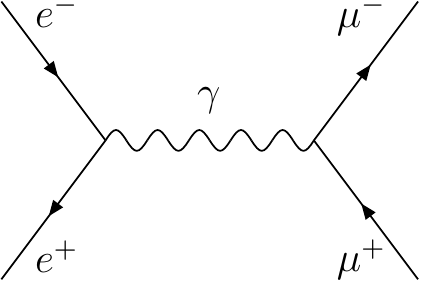}}}&\hspace{1cm}\subfigure[]{\s{\includegraphics{figs/fig24b}}}\\[-0.5cm]
\subfigure[]{\s{\includegraphics{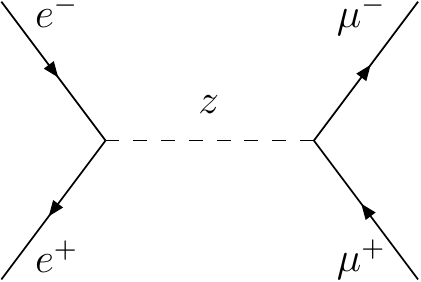}}}&\hspace{1cm}\subfigure[]{\s{\includegraphics{figs/fig24c}}}
\end{tabular}
\caption{Tree diagrams for $e^+e^- \rightarrow \mu^+\mu^-$ in an
$R$-gauge.} \label{fig25}\index{Feynman diagrams!for
$e^+e^-\rightarrow \mu^+\mu^-$!in the $R$-gauge}
\end{figure}

Now we are ready to examine the gauge independence of an
appropriate tree-level scattering amplitude. The particular
illustrative example we are going to discuss is the process
$e^+e^- \rightarrow \mu^+\mu^-$. In the $U$-gauge it is described
by the Feynman graphs shown in Fig.\,\ref{fig24}
and the relevant $R$-gauge diagrams are depicted in
Fig.\,\ref{fig25}.
To begin with, it should be remembered that fixing of the gauge
for the photon propagator is always done separately and does not
depend on whether we are working in the $U$-gauge or an $R$-gauge
for the $W$ and $Z$. In fact, the photon-exchange contribution
described by the graph in Fig.\,\ref{fig24}(a) is gauge
independent by itself (i.e. does not depend on the gauge-fixing
parameter $\alpha$); as we have noticed in the preceding section,
this is due to the conservation of the electromagnetic vector
current. Further, the contribution of the Higgs boson exchange is
the same in all gauges. Thus, it remains to be shown that the sum
of the diagrams (b), (c) in Fig.\,\ref{fig25} is independent of
the relevant gauge-fixing parameter and that it is equal to the
$U$-gauge graph in Fig.\,\ref{fig24}(b)\index{gauge-fixing
term|)}.

To see this, let us write down explicitly the matrix elements in
question. In an obvious notation, the contributions of the
$R_\xi$-gauge graphs in Fig.\,\ref{fig25} are given by
\begin{equation}
\begin{split}
i{\cal M}_\xi^{(Z)} = i^3
\frac{1}{4}\Bigl(\frac{g}{\cos\theta_W}\Bigr)^2 \bigl[
\bar{v}(p_2)\gamma_\mu(v-a\gamma_5) u(p_1) \bigr] \bigl[
\bar{u}(p_3)\gamma_\nu(v-a\gamma_5) v(p_4) \bigr] \\ \times
\frac{1}{q^2-m_Z^2}\bigl[ -g^{\mu\nu} +(1-\xi) \frac{q^\mu
q^\nu}{q^2-\xi m_Z^2} \bigl]
\end{split}
\label{eq7.161}
\end{equation}
and
\begin{equation}
\label{eq7.162}
\begin{split} i{\cal M}_\xi^{(z)} = i^3
\Bigl(-i\frac{g}{2}\frac{m_e}{m_W}\Bigr)
\Bigl(-i\frac{g}{2}\frac{m_\mu}{m_W}\Bigr)
\bigl[\bar{v}(p_2)\gamma_5 u(p_1)\bigr] \bigl[ \bar{u}(p_3)
\gamma_5 v(p_4) \bigr]\\ \times \frac{1}{q^2-\xi m_Z^2}
\end{split}
\end{equation}
respectively. Note that the vector and axial coupling parameters
$v$ and $a$ appearing in (\ref{eq7.161}) are\index{axial
vector!coupling}
\begin{align}
v &= \varepsilon_L + \varepsilon_R = -\frac{1}{2} +
2\sin^2\theta_W\notag\\
a &= \varepsilon_L - \varepsilon_R = -\frac{1}{2} \label{eq7.163}
\end{align}
(cf. (\ref{eq7.11}) and (\ref{eq5.58})). It is convenient to split
the amplitude (\ref{eq7.161}) as
\begin{equation}
{\cal M}^{(Z)}_\xi = {\cal M}^{(Z)}_\ti{diag.} + {\cal
M}^{(Z)}_\ti{long.} \label{eq7.164}
\end{equation}
where the two terms in (\ref{eq7.164}) correspond to the diagonal
and longitudinal part of the $Z$ propagator respectively, i.e.
\begin{multline}
{\cal M}^{(Z)}_\ti{diag.} =
-\frac{1}{4}\Bigl(\frac{g}{\cos\theta_W}\Bigr)^2 \bigl[
\bar{v}(p_2)\gamma_\mu(v-a\gamma_5) u(p_1) \bigr] \bigl[
\bar{u}(p_3)\gamma_\nu(v-a\gamma_5) v(p_4) \bigr]\\
\times \frac{-g^{\mu\nu}}{q^2-m_Z^2}
\end{multline}
and
\begin{multline}
\label{eq7.166}
{\cal M}^{(Z)}_\ti{long.} =
-\frac{1}{4}\Bigl(\frac{g}{\cos\theta_W}\Bigr)^2 \bigl[
\bar{v}(p_2)\slashed{q}(v-a\gamma_5) u(p_1) \bigr] \bigl[
\bar{u}(p_3)\slashed{q}(v-a\gamma_5) v(p_4) \bigr]\\
\times \frac{1-\xi}{(q^2-m_Z^2)(q^2-\xi m_Z^2)}
\end{multline}
Obviously, the ${\cal M}^{(Z)}_\ti{diag.}$ is obtained from
(\ref{eq7.161}) by setting there $\xi=1$, i.e. it coincides with
the ${\cal M}^{(Z)}_\xi$ value in the 't Hooft--Feynman gauge:
\begin{equation}
\label{eq7.167} {\cal M}^{(Z)}_\ti{diag.} = {\cal M}^{(Z)}_{\xi=1}
= {\cal M}^{(Z)}_{tHF}
\end{equation}
Now, taking into account that $q=p_1+p_2=p_3+p_4$, one can employ
the equations of motion for the Dirac spinors in (\ref{eq7.166}).
The vector component of the weak neutral current is conserved, so
a non-trivial contribution can only arise from its axial-vector
part. In particular, one has
\begin{align}
\bar{v}(p_2)\slashed{q}\gamma_5 u(p_1) &= -2m_e
\bar{v}(p_2)\gamma_5 u(p_1)\notag\\
\bar{u}(p_3)\slashed{q}\gamma_5 v(p_4) &= +2m_\mu
\bar{u}(p_3)\gamma_5 v(p_4) \label{eq7.168}
\end{align}
Thus, (\ref{eq7.166}) becomes
\begin{equation}\begin{split}
{\cal M}^{(Z)}_\ti{long.} = \frac{g^2}{4m_W^2}m_e m_\mu \bigl[
\bar{v}(p_2)\gamma_5 u(p_1)\bigr] \bigl[\bar{u}(p_3) \gamma_5
v(p_4) \bigr]\\
\times \frac{m_Z^2(1-\xi)}{(q^2-m_Z^2)(q^2-\xi m_Z^2)}
\end{split}
\end{equation}
where we have also taken into account (\ref{eq7.163}) and the
Weinberg mass relation $\cos^2\theta_W =m_W^2/m_Z^2$. With the
above results at hand, the sum ${\cal M}^{(Z)}_\ti{long.}+{\cal
M}^{(z)}_\xi$ can be written as
\begin{equation}\label{eq7.170}\begin{split}
{\cal M}^{(Z)}_\ti{long.}+{\cal M}^{(z)}_\xi = \frac{g^2}{4m_W^2}
m_e m_\mu \bigl[\bar{v}(p_2)\gamma_5 u(p_1)\bigr] \bigl[
\bar{u}(p_3)\gamma_5 v(p_4)\bigr] \\
\times \Bigl( \frac{m_Z^2 (1-\xi)}{(q^2-m_Z^2)(q^2-\xi m_Z^2)}
+\frac{1}{q^2-\xi m_Z^2} \Bigr)
\end{split}\end{equation}
It is clear that the $\xi$-dependent terms drop out from
(\ref{eq7.170}) and one gets
\begin{equation}
\label{eq7.171} {\cal M}^{(Z)}_\ti{long.}+{\cal M}^{(z)}_\xi =
\frac{g^2}{4m_W^2} \bigl[\bar{v}(p_2)\gamma_5 u(p_1)\bigr]
\bigl[\bar{u}(p_3)\gamma_5 v(p_4)\bigr] \frac{1}{q^2-m_Z^2}
\end{equation}
However, the last expression is seen to coincide with
(\ref{eq7.162}) for $\xi=1$; in other words, (\ref{eq7.171}) is
equal to the $z$-exchange contribution in the 't Hooft--Feynman
gauge. Then, taking into account also (\ref{eq7.164}) and
(\ref{eq7.167}), one can write
\begin{equation}
{\cal M}^{(Z)}_\xi + {\cal M}^{(z)}_\xi = {\cal M}^{(Z)}_{\xi=1} +
{\cal M}^{(z)}_{\xi=1} = {\cal M}^{(Z)}_{tHF} + {\cal
M}^{(z)}_{tHF}
\end{equation}
and the gauge independence of ${\cal M}_\xi(e^+e^-\rightarrow
\mu^+\mu^-)$ is thereby proved within the class of $R_\xi$-gauges.
The proof of an equivalence of 't Hooft--Feynman gauge and the
$U$-gauge for the considered process goes along the same lines as
the above derivation and we leave it to the reader as a
straightforward exercise.

Thus, we have been able to prove the gauge independence of the
tree-level amplitude in question in an elementary and rather
transparent way. In the present context, a cancellation mechanism
for the $\xi$-dependent terms seems to be clear: there is an
obvious correlation between the contribution of an unphysical
Goldstone boson and that of the longitudinal part of a massive
vector boson propagator\index{propagator!of massive vector boson},
which yields the desired compensation (recall that the location of
the $\xi$-dependent extra pole in an $R$-gauge vector propagator
coincides with that of the related unphysical scalar).
Technically, an essential ingredient of our calculation has been
the \qq{partial conservation} of weak currents (see
(\ref{eq7.168})), following simply from the equations of motion
for external particles. A generalization of such an analysis to
other tree-level scattering amplitudes would be quite
straightforward. However, let us emphasize that proving the gauge
independence of the $S$-matrix to all orders of perturbation
theory is a highly non-trivial task. In order to accomplish such a
goal, some advanced quantum field theory techniques are needed; in
particular, one has to employ all relevant Ward
identities\index{Ward identity} that express the contents of the
original gauge symmetry at quantum level (in fact, our elementary
calculation exemplifies how the Ward identities work in the lowest
order). Concerning these topics, an interested reader is referred
either to the original papers \cite{ref48}, \cite{ref72} or to the
monographs \cite{BaL}, \cite{Wei}, \cite{Pok}.

%\end{document}

%\input{kniha78}
%%%%%%%%%%%%%%%%%%%%%%%%%%%%%%%%%%%%%%%%%%%%%%%%%%%%%%%%%%%%%%%%%%%
%%%%%%%%%%%%%%%%%%%%%%%%%%%%%%%%%%%%%%%%%%%%%%%%%%%%%%%%%%%%%%%%%%%%%%%%%%%%%%%%%%%%%%%%%%%%%%%%%%%%%%%%%%%%%%%%%%%%%%%%%%%%%%%%%%%%%%%%
%\documentclass[12pt,tbtags]{report}
%\input{pream}
%\begin{document}
\section{Equivalence theorem for longitudinal vector bosons}
\index{equivalence theorem|(} We have already noted earlier (cf.
the end of Section~\ref{sec6.3}) that within a gauge theory with the Higgs
mechanism\index{Higgs mechanism} one should expect an intimate
dynamical connection between the unphysical Goldstone
bosons\index{Goldstone!boson} and longitudinally polarized vector
bosons\index{longitudinally polarized vector boson|ff}. In a
nutshell, such an expectation relies -- rather intuitively -- on
the fact that the physical longitudinal mode of a massive vector
boson emerges in place of a would-be Goldstone boson, \qq{eaten}
by the corresponding (originally massless) gauge field. Now we are
in a position to formulate the relevant \qq{equivalence theorem}
in more definite terms. Before doing it, we would like to discuss
two instructive examples involving some specific decay and
scattering processes.

\begin{figure}[h]
\centering \s{\includegraphics{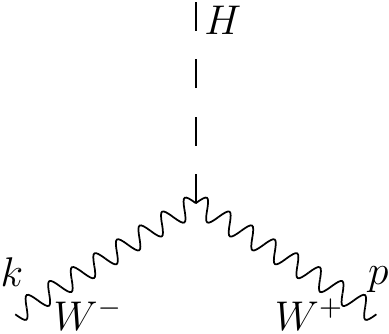}}
\caption{Tree-level graph for the decay $H\rightarrow W^+W^-.$}
\label{fig26}\index{Feynman diagrams!for $H\rightarrow W^+W^-$}
\end{figure}

\begin{figure}[h]
\centering \s{\includegraphics{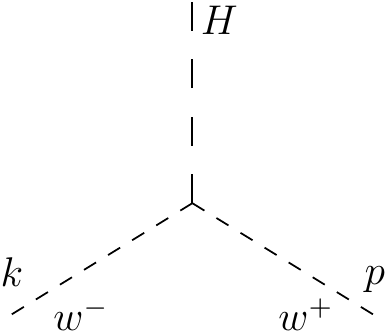}}
\caption{Tree-level Feynman graph for the unphysical process
$H\rightarrow w^+w^-$.} \label{fig27}\index{Feynman diagrams!for
$H\rightarrow w^+w^-$}
\end{figure}

First, let us consider the decay of a heavy Higgs boson ($m_H\gg
m_W$)\index{decay!of the Higgs boson} at rest into a pair of
longitudinally polarized vector bosons $W^\pm$. In lowest order,
the process is described by the diagram shown in
Fig.\,\ref{fig26}.
The corresponding amplitude is then
\begin{equation}
\label{eq7.173} {\cal M}(H\rightarrow W^+_L W^-_L) = g m_W
\varepsilon_L^\mu(k) \varepsilon_{L\mu}(p)
\end{equation}
where we have taken into account the form of the $WWH$ coupling
given by (\ref{eq6.76}) (plus the fact that the polarization
vectors are real\index{polarization!vector}). As we know, a longitudinal polarization vector
can be split as
\begin{equation}
\label{eq7.174} \varepsilon^\mu_L (p) = \frac{1}{m_W} p^\mu +
\Delta^\mu(p)
\end{equation}
with $\Delta^\mu(p)$ being of the order $\OO(m_W/E_W)$ in
high-energy limit (cf. (\ref{eq3.29})). In our case,
$E_W=\frac{1}{2}m_H$ and thus $\Delta^\mu(p)\ll 1$ according to
the above assumption. Decomposing the polarization vectors in
(\ref{eq7.173}) according to (\ref{eq7.174}), one gets
\begin{align}
{\cal M}(H\rightarrow W^+_L W^-_L) &= g m_W \Bigl(
\frac{1}{m_W}k^\mu +\Delta^\mu (k) \Bigr)
\Bigl(\frac{1}{m_W}p_\mu + \Delta_\mu(p)\Bigr)\notag\\
&= g m_W \Bigl( \frac{1}{m_W^2} k\cdot p + \OO(1) \Bigr)\notag\\
&=\frac{g}{2}\frac{m_H^2}{m_W} \Bigl(1+ \OO
\bigl(\frac{m_W^2}{E_W^2}\bigr)\Bigr) \label{eq7.175}
\end{align}
Having in mind the envisaged connection between the $W_L^\pm$ and
their scalar counterparts $w^\pm$, let us now calculate the
amplitude describing formally the unphysical process $H\rightarrow
w^+ w^-$. The corresponding lowest order diagram is shown in
Fig.\,\ref{fig27}.

\noindent Using our earlier results (\ref{eq7.156}) and
(\ref{eq7.157}), one can write immediately
\begin{equation}
{\cal M}(H\rightarrow w^+ w^- ) = -\frac{g}{2}\frac{m_H^2}{m_W}
\end{equation}
Comparing it with the last line in (\ref{eq7.175}), we thus have
\begin{equation}
\label{eq7.177} {\cal M}(H\rightarrow W^+_L W^-_L) = -{\cal
M}(H\rightarrow w^+ w^-)\times \Bigl[
1+\OO\bigl(\frac{m_W^2}{E_W^2}\bigr) \Bigr]
\end{equation}
This relation represents a simple explicit example of the
equivalence theorem for longitudinal vector bosons and unphysical
Goldstone bosons alluded to previously.

As a second example (that may be technically somewhat less
trivial), we shall discuss the process $e^+e^-\rightarrow W_L^+
W_L^-$. For simplicity, we set $m_e=0$ throughout the calculation.
Then, the relevant $U$-gauge tree diagrams are those shown in
Fig.\,\ref{fig19} (see p.\,\pageref{fig19}), since the
contribution of the Higgs exchange graph vanishes in the
considered approximation. In fact, the
$R$-gauge\index{R-gauge@$R$-gauge} graphs are the same as there is
no $zWW$ coupling. Gauge independence of the considered amplitude
can be proved readily -- it is a matter of a straightforward
application of the 't Hooft identity\index{t Hooft identity@'t
Hooft identity} (\ref{eq3.47}) for the vertices $WW\gamma$ and
$WWZ$. Further, it is easy to realize that in the chiral limit
$m_e\rightarrow 0$ we have in mind, the matrix element in question
can only be non-vanishing if the $e^+$ and $e^-$ have unlike
helicities. For definiteness, let the electron be left-handed and
the positron right-handed. Owing to the asymptotic behaviour of
longitudinal polarization vectors (cf. (\ref{eq7.174})),
contributions of the individual diagrams in Fig.\,\ref{fig19}(a),
(b), (c) diverge in the high-energy limit. As we have already seen
in Chapter~\ref{chap5}, the leading (quadratic)
divergences\index{high-energy divergences} cancel in their sum
(see (\ref{eq5.93}), (\ref{eq5.94}), (\ref{eq5.95})). The residual
(linear) divergences are proportional to $m_e$ and therefore
entirely disappear in the approximation considered here. Thus, the
graphs in Fig.\,\ref{fig19} yield an amplitude that is
asymptotically flat (i.e. satisfies the condition of tree
unitarity\index{tree unitarity|ff}); as a result of rather long
and tedious calculation, one gets
\begin{multline}%
\label{eq7.178} {\cal M}(e^+e^-\rightarrow W^+_L W^-_L) =
\frac{1}{s}\bar{v}_R(l)(\slashed{p}-\slashed{r})u_L(k)\\ \times
\Bigl[e^2 + g^2
(-\tfrac{1}{2}+\sin^2\theta_W)(\frac{m_Z^2}{2m_W^2}-1)\Bigr] +
\OO(\frac{m_W^2}{s})
\end{multline}
In arriving at the last expression, the computational tricks
employed earlier (see in particular Chapter~\ref{chap5}) are instrumental;
note also that another very useful relation is
\begin{equation}
p\cdot\Delta(p) = -m_W
\end{equation}
where the $\Delta(p)$ is the remainder in (\ref{eq7.174}) (this is
an immediate consequence of the identities $p\cdot
\varepsilon_L(p)=0$ and $p^2=m_W^2$). With these encouraging
remarks, we leave a derivation of (\ref{eq7.178}) as a challenge
for a seriously interested reader.

\begin{figure}[h]\centering
\begin{tabular}{ccc}
\subfigure[]{\s{\includegraphics{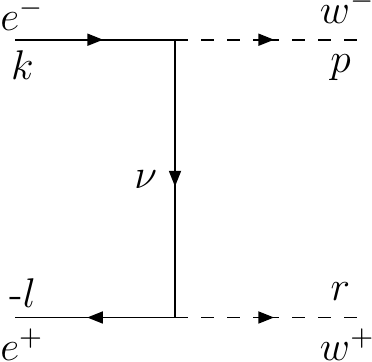}}}&\hspace{0.2cm}\subfigure[]{\s{\includegraphics{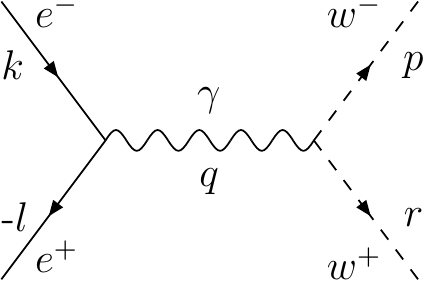}}}&\hspace{0.2cm}\subfigure[]{\s{\includegraphics{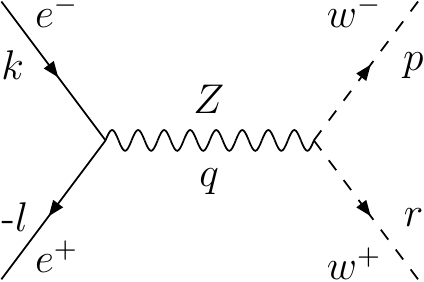}}}
\end{tabular}
\caption{Tree-level diagrams for the unphysical process
$e^+e^-\rightarrow w^+w^-$. The contribution of the diagram (a)
vanishes for $m_e=0$.} \label{fig28}\index{Feynman diagrams!for
$e^+e^-\rightarrow w^+w^-$}
\end{figure}

Let us now consider the unphysical counterpart of the above
process, namely $e^+e^-\rightarrow w^+w^-$. The corresponding
lowest-order diagrams are depicted in Fig.\,\ref{fig28}.

The relevant interactions of the $w^\pm$ are described by the
Lagrangian
\begin{align}
\lagr_\ti{int}^{(w^\pm)} = \phantom{+\
}&i\frac{g}{2\sqrt{2}}\frac{m_e}{m_W}\bar{\nu}_e(1+\gamma_5)ew^\pm
-i\frac{g}{2\sqrt{2}}\frac{m_e}{m_W}\bar{e}(1-\gamma_5)\nu_ew^-\notag\\
+\ &iew^- \partialvob_\mu w^+ A^\mu + i\frac{g}{\cos\theta_W}
(\tfrac{1}{2}-\sin^2\theta_W)w^-\partialvob_\mu w^+ Z^\mu
\end{align}
(see (\ref{eq7.153}) and (\ref{eq7.160})), which means that for
$m_e=0$ only the graphs (b) and (c) contribute. Obviously, the
Feynman rules for the interaction vertices $w^+w^-\gamma$ and
$w^+w^- Z$ are essentially those of the ordinary scalar
QED\index{quantum!electrodynamics (QED)}. Taking into account also
the other familiar rules, the amplitude in question can thus be
written as
\begin{multline}
{\cal M}(e^+e^-\rightarrow w^+w^-) = e^2 \bar{v}_R(l)\gamma_\mu
u_L(k) \frac{-g^{\mu\nu}}{q^2}(p_\nu - r_\nu)\\
-\Bigl(\frac{g}{\cos\theta_W}\Bigr)^2(-\tfrac{1}{2}+\sin^2\theta_W)
(\tfrac{1}{2}-\sin^2\theta_W) \bar{v}_R(l)\gamma_\mu u_L(k)
\\ \times \frac{-g^{\mu\nu}}{q^2-m_Z^2}(p_\nu-r_\nu)
\label{eq7.181}
\end{multline}
(we consider again a left-handed electron and right-handed
positron). Note that the gauge independence of such a matrix
element is proved easily, so in writing  down (\ref{eq7.181}) we
have used the Feynman gauge both for photon and for $Z$. From
(\ref{eq7.181}) one gets readily
\begin{multline}
{\cal M}(e^+e^-\rightarrow w^+w^-) = -\frac{1}{s} \bar{v}_R(l)
(\slashed{p}-\slashed{r}) u_L(k)\\
\times \Bigl[e^2 + \Bigl(\frac{g}{\cos\theta_W}\Bigr)^2
(\tfrac{1}{2}-\sin^2\theta_W)^2 \Bigr] + \OO(\frac{m_W^2}{s})
\end{multline}
where we have set $q^2=s$. Now it is clear that when the formula
$m_W/m_Z=\cos\theta_W$ is used in (\ref{eq7.178}), one has
\begin{equation}
\label{eq7.183} {\cal M}(e^+e^-\rightarrow W^+_L W^-_L)=-{\cal M}
(e^+e^-\rightarrow w^+w^-)\times \Bigl[
1+\OO\Bigl(\frac{m_W^2}{E_W^2}\Bigr)\Bigl]
\end{equation}
The analogy between the last relation and eq.\,(\ref{eq7.177}) is
striking. Of course, (\ref{eq7.183}) is another example of the
equivalence theorem (ET) mentioned above. Let us now formulate a
general statement of ET for tree-level matrix elements (for some
original papers see ref. \cite{ref49}).

{\bf Equivalence theorem:} Let us consider a process involving,
apart from other physical particles, a certain number of
longitudinally polarized vector bosons $V_L$ (i.e. $W_L^\pm$
and/or $Z_L$), with $n_1$ of them being in the initial state and
$n_2$ in the final state. Let $E_V$ denote generically the vector
boson energies; for $E_V\gg m_W$ one then has
\begin{multline}
\label{eq7.184} {\cal M}_{fi}\bigl( V_L(i_1), \ldots,
V_L(i_{n_1}), A\rightarrow V_L(f_1), \ldots, V_L(f_{n_2}), B
\bigr) = \\{\cal M}_{fi} \bigl( \varphi(i_1),
\ldots,\varphi(i_{n_1}), A\rightarrow \varphi(f_1), \ldots,
\varphi(f_{n_2}), B \bigr) \times i^{n_1} (-i)^{n_2}
\Bigl[1+\OO\Bigl(\frac{m_V}{E_V}\Bigr)\Bigr]\phantom{\Biggr)}
\end{multline}
where the $\varphi$'s stand for the unphysical Goldstone scalar
counterparts of the $V_L$'s and the $A$, $B$ symbolize all other
incoming and outgoing particles $\blacksquare$

Note that the precise form of the phase factor $i^{n_1}(-i)^{n_2}$
shown in (\ref{eq7.184}) is due to the conventional definition of
the Goldstone boson fields according to (\ref{eq7.131}). Looking
back at (\ref{eq7.177}) and (\ref{eq7.183}), the reader can see
immediately that the phase factor contained in the general formula
(\ref{eq7.184}) is indeed recovered in our previous two examples
(where $n_1=0$, $n_2=2$).

For completeness, a comment is in order here. As we noted, ET in
the above form is certainly valid at the tree level. When going to
higher orders of perturbation theory, the relation (\ref{eq7.184})
gets slightly modified by including a finite
renormalization\index{renormalization} factor $C=1+\OO(g^2)$ (with
$g$ denoting generically a gauge coupling constant) {\it
independent of the energies\/}. Such a generalization of the ET is
discussed in detail e.g. in \cite{ref73}, where also some further
references can be found. In fact, it turns out that in a suitable
gauge the $C$ can be made equal to unity in all orders.

The ET is a deep general result characteristic of any gauge theory
with the Higgs mechanism. Technically, it is a consequence of the
gauge symmetry\index{gauge invariance} expressed in terms of an
appropriate Ward identity\index{Ward identity}. A more detailed
commentary concerning the ET proof would go beyond the scope of
this text and the interested reader is referred to the literature.
In particular, an introductory treatment can be found in
\cite{ref74}, together with a comprehensive list of further
references. Among other things, the process $e^+e^- \rightarrow
W^+_L W^-_L$ is reconsidered in \cite{ref74} from the point of
view of a relevant Ward identity, which provides some insight into
the result (\ref{eq7.183}) obtained here by means of a
straightforward calculation. A brief discussion of the ET can also
be found in the books \cite{PeS} and \cite{Don}; for a more
sophisticated survey see e.g. \cite{Dob}.

It is important to realize that -- with the equivalence theorem at
hand -- general validity of the tree unitarity within the GWS
standard electroweak theory becomes quite clear. Indeed, as we
know, there are two sources of a possible \qq{bad} high-energy
behaviour of the tree-level amplitudes: the $U$-gauge massive
vector boson propagators (that contain a factor of $m_V^{-2}$) and
the polarization vectors of external massive vector bosons (each
bringing in a factor of $m_V^{-1}$ in the high-energy limit). Now,
one can rely on the gauge independence of physical scattering
amplitudes and pass from the $U$-gauge to an
$R$-gauge\index{R-gauge@$R$-gauge}, where the \qq{dangerous} parts
of vector boson propagators are absent. Subsequently, using ET in
the high-energy limit, one can replace the longitudinally
polarized vector bosons by the corresponding unphysical Goldstone
bosons. However, these are completely innocuous, as the Feynman
graphs with external scalars obviously respect the constraints of
tree unitarity (once there are no coupling constants with
dimension of a negative power of mass). {\bf In this way, one can
see that because of the general validity of ET, tree-level
unitarity is satisfied in any electroweak theory of
renormalizable\index{renormalizable theory} type.}

Finally, let us add a historical remark. A first proof of the tree
unitarity for gauge theories with Higgs mechanism has been given
by J. S. Bell \cite{ref75} (at a time when ET has not been known
yet) with the help of a different method. Bell's work has actually
been a precursor to the papers \cite{ref30}, \cite{ref35},
\cite{ref36}, where the GWS electroweak theory has been derived
from the constraints of tree-level unitarity (see also
\cite{Hor})\index{equivalence theorem}.
\newpage %!!!!!!!!!!!!!!!

%\input{kniha79}
%%%%%%%%%%%%%%%%%%%%%%%%%%%%%%%%%%%%%%%%%%%%%%%%%%%%%%%%%%%%%%%%%%%
%%%%%%%%%%%%%%%%%%%%%%%%%%%%%%%%%%%%%%%%%%%%%%%%%%%%%%%%%%%%%%%%%%%%%%%%%%%%%%%%%%%%%%%%%%%%%%%%%%%%%%%%%%%%%%%%%%%%%%%%%%%%%%%%%%%%%%%%
%\documentclass[12pt,tbtags]{report}
%\usepackage{mathrsfs,euscript,amsfonts, amssymb, mathbbol, bbm, dsfont,eufrak}
%\DeclareMathAlphabet{\mathpzc}{OML}{cmm}{m}{it}
\newcommand{\pA}{\text{\Large$a$}}
%\input{pream}
%\begin{document}
\section{Effects of ABJ anomaly}\label{sec7.9}
For complete understanding of structural properties of the
electroweak theory one has to take into account another important
concept of quantum field theory, namely the Adler--Bell--Jackiw
(ABJ) anomaly \cite{ref76}. \index{Adler--Bell--Jackiw (ABJ)
anomaly|(}This is a rather subtle phenomenon, which nevertheless
plays substantial role in the discussion of internal consistency
and renormalizability of the GWS standard model. The ABJ anomaly
reflects peculiar behaviour of closed fermionic loops involving
vector and axial-vector currents; from the technical point of view
it represents a violation of naive Ward identities\index{anomalous
Ward identity|(} for such Feynman graphs (for an introduction to
the subject, see e.g. \cite{ref77} and the monograph \cite{Ber}).

To elucidate the nature of possible effects due to the ABJ
anomaly, let us resume the investigation of our recurrent theme --
perturbative unitarity, or \qq{asymptotic softness} of scattering
amplitudes.\footnote{The reader may find it useful to look back
into the Section~\ref{sec3.1}; in what follows we are going to generalize
slightly our earlier considerations.} Throughout the present text
we stressed repeatedly that the tree-level unitarity is a
necessary condition for renormalizability in higher orders of
perturbation expansion. The crucial point is that power-like
growth of a scattering amplitude with energy (for a binary process
$1 + 2 \rightarrow 3 + 4$) would propagate into higher order
diagrams, leading to an uncontrollable proliferation of
divergences\index{ultraviolet divergences} and subsequent loss of
renormalizability \cite{ref30}. Such an argument is essentially
based on dispersion relations\index{dispersion relations|(} for
Feynman diagrams and we shall now recapitulate it briefly.

Using the technique of dispersion relations, a scattering
amplitude is evaluated through its imaginary part by means of a
Cauchy-type integral and -- depending on the asymptotic
(high-energy) behaviour of the integrand -- one eventually has to
employ an appropriate number of subtractions in order to get a
finite answer. As regards the imaginary parts, one should remember
that these are expressed (via the $S$-matrix unitarity\index{S
matrix unitarity@$S$-matrix unitarity}) in terms of products of
the amplitudes in lower perturbative orders. For example, the
imaginary part of a one-loop diagram\index{loop diagrams} can be
represented as a square of a tree-level graph (obtained by cutting
the internal lines of the closed loop), etc. All this means that
the leading asymptotic energy dependence of a higher order graph
is given, roughly speaking, by a product of contributions from
lower orders. Thus, if the tree-level unitarity is violated, one
can expect that the power-like growth of a considered scattering
amplitude will get worse at higher orders (barring some accidental
cancellations). Consequently, an indefinitely growing number of
subtractions is needed to make the dispersion integrals
convergent. In fact, the subtractions are tantamount to the
renormalization\index{renormalization} counterterms and their
infinite number means that the theory is not
renormalizable\index{renormalizable theory|(} in the usual
perturbative sense.

On the other hand, if the tree unitarity holds, one can expect
(naively) that the scattering amplitude in question remains
sufficiently \qq{soft} for $E \rightarrow \infty$ even at higher
orders of perturbation theory;\footnote{As usual, the $E$ is a
generic notation for a relevant energy variable, e.g. the total
centre-of-mass energy of the considered process,
$E_{c.m.}=s^{1/2}$.} in simple terms, the idea is (having in mind
e.g. a binary process) that through successive multiplication of
asymptotically flat matrix elements of lower order one should get
a result with the same high-energy behaviour. However, there is a
snag. When calculating the full one-loop amplitude, one must
perform -- apart from algebraic manipulations -- an integration
over an energy variable in the relevant dispersion relation, and
it might happen (in principle at least) that the result would
behave differently than the basic tree-level amplitude. In
particular, the real part of a one-loop amplitude could pick up a
contribution, scaling as a positive power of energy for $E
\rightarrow \infty$. This would mean that an originally expected
chain of well-behaved (i.e. asymptotically soft) perturbative
iterations breaks down: at the one-loop level and higher, one
would face a rapid violation of unitarity, ending up with
non-renormalizable\index{non-renormalizability} perturbation
series. In other words, while it seems to be true beyond any
reasonable doubt that the tree-level unitarity is a {\it
necessary\/} condition for renormalizability, one cannot be sure
whether it is a {\it sufficient\/} condition as well. {\bf It
turns out that the tree unitarity indeed does not, in general,
guarantee renormalizability.} The presaged \qq{pathological}
behaviour of one-loop scattering amplitudes is rather exceptional,
but it does occur within some field theory models. As we shall
explain below, its source is just the ABJ anomaly. Later on we
will also show how this potential problem is avoided within the
GWS standard model.\footnote{For the purpose of the preceding
heuristic discussion we have invoked the method of dispersion
relations, but we shall not pursue it any further. In our
subsequent calculations we simply utilize some particular results
of a direct evaluation of one-loop Feynman graphs. For the
dispersion-relation approach to the ABJ anomaly in the present
context see e.g. \cite{Hor}\index{dispersion relations|)}.}

\begin{figure}[h]
\centering \s{\includegraphics{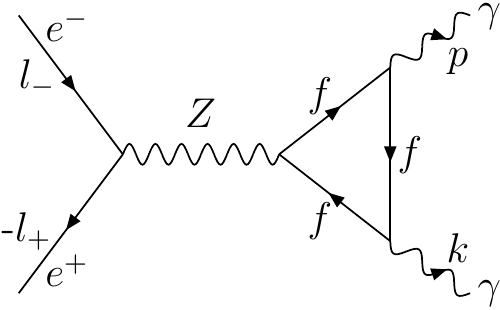}}
\caption{\qq{Anomalous} contribution to the process
$e^+e^-\rightarrow \gamma\gamma$ involving fermionic triangle
loop. Adding an analogous diagram with crossed external photon
lines is tacitly assumed.} \label{fig29}\index{Feynman
diagrams!for $e^+e^-\rightarrow \gamma\gamma$!in the $U$-gauge}
\end{figure}

As an illustrative example, let us consider the $e^+e^-$
annihilation into two photons and, for definiteness, we shall
first work in the $U$-gauge. As regards the high-energy behaviour
of the corresponding amplitude, such a process is completely
innocuous at the tree level, where it is represented by the
familiar QED diagrams (these, of course, do not involve any
\qq{dangerous} components since the massless photons can only have
transverse polarizations\index{polarization!of the
photon}\index{transverse polarization}). In the one-loop
approximation, there are many diagrams that contribute to the
process in question and most of them simply reproduce the decent
behaviour of tree-level matrix element (modified only by some
logarithmic corrections). However, there is one exception, namely
the graph shown in Fig.\,\ref{fig29}.
The corresponding matrix element can be written as
\begin{equation}
\label{eq7.185}
\begin{split}
i\mathcal{M}_\triangle = &i^8 (-1) \frac{1}{4}
\Bigl(\frac{g}{\cos\theta_W}\Bigr)^2 \bar{v}(l_+)\gamma_\lambda
(v_e - a_e \gamma_5) u(l_-)\\
&\times \frac{-g^{\lambda\alpha} + m_Z^{-2} q^\lambda
q^\alpha}{q^2 - m_Z^2}\cdot (-a_f) Q_f^2 e^2 T_{\alpha\mu\nu}(k,p)
\varepsilon^{*\mu}(k)\varepsilon^{*\nu}(p)
\end{split}
\end{equation}
where the coupling parameters for neutral currents have the usual
meaning (cf.\,(\ref{eq5.58}) or (\ref{eq7.163})), the $Q_f$ is
charge factor for the fermion circulating in the loop (e.g.
$Q_e=-1$ etc.) and the $T_{\alpha\mu\nu}(k,p)$ stands for the
triangle loop itself. Before specifying its form, note that in
(\ref{eq7.185}) we have already taken into account that only the
axial-vector ($A$) part of the weak neutral current entering the
triangle (in the vertex attached to the $Z$ propagator) can give a
non-vanishing contribution. The reason is that the electromagnetic
currents appearing in the other two vertices are of pure vector
($V$) nature and in combination with the vector part of the weak
neutral current one would get a $VVV$ triangle as in spinor QED --
but this is known to vanish identically according to\index{Furry's
theorem} the Furry's theorem.\footnote{Let us remind the reader
that purely fermionic closed loops with an (arbitrary) odd number
of vertices are discarded within spinor QED, because to any such
graph one can add the contribution of its counterpart with a
reverse orientation of internal lines (i.e. with fermion
circulating inside the loop in opposite direction) and this is
exactly opposite to the original one. For a triangle graph,
reverting the loop orientation is tantamount to the crossing of
two external photon lines attached to its vertices.} Thus, the
triangle loop appearing in Fig.\,\ref{fig29} is of the
$VVA$\index{VVA triangle@$VVA$ triangle graph|ff} type and it is
represented formally as
\begin{equation}\label{eq7.186}\begin{split}
T_{\alpha\mu\nu}(k,p) = &\int \frac{d^4 l}{(2\pi)^4} \text{Tr}
\Bigl(\frac{1}{\slashed{l}-\slashed{k}-m_f}\gamma_\mu
\frac{1}{\slashed{l}-m_f}\gamma_\nu
\frac{1}{\slashed{l}+\slashed{p}-m_f} \gamma_\alpha \gamma_5
\Bigr)\\ &+ \bigl[ (k,\mu) \leftrightarrow (p,\nu) \bigr]
\end{split}\end{equation}
where we have included the crossing of the external photon lines
(\qq{Bose symmetrization}) indicated in Fig.\,\ref{fig29}. The
usual factor of $(-1)$ associated with any purely fermionic closed
loop has already been incorporated into the overall factor in the
expression (\ref{eq7.185}) and the meaning of the integration
variable (loop momentum $l$) is obvious. At first sight, the
integral in (\ref{eq7.186}) has an ultraviolet divergence and must
be defined properly. Brief discussion of this issue, together with
a succinct summary of basic properties of the $T_{\alpha\mu\nu}$
can be found in the Appendix~\ref{appenE}; below we will only utilize some
key relations that are substantial for understanding of the
high-energy behaviour of the matrix element (\ref{eq7.185}).

Obviously, the only potentially dangerous term is that involving
the longitudinal part of the $Z$ boson propagator (because of the
factor $m_Z^{-2}$). Denoting the corresponding contribution to
(\ref{eq7.185}) as ${\cal M}_\triangle^\tiz{long.}$, one has
\begin{equation}\label{eq7.187}\begin{split}
\mathcal{M}_\triangle^\tiz{long.} = -&i
\frac{1}{4}\Bigl(\frac{g}{\cos\theta_W}\Bigr)^2
m_Z^{-2} a_f Q_f^2 e^2 
\bar{v}(l_+)
\slashed{q} (v_e -a_e\gamma_5) u(l_-) \\
& \times \frac{1}{s-m_Z^2} q^\alpha T_{\alpha\mu\nu} (k,p)
\varepsilon^{*\mu}(k)\varepsilon^{*\nu}(p)
\end{split}\end{equation}
where we have also set $s=q^2$. Taking into account that $q=l_-
+l_+$ and employing the equations of motion for the Dirac spinors,
one gets
\begin{align}
\bar{v}(l_+) \slashed{q} u(l_-) &= 0 \notag \\
\bar{v}(l_+) \slashed{q} \gamma_5 u(l_-) &= -2m_e \bar{v}(l_+)
\gamma_5 u(l_-)
\end{align}
and (\ref{eq7.187}) thus becomes
\begin{equation}\begin{split}
\mathcal{M}_\triangle^\tiz{long.} = &
\frac{i}{4}\Bigl(\frac{g}{\cos\theta_W}\Bigr)^2 a_f Q_f^2 e^2
\frac{m_e}{m_Z^2}\frac{1}{s-m_Z^2} \bar{v}(l_+)
\gamma_5 u(l_-)\\
& \times q^\alpha T_{\alpha\mu\nu} (k,p)
\varepsilon^{*\mu}(k)\varepsilon^{*\nu}(p) \label{eq7.189}
\end{split}\end{equation}
where we have already set $a_e=-1/2$ according to (\ref{eq7.163}).
It means that one factor of $m_Z^{-1}$ is effectively compensated
by the electron mass factorized from the four-divergence of the
axial-vector part of the weak neutral current and the matrix
element in question can therefore grow at worst linearly in the
high-energy limit. Now, we assume that the $T_{\alpha\mu\nu}(k,p)$
is defined in such a way that
\begin{equation}
k^\mu T_{\alpha\mu\nu} (k,p) = 0, \qquad p^\nu
T_{\alpha\mu\nu}(k,p)=0
\end{equation}
i.e. the vector Ward identities (cf. the Appendix~\ref{appenE}) are imposed
in order to maintain electromagnetic gauge invariance. Then
\begin{equation}
\label{eq7.191} q^\alpha T_{\alpha\mu\nu}(k,p) = 2 m_f
T_{\mu\nu}(k,p) +\frac{1}{2\pi^2}\epsilon_{\mu\nu\rho\sigma}
k^\rho p^\sigma
\end{equation}
where the $T_{\mu\nu}(k,p)$ is obtained from (\ref{eq7.186}) by
replacing the $\gamma_\alpha\gamma_5$ with $\gamma_5$ (let us
stress that the integral defining the $T_{\mu\nu}$ is perfectly
convergent). The relation (\ref{eq7.191}) represents an
\qq{anomalous axial-vector Ward identity}. The first contribution
in its right-hand side is usually called the \qq{normal term} and
the second one is the famous {\bf ABJ anomaly}. It is clear that
the two contributions in (\ref{eq7.191}) have substantially
different impact on the high-energy behaviour of the expression
(\ref{eq7.189}). The normal term is proportional to the fermion
mass and thus it compensates the remaining factor of $m_Z^{-1}$;
in other words, this yields a contribution that is asymptotically
flat for $E\rightarrow \infty$ (and proportional to $m_e
m_f/m_Z^2$). However, the ABJ anomaly represents a \qq{hard}
contribution that does not contain any compensating mass factor
and one is thus indeed left with a result for (\ref{eq7.189}) that
is linearly divergent in the limit $E\rightarrow \infty$.
Explicitly, the leading term in (\ref{eq7.189}) (and,
consequently, in (\ref{eq7.185})) has the form
\begin{equation}\begin{split}
\mathcal{M}_\ti{$\triangle$anomaly}^\tiz{long.} = &\frac{i}{4}
\Bigl(\frac{g}{\cos\theta_W}\Bigr)^2 a_f Q_f^2 e^2
\frac{m_e}{m_Z^2}\frac{1}{s-m_Z^2} \bar{v}(l_+)
\gamma_5 u(l_-)\\
&\times \frac{1}{2\pi^2} \epsilon_{\mu\nu\rho\sigma} k^\rho
p^\sigma \varepsilon^{*\mu}(k)\varepsilon^{*\nu}(p)
\end{split}\end{equation}
i.e. the high-energy asymptotics can be written schematically as
\begin{equation}
\label{eq7.193} \mathcal{M}_\triangle \simeq a_f Q_f^2 \
\OO\bigl(\frac{m_e}{m_Z^2}E\bigr)
\end{equation}
where we have singled out only the coefficients that depend on the
characteristics of the fermion circulating in the triangle loop.
Let us emphasize that it is the {\it real part\/} of the triangle
subgraph that yields the observed bad high-energy behaviour of
(\ref{eq7.185}); the corresponding imaginary part is sufficiently
\qq{soft} (for a more detailed discussion of this point and for
further explicit formulae see e.g. \cite{Hor}).

The lesson to be learned from the considered example is as
follows. If one considers e.g. the GWS electroweak theory with the
fermion sector restricted to a single lepton type, the tree-level
unitarity surely holds, but at the one-loop (and higher) level one
observes a rapid violation of unitarity induced by an effect of
the ABJ anomaly. In fact, the coefficient in (\ref{eq7.193}) is
the same for all lepton species (note that
$a_e=a_\mu=a_\tau=-\frac{1}{2}$ and $Q_e=Q_\mu=Q_\tau=-1$) and
this means that adding more \qq{standard} leptons to a single
generation does not make the situation any better. Thus, {\bf the
GWS model for a leptonic world would not be renormalizable}; in
particular, the original Weinberg model \cite{ref39} certainly
suffers from such an \qq{anomaly disease} and -- as we will show
later in this section -- this is cured only when the quark sector
is taken into account properly.

\begin{figure}[h]\centering
\begin{tabular}{cc}
\subfigure[]{\s{\includegraphics{figs/fig29}}}&\hspace{1.5cm}\subfigure[]{\s{\includegraphics{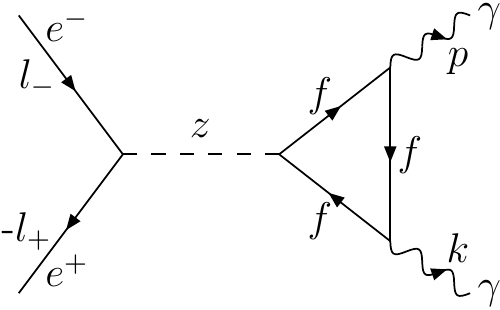}}}
\end{tabular}
\caption{Contributions of triangle fermion loops to the process
$e^+e^- \rightarrow \gamma\gamma$ in an $R$-gauge: (a) exchange of
the $Z$ boson (b) exchange of its unphysical counterpart $z$.}
\label{fig30}\index{Feynman diagrams!for $e^+e^-\rightarrow
\gamma\gamma$!in the $R$-gauge}
\end{figure}

In any case, the problem described above is characteristic of the
$U$-gauge (since only there one encounters vector boson
propagators containing pieces proportional to $m_V^{-2}$).
Irrespective of its ultimate solution within the full Standard
Model, it is interesting to know how the ABJ anomaly can manifest
itself in an $R$-gauge\index{R-gauge@$R$-gauge|(}, where all
propagators behave properly. To examine this, let us consider
again the process $e^+e^-\rightarrow \gamma\gamma$. Obviously, in
an $R_\xi$-gauge there is no power-like growth of the scattering
amplitude in question for $E\rightarrow \infty$, even at the level
of individual Feynman graphs. Instead, the crucial point now is
gauge invariance, i.e. the independence of the $S$-matrix element
on the gauge parameter $\xi$ (order by order in perturbation
theory). For tree graphs the problem is trivial, but it becomes
rather subtle already at the one-loop level. Motivated by our
previous experience, we shall focus our attention on the
\qq{suspect} diagrams depicted in Fig.\,\ref{fig30}.\footnote{Of
course, there are many other one-loop Feynman graphs contributing
to the considered process, that depend on the gauge parameter
$\xi$. It can be shown that the sum of all one-loop graphs, except
those in Fig.\,\ref{fig30}, is $\xi$-independent, but a
straightforward proof based on an explicit diagram calculations is
tedious.}

Normally, one would expect a cancellation of the $\xi$-dependent
part of the graph (a) against the contribution of (b), similarly
as in the tree-level example discussed in Section~\ref{sec7.7}. However,
the ABJ anomaly may violate such a mechanism, as it represents an
extra contribution to naive Ward identities (that simply \qq{copy}
classical relations for current divergences). We are going to show
that such an effect really occurs. For the purpose of our
discussion let us denote the contributions of Fig.\,\ref{fig30}(a)
and (b) (including the crossing of external photon lines) simply
as $\mathcal{M}^{(Z)}_\xi$ and $\mathcal{M}^{(z)}_\xi$
respectively, and identify the fermion inside the triangle loops
with a lepton, i.e. $f=e, \mu$ or $\tau$. According to the rules
established in Sections~\ref{sec7.6} and~\ref{sec7.7}, one has
\begin{equation}\label{eq7.194}\begin{split}
i\mathcal{M}_\xi^{(Z)} = i^8 &(-1) \frac{1}{4}
\Bigl(\frac{g}{\cos\theta_W}\Bigr)^2 (-a_f) Q_f^2 e^2
\bar{v}(l_+)\gamma_\lambda (v_e - a_e\gamma_5) u(l_-) \\
&\times \frac{-g^{\lambda\alpha} + (1-\xi) q^\lambda q^\alpha (q^2
- \xi m_Z^2)^{-1}}{q^2 - m_Z^2} T_{\alpha\mu\nu} (k,p)
\varepsilon^{*\mu}(k) \varepsilon^{*\nu}(p)
\end{split}
\end{equation}
and
\begin{equation}\begin{split}
i\mathcal{M}_\xi^{(z)} = i^8 (-1)
\Bigl(-i\frac{g}{2}\frac{m_e}{m_W}\Bigr)
\Bigl(-i\frac{g}{2}\frac{m_f}{m_W}\Bigr) Q_f^2 e^2 \bar{v}(l_+)
\gamma_5 u(l_-)\\
\times \frac{1}{q^2 - \xi m_Z^2} T_{\mu\nu}(k,p)
\varepsilon^{*\mu}(k) \varepsilon^{*\nu}(p) \label{eq7.195}
\end{split}\end{equation}
where the used symbols have the same meaning as before (note that
in writing (\ref{eq7.194}) we have automatically discarded a term
that would correspond to a loop of the $VVV$ type). Now, it is
convenient to compare
$\mathcal{M}_\xi^{(Z)}+\mathcal{M}_\xi^{(z)}$ with a \qq{reference
value}, corresponding e.g. to the 't Hooft--Feynman gauge (for
which $\xi=1$). The expression (\ref{eq7.194}) is naturally split
as
\begin{equation}
\mathcal{M}_\xi^{(Z)} = \mathcal{M}^{(Z)}_\ti{diag.} +
\mathcal{M}^{(Z)}_\ti{long.}
\end{equation}
in correspondence with the structure of $Z$ boson propagator (cf.
(\ref{eq7.164})). Obviously, $\mathcal{M}^{(Z)}_\ti{diag.}$
coincides with $\mathcal{M}^{(Z)}_{\xi=1}$, i.e.
\begin{equation}
\mathcal{M}^{(Z)}_\ti{diag.} = \mathcal{M}^{(Z)}_{tHF}
\label{eq7.197}
\end{equation}
Proceeding in the usual way, the $\mathcal{M}^{(Z)}_\ti{long.}$ is
recast as
\begin{equation}\begin{split}
\mathcal{M}^{(Z)}_\ti{long.} = -i \frac{1}{2}
\Bigl(\frac{g}{\cos\theta_W}\Bigr)^2 a_f Q_f^2 e^2 a_e m_e
\frac{1-\xi}{(q^2-\xi m_Z^2)(q^2-m_Z^2)}\\
\times \bar{v}(l_+) \gamma_5 u(l_-)\bigl[ 2m_f T_{\mu\nu}(k,p) + \mathcal{A}_{\mu\nu}(k,p)
\bigr] \varepsilon^{*\mu}(k) \varepsilon^{*\nu}(p) \label{eq7.198}
\end{split}\end{equation}
where we have also used the identity (\ref{eq7.191}) and denoted
\begin{equation}
\mathcal{A}_{\mu\nu}(k,p) = \frac{1}{2\pi^2}
\epsilon_{\mu\nu\rho\sigma} k^\rho p^\sigma
\end{equation}
The $\mathcal{M}^{(Z)}_\ti{long.}$  is thus divided into its
\qq{normal part} (corresponding to $2m_f T_{\mu\nu}$) and a
contribution of the ABJ anomaly. It is easy to show that by adding
the $\mathcal{M}^{(z)}_\xi$ shown in (\ref{eq7.195}) to the normal
part of (\ref{eq7.198}), one recovers the $z$-exchange
contribution in the 't Hooft--Feynman gauge (the reader is
recommended to verify this explicitly). Taking into account also
(\ref{eq7.197}), our results can be summarized as follows:
\begin{equation}
\mathcal{M}^{(Z)}_\xi + \mathcal{M}^{(z)}_\xi =
\mathcal{M}^{(Z)}_{tHF} +\mathcal{M}^{(z)}_{tHF} +
\mathcal{M}^{(Z)}_\ti{anomaly}
\end{equation}
where
\begin{equation}\label{eq7.201}\begin{split}
\mathcal{M}^{(Z)}_\ti{anomaly} = \frac{i}{4} &
\Bigl(\frac{g}{\cos\theta_W}\Bigr)^2 a_f Q_f^2 e^2 m_e
\frac{1-\xi}{(q^2 - \xi m_Z^2)(q^2 - m_Z^2)}\\
& \times\bar{v}(l_+) \gamma_5 u(l_-)\mathcal{A}_{\mu\nu} (k,p)
\varepsilon^{*\mu}(k)\varepsilon^{*\nu}(p)
\end{split}\end{equation}
In other words, the sum $\mathcal{M}^{(Z)}_\xi +
\mathcal{M}^{(z)}_\xi$ would be gauge independent, were it not for
the ABJ anomaly term in (\ref{eq7.198}). {\bf The anomaly effect
destroys gauge invariance} of the matrix element in question at
the one-loop level and this, of course, is a fatal blow to the
internal consistency of the GWS electroweak theory involving just
one fermion species. Notice that the resulting anomalous
contribution in (\ref{eq7.201}) is non-vanishing for any $\xi\neq
1$ and, as regards its dependence on the properties of the fermion
inside the triangle loop, this is carried by an overall factor
$a_f Q_f^2$ -- the same as in our previous result concerning the
violation of perturbative unitarity in $U$-gauge (cf.
(\ref{eq7.193})). Thus, in analogy with the observation following
the relation (\ref{eq7.193}), we can also conclude that the GWS
theory of leptons would be internally inconsistent (irrespective
of the number of lepton flavours), since the gauge independence of
the $S$-matrix would be lost because of the ABJ anomaly.

An upshot of the preceding discussion is as follows. If the
spectrum of elementary fermions is reduced to leptons alone, the
GWS electroweak theory suffers from serious problems due to the
ABJ anomaly. In the $U$-gauge, such a model is
non-renormalizable\index{non-renormalizability}, even though the
tree-level unitarity is satisfied. In renormalizable
$R_\xi$-gauges, some particular one-loop scattering amplitudes
depend explicitly on the gauge-fixing parameter $\xi$, i.e. there
is a manifest violation of gauge invariance. Both difficulties are
of similar technical origin, but the second issue is in fact more
fundamental: while one can imagine a perfectly consistent quantum
field theory model that is not perturbatively
renormalizable\footnote{It would only mean that an infinite number
of renormalization counterterms is needed and, consequently, such
a theory has less predictive power in comparison with models
renormalizable in the conventional sense.}, the loss of gauge
independence makes the considered perturbative approximation (and
thereby the whole perturbation expansion) totally meaningless.
Anyway, within the GWS theory, both problems are two sides of the
same coin, which is the anomalous Ward identity for a triangle
fermion loop\index{R-gauge@$R$-gauge|)}.

It is gratifying that within the full GWS standard model there is
a natural way out of the difficulties described above: it turns
out that the ABJ anomaly effects due to the lepton triangle loops
are exactly cancelled by an analogous contribution coming from
quarks. Let us now show, how such a simple mechanism works in the
example discussed above. As we have seen, the anomaly coefficient
corresponding to the $VVA$ triangle loop made of a fermion $f$ is
$a_f Q_f^2$, with $a_f$ and $Q_f$ having the usual meaning
explained above. According to the familiar rules for the weak
neutral current couplings, one has
\begin{equation}
a_f = T_{3L}^{(f)}
\end{equation}
where $T_{3L}^{(f)}$ is the weak isospin\index{weak!isospin|ff} of
the $f_L$. Thus, for $u$-quark ($Q_u=\frac{2}{3}$) and $d$-quark
($Q_d=-\frac{1}{3}$) one
gets\index{u-quark@$u$-quark}\index{dquark@$d$-quark}
\begin{equation}
\label{eq7.203} a_u Q_u^2 =
+\tfrac{1}{2}\bigl(\tfrac{2}{3}\bigr)^2, \qquad a_d Q_d^2 =
-\tfrac{1}{2}\bigl(-\tfrac{1}{3}\bigr)^2
\end{equation}
(obviously, the result (\ref{eq7.203}) is valid for any up-type
and down-type quark flavours). Now, it is important to realize
that any quark can occur in three colour \qq{copies} (more
precisely, a quark with a given flavour can exist in three states
distinguished by the colour quantum number). Colour charge is
substantial for strong interactions\index{strong interaction}
(described by quantum chromodynamics),
\index{quantum!chromodynamics (QCD)}but does not play any
dynamical role in electroweak interactions. This means that the
contribution of the considered triangle loop for a given quark
flavour should be simply multiplied by the number of colours $N_c
= 3$\index{colour}\index{colour|seealso{QCD}}. Thus, restricting
ourselves to the first generation of fermions ($\nu_e, e, u, d$),
the relevant results can be summarized as follows. Lepton
contribution to the full coefficient of the ABJ anomaly is
\begin{equation}
\pA_{ABJ}^\tiz{leptons} = -\tfrac{1}{2}(-1)^2 = -\tfrac{1}{2}
\end{equation}
while the quark loops yield
\begin{equation}
\pA_{ABJ}^\tiz{quarks $u,d$} = 3\times
\Bigl[\tfrac{1}{2}\bigl(\tfrac{2}{3}\bigr)^2
-\tfrac{1}{2}\bigl(-\tfrac{1}{3}\bigr)^2 \Bigr] = +\tfrac{1}{2}
\end{equation}
In this straightforward way, it is seen that the lepton and quark
anomalies indeed compensate each other. Obviously, for the other
two generations of elementary fermions the result must be the
same, since the pattern of the relevant parameters $a_f$, $Q_f$
repeats itself.

The algebraic condition for the anomaly cancellation\index{anomaly
cancellation|(} can be put in a more elegant form. To see this,
let us start with a general expression for the sum of the ABJ
anomaly coefficients (e.g. within the first generation) that can
be written as
\begin{equation}\label{eq7.206}
\pA_{ABJ} = \sum_f a_f Q_f^2 = T_{3L}^{(\nu)}Q_\nu^2 +
T_{3L}^{(e)} Q_e^2 + N_c (T_{3L}^{(u)}Q_u^2 + T_{3L}^{(d)}
Q_{d}^2)
\end{equation}
(for the sake of full symmetry we have included formally also the
neutrino with $Q_\nu=0$, though its contribution vanishes).
Utilizing the known values of the weak isospin for leptons and
quarks, (\ref{eq7.206}) is recast as
\begin{equation}
\pA_{ABJ} = \tfrac{1}{2}(Q_\nu^2 - Q_e^2) + \tfrac{1}{2}N_c (Q_u^2
-Q_d^2)
\end{equation}
However, electric charges of two fermions belonging to the same
isospin doublet differ by one unit, so the last expression becomes
\begin{equation}
\pA_{ABJ} = \tfrac{1}{2} \bigl[ Q_\nu +Q_e + N_c (Q_u + Q_d)
\bigr]
\end{equation}
Thus, the ABJ anomaly coefficient in question is seen to be
proportional to the sum of all fermion charges, with each quark
taken in $N_c$ colour mutations. The condition of vanishing of the
ABJ anomaly in the considered example thus reads
\begin{equation}\label{eq7.209}
\sum_f Q_f = 0
\end{equation}
where the sum in (\ref{eq7.209}) extends over all fermion species,
including the colour factor $N_c$ for quarks. From the above
discussion it is clear that for achieving such a cancellation of
anomalies it is essential that there are just three colours,
$N_c=3$. Equally obvious is that the presence of two quark
flavours in each generation (with electric charges $+\frac{2}{3}$
and $-\frac{1}{3}$ respectively) is necessary for this purpose.
Historically, this observation provided (among other things)
strong motivation for the $t$-quark\index{t-quark@$t$-quark}
searches after the discoveries of the $\tau$-lepton\index{tau
lepton@$\tau$ lepton} in 1975 and $b$-quark in 1977: at that
time\index{bquark@$b$-quark}, the top was desperately needed to
make the third generation complete\index{top}. As we noted earlier
in this chapter, this superheavy quark has been directly observed
only in mid 1990s. The simple algebraic condition (\ref{eq7.209})
represents indeed a highly remarkable result: it provides the only
known successful theoretical relation {\it between leptons and
quarks}, which otherwise form entirely independent sectors of the
spectrum of elementary fermions within SM.

In fact, within the GWS standard model one can find many other
examples of scattering amplitudes that could be affected by the
ABJ anomaly effects. The point is that any triangular fermion loop
of the type $VVA$ yields the anomaly, irrespective of the
assignments of the attached vector boson lines. Moreover, ABJ
anomalies also occur in $AAA$ fermion triangles (i.e. in those
made of three axial-vector currents). The potentially anomalous
cases can be simply classified in terms of labels for the external
vector boson lines entering vertices of the triangle loops in
question. Thus, apart from the $Z\gamma\gamma$ case discussed
previously, the other relevant configurations can be $ZZ\gamma,
ZZZ, ZWW$ and $\gamma WW$ (for example, the configuration
$ZZ\gamma$ can occur in the amplitude for the process $e^+e^-
\rightarrow Z\gamma$ etc.). Obviously, for combinations
$Z\gamma\gamma, ZZ\gamma$ and $\gamma WW$ only the $VVA$ anomalies
contribute, while in cases labelled as $ZZZ$ and $ZWW$ both $VVA$
and $AAA$ anomalies can play a role. When the whole collection of
triangle graphs is analyzed, a result that emerges is remarkably
simple: {\bf it turns out that satisfying the relation
(\ref{eq7.209}) already suffices for elimination of all anomalies
enumerated here} (a proof of this statement is left as a challenge
for a seriously interested reader). Therefore we can conclude that
the GWS standard model is completely free of ABJ anomalies, which
means that it is an internally consistent and perturbatively
renormalizable theory of electroweak
interactions\index{Adler--Bell--Jackiw (ABJ)
anomaly|)}\index{renormalizable theory|)}\index{anomalous Ward
identity|)}\index{anomaly cancellation|)}.

The possibility of a mutual compensation of ABJ triangle anomalies
due to different fermion species has been first observed -- within
the $SU(2)\times U(1)$ electroweak theory -- by C.~Bouchiat,
J.~Iliopoulos and Ph.~Meyer \cite{ref78}. Thus, the anomaly
cancellation mechanism outlined above should perhaps be
appropriately called the \qq{BIM mechanism}. Note also that almost
simultaneously with the work \cite{ref78}, the same issue was
analyzed independently in the papers \cite{ref79} and
\cite{ref80}.

%\end{document}

%\input{kniha7_10}
%%%%%%%%%%%%%%%%%%%%%%%%%%%%%%%%%%%%%%%%%%%%%%%%%%%%%%%%%%%%%%%%%%%
%%%%%%%%%%%%%%%%%%%%%%%%%%%%%%%%%%%%%%%%%%%%%%%%%%%%%%%%%%%%%%%%%%%%%%%%%%%%%%%%%%%%%%%%%%%%%%%%%%%%%%%%%%%%%%%%%%%%%%%%%%%%%%%%%%%%%%%%
%\documentclass[12pt,tbtags]{report}
%\usepackage{mathrsfs,euscript,amsfonts, amssymb, mathbbol, bbm, dsfont,eufrak}
%\input{pream}
%\begin{document}
\section{Synopsis of the GWS standard model}\label{sec7.10}
\index{Yang--Mills field|(} \index{SU(2) times U(1)
group@$SU(2)\times U(1)$ group} We have already described in
detail all relevant parts of the Glashow--Weinberg--Salam standard
model of electroweak interactions. For the reader's convenience,
we shall now summarize the whole construction, as well as the
corresponding interaction Lagrangian in the physical $U$-gauge.

As to the particle contents of the GWS standard model, there are
\begin{enumerate}
\item [i)] three generations of spin-1/2 elementary fermions, i.e. six
leptons ($e$, $\nu_e$, $\mu$, $\nu_\mu$, $\tau$, $\nu_\tau$) and
six quarks ($d$, $u$, $s$, $c$, $b$, $t$)
\item [ii)] four spin-1 bosons, namely the massive intermediate vector bosons $W^\pm$, $Z^0$
and massless photon $\gamma$
\item [iii)] a spin-0 Higgs boson $H$.
\end{enumerate}
The fermions are conventionally considered as constituting the
\qq{matter}, while the spin-1 bosons are \qq{carriers of
electroweak force} (since they mediate electroweak interactions).
The Higgs boson (not yet confirmed experimentally) is intimately
related to the mechanism of mass generation for the other
particles.

The fundamental dynamical principle is that of gauge invariance
(i.e. local internal symmetry\index{local symmetry}); the relevant
symmetry group is non-Abelian, namely $SU(2)\times U(1)$. The
$SU(2)$ is referred to as \qq{weak isospin} subgroup and the
factor $U(1)$ corresponds to \qq{weak
hypercharge}\index{weak!hypercharge}. Within this framework, the
$W^\pm$, $Z^0$ and $\gamma$ mediating electroweak interactions are
quanta of physical vector fields that are made of the four
original Yang--Mills fields associated with the four generators of
$SU(2)\times U(1)$. Basic building blocks of the fermion sector
are left-handed $SU(2)$ doublets
\begin{align}
L^{(e)}&=\bm{\nu_{eL}\\e_L},&L^{(\mu)}&=\bm{\nu_{\mu L}\\
\mu_L},&L^{(\tau)}&=\bm{\nu_{\tau L}\\ \tau_L}\notag\\
L^{(d)}_0&=\bm{u_{0L}\\d_{0L}},&\quad L^{(s)}_0&=\bm{c_{0L}\\
s_{0L}},&\quad L^{(b)}_0&=\bm{t_{0L}\\ b_{0L}} \label{eq7.210}
\end{align}
and right-handed singlets
\begin{gather}
e_R,\quad \mu_R,\quad \tau_R \notag\\
d_{0R},\quad u_{0R},\quad s_{0R},\quad c_{0R},\quad b_{0R},\quad
t_{0R} \label{eq7.211}
\end{gather}
where $e_L=\frac{1}{2}(1-\gamma_5)e$,
$e_R=\frac{1}{2}(1+\gamma_5)e$ etc. Note that we have not included
here the right-handed components of neutrino fields; for the
purpose of this overview we simply ignore neutrino masses (a more
detailed discussion concerning this issue can be found in Section~\ref{sec6.6}). The quark variables in (\ref{eq7.210}) and (\ref{eq7.211})
carry the label \qq{0} as they represent a set of \qq{primordial}
fields (or \qq{protofields}) that have yet to be transformed into
the physical ones, carrying definite masses.\footnote{In
comparison with the notation introduced in Sections~\ref{sec7.3} and~\ref{sec7.4} we
have changed the symbols for quark doublets; such a slight
modification makes the subsequent formulae more compact. The logic
of the labelling employed in (\ref{eq7.210}) should be obvious.}
For implementing the Higgs mechanism\index{Higgs mechanism}, which
yields mass terms of vector bosons, one makes use of an $SU(2)$
doublet of complex scalar fields
\begin{equation}
\label{eq7.212} \Phi = \bm{\varphi^+\\ \varphi^0}
\end{equation}
Furthermore, in order to generate masses of all fermions (in
particular, the Dirac masses of both down- and up-type quarks),
one has to employ both (\ref{eq7.212}) and a conjugate doublet
\begin{equation}
\widetilde{\Phi} = i \tau_2 \Phi^* \label{eq7.213}
\end{equation}
where the $\tau_2$ is the second Pauli matrix.

The GWS gauge invariant Lagrangian can be written as consisting of
four parts, namely
\begin{equation}
\label{eq7.214} \lagr_{GWS}=\lagr_\ti{gauge} + \lagr_\ti{fermion}
+ \lagr_\ti{Higgs} + \lagr_\ti{Yukawa}
\end{equation}
and the individual terms in (\ref{eq7.214}) are consecutively
defined below.

First, the $\lagr_\ti{gauge}$ is pure Yang--Mills Lagrangian
corresponding to the local symmetry $SU(2)\times U(1)$:
\begin{equation}
\lagr_\ti{gauge} = -\frac{1}{4} F_{\mu\nu}^a F^{a\mu\nu} -
\frac{1}{4}B_{\mu\nu}B^{\mu\nu} \label{eq7.215}
\end{equation}
where
\begin{align}
F^a_{\mu\nu} &= \partial_\mu A^a_\nu -\partial_\nu A^a_\mu + g
\epsilon^{abc} A^b_\mu A^c_\nu \notag \\
B_{\mu\nu} &= \partial_\mu B_\nu - \partial_\nu B_\mu
\label{eq7.216}
\end{align}
with $A^a_\mu$, $a=1,2,3$ and $B_\mu$ being the $SU(2)$ and $U(1)$
gauge fields respectively; the $\epsilon^{abc}$ stands for the
totally antisymmetric Levi-Civita symbol (structure constants of
the $SU(2)$ algebra) and $g$ is the $SU(2)$ gauge coupling
constant.

Next, the $\lagr_\ti{fermion}$ comprises kinetic terms for leptons
and quarks and their interactions with gauge fields. Making use of
the doublets and singlets (\ref{eq7.210}), (\ref{eq7.211}) it can
be written as
\begin{equation}\begin{split}
\lagr_\ti{fermion} &= \sum_{\ell=e,\mu,\tau} i \bar{L}^{(\ell)}
\gamma^\mu (\partial_\mu - ig A^a_\mu \frac{\tau^a}{2} - i
Y_L^{(\ell)} g' B_\mu ) L^{(\ell)} \\
&+\sum_{q=d,s,b} i \bar{L}_0^{(q)} \gamma^\mu (\partial_\mu - ig
A^a_\mu \frac{\tau^a}{2} - i Y_L^{(q)} g' B_\mu ) L_0^{(q)}\\
&+\sum_{\ell=e,\mu,\tau} i \bar{\ell}_R \gamma^\mu (\partial_\mu -
i Y_R^{(\ell)} g' B_\mu ) \ell_R \\
&+\sum_{q=d,u,s,c,b,t} i \bar{q}_{0R} \gamma^\mu (\partial_\mu - i
Y_R^{(q)} g' B_\mu ) q_{0R} \label{eq7.217}
\end{split}\end{equation}
where we have introduced standard covariant derivatives. Apart
from the parameter $g$ that has already appeared in
(\ref{eq7.216}), there is another independent coupling constant
$g'$, associated with the $U(1)$ subgroup\index{U(1) group@$U(1)$
group}. The weak hypercharge assignments for the matter fields are
conventionally defined by
\begin{equation}
Q=T_3 + Y \label{eq7.218}
\end{equation}
with $T_3$ being the third component of weak isospin and $Q$ the
corresponding particle charge (in units of positron charge).
Taking into account that
\begin{alignat}{3}
Q_e &= Q_\mu &&= Q_\tau &&= -1 \notag\\
Q_d &= Q_s &&= Q_b &&= -\frac{1}{3} \notag\\
Q_u &= Q_c &&= Q_t &&= +\frac{2}{3} \label{eq7.219}
\end{alignat}
the relation (\ref{eq7.218}) yields
\begin{alignat}{2}
Y_L^{(\ell)} &= -\frac{1}{2}\ ,&\qquad \quad \ell &= e,\mu,\tau \notag\\
Y_L^{(q)} &= +\frac{1}{6}\ ,&\qquad \quad q &= d,s,b \notag\\
Y_R^{(q)} &= -\frac{1}{3}\ ,&\qquad \quad q &= d,s,b \notag\\
Y_R^{(q)} &= +\frac{2}{3}\ ,&\qquad \quad q &= u,c,t
\label{eq7.220}
\end{alignat}
The $\lagr_\ti{Higgs}$ has the form
\begin{equation}
\lagr_\ti{Higgs} = \Phi^\dagger \bigl(\partialvb_\mu + i g A^a_\mu
\frac{\tau^a}{2} + \frac{i}{2} g' B_\mu \bigr)\bigl(
\partial^\mu - i g A^{b\mu} \frac{\tau^b}{2} - \frac{i}{2} g'
B^\mu \bigr) \Phi - \lambda \bigl( \Phi^\dagger \Phi -
\frac{v^2}{2} \bigr)^2 \label{eq7.221}
\end{equation}
where, in accordance with the rule (\ref{eq7.218}), we have set
$Y_\Phi = +\frac{1}{2}$ in the relevant covariant derivatives. The
$\lambda$ is a coupling constant for the Higgs scalar
self-interaction and the parameter $v$ (\qq{vacuum expectation
value}\index{vacuum expectation value} or \qq{vacuum
shift}\index{vacuum} of the Higgs field) provides an overall scale
for particle masses. Working out the scalar \qq{potential} in
(\ref{eq7.221}), one can see immediately that $\lambda v^2$ is the
coefficient of a wrong-sign mass term for the $\Phi$.

Finally, the Yukawa-type term reads\index{Yukawa coupling}
\begin{equation}\begin{split}
\lagr_\ti{Yukawa} = &- \sum_{\ell=e,\mu,\tau} h_\ell
\bar{L}^{(\ell)} \Phi \ell_R + \text{h.c.} - \sum_{\substack
{q=d,s,b \\ q'=d,s,b}} h_{qq'} \bar{L}_0^{(q)} \Phi q'_{0R} +
\text{h.c.} \\ &- \sum_{\substack{q=d,s,b\\q'=u,c,t}}
\tilde{h}_{qq'} \bar{L}_0^{(q)} \Phit q'_{0R} + \text{h.c.}
\label{eq7.222}
\end{split}\end{equation}
with $h_\ell$, $h_{qq'}$, $\tilde{h}_{qq'}$ being essentially
arbitrary (real) coupling constants; the overall minus sign is
purely conventional. If one takes into account (\ref{eq7.220}) as
well as
\begin{equation}
Y_{\Phit} = - Y_\Phi = -\frac{1}{2}
\end{equation}
the $U(1)$ invariance of (\ref{eq7.222}) can be checked easily
(its symmetry with respect to the $SU(2$) is obvious).

Now, the complex doublet (\ref{eq7.212}) embodies four real
scalars and three of them are would-be Goldstone
bosons\index{Goldstone!boson} associated with a spontaneously
broken $SU(2)$ symmetry\index{spontaneous symmetry breakdown} of
the potential in $\lagr_\ti{Higgs}$. These unphysical
scalars\index{unphysical scalars} can be eliminated by means of an
appropriate choice of gauge; such a procedure is formally
equivalent to an $SU(2)$ transformation within the Lagrangian
(\ref{eq7.214}) and amounts to replacing the $\Phi$ by
\begin{equation}
\Phi_U = \bm{0\\\frac{1}{\sqrt{2}}(v+H)} \label{eq7.224}
\end{equation}
where $H$ denotes the physical Higgs boson. Note also that
(\ref{eq7.213}) then immediately yields
\begin{equation}
\Phit_U = \bm{\frac{1}{\sqrt{2}}(v+H)\\0} \label{eq7.225}
\end{equation}
In what follows, we shall describe how the contents of the
original Lagrangian (\ref{eq7.214}) is disentangled in terms of
physical fields. We are going to concentrate first on the
structural aspects and a detailed form of the interaction
Lagrangian will be summarized later on. Let us start with the
leptonic part of $\lagr_\ti{fermion}$. The $A^a_\mu$ and $B_\mu$
are primordial gauge fields without a direct particle contents,
but the physical vector fields can be obtained from them by means
of appropriate linear combinations. In particular, the $W^\pm_\mu$
defined by
\begin{equation}
W^\pm_\mu = \frac{1}{\sqrt{2}} (A^1_\mu \mp i A^2_\mu)
\label{eq7.226}
\end{equation}
are coupled to weak charged currents (with coupling constant being
proportional to the $g$) and the $Z_\mu$, $A_\mu$ introduced via a
real orthogonal transformation\index{orthogonal transformation}
\begin{alignat}{2}
A^3_\mu &= \phantom{-}\cos\theta_W Z_\mu &&+\sin\theta_W A_\mu \notag \\
B_\mu &=-\sin\theta_W Z_\mu &&+\cos\theta_W A_\mu \label{eq7.227}
\end{alignat}
represent the $Z$ boson field and the electromagnetic
four-potential respectively. The mixing embodied in
(\ref{eq7.227}) represents the mathematical basis of the concept
of  \qq{electroweak unification} within the GWS theory. The
parameter $\theta_W$ is usually called the Weinberg angle, or
simply \qq{weak mixing angle}\index{weak!mixing angle}. The
requirement that the $A_\mu$ be coupled with equal strength to the
left-handed and right-handed leptons (in other words, that the
current coupled to the $A_\mu$ be a pure vector) leads to the
condition $\tan\theta_W =g'/g$, i.e.
\begin{equation}
\cos\theta_W = \frac{g}{\sqrt{g^2+g'^2}}, \qquad \sin\theta_W =
\frac{g'}{\sqrt{g^2+g'^2}} \label{eq7.228}
\end{equation}
and, subsequently, the electromagnetic coupling constant is
expressed as
\begin{equation}
e = \frac{gg'}{\sqrt{g^2+g'^2}}
\end{equation}
One thus arrives at a relation between $e$ and $g$, namely
\begin{equation}
e = g \sin\theta_W \label{eq7.230}
\end{equation}
Note that (\ref{eq7.230}) is sometimes called the \qq{unification
condition}\index{unification condition} (for an $SU(2) \times
U(1)$ electroweak theory). The $Z_\mu$ defined by (\ref{eq7.227})
is then coupled to a weak neutral current whose structure is fully
determined in terms of the parameter $\sin^2\theta_W$  (that has
to be fixed by experiments). As regards the electroweak
interactions of quarks, we shall discuss them a bit later; now let
us come back to the $\lagr_\ti{gauge}$.

Using the linear transformations (\ref{eq7.226}) and
(\ref{eq7.227}) in (\ref{eq7.215}), one gets kinetic terms for the
vector fields $W^\pm_\mu$, $Z_\mu$ and $A_\mu$, namely
\begin{equation}
\lagr^\tiz{kin.}_\ti{gauge} = -\frac{1}{2} W^-_{\mu\nu}
W^{+\mu\nu} - \frac{1}{4} Z_{\mu\nu}Z^{\mu\nu} - \frac{1}{4}
A_{\mu\nu}A^{\mu\nu} \label{eq7.231}
\end{equation}
where $W^-_{\mu\nu}=\partial_\mu W^-_\nu - \partial_\nu W^-_\mu$
etc., and a set of trilinear and quadrilinear\index{quadrilinear
vertex} vector boson self-interactions. These are of the following
types: $WW\gamma$, $WWZ$, $WWWW$, $WWZZ$, $WWZ\gamma$ and
$WW\gamma\gamma$. Notice that other types, such as e.g.
$Z\gamma\gamma$, $ZZZ$, $ZZZZ$, etc., are automatically excluded.

Next, we proceed to the $\lagr_\ti{Higgs}$. Substituting
(\ref{eq7.224}) into (\ref{eq7.221}), one identifies readily free
Lagrangian for the Higgs boson $H$, with a mass given by
\begin{equation}
m_H^2 = 2\lambda v^2
\end{equation}
As the most important item, mass terms for the $W^\pm$ and $Z$ are
obtained from the Higgs mechanism. To that end, one has to work
out the relevant quadratic form in variables $A^a_\mu$, $B_\mu$
(which is induced by the vacuum shift $v$ in (\ref{eq7.224})): the
$A^1_\mu$, $A^2_\mu$ are replaced by the $W^\pm_\mu$ defined
according to (\ref{eq7.226}) and a mass matrix for $A^3_\mu$,
$B_\mu$ is diagonalized in a straightforward way. Taking into
account the normalization of (\ref{eq7.231}), the vector boson
masses in question can then be identified as
\begin{equation}
m_W = \frac{1}{2}gv, \qquad m_Z = \frac{1}{2}(g^2+g'^2)^{1/2} v
\label{eq7.233}
\end{equation}
Thus, in view of (\ref{eq7.228}), one has
\begin{equation}
m_W = m_Z \cos\theta_W \label{eq7.234}
\end{equation}
It should be stressed that the $Z_\mu$ carrying the mass shown in
(\ref{eq7.233}) is given precisely by the expression following
from (\ref{eq7.227}), i.e.
\begin{equation}
Z_\mu = \cos\theta_W A^3_\mu - \sin\theta_W B_\mu
\end{equation}
(with $\cos\theta_W$ and $\sin\theta_W$ taking on the values
(\ref{eq7.228})). Similarly, the orthogonal combination
\begin{equation}
A_\mu = \sin\theta_W A^3_\mu + \cos\theta_W B_\mu
\end{equation}
corresponds to the massless photon.\footnote{In other words, it is
seen that the fields diagonalizing gauge boson mass terms in
$\lagr_\ti{Higgs}$ coincide with those descending from the
analysis of interactions in the leptonic sector of
$\lagr_\ti{fermion}$. Such a result is gratifying (as it manifests
the internal consistency of the electroweak theory), but one
should also keep in mind that we have actually anticipated it by
choosing carefully the relevant hypercharge values according to
(\ref{eq7.218}).} Invoking the familiar formula $G_F/\sqrt{2} =
g^2/(8m_W^2)$ and using (\ref{eq7.233}) one finds out immediately
that the $v$ is simply related to the Fermi constant:
\begin{equation}
v=(G_F \sqrt{2})^{-1/2} \doteq 246\ \GeV
\end{equation}
Further, utilizing the unification condition (\ref{eq7.230}) and
the relation between $e$ and the fine structure constant $\alpha$,
$\alpha=e^2/(4\pi)$, the mass formulae (\ref{eq7.233}) and
(\ref{eq7.234}) can be recast in a form most suitable for
practical purposes, namely
\begin{equation}
m_W = \Bigl(\frac{\pi\alpha}{G_F \sqrt{2}}\Bigr)^{1/2}
\frac{1}{\sin\theta_W},\quad m_Z =
\Bigl(\frac{\pi\alpha}{G_F\sqrt{2}}\Bigr)^{1/2}
\frac{1}{\sin\theta_W\cos\theta_W}
\end{equation}
Apart from the above-mentioned mass terms, the $\lagr_\ti{Higgs}$
yields also a set of interactions involving the $W^\pm$, $Z$ and
$H$. Schematically, the relevant couplings are $WWH$, $ZZH$,
$WWHH$, $ZZHH$, $HHH$ and $HHHH$.

Last but not least, let us consider the term $\lagr_\ti{Yukawa}$.
As for the leptonic part of (\ref{eq7.222}), this is worked out in
a straightforward way. Substituting there (\ref{eq7.224}), one
gets readily mass terms for charged leptons, with
\begin{equation}
m_\ell = \frac{1}{\sqrt{2}} h_\ell v \label{eq7.239}
\end{equation}
and pure scalar Yukawa interactions of the type $\ell\ell H$,
whose strength is -- in view of (\ref{eq7.239}) -- obviously
proportional to $m_\ell/v$. In the quark sector, one gets two
different (in general non-diagonal) $3\times 3$ mass matrices for
primordial fields (corresponding to the matrices of coupling
constants in (\ref{eq7.222})); in particular, the original Yukawa
interactions involving the $\Phi$ yield the down-type quark masses
while the up-type quarks gain masses from interactions with the
$\Phit$ displayed in (\ref{eq7.225}). The quark mass matrices are
diagonalized by means of appropriate biunitary
transformations\index{biunitary transformation} (involving
independent rotations of left-handed and right-handed fields).
Owing to the simple structure of  $\Phi_U$ and $\Phit_U$, Higgs
boson interactions are diagonalized simultaneously with mass terms
and one thus arrives at the same pattern of coupling constants as
in the case of leptons: the strength of a coupling $qqH$ ($q = d,
u, s, c, b, t$) is proportional to $m_q/v$.

Having diagonalized the mass matrices in question, one can return
to the quark sector of the $\lagr_\ti{fermion}$. Performing the
relevant unitary rotations of the primordial fields appearing in
(\ref{eq7.217}), the interaction Lagrangian is recast in terms of
variables corresponding to mass eigenstates. In this context, one
has to keep in mind that transformations of the up-type and
down-type quarks are completely independent; e.g. for the
left-handed fields one can write symbolically
\begin{equation}
\bm{d_L\\s_L\\b_L} = \U \bm{d_{0L}\\s_{0L}\\b_{0L}},\quad
\bm{u_L\\c_L\\t_L} = \Ut \bm{u_{0L}\\c_{0L}\\t_{0L}}
\end{equation}
where the $\U$ and $\Ut$ are in general different unitary $3\times
3$ matrices. Thus, an essentially arbitrary unitary matrix
\begin{equation}
\Ut\U^\dagger = V =
\bm{V_{ud}&V_{us}&V_{ub}\\V_{cd}&V_{cs}&V_{cb}\\V_{td}&V_{ts}&V_{tb}}
\label{eq7.241}
\end{equation}
shows up in the resulting weak interaction of charged currents.
The $V$ is the celebrated Cabibbo--Kobayashi--Maskawa (CKM) matrix
and can be ultimately parametrized in terms of just four
physically relevant parameters -- three \qq{Cabibbo-like} angles
and one $\cal CP$--violating phase (in case of two generations of
quarks one would end up with a real orthogonal $2 \times 2$ matrix
described in terms of the Cabibbo angle). In this way, one arrives
at a natural view of the origin of flavour
mixing\index{flavour!mixing} and $\cal CP$ violation: both these
phenomena are intimately related to diagonalization of quark
matrices descending from the most general Yukawa couplings
compatible with electroweak symmetry. Furthermore, one observes
that {\it three\/} is just the minimum number of fermion
generations for which $\cal CP$ violation can occur within the SM
scheme. On the other hand, neutral quark currents, that are
obviously flavour-diagonal in the primordial basis, remain
diagonal even after the transformation to the fields with definite
masses. In this way, SM provides a natural explanation for the
conspicuous absence of strangeness-changing (more generally,
flavour-changing) weak neutral currents\index{strangeness-changing
neutral current}.

Note finally that massive neutrinos and their eventual mixings can
be incorporated quite naturally into the SM scheme. Introducing
also the right-handed (singlet) components of neutrino fields from
the very beginning, the technique used for quarks can be
generalized in a straightforward way to the lepton sector of
$\lagr_\ti{Yukawa}$; one thus gets Dirac mass terms for neutrinos
(along with the corresponding Higgs boson Yukawa interactions) and
a leptonic analogue of the CKM matrix.\footnote{In fact, when
introducing neutrino mass terms one has more possibilities than in
the case of quarks; neutrinos are electrically neutral and this
opens the possibility that they could be Majorana particles. As we
said earlier, in the present text we do not pursue the issue of
neutrino masses in detail, although this currently represents one
of the hot topics of particle physics. The interested reader is
referred e.g. to the book \cite{Vog}.}

Let us now summarize the explicit form of the SM interaction
Lagrangian in the $U$-gauge. This can be written as
\begin{equation}\begin{split}
\lagr^{(GWS)}_{int} &= \sum_f Q_f e \bar{f} \gamma^\mu f A_\mu +
\lagr_{CC} + \lagr_{NC}\\ &-ig (W^0_\mu W^-_\nu \partialvob^\mu
W^{+\nu} + W^-_\mu W^+_\nu \partialvob^\mu W^{0\nu} + W^+_\mu
W^0_\nu \partialvob^\mu W^{-\nu})\\
&-g^2 \bigl[ \tfrac{1}{2}(W^-\cdot W^+)^2 - \tfrac{1}{2}(W^-)^2
(W^+)^2 + (W^0)^2 (W^-\cdot W^+) \\ &\hspace{5.7cm}- (W^-\cdot
W^0)(W^+\cdot W^0) \bigr]\\
&+gm_W W^-_\mu W^{+\mu} H + \frac{1}{2\cos\theta_W} g m_Z Z_\mu
Z^\mu H \\ &+\frac{1}{4} g^2 W^-_\mu W^{+\mu} H^2 +
\frac{1}{8}\frac{g^2}{\cos^2\theta_W} Z_\mu Z^\mu H^2 \\
&-\sum_f \frac{1}{2} g \frac{m_f}{m_W} \bar{f}f H - \frac{1}{4} g
\frac{m_H^2}{m_W} H^3 - \frac{1}{32} g^2 \frac{m_H^2}{m_W^2} H^4
\label{eq7.242}
\end{split}\end{equation}
where the indicated sums run over all elementary fermions (leptons
and quarks) and the relevant charge factors $Q_f$ are displayed in
(\ref{eq7.219}). In the self-interactions of vector bosons we have
used, for mnemonic convenience, the notation
\begin{equation}
W^0_\mu = \cos\theta_W Z_\mu + \sin\theta_W A_\mu \label{eq7.243}
\end{equation}
(of course, $W^0_\mu$ coincides with the $A^3_\mu$ used before).
The term $\lagr_{CC}$ describes the interactions of weak charged
currents and vector bosons $W^\pm$:
\begin{equation}\begin{split}
\lagr_{CC} = &\frac{g}{2\sqrt{2}} \sum_{\ell = e,\mu,\tau}
\bar{\nu}_\ell \gamma^\lambda (1-\gamma_5) \ell W^+_\lambda\\ &+
\frac{g}{2\sqrt{2}} \bm{\bar{u},\ \bar{c},\
\bar{t}}\gamma^\lambda(1-\gamma_5) V_{CKM}\bm{d\\s\\b}W^+_\lambda
+ \text{h.c.}
\end{split}\end{equation}
where $V_{CKM}$ is the CKM unitary matrix (\ref{eq7.241}). The
$\lagr_{NC}$ stands for the interaction of weak neutral currents
and the $Z$:
\begin{equation}
\lagr_{NC}=\frac{g}{\cos\theta_W} \sum_f \bigl(
\varepsilon_L^{(f)} \bar{f}_L \gamma^\lambda f_L +
\varepsilon_R^{(f)} \bar{f}_R \gamma^\lambda f_R \bigr) Z_\lambda
\end{equation}
where
\begin{align}
\varepsilon_L^{(f)} &= T^{(f)}_{3L} - Q_f \sin^2 \theta_W \notag \\
 \varepsilon_R^{(f)} &= T^{(f)}_{3R} - Q_f \sin^2 \theta_W
\end{align}
with $T^{(f)}_{3L}=+\frac{1}{2}$ for $f=\nu_e,\nu_\mu,\nu_\tau, u,
c, t$, $T^{(f)}_{3L} = -\frac{1}{2}$ for $f=e,\mu,\tau, d, s,b$
and $T^{(f)}_{3R} = 0$ for any $f$.

The neutral current interaction may alternatively be written in
the form
\begin{equation}
\lagr_{NC} = \frac{g}{2\cos\theta_W} \sum_f \bar{f}\gamma^\lambda
(v_f-a_f\gamma_5) f Z_\lambda
\end{equation}
with
\begin{align}
v_f &= \varepsilon_L^{(f)} + \varepsilon_R^{(f)} \notag \\
a_f &= \varepsilon_L^{(f)} - \varepsilon_R^{(f)}
\end{align}
that is
\begin{align}\label{eq7.249}
\left.
\begin{aligned}
v_f&=-\frac{1}{2}-2 Q_f \sin^2\theta_W\\
a_f&=-\frac{1}{2}
\end{aligned}
\right\} &\quad\text{for } f=e,\mu,\tau,d,s,b \notag \\
\left.
\begin{aligned}
v_f&=+\frac{1}{2}-2 Q_f \sin^2\theta_W\\ a_f&=+\frac{1}{2}
\end{aligned}
\right\} &\quad\text{for } f=\nu_e,\nu_\mu,\nu_\tau,u,c,t
\end{align}
As for the self-interactions of vector bosons, the compact form
shown in (\ref{eq7.242}) is worked out readily by using the
definition (\ref{eq7.243}) and one can thus identify the
individual couplings $WW\gamma$, $WWZ$, $WW\gamma\gamma$, $WWWW$,
$WWZZ$ and $WWZ\gamma$. For completeness, let us reiterate the
important relations
\begin{align}
e/g &= \sin\theta_W \notag\\
m_W/m_Z &= \cos\theta_W \notag\\
G_F/\sqrt{2} &= {g^2}/(8m_W^2) \label{eq7.250}
\end{align}

In summarizing the electroweak standard model, we should also
count the number of free parameters involved in its Lagrangian.
First of all, there are coupling constants $g$, $g'$, $\lambda$
and the mass scale $v$ (obviously, in view of (\ref{eq7.250}),
these basic four parameters can be traded e.g. for $\alpha$,
$\sin^2\theta_W$, $m_Z$, $m_H$ or $\alpha$, $G_F$, $m_Z$, $m_H$
etc.). The remaining free parameters come from the Yukawa sector.
Taking neutrinos as massless for the moment, one has three masses
of charged leptons, six quark masses and four parameters of the
CKM mixing matrix. Thus, within the original version of SM one has
$4 + 3 + 6 + 4 = 17$ free parameters. If one allows also for
neutrino masses and mixings, one has seven additional parameters
(three neutrino masses and four parameters of a leptonic CKM-like
mixing matrix). Thus, the total number of free parameters in a
\qq{realistic} present-day variant of electroweak SM is 24.

\begin{figure}[h]\centering
\begin{tabular}{ccc}
\s{\includegraphics{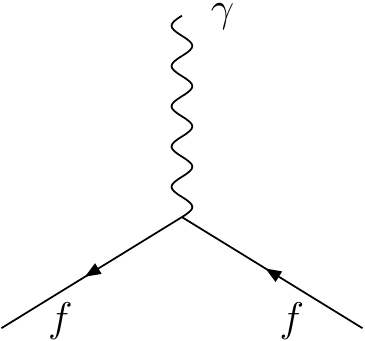}}&\hspace{0.5cm}\s{\includegraphics{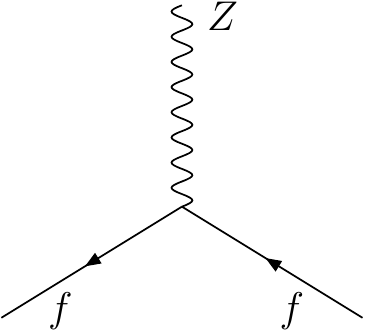}}
&\hspace{0.5cm}\s{\includegraphics{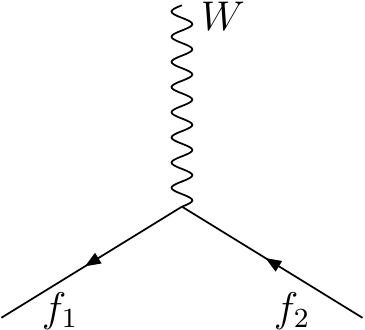}}\\[0.35cm]
\s{\includegraphics{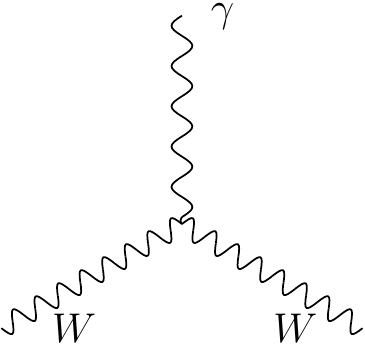}}&\hspace{0.5cm}\s{\includegraphics{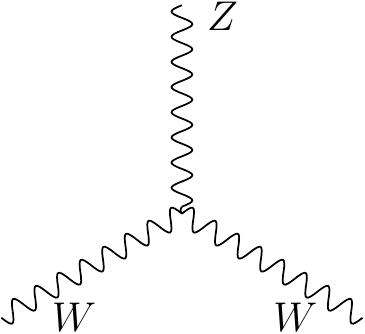}}
&\hspace{0.5cm}\s{\includegraphics{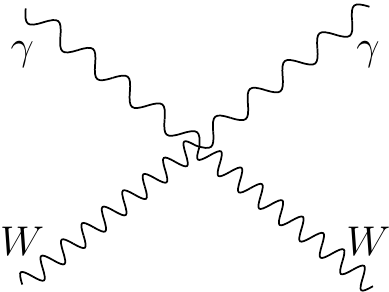}}\\[0.35cm]
\s{\includegraphics{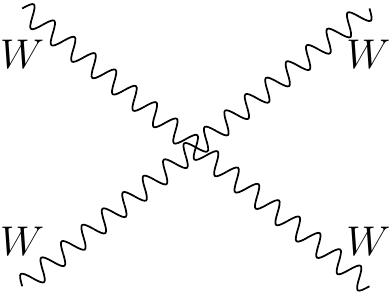}}&\hspace{0.5cm}\s{\includegraphics{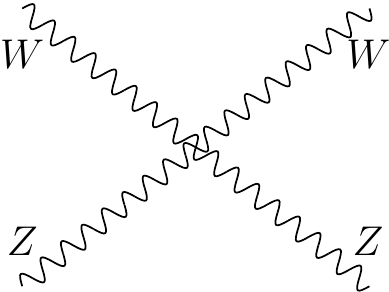}}
&\hspace{0.5cm}\s{\includegraphics{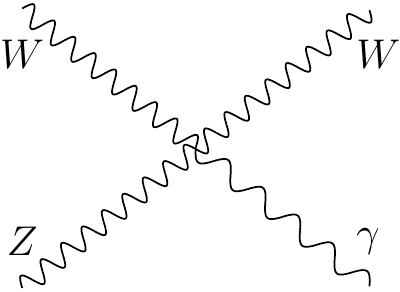}}\\[0.35cm]
\s{\includegraphics{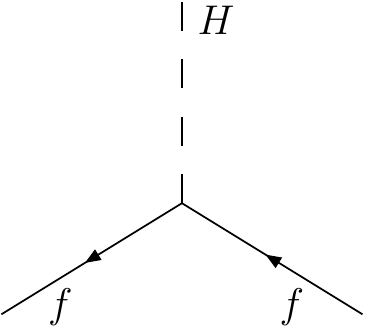}}&\hspace{0.5cm}\s{\includegraphics{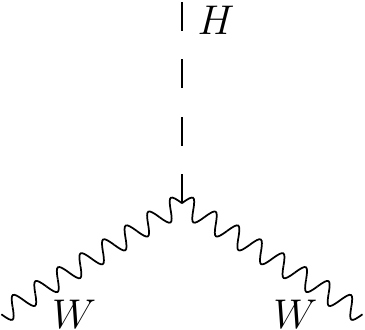}}
&\hspace{0.5cm}\s{\includegraphics{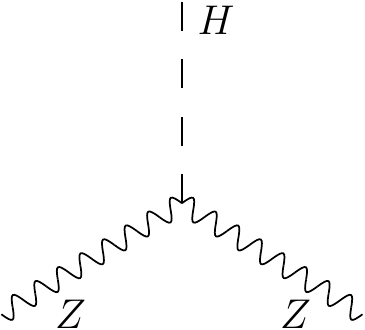}}\\[0.35cm]
\s{\includegraphics{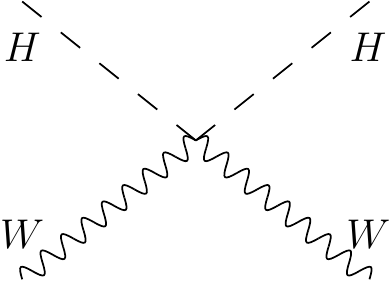}}&\hspace{0.5cm}\s{\includegraphics{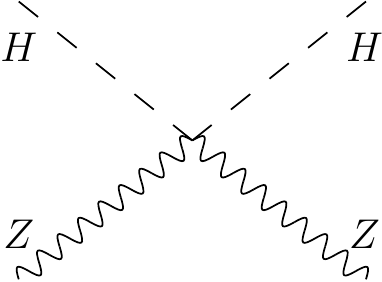}}
&\hspace{0.5cm}\s{\includegraphics{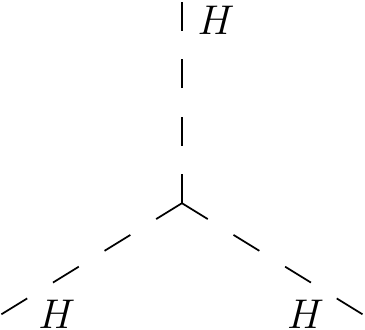}}\\[0.35cm]
\s{\includegraphics{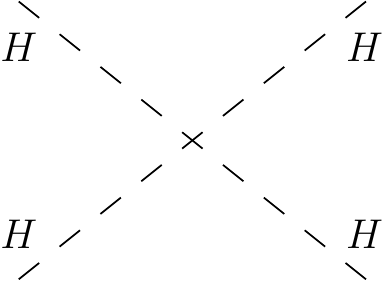}}
\end{tabular}
\caption{All types of interactions contained in the electroweak SM
Lagrangian. The set of vertices displayed here corresponds to the
physical $U$-gauge. The label $f$ refers generally to any relevant
fermion species. In the CC weak interaction vertex the $f_1$,
$f_2$ denote symbolically either $\nu_\ell$, $\ell$ or a pair of
quarks with charges differing by one
unit.}\label{fig31}\index{Feynman diagrams!vertices of the
Standard Model}
\end{figure}

Finally, the reader may find it instructive to see all the
interactions contained in (\ref{eq7.242}) depicted, schematically,
as the corresponding Feynman-graph vertices. Such a collection is
displayed in Fig.\,\ref{fig31}. In this context, it is useful to
realize that despite the common label \qq{Standard Model} used for
the GWS electroweak theory, detailed experimental tests are currently not available for all interaction vertices shown here. The present-day situation can be roughly summarized as follows. Interactions of $W$ and $Z$ with fermions (both leptons and quarks) are tested with good accuracy. Similarly, the couplings $WW\gamma$ and $WWZ$ have been already tested well at the facility LEP (Large Electron Positron Collider) at CERN. In contrast to this, the quartic self-interactions of vector bosons are tested rather poorly (some events corresponding to the $WW$ scattering have been detected only recently at LHC). Concerning the Higgs boson interactions, there are some experimental data for the couplings $WWH$, $ZZH$, $ttH$, $bbH$ and $\tau\tau H$. Couplings of $H$ to light fermions (including $c$-quark) are at present experimentally inaccessible, as well as the quartic interactions $WWHH$ and $ZZHH$. Higgs boson self-interactions $HHH$ and $HHHH$ are still entirely untested, but represent great challenge for the forthcoming collider experiments. The point is that such measurements would be crucial for a definitive identification of the currently known Higgs-like particle $H$ as the Higgs boson of SM. Thus, taking into account the full number of elementary fermions (and the related number of the elements of the CKM matrix) one may say that, roughly, about one third of the interaction vertices in Fig. 32 remain untested up to now.

Anyway, in view of the stunning phenomenological success of the
GWS theory since 1970s until present day, the label \qq{Standard
Model} is well justified and understandable (though the term
\qq{Standard Theory} would perhaps be more pertinent) and it seems
to be clear that SM will remain a \qq{textbook}
effective theory of electroweak interactions valid up to energy
scale of $O(100\ \GeV)$. An excellent survey of the SM physics (in
particular, with regard to the performance of the Large Electron
Positron collider (LEP) at CERN) can be found in \cite{Ven}. 
Further, as indicated above, during the past two decades great progress was made in the experiments on several facilities, most prominent being LHC at CERN. These achievements are covered in considerable detail e.g. in the monographs \cite{Alt}, \cite{Lan} and, of course, the full overview of the current data can be found in \cite{ref5}. It turns out that up to now all available experimental results confirm the validity of SM (though some ``smoking guns'' occur, encouraging the permanent quest for new physics beyond SM). Some popular theory models going far beyond present-day SM are described e.g. in \cite{Lan} and \cite{Pal}.

\index{Yang--Mills field|)}\index{U-gauge@$U$-gauge|)}
%\end{document}

\clearpage

%%%%%%%%%%%%%%%%%%%%%%%%%%%%%%%%%%%%%%%%%%%%%%%%%%%%%%%%%%%%%%%%%%%
%%%%%%%%%%%%%%%%%%%%%%%%%%%%%%%%%%%%%%%%%%%%%%%%%%%%%%%%%%%%%%%%%%%%%%%%%%%%%%%%%%%%%%%%%%%%%%%%%%%%%%%%%%%%%%%%%%%%%%%%%%%%%%%%%%%%%%%%

%\input{problems7}
%%%%%%%%%%%%%%%%%%%%%%%%%%%%%%%%%%%%%%%%%%%%%%%%%%%%%%%%%%%%%%%%%%%
%%%%%%%%%%%%%%%%%%%%%%%%%%%%%%%%%%%%%%%%%%%%%%%%%%%%%%%%%%%%%%%%%%%%%%%%%%%%%%%%%%%%%%%%%%%%%%%%%%%%%%%%%%%%%%%%%%%%%%%%%%%%%%%%%%%%%%%%

\begin{priklady}{11}

\item  
Evaluate lepton and hadron decay widths of the $W$ boson\index{decay!of the W@of the $W$ boson} within SM (in the tree approximation). Needless to say, one may assume complete hadronization of final-state quarks, so that the hadronic width is to be calculated by summing the $W$ decay rates involving all relevant quark-antiquark pairs. Note that to a good accuracy one may neglect masses of quarks $u$, $d$, $s$, $c$ and $b$ (as well as the lepton masses), since these are much smaller than $m_W$. Show that 
$$
\Gamma(W \to \text{leptons})  = 3\Gamma_0\,,\qquad
\Gamma(W \to \text{quarks}) = 6\Gamma_0
$$
where
$$
\Gamma_0  = \frac{1}{6\pi\sqrt2} G_F m_W^3
$$
(in this way, one sees that the ratio of the hadronic and leptonic widths is 2 : 1, in good agreement with experimental data). Using the known values of $G_F$ and $m_W$, one thus gets
$$
\Gamma_W  = \Gamma(W\to\text{all}) =  9 \Gamma_0 \doteq 2.1\,\text{GeV}
$$
{\it Hint:} For the calculation of the full decay width into quarks, the unitarity of the CKM mixing matrix is to be utilized. Further, one should not forget to include the colour factor $N_c = 3$.

\item Evaluate lepton and hadron decay widths of the $Z$ boson\index{decay!of the Z@of
the $Z$ boson} within SM (in tree approximation). Neglecting the relevant fermion masses, show first that for an individual decay $Z \to f \bar f$) one gets
$$
\Gamma(Z\to f\bar f) = \frac{G_F m_Z^3}{6\pi\sqrt2}(v_f^2+a_f^2)
$$
with the coupling factors $v_f$, $a_f$ being given by (\ref{eq7.249}) (cf. also the Problem \ref{pro52}).  Concerning the inclusive decay rates, one is supposed to recover the formulae (see also \cite{Pas})
\begin{align*}
    &\Gamma(Z\to \text{leptons}) = \frac{G_F m_Z^3}{3\pi\sqrt2}\Bigl(\frac32 - 3\sin^2\theta_W + 6 \sin^4\theta_W \Bigr)\\
    &\Gamma(Z\to \text{quarks}) = \frac{G_F m_Z^3}{3\pi\sqrt2}\Bigl(\frac{15}{4} - 7\sin^2\theta_W + \frac{22}{3} \sin^4\theta_W \Bigr)
\end{align*}
so that
$$
\Gamma_Z = \Gamma(Z\to \text{all}) = \frac{G_F m_Z^3}{3\pi\sqrt2}\Bigl(\frac{21}{4} - 10\sin^2\theta_W + \frac{40}{3}\sin^4\theta_W \Bigr)
$$
Employing the numerical value $\sin^2\theta_W \approx 0.23$, one may then check that the hadronic and leptonic decays constitute roughly 70\% and 30\% of the full $Z$ width respectively. Note that utilizing also the known values of $G_F$ and $m_Z$, one is led to an approximate prediction for $\Gamma_Z$, which reads $\Gamma_Z \approx 2.4\,\text{GeV}$ (an attentive reader may thus observe readily that the results for $\Gamma_W$ and $\Gamma_Z$  correspond to mean lifetimes of $W$ and $Z$ of the order of $10^{-25}$~s).

\item  Calculate the width (and mean lifetime) of the top quark\index{top}.
What is its dominant decay mode?

\item Concerning the decay $t \to b + W^+ $, it is also interesting to consider the production of longitudinal and transverse $W$ separately. Evaluate the ratio of the decay rates in question and show that
$$
\frac{\Gamma(t\to b\, W_L)}{\Gamma(t\to b\, W_T)} = \frac{m_t^2}{2 m_W^2}
$$
in the approximation $m_b = 0$.

\item  Compute the electron energy spectrum in the $b$-quark decay
$b\rightarrow c+e^- + \bar{\nu}_e$. Similarly, calculate the
energy spectrum for positron in $c\rightarrow s+e^+
+\bar{\nu}_e$.\\{\it Hint:} For such low-energy processes one can
employ an effective four-fermion Lagrangian involving charged
$V-A$ currents. Throughout the calculation, neglect the electron
mass.

\item Show that the SM tree-level amplitude for the process
$d+\bar{s} \rightarrow W^+ W^-$ behaves well in the high-energy
limit.

\item \label{pro7Compute} Compute the cross section of the process $e^+ e^- \rightarrow
\mu^+\mu^-$ in the vicinity of the $Z$ resonance, i.e. for
$s=E_{c.m.}^2$ close to $m_Z^2$. Calculate also the corresponding
forward-backward asymmetry\index{forward-backward asymmetry}
$A_{FB}$ (for relevant definitions, see the Problem~\ref{pro5.6} at the end
of Chapter~\ref{chap5}). Throughout the calculation, employ the
Breit--Wigner form for denominator of the $Z$ boson propagator,
i.e. replace the expression $(q^2-m_Z^2)^{-1}$ with $(q^2-m_Z^2+i
m_Z \Gamma)^{-1}$, where $\Gamma$ stands for the $Z$ total width
(concerning this, see e.g. \cite{PeS}, Section~\ref{sec7.3}). Show that the
interference cross section $\sigma_{\gamma Z}$ vanishes for
$s=m_Z^2$ and the $\sigma_Z$ (i.e. the $Z$-exchange contribution)
becomes
\begin{align*}
\sigma_Z \Bigl\rvert_{s=m_Z^2} &= 12\pi \frac
{\Gamma(Z\rightarrow e^+e^-) \Gamma(Z\rightarrow \mu^+ \mu^-)}{m_Z^2 \Gamma^2}\\
&= \frac{12\pi}{m_Z^2}\,\text{BR}(Z\rightarrow e^+e^-)
\text{BR}(Z\rightarrow \mu^+\mu^-)
\end{align*}
Using the familiar formula
$$
\sigma_\gamma \doteq \frac{4\pi\alpha^2}{3s}
$$
for the photon-exchange contribution at high energy, evaluate the
ratio $\sigma_Z/\sigma_\gamma$ for $s=m_Z^2$.

Further, one can define the cross section for $Z$ boson production
in $e^+e^-$ annihilation as
$$
\sigma(e^+e^-\rightarrow Z) = \sum_f \sigma_Z (e^+e^- \rightarrow
f\bar{f})
$$
where the sum runs over all fermions for which the decay channel
$Z\rightarrow f\bar{f}$ is open. Using the preceding results, it
is easy to see that
$$
\sigma(e^+e^-\rightarrow Z) = \frac{12\pi}{m_Z^2}
\text{BR}(Z\rightarrow e^+e^-)
$$
What is the numerical value of the ratio $\sigma(e^+e^-\rightarrow
Z)/\sigma_\gamma(s=m_Z^2)$? If the luminosity of an
electron-positron collider is $10^{32}
\text{cm}^{-2}\text{s}^{-1}$, how many $Z$ bosons are then
produced within one year?\\{\it Hint:} Remember that one year has
approximately $\pi\times 10^7\ \text{s}$\index{Z boson@$Z$
boson|)}. {\mbox{\fontencoding{U}\fontfamily{wasy}\selectfont
\char44}}%\epsfig{figure=smiley.eps,width=10pt}

\item  Imagine that you are in the position of a particle physics aficionado who on July 4, 2012 reads in news headlines about the discovery of a Higgs-like boson with mass of $125\ \GeV$. Would you then be able
to predict, at least roughly, its lifetime?

\item  Once again, suppose that the SM Higgs boson has the mass $m_H
\approx 125\ \GeV$. In analogy with contents of the Problem~\ref{pro7Compute}, perform
an analysis of the cross section for $\sigma(e^+e^-\rightarrow
f\bar{f})$ in the vicinity of the Higgs boson resonance, i.e. for
$s=E_{c.m.}^2$ close to $m_H^2$. Evaluate the ratios
$\sigma(e^+e^-\rightarrow H)/\sigma_\gamma$ and
$\sigma(e^+e^-\rightarrow H)/\sigma_Z$, with the $\sigma_\gamma$
and $\sigma_Z$ taken at $s=m_H^2$.

\item Consider the production of a pair of Higgs bosons in electron-positron annihilation, i.e. the process $e^+ e^- \to H H$. Identify the corresponding tree diagrams, single out the one giving the dominant contribution, and evaluate the cross section as a function of the collision energy in the c.m. system. For an explicit numerical illustration, choose e.g. the energy $E_{c.m.} = s^{1/2} = 500$~GeV as a reference point. It should be obvious {\it a priori\/} that the resulting value of such a tree-level cross section must be extremely small, because of the suppression factor $m_e/m_W$ due to the $eeH$ coupling. Indeed, the reader is supposed to find out, by means of an explicit calculation, that the value in question is of the order of $10^{-24}$ barn (i.e. yoctobarn). Actually, the considered process is a curious example of a situation where the higher order (one-loop) diagrams give much larger contribution than the tree-level ones. The interested reader is encouraged to figure out what the relevant SM one-loop diagrams could be. It turns out that, at the one-loop level, the relevant cross section may be of the order of $10^{-17}$ barn (i.e. $10^{-2}$ femtobarn). For details, see e.g. the papers \cite{Djouadi:1996hp}, \cite{Lopez-Villarejo:2008qii} and references therein.

\item Show that amplitudes of the processes $W^- W^+ \to H \gamma$ and $W^-  W^+ \to H Z$ satisfy the condition of tree unitarity.

\item The observation of the rare decay $H \to \gamma\gamma$ was one of the first experimental signals marking the discovery of the Higgs boson. As we know, within the SM Lagrangian there is no direct $H\gamma\gamma$ interaction, so that the process in question can only occur at the one-loop (and higher) level. The contribution of the relevant Feynman diagrams is free of the UV divergences, as one may anticipate in view of the perturbative renormalizability of SM. The reader is encouraged to demonstrate by means of an explicit calculation that the contribution of a purely fermionic triangle loop for $H \to \gamma\gamma$ is indeed UV finite (to this end, one may utilize the elementary techniques displayed in the Appendix~\ref{appenE}). Obviously, there are other two relevant loops that any observant reader is supposed to draw readily: a triangle and a bubble made of $W$ boson internal lines, involving the $WW\gamma$ and $WW\gamma\gamma$ couplings respectively. Their sum is UV finite as well, but the corresponding calculation is much more laborious; this may be left as a challenge for truly hard-working SM aficionados. Anyway, one may find a lot of detailed information concerning the decay process in question in the monograph \cite{Gun}. 

\end{priklady}

%%%%%%%%%%%%%%%%%%%%%%%%%%%%%%%%%%%%%%%%%%%%%%%%%%%%%%%%%%%%%%%%%%%
%%%%%%%%%%%%%%%%%%%%%%%%%%%%%%%%%%%%%%%%%%%%%%%%%%%%%%%%%%%%%%%%%%%%%%%%%%%%%%%%%%%%%%%%%%%%%%%%%%%%%%%%%%%%%%%%%%%%%%%%%%%%%%%%%%%%%%%%
\newpage\markboth{\MakeUppercase{Epilogue}}{\MakeUppercase{Epilogue}}
\setcounter{secnumdepth}{-1} 
\chapter{Epilogue}

The saga of the standard model of electroweak interactions is undoubtedly one of the most fascinating chapters of modern physics history. The road to the final form of SM had been a remarkable interplay of bold theoretical hypotheses and brilliant experiments that gradually confirmed the theory. Particularly impressive is the way how several new particles were successfully predicted: intermediate vector bosons $W$ and $Z$, the fourth quark $c$ (actually, also two extra quarks $b$ and $t$) and the enigmatic Higgs boson. In view of more than four decades of doubts and conceptual disputes concerning the nature of the electroweak symmetry breaking, the ultimate observation of the solitary scalar particle endowed with properties of the Higgs boson was perhaps one of the most astonishing discoveries of particle physics ever made. Thus, the GWS model of electroweak unification,  which was a highly speculative construction at the beginning of the 1970s, has finally become a widely recognized physically realistic theory of natural phenomena at a fundamental level. As we know, there are still some long-standing fundamental questions on the interface between particle physics and cosmology, which the standard model is not able to answer; another challenge is understanding a deeper unification of fundamental interactions (the time-honoured subject of ``grand unification''). Consequently, there is a lot of activity in the quest for physics beyond SM, both in theory and experiment. Anyway, the present-day SM represents one of the greatest achievements of modern physics, and for many years to come will certainly stay with us as a robust reference theory for evaluating the results of forthcoming experiments. 

\setcounter{secnumdepth}{2} 

\renewcommand{\TheAlphaChapter}{\thechapter}
\appendix

%\input{app_a}
%%%%%%%%%%%%%%%%%%%%%%%%%%%%%%%%%%%%%%%%%%%%%%%%%%%%%%%%%%%%%%%%%%%
%%%%%%%%%%%%%%%%%%%%%%%%%%%%%%%%%%%%%%%%%%%%%%%%%%%%%%%%%%%%%%%%%%%%%%%%%%%%%%%%%%%%%%%%%%%%%%%%%%%%%%%%%%%%%%%%%%%%%%%%%%%%%%%%%%%%%%%%
%\documentclass[11pt,tbtags]{book2}
%\input{pream}
%\begin{document}
%\appendix
\renewcommand{\chaptermark}[1]{\markright{\MakeUppercase{#1}}{}}
\chapter{Dirac equation and its solutions}\label{appenA}
\renewcommand{\sectionmark}[1]{\markboth{{\thechapter #1}}}
\fancyhead[CE]{\helv\MakeUppercase{\jmenokap\ \thechapter}}
\index{quantization of!Dirac field|appA}
\index{Weyl!spinor|appA}\index{Weyl!representation|appA}\index{trace
identities|appA} \index{Pauli matrices|appA}  \index{Majorana
representation|appA}
\index{Lorentz!group|appA}\index{Dirac!matrices|see{$\gamma$
matrices}|appA} \index{Dirac!equation|appA}\index{anticommutation
relations|appA}\index{metric tensor|appA}\index{gamma
matrices|appA}\index{Dirac!spinors|appA}\index{Levi--Civita
tensor|appA}\index{helicity|appA}\index{chirality|appA}\index{Gordon
identity|appA}\index{bilinear covariant
forms|appA}\index{Dirac!matrices|appA}

The use of Dirac equation in physics is twofold. Either it is
treated as relativistic quantum-mechanical equation for a
spin-$1/2$ particle, or it describes a classical bispinor field
(that is subsequently quantized in terms of spin-$1/2$ particles
and their antiparticles).\footnote{A terminological remark is
perhaps in order here. In general, bispinor (or Dirac spinor) is a
quantity that transforms according to the four-dimensional
representation $\bigl(\frac{1}{2},\,0\bigr)\oplus
\bigl(0,\,\frac{1}{2}\bigr)$ of the Lorentz group, i.e. it behaves
as a direct sum of two inequivalent two-component Weyl spinors. An
elementary introduction to the theory of relativistic spinors can
be found in most of the textbooks on field theory, see e.g.
\cite{Ryd}.} In what follows, we consider the case of free
particles (fields) -- this is just what is needed for the purposes
of perturbative quantum field theory (i.e. for the Feynman diagram
calculations).

The Dirac equation is written in the familiar covariant form as
\begin{equation}\label{eqA.1}
\bigl(i \gamma^\mu \partial_\mu - m\bigr) \psi(x) = 0
\end{equation}
where $m$ is a mass parameter and the coefficients $\gamma^\mu,\;
\mu=0,1,2,3$ are $4 \times 4$ matrices satisfying anticommutation
relations
\begin{equation}\label{eqA.2}
\{\gamma^\mu,\,\gamma^\nu\} \equiv \gamma^\mu\gamma^\nu +
\gamma^\nu \gamma^\mu = 2 g^{\mu\nu}
\end{equation}
Here $g^{\mu\nu}$ denotes a metric tensor in the flat
four-dimensional space-time; in our conventions, this is taken to
be
\begin{equation}\label{eqA.3}
g^{\mu\nu} = g_{\mu\nu} = \bm{1&\phantom{-}0&\phantom{-}0&\phantom{-}0\\
0&-1&\phantom{-}0&\phantom{-}0\\0&\phantom{-}0&-1&\phantom{-}0\\0&\phantom{-}0&\phantom{-}0&-1}
\end{equation}
(of course, the mixed components are
$g^\mu{}_\nu=\delta^\mu_\nu$). Needless to say, multiplication by
the $4\times 4$ unit matrix in the right-hand side of
(\ref{eqA.2}) is tacitly understood. Thus, (\ref{eqA.2}) means
that $\gamma^\mu\gamma^\nu=-\gamma^\nu\gamma^\mu$ for $\mu\neq
\nu$, $(\gamma^0)^2 =1$ and $(\gamma^j)^2=-1$ for $j = 1, 2, 3$
(we denote the $4 \times 4$ unit matrix simply as $1$).

It is natural to introduce also matrices $\gamma_\mu$,
defined by lowering formally the Lorentz labels of the $\gamma^\mu$, i.e.
\begin{equation}
\gamma_\mu = g_{\mu\nu} \gamma^\nu
\end{equation}
Using (\ref{eqA.3}), one thus has
\begin{equation}\label{eqA.5}
\gamma_0 = \gamma^0,\quad \gamma_j = -\gamma^j \qquad\quad
\text{for}\; j=1,2,3
\end{equation}
A standard ingredient of the relevant notation is the \qq{slash} symbol
$\slashed{a}$ defined for any four-vector $a$ as
\begin{equation}
\slashed{a}=a_\mu \gamma^\mu = a^\mu\gamma_\mu
\end{equation}
Employing this, the Dirac equation can be recast as
\begin{equation}
\bigl(i\slashed{\partial} -m\bigr) \psi(x) = 0
\end{equation}
with $\slashed{\partial}=\gamma^\mu \partial_\mu = \gamma_\mu \partial^\mu$.

In general, Dirac matrices must have some specific properties
under hermitean conjugation; for our conventional choice of the
$g^{\mu\nu}$ one has\footnote{For the metric with opposite
signature, i.e. for $g^{\mu\nu}=\text{diag}(-1,\,1,\,1,\,1)$, the
$\gamma^j$, $j=1,2,3$ would be hermitean and $\gamma^0$
anti-hermitean.}
\begin{equation}
\gamma_0^\dagger = \gamma_0,\qquad \gamma_j^\dagger = - \gamma_j
\end{equation}
Obviously, this can be written compactly as
\begin{equation}
\gamma_\mu^\dagger = \gamma_0 \gamma_\mu \gamma_0
\end{equation}
In this context, it is natural to introduce another standard
symbol, namely that of Dirac conjugation: for a $\psi$, the
conjugate spinor $\psib$ is defined by
\begin{equation}
\psib = \psi^\dagger \gamma_0
\end{equation}
A simple consequence of such a definition is that the $\psib(x)$ satisfies the equation
\begin{equation}
\psib(x)\bigl(i\slashed{\partial} + m\bigr) =0
\end{equation}
if $\psi(x)$ is a solution of (\ref{eqA.1}).

For various purposes, it is highly useful to introduce an
additional matrix denoted as $\gamma_5$, which anticommutes with
all $\gamma^\mu$, $\mu=0,1,2,3$. In view of the anticommutativity
of different $\gamma^\mu$, it is obvious that the product
$\gamma^0\gamma^1\gamma^2\gamma^3$ has the desired property;
conventionally, we shall define the $\gamma_5$ as
\begin{equation}
\label{eqA.12}
\gamma_5 = i \gamma^0 \gamma^1 \gamma^2 \gamma^3
\end{equation}
Then
\begin{equation}
\gamma^\dagger_5 = \gamma_5,\qquad (\gamma_5)^2 = 1
\end{equation}
Taking into account (\ref{eqA.5}), one can also recast (\ref{eqA.12}) as
\begin{equation}
\gamma_5 = -i \gamma_0 \gamma_1 \gamma_2 \gamma_3
\end{equation}
Let us stress that we do not introduce two different symbols $\gamma_5$ and $\gamma^5$, as it would make little practical sense.

Having displayed basic definitions and some elementary facts concerning the gamma matrices, we should also recall that the $\gamma^\mu$ are simply related to the matrices $\beta$, $\alpha^j$, $j = 1, 2, 3$ introduced originally by Dirac; one has
\begin{equation}
\gamma^0 = \beta,\qquad \gamma^j = \beta\alpha^j
\end{equation}
or, inverting the last relation,
\begin{equation}
\alpha^j = \gamma^0\gamma^j
\end{equation}
The \qq{old} matrices $\beta$, $\alpha^j$ are all hermitean and
appear in the Schr\"odinger-like form of the Dirac equation (in
which Lorentz covariance is not \qq{manifest}), namely
\begin{equation}\label{eqA.17}
i\partial_0\psi(x) = \bigl(-i\vec{\alpha}\cdot\vec{\nabla}+\beta
m\bigr) \psi(x)
\end{equation}
Under a Lorentz transformation of space-time coordinates $x'=\Lambda x$, the Dirac spinor $\psi$ in (\ref{eqA.1}) is transformed as
\begin{equation}
\label{eqA.18}
\psi'(x') = S(\Lambda)\psi(x)
\end{equation}
where $S(\Lambda)$ is a non-singular $4 \times 4$ matrix
fulfilling the condition\footnote{Note that (\ref{eqA.19})
reflects the relativistic covariance of Dirac equation, which
means that if  $\psi$ is a solution of eq. (\ref{eqA.1}), the
$\psi'$ defined by (\ref{eqA.18}) satisfies the same equation,
written in primed coordinates.}
\begin{equation}
\label{eqA.19} S^{-1}(\Lambda)\gamma^\mu S(\Lambda) =
\Lambda^\mu{}_\nu \gamma^\nu
\end{equation}
An explicit general form of the $S(\Lambda)$ can be found in any
textbook on relativistic quantum theory, but it will not be needed
for our present purposes. We recall here at least some of its
important properties. First, it holds\footnote{Thus,
(\ref{eqA.20}) indicates that, in general, the $S(\Lambda)$ is not
unitary. In fact, $S(\Lambda)$ is unitary for spatial rotations
and hermitean for pure Lorentz boosts.}
\begin{equation}
\label{eqA.20}
S^{-1} = \gamma_0 S^\dagger \gamma_0
\end{equation}
Its immediate consequence is a simple transformation law for
conjugate Dirac spinor: if $\psi'(x')=S\psi(x)$, then
\begin{equation}\label{eqA.21}
\psib'(x') = \psib(x)S^{-1}
\end{equation}
Further, note that the $S(\Lambda)$ has a particularly simple form
for the space inversion $\cal P$, i.e. for $x'=x_{\cal P} =
(x^0,\,-\vec{x})$ (obviously, the corresponding $\Lambda$ is
$\Lambda_{\cal P}=\text{diag}(1,\,-1,\,-1,\,-1)$). In that case,
one has
\begin{equation}
\psi_{\cal P}(x_{\cal P}) = \gamma_0 \psi(x)
\end{equation}
Finally, let us add that the covariance relation (\ref{eqA.19}) has a simple counterpart involving the $\gamma_5$, namely
\begin{equation}\label{eqA.23}
S^{-1}(\Lambda)\gamma_5 S(\Lambda) =\det{\Lambda}\ \gamma_5
\end{equation}
(remember that $\det\Lambda=+1$ for Lorentz boosts and spatial rotations, while $\det\Lambda = -1$ for space inversion).

There are infinitely many realizations of the anticommutation
relations (\ref{eqA.2}) in terms of $4 \times 4$ matrices, but it
turns out that they are all equivalent. If $\gamma^\mu$ and
$\gamma^\mu{}'$ are two sets satisfying (\ref{eqA.2}), then there
is a non-singular matrix $U$ such that $\gamma^\mu{}'=U\gamma^\mu
U^{-1}$; moreover, if both $\gamma^\mu$ and $\gamma^\mu{}'$ have
the above-mentioned hermiticity properties, the $U$ is unitary
(for a proof of this non-trivial statement, see e.g. \cite{Mes}).
For an illustration, let us display three frequently used
representations. The standard (or Dirac) representation reads
\begin{equation}\label{eqA.24}
\gamma^0 = \bm{\J&0\\0&-\J},\quad
\gamma^j = \bm{0&\sigma_j\\-\sigma_j&0}
\end{equation}
where $\J$ is the $2 \times 2$ unit matrix and $\sigma_j$, $j = 1, 2, 3$ are Pauli matrices
\begin{equation}
\sigma_1 = \bm{0&1\\1&0},\quad \sigma_2=\bm{0&-i\\i&0},\quad \sigma_3=\bm{1&0\\0&-1}
\end{equation}
Consequently,
\begin{equation}
\gamma_5 = \bm{0&\J\\ \J&0}
\end{equation}
and
\begin{equation}
\alpha^j = \bm{0&\sigma_j\\ \sigma_j&0}
\end{equation}
Next, the so-called chiral (also spinor, or Weyl) representation is defined as
\begin{equation}
\gamma^0_\ti{chiral} = \bm{0&\J\\ \J &0},\quad
\gamma^j_\ti{chiral} = \bm{0&-\sigma_j\\ \sigma_j&0}
\end{equation}
Then
\begin{equation}
(\gamma_5)_\ti{chiral} = \bm{\J&0\\0&-\J}
\end{equation}
Finally, the Majorana representation
\begin{alignat}{2}
\gamma^0_\ti{Majorana}&= \bm{0&\sigma_2\\ \sigma_2&0},\quad
&\gamma^1_\ti{Majorana}&= \bm{i\sigma_3&0\\ 0&i\sigma_3}\notag\\
\gamma^2_\ti{Majorana}&= \bm{0&-\sigma_2\\ \sigma_2&0},\quad
&\gamma^3_\ti{Majorana}&= \bm{-i\sigma_1&0\\0&-i\sigma_1}
\label{eqA.30}
\end{alignat}
consists of purely imaginary matrices (it means that Dirac
equation has only real coefficients in such a representation). In
fact, the explicit form (\ref{eqA.30}) corresponds to simple
expressions made of standard Dirac matrices (\ref{eqA.24}), namely
\begin{alignat}{2}
\gamma^0_\ti{Majorana}&=\gamma^0\gamma^2,\qquad
&\gamma^1_\ti{Majorana}&=-\gamma^1\gamma^2\notag\\
\gamma^2_\ti{Majorana}&=-\gamma^2,\qquad
&\gamma^3_\ti{Majorana}&=\gamma^2\gamma^3
\end{alignat}
For completeness, let us specify the equivalence transformations between the above-mentioned representations. One has
\begin{equation}
\gamma^\mu_\ti{chiral} = U\gamma^\mu U^{-1}
\end{equation}
where
\begin{equation}
U = U^\dagger = U^{-1} = \frac{1}{\sqrt{2}}(\gamma_0+\gamma_5) = \frac{1}{\sqrt{2}}\bm{\J&\J\\ \J&-\J}
\end{equation}
(the gamma matrices in the last expression are taken in the standard representation) and
\begin{equation}
\gamma^\mu_\ti{Majorana} = V\gamma^\mu V^{-1}
\end{equation}
where
\begin{equation}
V=V^\dagger = V^{-1} = \frac{1}{\sqrt{2}}\gamma^0 (1+\gamma^2) = \frac{1}{\sqrt{2}}\bm{\J&\sigma_2\\ \sigma_2&-\J}
\end{equation}
In the present text we employ only the standard representation (\ref{eqA.24}).

Dirac matrices are endowed with many remarkable properties that hold independently of a specific representation. Some of the relevant identities are summarized below. Let us start with a series of \qq{sandwich} relations
\begin{align}
\gamma_\alpha \gamma^\alpha &=4 \notag\\
\gamma_\alpha \gamma_\mu \gamma^\alpha &= -2\gamma_\mu \notag\\
\gamma_\alpha \gamma_\mu \gamma_\nu \gamma^\alpha &= 4 g_{\mu\nu} \notag\\
\gamma_\alpha \gamma_\mu \gamma_\nu \gamma_\rho \gamma^\alpha &= - 2\gamma_\rho \gamma_\nu \gamma_\mu
\end{align}
etc., which follow easily from the basic anticommutation relation
(\ref{eqA.2}) (when checking the above identities, don't forget
that $g^\alpha{}_\alpha=4$).

Further, there is a set of formulae for traces of products of gamma matrices. First of all, trace of the product of an arbitrary odd number of $\gamma^\mu$'s is identically zero, i.e.
\begin{equation}\label{eqA.37}
\Tr(\gamma_{\mu_1} \cdots \gamma_{\mu_{2k+1}}) = 0
\end{equation}
(for proving this, the existence of the fully anticommuting
$\gamma_5$ satisfying $(\gamma_5)^2=1$ is instrumental). For
products involving an even number of $\gamma^\mu$'s one has, in
particular,
\begin{align}
\Tr(\gamma_\mu\gamma_\nu) &= 4g_{\mu\nu} \notag\\
\Tr(\gamma_\mu\gamma_\nu\gamma_\rho\gamma_\sigma) &= 4 (g_{\mu\nu}
g_{\rho\sigma} - g_{\mu\rho}g_{\nu\sigma} + g_{\mu\sigma}
g_{\nu\rho}) \label{eqA.38}
\end{align}
These relations, as well as their eventual extensions for longer
chains of Dirac matrices, can be derived systematically by using
(\ref{eqA.2}) and the cyclicity property of traces, i.e. $\Tr(AB)
= \Tr(BA)$. Of course, the universal factor $4$ appearing in
(\ref{eqA.38}) is due to the trace of $4 \times 4$ unit matrix
(acting in the four-dimensional space of Dirac
spinors).\footnote{Note also that traces of products of Dirac
matrices behave, in general, as tensors under Lorentz
transformations (this is a simple consequence of trace cyclicity
and the covariance relation (\ref{eqA.19})). At the same time,
they consist of pure numbers and therefore can depend only on
components of metric tensor. Such an argument provides a useful
additional insight into the algebraic structure of
(\ref{eqA.38}).}

Similarly, there is a series of formulae for traces involving also
the $\gamma_5$. In particular,
\begin{align}
\Tr(\gamma_5) &= 0 \notag\\
\Tr(\gamma_\mu\gamma_\nu\gamma_5) &=0\notag\\
\Tr(\gamma_\mu\gamma_\nu\gamma_\rho\gamma_\sigma\gamma_5) &= 4 i
\epsilon_{\mu\nu\rho\sigma}\label{eqA.39}
\end{align}
where $\epsilon_{\mu\nu\rho\sigma}$ is the totally antisymmetric
Levi-Civita symbol; in our conventions, $\epsilon_{0123}=+1$.
Again, there is a tensor argument for the algebraic structure of
the last relation in (\ref{eqA.39}). Taking into account
(\ref{eqA.19}) together with (\ref{eqA.23}), it is seen that the
trace in question is a (purely numerical) pseudotensor under
Lorentz transformations; however, the only numerical four-index
pseudotensor in four space-time dimensions is just the Levi-Civita
symbol. Thus, the last trace in (\ref{eqA.39}) can only be
proportional to the $\epsilon_{\mu\nu\rho\sigma}$ (for the same
reason, the first two traces must vanish as there is no
possibility to make a pseudoscalar or a two-index pseudotensor out
of $\epsilon_{\mu\nu\rho\sigma}$ and the metric tensor). Traces of
longer chains of the type (\ref{eqA.39}) can be expressed as
linear combinations of appropriate products of the
$g_{\alpha\beta}$ and $\epsilon_{\mu\nu\rho\sigma}$ (one such
example is shown in (\ref{eqA.49})).

Finally, for the sake of completeness one should mention another
general trace identity, namely
\begin{equation*}
\Tr(\gamma_\alpha\gamma_\beta\cdots \gamma_\tau\gamma_\omega) =
\Tr(\gamma_\omega\gamma_\tau\cdots \gamma_\beta\gamma_\alpha)
\end{equation*}

In this context, let us list some general relations for products
of two Levi-Civita tensors, which are highly useful in
calculations involving Lorentz pseudotensors. The \qq{master
formula} reads
\begin{equation}\label{eqA.40}
\epsilon^{\iota\kappa\lambda\mu}\epsilon_{\rho\sigma\tau\omega} =
-\begin{vmatrix}
\delta^\iota_\rho&\delta^\iota_\sigma&\delta^\iota_\tau&\delta^\iota_\omega\\
\delta^\kappa_\rho&\delta^\kappa_\sigma&\delta^\kappa_\tau&\delta^\kappa_\omega\\
\delta^\lambda_\rho&\delta^\lambda_\sigma&\delta^\lambda_\tau&\delta^\lambda_\omega\\
\delta^\mu_\rho&\delta^\mu_\sigma&\delta^\mu_\tau&\delta^\mu_\omega\\
\end{vmatrix}
\end{equation}
(note that such a result is quite natural a priori, since the
product of two pseudotensors must be a true tensor and the
determinant maintains automatically the required antisymmetry).
Contractions of Lorentz indices in the left-hand side of
(\ref{eqA.40}) yield
\begin{equation}
\epsilon^{\iota\kappa\lambda\omega}\epsilon_{\rho\sigma\tau\omega}
= -\begin{vmatrix}
\delta^\iota_\rho&\delta^\iota_\sigma&\delta^\iota_\tau\\
\delta^\kappa_\rho&\delta^\kappa_\sigma&\delta^\kappa_\tau\\
\delta^\lambda_\rho&\delta^\lambda_\sigma&\delta^\lambda_\tau\\
\end{vmatrix}
\end{equation}
and, in particular,
\begin{equation}\label{eqA.42}
\epsilon^{\iota\kappa\tau\omega}\epsilon_{\rho\sigma\tau\omega} =
-2\begin{vmatrix}
\delta^\iota_\rho&\delta^\iota_\sigma\\
\delta^\kappa_\rho&\delta^\kappa_\sigma\\
\end{vmatrix}= -2 (\delta^\iota_\rho
\delta^\kappa_\sigma-\delta^\iota_\sigma \delta^\kappa_\rho)
\end{equation}
From (\ref{eqA.42}) one then gets readily
\begin{equation}
\epsilon^{\iota\sigma\tau\omega}\epsilon_{\rho\sigma\tau\omega} =
- 6\delta^\iota_\rho
\end{equation}
There is another useful relation (of a completely different type),
which is worth mentioning here:
\begin{equation}\label{eqA.44}
g_{\lambda\mu}\epsilon_{\nu\rho\sigma\tau}
-g_{\lambda\nu}\epsilon_{\mu\rho\sigma\tau}
+g_{\lambda\rho}\epsilon_{\mu\nu\sigma\tau}
-g_{\lambda\sigma}\epsilon_{\mu\nu\rho\tau}
+g_{\lambda\tau}\epsilon_{\mu\nu\rho\sigma}=0
\end{equation}
(note that this identity comes out easily when working out the
expression
$\Tr(\gamma_\lambda\gamma_\mu\gamma_\nu\gamma_\rho\gamma_\sigma\gamma_\tau\gamma_5)$
by using (\ref{eqA.2}), the $\gamma_5$ anticommutativity and trace
cyclicity).

A highly useful technical device of \qq{diracology} is a special
basis in the 16-dimensional space of all $4 \times 4$ matrices,
which is made of appropriate products of the $\gamma^\mu$'s. The
\qq{canonical} choice is
\begin{equation}\label{eqA.45}
\Gamma_S=1,\ \Gamma_V=\gamma_\mu,\ \Gamma_T=\sigma_{\mu\nu},\
\Gamma_A=\gamma_5\gamma_\mu,\ \Gamma_P=\gamma_5
\end{equation}
with $\mu,\nu=0,1,2,3$; the $\sigma_{\mu\nu}$ is defined as
\begin{equation}\label{eqA.46}
\sigma_{\mu\nu} = \frac{i}{2}[\gamma_\mu,\gamma_\nu]
\end{equation}
Total number of the matrices (\ref{eqA.45}) can be checked
immediately; one gets $1 + 4 + 6 + 4 + 1 = 16$ (obviously, there
are only six linearly independent matrices $\sigma_{\mu\nu}$
because of antisymmetry, $\sigma_{\mu\nu}=-\sigma_{\nu\mu}$). The
indices $S$, $V$, $T$, $A$, $P$ stand for scalar, vector, tensor,
axial-vector (pseudovector) and pseudoscalar respectively, and
they refer to the transformation properties of bilinear quantities
obtained by sandwiching the matrices (\ref{eqA.45}) between Dirac
spinors. In particular, let $\psi_1=\psi_1(x)$ and
$\psi_2=\psi_2(x)$ be two Dirac spinors; then the expressions
\begin{equation}
\psib_1\psi_2,\quad \psib_1\gamma_\mu\psi_2,\quad
\psib_1\sigma_{\mu\nu}\psi_2,\quad
\psib_1\gamma_5\gamma_\mu\psi_2,\quad \psib_1\gamma_5\psi_2
\end{equation}
behave consecutively as a scalar, vector, antisymmetric tensor,
axial vector and pseudoscalar under a Lorentz transformation. This
is proved easily if one takes into account the transformation laws
(\ref{eqA.18}), (\ref{eqA.21}) and the relations (\ref{eqA.19}),
(\ref{eqA.23}).

It is not difficult to see that the matrices $\Gamma_j$, $j = S,
V, T, A, P$ have the properties
\begin{equation}\label{eqA.48}
(\Gamma_j)^2=\pm 1, \quad \Tr(\Gamma_j\Gamma_k) =0 \qquad
\text{for}\ j\neq k
\end{equation}
Obviously, the relations (\ref{eqA.48}) are instrumental for
calculating the expansion coefficients of a general matrix in the
basis (\ref{eqA.45}). As a simple application, one can derive the
following formula for the product of three Dirac matrices:
\begin{equation}\label{eqA.49s}
\gamma_\lambda\gamma_\mu\gamma_\nu = (g_{\lambda\mu}g_{\nu\rho} -
g_{\lambda\nu}g_{\mu\rho} +g_{\lambda\rho}g_{\mu\nu}) \gamma^\rho
+ i \epsilon_{\lambda\mu\nu\rho} \gamma_5 \gamma^\rho
\end{equation}
(proving the last identity is left to the reader as an instructive
exercise). Note that using this and the identity (\ref{eqA.44}),
one obtains easily the trace formula
\begin{equation}\label{eqA.49}\begin{split}
\Tr(\gamma_\lambda\gamma_\mu\gamma_\nu\gamma_\rho\gamma_\sigma\gamma_\tau\gamma_5)
= 4i \bigl(&g_{\lambda\mu} \epsilon_{\nu\rho\sigma\tau} -
g_{\lambda\nu} \epsilon_{\mu\rho\sigma\tau} + g_{\mu\nu}
\epsilon_{\lambda\rho\sigma\tau}\\
+&g_{\sigma\tau}\epsilon_{\lambda\mu\nu\rho} -
g_{\rho\tau}\epsilon_{\lambda\mu\nu\sigma}
+g_{\rho\sigma}\epsilon_{\lambda\mu\nu\tau}\bigr)
\end{split}\end{equation}

When calculating scattering cross sections and decay probabilities
within perturbative quantum field theory, one often encounters
products of two traces of the type (\ref{eqA.38}) and/or
(\ref{eqA.39}), contracted over Lorentz indices. Here are some
practical formulae that improve greatly the efficiency of
algebraic manipulations:
\begin{align}
\Tr(\slashed{a}\gamma^\mu\slashed{b}\gamma^\nu)\cdot
\Tr(\slashed{c}\gamma_\mu\slashed{d}\gamma_\nu) &= 32\bigl[(a\cdot
c)(b\cdot d) + (a\cdot d)(b\cdot c)\bigr]\notag\\
\Tr(\slashed{a}\gamma^\mu\slashed{b}\gamma^\nu\gamma_5)\cdot
\Tr(\slashed{c}\gamma_\mu\slashed{d}\gamma_\nu\gamma_5) &=
32\bigl[(a\cdot c)(b\cdot d) - (a\cdot d)(b\cdot c)\bigr]\notag\\
\Tr(\slashed{a}\gamma^\mu\slashed{b}\gamma^\nu)\cdot
\Tr(\slashed{c}\gamma_\mu\slashed{d}\gamma_\nu\gamma_5) &=0
\label{eqA.50}
\end{align}
Note that these relations can be obtained in a straightforward
way, by using (\ref{eqA.38}), (\ref{eqA.39}) and the identity
(\ref{eqA.42}). Let us also give an analogous formula involving
the $\sigma_{\mu\nu}$:
\begin{equation*}
\Tr(\slashed{a}\sigma^{\alpha\beta}\slashed{b}\sigma^{\mu\nu})\cdot
\Tr(\slashed{c}\sigma_{\alpha\beta}\slashed{d}\sigma_{\mu\nu})
=128\bigl[2(a\cdot c)(b\cdot d) + 2(a\cdot d)(b\cdot c) - (a\cdot
b)(c\cdot d)\bigr]
\end{equation*}
(needless to say, a derivation of the last relation is much more
tedious than in the preceding case).

For completeness, we list some useful identities for Pauli
matrices:
\begin{align}\label{eqA.51}
\sigma_j\sigma_k &= \delta_{jk}\cdot \J +i \epsilon_{jkl}\sigma_l
\notag\\
\Tr(\sigma_j\sigma_k) &=2\delta_{jk} \notag\\
\sum_i (\sigma_i)_{ab}(\sigma_i)_{cd} &= 2\delta_{ad}\delta_{bc}
-\delta_{ab}\delta_{cd}
\end{align}

Let us now proceed further, to summarize some essential properties
of solutions of the free-particle Dirac equation (\ref{eqA.1}). We
consider plane waves, i.e. the solutions corresponding to a
definite energy and momentum. There are two independent Ans\"atze
for such a solution, namely
\begin{align}
\psi_+ (x) &= u(p)\,\text{e}^{-ipx}\notag\\
\psi_-(x) &= v(p)\,\text{e}^{ipx}\label{eqA.52}
\end{align}
with $px=p_0x_0-\vec{p}\cdot\vec{x}$, where we take $p_0>0$ by
definition. Substituting (\ref{eqA.52}) into (\ref{eqA.1}) one
gets
\begin{equation}\label{eqA.53}
(\slashed{p}-m)u(p) = 0
\end{equation}
and
\begin{equation}\label{eqA.54}
(\slashed{p}+m)v(p) =0
\end{equation}
Obviously, the $p$ must then satisfy $p^2=m^2$. Further, taking
into account (\ref{eqA.17}), it becomes clear that $\psi_+$
corresponds to a positive energy $E=p_0=\sqrt{\vec{p}\,^2 + m^2}$
while the $\psi_-$ carries negative energy
$-\sqrt{\vec{p}\,^2+m^2}$. It is useful to know that solutions of
(\ref{eqA.53}) and (\ref{eqA.54}) are interrelated through the
operation of charge conjugation\index{charge conjugation}, defined
in terms of the matrix $C=i\gamma^2\gamma^0$ (in standard
representation): if $u(p)$ is a solution of (\ref{eqA.53}), then
\begin{equation}\label{eqA.55}
u_c(p) = C\bar{u}(p)^T
\end{equation}
(with $T$ denoting matrix transposition) satisfies
eq.\,(\ref{eqA.54}).

A frequently used set of the $u(p)$ and $v(p)$, corresponding to
the standard representation of Dirac matrices in (\ref{eqA.53}),
(\ref{eqA.54}), can be described explicitly as
\begin{equation}\label{eqA.56}
u^{(r)}(p) = \sqrt{E+m} \begin{pmatrix}\chi^{(r)}\vspace{3pt}\\
\dfrac{\vec{\sigma}\cdot\vec{p}}{E+m}\chi^{(r)}\end{pmatrix},\qquad
r=1,2
\end{equation}
and
\begin{equation}\label{eqA.57}
v^{(r)}(p) =
\pm\sqrt{E+m}\begin{pmatrix}\dfrac{\vec{\sigma}\cdot\vec{p}}{E+m}
\chi^{(r)}\vspace{3pt}\\ \chi^{(r)}\end{pmatrix},\qquad r=1,2
\end{equation}
where $E=\sqrt{\vec{p}\,^2+m^2}$ and
\begin{equation}
\chi^{(1)}=\bm{1\\0},\qquad \chi^{(2)} = \bm{0\\1}
\end{equation}
The upper and lower sign in (\ref{eqA.57}) holds for $r = 1$ and
$r = 2$ respectively; note that the form of the $v^{(r)}(p)$
(including the overall sign) is determined by the
charge-conjugation transformation (\ref{eqA.55}). It is important
to stress that the solutions (\ref{eqA.56}), (\ref{eqA.57}) are
normalized according to
\begin{align}
\bar{u}(p)u(p) &= 2m\notag\\
\bar{v}(p)v(p) &= -2m
\end{align}
(such a normalization is most convenient for the discussion of
high-energy behaviour of scattering amplitudes represented by
Feynman graphs).

Let us also remark that in the non-relativistic limit, i.e. for
$|\vec{p}|\ll m$, the lower two components of the bispinor
(\ref{eqA.56}) become negligible and the Dirac particle is thus
effectively described by means of a two-component spinor,
proportional to $\chi^{(r)}$; this is a main advantage of working
in the standard representation.

The index $r=1,2$ in (\ref{eqA.56}) and (\ref{eqA.57}) labels spin
degrees of freedom. In particular, the $u^{(r)}(p)$ corresponds to
positive-energy solution with spin up ($r = 1$) or down ($r = 2$)
along the third axis of the coordinate system, in the particle
rest frame. In fact, spin states of a Dirac particle can be
described, quite generally, in an elegant covariant way. We are
now going to summarize briefly the contents of such a formalism,
as well as some relevant formulae. Unless stated otherwise, we
assume explicitly that $m\neq 0$.

A basic notion is that of the \qq{spin four-vector}. For a given
four-momentum $p$ one defines $s^\mu=s^\mu(p)$, $\mu=0,1,2,3$, so
that
\begin{equation}
s^\mu p_\mu =0
\end{equation}
and the $s^\mu$ behaves as a space-like Lorentz four-vector; its
normalization is conveniently fixed by
\begin{equation}
s^2 = -1
\end{equation}
(concerning the terminology, let us add that the spin four-vector
$s$ is sometimes also called \qq{polarization vector}). A remark
is in order here. For a conceptual construction of the spin vector
$s=s(p)$, one can start in the particle rest frame, where
$p=p^{(0)}=(m,\,0,\,0,\,0)$ and take $s=s^{(0)}=(0,\,\vec{s})$,
with $\vec{s}$ being a unit vector in three-dimensional space (the
$\vec{s}$ is to be understood as the spin direction in the rest
frame). Passing from $p^{(0)}$ to an arbitrary $p$, $p^2=m^2$, the
$s(p)$ is defined by means of the corresponding Lorentz
transformation of the $s^{(0)}$. Of course, given a spatial
direction $\vec{s}$ (in the rest frame), there are two independent
states of a Dirac particle, characterized by the spin projection
pointing up or down along the $\vec{s}$. Alternatively, one can
say that these two states correspond to the opposite directions
$\vec{s}$ and $-\vec{s}$ (i.e. to the spin parallel with either
$\vec{s}$ or $-\vec{s}$). In a general reference frame, this means
that for a given four-momentum $p$ (and, say, for a positive
energy) there are two independent states corresponding to the spin
four-vectors $s$ and $-s$ respectively.

Covariant description of the spin states in question can be
formulated as follows. With an $s=s(p)$ at hand, one considers
solutions of (\ref{eqA.53}) and (\ref{eqA.54})
satisfying\footnote{It should be noticed that the $\slashed{p}$
and $\gamma_5\slashed{s}$ commute; it is a simple consequence of
the relation $s\cdot p = 0$.}
\begin{align}
\gamma_5 \slashed{s} u(p,s) &= u(p,s)\notag\\
\gamma_5 \slashed{s} v(p,s) &= v(p,s)
\end{align}
In other words, $u(p,s)$ or $v(p,s)$ is obtained from an arbitrary
solution of (\ref{eqA.53}) or (\ref{eqA.54}) by means of the
projector
\begin{equation}
P_+(s) = \frac{1}{2}(1+\gamma_5\slashed{s})
\end{equation}
(the reader is recommended to verify explicitly that the $P_+(s)$
is indeed a projector, i.e. that it holds $(P_+(s))^2=P_+(s)$).
Similarly, using the projector
\begin{equation}
P_-(s)=\frac{1}{2}(1-\gamma_5\slashed{s})
\end{equation}
one obtains the remaining independent spin states, corresponding
to the spin vector $-s$. It can be shown that the four spinors
$u(p,\pm s)$ and $v(p,\pm s)$ determine a complete system of
solutions of the Dirac equation for a free particle. For routine
Feynman diagram calculations of cross sections or decay
probabilities (employing the familiar \qq{trace techniques}),
explicit expressions for the $u(p,s)$ and $v(p,s)$ are not
necessary; one really needs only the combinations like
$u(p,s)\bar{u}(p,s)$ etc. The relevant results are
\begin{align}
u(p,s)\bar{u}(p,s) &=
(\slashed{p}+m)\frac{1+\gamma_5\slashed{s}}{2}\notag\\
v(p,s)\bar{v}(p,s) &=
(\slashed{p}-m)\frac{1+\gamma_5\slashed{s}}{2}\label{eqA.65}
\end{align}
The corresponding formulae for $u(p,-s)\bar{u}(p,-s)$ and
$v(p,-s)\bar{v}(p,-s)$ are obtained from (\ref{eqA.65}) trivially
by replacing there $s$ with $-s$. Summing the expressions
(\ref{eqA.65}) over the individual spin states (or
\qq{polarizations}), one gets
\begin{align}
\sum_\ti{spin} u(p,s)\bar{u}(p,s) &= \slashed{p}+m \notag\\
\sum_\ti{spin} v(p,s)\bar{v}(p,s) &= \slashed{p}-m \label{eqA.66}
\end{align}

There is an important particular example of the spin vector that
deserves special attention. The specific spin states we have in
mind correspond to {\bf helicity} or \qq{longitudinal
polarization}, described in terms of an $s(p)$, whose spatial part
is directed along the three-momentum $\vec{p}$. For reasons that
are explained below, we denote the spin four-vector
$s=(s^0,\,\vec{s})$ having $\vec{s}$ parallel to $\vec{p}$ as
$s_R(p)$, indicating thus that it corresponds to right-handed
particle (with positive helicity); similarly, the left-handed
state (with negative helicity) is described by $s_L(p)=-s_R(p)$.
It is not difficult to find an explicit form of the $s_R(p)$.
Using the Ansatz $s_R(p)=(s^0,\,\lambda\vec{p})$ with $\lambda>0$
and taking into account the general relations $s\cdot p=0$,
$s^2=-1$, one gets readily
\begin{equation}\label{eqA.67}
s_R^\mu (p) =
\biggl(\frac{|\vec{p}|}{m},\,\frac{E}{m}\frac{\vec{p}}{|\vec{p}|}\biggr)
\end{equation}

Let us now comment on the connection between the above formal
description of helicity and its straightforward physical
definition (which may be more familiar to an average reader). The
helicity of a particle with definite momentum is generally defined
as the projection of spin on the direction of motion. Thus, for a
Dirac particle with momentum $\vec{p}$, helicity is identified
with an eigenvalue of the $4 \times 4$ matrix
\begin{equation}\label{eqA.68}
h(p)=\frac{\vec{\Sigma}\cdot\vec{p}}{|\vec{p}|}
\end{equation}
where
\begin{equation}
\vec{\Sigma}=\bm{\vec{\sigma}&0\\0&\vec{\sigma}}
\end{equation}
(strictly speaking, (\ref{eqA.68}) represents the spin projection
up to a factor of $1/2$, since the spin matrix for a Dirac
particle is $\frac{1}{2}\vec{\Sigma})$. Now, the crucial
observation is that
\begin{equation}
\label{eqA.70} \gamma_5\slashed{s}_R(p)u(p) =
\frac{\vec{\Sigma}\cdot\vec{p}}{|\vec{p}|}u(p)
\end{equation}
for any $u(p)$ satisfying eq.\,(\ref{eqA.53}). It is clear that
the identity (\ref{eqA.70}) establishes the aforementioned
equivalence between the two descriptions of helicity. A proof of
(\ref{eqA.70}) is not difficult; apart from some straightforward
algebraic manipulations, one has to take into account the identity
\begin{equation}\label{eqA.71}
\vec{\Sigma}=\gamma_5\vec{\alpha}
\end{equation}
that obviously holds in the standard representation (in fact,
defining generally
$\Sigma^j=\frac{1}{2}\epsilon^{jkl}\sigma^{kl}$, (\ref{eqA.71}) is
valid in any representation). Note that an identity analogous to
(\ref{eqA.70}) can be derived also for solutions of
(\ref{eqA.54}); however, when considering the helicity of a
$v(p)$, one should not forget that the corresponding plane wave
carries momentum $-\vec{p}$ (and negative energy).

For completeness, let us now discuss briefly the case of a
massless particle. Obviously, the expression (\ref{eqA.67}) makes
no sense for $m=0$; more generally, one can verify directly --
starting from the basic requirements -- that a space-like
longitudinal spin four-vector simply cannot be constructed in the
massless case. Nevertheless, the definition of helicity based on
(\ref{eqA.68}) is still applicable. Moreover, making use of the
identity (\ref{eqA.71}), the characterization of the helicity
states is greatly simplified: it turns out that helicity is
essentially reduced to {\bf chirality}, which is an eigenvalue of
the $\gamma_5$. In particular, for an $u(p)$ the helicity
coincides with chirality, while for an $v(p)$ helicity is equal to
chirality taken with minus sign. Thus, for $m=0$, the left-handed
and right-handed spinors $u_{L,R}(p)$, $v_{L,R}(p)$ satisfy
\begin{align}
\gamma_5 u_L(p) &= -u_L(p) \notag\\
\gamma_5 u_R(p) &= \phantom{+}u_R(p)
\end{align}
and
\begin{align}
\gamma_5 v_L(p) &= \phantom{+}v_L(p) \notag\\
\gamma_5 v_R(p) &= -v_R(p)
\end{align}
These relations can be recast in terms of appropriate projectors,
namely
\begin{align}
\frac{1}{2}(1-\gamma_5) u_L(p) &= u_L(p) \notag\\
\frac{1}{2}(1+\gamma_5) u_R(p) &= u_R(p) \label{eqA.74}
\end{align}
and
\begin{align}
\frac{1}{2}(1+\gamma_5) v_L(p) &= v_L(p) \notag\\
\frac{1}{2}(1-\gamma_5) v_R(p) &= v_R(p) \label{eqA.75}
\end{align}
Similarly as in the massive case, for practical calculations one
needs combinations like $u_L(p)\bar{u}_L(p)$ etc. For that
purpose, one cannot simply take the limit $m\rightarrow 0$ in
(\ref{eqA.65}) since it does not exist. On the other hand, the
summed expression (\ref{eqA.66}) is safe in the massless limit and
one has
\begin{alignat}{2}
\sum_\ti{spin} u(p)\bar{u}(p) &= u_L(p)\bar{u}_L(p) +
u_R(p)\bar{u}_R(p) &&= \slashed{p} \notag\\
\sum_\ti{spin} v(p)\bar{v}(p) &= v_L(p)\bar{v}_L(p) +
v_R(p)\bar{v}_R(p) &&= \slashed{p} \label{eqA.76}
\end{alignat}
The desired \qq{anatomy} of the relations (\ref{eqA.76}) can be
obtained from (\ref{eqA.74}), (\ref{eqA.75}) and (\ref{eqA.76}) by
means of simple algebraic tricks (among other things, one has to
utilize obvious relations like $\frac{1}{2}(1-\gamma_5)u_R=0$
etc.). Leaving a detailed derivation to the interested reader, we
give here only the result:
\begin{alignat}{2}
u_L(p)\bar{u}_L(p) &= \slashed{p}\ \frac{1+\gamma_5}{2},\qquad
u_R(p)\bar{u}_R(p) &&= \slashed{p}\ \frac{1-\gamma_5}{2}\notag\\
v_L(p)\bar{v}_L(p) &= \slashed{p}\ \frac{1-\gamma_5}{2},\qquad
v_R(p)\bar{v}_R(p) &&= \slashed{p}\ \frac{1+\gamma_5}{2}
\end{alignat}

When summarizing important properties of plane-wave solutions of
the Dirac equation, one should also mention the so-called Gordon
identity, that represents a practically useful decomposition of
the current $\bar{u}(p)\gamma_\mu u(p')$ into \qq{convective} and
\qq{spin} parts. To arrive at such a result, one can start with
the identity
\begin{equation}\label{eqA.78}
\bar{u}(p)\bigl[(\slashed{p}-m)\gamma_\mu +
\gamma_\mu(\slashed{p}'-m)\bigr] u(p') =0
\end{equation}
that obviously holds for solutions of eq.\,(\ref{eqA.53}) (we
suppress here the spin labels, since these are irrelevant in the
present context). Decomposing the matrix products in
(\ref{eqA.78}) into anticommutators and commutators, employing the
basic relation (\ref{eqA.2}) and the definition (\ref{eqA.46}),
one gets readily
\begin{equation}\label{eqA.79}
\bar{u}(p)\gamma_\mu u(p') = \frac{1}{2m} \bar{u}(p)\bigl[
(p_\mu+p'_\mu) + i\sigma_{\mu\nu}(p^\nu-p'^\nu)\bigr] u(p')
\end{equation}
In fact, it is easy to realize that there are three additional
identities of such a type, involving one or two spinors $v$
instead of $u$. Such generalizations of (\ref{eqA.79}) are derived
by modifying appropriately the \qq{master identity}
(\ref{eqA.78}): since a $v(p)$ satisfies eq.\,(\ref{eqA.54}), it
is sufficient to change the sign of the corresponding
four-momentum whenever the $v$ stands in place of a $u$. Thus, it
becomes clear that the resulting Gordon identities are obtained in
the same manner -- simply by changing signs of the relevant
four-momenta in (\ref{eqA.79}).

To conclude this appendix, let us now recapitulate briefly some
basic relations concerning the quantized free Dirac field. This is
represented by a four-component spinor operator in the Fock space,
written as
\begin{align}
\psi(x) &= \sum_{\pm s} \int \frac{d^3p}{(2\pi)^{3/2}(2p_0)^{1/2}}
\bigl[b(p,s)u(p,s)\,\text{e}^{-ipx} + d^+
(p,s)v(p,s)\,\text{e}^{ipx}
\bigr]\notag\\
\psib(x) &= \sum_{\pm s} \int
\frac{d^3p}{(2\pi)^{3/2}(2p_0)^{1/2}}
\bigl[b^+(p,s)\bar{u}(p,s)\,\text{e}^{ipx} +
d(p,s)\bar{v}(p,s)\,\text{e}^{-ipx} \bigr] \label{eqA.80}
\end{align}
Here $b(p,s)$, $d(p,s)$ are annihilation operators of the particle
and antiparticle respectively, and $b^+(p,s)$, $d^+(p,s)$ are the
corresponding creation operators. Of course, the annihilation and
creation operators are related through hermitean conjugation, i.e.
$b^+(p,s)=b^\dagger (p,s)$, $d^+(p,s)=d^\dagger(p,s)$. The
four-momenta in (\ref{eqA.80}) are on the mass shell, i.e. one
takes everywhere $p_0=E(p)=\sqrt{\vec{p}\,^2+m^2}$.

The field operators satisfy equal-time (E.T.) anticommutation
relations
\begin{align}
\bigl\{\psi_a(x),\,\psi_b(y)\bigr\}_{E.T.} &= 0\notag\\
\bigl\{\psi_a(x),\,\psi_b^\dagger(y)\bigr\}_{E.T.} &= \delta_{ab}
\delta^3(\vec{x}-\vec{y}) \label{eqA.82}
\end{align}
that yield the algebra of creation and annihilation operators
\begin{alignat}{3}
&\bigl\{b(p,s),\,b(p',s')\bigr\} &&=0,\qquad\qquad
\bigl\{d(p,s),\,d(p',s')\bigr\} &&=0\notag\\
&\bigl\{b(p,s),\,b^+(p',s')\bigr\}
&&=\delta_{ss'}\delta^3(\vec{p}-\vec{p}{\,}')\notag\\
&\bigl\{d(p,s),\,d^+(p',s')\bigr\} &&=\delta_{ss'}\delta^3(\vec{p}-\vec{p}{\,}')\notag\\
&\bigl\{b(p,s),\,d(p',s')\bigr\} &&=0,\qquad\qquad
\bigl\{b(p,s),\,d^+(p',s')\bigr\} &&=0 \label{eqA.83}
\end{alignat}
(other anticommutators are obtained by means of hermitean
conjugation). Note that a passage from (\ref{eqA.82}) to
(\ref{eqA.83}) (and vice versa) is guaranteed, among other things,
by the choice of the normalization factor $(2\pi)^{-3/2}
(2p_0)^{-1/2}$ introduced in the definition (\ref{eqA.80}). It
should be emphasized that the anticommutation relations
(\ref{eqA.83}) imply a specific normalization of one-particle
states; in particular, defining $|p,s\rangle = b^+(p,s)|0\rangle$
etc. (with $|0\rangle$ being the Fock vacuum state), one has
\begin{equation}
\langle p,s|p',s'\rangle =
\delta_{ss'}\delta^3(\vec{p}-\vec{p}{\,}')
\end{equation}
Although such a non-covariant normalization is not universally
accepted in current literature, we stick to this convention
(essentially corresponding to \cite{BjD}) throughout the present
text.
%\end{document}

%\input{app_b}
%%%%%%%%%%%%%%%%%%%%%%%%%%%%%%%%%%%%%%%%%%%%%%%%%%%%%%%%%%%%%%%%%%%
%%%%%%%%%%%%%%%%%%%%%%%%%%%%%%%%%%%%%%%%%%%%%%%%%%%%%%%%%%%%%%%%%%%%%%%%%%%%%%%%%%%%%%%%%%%%%%%%%%%%%%%%%%%%%%%%%%%%%%%%%%%%%%%%%%%%%%%%
%\documentclass[11pt,tbtags]{book2} \input{pream}
%\begin{document}
%\appendix
\chapter{Scattering amplitudes, cross sections and decay rates}\label{appenB}
\index{scattering amplitude|appB} \index{Lorentz!Invariant Phase
Space (LIPS)|\\appB} \index{Mandelstam variables|appB}
\index{partial-wave expansion|appB} \index{Legendre
polynomials|appB} \index{S matrix unitarity@$S$-matrix
unitarity|appB} \index{unitarity bound|appB} \index{angular
momentum|appB}\index{cross section|appB}\index{decay
rate|appB}\index{decay width|appB}\index{Jacob--Wick
expansion|appB}

Let us start with definition of the Lorentz invariant scattering
(or decay) amplitude $\mathcal{M}_{fi}$ in terms of an $S$-matrix
element $S_{fi}=\langle f| S |i\rangle$. This reads
\begin{equation}\label{eqB.1}
S_{fi}=\delta_{fi}+(2\pi)^4\delta^4(P_f-P_i)(i\mathcal{M}_{fi})\prod_{f,i}\frac{1}{(2\pi)^{3/2}(2E_{f,i})^{1/2}}
\end{equation}
where the $P_f$ and $P_i$ denote the total four-momenta of the
final and initial particles respectively. The normalization
factors under the product symbol correspond to the conventional
choice, exemplified by the formula (\ref{eqA.80}) for quantized
Dirac field; such a choice means that these factors have the same
form for bosons and fermions. Note also that our sign convention
for the $\mathcal{M}_{fi}$ differs from that adopted in some
standard textbooks: e.g. the definition used in \cite{BjD} is
obtained from (\ref{eqB.1}) by replacement
$i\mathcal{M}_{fi}\rightarrow -i\mathcal{M}_{fi}$.

In the context of practical calculations, the $\mathcal{M}_{fi}$
is often called simply \qq{matrix element} (for a given process).
Within perturbation theory this is evaluated by means of the
relevant covariant Feynman rules; in particular, the contributions
of external Dirac particles are represented by the corresponding
spinors $u$ or $v$, etc.\footnote{In other words, the definition
(\ref{eqB.1}) means that the non-covariant normalization factors
do not enter the routine Feynman diagram calculations.} Knowing
the matrix element $\mathcal{M}_{fi}$, one can compute physically
observable quantities for the considered process. In particular,
the differential cross section for a reaction $1 + 2 \rightarrow 3
+ 4 + \ldots + n$ is given by the general formula
\begin{equation}\label{eqB.2}
d\sigma=\frac{1}{|\vec{v}_1-\vec{v}_2|}\frac{1}{2E_1}\frac{1}{2E_2}|\mathcal{M}_{fi}|^2(2\pi)^4\delta^4\bigl(p_1+p_2-\sum_{j=3}^{n}p_j\bigr)
\frac{d^3p_3}{(2\pi)^3 2E_3} \cdots \frac{d^3p_n}{(2\pi)^3
2E_n}\; K
\end{equation}
(irrespectively of whether the particles $1, \ldots, n$ are bosons
or fermions). The $\vec{v}_1$, $\vec{v}_2$ denote velocities of
the initial particles (we assume that vectors $\vec{v}_1$,
$\vec{v}_2$ are parallel and  have opposite directions), the
$p_j=(E_j,\ \vec{p}_j),\; j=1,2,\ldots, n$ are four-momenta, i.e.
$E_j=\sqrt{\vec{p}_j^{\,2}+m_j^2}$, and the $K$ is a combinatorial
(\qq{statistical}) factor, which is different from $1$ only when
some of the final-state particles are identical:
\begin{equation}\label{eqB.3}
K=\prod_{r=1}^k \frac{1}{n_r!}
\end{equation}
where $n_r$ is the number of identical particles of the $r$th kind
in the final state $|f\rangle$ (of course, $n_1+\ldots+n_k=n-2$).

It is worth noticing here that with the result (\ref{eqB.2}) at
hand, one can determine the dimension of the matrix element
$\mathcal{M}_{fi}$ on quite general grounds. The argument goes as
follows. The dimension of the left-hand side of (\ref{eqB.2}) is
(length)$^2$, i.e. (mass)$^{-2}$ in the system of units where
$\hbar=c=1$. Thus,
\begin{equation}\label{eqB.4}
[d\sigma]=M^{-2}
\end{equation}
with $M$ being an arbitrary mass and in the right-hand side of
(\ref{eqB.2}) one has
\begin{equation}\label{eqB.5}
M^{-1}\cdot M^{-1}\cdot\bigl[|\mathcal{M}_{fi}|^2\bigr]\cdot
M^{-4}\cdot (M^2)^{n-2} = \bigl[|\mathcal{M}_{fi}|^2\bigr]\cdot
M^{2n-10}
\end{equation}
(recall that dimension of the $\delta^4(P_f-P_i)$ is $M^{-4}$ !).
Comparing (\ref{eqB.4}) and (\ref{eqB.5}) one gets immediately the
desired result:
\begin{equation}\label{eqB.6}
[\mathcal{M}_{fi}]=M^{4-n}
\end{equation}
In particular, (\ref{eqB.6}) shows that the matrix element for an
arbitrary {\it binary\/} process $1 + 2 \rightarrow 3 + 4$ is {\it
dimensionless}; this simple observation is quite useful in
estimating the high-energy behaviour of scattering amplitudes.

Before proceeding further, let us recall briefly some elementary
kinematics. Considering a binary process and using the above
notation for the corresponding four-momenta, one defines the
Lorentz invariant Mandelstam variables as
\begin{align}
s&=(p_1+p_2)^2 = (p_3+p_4)^2\notag\\
t&=(p_1-p_3)^2 = (p_2-p_4)^2\notag\\
u&=(p_1-p_4)^2 = (p_2-p_3)^2\label{eqB.7}
\end{align}
(in writing (\ref{eqB.7}) we have taken into account explicitly
the four-momentum conservation $p_1+p_2 = p_3+p_4$). It is not
difficult to show that the $s, t, u$ satisfy the identity
\begin{equation}
s+t+u = \sum_{j=1}^4 m_j^2
\end{equation}
One should also notice that the $s$ has a simple physical meaning:
it coincides with the square of total centre-of-mass (c.m.) energy
of the colliding particles. This is obvious, since $(p_1+p_2)^2$
has the same value in any Lorentz frame and in the c.m. system one
has $p_1+p_2=(E_1+E_2,\; \vec{0})$ by definition. Thus, one has
\begin{equation}\label{eqB.9}
s^{1/2}=\sqrt{|\vec{p}_{c.m.}|^2+m_1^2} +
\sqrt{|\vec{p}_{c.m.}|^2+m_2^2}
\end{equation}
with $\vec{p}_{c.m.}$ standing for the c.m. momentum of one of the
colliding particles (one can take e.g. $\vec{p}_{c.m.}=\vec{p}_{1
c.m.}=-\vec{p}_{2 c.m.}$). An explicit formula for the
$|\vec{p}_{c.m.}|$ then follows easily from (\ref{eqB.9}); one
gets
\begin{equation}\label{eqB.10}
|\vec{p}_{c.m.}| = \Bigl[
\frac{\lambda(s,m_1^2,m_2^2)}{4s}\Bigr]^{1/2}
\end{equation}
where
\begin{equation}\label{eqB.11}
\lambda(x,y,z)=x^2+y^2+z^2 - 2xy -2xz -2yz
\end{equation}
Integrating over an appropriate part of the phase space of final
states in (\ref{eqB.2}) one can derive special formulae that are
suitable for practical applications. A most frequently used result
is the expression for angular distribution of final-state
particles in a binary process $1 + 2 \rightarrow 3 + 4$,
considered in the c.m. frame. Below we quote the standard formula
for the corresponding differential cross section (its derivation
is rather straightforward and can be found in many places, see
e.g. Appendix C of the book \cite{Hor}). Assuming that the
final-state particles are not identical, one has
\begin{equation}\label{eqB.12}
\frac{d\sigma}{d\Omega_{c.m.}} =
\frac{1}{64\pi^2}\frac{1}{s}\frac{|\vec{p}{\,}'_{\!c.m.}|}{|\vec{p}_{c.m.}|}|\mathcal{M}_{fi}|^2
\end{equation}
where the $|\vec{p}_{c.m.}|$ has been defined in (\ref{eqB.9}) and
$|\vec{p}{\,}'_{\!c.m.}|$ has an analogous meaning for the
final-state particles; therefore,
\begin{equation}
|\vec{p}{\,}'_{\!c.m.}| =
\Bigl[\frac{\lambda(s,m_3^2,m_4^2)}{4s}\Bigr]^{1/2}
\end{equation}
The $d\Omega_{c.m.}$ is an element of solid angle corresponding to
the direction of $\vec{p}{\,}'_{\!c.m.}$; in spherical coordinates
this has the standard form
$$d\Omega_{c.m.}=\sin\vartheta_{c.m.}d\vartheta_{c.m.}d\varphi_{c.m.}$$
If the particles $3$ and $4$ were identical, the right-hand side
of (\ref{eqB.12}) would include the combinatorial factor $K =
1/2$.

For an elastic scattering process (where the final particles are
the same as those in the initial state), the formula
(\ref{eqB.12}) gets simplified: in such a case one has, obviously,
$|\vec{p}{\,}'_{\!c.m.}|=|\vec{p}_{c.m.}|$ and (\ref{eqB.12}) thus
becomes
\begin{equation}
\frac{d\sigma}{d\Omega_{c.m.}}\Bigl|_\ti{elast.} =
\frac{1}{64\pi^2}\frac{1}{s}|\mathcal{M}_{fi}|^2
\end{equation}
There is another frequently occurring situation, where this kind
of simplification is relevant. Considering a general binary
process $1 + 2 \rightarrow 3 + 4$ in the high-energy limit, i.e.
for $s^{1/2}\gg m_j,\; j=1,\ldots, 4$, the particle masses can be
safely neglected in kinematical relations and thus
$|\vec{p}{\,}'_{\!c.m.}|/|\vec{p}_{c.m.}|\doteq 1$ with good
accuracy. In fact, when all particles involved in a given process
are taken as effectively massless, one can further streamline the
cross section calculations by introducing a suitable new
kinematical variable. To comply with a traditional notation, let
us label the four-momenta of particles 1, 2, 3, 4 consecutively as
$k, p, k', p'$ and define
\begin{equation}
y=\frac{p\cdot q}{p\cdot k}
\end{equation}
where $q=k-k'$ (obviously, the $y$ is Lorentz invariant and
dimensionless). Then it is not difficult to see that for $m_j=0$,
$j = 1, \ldots , 4$, the Mandelstam variables $t$ and $u$ can be
expressed in terms of the $s$ and $y$ as
\begin{align}
t &= -sy\notag\\
u &= -s(1-y)\label{eqB.16}
\end{align}
Moreover, in such a case the $y$ is related simply to the
scattering angle in the c.m. system:
\begin{equation}
\label{eqB.17} y=\frac{1}{2}(1-\cos\vartheta_{c.m.})
\end{equation}
($\vartheta_{c.m.}$ is defined here as the angle between
$\vec{p}{\,}'_{\!c.m.}$ and $\vec{p}_{c.m.}$). From (\ref{eqB.17})
it is then clear that for vanishing masses the $y$ takes on values
between 0 and 1. The above observations make the practical
importance of the kinematical variable $y$ obvious. Thus, it is
also desirable to have an expression for differential cross
section, written directly with respect to the $y$. This is
achieved easily. Assuming that the matrix element squared
$|\mathcal{M}_{fi}|^2$ does not depend on the polar angle
$\varphi$ (which is usually the case), the integration of
(\ref{eqB.12}) over the $\varphi$ is done trivially and, taking
into account (\ref{eqB.17}), one gets readily
\begin{equation}\label{eqB.18}
\frac{d\sigma}{dy} =
\frac{1}{16\pi}\frac{1}{s}|\mathcal{M}_{fi}|^2
\end{equation}
When writing (\ref{eqB.18}), it is assumed implicitly that the
expression for $|\mathcal{M}_{fi}|^2$ has been recast in terms of
the $s$ and $y$ by using (\ref{eqB.16}) (this is certainly
possible when considering scattering of unpolarized particles,
i.e. when the $|\mathcal{M}_{fi}|^2$ is summed over the relevant
spin states). Let us emphasize again that (\ref{eqB.18}) is valid
as an approximate formula in the high-energy limit, or as an exact
formula in a strictly massless case.

Another important case that deserves a separate treatment is the
scattering on a fixed target, i.e. in the {\it laboratory\/}
(rest) system of one of the initial particles. In particular, let
us consider a binary process $1 + 2 \rightarrow 3 + 4$ in the rest
frame of the particle 2. One can calculate e.g. the angular
distribution of the particle 3 with respect to the direction of
the incident particle 1. We shall not present here a derivation of
the formula in question from the basic relation (\ref{eqB.2})
(though it is not a difficult task) and quote only the final
result for the corresponding differential cross section:
\begin{equation}\label{eqB.19}
\frac{d\sigma}{d\Omega} =
\frac{1}{64\pi^2}\frac{1}{|\vec{p}_1|m_2}|\mathcal{M}_{fi}|^2\frac{|\vec{p}_3|}{E_1+m_2-\frac{|\vec{p}_1|E_3}{|\vec{p}_3|}\cos\vartheta}
\end{equation}
where $d\Omega$ is an element of solid angle along the direction
of $\vec{p}_3$; the $\vartheta$ is the angle between $\vec{p}_3$
and $\vec{p}_1$, and the meaning of the other symbols should be
obvious (notice that for brevity we drop everywhere the labels
referring explicitly to the laboratory frame and write simply
$d\Omega$, $d\vartheta$ instead of $d\Omega_\ti{lab.}$,
$d\vartheta_\ti{lab.}$ etc.). Of course, for an evaluation of the
right-hand side of (\ref{eqB.19}) one has to take into account the
relevant energy-momentum constraints. Making use of the definition
$p_2=(m_2,\ \vec{0})$, one has $\vec{p}_4=\vec{p}_1-\vec{p}_3$ and
the energy conservation then yields
\begin{equation}\label{eqB.20}
\sqrt{|\vec{p}_3|^2+m_3^2} + \sqrt{|\vec{p}_1|^2 -
2|\vec{p}_1||\vec{p}_3|\cos\vartheta + |\vec{p}_3|^2 +m_4^2} =
E_1+m_2
\end{equation}
Eq. (\ref{eqB.20}) can be explicitly solved for the $|\vec{p}_3|$;
for a general combination of $m_1,\ldots, m_4$ the result is quite
complicated function of the scattering angle $\vartheta$, but it
can be considerably simplified when some of the masses vanish.
This, of course, is of practical interest, since such a
configuration occurs in some familiar physical processes, e.g. in
the Compton scattering $\gamma+e^-\rightarrow \gamma + e^-$ or in
the elastic scattering of (quasi)massless neutrino on a charged
lepton. Thus, let us consider the case of an elastic scattering
with $m_1=m_3=0$ and $m_2=m_4=m$. We shall label the $p_1$, $p_2$,
$p_3$, $p_4$ consecutively as $k$, $p$, $k'$, $p'$ and denote the
energies $E(k)$, $E(k')$ simply as $E$, $E'$ resp. (thus,
$E=|\vec{k}|$ and $E'=|\vec{k}'|$). The relation (\ref{eqB.20})
then becomes
\begin{equation}
E' + \sqrt{E'^2 - 2 E E' \cos\vartheta + E^2 +m^2} = E+m
\end{equation}
and this is easily reduced to
\begin{equation}\label{eqB.22}
\frac{1}{E'}-\frac{1}{E} = \frac{1}{m}(1-\cos\vartheta)
\end{equation}
From (\ref{eqB.22}) one gets immediately
\begin{equation}
E' = \frac{E}{1+\dfrac{E}{m}(1-\cos\vartheta)}
\end{equation}
(notice that the last result is precisely the famous Compton
relation for the change of frequency of a photon scattered off a
free electron). Using all kinematical relations shown above, the
formula (\ref{eqB.19}) is recast, after some simple manipulations,
as
\begin{equation}\label{eqB.24}
\frac{d\sigma}{d\Omega}= \frac{1}{64\pi^2} \frac{1}{m^2}
|\mathcal{M}_{fi}|^2 \Bigl(\frac{E'}{E}\Bigr)^2
\end{equation}
Thus, we have arrived at the desired result: eq.\,(\ref{eqB.24})
represents a relatively simple formula for angular distribution of
elastically scattered particles in the laboratory frame, which is
applicable whenever the incident particle is much lighter than the
target (so that its mass can be safely neglected).

Next, let us turn to the decay processes. In general, we consider
a particle with the mass $M$ decaying in its rest system into a
number ($n$) of lighter particles. The differential probability of
such a decay per unit of time (the differential decay rate) is
given by
\begin{equation}\label{eqB.25}
dw=\frac{1}{2M}|\mathcal{M}_{fi}|^2 (2\pi)^4
\delta^4\bigl(P-\sum_{j=1}^n p_j\bigr) \frac{d^3 p_1}{(2\pi)^3 2
E_1} \cdots \frac{d^3p_n}{(2\pi)^3 2 E_n}\ K
\end{equation}
where the $\mathcal{M}_{fi}$ is the corresponding Lorentz
invariant matrix element, $P$ denotes the four-momentum of the
decaying particle, i.e. (in the rest frame) $P=(M,\ \vec{0})$,
$p_j=(E_j,\ \vec{p}_j)$ for $j = 1, \ldots , n$ are the
four-momenta of the decay products and $K$ stands for the
combinatorial factor defined in (\ref{eqB.3}). The simplest
configuration is a two-body decay, i.e. $n = 2$ in (\ref{eqB.25}).
In such a case, the phase-space integration is particularly simple
and one can thus derive easily the formulae of immediate practical
interest. Below we summarize some relevant results (their detailed
derivation can be found in many places, see e.g. the Appendix C in
\cite{Hor}). For definiteness, we assume that the decay products
1, 2 are not identical particles, i.e. we set $K = 1$ in
(\ref{eqB.25}).

Thus, we start with the elementary 2-body differential decay rate
\begin{equation}\label{eqB.26}
dw=\frac{1}{2M}|\mathcal{M}_{fi}|^2 (2\pi)^4 \delta^4 (P-p_1-p_2)
\frac{d^3p_1}{(2\pi)^3 2E_1}\frac{d^3p_2}{(2\pi)^3 2E_2}
\end{equation}
When it makes sense to consider an angular distribution of the
decay products (e.g. when the decaying particle is polarized,
defining thus a preferred direction in space), one can just
integrate over the magnitudes of the final-state momenta (with the
energy-momentum constraint defined by the delta function in
(\ref{eqB.26})) and express the decay rate in question as
\begin{equation}\label{eqB.27}
dw=\frac{1}{2M}|\mathcal{M}_{fi}|^2 d(\LIPS_2)
\end{equation}
where $\LIPS_2$ is an acronym for \qq{2-body Lorentz Invariant
Phase Space}; its element $d(\LIPS_2)$ is given by
\begin{equation}
d(\LIPS_2) = \frac{|\vec{p}|}{M}\frac{d\Omega}{16\pi^2}
\end{equation}
with $\vec{p}$ denoting the momentum of a decay product (one can
take e.g. $\vec{p}=\vec{p}_1=-\vec{p}_2$) and $d\Omega$ stands for
an element of the solid angle along the direction of $\vec{p}$.
Obviously, the $|\vec{p}|$ can be calculated by means of the
formula (\ref{eqB.10}) with $s=M^2$ and one thus has
\begin{equation}\label{eqB.29}
|\vec{p}| = \frac{1}{2M}
\bigl[\lambda(M^2,m_1^2,m_2^2)\bigr]^{1/2}
\end{equation}
It is useful to notice that the expression (\ref{eqB.11}) for
$\lambda(M^2,m_1^2,m_2^2)$ can be recast, after some simple
manipulations, as
\begin{equation}\label{eqB.30}
\lambda(M^2,m_1^2,m_2^2) =
\bigl[M^2-(m_1+m_2)^2\bigr]\bigl[M^2-(m_1-m_2)^2\bigr]
\end{equation}
When the initial and final particles are unpolarized, the quantity
$|\mathcal{M}_{fi}|^2$ is summed (and averaged) over the relevant
spin states; it is easy to realize that the result can only depend
on $M^2, m_1^2, m_2^2$. The expression (\ref{eqB.27}) can then be
integrated trivially over the angles (one thus gets just a
multiplicative factor of $4\pi$) and the resulting decay rate
(decay width) $\Gamma$ becomes
\begin{equation}
\Gamma = \frac{1}{2M} \overline{|\mathcal{M}_{fi}|^2} \LIPS_2
\end{equation}
where the $\overline{|\mathcal{M}_{fi}|^2}$ stands for the
spin-averaged matrix element squared and
\begin{equation}\label{eqB.32}
\LIPS_2 = \frac{1}{4\pi}\frac{|\vec{p}|}{M} = \frac{1}{8\pi}
\Bigl[1-\frac{(m_1+m_2)^2}{M^2}\Bigr]^{1/2}
\Bigl[1-\frac{(m_1-m_2)^2}{M^2}\Bigr]^{1/2}
\end{equation}
(in writing the last expression, we have utilized the relations
(\ref{eqB.29}) and (\ref{eqB.30})). For completeness, let us
display two frequently used particular forms of (\ref{eqB.32}):
\begin{itemize}
\item[i)] For $m_1=m_2=m$, (\ref{eqB.32}) is reduced to
\begin{equation}
\LIPS_2\Bigl|_{m_1=m_2=m} =
\frac{1}{8\pi}\sqrt{1-\frac{4m^2}{M^2}}
\end{equation}
\item[ii)] For $m_1, m_2 \ll M$ one has the approximate relation
\begin{equation}
\LIPS_2\Bigl|_{m_1, m_2 \ll M} \doteq \frac{1}{8\pi}
\end{equation}
\end{itemize}

Finally, we shall discuss some general properties of relativistic
scattering amplitudes. In particular, below we summarize briefly
basic formulae concerning the partial-wave expansion (usually
called the Jacob--Wick expansion). More details can be found e.g.
in the textbook \cite{ItZ}. First, let us consider the elastic
scattering of particles 1, 2; as a reference frame, we always use
the corresponding c.m. system, but we suppress the label c.m. in
what follows. Initial and final states of both particles are
characterized by definite momenta ($\vec{p}_1=-\vec{p}_2=\vec{p}$,
$\vec{p}\,'\mspace{-7mu}_1=-\vec{p}\,'\mspace{-7mu}_2=\vec{p}\,'$)
and helicities (denoted as $h_1, h_2, h'_1, h'_2$); note that
$|\vec{p}| = |\vec{p}\,'|$ for elastic scattering. We identify the
third axis of our coordinate system with the direction of the
$\vec{p}$. For a scattering amplitude $f$, normalized with respect
to the differential cross section in such a way that
\begin{equation}\label{eqB.35}
\frac{d\sigma}{d\Omega} = |f|^2
\end{equation}
one can write the Jacob--Wick expansion
\begin{equation}\label{eqB.36}
f_{h'h}(s,\Omega) = \sum_j (2j+1)
f_{h'h}^{(j)}(s)\mathscr{D}^{(j)}_{\lambda'\lambda}(\Omega)
\end{equation}
where we denote collectively $h\equiv (h_1, h_2)$, $h'\equiv
(h'_1, h'_2)$, the angles $\Omega=(\vartheta, \phi)$ define the
direction of $\vec{p}\,'$ and the
$\mathscr{D}^{(j)}_{\lambda'\lambda}(\Omega)$ are Wigner functions
(known from the theory of angular momentum as matrix elements of
finite rotations, cf. e.g. \cite{Sak}). The indices $\lambda$,
$\lambda'$ are given by $\lambda=h_1-h_2$, $\lambda'=h'_1-h'_2$.
Some basic properties of the $\mathscr{D}$-functions are
summarized at the end of this appendix. The coefficients $f^{(j)}$
are the partial-wave amplitudes; the label $j$ stands for the
total angular momentum characterizing an individual partial wave.
The sum in (\ref{eqB.36}) runs over all non-negative integer or
half-integer values of the $j$, depending on whether there is an
even or odd number of fermions among the particles 1, 2. An
$f^{(j)}$ has the form
\begin{equation}
f^{(j)}_{h'h}(s) = \frac{1}{2i|\vec{p}|}
\bigl(S^{(j)}_{h'h}-1\bigr)
\end{equation}
where $S^{(j)}_{h'h}$ is an element of the $S$-matrix in the
angular momentum basis.\footnote{In such a basis, the $S$-matrix
has a block-diagonal form; for a fixed $j$, the $S^{(j)}$ is a
finite dimensional matrix living in a subspace spanned by the
helicity states.} The crucial point is that the $S$-matrix is
unitary; this implies an important constraint for the $f^{(j)}$,
namely
\begin{equation}
|f^{(j)}(s)|\leq \frac{1}{|\vec{p}|}
\end{equation}
(here and in what follows we usually suppress the indices $h,
h'$).

From eq.\,(\ref{eqB.36}) one can obtain easily a corresponding
expansion for the Lorentz invariant matrix element $\mathcal{M}$
entering the cross-section formula (\ref{eqB.12}). Indeed,
rescaling the $f$ normalized according to (\ref{eqB.35}) so as to
get an $\mathcal{M}$ satisfying (\ref{eqB.12}), and taking into
account that $|\vec{p}|=|\vec{p}\,'|$ for elastic scattering, one
can write
\begin{equation}\label{eqB.39}
\mathcal{M}(s,\Omega) = 16\pi \sum_j (2j+1) \mathcal{M}^{(j)}(s)
\mathscr{D}^{(j)}_{\lambda'\lambda}(\Omega)
\end{equation}
with
\begin{equation}\label{eqB.40}
\mathcal{M}^{(j)}(s) =
\frac{s^{1/2}}{4i|\vec{p}|}\bigl(S^{(j)}-1\bigr)
\end{equation}
This yields a unitarity bound for the $\mathcal{M}^{(j)}(s)$,
namely
\begin{equation}\label{eqB.41}
\bigl|\mathcal{M}^{(j)}(s)\bigr| \leq \frac{s^{1/2}}{2|\vec{p}|}
\end{equation}
In high-energy limit or for massless particles one has
$|\vec{p}|\doteq \frac{1}{2}\sqrt{s}$ and (\ref{eqB.41}) then
simplifies to
\begin{equation}\label{eqB.42}
\bigl|\mathcal{M}^{(j)}(s)\bigr| \leq 1
\end{equation}
Now, with the hindsight, it becomes clear that the choice of the
overall factor $16\pi$ in (\ref{eqB.39}) has been convenient, as
it leads to the simple constraint (\ref{eqB.42}).

A remark on practical evaluation of the partial-wave amplitudes is
in order here. For a given physical process, the matrix element
$\mathcal{M}(s,\Omega)$ can be calculated e.g. within standard
covariant perturbation theory (i.e. by means of Feynman diagrams).
Taking into account an appropriate orthogonality relation for the
Wigner $\mathscr{D}$-functions (see (\ref{eqB.56})), one can
evaluate an $\mathcal{M}^{j}(s)$ by means of the angular
integration
\begin{equation}\label{eqB.43}
\mathcal{M}^{(j)}(s)=\frac{1}{16\pi} \int \mathcal{M}(s,\Omega)
\mathscr{D}^{(j)*}_{\lambda'\lambda}(\Omega)\frac{d\Omega}{4\pi}
\end{equation}
In the particular case where $\lambda'=\lambda=0$ (i.e. for
$h_1=h_2$, $h'_1=h'_2$) the $\mathscr{D}$-functions are reduced to
Legendre polynomials (see (\ref{eqB.54})) and the formula
(\ref{eqB.43}) then becomes
\begin{equation}
\mathcal{M}^{(j)}(s) = \frac{1}{32\pi}\int_{-1}^1
\mathcal{M}(s,\vartheta) P_j(\cos\vartheta)d(\cos\vartheta)
\end{equation}
For an inelastic process $1 + 2 \rightarrow 3 + 4$ one can also
write a partial-wave expansion in the form (\ref{eqB.36}) or
(\ref{eqB.39}); however, in such a case only the purely
non-diagonal $S$-matrix elements are involved. Instead of
(\ref{eqB.40}) one then has
\begin{equation}\label{eqB.45}
\mathcal{M}^{(j)}_\ti{inel.}(s) =
\frac{s^{1/2}}{4i|\vec{p}|}S^{(j)}_\ti{inel.}
\end{equation}
where the symbol $S^{(j)}_\ti{inel.}$ again represents
collectively elements of the relevant unitary matrix and the index
\qq{inel.} denotes the inelastic channel $1 + 2 \rightarrow 3 +
4$. In high-energy limit, the relation (\ref{eqB.45}) implies the
bound
\begin{equation}
\bigl|\mathcal{M}^{(j)}_\ti{inel.}(s)\bigr| \leq \frac{1}{2}
\end{equation}
The constraints for partial-wave amplitudes following from
$S$-matrix unitarity can also be easily converted into
inequalities for partial cross sections (i.e. for cross sections
corresponding to the individual partial waves). Using the
expansion (\ref{eqB.39}) in the formula (\ref{eqB.12}) for
differential cross section, integrating (\ref{eqB.12}) over the
angles and utilizing the orthogonality relation (\ref{eqB.56}),
one gets
\begin{equation}
\sigma(s) = \sum_j \sigma^{(j)} (s)
\end{equation}
(for a given set of the initial and final helicities), where
\begin{equation}\label{eqB.48}
\sigma^{(j)}(s) = \frac{16\pi}{s} (2j+1)
\bigl|\mathcal{M}^{(j)}(s)\bigr|^2
\end{equation}
For elastic scattering, the inequality (\ref{eqB.41}) then implies
a bound for the partial cross sections (\ref{eqB.48}), namely
\begin{equation}
\sigma^{(j)}(s)\leq (2j+1) \frac{4\pi}{|\vec{p}|^2}
\end{equation}
which in the high-energy limit becomes
\begin{equation}
\sigma^{(j)}(s)\leq (2j+1) \frac{16\pi}{s}
\end{equation}
In the case of an inelastic process it is easy to derive analogous
inequalities; in high-energy limit (or for massless particles) one
gets
\begin{equation}
\sigma^{(j)}_\ti{inel.}(s) \leq (2j+1)\frac{4\pi}{s}
\end{equation}
We close this appendix by collecting some important formulae for
the Wigner $\mathscr{D}$-functions appearing in the Jacob--Wick
expansion. For a non-negative integer or half-integer $j$ one
defines
\begin{equation}
\mathscr{D}^{(j)}_{m'm}(\Omega) = \text{e}^{im\varphi}
d^{(j)}_{m'm}(\vartheta)
\end{equation}
where the indices $m, m'$ may only take on values $-j, -j+1,
\ldots, j-1, j$, and the functions $d^{(j)}_{m'm}(\vartheta)$ are
given by the general formula
\begin{multline}\label{eqB.53}
d^{(j)}_{m'm}(\vartheta) = \sum_k (-1)^{k-m+m'}
\frac{\bigl[(j+m)!(j-m)!(j+m')!(j-m')!\bigr]^{1/2}}{(j+m-k)!k!(j-k-m')!(k-m+m')!}\\
\times \Bigl(\cos\frac{\vartheta}{2}\Bigr)^{2j-2k+m-m'}
\Bigl(\sin\frac{\vartheta}{2}\Bigr)^{2k-m+m'}
\end{multline}
where the sum runs over the integers $k$ such that the arguments
of all factorials in (\ref{eqB.53}) are non-negative.

When $m=m'=0$, for an arbitrary integer $l\geq 0$ one has
\begin{equation}\label{eqB.54}
\mathscr{D}^{(l)}_{00}(\Omega) = P_l (\cos\vartheta)
\end{equation}
where $P_l$ is Legendre polynomial.

As another example, let us show the explicit form of the functions
$d^{(j)}_{m'm}(\vartheta)$ for $j = 1$:
\begin{align}
d_{11}^{(1)}(\vartheta) &= d_{\m1\m1}^{(1)}(\vartheta) =
\frac{1}{2}(1+\cos\vartheta)\notag\\
d_{00}^{(1)}(\vartheta) &= \cos\vartheta\notag\\
d_{1\m1}^{(1)}(\vartheta)&= d_{\m11}^{(1)}(\vartheta)
=\frac{1}{2}(1-\cos\vartheta)\notag\\
d_{10}^{(1)}(\vartheta) &= -d_{01}^{(1)}(\vartheta) =
d_{0\m1}^{(1)}(\vartheta) = - d_{\m10}^{(1)}(\vartheta) =
\frac{1}{\sqrt{2}}\sin\vartheta\label{eqB.55}
\end{align}
An orthogonality relation for the $\mathscr{D}$-functions reads:
\begin{equation}\label{eqB.56}
\int\mathscr{D}^{(j_1)*}_{m^{\prime}_1
m^{\phantom{\prime}}_1}(\Omega)\mathscr{D}^{(j_2)*}_{m^\prime_2
m^{\phantom{\prime}}_2}(\Omega) \frac{d\Omega}{4\pi} =
\frac{1}{2j_1+1}\delta_{j_1j_2}\delta_{m_1m_2}
\end{equation}
%\end{document}

%\input{app_c}
%%%%%%%%%%%%%%%%%%%%%%%%%%%%%%%%%%%%%%%%%%%%%%%%%%%%%%%%%%%%%%%%%%%
%%%%%%%%%%%%%%%%%%%%%%%%%%%%%%%%%%%%%%%%%%%%%%%%%%%%%%%%%%%%%%%%%%%%%%%%%%%%%%%%%%%%%%%%%%%%%%%%%%%%%%%%%%%%%%%%%%%%%%%%%%%%%%%%%%%%%%%%
%\documentclass[11pt,tbtags,kniha]{book2} \input{pream}
%\begin{document}
%\appendix
\chapter{Beta decay of polarized neutron}\label{appenC}
Our starting point is the beta-decay matrix element
\begin{align}
\mathcal{M}=\ &C_V (U^\dagger_p U_n)\bigl[\bar{u}_e(1+\alpha_V
\gamma_5)\gamma_0 v_\nu \bigr] \notag\\ +\ &C_A (U^\dagger_p
\sigma_j U_n) \bigl[\bar{u}_e (1+\alpha_A
\gamma_5)\gamma_5\gamma^j v_\nu\bigr]\label{eqC.1}
\end{align}
(cf. (\ref{eq1.111})), where the neutron is assumed to be
polarized along the third axis. It is easy to see that such an
assumption can be technically implemented by writing
\begin{equation}\label{eqC.2}
U_n U_n^\dagger = 2 M \frac{1+\sigma_3}{2}
\end{equation}
where the $M$ denotes, in accordance with conventions of Chapter~\ref{chap1}, the average nucleon mass.

First, we are going to calculate the matrix element (\ref{eqC.1})
squared, summing eventually over the spin states of $p, e,
\bar{\nu}$. Employing some familiar properties of the Dirac
matrices and introducing spinor traces in the usual way,
$|\mathcal{M}|^2$ can be written as
\begin{align}
&|\mathcal{M}|^2 = C_V^2 \Tr\bigl(U_p U^\dagger_p U_n
U^\dagger_n\bigr) \cdot \Tr\bigl[ (1-\alpha_V\gamma_5)u\bar{u}
(1+\alpha_V\gamma_5)\gamma_0 v\bar{v}\gamma_0 \bigr]\notag\\
&+C_A C_V \Tr\bigl(U_p U^\dagger_p U_n U^\dagger_n\sigma_j\bigr)
\cdot \Tr\bigl[ (\alpha_A-\gamma_5)u\bar{u}
(1+\alpha_V\gamma_5)\gamma_0 v\bar{v}\gamma^j \bigr] + \text{c.c.}\notag\\
&+C_A^2 \Tr\bigl(U_p U^\dagger_p \sigma_j U_n
U^\dagger_n\sigma_k\bigr) \cdot
\Tr\bigl[(\alpha_A-\gamma_5)u\bar{u} (\alpha_A+\gamma_5)\gamma^j
v\bar{v}\gamma^k \bigr]
\end{align}
(notice that we have suppressed here the labels $e, \nu$ at the
corresponding spinors, but this cannot lead to any confusion). The
spin summation indicated above is carried out in several steps.
Using (\ref{eqC.2}) and the relation
\begin{equation*}
\sum_\ti{spin} U_p U^\dagger_p = 2M\cdot \J
\end{equation*}
(with $\J$ denoting the $2 \times 2$ unit matrix, cf.
(\ref{eq1.39})), as well as the identities (\ref{eqA.66}), one
gets
\begin{align}
&\sum_{spin\ p, e, \bar{\nu}}|\mathcal{M}|^2 =\notag\\
&\phantom{+}4 M^2 C_V^2 \Tr\Bigl(\frac{1+\sigma_3}{2}\Bigr) \cdot
\Tr\bigl[ (1-\alpha_V\gamma_5)(\slashed{p}_e+m_e)
(1+\alpha_V\gamma_5)\gamma_0 \slashed{p}_{\bar{\nu}}\gamma_0 \bigr]\notag\\
&+4M^2 C_A C_V \Tr\bigl(\frac{1+\sigma_3}{2}\sigma_j\bigr) \cdot
\Tr\bigl[(\alpha_A-\gamma_5)(\slashed{p}_e+m_e)
(1+\alpha_V\gamma_5)\gamma_0 \slashed{p}_{\bar{\nu}}\gamma^j \bigr] + \text{c.c.}\notag\\
&+4 M^2 C_A^2 \Tr\bigl(\sigma_j \frac{1+\sigma_3}{2}\sigma_k\bigr)
\cdot \Tr\bigl[(\alpha_A-\gamma_5)(\slashed{p}_e+m_e)
(\alpha_A+\gamma_5)\gamma^j \slashed{p}_{\bar{\nu}}\gamma^k
\bigr]\label{eqC.4}
\end{align}
Next, with the help of standard identities for the Dirac and Pauli
matrices (see Appendix~\ref{appenA}, in particular (\ref{eqA.51})), eq.
(\ref{eqC.4}) is recast as
\begin{align}
\sum_{spin\ p, e, \bar{\nu}} |\mathcal{M}|^2 =\ &4 M^2 C_V^2 \cdot
\Tr\bigl[(1+\alpha_V^2 - 2\alpha_V\gamma_5)\slashed{p}_e\gamma_0
\slashed{p}_{\bar{\nu}}\gamma_0\bigr]\notag\\[-0.4cm]
+\ &4M^2 C_A C_V \cdot \Tr\bigl[\bigl(\alpha_A +\alpha_V
-(1+\alpha_A\alpha_V)\gamma_5\bigr) \slashed{p}_e \gamma_0
\slashed{p}_{\bar{\nu}} \gamma^3\bigr] + \text{c.c.}\notag\\
+\ &4M^2 C_A^2 \cdot \Tr\bigl[(1+\alpha_A^2 - 2\alpha_A \gamma_5)
\slashed{p}_e \gamma^j \slashed{p}_{\bar{\nu}} \gamma^j \bigr]\notag\\
-\ &4M^2 C_A^2 \cdot i \epsilon_{jk3} \Tr\bigl[(1+\alpha_A^2 -
2\alpha_A \gamma_5)\slashed{p}_e \gamma^j \slashed{p}_{\bar{\nu}}
\gamma^k \bigr]\label{eqC.5}
\end{align}
This expression can be simplified considerably just on the basis
of symmetry arguments. Indeed, in the first line on the right-hand
side of (\ref{eqC.5}), the term involving $\gamma_5$ vanishes
because of antisymmetry of the Levi-Civita symbol
$\epsilon_{\mu\nu\rho\sigma}$; the same is true for the third
line. In the second line, the term with $\gamma_5$ gives
effectively zero, since the corresponding trace is purely
imaginary and thus it gets cancelled when combined with its c.c.
counterpart. Finally, in the fourth line, the term without
$\gamma_5$ does not contribute, as the trace in question is
symmetric under $j\leftrightarrow k$ and the $\epsilon_{jk3}$ is
antisymmetric. Thus, (\ref{eqC.5}) is reduced to
\begin{align}
\sum_{spin\ p, e, \bar{\nu}} |\mathcal{M}|^2 =\ &4 M^2
C_V^2(1+\alpha_V^2) \cdot \Tr(\slashed{p}_e\gamma_0
\slashed{p}_{\bar{\nu}}\gamma_0)\notag\\[-0.4cm]
+\ &8M^2 C_A C_V (\alpha_A+\alpha_V)\cdot\Tr(\slashed{p}_e
\gamma_0\slashed{p}_{\bar{\nu}} \gamma^3)\notag\\
+\ &4M^2 C_A^2 (1+\alpha_A^2)\cdot \Tr(\slashed{p}_e \gamma^j \slashed{p}_{\bar{\nu}} \gamma^j)
\notag\\
+\ &16 i M^2 C_A^2 \alpha_A \cdot \Tr(\slashed{p}_e \gamma_1
\slashed{p}_{\bar{\nu}} \gamma_2 \gamma_5) \label{eqC.6}
\end{align}
Evaluating the traces in (\ref{eqC.6}) one gets, after some simple
manipulations
\begin{align}
\sum_{spin\ p, e, \bar{\nu}} |\mathcal{M}|^2 =\ &16 M^2
C_V^2(1+\alpha_V^2) (E_e E_{\bar{\nu}}+\vec{p}_e\cdot \vec{p}_{\bar{\nu}})\notag\\[-0.4cm]
+\ &32M^2 C_A C_V (\alpha_A+\alpha_V)(E_e p^3_{\bar{\nu}} + E_{\bar{\nu}}p^3_e))\notag\\
+\ &16M^2 C_A^2 (1+\alpha_A^2)(3E_e E_{\bar{\nu}}-\vec{p}_e\cdot
\vec{p}_{\bar{\nu}}) \notag\\
+\ &64 M^2 C_A^2 \alpha_A (E_e p^3_{\bar{\nu}} - E_{\bar{\nu}}
p_e^3) \label{eqC.7}
\end{align}
Now, components of the vectors $\vec{p}_e$, $\vec{p}_{\bar{\nu}}$
can be parametrized in terms of spherical angles as
\begin{align}
\vec{p}_e &= \bigl(|\vec{p}_e|\sin{\vartheta_e}\cos\varphi_e,\;
|\vec{p}_e|\sin\vartheta_e \sin\varphi_e,\; |\vec{p}_e|
\cos\vartheta_e\bigr)\notag\\
\vec{p}_{\bar{\nu}} &=
\bigl(|\vec{p}_{\bar{\nu}}|\sin{\vartheta_{\bar{\nu}}}\cos\varphi_{\bar{\nu}},\;
|\vec{p}_{\bar{\nu}}|\sin\vartheta_{\bar{\nu}}
\sin\varphi_{\bar{\nu}},\; |\vec{p}_{\bar{\nu}}|
\cos\vartheta_{\bar{\nu}}\bigr)\label{eqC.8}
\end{align}
Then the scalar product $\vec{p}_e\cdot \vec{p}_{\bar{\nu}}$ is
expressed as
\begin{equation}
\vec{p}_e\cdot \vec{p}_{\bar{\nu}} = |\vec{p}_e|.
|\vec{p}_{\bar{\nu}}| \sin\vartheta_e \sin\vartheta_{\bar{\nu}}
\cos(\varphi_e-\varphi_{\bar{\nu}}) +
|\vec{p}_e|.|\vec{p}_{\bar{\nu}}| \cos\vartheta_e
\cos\vartheta_{\bar{\nu}}\label{eqC.9}
\end{equation}
We are interested in the angular distribution of the electron with
respect to the direction of neutron polarization, which means that
the relevant variable is the $\vartheta_e$. To obtain the quantity
in question, one has to integrate the expression (\ref{eqC.7})
over directions of the $\vec{p}_{\bar{\nu}}$. Obviously, for a
fixed $\vec{p}_e$, one has
\begin{align}
\int_0^{2\pi} \cos (\varphi_e - \varphi_{\bar{\nu}})
d\varphi_{\bar{\nu}} &=0\notag\\
\int_0^{\pi} \cos \vartheta_{\bar{\nu}} \sin\vartheta_{\bar{\nu}}
d\vartheta_{\bar{\nu}} &=0\label{eqC.10}
\end{align}
(recall that the element of the relevant solid angle is
$d\Omega_{\bar{\nu}} =
\sin\vartheta_{\bar{\nu}}d\vartheta_{\bar{\nu}}d\varphi_{\bar{\nu}}$).
Thus, relations (\ref{eqC.8}), (\ref{eqC.9}) and (\ref{eqC.10})
make it clear that an integration over the directions of
antineutrino momentum eliminates from (\ref{eqC.7}) all terms
involving the scalar product $\vec{p}_e \cdot \vec{p}_{\bar{\nu}}$
or the $p^3_{\bar{\nu}}$. As a result, one obtains
\begin{align}
\int &\frac{d\Omega_{\bar{\nu}}}{4\pi} \sum_{spin\ p,e,\bar{\nu}}
|\mathcal{M}|^2 = 16 M^2 E_e E_{\bar{\nu}} \\ &\times\bigl[ C_V^2
(1+\alpha_V^2) + 3 C_A^2 (1+\alpha_A^2) + \bigl(2C_A C_V (\alpha_A
+ \alpha_V) - 4 C_A^2 \alpha_A\bigr)\beta\cos\vartheta_e \bigr]
\notag
\end{align}
with $\beta=|\vec{p}_e|/E_e$. This is precisely the formula
(\ref{eq1.113}) of Chapter~\ref{chap1} (there we have set, conventionally,
$E=E_e$).
%\end{document}

%\input{app_d}
%%%%%%%%%%%%%%%%%%%%%%%%%%%%%%%%%%%%%%%%%%%%%%%%%%%%%%%%%%%%%%%%%%%
%%%%%%%%%%%%%%%%%%%%%%%%%%%%%%%%%%%%%%%%%%%%%%%%%%%%%%%%%%%%%%%%%%%%%%%%%%%%%%%%%%%%%%%%%%%%%%%%%%%%%%%%%%%%%%%%%%%%%%%%%%%%%%%%%%%%%%%%
%\documentclass[11pt,tbtags,kniha]{book2} \input{pream}
%\begin{document}
%\appendix
\chapter{Massive vector bosons}\label{appenD}
\index{quantization of!massive vector field|appD} \index{Proca
equation|appD}
\index{polarization!vector|appD}\index{polarization!sum|appD}
\index{polarization!of massive vector boson|appD}\index{massive
vector boson|appD} \index{Lorenz!condition|appD}\index{transverse
polarization|appD} \index{propagator!of massive vector
boson|appD}\index{polarization|appD} Relativistic theory of free
massive particles with spin 1 is based on the Proca equation
\begin{equation}\label{eqD.1}
\partial_\mu F^{\mu\nu} + m^2 B^\nu = 0
\end{equation}
where
\begin{equation}\label{eqD.2}
F^{\mu\nu} = \partial^\mu B^\nu - \partial^\nu B^\mu
\end{equation}
and $B^\mu=B^\mu(x)$ is a four-vector under Lorentz
transformations of space-time coordinates. Similarly as any other
relativistic wave equation, (\ref{eqD.1}) can either be used as
one-particle equation of relativistic quantum mechanics, or it is
treated as the equation of motion of a classical vector field that
is subsequently quantized in terms of spin-1 particles with
non-zero mass.

Before discussing solutions and other properties of
eq. (\ref{eqD.1}), the following important comment is in order
here. Substituting (\ref{eqD.2}) into (\ref{eqD.1}), one gets
\begin{equation}\label{eqD.3}
(\Box + m^2) B^\nu - \partial^\nu (\partial \cdot B) = 0
\end{equation}
where we have denoted $\partial\cdot B\equiv \partial_\mu B^\mu$.
Acting on (\ref{eqD.3}) with $\partial_\nu$, it is easy to see
that one is ultimately left with $m^2 \partial\cdot B=0$ and,
since $m\neq 0$, this yields
\begin{equation}\label{eqD.4}
\partial_\mu B^\mu = 0
\end{equation}
In other words, a \qq{Lorenz condition} follows directly from the
equation of motion (\ref{eqD.1}); obviously, the crucial point in
this respect is that $m\neq 0$ (remember that for Maxwell
equations the Lorenz condition has to be added by hand). Thus,
looking back at (\ref{eqD.3}), it becomes clear that instead of
(\ref{eqD.1}), one can write a pair of equations
\begin{equation}\label{eqD.5}
(\Box + m^2) B^\mu = 0, \qquad \partial_\mu B^\mu =0
\end{equation}
(i.e. one has the Klein--Gordon equation for each component
$B^\mu$, supplemented with the Lorenz condition). Now, since a
passage in the reverse direction, i.e. from (\ref{eqD.5}) to
(\ref{eqD.1}), is quite obvious, one can conclude that
eq.\,(\ref{eqD.1}) is {\it equivalent\/} to (\ref{eqD.5}). The
condition (\ref{eqD.4}) involves only the first time derivative
and represents, in fact, a constraint on the components of the
four-vector $B^\mu$: only three of them are thus independent,
corresponding to the three internal degrees of freedom of a
massive spin-1 particle.

Let us now describe briefly the plane-wave solutions of eq.
(\ref{eqD.5}). For such a solution we use an Ansatz
\begin{equation}\label{eqD.6}
B_\mu(x) = \varepsilon_\mu(k) \text{e}^{-ikx}
\end{equation}
with $k=(k^0,\ \vec{k})$. The $\varepsilon_\mu(k)$ is called, in
analogy with an electromagnetic plane wave, a \qq{polarization
vector}; its components are, in general, complex. We omit here a
usual normalization factor, as this is inessential for the present
purpose. Note that the $\varepsilon_\mu(k)$ plays a similar role
as the $u(k)$ in a plane-wave solution of the Dirac equation.
Obviously, another independent solution of eq. (\ref{eqD.5}) is
obtained by complex conjugation of (\ref{eqD.6}). Substituting
(\ref{eqD.6}) into (\ref{eqD.5}), the Klein--Gordon equation
yields immediately the mass-shell condition for the $k$, i.e.
\begin{equation}
k^2 = (k_0)^2 -\vec{k}^2 = m^2
\end{equation}
(we shall assume, conventionally, that $k_0>0$) and the Lorenz
condition turns into the requirement of transversality of the
polarization vector in the four-dimensional momentum space:
\begin{equation}\label{eqD.8}
k^\mu \varepsilon_\mu(k) = 0
\end{equation}
It is not difficult to realize that for a given $k$ there are
three independent four-vectors $\varepsilon(k)$ satisfying the
condition (\ref{eqD.8}) and, moreover, that they are {\it
space-like}. The argument goes as follows. Because of the
four-vector character of the $k$ and $\varepsilon(k)$, the scalar
product $k\cdot \varepsilon(k)$ has the same value in any Lorentz
frame. In particular, one can pass to the rest system of the $k$,
in which $k=k^{(0)}=(m,\ \vec{0})$; from (\ref{eqD.8}) it is then
obvious that the time component of any polarization vector must
vanish in such a system. This means, generally, that the
$\varepsilon(k)$ is of space-like character. It is also clear that
there are just three such vectors -- these correspond to the three
linearly independent spatial directions in the rest frame. Note
also that the normalization of an $\varepsilon(k)$ is
conventionally fixed by
\begin{equation}\label{eqD.9}
\varepsilon(k) \cdot \varepsilon^*(k)=-1
\end{equation}
(here we take into account that the $\varepsilon(k)$ may be
complex).

The polarization vectors in question are labelled, for a given
$k$, as $\varepsilon^\mu(k,\lambda)$, with $\lambda = 1, 2, 3$. A
particularly useful triad can be defined in the following manner.
The $\varepsilon(k,1)$ and $\varepsilon(k,2)$ are taken in the
form
\begin{align}
\varepsilon^\mu(k,1) &= \bigl(0,\ \vec{\varepsilon}{\,}^{(1)}(\vec{k})\bigr)\notag\\
\varepsilon^\mu(k,2) &= \bigl(0,\
\vec{\varepsilon}{\,}^{(2)}(\vec{k})\bigr) \label{eqD.10}
\end{align}
where the $\vec{\varepsilon}{\,}^{(\lambda)}$, $\lambda= 1,2$ are
two linearly independent vectors lying in the plane perpendicular
to the $\vec{k}$ (it means that
$\vec{k}\cdot\vec{\varepsilon}{\,}^{(1)}=0$,
$\vec{k}\cdot\vec{\varepsilon}{\,}^{(2)}=0$). As for the
$\varepsilon(k,3)$, this is chosen to have its spatial part
directed along the $\vec{k}$. Thus, it can be written as
\begin{equation}
\varepsilon^\mu(k,3) = (\varepsilon^0,\
\alpha\frac{\vec{k}}{|\vec{k}|})
\end{equation}
with $\alpha>0$. The parameters $\varepsilon^0$ and $\alpha$ are
determined uniquely by making use of the conditions (\ref{eqD.8}),
(\ref{eqD.9}) and one gets
\begin{equation}
\varepsilon^\mu(k,3) = \Bigl(\frac{|\vec{k}|}{m},\
\frac{k_0}{m}\frac{\vec{k}}{|\vec{k}|}\Bigr)
\end{equation}
where $k_0=\sqrt{\vec{k}^2+m^2}$.

In usual terminology, the $\varepsilon(k,3)$ is called {\bf
longitudinal polarization} vector, while the $\varepsilon(k,1)$
and $\varepsilon(k,2)$ correspond to two independent {\bf
transverse polarizations}. For practical purposes, it is
convenient to introduce a specific symbol for longitudinal
polarization: thus, we will usually denote the $\varepsilon(k,3)$
as $\varepsilon_L(k)$. As regards the transverse polarizations
shown in (\ref{eqD.10}), the $\vec{\varepsilon}{\,}^{(1)}$,
$\vec{\varepsilon}{\,}^{(2)}$ can be chosen e.g. as two real (and
mutually orthogonal) vectors; in such a case we speak of
\qq{linear polarizations}. Next, one can also form complex vectors
\begin{equation}\label{eqD.13}
\vec{\varepsilon}_\pm =
\frac{1}{\sqrt{2}}(\vec{\varepsilon}{\,}^{(1)}\pm i
\vec{\varepsilon}{\,}^{(2)})
\end{equation}
corresponding to \qq{circular polarizations}. Needless to say,
such a terminology is based on a straightforward analogy with
electromagnetic plane waves. Notice that the considered
polarization vectors obviously satisfy orthonormality relations
\begin{equation}
\varepsilon(k,\lambda)\cdot \varepsilon^*(k,\lambda') =
-\delta_{\lambda\lambda'}
\end{equation}
In the context of relativistic quantum mechanics of spin-1 bosons,
it is important to note that the plane waves specified above
describe the states with definite helicities (and fixed
energy-momentum): in particular, the circular transverse
polarizations
\begin{equation}
\varepsilon(k,\pm) = (0,\ \vec{\varepsilon}_\pm)
\end{equation}
correspond to helicities $\pm 1$ (right-handed and left-handed
motion respectively) and the longitudinally polarized plane wave
carries the helicity zero. For more details concerning this issue,
see e.g. the Appendix H in \cite{Hor}. Thus, a \qq{canonical} set
of polarization vectors can be taken as consisting of the
$\varepsilon(k,\pm)$ given by (\ref{eqD.13}) and the
$\varepsilon_L(k)$,
\begin{equation}\label{eqD.16}
\varepsilon_L^\mu (k) = \Bigl(\frac{|\vec{k}|}{m},\
\frac{k_0}{m}\frac{\vec{k}}{|\vec{k}|}\Bigr)
\end{equation}
An astute reader may observe that the longitudinal polarization
vector of spin-1 boson coincides with the spin four-vector
describing helicity of a spin-$\frac{1}{2}$ fermion. Of course,
this is not surprising as the relevant requirements are formally
the same in both cases; however, the physical meaning of the two
quantities is different: as we noted before, the $\varepsilon(k)$
plays the role of a one-particle wave function in momentum space.

In practical calculations, one needs some further particular
properties of the polarization vectors $\varepsilon(k,\lambda)$.
First, from (\ref{eqD.16}) one can infer quite easily that in the
high-energy limit, components of the $\varepsilon_L(k)$ behave
essentially as the four-momentum $k$ itself; in explicit terms,
the relevant statement reads
\begin{equation}\label{eqD.17}
\varepsilon_L^\mu(k) = \frac{k^\mu}{m} +
O\Bigl(\frac{m}{|\vec{k}|}\Bigr),\qquad \text{for } |\vec{k}|\gg m
\end{equation}
(of course, the remainder in (\ref{eqD.17}) could also be written
as $O(m/k_0)$). On the other hand, components of a transverse
polarization vector $\varepsilon_T(k)$ cannot grow indefinitely:
there is an obvious bound $|\varepsilon_T^\mu(k)|\leq 1$ set by
the euclidean norm of the $\vec{\varepsilon}{\,}^{(\lambda)}$,
$\lambda = 1, 2$ in (\ref{eqD.10}). Another important formula is
the \qq{completeness relation}
\begin{equation}\label{eqD.18}
\sum_{\lambda=1}^3
\varepsilon_\mu(k,\lambda)\varepsilon_\nu^*(k,\lambda) =
-g_{\mu\nu} + \frac{1}{m^2} k_\mu k_\nu
\end{equation}
One should notice that this is an analogue of the identities
(\ref{eqA.66}) for Dirac spinors. A straightforward proof of eq.
(\ref{eqD.18}) goes as follows. For a given $k$ satisfying
$k^2=m^2$ one considers the unit time-like vector
\begin{equation}
\varepsilon^\mu(k,0) = \frac{1}{m} k^\mu
\end{equation}
together with the space-like polarization vectors
$\varepsilon(k,\lambda)$ described above. The
$\varepsilon(k,\lambda)$, $\lambda = 0, 1, 2, 3$ obviously satisfy
an orthonormality relation
\begin{equation}\label{eqD.20}
\varepsilon(k,\lambda)\cdot \varepsilon^*(k,\lambda') =
g_{\lambda\lambda'}
\end{equation}
and form a basis in the four-dimensional space endowed with the
usual metric. The latter statement means that
\begin{equation}\label{eqD.21}
\varepsilon_\mu(k,0)\varepsilon^*_\nu(k,0) - \sum_{\lambda=1}^3
\varepsilon_\mu(k,\lambda)\varepsilon^*_\nu (k,\lambda)=g_{\mu\nu}
\end{equation}
(this can be verified easily by multiplying both sides of
(\ref{eqD.21}) with $\varepsilon^\nu(k,\lambda')$, taking
consecutively $\lambda' = 0, 1, 2, 3$ and utilizing
(\ref{eqD.20})). From (\ref{eqD.21}) then immediately follows the
result (\ref{eqD.18}) for the polarization sum in question.

For reader's convenience, let us also add that there is an
independent and frequently used argument for (\ref{eqD.18}), which
can be formulated in the following way. Since the
$\varepsilon(k,\lambda)$, $\lambda = 1, 2, 3$ are supposed to be
four-vectors, the polarization sum on the left-hand side of
(\ref{eqD.18}) should be a 2nd rank Lorentz tensor depending on
the four-momentum $k$. Thus, on general grounds, one can write
\begin{equation}\label{eqD.22}
\sum_{\lambda=1}^3 \varepsilon_\mu(k,\lambda) \varepsilon_\nu^*
(k,\lambda) = A g_{\mu\nu} + B k_\mu k_\nu
\end{equation}
where the coefficients $A$ and $B$ may only depend on $k^2$;
however, one has $k^2=m^2$ and, therefore, $A$, $B$ are simply
constants. Now, multiplying (\ref{eqD.22}) with $k^\mu$ and
utilizing (\ref{eqD.8}), one gets the constraint
\begin{equation}\label{eqD.23}
A + B m^2 = 0
\end{equation}
Further, one can raise e.g. the index $\nu$ in both sides of
(\ref{eqD.22}) and take then the corresponding trace; this yields
\begin{equation}\label{eqD.24}
4 A + B m^2 = -3
\end{equation}
Solving (\ref{eqD.23}) and (\ref{eqD.24}) one obtains
\begin{equation}
A=-1, \qquad B=\frac{1}{m^2}
\end{equation}
and the result (\ref{eqD.18}) is thus recovered.

Let us now discuss quantization of the free massive vector field.
For simplicity, we shall consider the case of a real (hermitean)
field. The relevant Lagrangian density can be written as
\begin{equation}\label{eqD.26}
\lagr = -\frac{1}{4}F_{\mu\nu}F^{\mu\nu} + \frac{1}{2}m^2 B_\mu
B^\mu
\end{equation}
(here and in what follows we usually write simply $B_\mu$ instead
of $B_\mu(x)$ etc.). It is not difficult to verify that
(\ref{eqD.26}) yields (\ref{eqD.1}) as the corresponding equation
of motion. Indeed, from (\ref{eqD.26}) one gets
\begin{equation}\label{eqD.27}
\frac{\delta\lagr}{\delta(\partial_\mu B_\nu)} = -F^{\mu\nu}
\end{equation}
and
\begin{equation}
\frac{\delta\lagr}{\delta B_\nu} = m^2 B^\nu
\end{equation}
Using now these results in the Euler--Lagrange equation
\begin{equation}
\partial_\mu \frac{\delta\lagr}{\delta(\partial_\mu B_\nu)} -
\frac{\delta\lagr}{\delta B_\nu} = 0
\end{equation}
one recovers immediately eq. (\ref{eqD.1}).

According to our previous analysis, only three of the four field
components $B_\mu$ are to be taken as independent, since they are
constrained by
\begin{equation}\label{eqD.30}
\partial^\mu B_\mu = 0
\end{equation}
(cf. the remark following eq. (\ref{eqD.5})). For the purpose of
canonical quantization, we take the $B^j$, $j = 1, 2, 3$ as the
relevant independent variables (\qq{generalized coordinates}) and
the $B_0$ is understood as a solution of the constraint
(\ref{eqD.30}). The corresponding canonically conjugate momenta
are defined in the usual way:
\begin{equation}\label{eqD.31}
\pi_j \equiv \frac{\delta\lagr}{\delta(\partial_0 B_j)}
\end{equation}
(the reader should not be confused by the seemingly non-covariant
position of the indices -- the $\pi_j$ is simply a convenient
notation for canonical momentum associated with the $B_j$). Using
(\ref{eqD.27}), the definition (\ref{eqD.31}) yields
\begin{equation}\label{eqD.32}
\pi_j = F_{0j} = \partial_0 B_j - \partial_j B_0
\end{equation}
(note that (\ref{eqD.27}) also makes it clear that the canonical
momentum conjugate to the $B_0$ would be identically zero).

Let us see how the constraint (\ref{eqD.30}) can be solved, i.e.
whether and how the $B_0$ can be expressed in terms of our
canonical variables. To this end, it is convenient to utilize
directly the original form (\ref{eqD.1}) of the equation of
motion. One thus gets
\begin{equation}\label{eqD.33}
B_0 = -\frac{1}{m^2} \partial_j F_{0j}
\end{equation}
(of course, the symbol $\partial_j$ ( $=-\partial^j$) stands for
$\partial/\partial x^j$ as usual). Taking now into account
(\ref{eqD.32}), eq. (\ref{eqD.33}) is recast as
\begin{equation}\label{eqD.34}
B_0 = -\frac{1}{m^2} \partial_j \pi_j
\end{equation}
and this, of course, is a crucial result since the last expression
involves only derivatives of canonical momenta with respect to the
space coordinates.

For canonical quantization, one postulates the equal-time (E.T.)
commutation relations
\begin{align}
[B_j(x),\,B_k(y)]_{E.T.} &= 0\notag\\
[\pi_j(x),\,\pi_k(y)]_{E.T.} &= 0\notag\\
[B_j(x),\,\pi_k(y)]_{E.T.} &= i \delta_{jk}
\delta^3(\vec{x}-\vec{y}) \label{eqD.35}
\end{align}
Any component $B_\mu(x)$ is a solution of (\ref{eqD.5}) and thus
can be written in terms of a plane-wave expansion
\begin{align}
B_\mu(x) = \sum_{\lambda=1}^3 \int \frac{d^3
k}{(2\pi)^{3/2}(2k_0)^{1/2}} \bigl[&\,a(k,\lambda)\varepsilon_\mu
(k,\lambda) \text{e}^{-ikx}\notag\\ +\,&\,a^+(k,\lambda)
\varepsilon_\mu^*(k,\lambda) \text{e}^{ikx}\bigr]\label{eqD.36}
\end{align}
where $k_0=\sqrt{\vec{k}^2+m^2}$ and the
$\varepsilon_\mu(k,\lambda)$ are polarization vectors described
above. We assume that the field operator $B_\mu$ is hermitean, so
that the $a^+(k,\lambda)$ is hermitean conjugate of
$a(k,\lambda)$, i.e. $a^+(k,\lambda)=a^\dagger(k,\lambda)$. Of
course, the $a(k,\lambda)$, $a^+(k,\lambda)$ are to be identified
with annihilation and creation operators corresponding to
particles (vector bosons) with definite energy-momentum and spin
(polarization).

For convenience, we also introduce the linear combinations
\begin{align}
a_\mu(k) &\equiv \sum_{\lambda=1}^3 a(k,\lambda)
\varepsilon_\mu(k,\lambda)\notag\\
a^+_\mu(k) \equiv a_\mu^\dagger (k) &= \sum_{\lambda=1}^3
a^+(k,\lambda)\varepsilon^*_\mu(k,\lambda)\label{eqD.37}
\end{align}
The $a_\mu(k)$ and $a^+_\mu(k)$ can be calculated from
(\ref{eqD.36}) and expressed in terms of the $B_\mu(x)$ and time
derivatives $\dot{B}_\mu(x)=\partial_0 B_\mu(x)$ (for the relevant
technique, see e.g. \cite{BjD}). Employing canonical commutation
relations (\ref{eqD.35}) and the result (\ref{eqD.34}) (as well as
eq.\,(\ref{eqD.30})), one can evaluate all possible commutators of
the $B_\mu$ and $\dot{B}_\nu$ for $\mu, \nu = 0, 1, 2, 3$ and thus
one is also able to determine all commutators involving the
momentum-space operators $a_\mu(k)$ and $a_\nu^+(k')$. The
calculation is rather tedious, but the result is rewarding and
easy to remember:
\begin{align}
[a_\mu(k),\,a_\nu(k')] &= 0 \notag\\
[a^+_\mu(k),\,a^+_\nu(k')] &= 0 \notag\\
[a_\mu(k),\,a^+_\nu(k')] &= (-g_{\mu\nu} + \frac{1}{m^2}k_\mu
k_\nu) \delta^3 (\vec{k}-\vec{k}')\label{eqD.38}
\end{align}
(let us stress again that the four-momenta labelling the operators
$a$, $a^+$ are on the mass shell, i.e.
$k_0=\sqrt{\vec{k}^2+m^2}$). Next, making use of the
orthonormality properties of the polarization vectors
$\varepsilon(k,\lambda)$, one can solve eq.\,(\ref{eqD.37}) and
turn subsequently the relations (\ref{eqD.38}) into an algebra of
the operators $a(k,\lambda)$ and $a^+(k,\lambda)$. The calculation
is straightforward and the result is, as expected,
\begin{align}
[a(k,\lambda),\,a(k',\lambda')] &= 0 \notag\\
[a^+(k,\lambda),\,a^+(k',\lambda')] &= 0 \notag\\
[a(k,\lambda),\,a^+(k',\lambda')] &= \delta_{\lambda\lambda'}
\delta^3 (\vec{k}-\vec{k}')
\end{align}
Of course, for interpretation of the $a(k,\lambda)$ and
$a^+(k,\lambda)$ as annihilation and creation operators one has to
calculate the relevant physical quantities (energy, momentum,
etc.) for the considered quantized field. Here we take the
connection of the operators $a$, $a^+$ with one-particle vector
boson states for granted; the main purpose of the preceding
discussion was to emphasize that massive vector field is a
constrained system that can be canonically quantized in a
straightforward way -- by solving explicitly the constraint in
terms of canonical variables (see eq. (\ref{eqD.34})).

Now we are going to discuss briefly the Feynman propagator. One
can start with the definition
\begin{equation}\label{eqD.40}
i\mathcal{D}_{\mu\nu}(x-y) = \langle 0 |
T\bigl(B_\mu(x)B_\nu(y)\bigr) |0\rangle
\end{equation}
where the time-ordered operator product (or simply $T$-product) in
(\ref{eqD.40}) is conventionally defined by means of the Heaviside
step function; let us recall that such a definition reads, in
general
\begin{equation}\label{eqD.41}
T\bigl(A(x)B(y)\bigr) = \theta(x_0-y_0) A(x)B(y) +
\theta(y_0-x_0)B(y)A(x)
\end{equation}
if one considers two bosonic operators $A$, $B$ depending on
space-time coordinates. Using in (\ref{eqD.40}) the decomposition
(\ref{eqD.36}), one obtains (after a somewhat tedious calculation)
the result for $\mathcal{D}_{\mu\nu}$ in the usual form of Fourier
integral:
\begin{equation}\label{eqD.42}
\mathcal{D}_{\mu\nu}(x-y) = \int\frac{d^4 q}{(2\pi)^4} \Bigl(
\frac{P_{\mu\nu}(q)}{q^2-m^2+i\varepsilon} - \frac{1}{m^2}
g_{0\mu}g_{0\nu}\Bigr) \text{e}^{iq(x-y)}
\end{equation}
where
\begin{equation}
P_{\mu\nu}(q) = -g_{\mu\nu} + \frac{1}{m^2}q_\mu q_\nu
\end{equation}
and the $+i\varepsilon$ prescription has the usual meaning as in
any other Feynman propagator.

A remarkable feature of the expression (\ref{eqD.42}) is that,
apart from the \qq{normal} covariant term involving the tensor
$P_{\mu\nu}(q)$, there is a non-covariant contribution
proportional to $g_{0\mu}g_{0\nu}$; obviously, this has contact
character, since the integration of the exponential factor yields
four-dimensional delta function. Note that the appearance of such
a term is related to the non-covariant nature of the conventional
$T$-product (\ref{eqD.41}). Thus, (\ref{eqD.42}) can be written as
\begin{equation}\label{eqD.44}
\mathcal{D}_{\mu\nu}(x-y) = \mathcal{D}_{\mu\nu}^\tiz{covar.}(x-y)
- \frac{1}{m^2} g_{0\mu} g_{0\nu} \delta^4(x-y)
\end{equation}
where
$$
\mathcal{D}_{\mu\nu}^\tiz{covar.}(x-y) = \int
\frac{d^4q}{(2\pi)^4} D_{\mu\nu}^\tiz{covar.}(q)
\text{e}^{iq(x-y)}
$$
with
\begin{equation}\label{eqD.45}
D_{\mu\nu}^\tiz{covar.}(q) = \frac{-g_{\mu\nu}+ m^{-2}q_\mu
q_\nu}{q^2-m^2 + i\varepsilon}
\end{equation}

The presence of the non-covariant term in the propagator
(\ref{eqD.44}) seems to be a disturbing feature of the theory of
massive vector bosons. Nevertheless, in Feynman diagram
calculations within common field theory models one does employ the
familiar form (\ref{eqD.45}), simply omitting the non-covariant
terms. A basic reason for that is, briefly, the following. To
develop the perturbation expansion in the usual Dirac picture, one
passes from an interaction Lagrangian $\lagr_\ti{int}$ to a
corresponding Hamiltonian $\mathscr{H}_\ti{int}$. It turns out
that the $\mathscr{H}_\ti{int}$ differs from $-\lagr_\ti{int}$ by
an additional term, which cancels exactly the contribution of the
contact non-covariant term in the propagator (\ref{eqD.44}). A
technical discussion of this issue would go beyond the scope of
this appendix; we have mentioned it here in order to make the
reader aware of subtleties and possible pitfalls of the canonical
operator quantization of the massive vector field. For a detailed
exposition, see e.g. \cite{Chg}.

It is useful to know that there is another independent way how to
arrive at the covariant form (\ref{eqD.45}). It is based on the
observation that the propagator in question can also be understood
as the (causal) Green's function\index{Green's function} of the
Proca equation (\ref{eqD.1}); as we shall see, such an approach is
well suited for practical calculations. Let us show how this
works. To find the covariant propagator function
$\mathcal{D}_{\mu\nu}(x)$, one has to solve the equation
\begin{equation}\label{eqD.46}
(\Box + m^2) \mathcal{D}^\mu{}_\nu(x)-\partial^\mu
\bigl(\partial_\lambda \mathcal{D}^\lambda{}_\nu (x)\bigr) =
g^\mu{}_\nu \delta^4(x)
\end{equation}
(let us recall again that $g^\mu{}_\nu=\delta^\mu_\nu$).
Performing Fourier transformation, i.e. defining a function
$D_{\mu\nu}(q)$ through
\begin{equation}
\mathcal{D}_{\mu\nu}(x) =\int \frac{d^4q}{(2\pi)^4} D_{\mu\nu}(q)
\text{e}^{iqx}
\end{equation}
one gets from (\ref{eqD.46}) the system of linear algebraic
equations
\begin{equation}
(-q^2 + m^2) D^\mu{}_\nu(q) + q^\mu q_\lambda D^\lambda{}_\nu(q) =
g^\mu{}_\nu
\end{equation}
that can be written compactly as
\begin{equation}\label{eqD.49}
L^\mu{}_\lambda D^\lambda{}_\nu = g^\mu{}_\nu
\end{equation}
with
\begin{equation}\label{eqD.50}
L^\mu{}_\lambda = (-q^2+m^2)g^\mu{}_\lambda + q^\mu q_\lambda
\end{equation}
The $D^{\mu\nu}(q)$ is a 2nd rank tensor depending on a
four-vector $q$ and, therefore, its most general form reads
\begin{equation}
\label{eqD.51} D^{\mu\nu}(q) = D_T (q^2) P_T^{\mu\nu} (q) +
D_L(q^2)P_L^{\mu\nu}(q)
\end{equation}
where
\begin{align}
P_T^{\mu\nu} &= g^{\mu\nu} - \frac{q^\mu q^\nu}{q^2}\notag\\
P_L^{\mu\nu} &= \frac{q^\mu q^\nu}{q^2}\label{eqD.52}
\end{align}
Denoting as $P_T$ and $P_L$ the $4 \times 4$ matrices whose
elements coincide with the mixed components of tensors
(\ref{eqD.52}), one finds easily that
\begin{equation}\label{eqD.53}
P_T^2 = P_T,\quad P_L^2 = P_L,\quad P_T P_L = P_L P_T =0
\end{equation}
Thus, the matrices $P_T$, $P_L$ represent orthogonal projectors
(this is the main advantage of the form (\ref{eqD.51}) over a
parametrization in terms of the basis made simply of $g_{\mu\nu}$
and $q_\mu q_\nu$). The matrix $L$ defined in (\ref{eqD.50}) can
be recast, accordingly, as
\begin{equation}
L = (-q^2 + m^2) P_T + m^2 P_L
\end{equation}
The matrix equation (\ref{eqD.49}) can now be solved easily by
utilizing the relations (\ref{eqD.53}); taking into account that
the unit matrix on the right-hand side of (\ref{eqD.49}) can be
decomposed as $P_T + P_L$, one gets readily
\begin{equation}\label{eqD.55}
D_T = \frac{1}{-q^2 + m^2}, \qquad D_L =\frac{1}{m^2}
\end{equation}
for $q^2\neq m^2$. This is the desired answer; substituting
(\ref{eqD.55}) into (\ref{eqD.51}) one recovers the result
(\ref{eqD.45}).

Of course, within the approach described above one has to make the
replacement $m^2 \rightarrow m^2 - i\varepsilon$ in the propagator
denominator by hand (relying on the general knowledge of
properties of causal Green's functions). In this context, one
should keep in mind that there are infinitely many Green's
functions, which are all solutions of the original eq.
(\ref{eqD.46}) (to any particular solution of the inhomogeneous
equation (\ref{eqD.46}) one may add an arbitrary solution of the
corresponding homogeneous equation); by removing the singularity
at $q^2=m^2$ in a specific way (e.g. through the $i\varepsilon$
prescription), the ambiguity is fixed. In any case, the nice
feature of the tensor method explained above is that it provides a
very efficient tool for finding the algebraic form of the
propagator in momentum space; such an approach can be used
conveniently in many other situations.

In closing this appendix, let us add that most of the previous
results can be generalized almost without changes to the case of a
complex (non-hermitean) vector field. Denoting, for convenience,
the four components of such a field as $W^-_\mu$, a corresponding
free Lagrangian can be written as
\begin{equation}\label{eqD.56}
\lagr = -\frac{1}{2}(\partial_\mu W^-_\nu - \partial_\nu
W^-_\mu)(\partial^\mu W^{+\nu} - \partial^\nu W^{+\mu}) + m^2
W^-_\mu W^{+\mu}
\end{equation}
where $W^+_\mu = (W^-_\mu)^*$ in a classical theory or
$W^+_\mu=(W^-_\mu)^\dagger$ in the quantum case. The $W^-_\mu$ and
$W^+_\mu$ are treated as independent dynamical variables (for this
reason, the coefficients in (\ref{eqD.56}) differ from those in
(\ref{eqD.26})). The plane-wave expansion of a quantized field
$W^\pm_\mu$ is written as
\begin{align}W_\mu^-(x) &= \sum_{\lambda=1}^3 \int
\frac{d^3k}{(2\pi)^{3/2}(2k_0)^{1/2}} \bigl[ b(k,\lambda)
\varepsilon_\mu (k,\lambda) \text{e}^{-ikx}
\notag\\&\hspace{4.5cm}+
d^+(k,\lambda) \varepsilon_\mu^* (k,\lambda) \text{e}^{ikx}\bigr] \notag\\
W_\mu^+(x) &= \sum_{\lambda=1}^3 \int
\frac{d^3k}{(2\pi)^{3/2}(2k_0)^{1/2}} \bigl[ b^+(k,\lambda)
\varepsilon^*_\mu (k,\lambda) \text{e}^{ikx}
\notag\\&\hspace{4.5cm}+ d(k,\lambda) \varepsilon_\mu(k,\lambda)
\text{e}^{-ikx}\bigr]\label{eqD.57}
\end{align}
where $b$, $b^+$ are the annihilation and creation operators of
particles (conventionally taken to be the $W^-$ bosons), and the
$d$, $d^+$ play an analogous role for the antiparticles ($W^+$).
The other symbols have the same meaning as in (\ref{eqD.36}). The
algebra of creation and annihilation operators now reads
\begin{align}
&[b(k,\lambda),\,b(k',\lambda')] =
[d(k,\lambda),\,d(k',\lambda')] =0\notag\\
&[b(k,\lambda),\,b^+(k',\lambda')]=[d(k,\lambda),\,d^+(k',\lambda')]
=\delta_{\lambda\lambda'}\delta^3(\vec{k}-\vec{k}')\notag\\
&[b(k,\lambda),\,d(k',\lambda')]=0,\qquad
[b(k,\lambda),\,d^+(k',\lambda')]=0\label{eqD.58}
\end{align}
where we have omitted commutators that follow from (\ref{eqD.58})
by hermitean conjugation. Note that from the representation
(\ref{eqD.57}) one can infer the following rule for external lines
corresponding to vector bosons: an incoming line always
contributes a factor of $\varepsilon_\mu(k,\lambda)$ and the
outgoing line a factor of $\varepsilon^*_\mu(k,\lambda)$,
independently of whether the line in question represents a
particle or antiparticle. The Feynman propagator can be defined
through the time-ordered product of  the $W^-_\mu(x)$ and
$W^+_\nu(y)$; the formula (\ref{eqD.45}) remains unchanged.
%\end{document}

%\input{app_e}
%%%%%%%%%%%%%%%%%%%%%%%%%%%%%%%%%%%%%%%%%%%%%%%%%%%%%%%%%%%%%%%%%%%
%%%%%%%%%%%%%%%%%%%%%%%%%%%%%%%%%%%%%%%%%%%%%%%%%%%%%%%%%%%%%%%%%%%%%%%%%%%%%%%%%%%%%%%%%%%%%%%%%%%%%%%%%%%%%%%%%%%%%%%%%%%%%%%%%%%%%%%%
%\documentclass[11pt,tbtags,kniha]{book2} \input{pream}
%\begin{document}
%\appendix
\chapter{Basics of the ABJ anomaly}\label{appenE}
\index{intermediate vector boson|seealso{$W$, $Z$ bosons}}
\index{Pauli--Villars
method|appE}\index{regularization|appE}\index{anomalous Ward
identity|appE}\index{VVA triangle@$VVA$ triangle graph}

 In this appendix we derive the basic formula for the
Adler--Bell--Jackiw (ABJ) axial anomaly, employed in Section~\ref{sec7.9}.
The anomaly has many facets and there are many different ways how
to derive it; accordingly, the relevant literature is vast. Our
discussion is aimed at an uninitiated reader and for this purpose
we adopt here a traditional elementary approach, which
nevertheless provides substantial insight into the nature and
origin of the axial anomaly.

\begin{figure}[h]
\centering \includegraphics{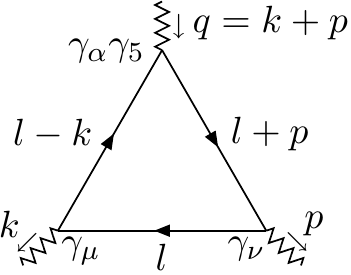} \caption{$VVA$
triangle graph: a closed fermionic loop with two vector ($V$)
vertices and one axial-vector ($A$) vertex. External lines
attached to the vertices merely symbolize the incoming and
outgoing momenta. Additional contribution of the crossed graph
with $(k,\mu)\leftrightarrow (p,\nu)$ has to be included in the
full $VVA$ amplitude.} \label{figapp}\index{Feynman diagrams!for
$VVA$ triangle}
\end{figure}

Let us start with the $VVA$ triangle graph (see
Fig.\,\ref{figapp}), which represents a correlation function of
two vector currents and one axial-vector current made of a single
fermion (Dirac) field.\footnote{More precisely, the quantity in
question is a Fourier transform of the vacuum expectation
value\index{vacuum expectation value} of the above-mentioned three
currents. To keep our discussion as general as possible, we do not
impose any particular restrictions on the external four-momenta
$k$ and $p$.}

A formal expression for the $VVA$ amplitude can be written as
\begin{multline}\label{eqE.1}
T_{\alpha\mu\nu}(k,p;m) = \int \frac{d^4 l}{(2\pi)^4}\Tr\Bigl(
\frac{1}{\slashed{l}-\slashed{k}-m}\gamma_\mu
\frac{1}{\slashed{l}-m}\gamma_\nu
\frac{1}{\slashed{l}+\slashed{p}-m}\gamma_\alpha\gamma_5\Bigr) +
\bigl[(k,\mu)\leftrightarrow (p,\nu)\bigr]
\end{multline}
where $m$ stands for the mass of the fermion circulating in the
loop. Another relevant quantity, closely related to (\ref{eqE.1}),
is
\begin{multline}\label{eqE.2}
T_{\mu\nu}(k,p;m) = \int \frac{d^4 l}{(2\pi)^4}\Tr\Bigl(
\frac{1}{\slashed{l}-\slashed{k}-m}\gamma_\mu
\frac{1}{\slashed{l}-m}\gamma_\nu
\frac{1}{\slashed{l}+\slashed{p}-m}\gamma_5\Bigr) +
\bigl[(k,\mu)\leftrightarrow (p,\nu)\bigr]
\end{multline}
Note that (\ref{eqE.2}) corresponds to a triangle loop obtained
from the original $VVA$ graph by replacing the axial-vector vertex
with a pseudoscalar one (i.e. by $\gamma_\alpha\gamma_5\rightarrow
\gamma_5$). Thus, the quantity (\ref{eqE.2}) can be naturally
called a $VVP$ amplitude. When speaking of the expression for the
$T_{\alpha\mu\nu}$, we stress the adjective {\it formal\/}: the
integral in (\ref{eqE.1}) has in fact an ultraviolet (UV)
divergence and its proper definition requires a special care. We
shall discuss this issue later on, and now let us focus on the
$T_{\mu\nu}$.

At first sight, the degree of divergence of the integral in
(\ref{eqE.2}) would seem to be the same as that of the
(\ref{eqE.1}). However, it turns out that -- for purely algebraic
reasons -- the integral (\ref{eqE.2}) is perfectly convergent! To
see this, let us recast the expression (\ref{eqE.2}) in the usual
manner as
\begin{equation}\label{eqE.3}
T_{\mu\nu}(k,p;m) = \int\frac{d^4 l}{(2\pi)^4}
\frac{\Tr\bigl[(\slashed{l}-\slashed{k}+m)\gamma_\mu
(\slashed{l}+m)\gamma_\nu
(\slashed{l}+\slashed{p}+m)\gamma_5\bigr]}{[(l-k)^2-m^2](l^2-m^2)[(l+p)^2-m^2]}
\end{equation}
Working out the trace in (\ref{eqE.3}), one finds out that this is
simplified drastically, and the result is
\begin{multline}
\label{eqE.4} \Tr\bigl[(\slashed{l}-\slashed{k}+m)\gamma_\mu
(\slashed{l}+m)\gamma_\nu
(\slashed{l}+\slashed{p}+m)\gamma_5\bigr]
=-m \Tr(\gamma_\mu\gamma_\nu\slashed{k}\slashed{p}\gamma_5) =
-4im\epsilon_{\mu\nu\rho\sigma}k^\rho p^\sigma
\end{multline}
(the reader is recommended to verify this independently, utilizing
the familiar properties of traces of the Dirac matrices,
summarized in Appendix~\ref{appenA}). Thus, we see that the $l$-dependence of
the integrand in (\ref{eqE.2}) is entirely due to its denominator
and, consequently, the whole integrand behaves as $l^{-6}$ for
$l\rightarrow \infty$; this means that the integral (\ref{eqE.2})
is actually even more convergent than necessary.\footnote{Note
that in (hyper)spherical coordinates one can write the $d^4 l$
schematically as $l^3dld\Omega$ (with $d\Omega$ denoting the
angular part); in this way, (\ref{eqE.2}) is eventually reduced to
a radial integral involving, asymptotically, $l^{-6}\cdot l^3 dl =
l^{-3} dl$.}

Now, taking into account (\ref{eqE.4}), it is not difficult to
realize that the contribution of the crossed term in (\ref{eqE.3})
is the same as that of the direct one. Thus, we have
\begin{equation}\label{eqE.5}
T_{\mu\nu}(k,p;m)=-8im\epsilon_{\mu\nu\rho\sigma}k^\rho p^\sigma
\int\frac{d^4l}{(2\pi)^4}\frac{1}{[(l-k)^2-m^2][(l+p)^2-m^2](l^2-m^2)}
\end{equation}
For the purpose of later discussion, we are going to recast the
last expression in terms of an integral over Feynman parameters.
This is done as follows. First, one introduces an integral
representation of the integrand in (\ref{eqE.5}) by means of the
general formula
\begin{equation}
\frac{1}{ABC}=2\int_0^1 dx \int_0^{1-x} dy \frac{1}{\bigl[Ax+By
+C(1-x-y)\bigr]^3}
\end{equation}
Then, after some simple manipulations, the expression
(\ref{eqE.5}) is rewritten as
\begin{multline}\label{eqE.7}
T_{\mu\nu}(k,p;m)=-16 i m \epsilon_{\mu\nu\rho\sigma} k^\rho
p^\sigma \int_0^1 dx \int_0^{1-x} dy \int \frac{d^4l}{(2\pi)^4}\\
\times \frac{1}{\bigl[(l-xk+yp)^2+x(1-x)k^2 +y(1-y)p^2 + 2xy
k\cdot p -m^2 \bigr]^3}
\end{multline}
As a next step, one performs the shift $l-xk+yp\rightarrow l$ in
the loop-momentum integral; (\ref{eqE.7}) thus becomes
\begin{equation}\label{eqE.8}
T_{\mu\nu}(k,p;m)=-16im\epsilon_{\mu\nu\rho\sigma}k^\rho p^\sigma
\int_0^1 dx \int_0^{1-x} dy \int \frac{d^4 l}{(2\pi)^4}
\frac{1}{\bigl[l^2-C(x,y;k,p,m^2)\bigr]^3}
\end{equation}
where we have denoted
\begin{equation}\label{eqE.9}
C(x,y;k,p,m^2) = m^2 -x(1-x)k^2 -y(1-y)p^2 -2xyk\cdot p
\end{equation}
Now, integration over the loop momentum can be carried out by
means of the general formula
\begin{equation}\label{eqE.10}
\int\frac{d^nl}{(2\pi)^n}\frac{(l^2)^r}{(l^2-C+i\varepsilon)^s} =
\frac{i}{(4\pi)^{\frac{n}{2}}}(-1)^{r-s} C^{r+\frac{n}{2}-s}
\,\frac{\Gamma(r+\frac{n}{2})\Gamma(s-r-\frac{n}{2})}{\Gamma(\frac{n}{2})\Gamma(s)}
\end{equation}
valid in $n$ dimensions, for values of the $r$ and $s$ such that
the integral converges (the $C$ is an essentially arbitrary real
parameter and we may suppose, for convenience, that $C>0$; note
that we have also retrieved the $i\varepsilon$ term, omitted in
(\ref{eqE.8}) for brevity). Using (\ref{eqE.10}) in (\ref{eqE.8}),
one gets the desired Feynman-parametric representation of the
$VVP$ amplitude in question:
\begin{equation}\label{eqE.11}
T_{\mu\nu}(k,p;m)=-\frac{1}{2\pi^2} \epsilon_{\mu\nu\rho\sigma}
k^\rho p^\sigma \int_0^1 dx \int_0^{1-x}\hspace{-0.3cm} dy \,
\frac{m}{C(x,y;k,p,m^2)}
\end{equation}

Let us now examine the relevant Ward identities for the $VVA$
amplitude. The preliminary discussion that follows is heuristic
and \qq{naive} (i.e. non-rigorous), in the sense that we ignore
temporarily the divergent nature of the considered integrals; a
regularization of the UV divergences will be taken into account in
a second step of our investigation.

We start with an evaluation of the quantity $k^\mu
T_{\alpha\mu\nu}$, which corresponds to the four-divergence of one
of the vector currents involved in the $VVA$ triangle graph. Using
the formal representation (\ref{eqE.1}), one has
\begin{multline}\label{eqE.12}
k^\mu T_{\alpha\mu\nu}(k,p;m) = \int\frac{d^4l}{(2\pi)^4}
\Tr\Bigl(
\frac{1}{\slashed{l}-\slashed{k}-m}\slashed{k}\frac{1}{\slashed{l}-m}
\gamma_\nu \frac{1}{\slashed{l}+\slashed{p}-m} \gamma_\alpha
\gamma_5 \Bigr) \\
+\int\frac{d^4l}{(2\pi)^4} \Tr\Bigl(
\frac{1}{\slashed{l}-\slashed{p}-m}\gamma_\nu\frac{1}{\slashed{l}-m}
\slashed{k} \frac{1}{\slashed{l}+\slashed{k}-m} \gamma_\alpha
\gamma_5 \Bigr)
\end{multline}
To simplify the expression (\ref{eqE.12}), one employs the
following simple algebraic trick. In the first integral, the
$\slashed{k}$ is recast as
\begin{equation}
\slashed{k}=(\slashed{l}-m)-(\slashed{l}-\slashed{k}-m)
\end{equation}
and, similarly, in the second integral one writes
\begin{equation}
\slashed{k}=(\slashed{l}+\slashed{k}-m)-(\slashed{l}-m)
\end{equation}
These substitutions result in a partial cancellation of propagator
denominators and (\ref{eqE.12}) is rewritten as a sum of four
integrals, namely
\begin{align}
k^\mu T_{\alpha\mu\nu}(k,p;m) &= \int\frac{d^4l}{(2\pi)^4}
\Tr\Bigl(
\frac{1}{\slashed{l}-\slashed{k}-m}\gamma_\nu\frac{1}{\slashed{l}+\slashed{p}-m}
\gamma_\alpha\gamma_5 \Bigr) \notag\\
&-\int\frac{d^4l}{(2\pi)^4} \Tr\Bigl(
\frac{1}{\slashed{l}-m}\gamma_\nu\frac{1}{\slashed{l}+\slashed{p}-m}
\gamma_\alpha \gamma_5 \Bigr)\notag\\
&+\int\frac{d^4l}{(2\pi)^4} \Tr\Bigl(
\frac{1}{\slashed{l}-\slashed{p}-m}\gamma_\nu\frac{1}{\slashed{l}-m}
\gamma_\alpha\gamma_5 \Bigr) \notag\\
&-\int\frac{d^4l}{(2\pi)^4} \Tr\Bigl(
\frac{1}{\slashed{l}-\slashed{p}-m}\gamma_\nu\frac{1}{\slashed{l}+\slashed{k}-m}
\gamma_\alpha \gamma_5 \Bigr)\label{eqE.15}
\end{align}
Now, it is easy to see that the first and the fourth integral in
(\ref{eqE.15}) mutually cancel -- this becomes clear when one
performs the shift $l\rightarrow l+k-p$ in the first integral. As
for the second and the third integral, these can be shown to
vanish (separately) on symmetry grounds: it is not difficult to
realize that each of them would be a 2nd rank {\it pseudotensor},
depending on a single four-vector $p$. However, one obviously
cannot construct such an object, because of full antisymmetry of
the Levi-Civita pseudotensor (that would have to be involved in a
corresponding expression). Thus, we arrive at the identity
\begin{equation}\label{eqE.16}
k^\mu T_{\alpha\mu\nu}(k,p;m) = 0
\end{equation}
In view of the symmetry of the $VVA$ amplitude (\ref{eqE.1}) under
$(k,\mu)\leftrightarrow (p,\nu)$, eq. (\ref{eqE.16}) implies
immediately also
\begin{equation}\label{eqE.17}
p^\nu T_{\alpha\mu\nu} (k,p;m) = 0
\end{equation}
The identities (\ref{eqE.16}), (\ref{eqE.17}) are in fact
anticipated results, since the vector current (made of a single
free Dirac field) is conserved.

In the same manner, we can calculate the quantity $q^\alpha
T_{\alpha\mu\nu}$ that expresses four-divergence of the
axial-vector current within the $VVA$ triangle graph. Using in
(\ref{eqE.1}) the trace cyclicity, it is convenient to start with
\begin{multline}
q^\alpha T_{\alpha\mu\nu}(k,p;m) = \int\frac{d^4l}{(2\pi)^4}
\Tr\Bigl(\gamma_\mu\frac{1}{\slashed{l}-m}\gamma_\nu\frac{1}{\slashed{l}+\slashed{p}-m}
\slashed{q}\gamma_5\frac{1}{\slashed{l}-\slashed{k}-m}\Bigr)\\
+\int\frac{d^4l}{(2\pi)^4}
\Tr\Bigl(\gamma_\nu\frac{1}{\slashed{l}-m}\gamma_\mu\frac{1}{\slashed{l}+\slashed{k}-m}
\slashed{q}\gamma_5\frac{1}{\slashed{l}+\slashed{p}-m}\Bigr)
\end{multline}
Taking into account that $q=k+p$, the $\slashed{q}\gamma_5$ in the
first integral can be recast, for obvious reasons, as
\begin{equation}
\slashed{q}\gamma_5 = (\slashed{l}+\slashed{p}-m)\gamma_5 +
\gamma_5(\slashed{l}-\slashed{k}-m) + 2m\gamma_5
\end{equation}
and, similarly, in the second integral one writes
\begin{equation}
\slashed{q}\gamma_5 = (\slashed{l}+\slashed{k}-m)\gamma_5 +
\gamma_5(\slashed{l}-\slashed{p}-m) + 2m\gamma_5
\end{equation}
Then, using the symmetry argument explained above and remembering
the definition (\ref{eqE.2}), one arrives at the result
\begin{equation}\label{eqE.21}
q^\alpha T_{\alpha\mu\nu}(k,p;m) = 2m T_{\mu\nu}(k,p;m)
\end{equation}
Again, this looks as an expected result, since it corresponds to a
\qq{partial conservation}\index{partially conserved axial current}
of the axial-vector current made of a massive Dirac field.

The relations (\ref{eqE.16}), (\ref{eqE.17}) and (\ref{eqE.21})
represent \qq{naive} (or \qq{canonical}) Ward identities (WI) for
the $VVA$ amplitude. The corresponding nomenclature sounds quite
naturally: the equations (\ref{eqE.16}), (\ref{eqE.17}) are called
{\bf vector WI}, while eq. (\ref{eqE.21}) is the {\bf axial WI}.
As we stressed earlier, in deriving them we have entirely ignored
all possible complications that could be due to the UV divergences
in the considered loop-momentum integrals. Now, we are going to
make up for this flaw. In particular, we will regularize the
formal expression for the contribution of the $VVA$ diagram by
means of the Pauli--Villars (PV) method (see e.g. \cite{ItZ}).
This consists in subtracting from (\ref{eqE.1}) the contribution
of an analogous loop, in which the original fermion mass is
replaced by an auxiliary regulator mass $M$ (of course, the
subtraction is made at the level of the corresponding integrand).
Since the integral in (\ref{eqE.1}) is only linearly divergent, one such
PV subtraction is sufficient. In explicit terms, the
PV-regularized $VVA$ amplitude reads
\begin{multline}\label{eqE.22}
T^\ti{reg.}_{\alpha\mu\nu}(k,p;M) = \int\frac{d^4l}{(2\pi)^4}
\Bigl\{ \Tr\Bigl( \frac{1}{\slashed{l}-\slashed{k}-m}\gamma_\mu
\frac{1}{\slashed{l}-m}\gamma_\nu
\frac{1}{\slashed{l}+\slashed{p}-m} \gamma_\alpha \gamma_5
\Bigr)\\
-\Tr\Bigl( \frac{1}{\slashed{l}-\slashed{k}-M}\gamma_\mu
\frac{1}{\slashed{l}-M}\gamma_\nu
\frac{1}{\slashed{l}+\slashed{p}-M} \gamma_\alpha \gamma_5
\Bigr)\Bigr\} + \bigl[(k,\mu)\leftrightarrow (p,\nu) \bigr]
\end{multline}
The salient feature of this regularization procedure is that it
preserves automatically the vector WI (note that precisely the
same effect occurs in the familiar example of the vacuum
polarization graph in spinor QED).\footnote{The point is that
within such a scheme, the internal fermion lines entering the
vector vertex carry the same mass $M$ and the vector current
conservation is thus maintained.}

For a general Feynman graph, one cannot simply remove the UV
cut-off by performing the limit $M\rightarrow \infty$ in the
regulated expression (before doing that, the quantity in question
has to be renormalized properly). However, the convergence
properties of the $VVA$ triangle graph are subtle and rather
amusing. In particular, although the integral in (\ref{eqE.1}) is
certainly UV divergent in a strict mathematical sense, it turns
out that the limit $M \rightarrow \infty$ for the PV-regularized
expression (\ref{eqE.22}) does exist! This statement is
non-trivial and will not be proved here; the interested reader can
find a very detailed treatment of the convergence properties of
the $VVA$ diagram e.g. in a paper by the present author and O. I.
Zavialov, published in Czech. J. Phys. B39 (1989), p. 478. An
upshot of all this is as follows. A \qq{renormalized} contribution
of the $VVA$ graph can be defined in a straightforward way as
\begin{equation}\label{eqE.23}
T^\ti{ren.}_{\alpha\mu\nu}(k,p;m) = \lim_{M\rightarrow \infty}
T^\ti{reg.}_{\alpha\mu\nu} (k,p;m,M)
\end{equation}
Moreover, since the vector WI hold for any value of the
regularization parameter $M$, the $T^\ti{ren.}_{\alpha\mu\nu}$
must obviously satisfy them as well, i.e., one has
\begin{equation}
k^\mu T^\ti{ren.}_{\alpha\mu\nu}(k,p;m) =0, \qquad p^\nu
T^\ti{ren.}_{\alpha\mu\nu} (k,p;m) = 0
\end{equation}

Let us now focus on the axial WI. All manipulations that led to
the naive identity (\ref{eqE.21}) are now legal for regularized
quantities. Then, keeping in mind the simple structure of the
definition (\ref{eqE.22}), it is easy to realize that the
\qq{intermediate} identity
\begin{equation}\label{eqE.25}
q^\alpha T^\ti{reg.}_{\alpha\mu\nu}(k,p;m,M) = 2m
T_{\mu\nu}(k,p;m) - 2 M T_{\mu\nu}(k,p;M)
\end{equation}
must be valid for any finite value of the $M$. In view of
(\ref{eqE.23}), the limit $M\rightarrow \infty$ can be performed
in (\ref{eqE.25}) and one thus obtains
\begin{equation}\label{eqE.26}
q^\alpha T^\ti{ren.}_{\alpha\mu\nu}(k,p;m) = 2 m T_{\mu\nu}(k,p;m)
- \lim_{M\rightarrow \infty} 2M T_{\mu\nu}(k,p;M)
\end{equation}
Now we come to the crucial point of our discussion. The appearance
of the second term in the right-hand side of eq. (\ref{eqE.26})
indicates a possible deviation from the naive identity
(\ref{eqE.21}); it only remains to be seen whether such an extra
term is indeed non-vanishing. Using our previous result for the
$T_{\mu\nu}$, one finds easily that the answer is {\it yes}; from
(\ref{eqE.11}) and (\ref{eqE.9}) one gets readily
\begin{align}
\lim_{M\rightarrow \infty} 2 M T_{\mu\nu} (k,p;M) &=
-\frac{1}{\pi^2}\epsilon_{\mu\nu\rho\sigma}k^\rho p^\sigma \times
\notag\\ \times\lim_{M\rightarrow\infty} \int_0^1 dx &\int_0^{1-x}
\hspace{-0.3cm}
dy \frac{M^2}{M^2 - x(1-x)k^2 -y(1-y)p^2 -2xy k\ccdot p}\notag\\
&=-\frac{1}{2\pi^2}\epsilon_{\mu\nu\rho\sigma}k^\rho p^\sigma
\end{align}
Thus, we arrive at an identity
\begin{equation}\label{eqE.28}
q^\alpha T^\ti{ren.}_{\alpha\mu\nu} (k,p;m) = 2mT_{\mu\nu} (k,p;m)
+\frac{1}{2\pi^2} \epsilon_{\mu\nu\rho\sigma} k^\rho p^\sigma
\end{equation}
which is precisely eq. (\ref{eq7.191}) quoted in the main text.

The second term on the right-hand side of (\ref{eqE.28}) is the
celebrated {\bf ABJ axial anomaly} \cite{ref75, ref76}.
Accordingly, the relation (\ref{eqE.28}) is often called the {\bf
anomalous axial Ward identity}. Clearly, the labels \qq{anomaly}
and \qq{anomalous} are of historical origin: they reflect the fact
that the discovery of the ABJ anomaly was indeed a kind of
surprise, taking into account that in many other situations, naive
results (in the above sense) often prove to be correct, i.e. they
are recovered when an appropriate regularization is included. On
the other hand, our preceding discussion should have made it clear
that, as a matter of fact, there is nothing anomalous about the
anomaly: it emerges as a result of a proper definition of the
$VVA$ amplitude in question (whereas the naive Ward identities are
derived by sloppy manipulations with ill-defined quantities). The
mechanism, by which the anomaly is generated, becomes quite
transparent within our approach. If one wants to maintain the
vector current conservation, the corresponding pair of internal
fermion lines must carry the same regulator mass $M$ and this must
be so for the two neighbouring vector vertices. Because of the
extremely simple topology of the triangle graph, the $M$ thus
automatically appears in both internal lines entering the
axial-vector vertex and, consequently, the axial WI gets modified.

For completeness, let us add that the evaluation of the anomaly
can be generalized in such a way that one need not rely on a
particular regularization procedure; we adopted here the PV method
since it is instructive and transparent for a first reading. The
essence of the anomaly phenomenon can be described succinctly as
follows. {\bf There is no consistent way of defining the
contribution of the $\boldsymbol{VVA}$ graph, such that the naive
vector and axial WI would hold simultaneously; in particular, when
the vector WI are imposed, the axial WI inevitably picks up the
extra term shown in (\ref{eqE.28}).} As we noted before, the ABJ
anomaly has many interesting aspects and the relevant literature
is rich. The reader seeking a deeper knowledge of the subject can
find an appropriate introduction e.g. in the review article
\cite{ref77} or in the comprehensive monograph \cite{Ber};
needless to say, these sources contain many other relevant
references.

%\end{document}

%\input{kniha_references}
%%%%%%%%%%%%%%%%%%%%%%%%%%%%%%%%%%%%%%%%%%%%%%%%%%%%%%%%%%%%%%%%%%%
%%%%%%%%%%%%%%%%%%%%%%%%%%%%%%%%%%%%%%%%%%%%%%%%%%%%%%%%%%%%%%%%%%%%%%%%%%%%%%%%%%%%%%%%%%%%%%%%%%%%%%%%%%%%%%%%%%%%%%%%%%%%%%%%%%%%%%%%
%\setcounter{secnumdepth}{-1} 
\phantomsection\label{references1}
\renewcommand\bibname{References}

\phantomsection\label{references2}
\renewcommand\bibname{Bibliography}

\phantomsection\label{index1}
\printindex

\end{document}